\RequirePackage{lineno} 

\documentclass{cernrep}
\usepackage{texnames,subfigure,epsfig}
\usepackage[T1]{fontenc}
\usepackage{geometry,graphicx}
\usepackage{color}
\usepackage{pslatex}
\usepackage{hyperref} 

\def\ra{ \rightarrow }
\def\bb{{b\bar{b}}}
\def\Mhmax{M_h^{\mathrm{max}}}

\def\tb{\tan\beta}

\def\pom{I\!\!P}

\def\GeVc{GeV/c}
\def\GeVcc{GeV/c$^{2}$}

\def\mean#1{\ensuremath{\left<#1\right>}}
\def\ttt#1{\texttt{\small #1}}

\begin{document}



\vspace*{-25mm}
\begin{flushright}
{\large \today}
\end{flushright}

\vspace*{-0.5cm}


\title{\includegraphics[width=0.15\textwidth]{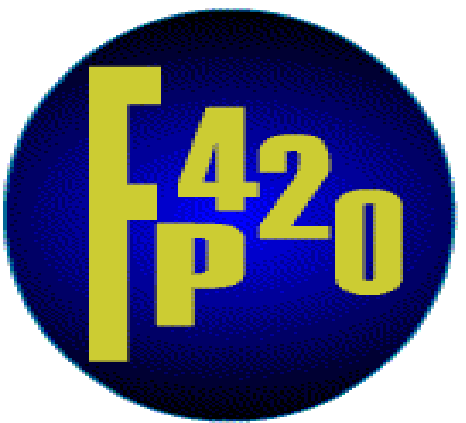}
{\Large The FP420 R\&D Project: Higgs and New Physics with forward protons at the LHC}}
\author{
M.~G.~Albrow$^{1}$, R.~B.~Appleby$^{2}$, M.~Arneodo$^{3}$, G.~Atoian$^{4}$, 
I.L.~Azhgirey$^{5}$, R.~Barlow$^{2}$, I.S.~Bayshev$^{5}$, W.~Beaumont$^{6}$, L.~Bonnet$^{7}$, 
A.~Brandt$^{8}$, P.~Bussey$^{9}$, C.~Buttar$^{9}$, J.~M.~Butterworth$^{10}$, 
M.~Carter$^{11}$, B.E.~Cox$^{2,}$\footnote{Contact persons: Brian.Cox@manchester.ac.uk, Albert.de.Roeck@cern.ch}, 
D.~Dattola$^{12}$, C.~Da~Via$^{13}$, J.~de~Favereau$^{7}$, D.~d'Enterria$^{14}$, P.~De~Remigis$^{12}$, 
A.~De~Roeck$^{14,6,*}$, E.A.~De~Wolf$^{6}$, P.~Duarte$^{8,}$\footnote{Now~at~Rice~University}, 
J.~R.~Ellis$^{14}$, B.~Florins$^{7}$, J.~R.~Forshaw$^{13}$, J.~Freestone$^{13}$, K.~Goulianos$^{15}$, 
J.~Gronberg$^{16}$, M.~Grothe$^{17}$, J.~F.~Gunion$^{18}$, J.~Hasi$^{13}$, ~S.~Heinemeyer$^{19}$, 
J.~J.~Hollar$^{16}$, S.~Houston$^{9}$, V.~Issakov$^{4}$, R.~M.~Jones$^{2}$, M.~Kelly$^{13}$, 
C.~Kenney$^{20}$, V.A.~Khoze$^{21}$, S.~Kolya$^{13}$, N.~Konstantinidis$^{10}$, H.~Kowalski$^{22}$, 
H.E.~Larsen$^{23}$, V.~Lemaitre$^{7}$, S.-L. Liu$^{24}$, A.~Lyapine$^{10}$, F.K.~Loebinger$^{13}$, 
R.~Marshall$^{13}$, ~A.~D.~Martin$^{21}$, J.~Monk$^{10}$, I.~Nasteva$^{13}$, P.~Nemegeer$^{7}$, 
M.~M.~Obertino$^{3}$, R.~Orava$^{25}$, V.~O'Shea$^{9}$, S.~Ovyn$^{7}$, A.~Pal$^{8}$, S.~Parker$^{20}$, 
J.~Pater$^{13}$, A.-L.~Perrot$^{26}$, T.~Pierzchala$^{7}$, A.~D.~Pilkington$^{13}$, J.~Pinfold$^{24}$, 
K.~Piotrzkowski$^{7}$, W. Plano$^{13}$, A.~Poblaguev$^{4}$, V.~Popov$^{27}$, K.~M.~Potter$^{2}$, 
S.~Rescia$^{28}$, F.~Roncarolo$^{2}$, A.~Rostovtsev$^{27}$, X.~Rouby$^{7}$, ~M.~Ruspa$^{3}$, 
M.G.~Ryskin$^{21}$, A.~Santoro$^{29}$, N.~Schul$^{7}$, G.~Sellers$^{2}$, 
A.~Solano$^{23}$, S.~Spivey$^{8}$, W.J.~Stirling$^{21}$, D.~Swoboda$^{26}$, 
M.~Tasevsky$^{30}$, R.~Thompson$^{13}$, T.~Tsang$^{28}$, P.~Van~Mechelen$^{6}$, A.~Vilela Pereira$^{23}$, S.J.~Watts$^{13}$, 
M.~R.~M.~Warren$^{10}$, G.~Weiglein$^{21}$, T.~Wengler$^{13}$, S.N.~White$^{28}$, B.~Winter$^{11}$,
Y.~Yao$^{24}$, D.~Zaborov$^{27}$, A.~Zampieri$^{12}$, M.~Zeller$^{4}$, A.~Zhokin$^{6,27}$
\\ \hspace*{\fill} 
\\ \hspace*{\fill}
\protect \centerline{\large FP420 R\&D Collaboration} 
}
\institute{\vspace{1.cm}
$^{1}$Fermilab, 
$^{2}$University of Manchester and the Cockcroft Institute, 
$^{3}$Universit\a`a del Piemonte Orientale, Novara, and INFN, Torino,
$^{4}$Yale University,
$^{5}$State Research Center of Russian Federation, Institute for High Energy Physics, Protvino, 
$^{6}$Universiteit Antwerpen,
$^{7}$Universit\a'e Catholique de Louvain,
$^{8}$University of Texas at Arlington,
$^{9}$University of Glasgow, 
$^{10}$University College London (UCL), 
$^{11}$Mullard Space Science Laboratory (UCL),
$^{12}$INFN Torino, 
$^{13}$University of Manchester,
$^{14}$CERN, PH Department,
$^{15}$Rockefeller University, NY,
$^{16}$Lawrence Livermore National Laboratory (LLNL),
$^{17}$University of Wisconsin, Madison,
$^{18}$UC Davis,
$^{19}$IFCA (CSIC-UC, Santander),
$^{20}$Molecular Biology Consortium, Stanford University,
$^{21}$Institute for Particle Physics Phenomenology, Durham,
$^{22}$DESY, 
$^{23}$Universit\a`a di Torino and INFN, Torino,
$^{24}$University of Alberta,
$^{25}$Helsinki Institute of Physics, 
$^{26}$CERN, TS/LEA,
$^{27}$ITEP Moscow,
$^{28}$Brookhaven National Lab (BNL),
$^{29}$Universidade do Estado do Rio De Janeiro (UERJ),
$^{30}$Institute of Physics, Prague
}

\maketitle 

\newpage

\begin{abstract}
We present the FP420 R\&D project, which has been studying the key aspects of the 
development and installation of a silicon tracker and fast-timing detectors in the LHC tunnel 
at 420~m from the interaction points of the ATLAS and CMS experiments. These detectors 
would  measure precisely very forward protons in conjunction with the corresponding 
central detectors as a means to study Standard Model (SM) physics, and to search for 
and characterise New Physics signals. This report includes a detailed description of the physics case 
for the detector and, in particular, for the measurement of Central Exclusive Production,
$pp \rightarrow p + \phi + p$, in which the outgoing protons remain intact and 
the central system $\phi$ may be a single particle such as a SM or MSSM Higgs boson. 
Other physics topics discussed are $\gamma \gamma$ and $\gamma p$ interactions,
and diffractive processes. The report includes a detailed study of the trigger strategy, 
acceptance, reconstruction efficiencies, and expected yields for a particular 
$p\,p \to p\, H\, p$ measurement with Higgs boson decay in the $b\bar{b}$ mode.
The document also describes the detector acceptance as given by the LHC beam optics 
between the interaction points and the FP420 location, the machine backgrounds, 
the new proposed connection cryostat and the moving (``Hamburg'') beam-pipe at 420~m, 
and the radio-frequency impact of the design on the LHC. The last part of the document 
is devoted to a description of the 3D silicon sensors and associated tracking performances, 
the design of two fast-timing detectors capable of accurate vertex reconstruction for background 
rejection at high-luminosities, and the detector alignment and calibration strategy.
\end{abstract}


\tableofcontents

\newpage

\section{Introduction}
\label{sec:Introduction}

\subsection{Executive summary}

Although forward proton detectors have been used to study Standard Model (SM) physics for a 
couple of decades, the benefits of using proton detectors to search for New Physics at the LHC have
only been fully appreciated within the last few years~\cite{alb,KMRtag,cox1,KKMRS,jeffrev,kp}. 
By detecting both outgoing protons that have lost less than 2\%  of their longitudinal
momentum~\cite{fp420}, in conjunction with a measurement of the associated centrally produced
system using the current ATLAS and/or CMS detectors, a rich programme of studies in QCD, electroweak, 
Higgs and Beyond the Standard Model physics becomes accessible, with the potential to make 
unique measurements at the LHC. 
A prime process of interest is Central Exclusive Production (CEP), $pp \rightarrow p + \phi + p$, 
in which the outgoing protons remain intact and the central system $\phi$ may be a single particle 
such as a Higgs boson. In order to detect both outgoing protons in the range of momentum loss appropriate 
for central systems in the $\sim 100$~\GeVcc\ mass range during nominal high-luminosity running, proton tagging 
detectors must be installed close to the outgoing beams in the high-dispersion region 420~m from the 
interaction points on each side of the ATLAS and CMS experiments. The FP420 R\&D project is a 
collaboration including members from ATLAS, CMS, TOTEM and the accelerator physics 
community, with support from theorists, aimed at assessing the feasibility of installing such detectors. 

The proposed FP420 detector system is a magnetic spectrometer. The LHC magnets between the interaction 
points and the 420~m regions bend protons that have lost a small fraction of their initial momentum out of 
the beam envelope. The FP420 detector consists of a silicon tracking system that can be
moved transversely and measures 
the spatial position of these protons relative to the LHC beam line and their arrival times at several points 
in a 12 m region around 420~m. The proposed instrumentation of the 420 m region includes the replacement 
of the existing 14 m long connection cryostat with a warm beam-pipe section and a cryogenic bypass. 
To this purpose, a new connection cryostat has been designed, based on a modified arc termination module, 
so as to minimise the impact on the machine. The FP420 detector must be moveable because it should be parked at a large 
distance from the beams during injection and luminosity tuning, but must operate at distances between 
4~mm and 7~mm from the beam centre during data taking, depending on the beam conditions. 
A measurement of the displacement and angle of the 
outgoing protons relative to the beam allows the momentum loss and transverse momentum of the 
scattered protons to be reconstructed. This in turn allows the mass of the centrally produced 
system $\phi$ to be reconstructed by the missing mass method~\cite{alb} with a resolution 
($\sigma$) between 2~GeV/c$^{2}$ and 3~GeV/c$^{2}$ {\it per event}
irrespective of the decay products of the central system.

The detector position relative to the beam can be measured both by employing beam position 
monitors and by using a high-rate physics process which produces protons of a known momentum loss 
(from a central detector measurement of the central system) in the FP420 acceptance range. 
The second method has the advantage that the magnetic field between the central 
detectors and FP420 does not have to be precisely known {\it a priori}.  

The cross sections for CEP of the SM Higgs boson and other new physics scenarios are expected 
to be small, on the femtobarn scale. FP420 must therefore be designed to operate up
to the highest LHC instantaneous 
luminosities of $10^{34}$cm$^{-2}$s$^{-1}$, where there will be on average 
35 overlap interactions per bunch crossing (assuming $\sigma_{tot}$~=~110~mb). 
These overlap events can result in a large fake background, consisting of a central system from one interaction 
and protons from other interactions in the same bunch crossing. Fortunately, there are many kinematic and 
topological constraints which offer a large factor of background rejection. In addition, a
measurement of the difference in the arrival times of the two protons at FP420 in the 10 picosecond range allows 
for matching of the detected protons with a central vertex within $\sim$2~mm, which will enable the 
rejection of most of the residual overlap background, reducing it to a manageable level.

Studies presented in this document show that it is possible to install detectors in the 420~m 
region with no impact on the operation or luminosity of the LHC (Section~\ref{sec:silicon}). 
These detectors can be calibrated to the accuracy required to 
measure the mass of the centrally produced system to between $2$ and $3$~GeV/c$^{2}$. 
This would allow an observation of new particles in the $60 - 180$~GeV/c$^{2}$  mass range in 
certain physics scenarios during 3 years of LHC running at instantaneous luminosities of 
$2 \times 10^{33}$ cm$^{-2}$ s$^{-1}$, and in many more scenarios at instantaneous luminosities 
of up to $10^{34}$ cm$^{-2}$ s$^{-1}$.   Events can be triggered using the central detectors alone at 
Level 1, using information from the 420~m detectors at higher trigger levels to reduce the event 
rate. Observation of new particle production in the CEP channel 
would allow a direct measurement of the quantum numbers of the particle and an 
accurate determination of the mass, irrespective of the decay channel of the particle. In some scenarios, 
these detectors may be the primary means of discovering new particles at the LHC, with unique ability 
to measure their quantum numbers. There is also an extensive, high-rate $\gamma \gamma$ 
and $\gamma p$ baseline physics program.

We therefore conclude that the addition of such detectors will, 
for a relatively small cost, enhance the discovery and physics potential of the ATLAS and CMS experiments.   

\subsection{Outline}

The outline of this document is as follows. In Section~\ref{sec:physics} we provide a brief overview of the 
physics case for FP420. In Section~\ref{sec:pilko} we describe in detail a physics and detector simulation of 
a particular scenario which may be observable if 420~m detectors are installed. The acceptance and mass
resolutions used in this analysis are presented in Section~\ref{sec:optics}. In Section~\ref{sec:backgrounds} 
we describe the machine-induced backgrounds at 420~m such as beam-halo and beam-gas backgrounds. 
We then turn to the hardware design of FP420. Section~\ref{sec:cryostat} describes the new 420~m connection 
cryostat which will allow moving near-beam detectors with no effects on LHC operations. The design of the 
beam pipe in the FP420 region and the movement mechanism are described in Section~\ref{sec:hhpipe}, 
and the studies of the radio-frequency impact of the design on the LHC are described in Section~\ref{sec:RF}.
Section~\ref{sec:silicon} describes the design of the FP420 3D silicon sensors, detectors and detector 
housings and off-detector services such as cabling and power supplies. 
Section~\ref{sec:timing} describes two complementary fast timing detector designs, both of which are likely to 
be used at FP420. Section~\ref{sec:alignment} describes the 
alignment and calibration strategy, using both physics and beam position monitor techniques. We present our 
conclusions and future plans
in Section~\ref{sec:conclusions}.     

\subsection{Integration of 420~m detectors into ATLAS and CMS forward physics programs}
\label{sec:integration}

This report focuses primarily on the design of 420~m proton tagging detectors. 
CMS will have proton taggers installed at 220~m around its IP at startup, provided by the TOTEM 
experiment and for which common data taking with CMS is planned~\cite{Albrow:2006xt}. 
ATLAS also has an approved forward physics experiment, ALFA, with proton taggers at 240m designed to measure 
elastic scattering in special optics runs~\cite{ALFA}. 

There are ideas to upgrade the currently approved TOTEM detectors and a proposal to 
install FP420-like detectors at 220~m around ATLAS~\cite{ATLAS220m}. Adding detectors at 220~m
capable of operating at high luminosity increases the acceptance of FP420 for central masses of 
$\sim$120~\GeVcc\ and upwards, depending on the interaction point\footnote{For 220~m detectors, 
the acceptance is different around IP1 (ATLAS) and IP5 (CMS).} 
and the distance of approach of both the 220~m and 420~m detectors to the beam
(see chapter~\ref{sec:optics}). Throughout this document we present results for 420~m detectors alone 
and where appropriate for a combined 220~m + 420~m system. It is envisaged that FP420  
collaboration members will become parts of the already existing ATLAS and CMS forward physics groups, 
and will join with them to propose forward physics upgrade programmes that will be developed separately by ATLAS and CMS, 
incorporating the findings of this report.    
\newpage
\section{The Physics Case for Forward Proton Tagging at the LHC}
\label{sec:physics} 


\subsection{Introduction}

A forward proton tagging capability can enhance the ability of the ATLAS and CMS detectors to
carry out the primary physics program of the LHC. This 
includes measurement of the mass and quantum numbers of the Higgs boson,
should it be discovered via traditional searches, and augmenting the discovery 
reach if nature favours certain plausible beyond the Standard Model scenarios, 
such as its minimal supersymmetric extension (MSSM). In this context, the central exclusive production (CEP) of new particles offers unique possibilities, 
although the rich photon-photon and photon-proton physics program also delivers promising search 
channels for new physics. These channels are described in Section~\ref{sec:photon_phys}.
 
\begin{figure}[htbp]
\centering
\includegraphics[width=0.3\textwidth]{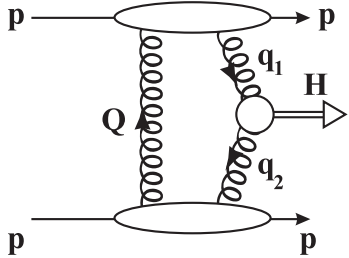}
\caption{Central Exclusive Production (CEP): $pp \to p+H+p$.} 
\label{fig:H}
\end{figure}

By central exclusive production we refer to the process $pp\rightarrow p + \phi + p$, where 
the `+' signs denote the absence of hadronic activity 
(that is, the presence of a rapidity gap) between the outgoing protons and the 
decay products of the central system $\phi$. The final state therefore consists 
{\it solely} of the two outgoing protons, which we intend to 
detect in FP420, and the decay products of the central system which will be detected in the 
ATLAS or CMS detectors. We note that gaps will not typically be part of the experimental signature 
due to the presence of minimum bias pile-up events, which fill in the gap but do not affect our ability 
to detect the outgoing protons. Of particular interest is the production of Higgs bosons, 
but there is also a rich and more exotic physics menu that includes the production of many kinds of 
supersymmetric particles, other exotica, and indeed any new
object which has $0^{++}$  (or $2^{++}$) quantum 
numbers and couples strongly to gluons~\cite{KMRtag,jeffg} or to photons~\cite{louvain}. 
The CEP process is illustrated for Higgs boson 
production in Fig.~\ref{fig:H}. The Higgs boson is produced as usual through gluon-gluon fusion, 
while another colour-cancelling gluon is exchanged, and no other particles are produced.

There are three important reasons why CEP is especially attractive for studies of
new heavy objects. Firstly, if the outgoing protons remain intact and scatter through 
small angles then, to a very good approximation, the primary active di-gluon system obeys 
a $J_z$ = 0, C-even, P-even, selection rule~\cite{KMRmm}. Here $J_z$ is the projection of 
the total angular momentum along the proton beam axis. This selection rule readily permits 
a clean determination of the quantum numbers of any new resonance, which is predominantly 
$0^{++}$  in CEP.~Secondly, because the process is exclusive, the energy loss of the 
 outgoing protons is directly 
related to the invariant mass of the central system, allowing an excellent mass measurement 
irrespective of the decay mode of the central system. Even final states containing jets and/or one or 
more neutrinos are measured with $\sigma_{M}\sim 2$~GeV/c$^{2}$. Thirdly, in many topical cases and in particular
for Higgs boson production, a signal-to-background ratio of order 1 or better is achievable 
\cite{DKMOR,Kaidalov:2003ys,cox2,krs2,kmrN}. This ratio becomes significantly larger for 
Higgs bosons in certain regions of MSSM parameter space~\cite{Kaidalov:2003ys,Heinemeyer:2007tu,Cox:2007sw}. 

There  is also a broad, high-rate QCD and electro-weak physics program;  by tagging both of the 
outgoing protons, the LHC is effectively turned into a gluon-gluon, photon-proton and photon-photon 
collider~\cite{kp,KMRphot}. In the QCD sector, detailed studies of 
 diffractive scattering, skewed, unintegrated gluon densities and the rapidity gap survival 
probability~\cite{KMRtag,Ryskin:2007qx,KMRsoft,Gotsman:2007ac} can be carried out.  In addition, 
CEP would provide a source of practically
pure gluon jets, turning the LHC into a `gluon factory'~\cite{KMRmm} and providing a unique 
laboratory in which to study the detailed properties of gluon jets, 
especially in comparison with quark jets. Forward proton tagging also provides unique capabilities to 
study photon-photon and photon-proton interactions at centre-of-mass energies never reached before. 
Anomalous top production, anomalous gauge boson couplings, exclusive dilepton production, or 
quarkonia photoproduction, to name a few, can be studied in the clean environment of photon-induced 
collisions.

In what follows we will give a brief overview of the theoretical predictions including a survey of the 
uncertainties in the expected cross sections. We will then review the possibilities of observing Higgs 
bosons in the Standard Model, MSSM and NMSSM for $W$, $\tau$ and $b$-quark decay channels.
A major potential contribution of FP420 to the LHC program is the
possibility to exploit the $b\bar{b}$ decay channel of the Higgs particle, which
is not available to standard Higgs analyses due to overwhelming backgrounds.
The combination of the suppression of the $b\bar{b}$ background, due to the $J_z$ = 0
selection rule, and the superior mass resolution of the FP420 detectors opens
up the possibility of exploiting this high branching ratio channel. Although the penalty for demanding 
two forward protons makes the discovery of a Standard Model Higgs boson in the $b\bar{b}$ channel 
unlikely despite a reasonable signal-to-background ratio, the cross section enhancements in other
scenarios indicate that this could be a discovery channel. 
For example, it has recently been shown that the heavy  CP-even MSSM Higgs boson, $H$, could 
be detected over a large region of the $M_A - \rm{tan}\beta$ plane; for $M_A \sim 140$ GeV/c$^{2}$,
discovery of $H$ should be possible for all values of tan $\beta$. The $5 \sigma$ discovery reach 
extends beyond $M_A = 200$ GeV/c$^{2}$ for tan $\beta > 30$~\cite{Weiglein:2007sc,Heinemeyer:2007tu}. 
We discuss the MSSM Higgs bosons measurements in the $b\bar{b}$ decay channel in detail in Section~\ref{sec:mssm}. 

In addition, for certain MSSM scenarios, FP420 provides an opportunity for a detailed lineshape 
analysis~\cite{Kaidalov:2003ys,je}. In the NMSSM, the complex decay chain $h \to a a \to 4 \tau$ 
becomes viable in CEP, and even offers the possibility to measure the mass of the pseudoscalar Higgs boson~\cite{Forshaw:2007ra}.
Another attractive feature of the FP420 programme is the ability to 
probe the CP-structure of the Higgs sector either by measuring directly the azimuthal asymmetry 
of the outgoing tagged protons~\cite{Khoze:2004rc} or by studying the correlations between
the decay products~\cite{je}. 
 

\subsection{The theoretical predictions}
\label{sec:calculation}

In this section we provide a very brief overview of the theoretical calculation involved in making
predictions for CEP.
~We shall, for the sake of definiteness, focus upon Higgs boson
production. A more detailed review can be found in~\cite{jeffrev}. 
Referring to Fig.~\ref{fig:H}, the dominant contribution comes from the region 
$\Lambda_{\rm QCD}^2\ll Q^2\ll M_h^2$ and hence the amplitude may
be calculated using perturbative QCD techniques~\cite{KMR,KMRmm}. The result is
\begin{equation}
{\cal A}_h \simeq N\int\frac{dQ^2}{Q^6} V_h \: f_g(x_1, x_1', Q^2, \mu^2)f_g(x_2,x_2',Q^2,\mu^2),
\label{eq:M}
\end{equation}
where the $gg\to h$ vertex factor for the $0^+$ Higgs boson production is (after azimuthal-averaging)
$V_h \simeq  Q^2$ and the normalization constant $N$ can be written in terms of the $h \to gg$ 
decay width~\cite{KMRtag,KMR}.~Equation~(\ref{eq:M}) holds for small transverse momenta of 
the outgoing protons, although including the full transverse momentum dependence is 
straightforward~\cite{Khoze:2002nf,Kaidalov:2003ys}.

The $f_g$'s are known as `skewed unintegrated gluon densities'~\cite{Ji:1996nm,Radyushkin:1996ru}. 
They are evaluated at the scale $\mu$, taken to be $\sim M_h/2$. Since $(x'\sim Q/\sqrt s)\ll (x\sim M_h/\sqrt s)\ll 1$, 
it is possible to express $f_g(x,x',Q^2,\mu^2)$, to single logarithmic accuracy, in terms of the gluon
distribution function  $g(x,Q^2)$. The $f_g$'s each contain a Sudakov suppression factor, which is the probability that 
the gluons which fuse to make the central system do not radiate in their evolution from $Q$ up to the hard scale.  
The apparent infrared divergence of Equation~(\ref{eq:M}) is nullified by these Sudakov factors and, for the 
production of $J_z=0$ central systems with invariant mass above 50 GeV/c$^{2}$, there is good control of the 
unknown infrared region of QCD.

Perturbative radiation associated with the $gg\ra h$ subprocess, which is vetoed by the Sudakov factors,
is not the only way to populate and to destroy the rapidity gaps. There is also the possibility of soft rescattering 
in which particles from the underlying proton-proton event (i.e. from other parton interactions) populate the gaps. 
The production of soft secondaries caused by the rescattering is expected to be almost independent of the 
short-distance subprocess and therefore can be effectively accounted for by a multiplicative factor
$S^2$, usually termed the soft gap survival factor or survival probability~\cite{Bjorken:1992er}. The value of
$S^2$ is not universal and depends on the centre-of-mass energy of the collision and the transverse momenta, $p_T$, of the
outgoing forward protons; the most sophisticated of the models for gap survival use a two~\cite{KMRsoft} 
and three-channel ~\cite{Ryskin:2007qx} eikonal model incorporating high mass diffraction. To simplify the discussion it is common to use a fixed value corresponding to the average over the $p_T$ 
acceptance of the forward detectors (for a 120~GeV/c$^{2}$ Higgs boson, $S^2$ is about 0.03 at the LHC).  
Taking this factor into account, the calculation of the production cross section for a 120~GeV/c$^{2}$ 
Standard Model Higgs boson via the CEP process at the LHC yields a central value of 3~fb.

\begin{figure}
\centering
\includegraphics[width=0.5\textwidth]{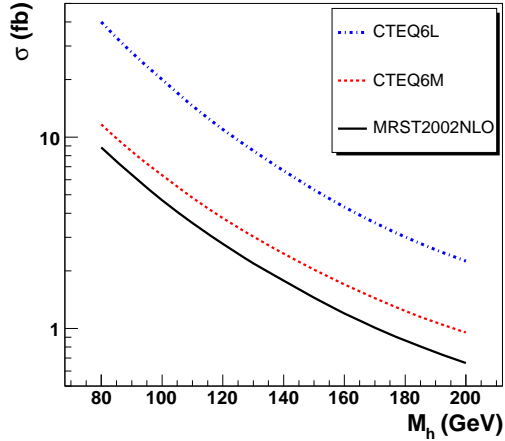}
\caption{The cross section for the central exclusive production of a Standard Model 
Higgs boson as a function of $M_h$, for three different proton parton distribution functions. 
\label{fig:sigmamh}}
\end{figure}

The primary uncertainties in the predicted cross section come from two sources. Firstly, since the gluon 
distribution functions $g(x,Q^2)$ enter to the fourth power, the predictions are sensitive to the choice 
of parton distribution function (PDF) in the proton and in particular to the gluon densities at $x$ = $\mathcal{O}$(0.01).
These are currently obtained from fits to data from HERA and the Tevatron.
Figure~\ref{fig:sigmamh} shows the prediction for the cross section for the CEP of a SM Higgs boson as a 
function of $M_h$ for three different choices of PDF at the LHC~\cite{Cox:2007sw}. The cross section varies 
from 2.8~fb to 11~fb for a 120~GeV/c$^{2}$ SM Higgs boson, although the highest prediction comes from 
a leading order PDF choice and, since the calculation includes an NLO K-factor (K$=$1.5), one might conclude 
that this choice is the least favoured. 
Secondly, there is some uncertainty in the calculation of the soft survival factor $S^2$. Until recently, 
the consensus was that $S^2$ has a value between 2.5\% and 4\% at LHC energies~\cite{Alekhin:2005dx}, 
but a lower value has been discussed~\cite{Gotsman:2007ac,Frankfurt:2006jp}
(although these have been challenged in~\cite{Ryskin:2007qx}). 
Early LHC data on various diffractive processes should settle this issue~\cite{Khoze:2008cx}.

The reliability of the theoretical calculations can be checked to some extent at the Tevatron. The CDF collaboration 
has observed a 6$\sigma$ excess of events in the exclusive dijet sample, $p\bar{p} \ra p + jj + \bar{p}$~\cite{Aaltonen:2007hs}, 
which is well described 
by the theory. 
CDF has also observed several candidates for central exclusive di-photon production, $p\bar{p} \ra p + \gamma \gamma + \bar{p}$, 
at the predicted rates~\cite{:2007na}, although the invariant mass of the di-photon pair is approximately 10~GeV/c$^{2}$ 
and the infrared region may not be under good control. Both of these predictions include calculations for the soft survival 
factor at Tevatron energies. 

The CDF measurements give some confidence in the predicted cross sections at the LHC. However, the theoretical uncertainties are approximately a factor of three, giving a predicted cross section range for a 120~GeV/c$^{2}$ SM Higgs boson of 1 to 9~fb.


\subsection{Standard Model Higgs boson}
\label{sec:sm}

The calculations of the previous section give a central cross section value of 3~fb for a 120~GeV/c$^{2}$ 
SM Higgs boson, falling to 1 fb for a mass of 200 GeV/c$^{2}$ (Fig.~\ref{fig:sigmamh}, where we take the
more conservative case obtained with the MRST PDFs).
Out of the two dominant decay channels ($h\rightarrow b \bar{b}$, $WW^*$), the $WW^*$ channel is the simplest way 
to observe the SM Higgs boson in CEP because the events are easy to trigger for the semi-leptonic and fully leptonic decay modes. 
A study taking into account basic experimental cuts was performed in~\cite{cox2} assuming that forward proton detectors 
were operational at 220~m and 420~m from the interaction point. With Level 1 trigger thresholds of $p_T > 25$ GeV/c 
for single electrons and $p_T > 20$ GeV/c for single muons, and reduced thresholds for dileptons, it was found that 
there should be $\sim$3 events in 30~fb$^{-1}$ for 140~GeV/c$^{2}$$< M_h <$ 200~GeV/c$^{2}$. 
For a lighter Standard Model Higgs boson, $M_h = 120$ GeV/c$^{2}$, there would be $\sim$0.5 events per 
30~fb$^{-1}$, and it was concluded that the event rate is marginal at low luminosity for $M_h < 140$~GeV/c$^{2}$. 
The event yields are similar for ATLAS and CMS.
All background processes, primarily from either photon-photon fusion or $W$-strahlung from the CEP of light-quark dijets, 
were studied and the conclusion was that signal-to-background ratio of one (or better) should be achievable, although 
below the 2-$W$ threshold there is a potentially dangerous background in the case where the off-shell $W^*$ from the 
Higgs boson decays hadronically. For the gold-plated doubly-leptonic decay modes, there would be approximately one event 
per 30~fb$^{-1}$ with no appreciable backgrounds.

Since above analysis was published, it has become clear that it will not be necessary to impose such high leptonic trigger thresholds because forward proton 
detector information can be employed at higher trigger levels to reduce the rates significantly, allowing for higher Level 1 
input rates (as discussed in Section~\ref{sec:trigger}). If the trigger thresholds are reduced to 15~GeV/c for both electrons and 
muons (which could also be achieved by demanding a coincidence with one or two jets) then the signal rates double. 
Detector effects have been investigated using the fast simulations of CMS (ATLAS) for the CEP of a 165~(160)~GeV/c$^{2}$ 
Higgs boson~\cite{vilela}. These studies showed that the experimental efficiency of the fully leptonic channel is in good agreement with the 
analysis presented in~\cite{cox2}, but that the semi-leptonic event rates may be reduced by up to a factor of four during data taking at instantaneous luminosities in excess of $5 \times 10^{33}$cm$^{-2}$ s$^{-1}$ due to kinematic cuts necessary to reduce the luminosity dependent `overlap' backgrounds, which are discussed in Section~\ref{sec:pilko}. Taking into account the increase in integrated luminosity, it is expected that ~10 events could be observed
in the gold-plated fully leptonic decay channel for 300~fb$^{-1}$ of data. Note that these events have the striking characteristic of a dilepton vertex with 
no additional tracks allowing for excellent background suppression and will afford a measurement of the Higgs mass to within $\sim$2 \GeVcc\ (the mass measurement by FP420 is not affected by the two undetected neutrinos).
For a 120~GeV/c$^{2}$ Higgs boson, there will be a total of 5 events for 300~fb$^{-1}$.

The conclusion is that the CEP of a SM Higgs boson should be observable in the $WW^{*}$ decay channel for all masses in 
300 fb$^{-1}$ with a signal to background ratio of one or better. This will provide confirmation that any observed resonance is 
indeed a scalar with quantum numbers $0^{++}$, and allow for a mass measurement\footnote{The mass resolution of FP420 
is discussed in detail in Section~\ref{sec:optics}.} on an event-by-event basis of better than 3 GeV/c$^{2}$ even in the 
doubly-leptonic decay channels in which there are two final state neutrinos. This will be a vitally important measurement at the LHC, 
where determining the Higgs quantum numbers is extremely difficult without CEP.~Furthermore, in certain regions of MSSM parameter 
space, in particular for $140$~GeV/c$^{2}$$ < M_A < 170$~GeV/c$^{2}$ and intermediate tan$\beta$, the CEP rate for 
$h \rightarrow WW^{*}$ may be enhanced by up to a factor of four~\cite{Heinemeyer:2007tu}. We discuss the MSSM in 
more detail in the following section for the $b\bar{b}$ decay channel.  
  
For the Standard Model Higgs boson, the $b \bar b$ decay channel is more challenging. It is the conclusion of~\cite{Albrow:2006xt,Cox:2007sw} 
that this channel will be very difficult to observe for $M_h = 120$~GeV/c$^{2}$ using FP420 alone, but may be observable 
at the $3 \sigma$ level if 220~m detectors are used in conjunction with FP420 and the cross sections are 
at the upper end of the theoretical expectations and/or the experimental acceptance and trigger and $b$-tagging efficiencies are 
improved beyond the currently assumed values. This should not be dismissed, because such an observation would be extremely 
valuable, since there may be no other way of measuring the $b$-quark couplings of the SM Higgs at the LHC. We discuss the 
experimental approach to observing Higgs bosons in the $b \bar b$ decay channel in detail in Section~\ref{sec:pilko}.       


\subsection{$h,H$ in the MSSM}
\label{sec:mssm}

In many MSSM scenarios, the additional capabilities brought to the LHC detectors by FP420 would be vitally 
important for the discovery of the Higgs bosons\footnote{Here we are dealing with the lightest 
MSSM Higgs boson $h$ and the heavier state $H$. Note that production of the pseudo-scalar Higgs, $A$, 
is suppressed in CEP due to the $J_z=0$ selection rule.} 
and the measurement of their properties. The coupling of the lightest MSSM Higgs boson to $b$ quarks and $\tau$ 
leptons may be strongly enhanced at large tan $\beta$ and small $M_A$, opening up both modes to FP420. 
The cross sections may become so large in CEP that one could carry out a lineshape analysis to 
distinguish between different models~\cite{Kaidalov:2003ys,je} and to make direct observations of CP 
violation in the Higgs sector~\cite{je,Khoze:2004rc}. If the widths are a few \GeVcc, a direct width measurement 
may be possible, a unique capability of FP420.


\subsubsection{$h,H\ra\bb$ decay modes}

\begin{figure}
\centering
\includegraphics[width=1.\textwidth]{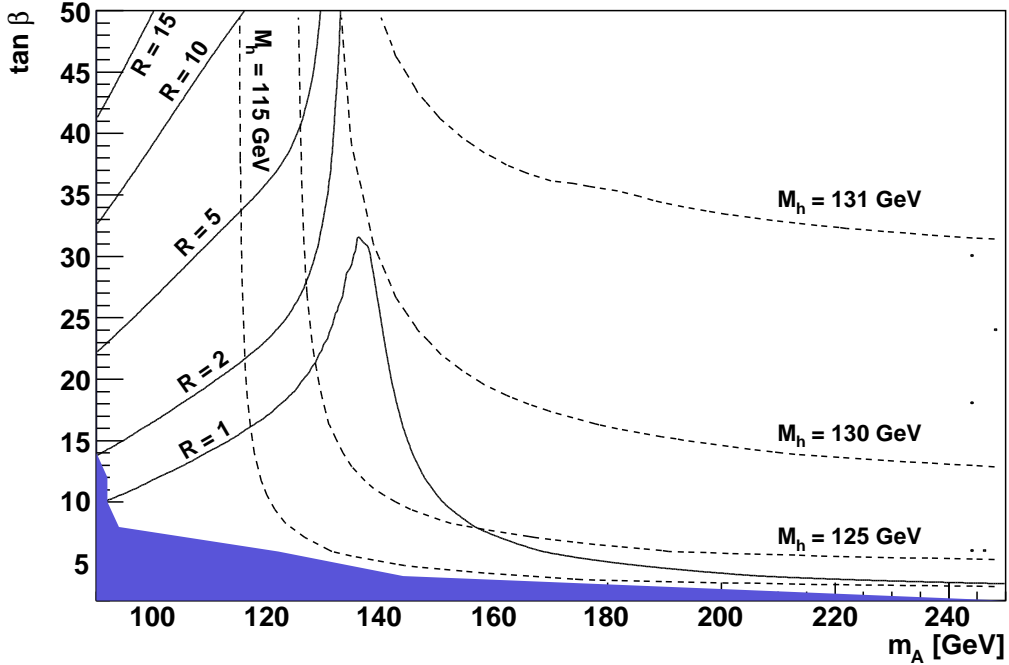}
\caption{The ratio, $R$, of cross section $\times$ branching ratio in the CEP $h \to b \bar b$ channel 
in the $M_A$ - tan$\beta$ plane of the MSSM within the $M_h^{max}$ benchmark scenario 
(with $\mu = +200$ GeV) to the SM Higgs cross section~\cite{Heinemeyer:2007tu}. 
The dark shaded (blue) region corresponds to the parameter region that is excluded by the LEP Higgs 
boson searches~\cite{LEPHiggsSM,LEPHiggsMSSM}.}
\label{fig:heinemeyer1}
\end{figure}

In~\cite{Heinemeyer:2007tu} (Heinemeyer {\it et al.}) a detailed study of the additional coverage 
in the $M_A - \rm{tan}\beta$ plane afforded by FP420 and 220~m detectors was carried out for several 
benchmark MSSM scenarios.  In particular, the observation of the CP-even Higgs bosons ($h$, $H$) in 
the $b$-quark decay channel was investigated. Figure~\ref{fig:heinemeyer1} shows the ratio of the MSSM 
to SM cross sections $\times$ the branching-ratio for the $h \to b \bar b$ channel within the $M_h^{max}$ 
scenario~\cite{sven3} as a function of $M_A$ and tan$\beta$. 
For example, at tan$\beta=33$ and $M_A=120$~GeV/c$^{2}$, the cross section for $h\rightarrow b\bar{b}$ 
in the MSSM is enhanced by a factor of five with respect to the Standard Model.
The results shown are for $\mu = +200$~GeV, where the parameter $\mu$ determines the size and effect 
of higher order corrections; negative (positive) $\mu$ leads to enhanced (suppressed) bottom Yukawa couplings. 

\begin{figure}
\begin{center}
\includegraphics[width=14cm,height=8.8cm]
                {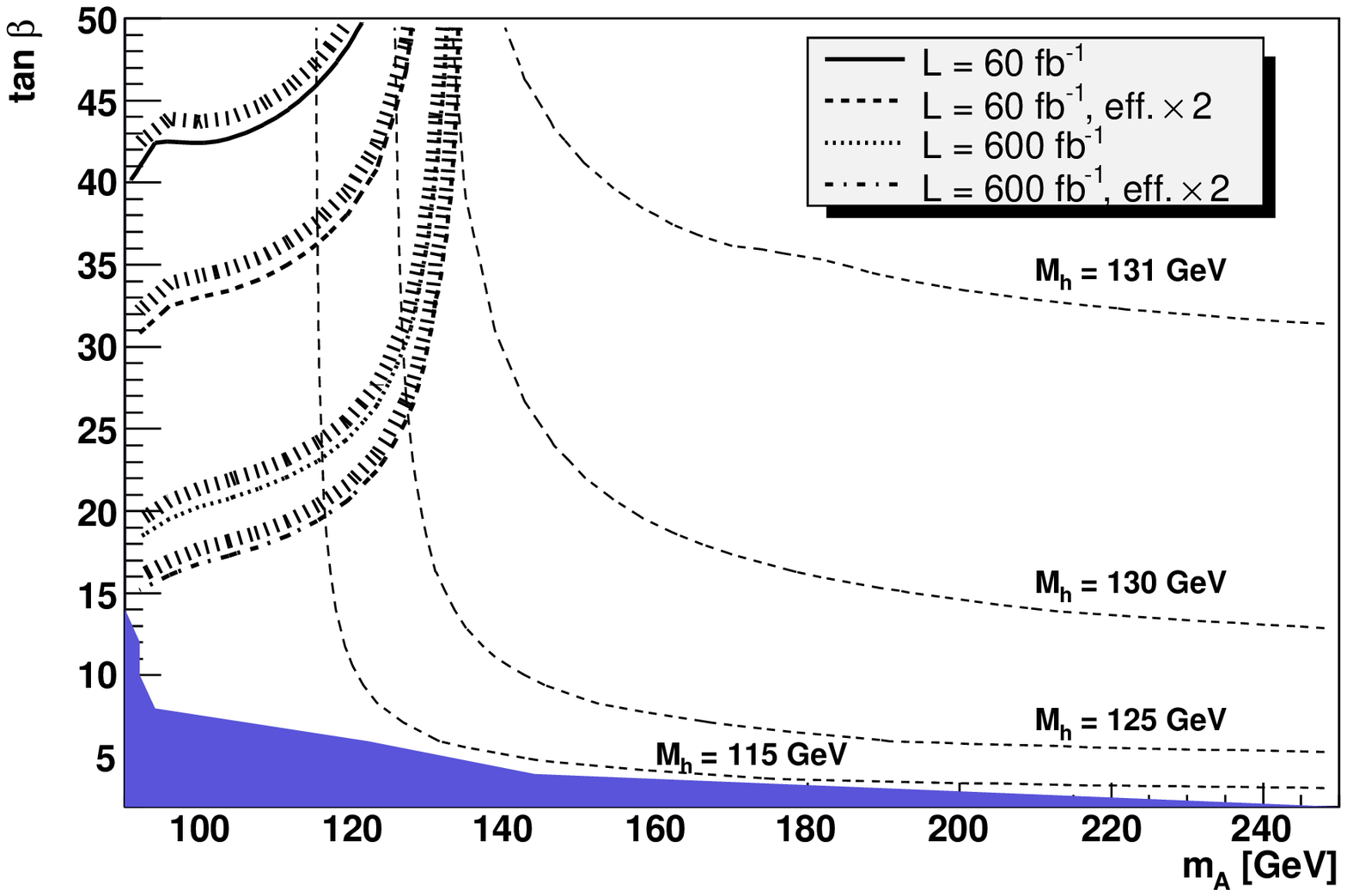}
\includegraphics[width=14cm,height=8.8cm]
                {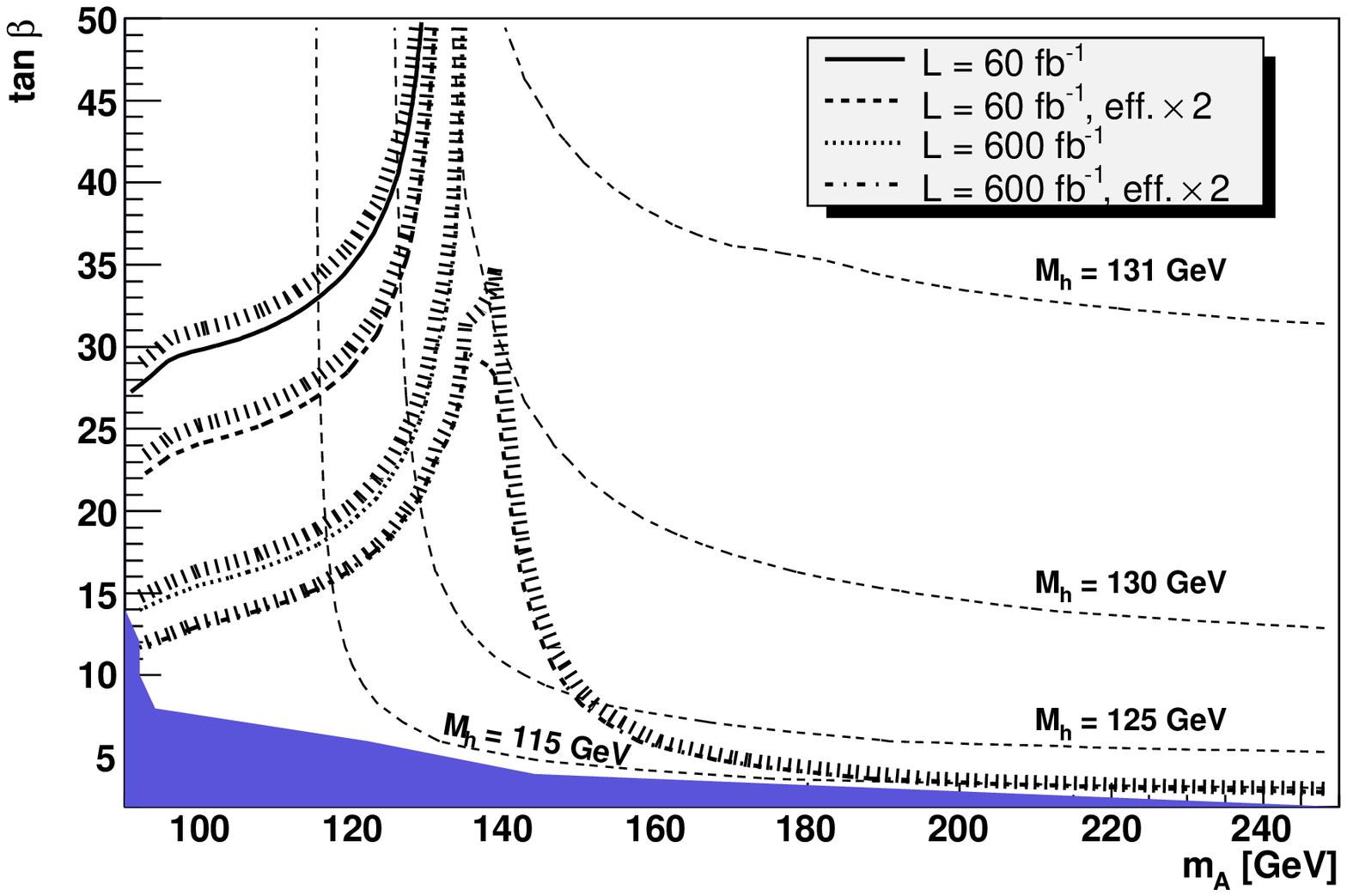}
\caption{
$5 \sigma$ discovery contours (upper plot) and contours of $3 \sigma$
statistical significance (lower plot) for the $h \to b \bar b$ channel in
CEP in the $M_A$ - \rm{tan}$\beta$ plane of the MSSM within the $\Mhmax$
benchmark scenario for different luminosity scenarios as described in the text~\cite{Heinemeyer:2007tu}.  
The values of the mass of the light CP-even Higgs boson, $M_h$, are
indicated by contour lines. No pile-up background assumed. 
The dark shaded (blue) region corresponds to the parameter region that
is excluded by the LEP Higgs boson searches~\cite{LEPHiggsSM,LEPHiggsMSSM}. 
}
\label{fig:heinemeyer2}
\end{center}
\end{figure}

\begin{figure}
\begin{center}
\includegraphics[width=14cm,height=8.8cm]
                {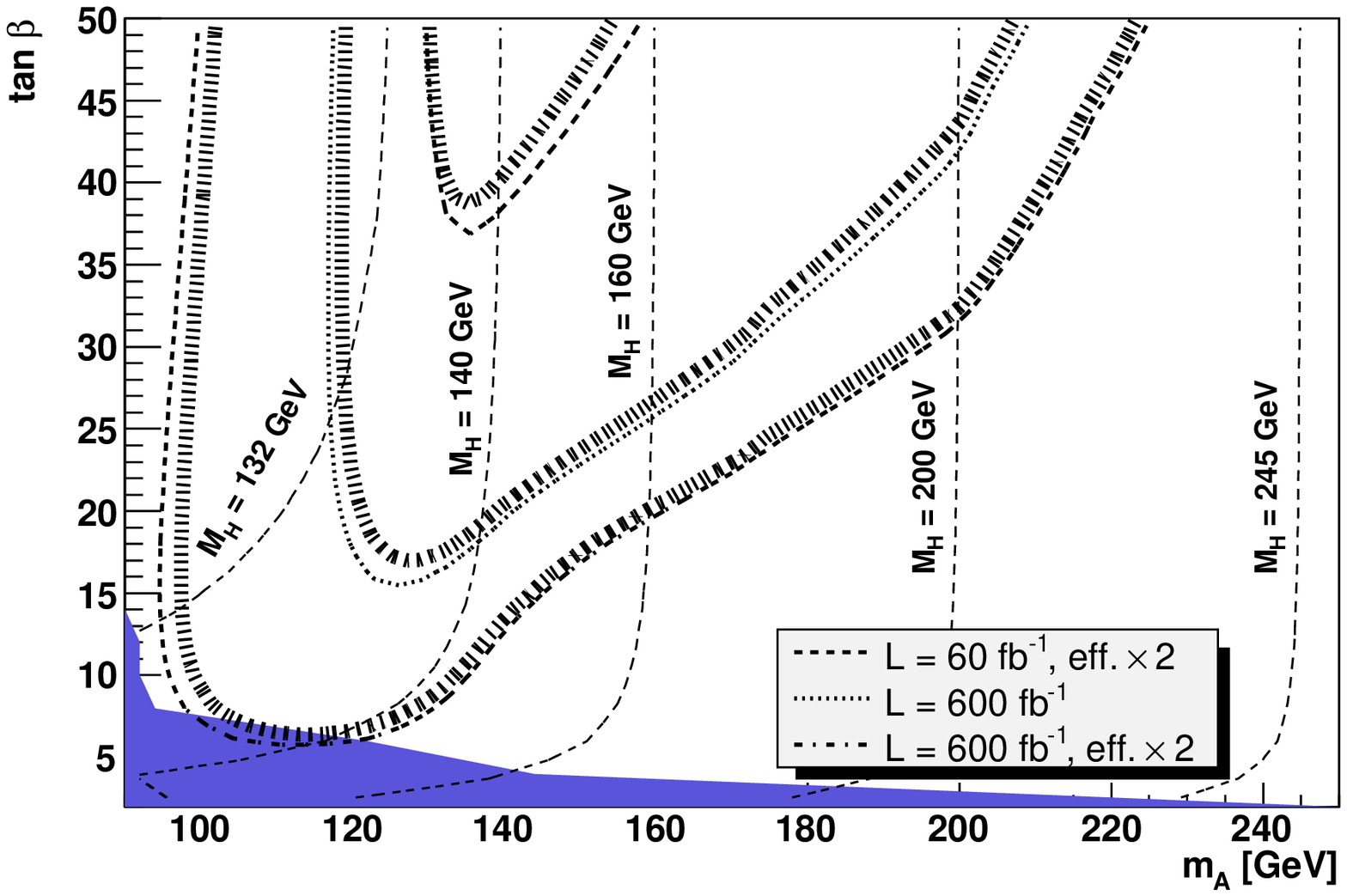}
\includegraphics[width=14cm,height=8.8cm]
                {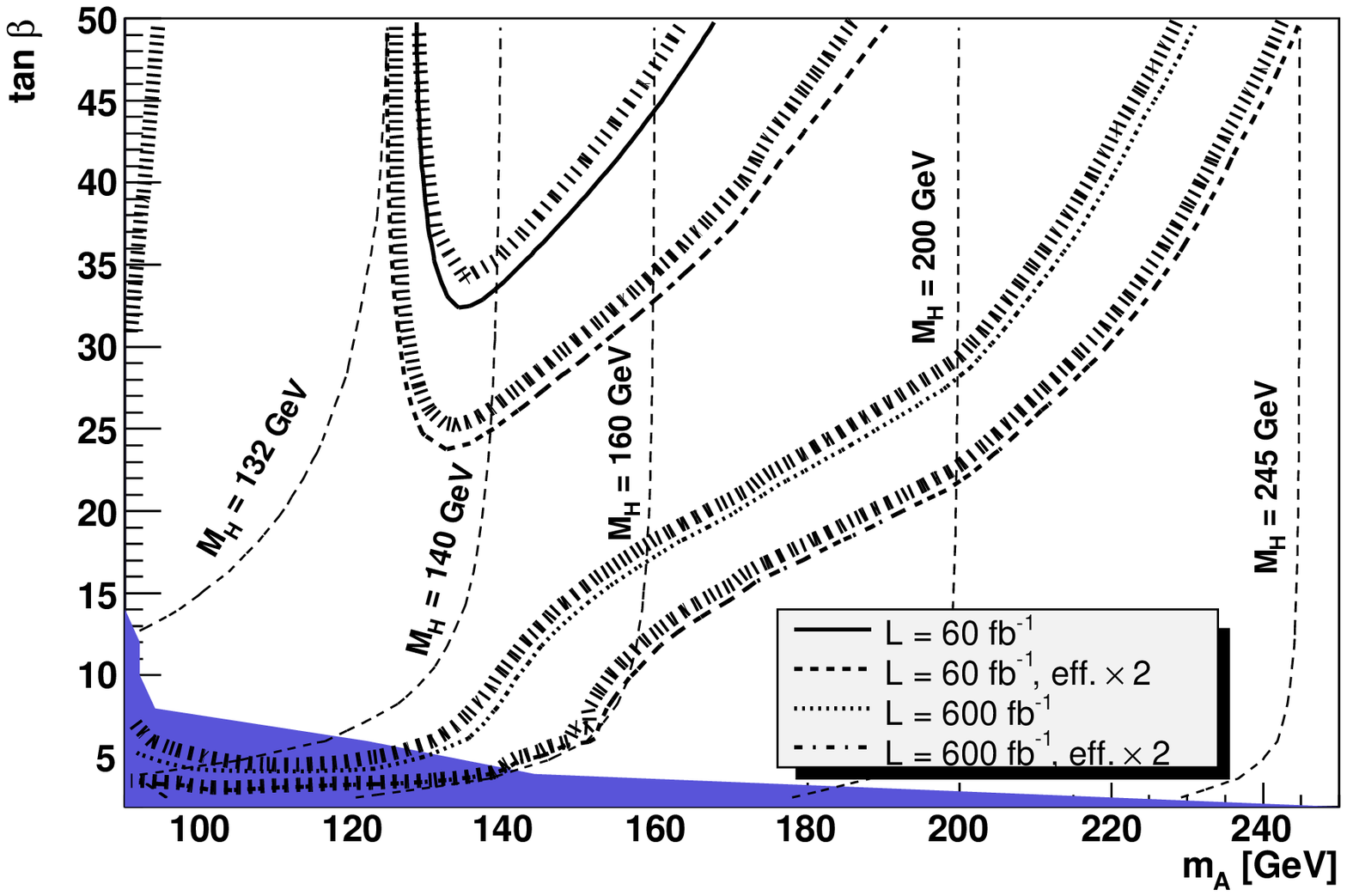}
\caption{
$5 \sigma$ discovery contours (upper plot) and contours of $3 \sigma$
statistical significance (lower plot) for the CEP $H \to b \bar b$ channel
in the $M_A$ - \rm{tan}$\beta$ plane of the MSSM within the $M_h^{max}$
benchmark scenario (with $\mu = +200$GeV) for different luminosity scenarios as 
described in the text~\cite{Heinemeyer:2007tu}. 
The values of the mass of the heavier CP-even Higgs boson, $M_H$, are
indicated by contour lines. No pile-up background assumed.
The dark shaded (blue) region corresponds to the parameter region that
is excluded by the LEP Higgs boson searches~\cite{LEPHiggsSM,LEPHiggsMSSM}.
}
\label{fig:heinemeyer3}
\end{center}
\end{figure}

Figure~\ref{fig:heinemeyer2} shows the $5 \sigma$ discovery contours (upper plot) and the $3 \sigma$ 
contours (lower plot) for this scenario. The discovery contours were calculated using an experimental 
efficiency based on the simulated analysis in the CMS-TOTEM studies~\cite{Albrow:2006xt}, 
with a full simulation of the acceptance of both FP420 and 220~m forward proton detectors. The Level 1 
trigger strategy was based on information only from the central detectors and 220~m detectors. Full 
details can be found in~\cite{Heinemeyer:2007tu}. Curves are shown for several luminosity scenarios; 
$\int\mathcal{L} = 60$ fb$^{-1}$ corresponds to 3 years of data taking by ATLAS and CMS at 
$10^{33}$~cm$^{-2}$~s$^{-1}$, and $\int\mathcal{L}=600$~fb$^{-1}$ corresponds to 
3 years of data taking by both experiments at 10$^{34}$~cm$^{-2}$~s$^{-1}$.  
For example, if tan$\beta=40$ and $M_A=120$~GeV/c$^{2}$,~$h \ra b\bar{b}$ would be observed 
with more than 3$\sigma$ confidence with 60~fb$^{-1}$ of data (lower plot), but would require twice the 
experimental efficiency or  more integrated luminosity to be observed with 5$\sigma$ confidence (upper plot). 
Figure~\ref{fig:heinemeyer3} shows the $5 \sigma$ discovery  contours (upper plot) and the $3 \sigma$ 
contours (lower plot) for the heavy scalar, $H$, in the same scenario. With sufficient integrated luminosity 
(few hundreds fb$^{-1}$), all values of tan$\beta$ are covered for $M_A \sim 140$~GeV/c$^{2}$ and at high 
tan$\beta$ observation remains possible for Higgs bosons with masses in excess of 200~GeV/c$^{2}$.

An important challenge of the $b \bar b$ channel measurement at the LHC is the combinatorial 'overlap' 
background caused by multiple proton-proton interactions in the same bunch crossing. The analysis presented 
above uses the selection efficiencies discussed in~\cite{Albrow:2006xt} which are based on stringent cuts that 
are expected to reduce such pile-up contributions. This background is indeed negligible at low luminosities 
($\sim$10$^{33}$~cm$^{-2}$~s$^{-1}$), but becomes more problematic at the highest luminosities. 
For the latter cases, additional software as well as hardware 
improvements in rejecting the background have been assumed. Such improvements are presented in the analysis 
of~\cite{Cox:2007sw} (Cox et al.) which examines the MSSM point given by tan$\beta$~=~40 and $M_A$~=~120 \GeVcc\
in detail. Figure~\ref{fig:heinemeyer2} indicates that, for this choice of parameters, $h \ra b\bar{b}$ 
should be observable with a significance close to 4$\sigma$ for 60 fb$^{-1}$ of data. Section~\ref{sec:pilko} 
summarises the results obtained in~\cite{Cox:2007sw} and demonstrates the experimental procedure 
and hardware requirements needed to reduce the overlap backgrounds. We compare the results of the
two independent $h \ra b\bar{b}$ analyses in Section~\ref{sec:Hbbar_comparison}.



\subsubsection{$h,H \rightarrow \tau \tau$ decay modes}

In the standard (non-CEP) search channels at the LHC, the primary means of detecting the 
heavy CP-even Higgs boson $H$ (and the CP-odd $A$)  in the MSSM is in the $b$-quark associated 
production channel, with subsequent decay of the Higgs boson in the $\tau \tau$ decay mode.
This decay mode is also open to CEP and was studied in~\cite{Heinemeyer:2007tu}. The branching ratio 
of the Higgs bosons to $\tau \tau$ is approximately 10\% for $M_{H/A} > 150$ GeV/c$^{2}$ and 
90\% to $b \bar b$, if the decays to light SUSY particles are not allowed. Note that $\tau$'s decay to 
1-prong (85\%) or 3-prong (15\%); requiring no additional tracks on the $\tau\tau$ vertex is very 
effective at reducing non-exclusive background.

\begin{figure}
\begin{center}
\includegraphics[width=14cm,height=8.8cm]
                {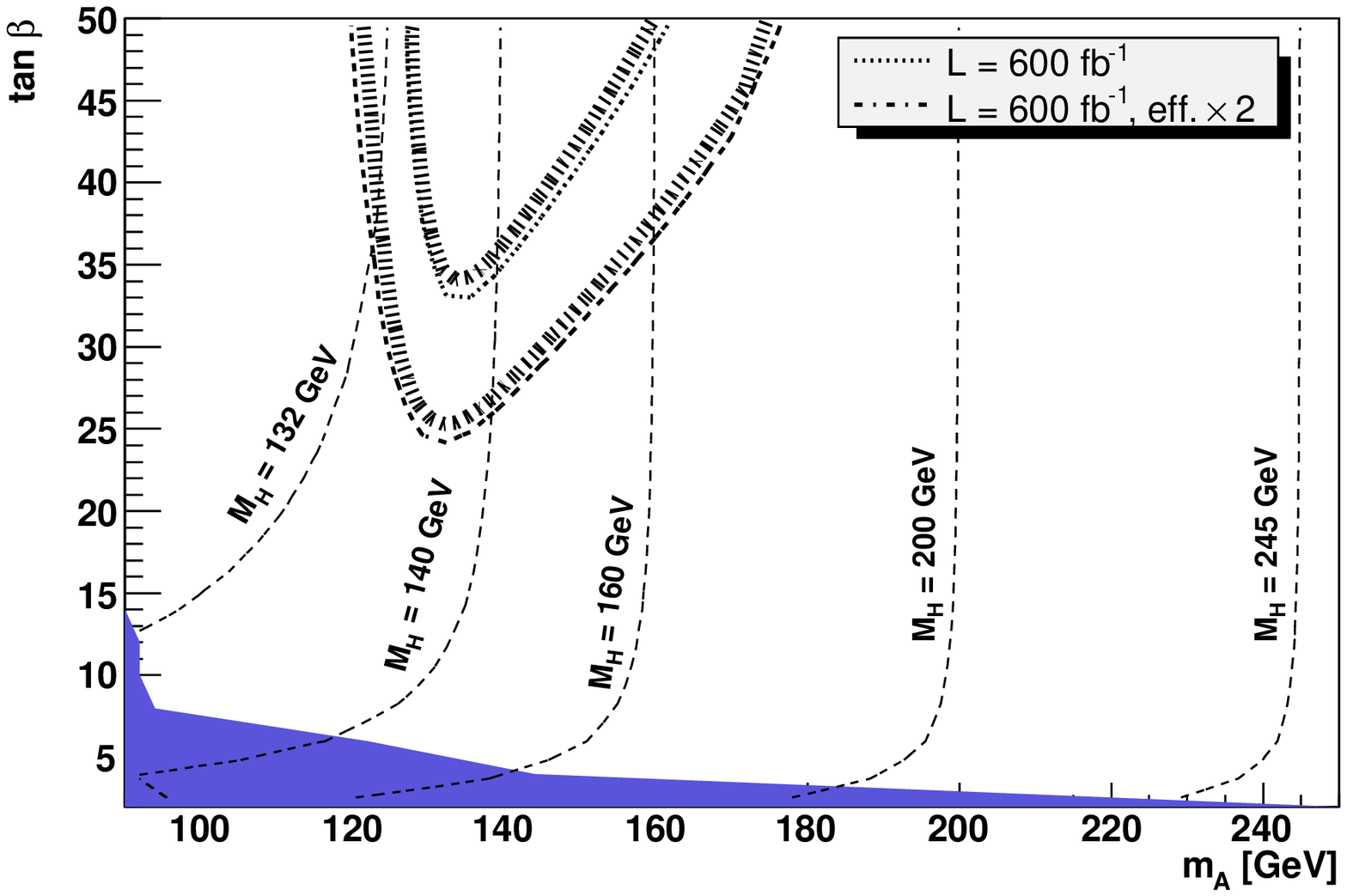}
\includegraphics[width=14cm,height=8.8cm]
                {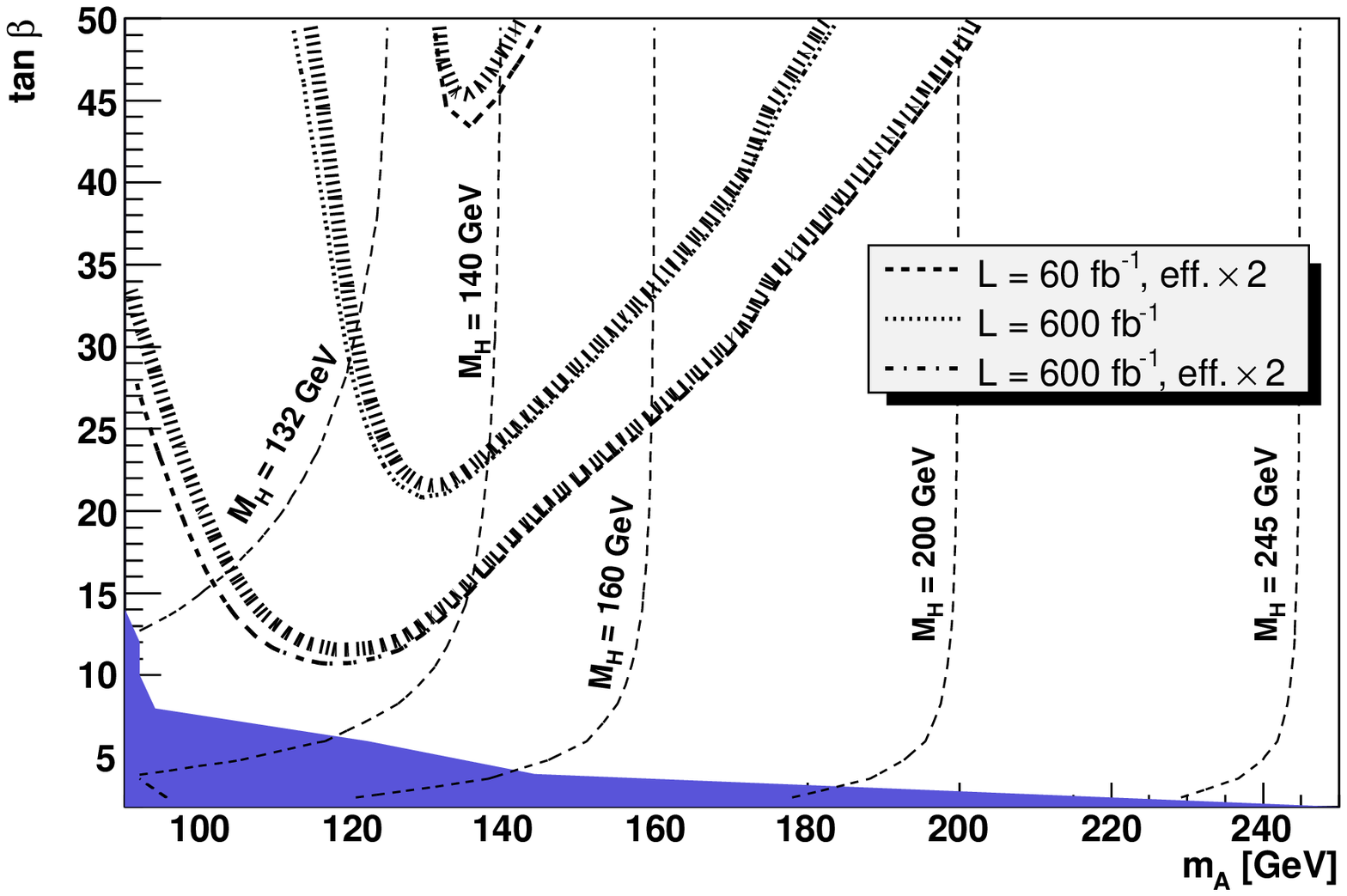}
\caption{
$5 \sigma$ discovery contours (upper plot) and contours of $3 \sigma$
statistical significance (lower plot) for the $H \to \tau^+\tau^-$ channel
in CEP in the $M_A$--tan$\beta$ plane of the MSSM within the $M_h^{max}$
benchmark scenario (with $\mu = +200$GeV) for different 
luminosities.
The values of the mass of the heavier CP-even Higgs boson, $M_H$, are
indicated by contour lines. No pile-up background assumed.
The dark shaded (blue) region corresponds to the parameter region that
is excluded by the LEP Higgs boson searches.}
\label{fig:heinemeyer4}
\end{center}
\end{figure}

Figure~\ref{fig:heinemeyer4} shows the $5 \sigma$ discovery contours and the $3 \sigma$ contours 
in the $M_A - \rm{tan}\beta$ plane for the $M_h^{max}$ benchmark scenario for different luminosity 
scenarios. The discovery region is significantly smaller than for the $b \bar b$ case, although the decay 
channel can be observed at 3$\sigma$ across a large area of parameter space. This would be an important 
complementary measurement to the standard search channels, affording a direct measurement of the 
quantum numbers of the $H$. Furthermore, in this region of parameter space, the $A$ is very close in 
mass to the $H$ and, since the $A$ is heavily suppressed in CEP, a clean high-precision measurement 
of the $H$ mass in the $\tau \tau$ channel will be possible using forward proton tagging. 
Heinemeyer {\it et al.} \cite{Heinemeyer:2007tu} also investigated the coverage for the di-tau decay channel of the light $h$, 
and found that a $3 \sigma$ observation could be made in the region tan$\beta \geq 15$, 
$M_h < 130$~GeV/c$^{2}$ at high luminosity.      

\subsection{Observation of Higgs bosons in the NMSSM}


The Next-to-Minimal Supersymmetric Standard Model (NMSSM) extends the MSSM by the inclusion of a singlet superfield, $\hat{S}$~\cite{Ellis:1988er}. 
This provides a natural solution to the $\mu$ problem through the $\lambda\hat{S}\hat{H_u}\hat{H_d}$ superpotential term when the 
scalar component of $\hat{S}$ acquires a vacuum expectation value. The Higgs sector of the NMSSM contains three CP-even and two 
CP-odd neutral Higgs bosons, and a charged Higgs boson. According to~\cite{dermisekgunion}  the part of parameter space that has 
no fine-tuning problems results in the lightest scalar Higgs boson decaying predominantly via $h\rightarrow aa$, where $a$ is the lightest 
pseudo-scalar. The scalar Higgs boson has a mass of $\sim$100~GeV/c$^{2}$. If the $a$ has a mass 
of $2m_{\tau} \lesssim m_a \lesssim 2m_b$, which is in fact preferred, then the decay channel $h\rightarrow aa \rightarrow 4\tau$ 
would become the dominant decay chain. This is not excluded by LEP data.  In such a scenario the LHC could fail to discover any 
of the Higgs bosons~\cite{dermisekgunion}.

\begin{figure}
\centering
\mbox{
	\subfigure[]{\includegraphics[width=.55\textwidth]{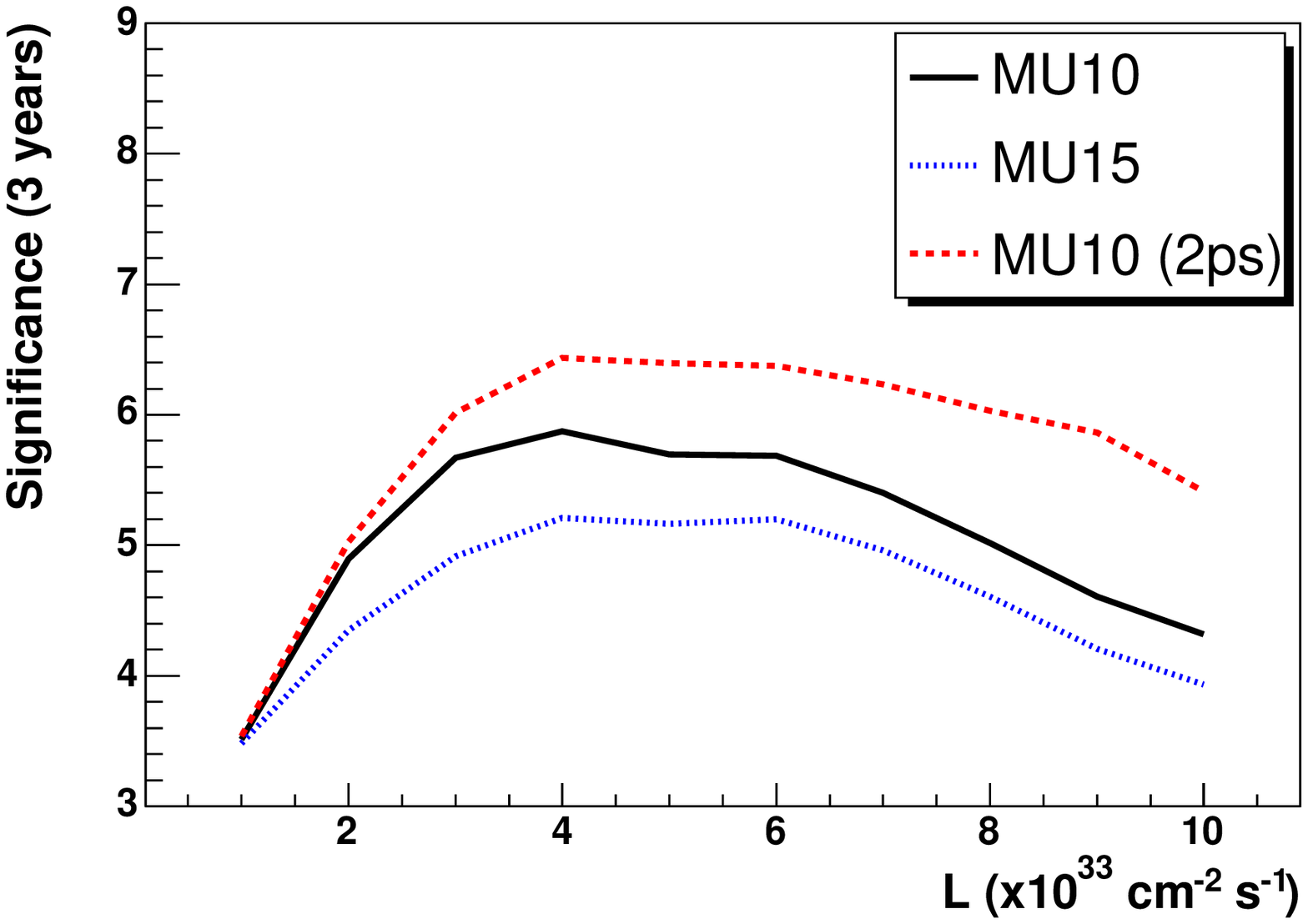}}
	\subfigure[]{\includegraphics[width=.55\textwidth]{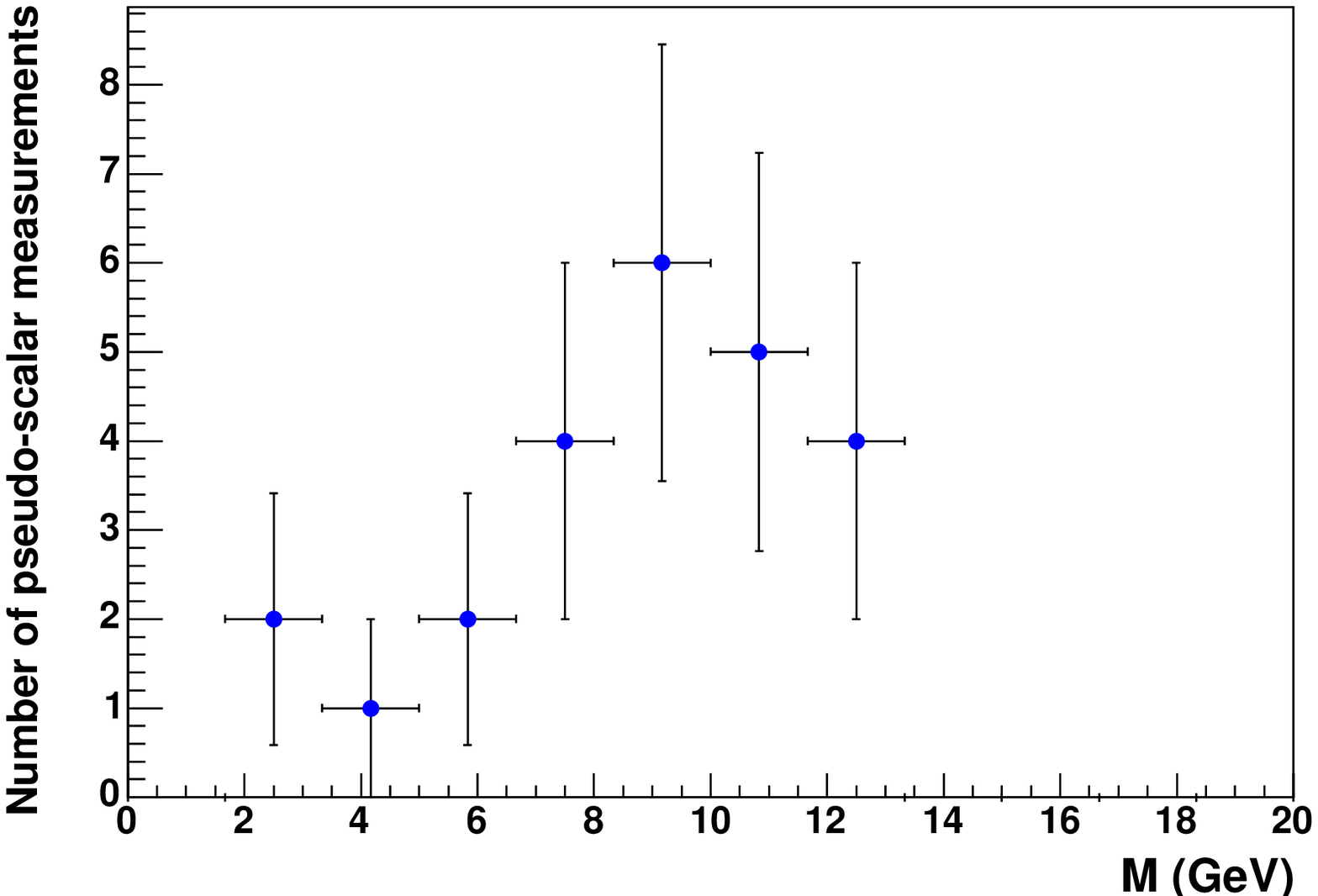}}
}

\caption{(a) The significance of observation of $h \rightarrow aa \rightarrow 4\tau$ using a muon $p_T$ trigger 
threshold of 10~GeV/c (or 15~GeV/c) for three years of data taking at ATLAS and CMS. Also shown is the increase 
in the significance due to a factor of five improvement in background rejection from a 2~ps proton time-of-flight measurement, see Sections~\ref{sec:pilko} 
and~\ref{sec:timing}, or a comparable gain across all of the rejection variables \cite{Forshaw:2007ra}. 
(b) A typical $a$ mass measurement for 150~fb$^{-1}$ of data.
\label{fig:nmssmhiggs}}
\end{figure}

Subsequently, however, it was shown in~\cite{Forshaw:2007ra} that the lightest Higgs boson could be 
discovered in CEP using FP420. The parameter point chosen was similar to scenario 1 in~\cite{Gunion1} 
and resulted in $M_h = 92.9$~GeV/c$^{2}$ and $m_a = 9.7$~GeV/c$^{2}$, with BR$(h\ra aa) = 92\%$ 
and BR$(a\ra \tau\tau) = 81\%$.~The analysis uses mainly tracking information to define the $4\tau$ final 
state and triggers on a single muon with a transverse momentum greater than 10 GeV/c, although the analysis 
still works for an increased muon threshold of 15~GeV/c.~The final event rates are low, approximately 3-4 
events after all cuts at ATLAS or CMS over three years of data taking if the instantaneous luminosity is greater 
than 10$^{33}$~cm$^{-2}$~s$^{-1}$. There is however no appreciable background. Figure~\ref{fig:nmssmhiggs}(a) 
shows the combined significance of observation at ATLAS and CMS after three years of data taking at a specific 
instantaneous luminosity. The mass of the $h$ is obtained using FP420 to an accuracy of $2-3$~GeV/c$^{2}$ 
(per event). Furthermore, using FP420 and the tracking information from the central detector, it is possible to 
make measurements of the $a$ mass on an event-by-event basis. This is shown in Figure~\ref{fig:nmssmhiggs}(b) 
for an example pseudo-data set corresponding to 150~fb$^{-1}$ of integrated luminosity.~ From examining 
many such pseudo-data sets, the mass of the $a$ in this scenario would be measured as $9.3\pm 2.3$~GeV/c$^{2}$.

A complementary, independent trigger study has also been performed for this decay channel using the CMS fast 
simulation. Using only the standard CMS single muon trigger of 14~GeV/c, a trigger efficiency of 13\% for the 
$h \ra aa \ra 4\tau$ was observed. This is in reasonably good agreement with the study presented above, which 
observed a 12\% efficiency for a 15~GeV/c trigger (assuming ATLAS efficiencies). Furthermore, the study also 
observed that the analysis presented above would benefit from additional triggers, which were not considered 
in~\cite{Forshaw:2007ra}. The total trigger efficiency increases to $\sim$28\% if a combination of lepton triggers 
are used. It is likely that the majority of these events will pass the analysis cuts presented in~\cite{Forshaw:2007ra} 
and so would boost the event rate by up to a factor of two. If the lepton trigger thresholds can be reduced, 
which could be possible at low luminosities, the trigger efficiency increases to 45\% resulting in a factor of 3.5 increase in the event rate.


\subsection{Invisible Higgs boson decay modes}

In some extensions of the SM, the Higgs boson decays dominantly
into particles which cannot be directly detected, the so called
invisible Higgs. The prospects of observing such Higgs boson
via the forward proton mode are quite promising~\cite{BKMR}
assuming that the overlap backgrounds can be kept under control.
Note that contrary to the conventional parton-parton inelastic production,
the mass of such invisible Higgs boson can be accurately measured
by the missing mass method.

 
\subsection{Conclusion of the studies of the CEP of $h,H$}   

It is a general feature of extended Higgs sectors that the heavy Higgs bosons decouple from the gauge bosons 
and therefore decay predominantly to heavy SM fermions. Adding the possibility to detect the $b \bar b$ decay 
channel and enhancing the capacity to detect the $\tau \tau$ channel would therefore be of enormous value. 
In the $M_h^{max}$ scenario of the MSSM, if forward proton detectors are installed at 420~m and 220~m and 
operated at all luminosities, then nearly the whole of the 
$M_A - \rm{tan}\beta$   plane can be covered at the $3 \sigma$ level. Even with only 60 fb$^{-1}$ of luminosity 
the large tan$\beta$ / small $M_A$ region can be probed. For the heavy CP-even MSSM Higgs boson with a mass 
of approximately 140 GeV/c$^{2}$,  observation should be guaranteed for all values of tan$\beta$ with sufficient 
integrated luminosity. At high tan$\beta$, Higgs bosons of masses up to $\sim 240$ GeV/c$^{2}$ should be 
observed with 220~m proton taggers. The coverage and significance are further enhanced for 
negative values of the $\mu$ parameter.  For scenarios in which the light (heavy) Higgs boson and the $A$ boson are nearly degenerate in mass, 
FP420 (together with the 220 proton tagger) will allow for a clean separation of the states since the $A$ cannot be produced in central exclusive production. 
In the NMSSM, forward proton tagging could become the discovery channel in the area of parameter space in which 
there are no fine-tuning issues through the decay chain $h\ra aa \ra 4\tau$. Using the information from FP420, 
the mass of both the $h$ and $a$ can be obtained on an event-by-event basis.

Observation of any Higgs state in CEP allows for direct observation of its quantum numbers and a high-precision 
mass measurement. As we shall see in Section~\ref{sec:pilko}, it will be possible in many scenarios to measure the 
mass with a precision of better than $1$ GeV/c$^{2}$ and a width measurement may also be possible.
Installation of FP420 would therefore provide a significant enhancement in the discovery potential of the current baseline LHC detectors.

\nopagebreak 

\subsection{Photon-photon and photon-proton physics}
\label{sec:photon_phys}


\subsubsection{Introduction}

Photon-induced interactions have been extensively studied 
in electron-proton and electron-positron collisions at HERA and LEP, respectively. 
A significant fraction of $pp$ collisions at the LHC will also involve 
quasi-real (low-$Q^2$) photon interactions, occurring for the first time at centre-of-mass energies well 
beyond the electroweak scale. The LHC will thus offer a unique possibility for novel 
research -- complementary to the standard parton-parton interactions -- via photon-photon and 
photon-proton processes in a completely unexplored regime. The much larger effective  
luminosity available in parton-parton scatterings will be compensated by the better known initial 
conditions and much simpler final states in photon-induced interactions. 
The distinct experimental signatures of events involving photon exchanges are the presence of very 
forward scattered protons and of large rapidity gaps (LRGs) in forward directions. 
Dedicated very forward detectors are thus required in order to efficiently {\it tag} photon-induced events 
and keep the backgrounds under control~\cite{kp}. Very recently, exclusive two-photon production 
of lepton pairs~\cite{Abulencia:2006nb} and diffractive photoproduction of quarkonia~\cite{tev2} have 
been successfully measured in $p\bar{p}$ collisions at Tevatron (and also in heavy-ion collisions at 
RHIC~\cite{d'Enterria:2006ep}) using LRGs. In both measurements, clear signals were obtained 
with low backgrounds. Apart from their intrinsic interest, these exclusive processes 
p + p $\rightarrow$ p + $e^+e^-$ + p, p + $\mu^+\mu^-$ + p, both through $\gamma\gamma \rightarrow l^+l^-$ 
and $\gamma + p \rightarrow \Upsilon + p$ provide a source of forward protons with momenta 
known to better than 1 GeV/c (dominated by the incoming beam momentum spread $dp/p\sim$~10$^{-4}$). 
They therefore give an important calibration of the FP420 momentum scale and resolution, even 
though usually only one proton is detected (see Section~\ref{sec:align_physics}).

\begin{figure}[htbp]  
\begin{center}
\epsfig{file=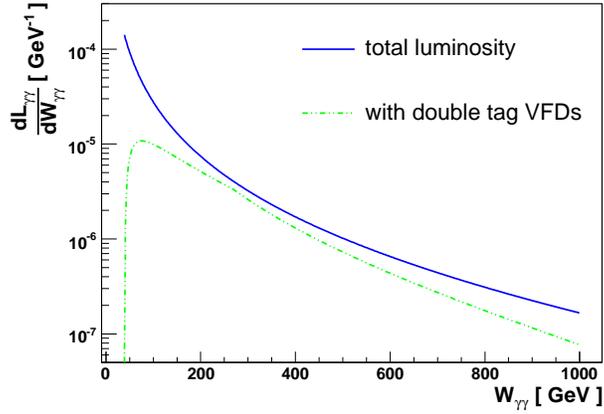,width=0.55\linewidth} 
\caption{Relative elastic luminosity spectrum of photon-photon collisions at the LHC 
in the range $Q^2_{min}<Q^2<2$~GeV$^2$ (solid blue line) compared to the 
corresponding luminosity if the energy of each photon is restricted to the forward detector (VFD) tagging range 
20 GeV $< E_\gamma <$ 900 GeV (dashed green curve)~\cite{louvain}.}
\label{fig:GGlumi} 
\end{center}
\end{figure}

The equivalent photon (or Weizs\"acker-Williams) approximation (EPA)~\cite{EPA} provides the appropriate
framework to describe processes involving photon exchange with proton beams at the LHC.
In this approximation, one photon is emitted by one (or both) incoming proton(s) which then
subsequently collides with the other proton (photon) producing a system $X$. Here, we will
only consider\footnote{A third class of events where the two colliding protons dissociate is
not considered here.} {\it elastic} photon-photon collisions, $\gamma\gamma\rightarrow X$, where both proton ``emitters'' 
remain intact (i.e. $pp\rightarrow ppX$) and 
{\it inelastic} photoproduction, $\gamma p\rightarrow X$, where the ``target'' proton dissociates into a state $Y$ (i.e. $pp\rightarrow p\,X\,Y$). 
In both cases, we ignore additional parton interactions which destroy the rapidity gaps. The probability that the gaps 
survive (gap survival probability, see Section~\ref{sec:calculation}) is much larger in the case 
of photon-photon  interactions -- which occur at relatively large impact parameters -- compared 
to exclusive Pomeron- or gluon- induced processes~\cite{KMRtag,KMRphot}.
In the EPA, the photon spectrum is a function of the photon energy 
$E_{\gamma}$ and its virtuality $Q^2$~\cite{EPA}:
\begin{equation}
{\rm d}N=\frac{\alpha}{\pi}\frac{{\rm d}E_{\gamma}}{E_{\gamma}}\frac{{\rm d}Q^2}{Q^2}
\left[\left(1-\frac{E_{\gamma}}{E}\right)
\left(1-\frac{Q^2_{min}}{Q^2}\right)F_E
+\frac{E_{\gamma}^2}{2E^2}F_M\right],
\label{eq:epa}
\end{equation}
where $\alpha$ is the fine-structure constant, $E$ is the incoming proton energy and the 
minimum photon virtuality $Q^2_{min}\simeq[M_Y^2E/(E-E_{\gamma})-M_p^2]E_{\gamma}/E$, where $M_p$ 
is the proton mass and $M_Y$ is the invariant mass of the final state $Y$, and $F_M$ and $F_E$
for the elastic production are given by the proton form factors, in the dipole approximation: 
$F_M=G_M^2$ and $F_E=(4M_p^2G_E^ 2+Q^2G_M^2)/(4M_p^2+Q^2)$,
where $G_E^2 = G_M^2 /7.78 = (1+Q^2/0.71 \mbox{GeV}^2)^{-4}$.
The spectrum is strongly peaked at low $E_{\gamma}$, therefore photon-photon 
centre-of-mass energies $W\simeq 2\sqrt{E_{\gamma_1}E_{\gamma_2}}$ are 
usually much smaller than the total centre-of-mass energy of $2E=14$~TeV. 
In the elastic case, the photon virtuality is usually low, 
$\langle Q^2\rangle\approx0.01$~GeV$^2$, and therefore
the proton scattering angle is very small, $\lesssim 20~\mu$rad.
The luminosity spectrum of photon-photon collisions, $d\mathcal{L}_{\gamma\gamma}/dW_{\gamma\gamma}$, 
can be obtained by integration of the product of the photon spectra, given by Eq.~(\ref{eq:epa}),
over the photon virtualities and energies keeping fixed $W$. The elastic $\gamma\gamma$
luminosity spectrum (Fig.~\ref{fig:GGlumi})
peaks strongly at low values of $W$, but extends to large values, even 
beyond\footnote{Note that the $W$-pair invariant mass coverage in $\gamma\gamma$ reactions is much 
larger than in the hard exclusive central diffractive processes where the cross sections at large masses 
are strongly suppressed by the QCD Sudakov form factor~\cite{KMRtag}.} 1~TeV.
The integrated spectrum directly gives a fraction of the $pp$ LHC luminosity
available in $\gamma\gamma$ collisions at $W>W_0$. 
The relative photon-photon effective luminosity amounts to 1\% for $W_{\gamma \gamma}>23$~GeV 
and to 0.1\% for $W_{\gamma \gamma}>225$~GeV. Given the very large LHC 
luminosity, this leads to significant event rates for high-energy processes with relatively 
small photon-photon cross-sections. This is even more true for $\gamma p$ 
interactions, where both energy reach and effective luminosities are much higher 
than for the $\gamma \gamma$ case.
Finally, photon physics can be studied also in ion collisions at the LHC~\cite{Baltz:2007kq}, 
where the lower ion luminosities are largely compensated by the high photon fluxes due to the 
$Z^2$ enhancement (for each nucleus), where $Z$ is the ion charge.\\

In this section, we will consider the following exclusive photon-induced processes accessible to measurement 
at the LHC with very forward proton tags:
\begin{enumerate}
\item two-photon production of lepton pairs (an excellent LHC ``luminometer'' process),
\item two-photon production $W$ and $Z$ pairs (as a means to investigate anomalous triple and quartic
gauge couplings),
\item two-photon production of supersymmetric pairs; as well as 
\item associated $WH$ photoproduction, and
\item anomalous single top photoproduction.
\end{enumerate}
Realistic studies of all these processes -- computed with dedicated packages 
({\sc madgraph} / {\sc madevent}~\cite{mad}, {\sc calchep}~\cite{Puk03}, {\sc lpair}~\cite{lpair}) 
including typical ATLAS/CMS acceptance cuts and a modified version of the Pythia
generator~\cite{pythia} for all processes involving final-state partons -- are discussed in 
detail in a recent review on photon-induced interactions at the LHC~\cite{louvain}. 
A summary of this work is presented in the following subsections.


\subsubsection{Two-photon processes}
\label{sec:GG}

\begin{figure}[htbp]  
\begin{center}
\epsfig{file=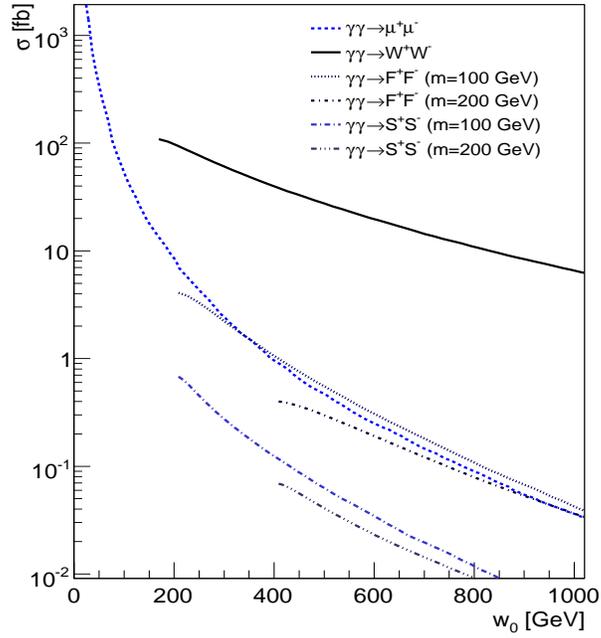,width=8.5cm,height=9.cm} 
\caption{Cross sections for various $\gamma \gamma$ processes at the LHC as a function
 of the minimal $\gamma \gamma$ centre-of-mass energy $W_0$~\cite{louvain}.}
\label{fig:IntGG} 
\end{center}
\end{figure}

Elastic two photon interactions yield very clean event topologies at the LHC: two very forward protons 
measured far away from the IP plus some centrally produced system. In addition, the photon momenta 
can be precisely measured using the forward proton taggers, allowing the reconstruction of the 
event kinematics. To illustrate the photon physics potential of the LHC, various pair production cross sections 
in two-photon collisions have been computed using a modified version~\cite{louvain} of 
{\sc madgraph}/{\sc madevent}~\cite{mad}.  The corresponding production cross sections are 
summarised in Table~\ref{tab:backGG}. Since the cross sections for pair production depend 
only on charge, spin and mass of the produced particles, the results are shown for charged 
and colourless fermions and scalars of two different masses. These cross sections are shown as 
a function of the minimal $\gamma \gamma$ centre-of-mass energy $W_0$ in Fig.~\ref{fig:IntGG}.\\

\begin{table}[htbp]
\begin{center}
\begin{tabular}[htbp]{ll||c |c}
\hline
\multicolumn{2}{l||}{\textbf{Processes}}  & $\mathbf{\sigma}$ (fb) & \textbf{Generator}   \\\hline
$\gamma \gamma \rightarrow$ & $\mu^+ \mu^-$ ($p_T^{\mu}>$ 2 \GeVc, $|\eta^\mu|<$3.1) &  72 500 & { \sc lpair}~\cite{lpair}\\
& $W^+ W^-$ & 108.5 & { \sc mg/me}~\cite{mad} \\ 
& $F^+ F^-$ (M = 100 \GeVcc) & 4.06 & $\scriptscriptstyle{//}$ \\
& $F^+ F^-$ (M = 200 \GeVcc) & 0.40 & $\scriptscriptstyle{//}$ \\
& $S^+ S^-$ (M = 100 \GeVcc) & 0.68 & $\scriptscriptstyle{//}$ \\
& $S^+ S^-$ (M = 200 \GeVcc) & 0.07 & $\scriptscriptstyle{//}$ \\
& $H \rightarrow b\overline{b}$ (M = 120 \GeVcc) & 0.15 & { \sc mg/me}~\cite{mad} \\
\hline
\end{tabular}
\end{center}
\caption{Production cross sections for $pp \rightarrow p p X$ (via $\gamma \gamma$ exchange) for various 
processes ($F$ for fermion, $S$ for scalar) computed with various generators~\cite{louvain}.}
\label{tab:backGG}
\end{table}

Clearly, interesting $\gamma \gamma$ exclusive cross sections at the LHC are 
accessible to measurement. In particular, the high expected statistics for exclusive 
$W$~pair production should allow for precise measurements of the $\gamma\gamma WW$ quartic 
couplings. The production of new massive charged particles such as supersymmetric pairs~\cite{bib:Ohnemus}, 
is also an intriguing possibility. Similarly, the exclusive production of the Higgs boson -- which has 
a low SM cross section~\cite{higgs} -- could become interesting in the case of an enhanced $H\gamma\gamma$ 
coupling. Last but not least, the two-photon exclusive production of muon pairs will provide an excellent 
calibration of luminosity monitors~\cite{lumi,kp}. 


\subsubsubsection*{Lepton pairs}
\label{sec:muonpairs}

Two-photon exclusive production of muon pairs has a well known QED cross section, including 
very small hadronic corrections~\cite{Khoze:2000db}. Small theoretical uncertainties and a large cross section 
at LHC energies ($\sigma = 72.5$ pb, Table~\ref{tab:backGG}) makes this process a perfect candidate for 
the measurement of the LHC absolute luminosity~\cite{kp}. Thanks to its distinct signature the 
selection procedure is very simple: two muons within the central detector acceptance ($|\eta|<2.5$), 
with transverse momenta above two possible thresholds ($p_T^{\mu}>3$ or $10$ \GeVc). 
As the forward protons have very low $p_T$, the muons have equal and opposite (in $\phi$) momenta.
The effective cross sections after the application of these acceptance cuts ($\sigma_\textrm{acc}$), 
with or without the requirement of at least one FP420 tag, are presented in Table~\ref{tab:lpaircuts}. 
About $800$ muon pairs should be detected in 12 hour run at the average luminosity of $10^{33}~\textrm{cm}^{-2} \textrm{s}^{-1}$.

\begin{table}[h!]
\begin{center}
\begin{tabular}{c||c|c}
      \hline  cross section [fb]     & $\sigma_{acc}$ &  $\sigma_{acc}$ (with forward proton tag) \\ \hline
        $p_T^{\mu} > 3$ \GeVc,  $|\eta^{\mu}|<2.5$  & $21~600$  & $1~340$  \\
        $p_T^{\mu} > 10$ \GeVc, $|\eta^{\mu}|<2.5$ & $7~260$  & $1~270$  \\
        \hline
\end{tabular}
\end{center}
\caption{Cross sections for $pp(\gamma \gamma \rightarrow \mu \mu)pp$ after application of typical ATLAS/CMS 
muon acceptance cuts, and coincident requirement of a forward proton~\cite{louvain}.}
\label{tab:lpaircuts}
\end{table}

An important application of these exclusive events is the absolute calibration of the very forward 
proton detectors. As the energy of the produced muons is well measured in the central detector, the forward proton 
energy can be precisely predicted using the kinematics constrains. This allows for precise calibration of 
the proton taggers, both momentum scale and resolution, in case of e.g. misalignment of the LHC beam-line elements, and leads to a 
good control of the reconstructed energy of the exchanged photon~\cite{hector}. The large cross sections 
could even allow for run-by-run calibration, as the requirement of at least one forward proton tag results 
in more than $300$ events per run. As the momenta of both forward protons are known from the central leptons, 
it is only necessary to measure one of them. This is fortunate as it allows low mass ($\sim$10 \GeVcc) forward pairs to be used, 
with rates much higher than in the FP420 double-arm acceptance. Finally, it is worth noting that the two-photon exclusive 
production of $e^+e^-$ pairs can also be studied at the LHC, though triggering of such events 
is more difficult. Electron pair reconstruction, e.g. in the CMS CASTOR forward calorimeter, has been discussed 
in~\cite{Albrow:2006xt}.


\subsubsubsection*{$W$ and $Z$ boson pairs}
\label{sec:WWZZ}

A large cross section of about 100 fb is expected for the exclusive two-photon production of $W$ boson pairs 
at the LHC. The very clean event signatures offer the possibility to study the properties of the $W$ 
gauge bosons and to make stringent tests of the Standard Model at average centre-of-mass energies of 
$\mean{W_{\gamma\gamma\rightarrow WW}}\approx$ 500~GeV. 
The cross section for events where both $W$ bosons decay into a muon and a neutrino -- resulting in 
events with two muons with large transverse momentum within the typical $|\eta|<2.5$ 
ATLAS/CMS muon acceptance range -- are large and only slightly reduced after adding the 
requirement of at least one forward proton tag (Table~\ref{tab:wwcut}).\\

\begin{table}[h!]
\begin{center}
\begin{tabular}{c||c|c}
        \hline cross section [fb]   & $\sigma_{acc}$   & $\sigma_{acc}$ (with forward proton tag) \\ \hline
        $p_T^{\mu} > 3$ \GeVc,  $|\eta^{\mu}|<2.5$   & $0.80$  & $0.76$  \\
        $p_T^{\mu} > 10$ \GeVc,  $|\eta^{\mu}|<2.5$ & $0.70$  & $0.66$  \\
        \hline
\end{tabular}
\end{center}
\caption{Cross section $\sigma_{acc}$ for $\gamma \gamma \rightarrow W^+ W^- \rightarrow \mu^+ \mu^- \bar{\nu_{\mu}}\nu_{\mu}$ after  
application of typical ATLAS/CMS muon acceptance cuts, and coincident requirement of a forward proton~\cite{louvain}.}
\label{tab:wwcut}
\end{table}

The unique signature of $WW$ pairs in the fully leptonic final state, no additional tracks on the $l^+l^-$ vertex, 
large lepton acoplanarity and large missing transverse momentum strongly reduces the backgrounds. 
The two-photon production of tau-lepton pairs, having in 
addition low cross-section at large invariant masses, can then be completely neglected. Moreover, the double 
diffractive production of the $W$ boson pairs is also negligible, and the inclusive partonic production (about 1~pb, 
assuming fully leptonic decays, and both leptons passing the acceptance cuts) can be 
very efficiently suppressed too by applying either the double tagging in the forward proton detectors, or the double 
LRG signature. Similar conclusions can be reached for the exclusive two-photon production of $Z$ boson pairs,
assuming fully leptonic, or semi-leptonic decays. In the SM, $\gamma\gamma \rightarrow ZZ$ is negligible; 
this would be a test of anomalous $\gamma ZZ$ couplings. The dominant SM source of exclusive $ZZ$ is 
$H \rightarrow ZZ$ if the Higgs boson exists, so the background in this channel is very small.\\

Two-photon production of $W$ pairs provides a unique opportunity to investigate anomalous gauge boson couplings, in particular the 
quartic gauge couplings (\textsc{qgc}s), $\gamma\gamma WW$~\cite{anom}. 
The sensitivity to the anomalous quartic vector boson couplings has been investigated~\cite{louvain} in the processes  
$\gamma\gamma\rightarrow W^+W^-\rightarrow l^+l^-\nu\bar{\nu}$ and $\gamma\gamma\rightarrow ZZ\rightarrow l^+l^- j j$
using the signature of two leptons ($e$ or $\mu$) within the acceptance cuts  $|\eta|<2.5$ and $p_T>10$~\GeVc.
The upper limits  $\lambda^{up}$ on the number of events at the 95\% confidence level have been calculated assuming 
that the number of observed events equals that of the SM prediction (corresponding to all anomalous couplings equal to zero). 
The calculated cross section upper limits can then be converted to one-parameter limits (when the other anomalous 
coupling is set to zero) on the anomalous quartic couplings. The obtained limits (Table~\ref{tab:CL}) are about 
10000 times better than the best limits established at LEP2~\cite{OPAL.limits} clearly showing the large 
and unique potential of such studies at the LHC.
A corresponding study of the anomalous triple gauge couplings can also be performed~\cite{royon}. However, in this
case the expected sensitivities are not as favourable as for the anomalous \textsc{qgc}s.


\begin{table}[h!]
\begin{center}
\begin{tabular}{c||c|c}
\hline
Coupling limits &  $\mathrm{\int\mathcal{L}dt = 1~fb^{-1}}$ & $ \mathrm{\int\mathcal{L}dt = 10~fb^{-1}}$ \\
$\mathrm{[10^{-6}~GeV^{-2}~]}$ & & \\
\hline
 $ \mathrm{|a^Z_0/\Lambda^2|} $& $ 0.49 $ & $ 0.16 $ \\
 $ \mathrm{|a^Z_C/\Lambda^2|} $& $ 1.84 $ & $ 0.58 $ \\
 $ \mathrm{|a^W_0/\Lambda^2|} $& $ 0.54 $ & $ 0.27 $ \\
 $ \mathrm{|a^W_C/\Lambda^2|} $& $ 2.02 $ & $ 0.99 $ \\
\hline
\end{tabular}
\end{center}
\caption{Expected one-parameter limits for anomalous quartic vector boson couplings at 
95$\%$ CL~\cite{louvain}.}
\label{tab:CL}
\end{table}


\subsubsubsection*{Supersymmetric pairs}
\label{sec:SusyBSM}

The interest in the two-photon exclusive production of pairs of new charged particles is three-fold: (i) it provides a new 
and very simple production mechanism for physics beyond the SM, complementary to the standard parton-parton processes; 
(ii) it can significantly constrain the masses of the new particles, using double forward-proton tagging information; 
(iii) in the case of SUSY pairs, simple final states are usually produced without cascade decays, characterised by a
fully leptonic final state composed of two charged leptons with large missing energy (and large lepton acoplanarity) 
with low backgrounds, and large high-level-trigger efficiencies.\\

The two-photon production of supersymmetric leptons or other heavy non-Standard Model leptons has been investigated in
~\cite{bib:Ohnemus,bib:Bhattacharya,bib:Drees,bib:KP}. The total 
cross-section at the LHC for the process $\gamma \gamma \rightarrow \tilde{l}^{+} \tilde{l}^{-}$ 
can be as large as $\sim 20$~fb ($\mathcal{O}(1 fb)$ for the elastic case alone), while still being consistent 
with the model-dependent direct search limits from LEP~\cite{bib:Delphi,bib:Opal}. 
While sleptons are also produced in other processes (Drell-Yan or squark/gluino decays), $\gamma \gamma$ production 
has the advantage of being a direct QED process with minimal theoretical uncertainties.


\begin{figure}[hhh]
\begin{center}
\epsfig{file=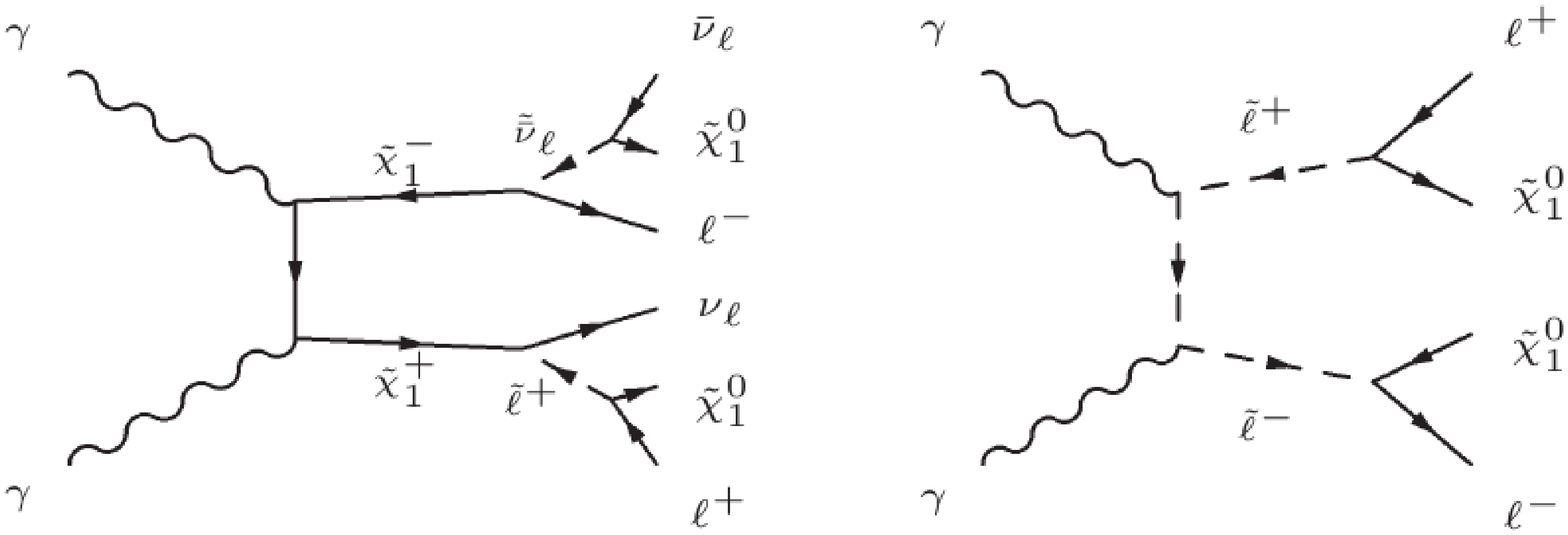,width=0.8\linewidth}
\caption{Relevant Feynman diagrams for SUSY pair production with leptons in
 the final state: chargino disintegration in a charged/neutral scalar and a
 neutral/charged fermion (left); slepton disintegration (right)~\cite{louvain}.}
\label{fig:diag.susy}
\end{center}
\end{figure}

In~\cite{louvain}, three benchmark points in mSUGRA/CMSSM parameter space constrained 
by the post-\textsc{wmap} research~\cite{wmap} have been chosen:
\begin{itemize}                                               
\item LM1: very light LSP, light $\tilde{\ell}$, light $\tilde{\chi}$ and $\tb$=10;
\item LM2: medium LSP, heavy $\tilde{\ell}$, heavy $\tilde{\chi}$ and $\tb$=35;    
\item LM6: heaviest LSP, light right $\tilde{\ell}$, heavy left $\tilde{\ell}$, heavy $\tilde{\chi}$ and $\tb$=10.
\end{itemize}
The masses of the corresponding supersymmetric particles are listed in Table~\ref{tab:mass.susy}.\\
\begin{table}[htpb]                                                                                                   
\begin{center}
\begin{tabular}[htpb]{c||c c c}                                                                                             
\hline
m [GeV/c$^2$] & LM1 & LM2 & LM6 \\                                                                            
\hline
 $\tilde{\chi}_1^0$ & 97 & 141 & 162 \\  
 $\tilde{\ell}_R^+$ & 118 & 229 & 175 \\
 $\tilde{\ell}_L^+$ & 184 & 301 & 283 \\                                     
 $\tilde{\tau}_1^+$ & 109 & 156 & 168 \\
 $\tilde{\tau}_2^+$ & 189 & 313 & 286 \\                                      
 $\tilde{\chi}_1^+$ & 180 & 265 & 303 \\
 $\tilde{\chi}_2^+$ & 369 & 474 & 541 \\
 $H^+$              & 386 & 431 & 589 \\
\hline
\end{tabular}
\caption{\small{Masses of \textsc{susy} particles, in GeV/c$^2$, for different benchmarks (here $\ell=e,\mu$)}}
\label{tab:mass.susy}
\end{center} 
\end{table}

The study concentrates on the fully leptonic SUSY case only. The corresponding Feynman diagrams 
are shown in Figure~\ref{fig:diag.susy}.
Signal and background samples coming from SUSY and SM pairs were produced using a modified
version of {\sc calchep}~\cite{Puk03}. The following acceptance cuts have been applied: two leptons with 
$p_T >$ 3 \GeVc\ or 10 \GeVc\ and $\vert \eta \vert < 2.5$. The only irreducible background for this type of processes 
is the exclusive $W$ pair production since direct lepton pairs $pp(\gamma \gamma \to \ell^+ \ell^-)pp$ 
can be suppressed by applying large acoplanarity cuts. 
Standard high-level-trigger (HLT) efficiencies are high for all these types of events. In typical 
mSUGRA/CMSSM scenarios, a light right-handed slepton will have a branching fraction of 
$B(\tilde{l}^{\pm} \rightarrow \chi_{1}^{0} l^{\pm}) = 100\%$. This results in a final state with two same-flavor 
opposite-sign leptons, missing energy, and two off-energy forward protons. Assuming a trigger threshold of 
$7$ \GeVc~for two isolated muons, the efficiency would be $71 - 74\%$ for smuons
in the range of typical light mSUGRA/CMSSM benchmark points (LM1 or SPS1a). With an integrated luminosity of $100$~fb$^{-1}$, this
would result in a sample of $15 - 30$ triggered elastic-elastic smuon pairs, plus a slightly smaller number of selectron
pairs. Including the less clean singly-elastic events would increase these yields by roughly a factor of 5.
The irreducible $\gamma\gamma\to WW$ background can be suppressed by a factor of two by selecting only same lepton-flavour 
($ee$, $\mu\mu$) final states. 
The measured energy of the two scattered protons in forward proton taggers could allow for the distinction between various contributions 
to the signal by looking at the distribution of the photon-photon invariant mass $W_{\gamma \gamma}$. 
HECTOR~\cite{hector} simulations of forward protons from slepton events consistent with LM1 benchmark point indicate that
the TOTEM 220 m detectors will have both protons tagged for only $30\%$ of events.  Addition of detectors at 420 m increases
that to $90\%$ of events. 

\begin{figure}[htbp]
\begin{center}
\begin{tabular}{c}
\epsfig{file=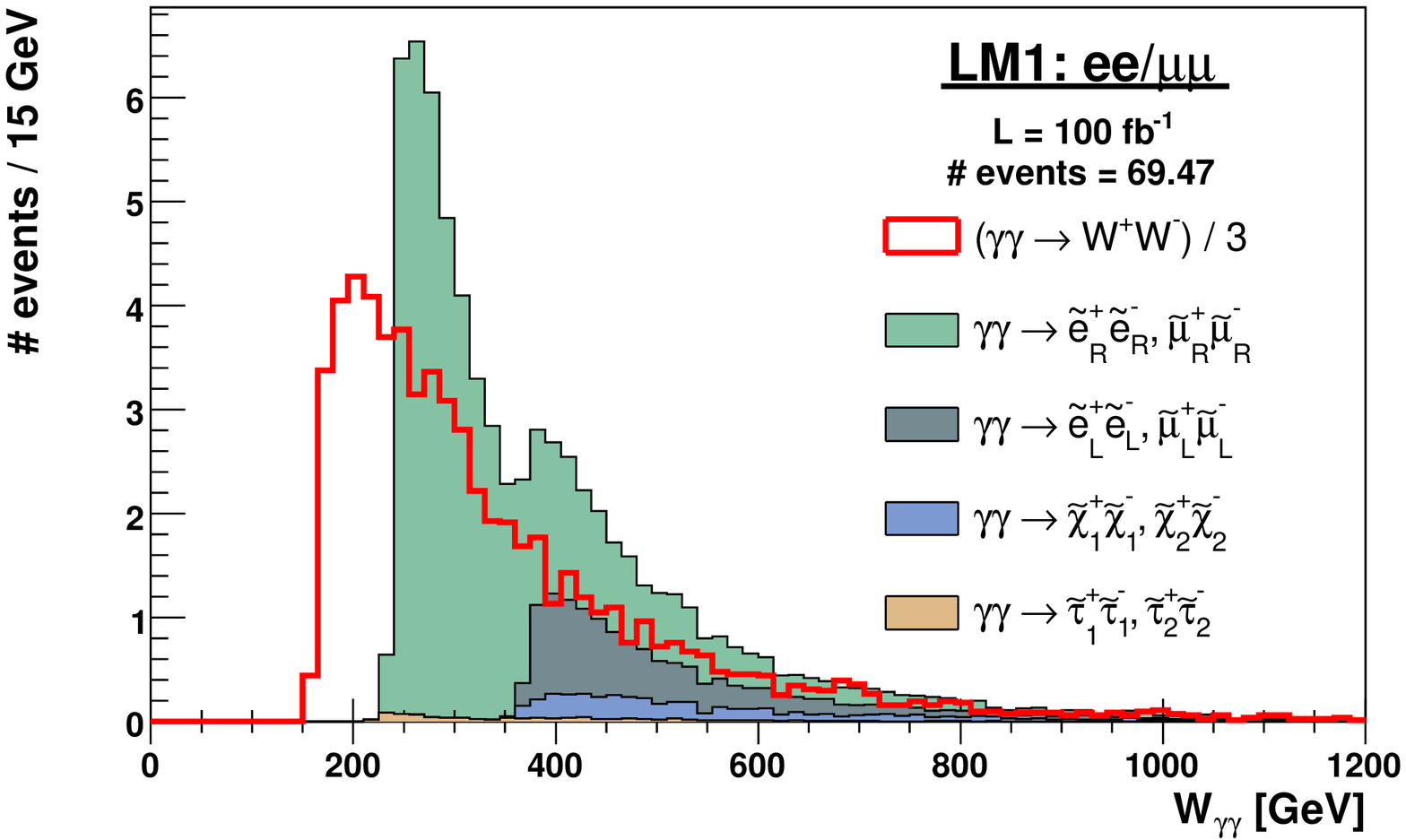, width=0.49\linewidth}
\epsfig{file=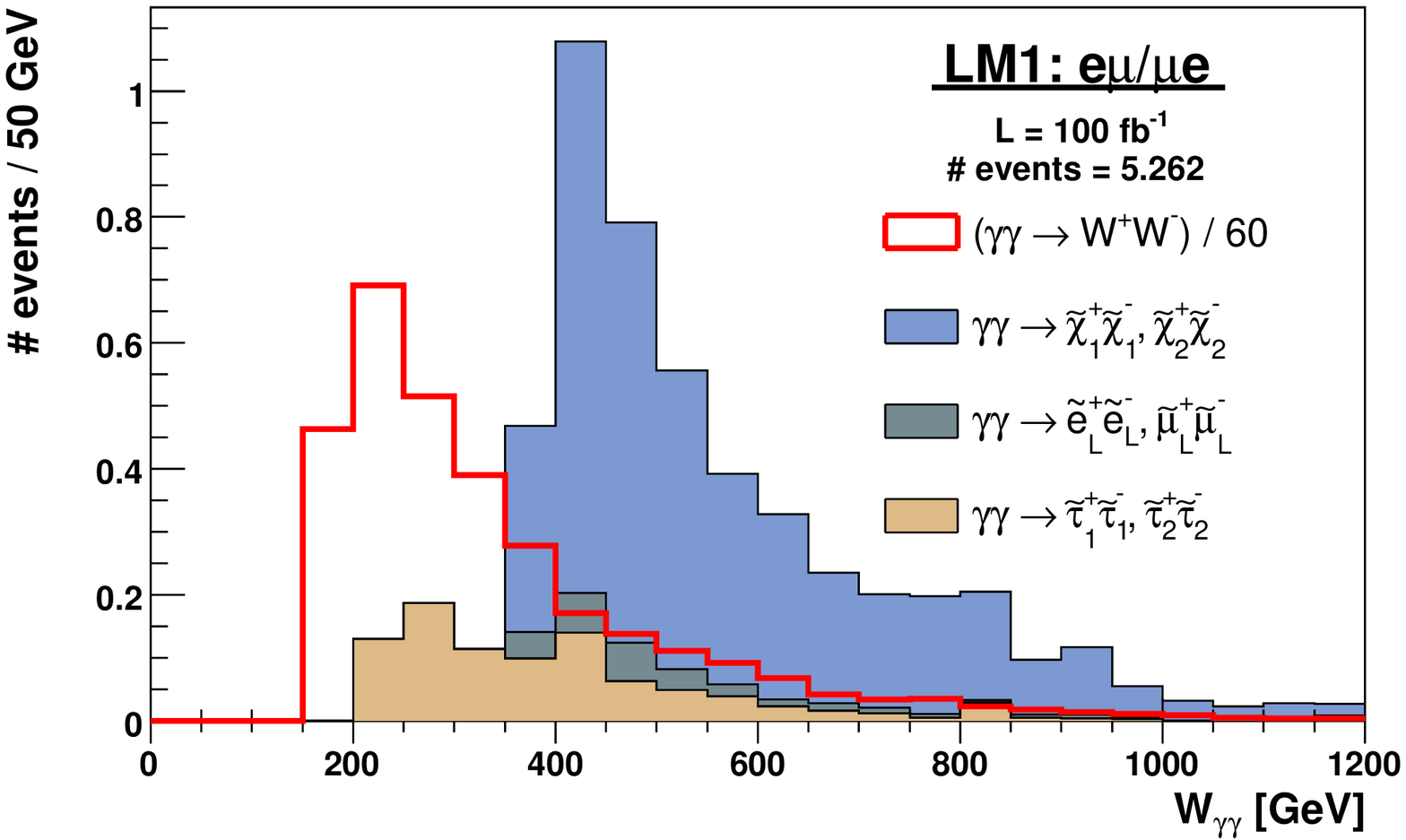, width=0.49\linewidth} 
\end{tabular}
\end{center}
\caption{Photon-photon invariant mass for benchmark point LM1 with $\int\mathcal{L}dt$ = 100 fb$^{-1}$.
Cumulative distributions for signal with two detected leptons ($p_T > 3$ \GeVc, $|\eta| <$ 2.5), 
two detected protons, with same (left) or different flavour (right). The $WW$ background 
has been down-scaled by the quoted factor~\cite{louvain}.}
\label{fig:gamma-gamma-inv-mass}
\end{figure}

The expected cumulative $W_{\gamma \gamma}$ distributions for LM1 events with two centrally measured 
leptons and two forward detected protons are illustrated in Figure~\ref{fig:gamma-gamma-inv-mass}.
With this technique and sufficient statistics, masses of supersymmetric particles could be measured with precision of a 
few \GeVcc~by looking at the minimal centre-of-mass energy required to produce a pair of SUSY particles. In the same way, 
missing energy can be computed by subtracting the detected lepton energies from the measured two-photon centre-of-mass energy. 
For backgrounds missing energy distributions start at zero missing energy, while in SUSY cases they start only at 
two times the mass of the LSP.


\subsubsection{Photon-proton processes}
\label{sec:GP}

\begin{figure}[htbp]
\begin{center}
 \epsfig{file=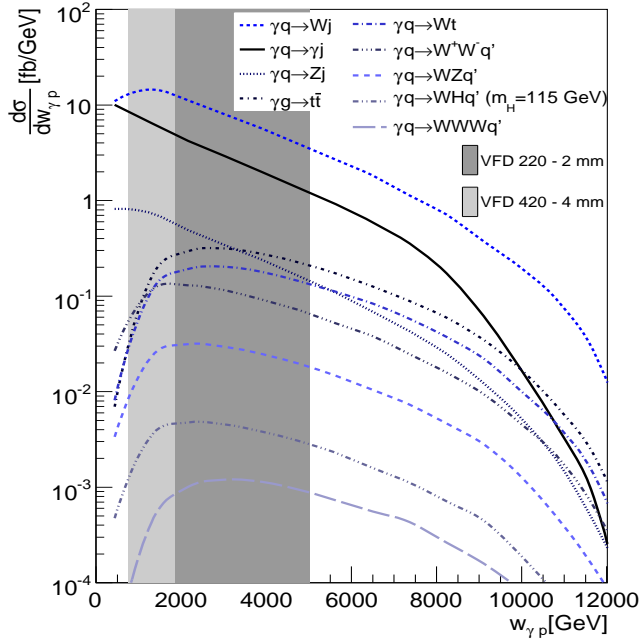,width=8.5cm,height=9.cm}
 \caption{Differential cross-sections for $pp (\gamma q/g \rightarrow X) p Y$ processes as 
a function of the c.m.s. energy in photon-proton collisions, $W_{\gamma p}$. The acceptance of roman pots 
(220~m at 2~mm from the beam axis and 420~m at 4~mm from the beam axis) is also sketched~\cite{louvain}.}
\label{fig:Int}
\end{center}
\end{figure}

The high luminosity and the high centre-of-mass energies of photo-production processes at 
the LHC offer very interesting possibilities for the study of electroweak interaction and 
for searches Beyond the Standard Model (BSM) up to the TeV scale~\cite{louvain}. 
Differential cross sections for $pp (\gamma q/g \rightarrow X) p Y$ reactions, as a function of the photon-proton 
centre-of-mass energy, are presented in Figure~\ref{fig:Int} together with the acceptance region of forward proton taggers.
A large variety of processes have sizeable cross section up to the electroweak scale and could therefore be studied 
during the very low and low luminosity phases of the LHC.
Interestingly, potential Standard Model background processes with hard leptons, missing energy and jets 
coming from the production of gauge bosons, have cross sections only one or two orders of magnitude higher than 
those involving top quarks. 
The large top quark photo-production cross sections, $\mathcal{O}$(pb), are particularly interesting for measuring 
top quark related SM parameters, such as the top quark mass and its electric charge. In addition, and in contrast 
to parton-parton top production, photo-production of top quark pairs and of single top in association 
with a $W$ boson have similar cross sections. This will certainly be advantageous in analyses aiming at measuring
the Cabibbo-Kobayashi-Maskawa (\textsc{ckm}) matrix element $|V_{tb}|$ in associated $Wt$ production. 



In order to illustrate the discovery potential of photon-proton interactions at the LHC, we discuss in the
next two subsections the possibility to observe: (i) the SM Higgs boson produced in association with a 
$W$ ($\sigma_{WH}\approx$ 20 fb for $M_H$ = 115 \GeVcc, representing more than $2\%$ 
of the total inclusive $WH$ production at the LHC), (ii)  the anomalous production 
of single top, which could reveal BSM phenomena via Flavour Changing Neutral Currents (FCNC).\\


\subsubsubsection*{Associated $WH$ production}
\label{sec.WH}

\begin{figure}[htbp]
\begin{center}
\epsfig{file=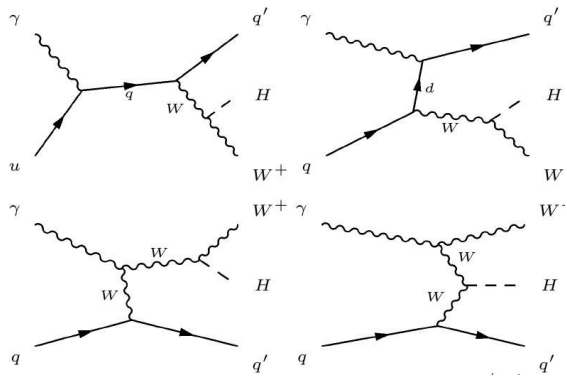,width=0.5\linewidth}
\caption{The Feynman diagrams for $\gamma q\rightarrow H W^{+} q'$ 
associated production at \textsc{lo}~\cite{louvain}.}
\label{prod}
\end{center} 
\end{figure}

The search for $WH$ associate production at the LHC will be challenging due to the 
large $W$+jets, $t\overline{t}$ and $WZ$ cross sections. Indeed, although Standard Model 
cross sections for the process $pp \rightarrow WH\,X$ range from 1.5 pb to 425 fb for Higgs 
boson masses of 115 \GeVcc~and 170 \GeVcc~respectively, this reaction is generally not considered 
as a Higgs discovery channel. This production mechanism however, is sensitive to 
$WWH$ coupling which might be enhanced when considering fermiophobic models, 
and might also give valuable information on the $Hb\overline{b}$ coupling, which
is particularly difficult to determine at the LHC.
The possibility of using $\gamma p$ collisions to search for $WH$ associate production was 
already considered at electron-proton colliders \cite{WH}. At the LHC the cross section 
for $pp(\gamma q \rightarrow WHq')pY$ reaction reaches 23 (17.5) fb for a Higgs boson 
mass of 115 \GeVcc~(170 \GeVcc). The dominant Feynman diagrams are shown in Figure~\ref{prod}. 
Although cross sections are smaller than the ones initiated by quarks, 
the signal-to-background ratio is improved by more than one order of magnitude~\cite{louvain}.\\

\subsubsubsection*{Anomalous top production}
\label{sec:anomtop}

\begin{figure}[htbp]
\begin{center}
\epsfig{file=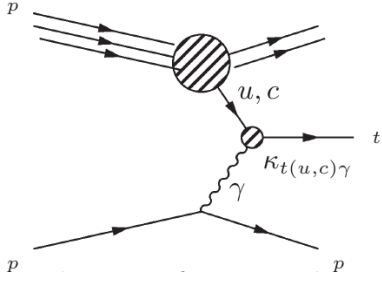,width=0.35\linewidth}
\caption{Photo-production of top quarks at LHC through FCNC~\cite{louvain}.}
\label{fig:anotop_diag}
\end{center}
\end{figure}

In the Standard Model, exclusive single top photo-production at LHC
energies is only possible for higher order electroweak interactions, since neutral 
currents preserve quarks flavour at tree level. The observation of a large number 
of single top events would hence be a sign of FCNC induced by processes 
beyond the Standard Model. FCNC appear in many extensions of the Standard Model, 
such as two Higgs-doublet models or R-parity violating supersymmetry. The dominant 
Feynman diagram contributing to photo-production of top quarks via FCNC, 
can be seen in Fig.~\ref{fig:anotop_diag}. The effective Lagrangian for this anomalous 
coupling can be written as \cite{eff_lag_anotop} : 
$$ L = iee_t\bar{t}\frac{\sigma_{\mu\nu}q^{\nu}}{\Lambda}k_{tu\gamma}uA^{\mu} + iee_t\bar{t}\frac{\sigma_{\mu\nu}q^{\nu}}{\Lambda}k_{tc\gamma}cA^{\mu} + h.c., $$
where $\sigma_{\mu\nu}$ is defined as $(\gamma^{\mu} \gamma^{\nu} - \gamma^{\nu} \gamma^{\mu})/2$, $q^{\nu}$ 
being the photon 4-vector and $\Lambda$ an arbitrary scale, conventionally taken as the top mass. 
The couplings $k_{tu\gamma}$ and $k_{tc\gamma}$ are real and positive such that the cross section 
takes the form : $$\sigma_{pp \rightarrow t} = \alpha_u\ k^2_{tu\gamma} + \alpha_c\ k^2_{tc\gamma}. $$
The computed $\alpha$ parameters using {\sc calchep} are $ \alpha_u = 368\ \textrm{pb} ,\alpha_c = 122\ \textrm{pb}.$
The best limit on $k_{tu\gamma}$ is around 0.14, depending on the top mass \cite{zeus_st} 
while the anomalous coupling $k_{tc\gamma}$ has not been probed yet.\\

The single top final state is composed of a $b$-jet and a $W$ boson. The main irreducible backgrounds 
for the considered topology, $\ell E_T^{miss} b$, come from $\gamma p$ interactions producing a 
$W$ boson and a jet, especially $c$-jets which can be miss-tagged as a $b$-jets. Limits on the 
anomalous couplings $k_{tu\gamma}$ and $k_{tc\gamma}$ have been extracted after application
of acceptance cuts in~\cite{louvain}. These results appear on Table~\ref{tab:cl.anotop} for two integrated
luminosities.

\renewcommand{\arraystretch}{1.1}
\begin{table}[h!]
\centering
\begin{tabular}{c||c|c}
\hline
Coupling limits & $\int\mathcal{L}dt$ = 1 fb$^{-1}$	& $\int\mathcal{L}dt$ = 10 fb$^{-1}$ 	\\
\hline
$k_{tu\gamma}$	& 0.043 & 0.024		\\
$k_{tc\gamma}$	& 0.074 & 0.042  		\\
\hline
\end{tabular}
\label{tab:cl.anotop}
\caption{Expected limits for anomalous couplings at 95$\%$ CL~\cite{louvain}.}
\end{table}


\subsubsection{Photon-photon and photon-proton physics summary}
\label{sec:Summ}

A summary of various unique photon-photon and photon-proton interactions accessible 
to measurement at the LHC, and discussed in detail in~\cite{louvain}, has been 
presented in this section. Interesting studies and searches can be performed for initial integrated luminosities 
of about 1~fb$^{-1}$, such as exclusive dimuon production in two-photon
collisions tagged with forward large rapidity gaps. At higher luminosities, the efficient selection 
of photon-induced processes is greatly enhanced with dedicated forward proton taggers such as
FP420. Photon induced reactions can provide much higher sensitivity than partonic reactions 
for various BSM signals such as e.g. anomalous quartic $\gamma\gamma WW$ gauge 
couplings. The associated photoproduction of a top quark or a $W$ boson is also very large, 
offering a unique opportunity to measure the fundamental Standard Model parameters, 
such as the top quark charge or the $V_{tb}$ element of the quark mixing matrix. 
Anomalous $\gamma qt$ couplings might also be uniquely revealed in single top photoproduction.
Larger integrated luminosity, of about hundred inverse femtobarns, will open complementary 
ways to search for production of supersymmetric particles in photon-photon interactions. 
Even larger luminosities might help to access important information on the Higgs 
boson coupling to $b$ quarks and $W$ bosons.
FP420 detectors are mandatory for the determination of the masses of the centrally produced particles, 
and to increase the sensitivity to new anomalous couplings contributions in two-photon interactions.

Last but not least, studying the photon-induced processes in the early LHC runs can provide valuable checks 
of the various components of the general formalism used to predict the cross sections of central 
exclusive reactions~\cite{Khoze:2008cx}. Thus, the photon-exchange dominated $W$-boson production 
with rapidity gaps on either side provides information on the gap survival factor $S^2$. As discussed in~\cite{Khoze:2008cx}, 
such studies can be performed even without tagging of the forward proton. Another example is 
exclusive $\Upsilon$ photoproduction induced by the process $ \gamma p \to \Upsilon p$~\cite{CMS_ExclUps}, now observed by CDF~\cite{tev2}. 
The study of such processes will not only reduce the theoretical uncertainties associated with the generalised, 
unintegrated gluon distributions $f_g$, e.g., by testing models based on diffusion in transverse momentum as 
incarnated in the Balitsky-Fadin-Kuraev-Lipatov (BFKL) equation~\cite{BFKL},
but will be of help to calibrate and align the forward proton detectors.


\nopagebreak

\nopagebreak

\subsection{Diffractive physics}
\label{sec:diffraction}

\begin{figure}[htb]
\begin{center}
\includegraphics[angle=270,width=0.6\textwidth]{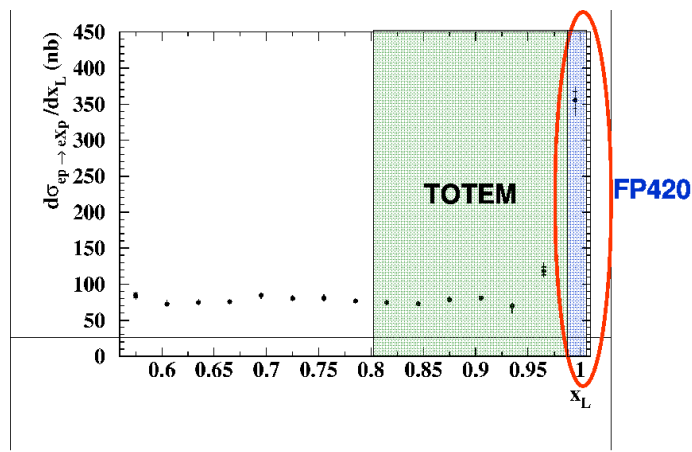}
\caption{\label{fp420-totem}
$x_L=1-\xi$ coverage of FP420 and TOTEM (or any near-beam detectors at 
220~m from the interaction point). The data points are for the reaction
$ep \to eXp$~\protect\cite{Chekanov:2002yh} and are only meant to 
illustrate the position of the diffractive peak at $x_L \approx 1$.}  
\end{center} \end{figure}

Proton tagging with FP420 will allow a continuation of the study of hard
diffraction, expanding and extending the investigations carried out at
CERN by UA8~\cite{Brandt:1992zu}, and more recently at HERA by H1 and ZEUS
and at Fermilab by CDF and D0 (see,
e.g.,~\cite{BaronePredazzi,Alekhin:2005dy,Albrow:2006xt,Arneodo:2005kd} and
references therein). The coverage of FP420, $0.002<\xi<0.02$, is centred
on the diffractive-peak region where the contribution from mesonic
exchanges (Reggeons) is negligible, and is thus complementary to that of
TOTEM (or of any near-beam detectors at 220~m from the interaction point),
which is $0.02<\xi<0.2$ with high-luminosity LHC optics (see
Fig.~\ref{fp420-totem}).

The following reactions can be studied:
\begin{enumerate}
\item at instantaneous luminosities where pile-up is negligible, single
diffractive (SD) dissociation of the proton, $pp \to Xp$, where one proton
is measured in FP420 and the other dissociates into a state $X$ which
contains high-$E_T$ jets, vector bosons or heavy flavours: the limitation
to low luminosities is due to the fact that the timing constraint cannot
be applied when only one proton is measured;
\item at all luminosities, double Pomeron exchange (DPE), $pp \to pXp$, 
where both protons are tagged by FP420, and again $X$ includes high-$E_T$ 
jets, vector bosons or heavy flavours;
\item also at all luminosities, central exclusive production of di-jets,
$pp \to p jjp$.
\end{enumerate}

Processes 1 and 2 are sensitive to the low-$x$ structure of the proton
and the diffractive parton distribution functions (dPDFs), which can be
interpreted as conditional probabilities to find a parton in the proton
when the final state of the process contains a fast proton of given
four-momentum. Process 3  is sensitive to the generalised (skewed) parton
distribution functions (GPD), which are crucial for the estimate of the
cross section for central-exclusive Higgs production.

Inclusive jet and heavy quark production are mainly sensitive to the gluon
component of the dPDFs, while vector boson production is sensitive to
quarks. The kinematic region covered expands that explored at HERA and
Tevatron, with values of $\beta$ (the fractional momentum of the struck 
parton in the diffractive exchange) as low as $10^{-4}$ and of $Q^2$  
up to tens of thousands of GeV$^2$. 

The extraction of the dPDFs and the GPDs is complicated by the breakdown
of QCD diffractive factorisation in hadron-hadron collisions: to determine the
dPDFs and GPDs, it is necessary to establish by how much diffractive
interactions are suppressed because of soft interactions of the
spectator partons from the interacting
hadrons~\cite{Bjorken:1992er,Kaidalov01}. This is quantified by the
so-called rapidity-gap survival probability, a critical ingredient for the
calculation of the cross section for central-exclusive Higgs production.
The rapidity-gap survival probability is interesting in its own right
because of its relationship with multiple scattering effects and hence the
structure of the underlying event in hard collisions. All three processes
listed above can be used to determine the rapidity-gap survival
probability. For example, as a consequence of the factorisation breakdown,
the diffractive structure function extracted from SD jet production
will differ from that obtained from DPE jet production. The ratio of these
two structure functions is sensitive to the rapidity-gap survival
probability. A rather unique additional possibility which arises with
FP420 is to observe events with three (or more) large rapidity gaps; two
gaps fixed by the forward protons and the third gap selected in the
central detector.  This may help shed further light on the dynamics behind
the rapidity-gap survival probability.

Also of interest is the fact that good data on single diffractive
dissociation at high energies could prove very important for a
better understanding of the nature of ultra high-energy cosmic ray
interactions, see e.g. Chapter 10 of ref.~\cite{Albrow:2006xt}.

Finally, it is natural to expect that the secondaries produced by an
`incoming' pomeron ($\pom$) will be enriched with glueballs ($G$). With
tagged protons, one could look for the quasi-elastic diffractive $\pom p
\to G X$ process. Similarly, tagging both protons allows one to observe
$\pom-\pom$ interactions at much larger energies, $\sqrt{s_{\pom \pom}}
\sim 100-200$~GeV, than have been explored so far.

Cross sections for hard-diffractive processes can be large, as shown in
Table~\ref{tab:hard-diffr-crossections}. In the following, we summarise 
some of the studies that have been performed.

\begin{table}[htbp]
\begin{center}
\begin{tabular}{|l|c|}
\hline
\multicolumn{1}{|c|}{Process} & Cross section \\ \hline
$pp \to Xp$, with $X$ including a $W$ boson & 70~pb \\ \hline
$pp \to Xp$, with $X$ including a di-jet ($E_T>50$~GeV) & 30~nb \\ \hline
$pp \to pXp$, with $X$ including a di-jet ($E_T>50$~GeV) & 1.5~nb \\ 
\hline
\end{tabular}
\caption{Cross sections for a few hard-diffractive processes, as obtained 
with the POMWIG generator~\protect\cite{Cox:2000jt}.\label{tab:hard-diffr-crossections}}
\end{center}
\end{table}

\subsubsection{Single-diffractive production of $W$, $Z$ bosons or di-jets}

\begin{figure}[bth]
\begin{center}
\begin{tabular}{cc}
\includegraphics[width=0.48\textwidth]{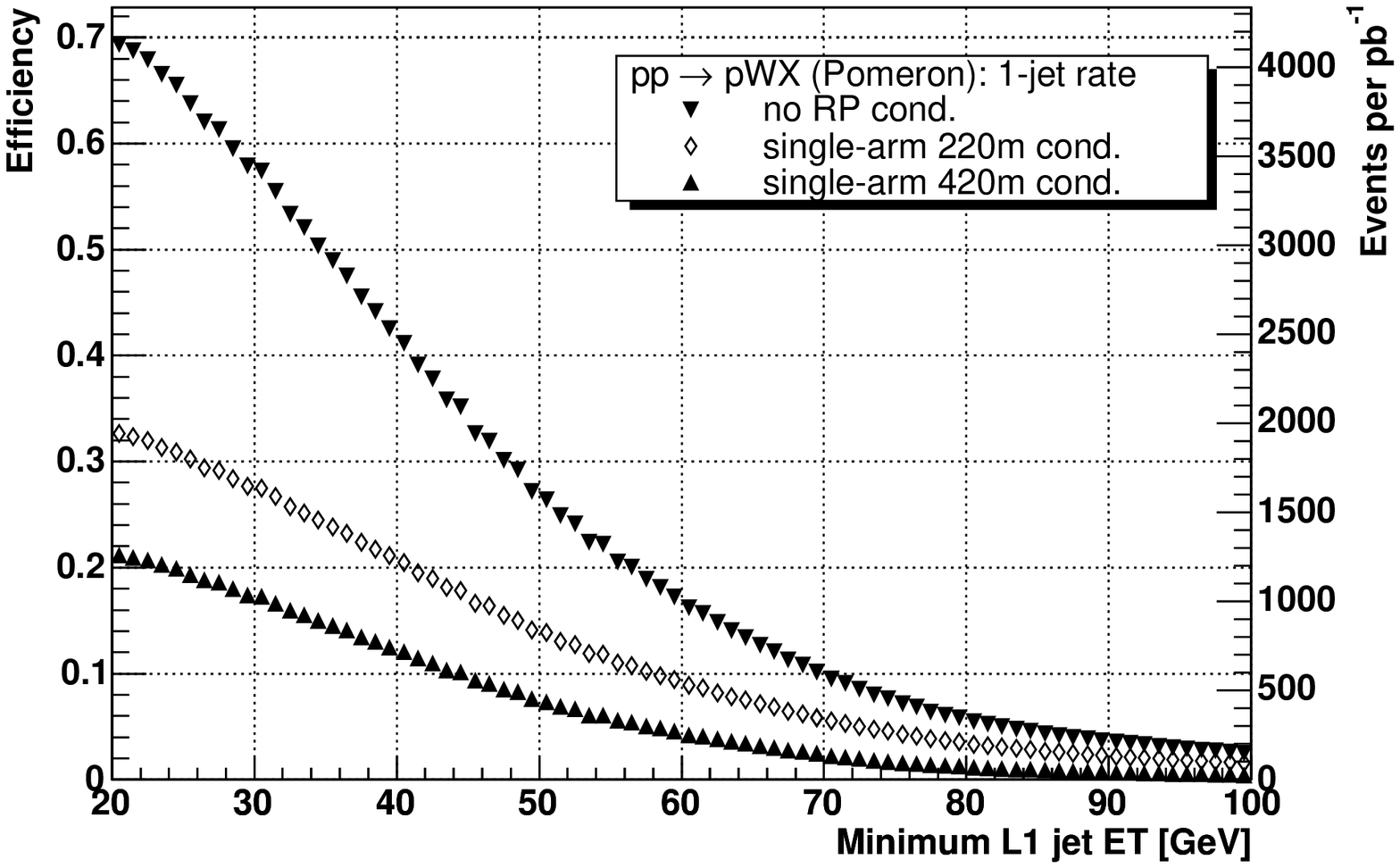} &
\includegraphics[width=0.48\textwidth]{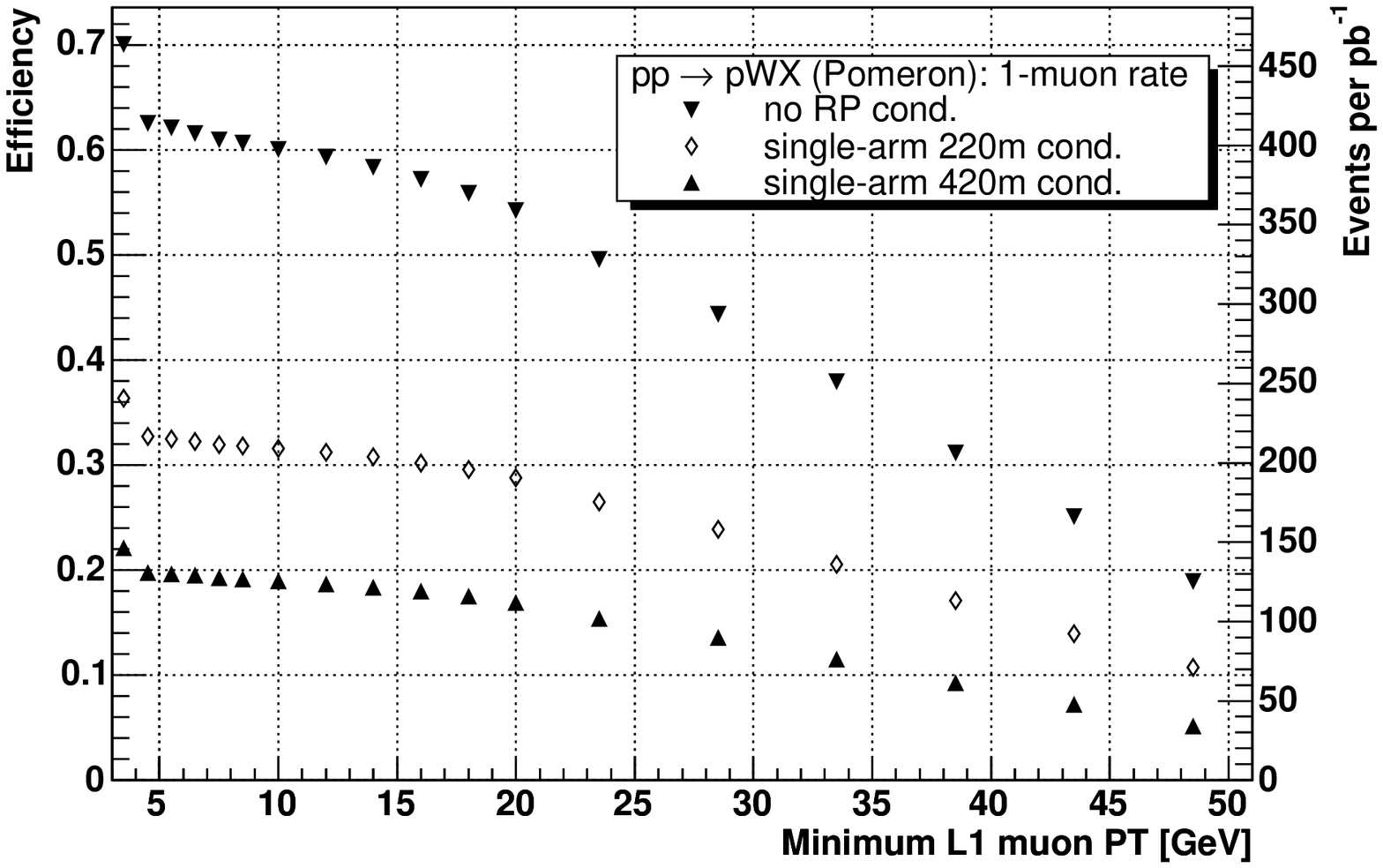}\\
\includegraphics[width=0.48\textwidth]{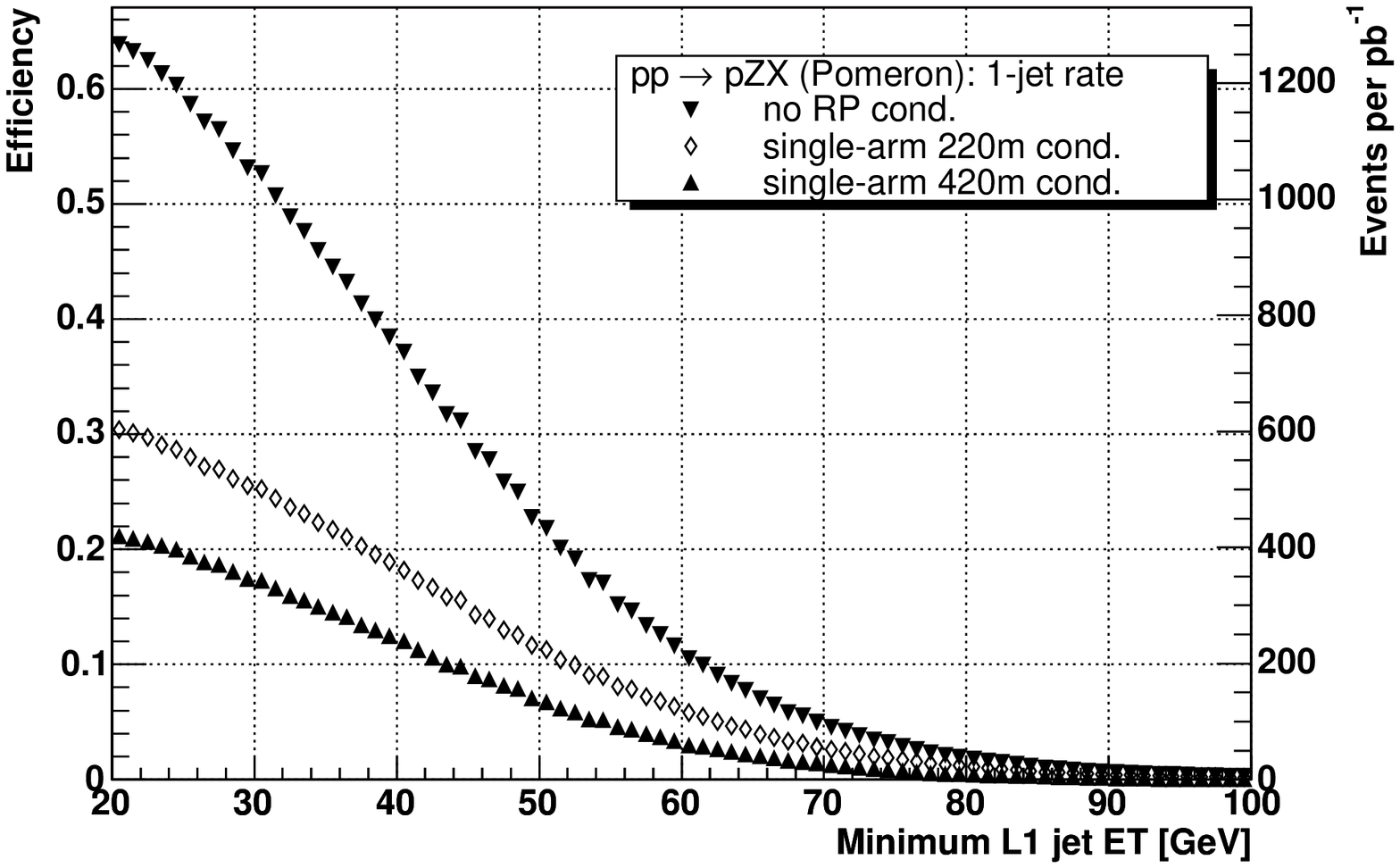} &
\includegraphics[width=0.48\textwidth]{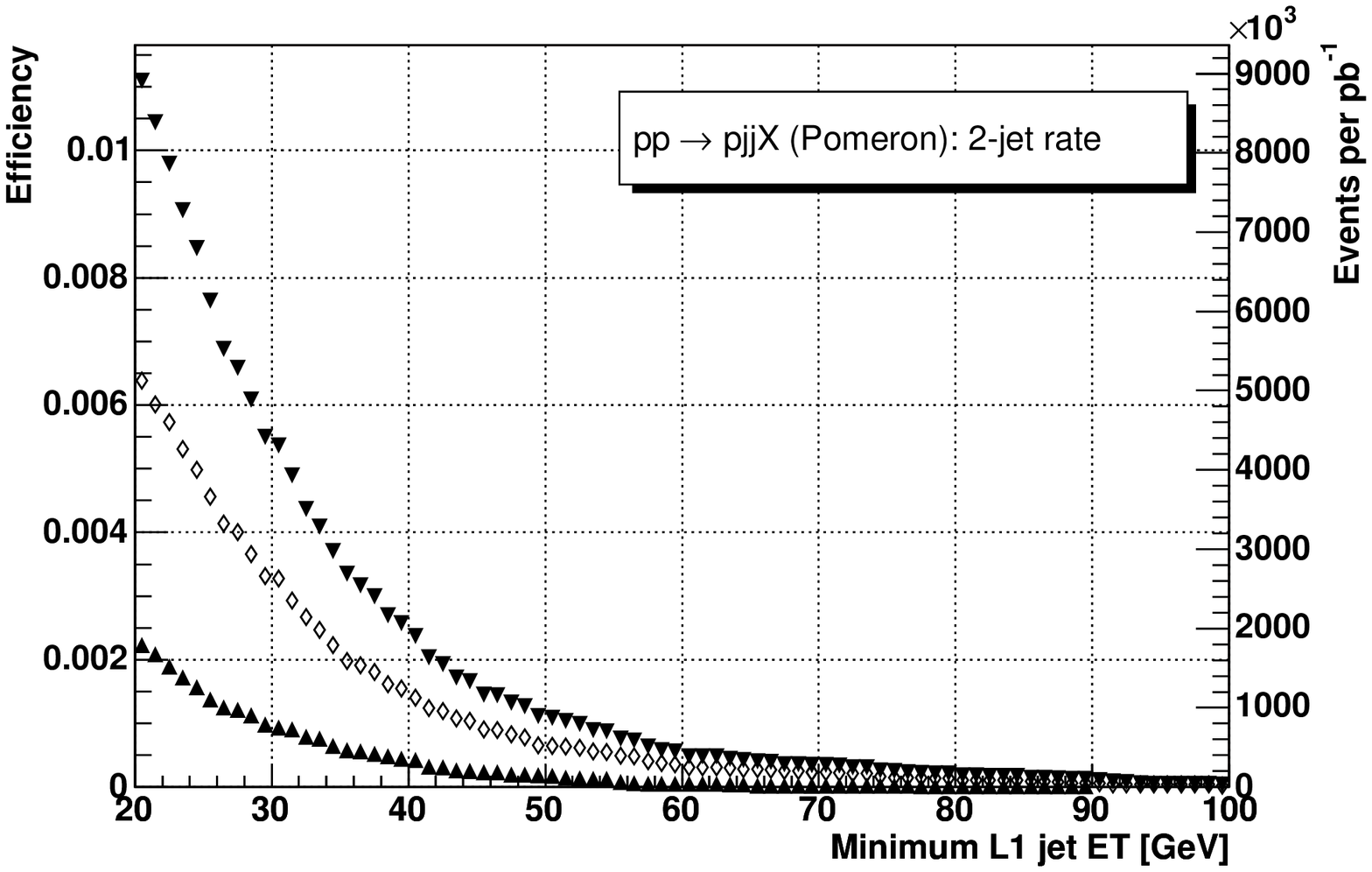}
\end{tabular}
\caption{Selection efficiency as function of the threshold value for $pp
\to p W X$ (upper left and upper right), $pp \to p Z X$ (lower left), $pp
\to p jj X$ (lower right). At least one L1 jet with $E_T$ above threshold
is required (upper and lower left), at least two L1 jets with $E_T$ above
threshold are required (lower right), at least one L1 muon with $p_T$
above threshold is required (upper right). The normalization of the
efficiency curves (left y-axis) is explained in the text. The number of
events expected to pass the L1 selection per pb$^{-1}$ of LHC data (right
y-axis) does not take into account the gap survival probability which at
the LHC is expected to be $\cal O$(0.1). All plots are for the non-pile-up
case. From~\protect\cite{Albrow:2006xt}. \label{fig:SD}}
\end{center}
\end{figure}

Selection efficiencies were studied in~\cite{Albrow:2006xt,PTDR2} for $pp
\to pX$, with $X$ containing a $W$ or a $Z$ boson that decays to jets or
to muons, as well as with $X$ containing a di-jet system. Samples of
100,000 signal events each were generated with the POMWIG Monte Carlo
generator~\cite{Cox:2000jt} (version 1.3). For these studies, the CMS
detector response was simulated using the OSCAR~\cite{oscar} package. The
digitisation (simulation of the electronic response), the emulation of the
Level-1 and High-Level Triggers (HLT), and the offline reconstruction of
physics objects were performed with the CMS full-reconstruction ORCA
package~\cite{orca}. For four example processes, Fig.~\ref{fig:SD} shows
the efficiency as a function of the L1 threshold value, normalised to the
number of events (in the muon rate case to the number of events with a
muon in the final state) with $0.001 < \xi < 0.2$. 
Three different trigger
conditions are considered: 
(i) only central detector information,
(ii) central detector information in conjunction with a single arm track at 220~m and 
(iii) central detector information in  conjunction with a single arm track at 420~m.
Also shown is the number of events expected to pass the L1 selection per
pb$^{-1}$ of LHC running. In~\cite{Albrow:2006xt,PTDR2}, a gap survival
probability of unity was assumed.  However, at the LHC this factor is
expected to be $\cal O$(0.1)~\cite{Khoze:2006gg}.

\subsubsection{Single-diffractive and double-Pomeron exchange production
of $B$ mesons}

Inclusive SD and DPE production of $B$ mesons, with $B \to J/\psi X$ and
$J/\psi \to \mu^+ \mu^-$, was studied in~\cite{Albrow:2006xt} using the
generator DPEMC 2.4~\cite{Boonekamp:2003ie} in conjunction with the fast
CMS simulation code FAMOS, version 1.3.1~\cite{PTDR1_FAMOS}. As discussed
earlier, this process is sensitive to the dPDFs of the proton. Events were
selected which had at least one pair of oppositely charged muons. If two
pairs were found, the one with invariant mass closer to that of the
$J/\psi$ meson was taken to be the one originating from the $J/\psi$
decay. Events were selected if $2.7 < M_{\mu \mu}< 3.5$~\GeVcc, with
$M_{\mu\mu}$ the invariant mass of the muon pair, $p_T^{\mu} > 3$~\GeVc\ (at
L1) and $p_T^{\mu} > 7$~\GeVc\ (HLT).  In addition, the detection of a proton
on either side of the interaction point was required for the SD events and
on both sides for the DPE events. The estimated event yield, after the
cuts, for an integrated luminosity of 1~fb$^{-1}$ is of hundreds of SD
events and a few DPE events. 



\subsubsection{Double-Pomeron exchange production of $W$ bosons}

\begin{figure} [htb]
\begin{tabular}{cc}
\includegraphics[width=0.48\textwidth]{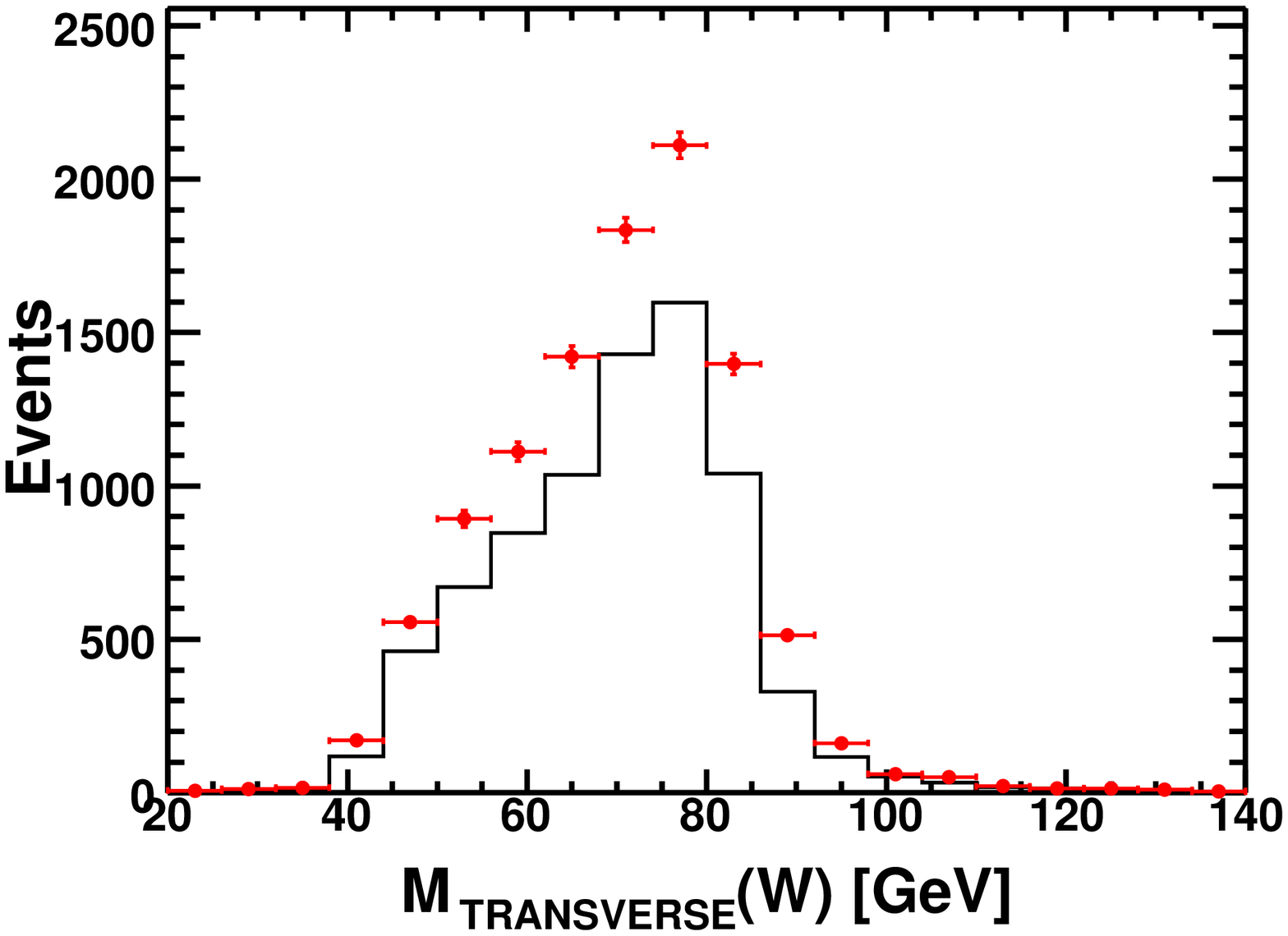} &
\includegraphics[width=0.48\textwidth]{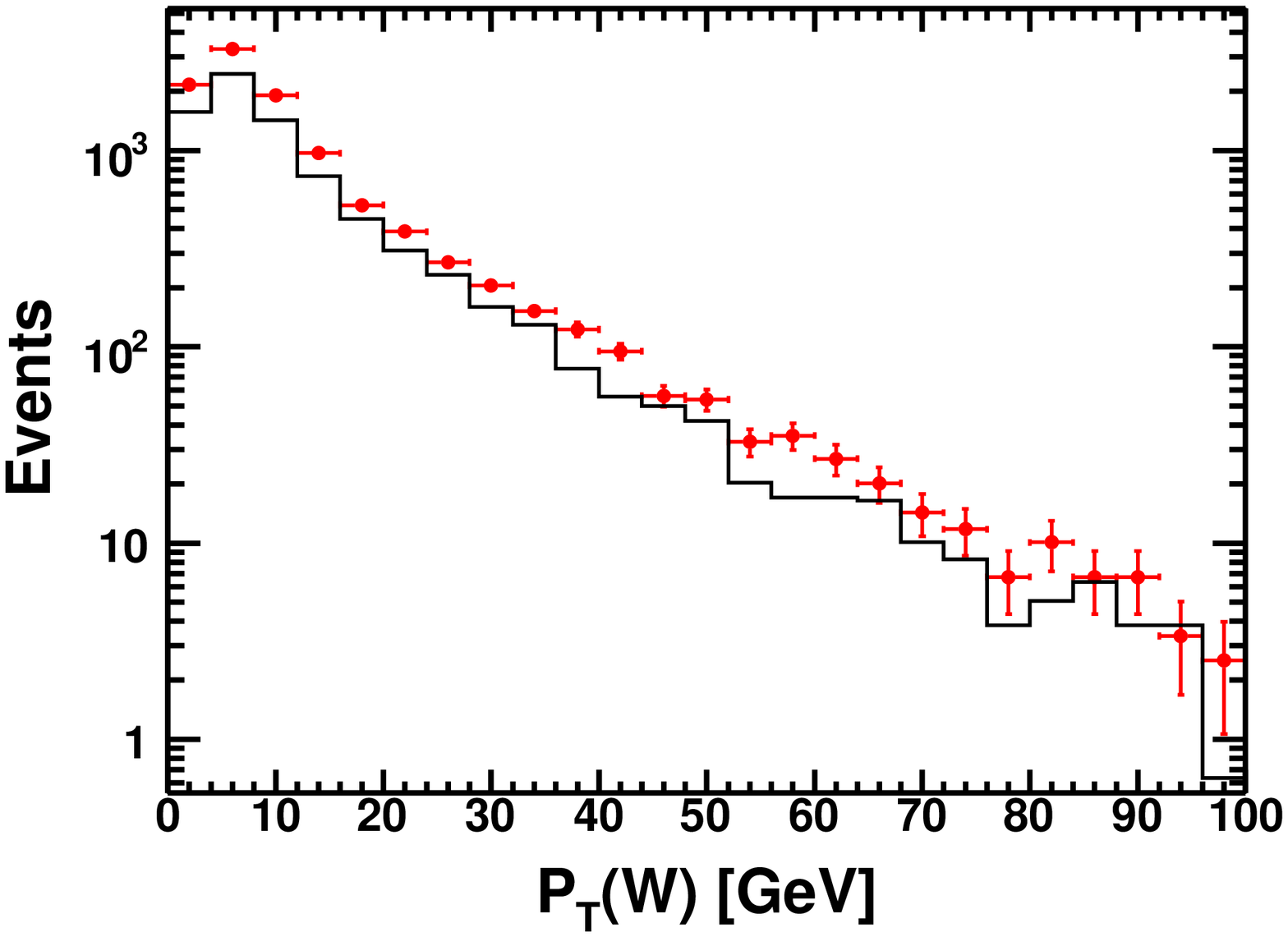} \\
\includegraphics[width=0.48\textwidth]{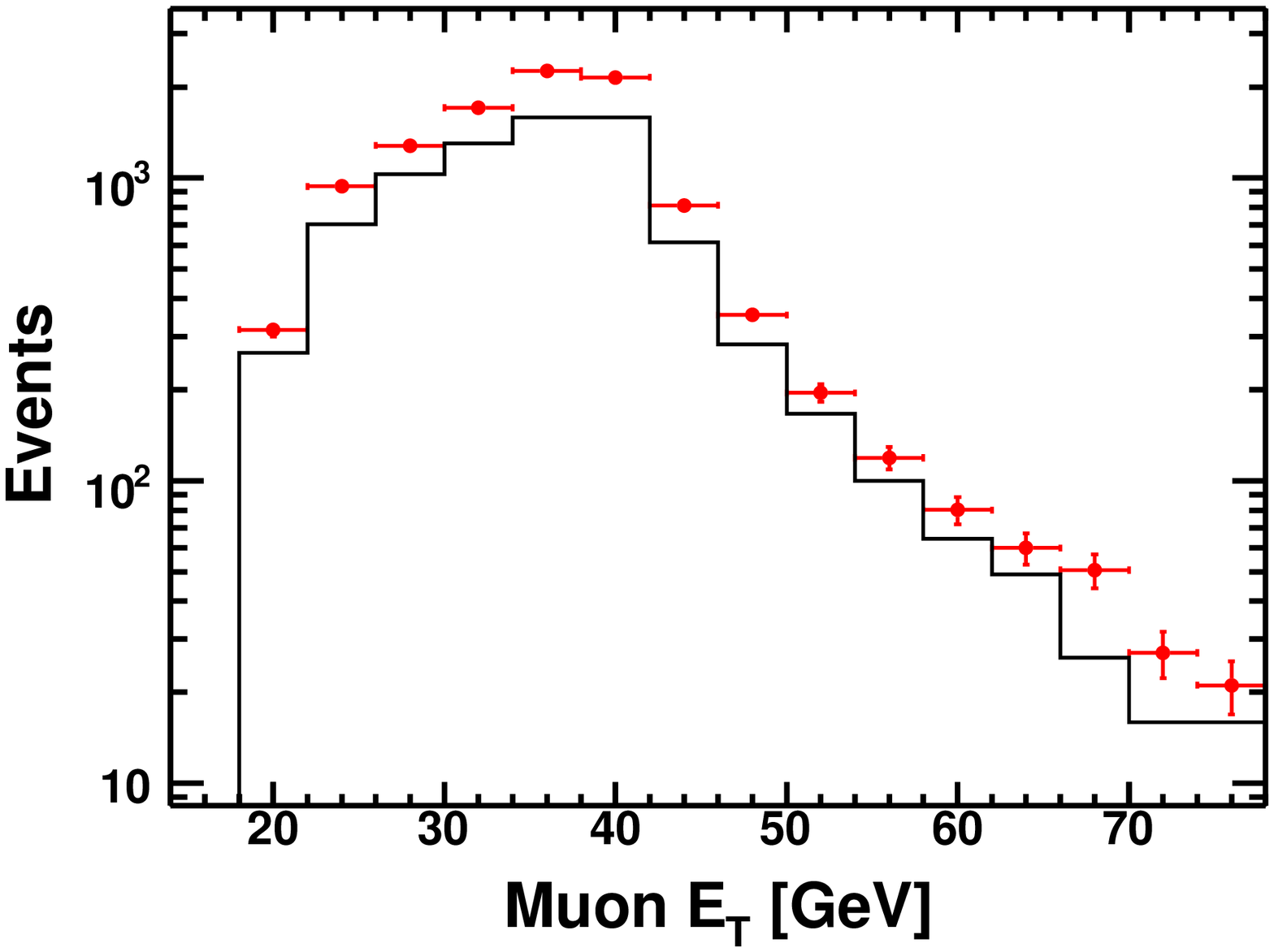} &
\includegraphics[width=0.48\textwidth]{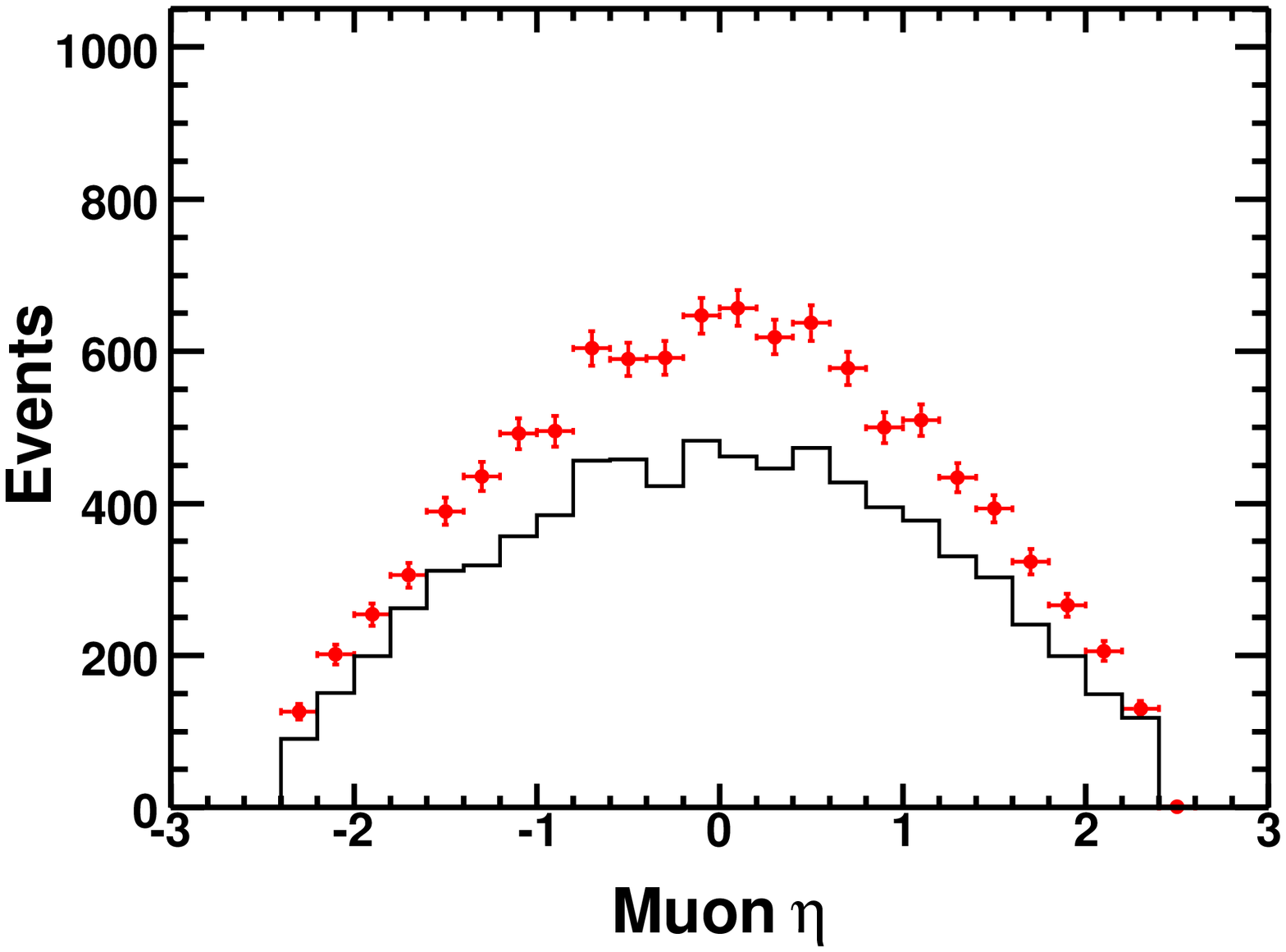}
\end{tabular}
\caption {Distributions, for $\int\mathcal{L}dt$~=~1~fb$^{-1}$, of 
(a) transverse mass of the $W^{\pm}$ boson, 
(b) transverse momentum of the $W^{\pm}$,
(c) transverse momentum of the muon, 
(d) pseudorapidity of the muon for $W \to \mu \nu$. 
Full points: approximately flat diffractive 
gluon density (H1 fit 2~\protect\cite{Adloff:1997sc});
histograms: more peaked diffractive gluon density (H1 fit 3~\protect\cite{Adloff:1997sc}). 
From~\protect\cite{Albrow:2006xt}.
\label{fig:wmunu}}
\end{figure}

Also studied in~\cite{Albrow:2006xt} is inclusive DPE production of $W$
bosons, $pp \to pXWp$, which probes the dPDFs of the proton. The reaction
was simulated with the DPEMC generator v2.4~\cite{Boonekamp:2003ie}. The
generated events were passed through the fast simulation of the CMS
detector, FAMOS version 1.2.0~\cite{PTDR1_FAMOS}. Events in the electron
channel, $W \to e \nu$, were selected by requiring an electron with
$E_T>30$~GeV and missing $E_T$ larger than 20~GeV.  These cuts are tighter
than the CMS L1 trigger thresholds. Several thousand events are expected
after the selection cuts, which include the demand of a tagged proton, for
1~fb$^{-1}$. Events in the muon channel, $W \to \mu\nu$, were selected by
requiring a muon with $E_T>20$~GeV and missing $E_T$ $> 20$~GeV. Also
these cuts are tighter than the CMS L1 trigger thresholds. The expected
distributions of the $W$ and muon variables for 1~fb$^{-1}$ are shown in
Fig.~\ref{fig:wmunu} for different choices of the diffractive PDFs. Here
again, several thousand events are expected after the selection cuts.

\nopagebreak

\subsection{Physics potential of $p_T$ measurements in FP420}
\label{sec:pt}

A study of the correlations between the proton transverse momenta $p_{it}$ 
in the CEP processes will provide us with extra leverage 
in the the forward physics programme. 
First of all, such measurements are important for
testing the underlying physics of diffraction~\cite{KMRtag}.
The absorptive re-scattering effects present in inelastic diffraction clearly violate
Regge factorization and lead to nontrivial
correlations between proton transverse momenta $p_{1t}$
and $p_{2t}$ in the process $pp\rightarrow p + M + p$. 
Measuring the transverse momenta and the azimuthal angle $\varphi$
distribution for different values  $p_{it}$,
allows a detailed probe of the opacity of the incoming
proton, and more generally, testing the dynamics of soft  
survival. One of the best examples to study such effects is exclusive
high $E_T$ dijet production~\cite{KMRtag,Aaltonen:2007hs}, where the 
cross section for the hard subprocess is large and well known.

Another important feature of the
correlation study is that it offers a unique possibility
 for direct observation of a CP-violating signal in the Higgs sector
by measuring the azimuthal asymmetry 
of the outgoing tagged protons~\cite{Khoze:2004rc,Accomando:2006ga}. 
In some MSSM scenarios the azimuthal asymmetry 
\begin{equation} A=\frac{\sigma(\varphi<\pi)-\sigma(\varphi>\pi)} {\sigma(\varphi<\pi)+\sigma(\varphi>\pi)}
\label{eq:AA}
\end{equation}
is expected~\cite{Khoze:2004rc} to be quite sizable.
For instance, $A\simeq 0.07$, in  a benchmark scenario of maximal CP-violation (\cite{CEPW}) or
in  the tri-mixing scenario of Ref.~\cite{je}. 


\subsection{Other physics topics}

\subsubsection{Pomeron/Graviton duality in AdS/CFT}

Another motivation for further study of central exclusive production
is as testing ground of possible connections with string theory, through the
so-called Anti-de-Sitter/Conformal-Field-Theory (AdS/CFT) or ``gauge/string'' 
correspondence~\cite{ads_cft}. The application of AdS/CFT correspondence 
between strongly coupled QCD and weakly coupled gravity has recently
successfully applied to the computation of various observables 
in high-energy heavy-ion physics (see e.g.~\cite{Liu:2007ab} and refs. therein).
Diffractive scattering and the Pomeron represent another
area where a connection with the string-theory-based techniques may
well be ripe.  Like heavy-ion physics, the physics of diffraction and
the Pomeron lies largely outside the regime where perturbative field
theory computations can be performed with confidence.  Indeed they are
known not to fully describe HERA data.  Thus as in heavy-ion physics,
there is much interest in approaching these phenomena with a tool
which replaces non-perturbative field theory with perturbative string
theory.  
The connection with the stringy aspects of the five-dimensional
description 
is  indeed very direct in the case of Regge phenomenology.

A number of papers by string theorists and recently even QCD/nuclear
theorists have studied aspects of the Pomeron, which in the string
theory description is related to the graviton and its higher spin
partners on the leading (five-dimensional) Regge trajectory (see 
e.g.~\cite{Brower:2007xg} and refs. therein).  
The physics of the Pomeron has been described with considerable technical
success, allowing insights into various aspects of Regge phenomenology
in the corresponding four-dimensional gauge theories.  Few attempts
have been made so far to connect these technical results with QCD
data, and the question of whether this connection will be as
suggestive as in the nucleus-nucleus case remains open at present.  Also,
important problems of relevance to the current proposal, such as
central hadron production, rapidity gap suppression, and Higgs boson
production are just now receiving some (as yet unpublished) technical
consideration from theorists.  But there is a real opportunity for
growth of an interdisciplinary research area out of diffractive
physics in general and central diffractive production in particular.

\subsubsection{Exotic new physics scenarios in CEP}

A.R.~White has developed~\cite{white1,white2} a theory of the pomeron 
which requires the existence of new particles in the LHC domain, and would 
give rise to dramatic effects in diffraction. If correct, exclusive processes such as
 $p+p\rightarrow p + W^+W^-+p$ and $p+p\rightarrow p+ZZ+p$ could be 
orders of magnitude higher than in the Standard Model. In the Standard Model, 
exclusive $W^+W^-$ production occurs mainly through $\gamma\gamma \rightarrow W^+W^-$ and
 $h\rightarrow W^+W^-$, if the Higgs boson exists with $M_h \sim$ 135 GeV/c$^{2}$. 
Exclusive $ZZ$ production only proceeds, to a good approximation, through $h$ decay. 
In White's theory the pomeron is approximately a reggeised gluon together with a sea
 of `wee' gluons, with the unitary Critical Pomeron produced via reggeon field theory 
interactions. A special version of QCD, $QCD_S$, is required in which the asymptotic freedom 
constraint is saturated, a requirement naturally satisfied by $QCD_S$, which contains the known 
six colour triplet quarks plus a doublet, $[U,D]$, of heavy (hundreds of GeV) colour 
sextet quarks. The Higgs mechanism is provided, not by a fundamental scalar Higgs boson, 
but by sextet pion composites, i.e. $[U\bar{U} - D\bar{D}]$. This results in a relatively
 strong coupling between the pomeron and vector bosons, with large cross sections for 
$I\!\!PI\!\!P \rightarrow W^+W^-$ and $I\!\!PI\!\!P\rightarrow ZZ$. The enhancement 
in diffractive $W^+W^-$ and $ZZ$ production (but not $WZ$ production) should be
large enough to see without forward proton tagging, with or without requiring large rapidity gaps. 
However, determining that the pomeron - vector boson coupling is responsible and studying 
it in detail will require forward proton measurements. $QCD_S$ provides a natural dark 
matter candidate, the sextet neutron, $N_6 = [UDD]$, which should be stable and have a 
mass in the TeV range. $QCD_S$ also embeds uniquely in an underlying SU(5) theory, 
called $QUD$, which potentially describes the full Standard Model.

\newpage

\section{Simulated measurement of $h \rightarrow b \bar b$ in the MSSM}
\label{sec:pilko}

As a more detailed example of our proposed methodology, we describe in this section how we 
intend to study the production of the MSSM $h\rightarrow b\bar{b}$ channel. Full details can be found in~\cite{Cox:2007sw}. 
Similar cuts to reduce the backgrounds are found in the analysis for the CMS-TOTEM document~\cite{Albrow:2006xt} 
and also in Ref.~\cite{Heinemeyer:2007tu}.
There are two properties of FP420 that are critical to the detection of Higgs bosons in any decay channel. The first is the acceptance, 
described in detail in Section~\ref{sec:optics}, which for a fixed LHC optics depends primarily on the distance of approach of the active 
edge of the silicon detectors to the beam. We will focus on those events in which both protons are tagged at 420~m, although we 
comment on the inclusion of forward detectors at 220~m in Section~\ref{sec:twotwenty}. For 120 GeV/c$^{2}$ central systems, 
the acceptance is independent of the distance of approach out to approximately 7~mm (10$\sigma$ is 2.5~mm at 420~m). 
Here we assume that the active edge of the 420~m detectors is 5~mm from the beam, which gives an acceptance of 28\% for both protons to be detected. 

The second important property of FP420 is its ability to measure the difference in arrival time of the forward 
protons on opposite sides of the central detector. This allows a measurement, from timing information alone, 
of the vertex position of the Higgs candidate event in the central detector, under the assumption that the 
detected protons are from the same proton-proton collision as the Higgs candidate. This vertex-matching 
requirement -- between the vertex determined with the central detectors and that obtained with the fast-timing forward
detectors -- is vitally important at the high LHC luminosities, where the large number of proton-proton 
collisions per bunch crossing (often referred to as pile-up) leads to a high probability that forward protons 
from single diffractive or double pomeron (DPE) collisions not associated with the Higgs candidate event 
will enter the forward detectors during the same bunch crossing. The design goal is to achieve a timing 
resolution of 10~ps in the detectors with negligible jitter in the reference timing system. This corresponds 
to a vertex measurement accurate to 2.1~mm from the tagged protons. The FP420 fast timing system 
is described in detail in Section~\ref{sec:timing}. 

The central exclusive signal events were generated using the ExHuME Monte Carlo v1.3.4~\cite{exhume}, which contains a direct implementation of the calculation described in Section~\ref{sec:calculation}. Using CTEQ6M PDFs and soft survival factor $S^2 = 0.03$, the cross section $\times$ branching ratio to $b \bar b$ for the CEP of a Higgs boson of mass $M_h = 119.5$~GeV/c$^{2}$ in the $M_h^{max}$ scenario of the MSSM is predicted to be 20~fb. There are three primary sources of background;
\begin{enumerate}
\item Central exclusive dijet backgrounds. Central exclusive $b \bar b$ production is suppressed by the $J_z = 0$ selection rule, but will still be present at a reduced rate and forms an irreducible continuum beneath the Higgs boson mass peak. Central exclusive glue-glue production is not suppressed, and contributes to the background when the gluon jets are mis-identified as $b$-jets. The mis-tag rate at ATLAS for gluon jets is 1.3\%, leading to a mis-tag rate for di-gluons of $1.69 \times 10^{-4}$. These are the dominant central exclusive backgrounds; the other CEP background contributions, such as $gHg$ and $b\bar{b}g$ discussed in~\cite{DKMOR,Khoze:2006um}, are either small from the beginning or could be suppressed due to the experimental cuts outlined in Section~\ref{sec:pilkoexp}.
 \item Double pomeron backgrounds. Double pomeron exchange (DPE) is defined as the process $pp \rightarrow p + X + p$ where $X$ is a central system produced by pomeron-pomeron fusion. In this picture the pomeron has a partonic structure and the system $X$ therefore always contains pomeron remnants in addition to the hard scatter. DPE events are simulated using the POMWIG v2.0 Monte Carlo~\cite{Cox:2000jt} with the H1 2006 fit B diffractive PDFs~\cite{Aktas:2006hy} and $S^2 = 0.03$. With this choice of PDF, the DPE background is expected to be small~\cite{Khoze:2007hx}. The effect of different choices of diffractive PDFs is studied in 
 ~\cite{Cox:2007sw} and found to make little difference to the overall conclusions. 
  \item Overlap backgrounds. Overlap events (as discussed in~\cite{Albrow:2006xt,Cox:2007sw}) are defined as a coincidence between an event that produces a Higgs boson candidate in the central detector and one or more single diffractive or DPE events which produce protons in the acceptance range of the forward detectors. Note that non-diffractive protons become important only for detectors at 220~m \cite{Cox:2007sw} and that protons from photon-induced processes are negligible in comparison to single diffraction. 
At a luminosity of  $10^{33}$~cm$^{-2}$~s$^{-1}$ (low luminosity) there will be on average 3.5 interactions per bunch crossing including elastic scattering, and 35 interactions per bunch crossing at $10^{34}$~cm$^{-2}$~s$^{-1}$ (high luminosity). There are three possible types of overlap background, for which we use the following notation: [p][X][p] for events in which there is a coincidence of three overlapping events, the detected protons coming overwhelmingly from soft single diffractive events; [pp][X] where the detected protons come from a single double pomeron exchange event; [pX][p] for events in which a single diffractive event produces a hard central system which fakes a Higgs candidate, and a second event produces a proton on the opposite side. 

These backgrounds are approximately 10$^{7}$ times larger than the signal. The majority of the rejection is achieved through kinematic and topological variables as demonstrated in the following sections. However, the proton time-of-flight (TOF) information from FP420 provides an additional reduction. As described above, a 10~ps resolution in the proton time-of-flight gives a vertex measurement accurate to 2.1~mm. For the overlap backgrounds however, the protons tagged by FP420 do not come from the same interaction as the dijets and therefore the event vertex implied from proton TOF will not, in general, match the dijet vertex measured by the inner tracking detectors. A TOF measurement accurate to 10~ps gives a rejection factor of 18 at low luminosity and 14 at high luminosity
\footnote{The luminosity dependence arises due more than one proton occurring in an arm of FP420. In this case, the event is retained if any of the predicted vertices from $\Delta$TOF matched the dijet vertex.} for [p][X][p] events, 
if we require that the two vertex measurements differ by no more than 4.2~mm (2$\sigma$) and the spread in interaction points is $\sim$4.5~cm. This rejection factor is used as a default in the following sections. 
Results are also presented in the scenario that the overlap background can be effectively removed, for example by improved efficiency of the kinematic and topological rejection variables (discussed in Section~\ref{sec:pilkoexp}) and/or an improvement in the fast-timing system - i.e. a TOF measurement accurate to 2~ps results in a factor 5 increase in the overlap rejection factor (see Sec. \ref{sec:timing})\footnote{While this is beyond present-day performance, it may be achievable on a few-years timescale and there is an active R\&D programme. 
Note that the detectors have very small area $\sim$1~cm$^2$.}.
 \end{enumerate}

Background events are constructed using ExHuME for the exclusive events, POMWIG for the hard single diffractive and DPE events, and HERWIG + JIMMY~\cite{herwig,jimmy} for the hard dijet system [X] in [p][X][p] and [pp][X] events. The soft single diffractive protons are generated according to a parameterisation of the single diffractive cross section at LHC energies given in~\cite{Khoze:2006gg}, which has been normalised to CDF data, and added into the event record. 
The forward proton momenta are smeared by the expected resolution of FP420 and the central particles are smeared to simulate the response of the ATLAS detectors. 
Full details are given in ~\cite{Cox:2007sw}. 


\subsection{Trigger strategy for $h \rightarrow b \bar b$}
\label{sec:trigger}

FP420 is too far away from the central detector to be included in the current level 1 (L1) trigger systems of ATLAS and CMS, which have a latency of 2.5~$\mu$s and 3.5~$\mu$s respectively. However, for all CEP analyses, information from FP420 can be used at level 2 (L2) and/or high-level-trigger (HLT) to substantially reduce the rate. The requirement that there be two in-time protons detected at 420~m would reduce the rate at L2 by a factor of $\sim$20000 (140) at a luminosity of $10^{33}$ ($10^{34}$)~cm$^{-2}$~s$^{-1}$. In addition, cuts on basic topological variables, such as those outlined in Section~\ref{sec:pilkoexp}, which compare the kinematics of the central system measured by ATLAS/CMS to that measured by FP420, would reduce the rate further.

The challenge therefore is to design a trigger strategy, based on central detector information, that is capable of retaining CEP events at L1. The situation for a light Higgs boson decaying to $b$-jets is especially  difficult because the un-prescaled threshold for dijets at ATLAS is foreseen to be 180 GeV at low luminosity and 290 GeV at high luminosity due to the large rate for QCD $2 \to 2$ scatters at hadron colliders. In this analysis, we consider three possible L1 triggers. The first is a low $p_T$ muon trigger of 6~GeV/c in addition to a 40~GeV jet, which is labeled MU6 in the analysis that follows. The jet requirement is required to reduce the rate for low $p_T$ muons at high luminosity. We also consider a higher muon threshold of 10~GeV/c (MU10). The MU6 (MU10) trigger at ATLAS has an efficiency of 10\% (6\%) for a $b\bar{b}$ system. A similar trigger was considered in the CMS-TOTEM studies~\cite{Albrow:2006xt}, that is, a 40~GeV jet with a 3~GeV/c muon, which was found to have an efficiency of 9\%.

The second trigger is to require a rapidity gap in addition to the 40~GeV jet. Such a trigger requires a central jets with $E_T>40$~GeV 
and a lack of hadronic activity in the forward region. The gap would be defined in the forward calorimeters of ATLAS/CMS, which approximately 
cover $3 < |\eta| < 5$. At ATLAS, an additional gap could be defined in the LUCID detectors, which cover $5.4<|\eta|<6.1$, and the 
Zero Degree Calorimeter (ZDC)~\cite{zdc_atlas}, which covers $8.3<|\eta|<9.2$. At CMS IP, the gap could be extended to cover 
$3.1< |\eta| < 4.7$ by the TOTEM T1 detector, $ 5.1 < |\eta| < 6.5 $, by the CASTOR~\cite{castor} and TOTEM T2 detectors, 
$5.2 \lesssim |\eta| \lesssim 6.6$, and $|\eta| > 8.1$ for neutral particles in the ZDC~\cite{zdc_cms}. 
It was found in~\cite{Albrow:2006xt} that the L1 rate for the QCD production of jets was reduced by several orders of magnitude 
by requiring that the T1 and T2 detectors be devoid of activity. The CEP process, however, would have 90\% efficiency in the absence 
of pile-up events in the same bunch crossing. This means that the rapidity gap trigger is self pre-scaling with luminosity; at 
$10^{33}$~cm$^{-2}$~s$^{-1}$ the probability for no pile-up events is 17\%, which drops to  2\% at $2\times10^{33}$~cm$^{-2}$~s$^{-1}$.

The final trigger is to allow a high, fixed L1 rate for 40~GeV jets, which is then substantially reduced at L2 by utilizing information from FP420 as outlined above. In this analysis, we consider a 25~kHz (J25) and a 10~kHz (J10) fixed L1 rate. The J25 trigger would not be pre-scaled at a luminosity of $10^{33}$~cm$^{-2}$~s$^{-1}$ and would be pre-scaled by a factor of 10 at $10^{34}$~cm$^{-2}$~s$^{-1}$. At L2, requiring two in-time proton hits would reduce the J25 rate to less than 200~Hz at high luminosity and could be reduced further to a few Hz by using the basic topological requirements outlined in Section~\ref{sec:pilkoexp}.

A complementary L1 trigger has been considered in~\cite{Albrow:2006xt} for the CMS-TOTEM system, which was not considered in the analysis presented here. The trigger strategy utilises the scalar sum, $H_T$, of all jets. The requirement that essentially all of the transverse energy be concentrated in two central jets, i.e that $(E_{T}^{1}+E_T^2)/H_T>0.9$, reduces the QCD rate by a factor of two but barely affects the signal. Thus the J25 trigger, which is considered to be a fixed rate of 25~kHz, could in fact have a final L1 output rate of 12.5~kHz. 
Another way to tag events with protons in FP420 proposed in~\cite{Albrow:2006xt} makes use of a diffractive type of trigger sensitive to
asymmetric events where one proton is detected in one FP420 detector and the other proton in the 220~m Roman Pot on the other side. 
This is briefly discussed in Section~\ref{sec:twotwenty}


\subsection{Experimental cuts on the final state}
\label{sec:pilkoexp}

\begin{figure}
\centering
\mbox{
	\subfigure[]{\includegraphics[width=.5\textwidth]{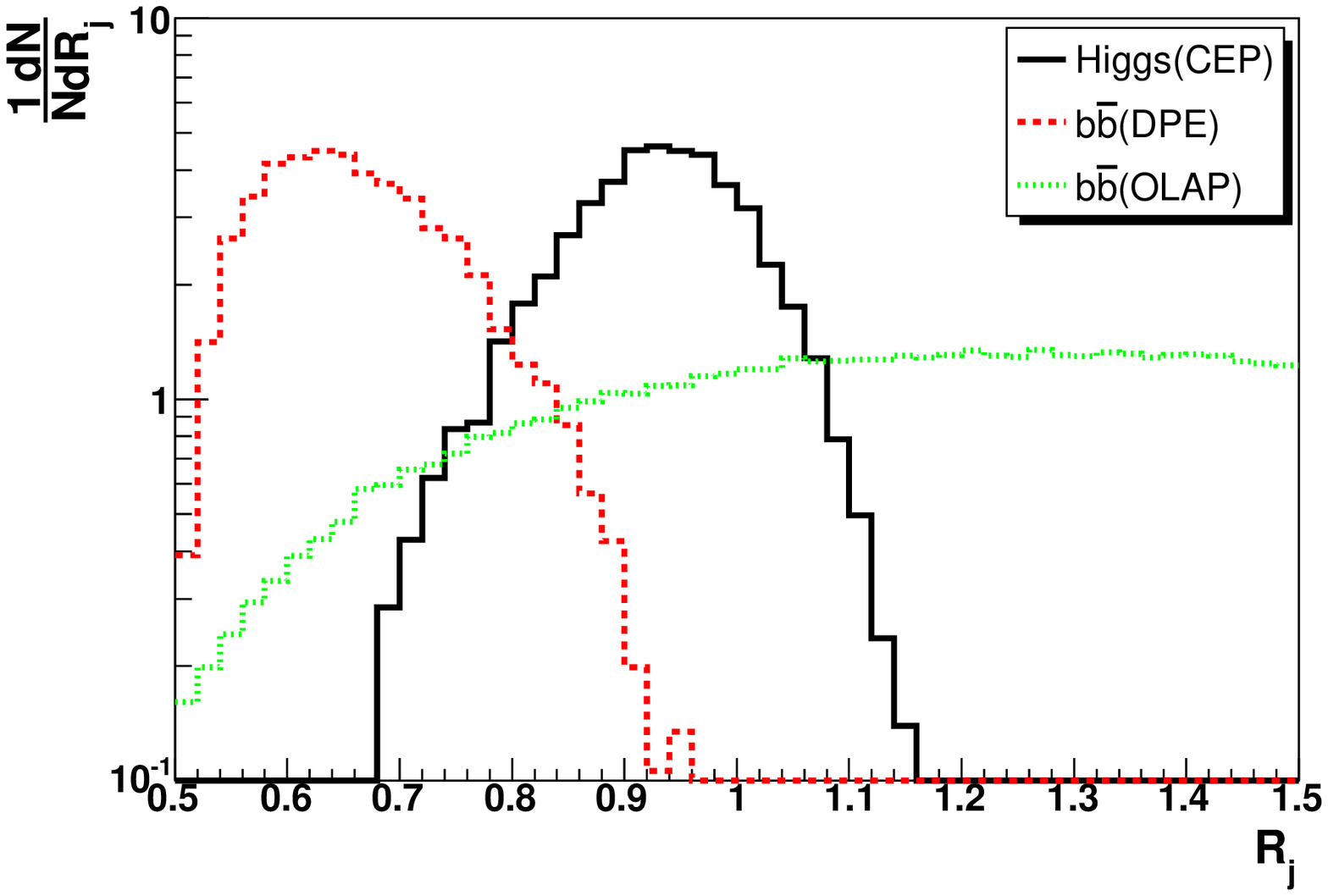}} \quad
	\subfigure[]{\includegraphics[width=.5\textwidth]{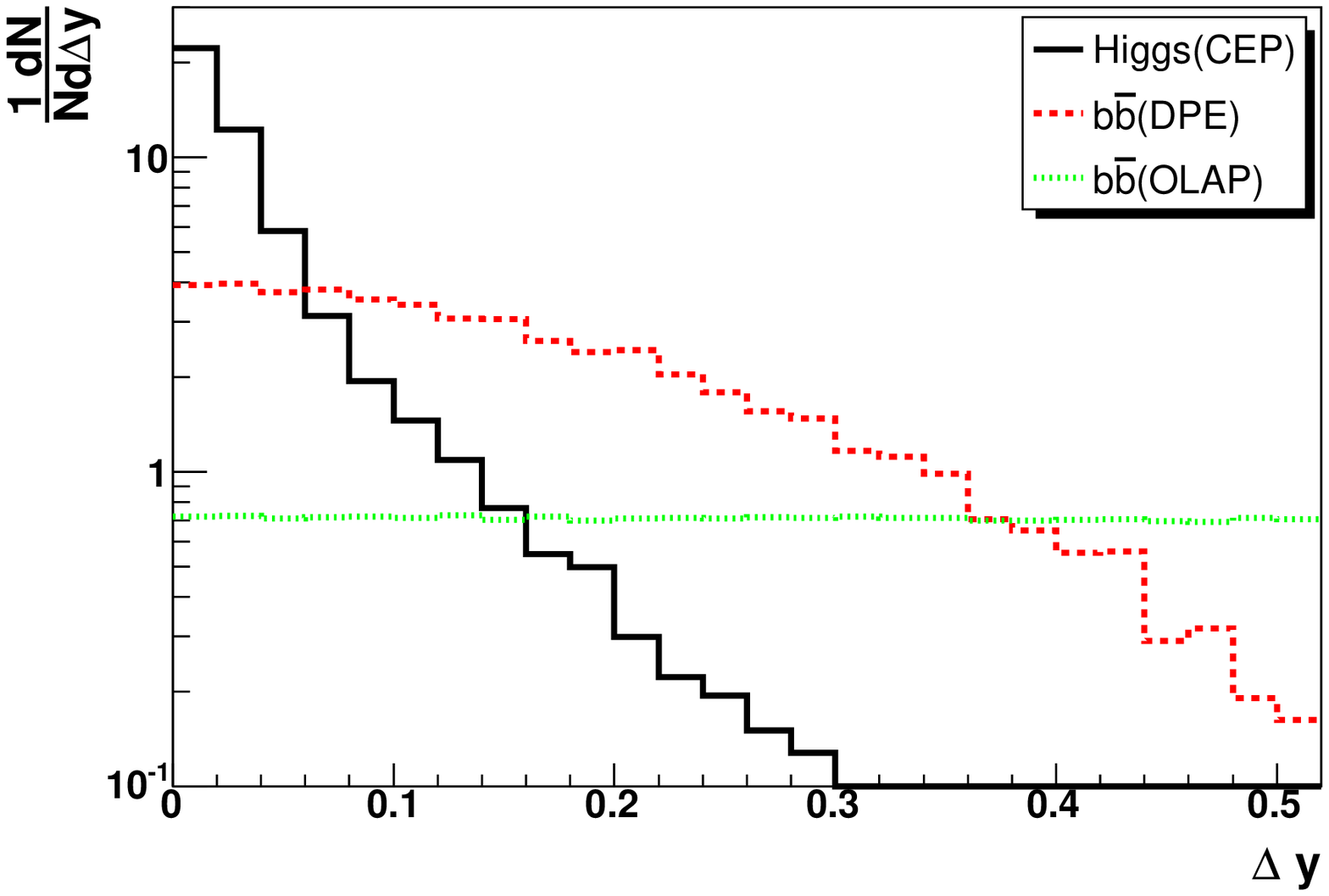}}
	}
\caption{The $R_j$ and $\Delta y$ distributions are shown in (a) and (b) respectively for the signal, 
[p][$b\bar{b}$][p] and DPE [p $b\bar{b}$ p] backgrounds . The distributions were reconstructed 
using a cone radius of 0.7 after smearing the particles with detector resolution. 
\label{fig:signalvbackgroundrjdely}}
\end{figure}

\begin{figure}
\centering
\mbox{
	\subfigure[]{\includegraphics[width=.5\textwidth]{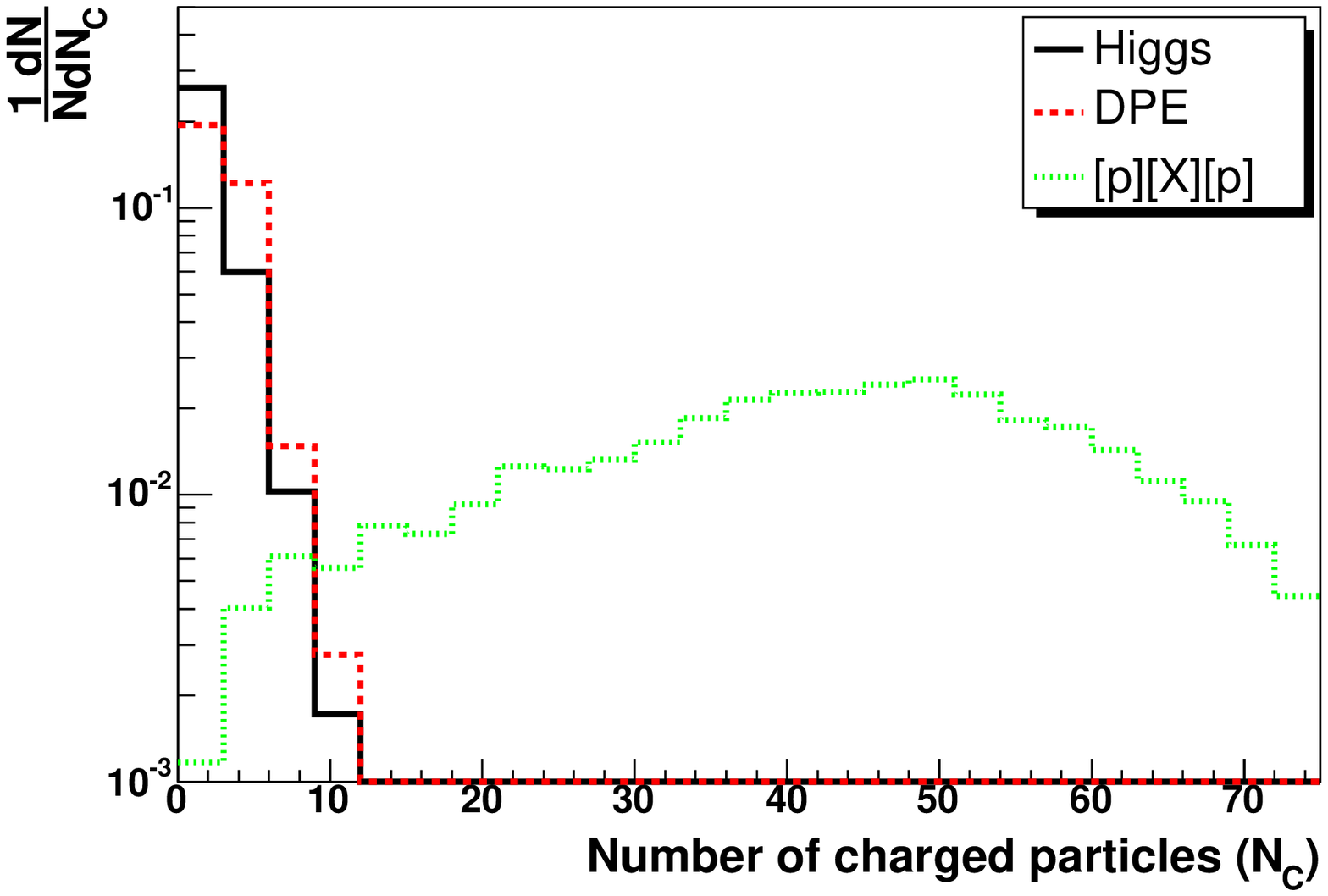}} \quad
	\subfigure[]{\includegraphics[width=.5\textwidth]{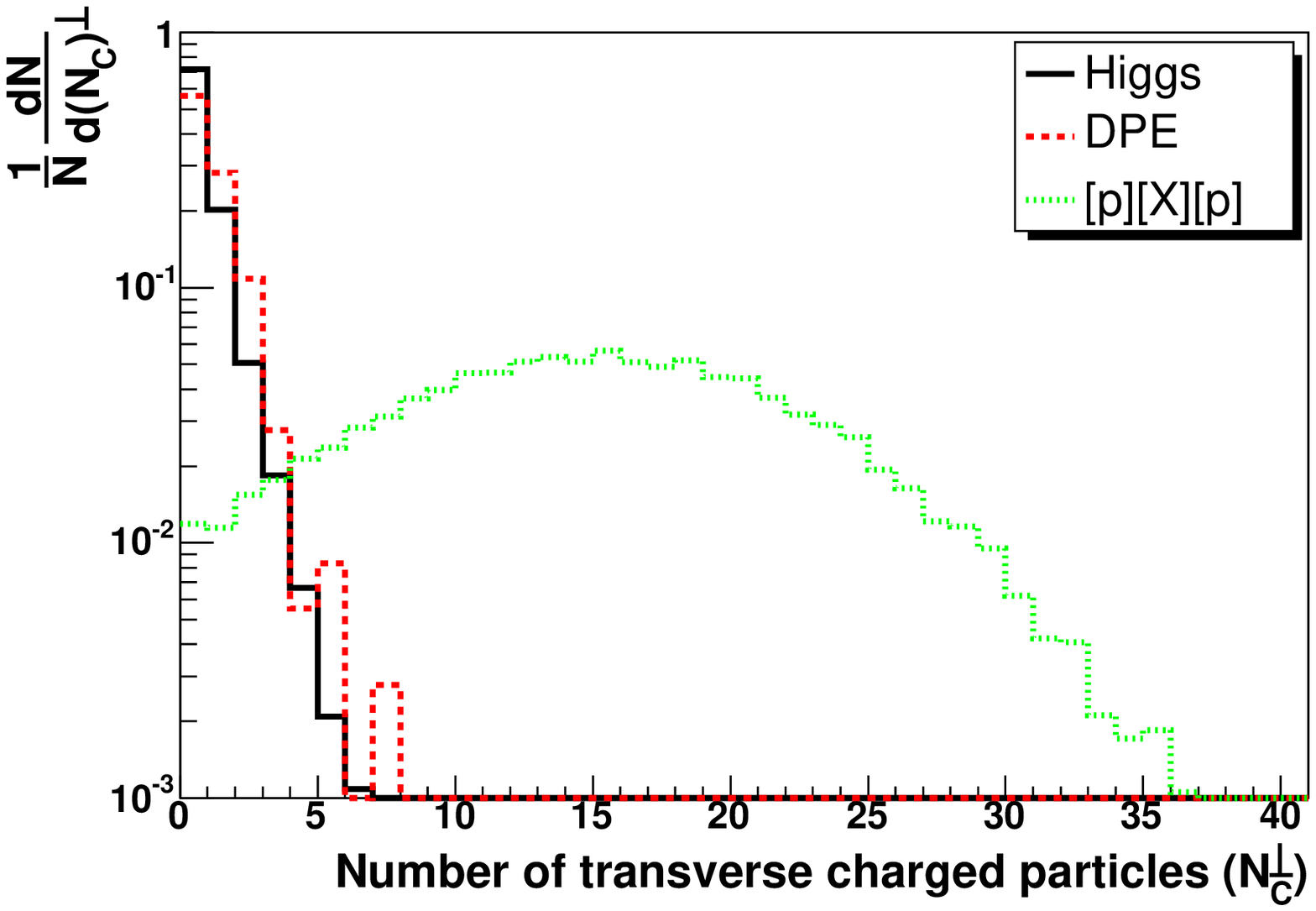}}
	}
\caption{(a) The charged track multiplicity outside of the dijet system, $N_C$. (b) The number of 
charged particles that are transverse to the leading jet as defined by equation~(\ref{eq:transverse}). 
In both cases the particles must satisfy $p_T>0.5$ GeV/c and $|\eta|\leq1.75$. Only 80\% of the 
particles are used, which replicates ATLAS reconstruction efficiency for low $p_T$ tracks. \label{fig:ncharged}}
\end{figure}

The Monte Carlo samples are initially standardised by requiring that there are two jets, one with $E_T>45$~GeV and one with $E_T>30$~GeV; the jets are reconstructed using the cone algorithm with cone radius of 0.7. Furthermore, the outgoing protons are required to lie within the acceptance of FP420 as defined in Section~\ref{sec:optics}. This corresponds approximately to the kinematic range $0.005 \leq \xi_1 \leq 0.018$, $0.004 \leq \xi_2 \leq 0.014$ and unrestricted in~$t$, where $\xi$ is the fractional longitudinal momentum loss of the outgoing proton and $t$ is the squared 4-momentum transfer at the proton vertex. Full details are given in~\cite{Cox:2007sw}. The following variables are then useful to characterise CEP events:
\begin{itemize}
\item The difference in rapidity, $\Delta y$, of the central system measured by FP420 to that measured from the average pseudo-rapidity of the dijets, i.e.
\begin{equation} 
\Delta y = \left| y - \left( \frac{\eta_1 + \eta_2}{2} \right) \right|
\end{equation}
where $y$ is the rapidity of the central system measured by FP420 and is given by
\begin{equation}
y= \frac{1}{2} \textrm{ln} \left( \frac{\xi_1}{\xi_2} \right).
\end{equation}
\item The dijet mass fraction, $R_j$, which is the fraction of the mass of the centrally produced system carried by the dijets. $R_j$ is an improved definition~\cite{Khoze:2006iw} of the dijet mass fraction variable $R_{jj}$, which has been used to identify exclusive events at CDF~\cite{Aaltonen:2007hs}. $R_j$ is defined as 
\begin{equation}
R_j = \frac{2E_{T}^{1}}{M} \textrm{cosh} (\eta_1 - y),
\end{equation}
where $E_{T}^{1}$ and $\eta_1$ are the transverse energy and pseudo-rapidity of the leading jet in the event and $M$ is the mass of the central system measured by FP420, given by 
\begin{equation}
M^2 \approx \xi_1 \, \xi_2 \, s
\end{equation}
where $\sqrt{s}$ is the centre-of-mass energy of the proton-proton interaction. 
For a true CEP event with no out-of-cone and detector smearing effects, $R_{j} = 1$.
\item The multiplicities of charged tracks, $N_C$ and $N_C^{\perp}$, with $p_T \geq 0.5$ GeV/c 
and $|\eta| \leq 1.75$ that are associated with (i.e. within $\pm$2.6~mm of) the dijet vertex. 
$N_C$ is the number of charged particles in the event that are not associated with the hard scatter, 
i.e. not contained within the jet cones. It is of course dependent on the jet algorithm used to reconstruct 
the jets. $N_C^{\perp}$, defined as the number of charged tracks that are perpendicular in azimuth 
to the leading jet, provides a measure of the particle multiplicity associated with the underlying event. 
Both $N_C$ and $N_C^{\perp}$ should vanish for CEP processes with negligible final-state radiation and underlying event.
We use the definition adopted in~\cite{Field:2006gq}, which assigns charged particles to the underlying 
event if they satisfy
\begin{equation}
\frac{\pi}{3} \leq | \phi_k - \phi_1 | \leq \frac{2\pi}{3} \quad \textrm{and} \quad 
\frac{4\pi}{3} \leq | \phi_k - \phi_1 | \leq \frac{5\pi}{3},
\label{eq:transverse}
\end{equation}
where $\phi_k$ is the azimuthal angle of a given charged particle and $\phi_1$ is the azimuthal angle of the highest transverse energy jet.
We also choose to not use the full inner detector tracking coverage ($|\eta| \leq 2.5$) to count charged tracks so that a small vertex window can be used; particles at large pseudo-rapidity have the poorest vertex reconstruction and would require a larger vertex window, which would increase the probability of tracks from pile-up events contaminating the signal (and background).
\end{itemize}
These variables are extremely efficient at separating the overlap and DPE backgrounds from the CEP events. For overlap events, the central system kinematics predicted by FP420 do not, in general, match the observed dijet kinematics. Figure~\ref{fig:signalvbackgroundrjdely}(a) shows the $R_j$ distribution for signal, DPE and overlap events and Fig.~\ref{fig:signalvbackgroundrjdely}(b) shows the $\Delta y$ distribution. To a good approximation, the overlap background is flat over a very large region of $R_j$ and $\Delta y$, whereas the signal forms a well defined and narrow peak.
Figures~\ref{fig:ncharged}(a) and~\ref{fig:ncharged}(b) show the $N_C$ and $N_C^{\perp}$ distributions respectively. As expected, the central exclusive events have few charged particles outside of the jet cones. In contrast, the overlap events have many charged particles due to the break up of the protons and the underlying event activity associated with standard QCD events at the LHC. The final exclusive candidate sample is defined by the following cuts:
\begin{itemize}
\item The dijet mass fraction, $0.75 \leq R_j \leq 1.1$.
\item The difference in rapidity of the central system measured by FP420 to that measured from the dijets, $\Delta y \leq 0.06$.
\item The jets are back-to-back  , i.e $\pi - |\Delta \phi| \leq 0.15$.
\item The charged track multiplicity associated with the dijet vertex, $N_C \leq 3$ and $N_C^{\perp} \leq 1$.
\end{itemize}

\subsection{Results and significances}
\label{sec:results}

The cross sections for the signal and the dominant backgrounds, excluding the trigger efficiency, are shown in Table~\ref{tab:cuteff}. 
The final cross sections are defined in a mass window around the Higgs boson mass of $\Delta M = \pm 5.2$ GeV/c$^{2}$. 
This is significantly larger than the projected resolution of the forward detectors because the width of the Higgs boson with 
this choice of MSSM parameters ($M_A$~=~120~GeV, tan$\beta=40$, $\mu$~=~200~GeV) is 3.3 GeV/c$^{2}$. 
The overlap backgrounds are defined at a luminosity of 10$^{34}$~cm$^{-2}$~s$^{-1}$, which is the worst-case scenario because even when the LHC is operating at peak design luminosity 
the average luminosity over a fill will be lower than 10$^{34}$~cm$^{-2}$~s$^{-1}$.  
Table~\ref{tab:cuteff} shows that the 
dominant background at high luminosity is the [p][X][p] overlap background. 
 
\begin{table}[t]
\centering
\begin{tabular}{|c||c|c|c||c||c|c|c|}\hline
& \multicolumn{7}{|c|}{Cross section (fb)} \\\hline
Cut & \multicolumn{3}{|c||}{CEP} & DPE & [p][X][p] & [p][pX] & [pp][X] \\\hline
& $h \rightarrow b \bar b$ & $b\bar{b}$ & $gg$ & $b\bar{b}$ & $b\bar{b}$  & $b\bar{b}$ & $b\bar{b}$ \\\hline
 $E_T$, $\xi_1$, $\xi_2$, $M$ & 1.011 & 1.390 & 2.145 & 0.666 & 5.42$\times10^{6}$& 8.98$\times10^{3}$ & 1.16$\times10^{6}$  \\
 TOF (2$\sigma$,10~ps) & 0.960 & 1.320 & 2.038 & 0.633 & 3.91$\times10^{5}$& 7.33$\times10^{2}$ & 6.29$\times10^{4}$  \\
 $R_j$ &  0.919 & 1.182 & 1.905 & 0.218 & 4.73$\times10^{4}$ & 85.2 & 7.59$\times10^{3}$ \\
 $\Delta y$ & 0.774 & 1.036 & 1.397 & 0.063 & 2.16$\times10^{3}$ & 1.38 & 3.50$\times10^{2}$ \\
 $\Delta \Phi$ & 0.724 & 0.996 & 1.229 & 0.058 & 6.66$\times10^{2}$ & 0.77 & 1.07$\times10^{2}$\\
 $N_C$, $N_C^{\perp}$ & 0.652 & 0.923 & 0.932 & 0.044 & 6.49 & 0.45 & 1.35 \\
 $\Delta$M & 0.539 & 0.152 & 0.191 & 0.009 & 1.28 & 0.06 & 0.28 \\\hline
\end{tabular}
\caption{Cross section (fb) for the CEP Higgs boson signal and associated backgrounds after applying 
each one of the cuts in the text. The first cut requires that both protons are 
tagged at 420~m, the mass measured by the forward detectors is between 80 and 160 GeV/c$^{2}$ 
and the transverse energy of the leading jet is greater than 40 GeV. The second cut is the requirement that the di-jet vertex is within $\pm4.2$~mm of the vertex predicted by proton TOF. The overlap backgrounds are defined 
at high luminosity (10$^{34}$~cm$^{-2}$~s$^{-1}$).}
\label{tab:cuteff}
\end{table}

\begin{figure}
\centering
\mbox{
	\subfigure[]{\includegraphics[width=.5\textwidth]{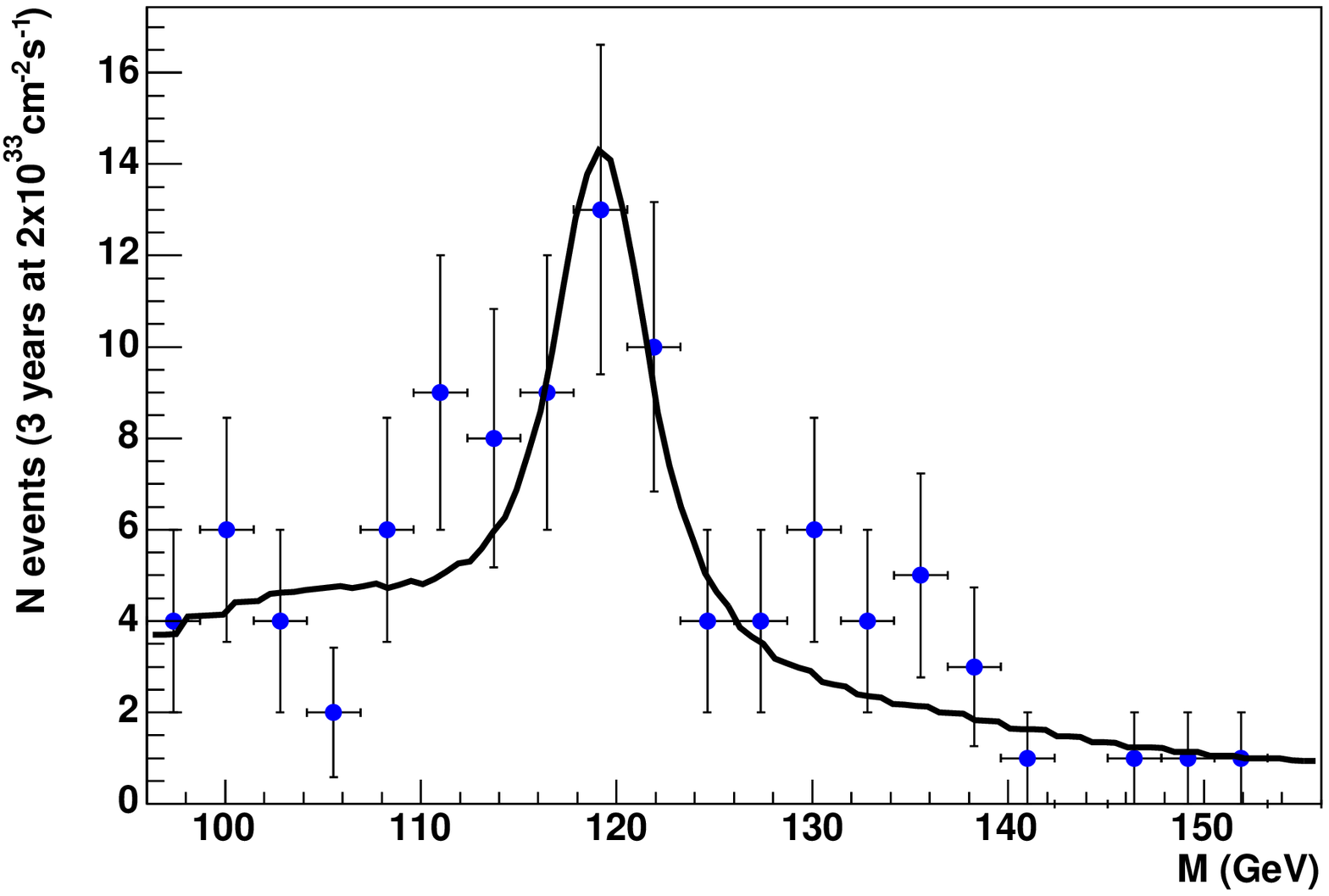}}
	\subfigure[]{\includegraphics[width=.5\textwidth]{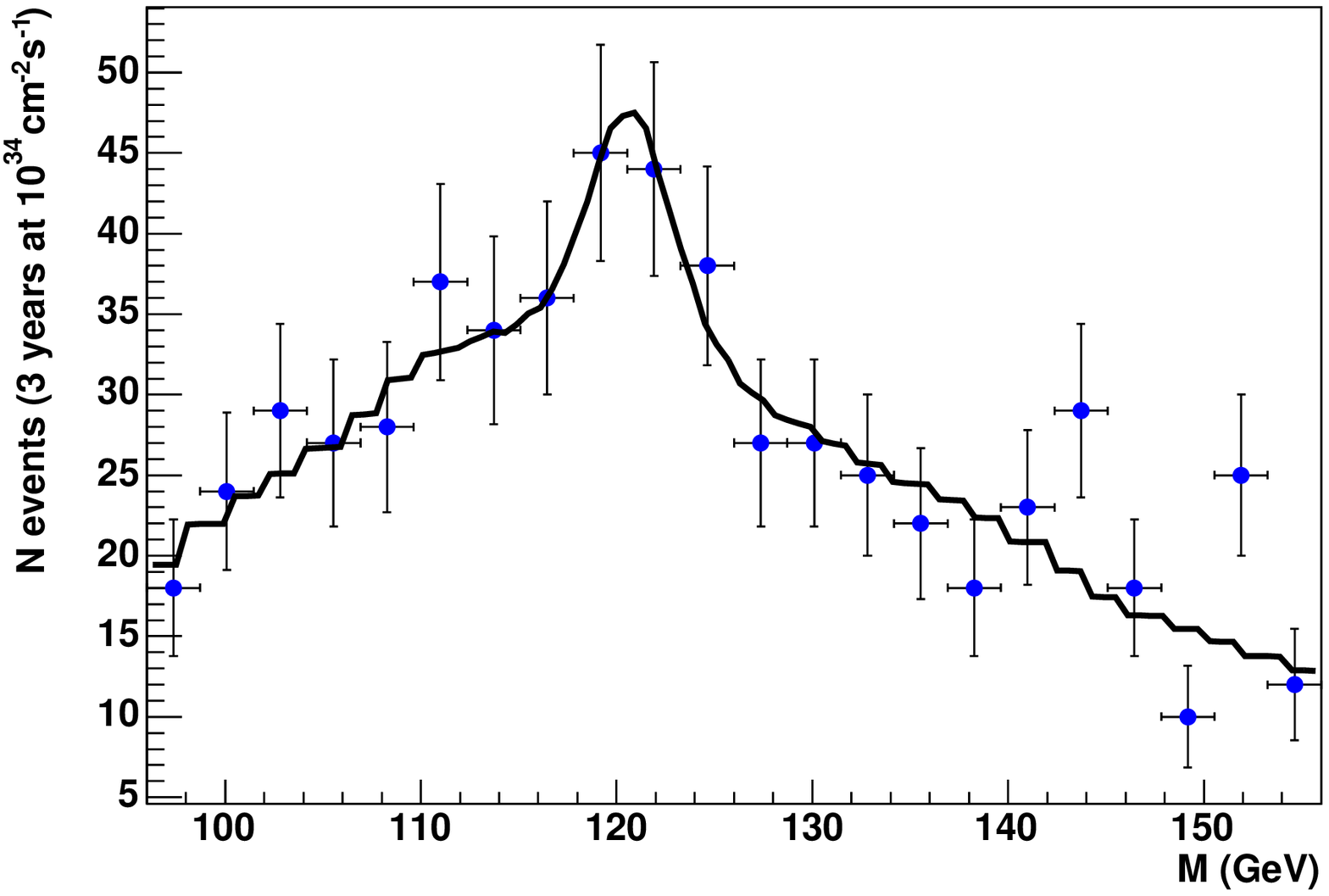}}
}

\caption{ Typical mass fits for the 120~GeV/c$^{2}$ MSSM $h\rightarrow b\bar{b}$, 
with the L1 trigger and analysis cuts discussed in the text, for 3 years of data taking at 
2$\times~10^{33}$~cm$^{-2}$~s$^{-1}$ (60~fb$^{-1}$ 3.5$\sigma$, left plot) 
and at 10$^{34}$~cm$^{-2}$~s$^{-1}$ (300 fb$^{-1}$, $3 \sigma$, right plot).}
\label{fig:peak}
\end{figure}

In order to determine the significance of the signal, a pseudo-data sample was constructed using the generators described above and a full analysis was performed including various L1 trigger strategies  and applying the aforementioned selection cuts. Figure~\ref{fig:peak}(a) shows a simulated mass fit after 3 years of data taking at $2 \times 10^{33}$~cm$^{-2}$~s$^{-1}$, corresponding to an integrated luminosity of 60 fb$^{-1}$. 
The L1 trigger strategy is J25 + MU6 + rapidity gap trigger (see definitions in Sec~\ref{sec:trigger}). 
The peak is fitted with a Gaussian function, which represents the known mass resolution of FP420, convoluted with a Lorentzian function. The shape of the background is assumed to be well known, as it can be measured with high statistics using the forward detectors; in our case, we use all the MC events (in the correct ratio) to determine the shape\footnote{We have also checked the possibility of using a quadratic background and reach the same results.}. 
The significance of this fit is $3.5 \sigma$. 

\begin{figure}
\centering
\mbox{
	\subfigure[]{\includegraphics[width=.5\textwidth]{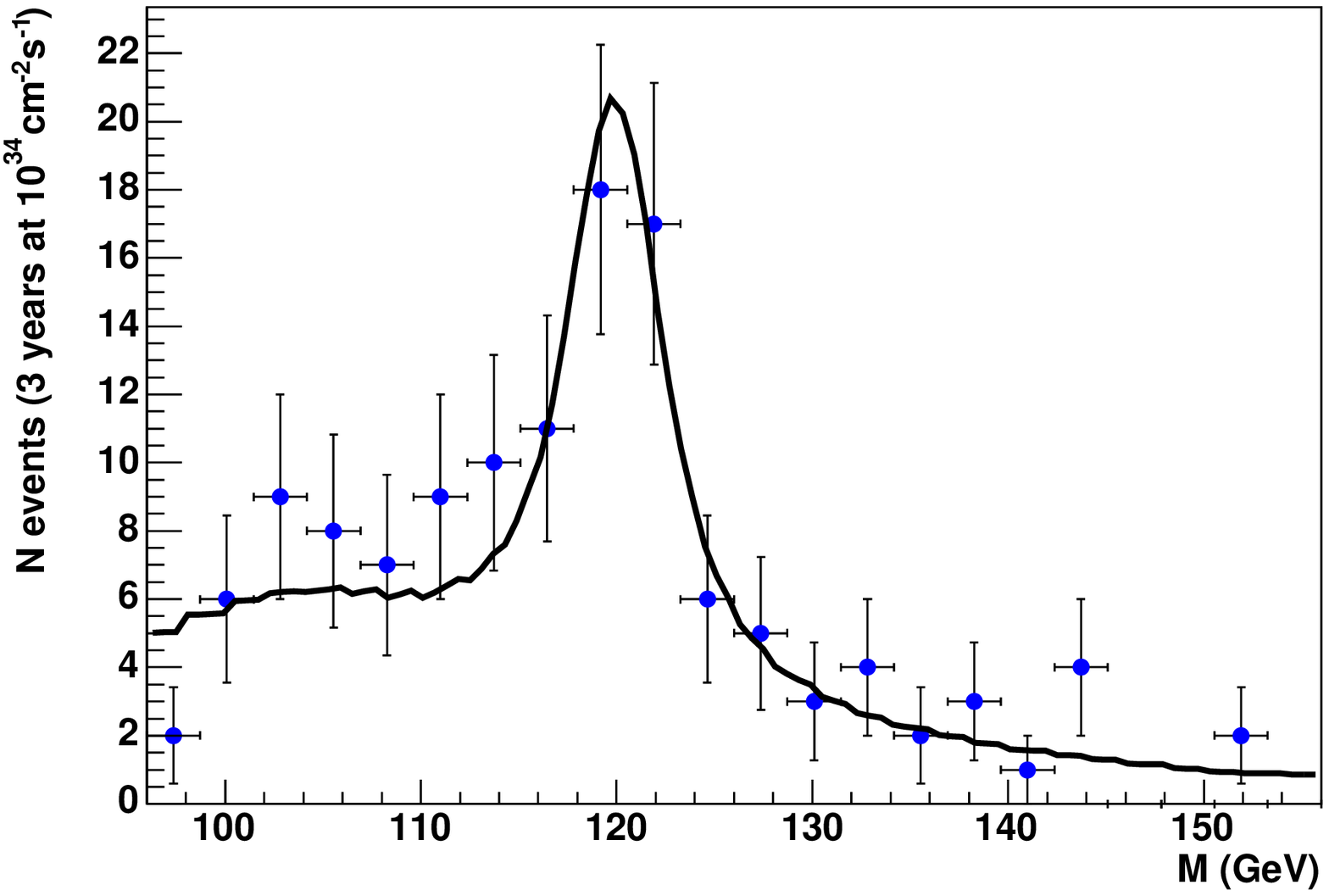}}
	\subfigure[]{\includegraphics[width=.5\textwidth]{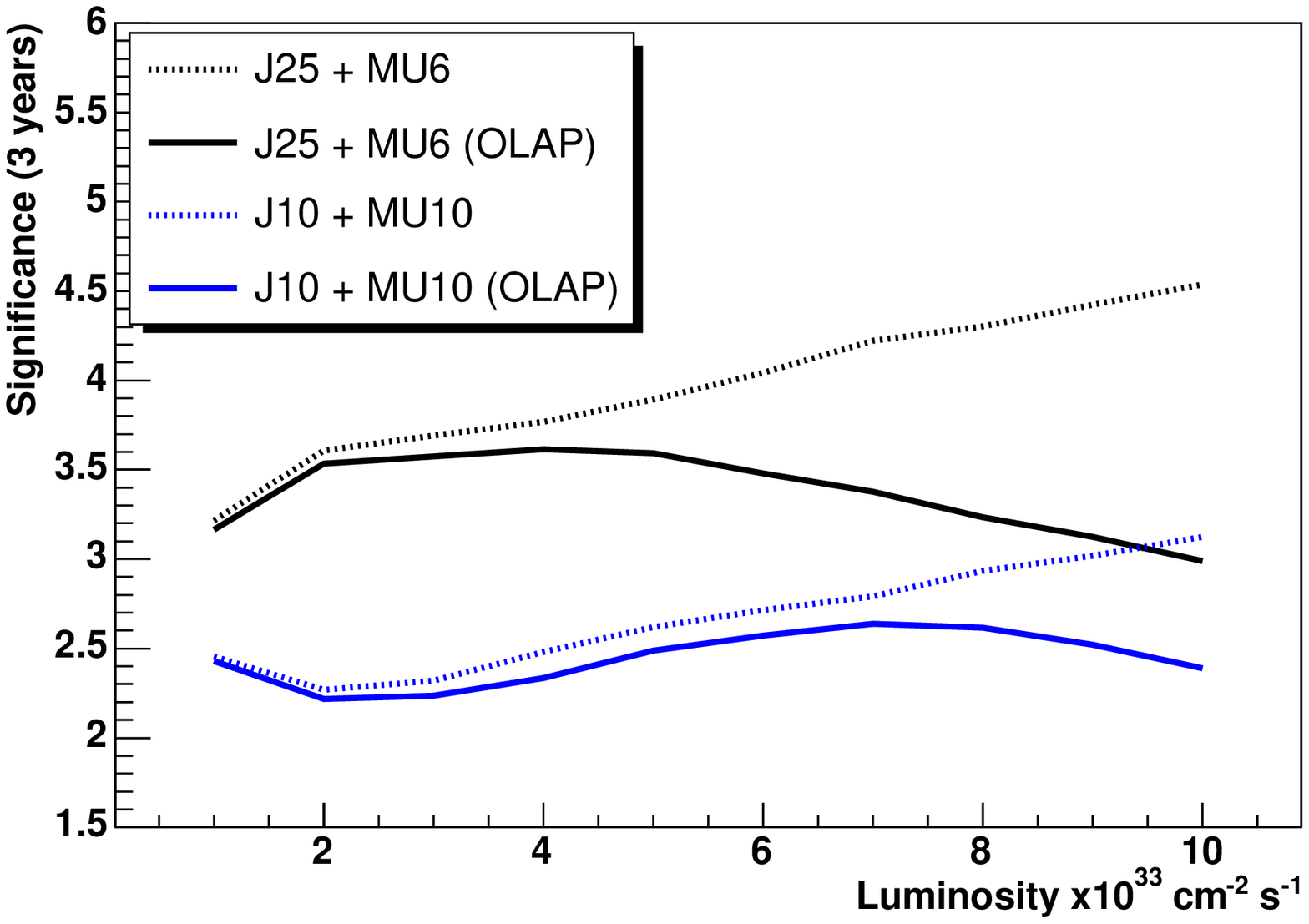}}
}

\caption{(a) Typical mass fit for the 120~GeV/c$^{2}$ MSSM $h\rightarrow b\bar{b}$ 
for 3 years of data taking at $10^{34}~$cm$^{-2}~$s$^{-1}$ after removing the overlap 
background contribution completely with improved timing detectors. The significance is 
$5 \sigma$ for these data. (b) Significance of the measurement of the 120 GeV/c$^{2}$ MSSM 
Higgs boson versus luminosity, for two different combinations of muon -- MU6, MU10 -- and 
jet-rate -- J25, J10 -- triggers, see Sec.~\ref{sec:trigger}, and with an improved (baseline) FP420 
timing design (OLAP labels).}
\label{fig:peak2}
\end{figure}

Figure~\ref{fig:peak}(b) shows a mass fit for the same experimental conditions for 3 years of data taking at  $10^{34}$~cm$^{-2}$~s$^{-1}$ (300 fb$^{-1}$). Because of the increase in overlap backgrounds, the significance falls slightly to $3 \sigma$ and 
improvements in the overlap rejection are required to take full advantage of the high luminosity. This could be achieved through an upgrade to the fast-timing system, as discussed in Section~\ref{sec:timing}, or an improvement in the background rejection variables. Figure~\ref{fig:peak2}(a) shows the same mass fit under the assumption that the overlap backgrounds can be effectively eliminated; the significance is now $5 \sigma$. 
Figure~\ref{fig:peak2}(b) shows the significance as a function of luminosity for two different L1 trigger strategies, J25 + MU6 and a more conservative J10 + MU10. The curves labelled OLAP are for the baseline rejection factors shown in Table \ref{tab:cuteff}. Curves are also shown for the improved overlap rejection and above luminosities of $5 \times 10^{33}$~cm$^{-2}$~s$^{-1}$ it becomes valuable to push for additional rejection and improved timing. This suggests a possible upgrade strategy for the FP420 timing system. 

The largest loss of events at high luminosity comes from the L1 trigger efficiency, which is at best around 20\% at $10^{34}$~cm$^{-2}$~s$^{-1}$  for the L1 strategies considered here, although it is close to 100\% at $10^{33}$~cm$^{-2}$~s$^{-1}$ . If, in a future upgrade to the central detectors, the L1 trigger latency were increased beyond 4 $\mu$s, a trigger efficiency of close to 100\% could be achieved by requiring two forward protons tagged by FP420. Coupled with improved fast timing, a $5 \sigma$ observation with a mass measurement better than 1 GeV/c$^{2}$ could be achieved for 100 fb$^{-1}$ of data taken at $10^{34}$~cm$^{-2}$~s$^{-1}$.   

For the benchmark scenario discussed above, the fits to the simulated data were relatively insensitive to the width of the Higgs state, which was 3.3 GeV/c$^{2}$. For Higgs bosons of decay width $\sim 5$ GeV/c$^{2}$ and greater, a measurement of the width should be possible with the standard FP420 experimental configuration for those regions of MSSM parameter space in which the cross sections are 10 times larger than the Standard Model cross section.  

\subsection{Inclusion of forward detectors at 220~m}
\label{sec:twotwenty}

Adding forward detectors at 220~m, in addition to FP420, has a number of benefits for this analysis.  Firstly, for a 120~GeV/c$^{2}$ central system, there can be a large acceptance for asymmetrically tagged protons, i.e. one tagged at 420~m and one at 220~m. The exact acceptance depends on the distance of the detectors from the beam (see Section~\ref{sec:optics}) and is approximately 16\% if the 220~m  detectors are 2~mm from the beam and the FP420 detectors are 5~mm from the beam\footnote{The beam spot is smaller at 220~m and the detectors can be placed much closer to the beam}. 
If the analysis is repeated for both symmetric and asymmetric tagged events, the significance increases to 4.4$\sigma$ for the J25$+$MU6 trigger for 60~fb$^{-1}$ collected at $2\times10^{33}$~cm$^{-2}$~s$^{-1}$. If improvements in fast-timing or rejection techniques result in the removal of the overlap backgrounds, the significance for 300~fb$^{-1}$ of data collected at $10^{34}~$cm$^{-2}~$s$^{-1}$ increases to 5.5$\sigma$ (3.6$\sigma$) for the J25$+$MU6 (J10$+$MU10) trigger strategy. The combined significances increases further if the detectors are moved closer to the beam.

It is also possible to devise a L1 trigger strategy for the asymmetric events incorporating information from the 220~m detectors. The trigger would require a proton hit, with a momentum loss measurement that is compatible with an opposite side proton hit at 420~m, and that the jet energies contained the majority of the energy deposited in the calorimeters. Such a trigger would have a rate below 1kHz up to an instantaneous luminosity of 2$\times10^{33}$~cm$^{-2}$~s$^{-1}$~\cite{Albrow:2006xt}. Thus at low luminosities, all of the asymmetric tagged events could be retained for little bandwidth and at high luminosities this approach would act as another method to reduce the prescale for events with low $E_T$ jets. It is demonstrated in~\cite{Cox:2007sw} that the significance of the asymmetric events using this trigger strategy is 3.2$\sigma$ for 60~fb$^{-1}$ of data collected at $2\times10^{33}$~cm$^{-2}$~s$^{-1}$, if the forward detectors at 220~m (420~m) are placed at 2~mm (5~mm) from the beam. The significance increases to 5$\sigma$ if the detectors are moved to 1.5~mm (3~mm) from the beam.


\subsection{Comparison of the $h,H \rightarrow b \bar b$ analyses}
\label{sec:Hbbar_comparison}

In this section, we compare the results of Heinemeyer et al.~\cite{Heinemeyer:2007tu} (Sec.~\ref{sec:mssm}) to the results of Cox et al.~\cite{Cox:2007sw} (presented in Section~\ref{sec:results}) for a 120~GeV/c$^{2}$ Higgs boson. The overall signal efficiency, excluding the trigger, assumed by Heinemeyer et al. is 2.0\% for  protons tagged at 420~m. This efficiency is found using the fast simulation of CMS and is very similar to the analysis published in~\cite{Albrow:2006xt} for SM $h\ra b \bar{b}$. The corresponding efficiency observed by Cox et al. is 2.7\%. Note that Cox et al. use a larger mass window, which results in the a factor $\sim$1.3 more events. After normalizing to the larger mass window, the Heinemeyer et al. efficiency increases to 2.5\%. There are two ongoing analysis using the ATLAS fast detector simulation that show similar experimental efficiencies.

The expected number of overlap events, for the combined 420/220 detector acceptance, is found to be 1.8  by Cox et al. for 30~fb$^{-1}$ of data taken at 10$^{33}$~cm$^{-2}$~s$^{-1}$. This includes a mass window around the Higgs boson peak as outlined in Section~\ref{sec:results}. Very large rejection factors are obtained using the exclusivity variables, $R_j$, $\Delta y$, $N_C$ and $N_C^{\perp}$, as shown in Table~\ref{tab:cuteff}.
These rejection factors are also being studied using the ATLAS fast detector simulation. Preliminary results are consistent with Cox et al.~\cite{Cox:2007sw}. The effects of using different event generators for the inclusive QCD event background has also been studied. Using Pythia~\cite{Sjostrand:2001yu}, with the ATLAS/DWT tunes to Tevatron data~\cite{Field:2006gq} predicts less underlying event activity at the LHC than HERWIG$+$JIMMY. The corresponding rejection factor of the $N_C$ and $N_C^{\perp}$ cuts is at least a factor of two smaller when using Pythia~\cite{Cox:2007sw}. The nature of the underlying event at the LHC will be determined with very early data. Despite this uncertainty due to the different underlying event models, it has been demonstrated that the overlap background rejection from the charged track multiplicity cut is largely unaffected by changes in luminosity.

Different trigger strategies are employed in the analyses presented in Secs.~\ref{sec:mssm} and~\ref{sec:pilko}. Heinemeyer et al. do not consider a pre-scaled jet rate trigger - the majority of the events  in the analysis are triggered at L1 by a proton tag at 220~m. The significance of the measurement, given a 120~GeV/c$^{2}$ Higgs boson at tan$\beta=40$, is slightly larger than 3$\sigma$ given 60~fb$^{-1}$ of data (Fig~\ref{fig:heinemeyer2}, 60~fb$^{-1}$). As discussed in Section~\ref{sec:twotwenty}, Cox et al. find that the asymmetric tagging alone achieves a significance in the region of 3.2$\sigma$ to 5$\sigma$ for 60~fb$^{-1}$ of data collected at 2$\times10^{33}$~cm$^{-2}$~s$^{-1}$; the exact value is dependent on the distance of the active detector edge to the beam. Furthermore, the jet-rate trigger could retain up to 50\% of the symmetric events at this luminosity, with a significance of up to 3.5$\sigma$ as shown in Fig.~\ref{fig:peak2}(b). Thus it is likely that, at low luminosity, the efficiency curves in Figs.~\ref{fig:heinemeyer2} and~\ref{fig:heinemeyer3} (labelled 60~fb$^{-1}$) are a conservative estimate.
Heinemeyer et al. do not consider the contribution of the overlap backgrounds, however, which become the dominant background at high luminosity. Thus, the high luminosity curves in Figs~\ref{fig:heinemeyer2} and~\ref{fig:heinemeyer3} (600~fb$^{-1}$) are only valid if the overlap background can be effectively eliminated. This could be achieved through improved efficiency of the rejection variables, outlined in Sec.~\ref{sec:pilkoexp}, or if the time-of-flight system is upgraded, as discussed in Sec.~\ref{sec:timing}.

\subsection{Recent improvements in background estimation}
\label{sec:updatedanalysis}

Recently, there have been a number of improvements in the calculations of the backgrounds in the 
$h\rightarrow b\bar{b}$ channel. Firstly, NLO calculations~\cite{kmrheralhc} indicate that the central 
exclusive production of $gg \rightarrow b\bar{b}$ is a factor of two (or more) smaller than the LO
values assumed in the estimates in Sec.~\ref{sec:results}. Secondly, the overlap backgrounds presented 
in the previous sections were calculated assuming fixed instantaneous luminosity for a given integrated 
luminosity. This is a very conservative estimate as the luminosity decreases during a store and the largest 
overlap background cross section scales with $\mathcal{L}^2$. For 300~fb$^{-1}$ of data, it is maybe 
more realistic to assume that half of the data was collected at a luminosity of 
10$^{34}$~cm$^{-2}$~s$^{-1}$ and half of the data was collected at 7.5$\times$10$^{33}$~cm$^{-2}$~s$^{-1}$. 
Although crude, this approximates the luminosity profile typical of a LHC store. Such a choice would 
reduce the dominant [p][X][p] background by 25\%. 

Further improvements related to the experimental efficiency with respect to reducible backgrounds have 
been investigated. Firstly, recent studies suggest that an improvement in b-tagging efficiency could be obtained 
with respect to gluons, improving the rejection of the CEP $gg \rightarrow gg$ background. Secondly, 
it is expected that the [pp][X] background contribution is overestimated. The calculation of this background depends 
crucially on the fraction of events at the LHC that produce two forward protons, $f_{[pp]}$. The cross 
section presented in Sec.~\ref{sec:results} uses the value of $f_{[pp]}$ predicted by the PHOJET event 
generator~\cite{phojet}, but other theoretical predictions result in a value of $f_{[pp]}$ which is more than an order 
of magnitude smaller~\cite{kmrN,Ryskin:2007qx}. In addition, the DPE central system must be about 100~GeV as the protons 
have to lose enough energy to produce a `missing mass' that is approximately the same size as the signal. 
In the analysis presented above the charged tracks from the [p] and [pp] vertices were not simulated. 
However, after fast-timing constraints, the [pp] vertex will be within 4.2~mm of the di-jet vertex and it 
is likely that additional charged tracks will cause the whole event to fail the charged multiplicity cuts 
outlined in Sec. \ref{sec:pilkoexp}. 

\begin{figure}
\centering
\mbox{
\subfigure[]{\includegraphics[width=.5\textwidth]{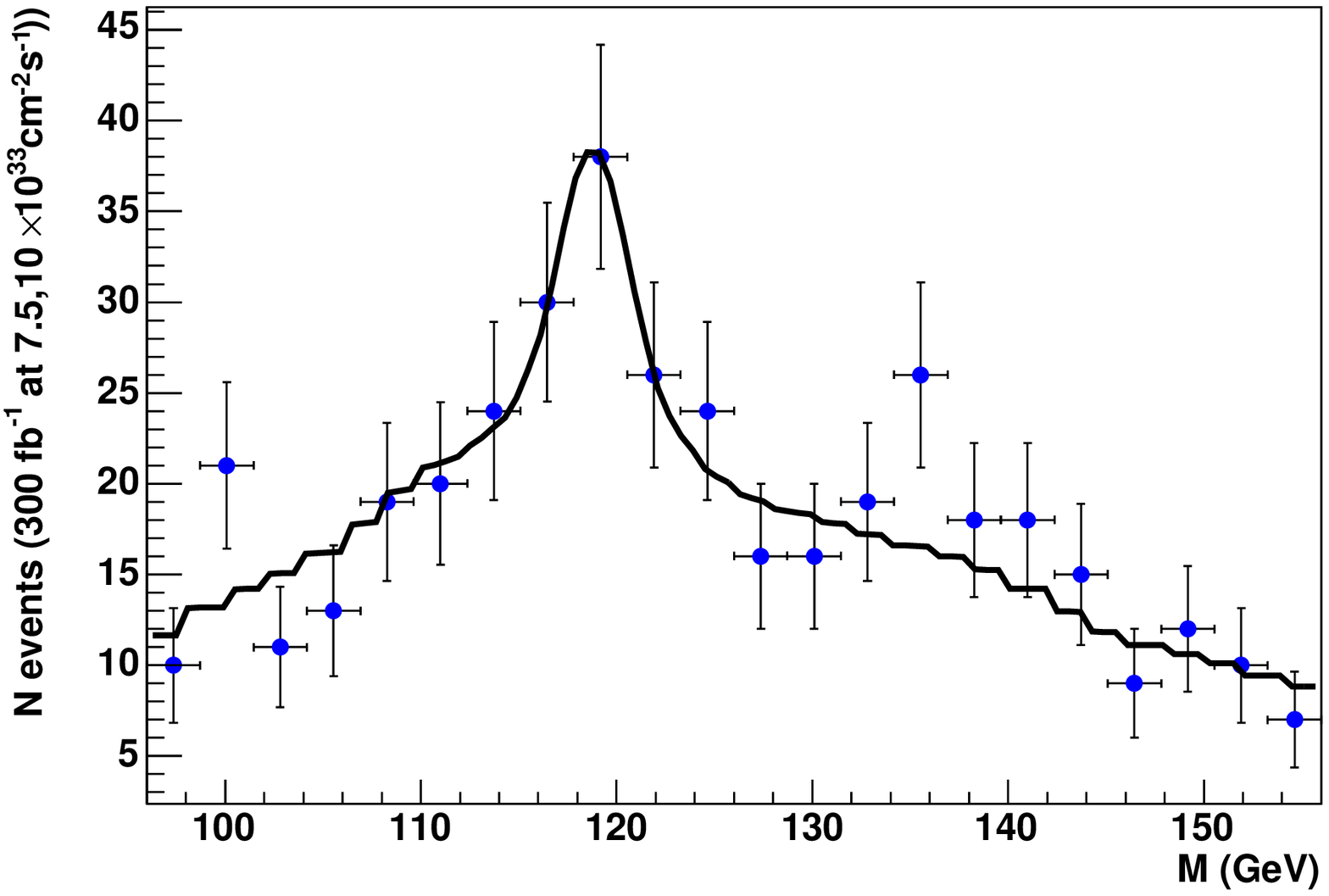}}
\subfigure[]{\includegraphics[width=.5\textwidth]{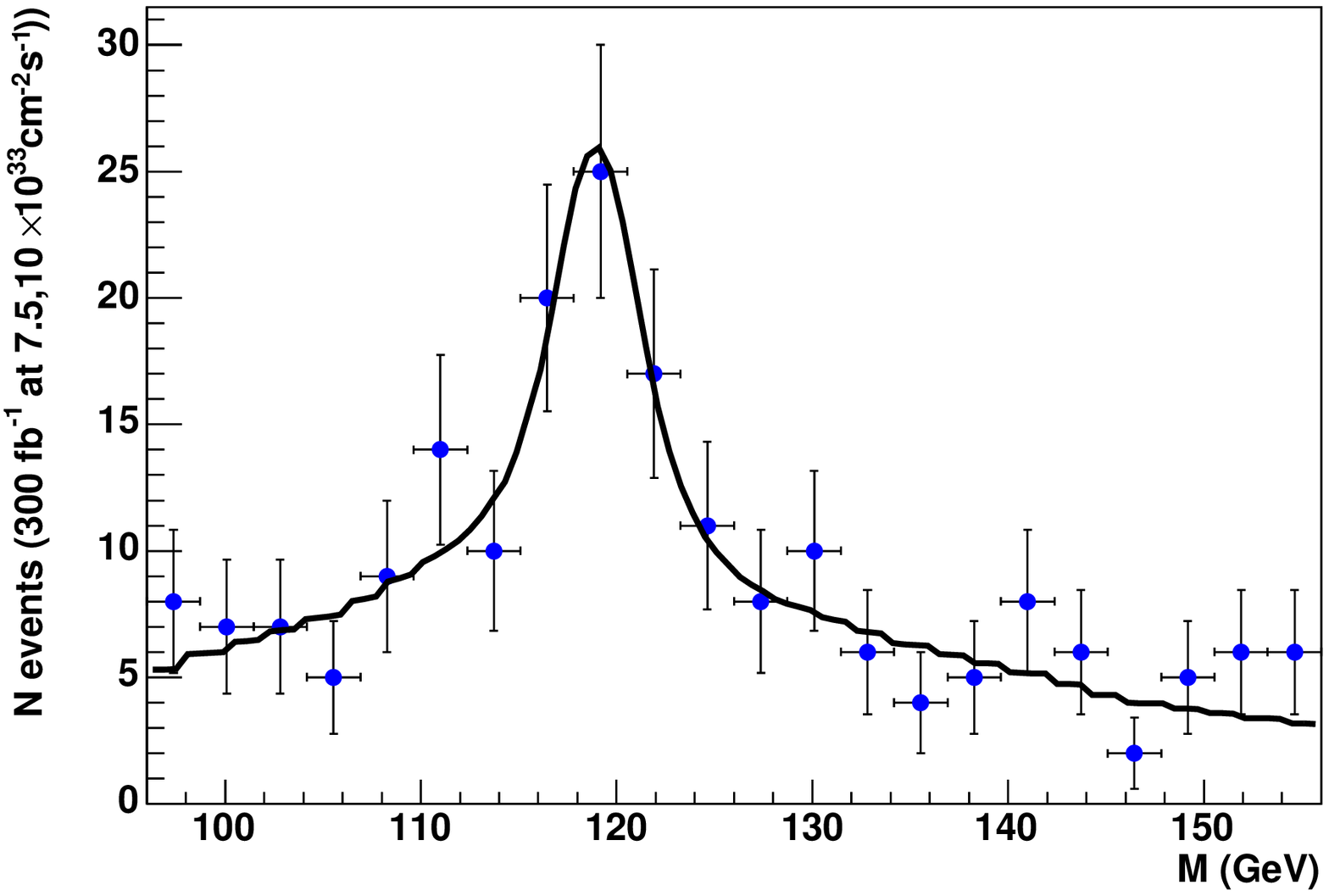}}
}
\caption{Mass fit for 300~fb$^{-1}$ of data for the improved background estimates described in the text 
(reduced CEP backgrounds, negligible [pp][X] and a luminosity profile consisting of half the data collected 
at 7.5$\times$10$^{33}$~cm$^{-2}$~s$^{-1}$ and half at 10$^{34}$~cm$^{-2}$~s$^{-1}$).
The plots are made assuming (a) baseline timing of 10ps and (b) improved timing of 5~ps or central timing.}
\label{fig:peak3}
\end{figure}

To estimate the effects of these improvements we have repeated the analysis detailed above, with the 
following modifications: (i) The CEP $b\bar{b}$ background is reduced by a factor of two, (ii) the gluon 
mis-tag probability is reduced from 1.3\% to 0.5\%, (iii) the [pp][X] background is assumed to be negligible, 
(iv) the luminosity profile is not fixed: For example for 300~fb$^{-1}$, half the data is assumed to be 
collected at 7.5$\times$10$^{33}$~cm$^{-2}$~s$^{-1}$ and half at 10$^{34}$~cm$^{-2}$~s$^{-1}$. 
Figure~\ref{fig:peak3} (a) shows the effect of these improvements given the baseline 10~ps fast-timing 
resolution and Fig.~\ref{fig:peak3} (b) shows the effects given a factor of two improvement in the fast-timing system (central timing or 5~ps forward timing resolution). The significance is increased to 
3.7$\sigma$ and 4.5$\sigma$ respectively from the 3$\sigma$ significance of Fig.~\ref{fig:peak}(b).

\newpage

\section{LHC Optics, acceptance, and resolution}
\label{sec:optics}

\subsection{Introduction}

\begin{figure}[htpb]
\centering
\includegraphics[width=13cm]{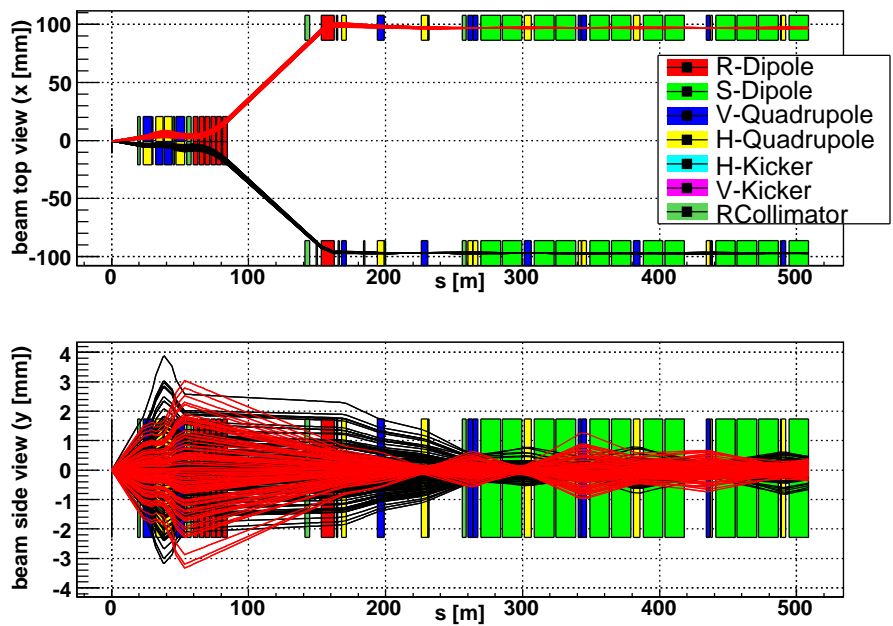}
\caption{Schematic plan view and side view of the beamline at IP5 (CMS); 
the IP1 configuration (ATLAS) is similar except that the 
kicker magnets are vertical (see also Fig.~\ref{fig:beampaths})~\cite{hector}.  
The horizontal curvature of the beamline has been straighted out 
for purposes of simplification here. }
\label{fig:beamline}
\end{figure}

The configuration of the LHC beamline around the interaction points is
shown schematically in Figure~\ref{fig:beamline}.  The proposed
forward detector stations are to be installed in the regions located
at approximately 220~m and 420~m from the IP1 and IP5 interaction
points in both beamlines downstream of the central detector.  Here
protons that have lost energy in the primary interaction are able to
emerge from the beamline. The acceptance and the ultimately achievable
experimental resolution of the forward detectors depends on the LHC
beam optics and on the position of the detectors relative to the beam.

\begin{figure}[htpb]
\centering
\includegraphics[width=6.5cm]{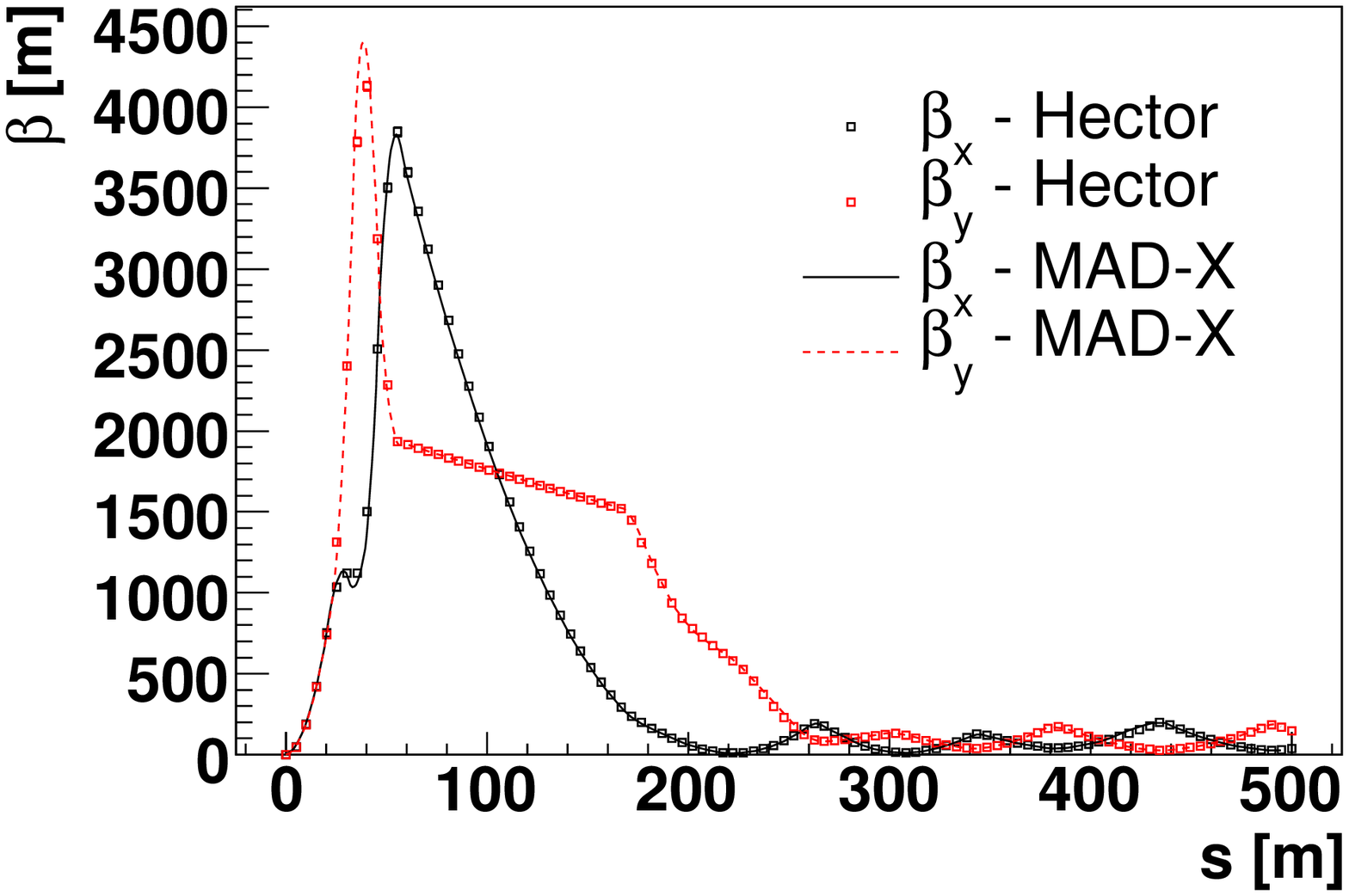}
\includegraphics[width=6.5cm]{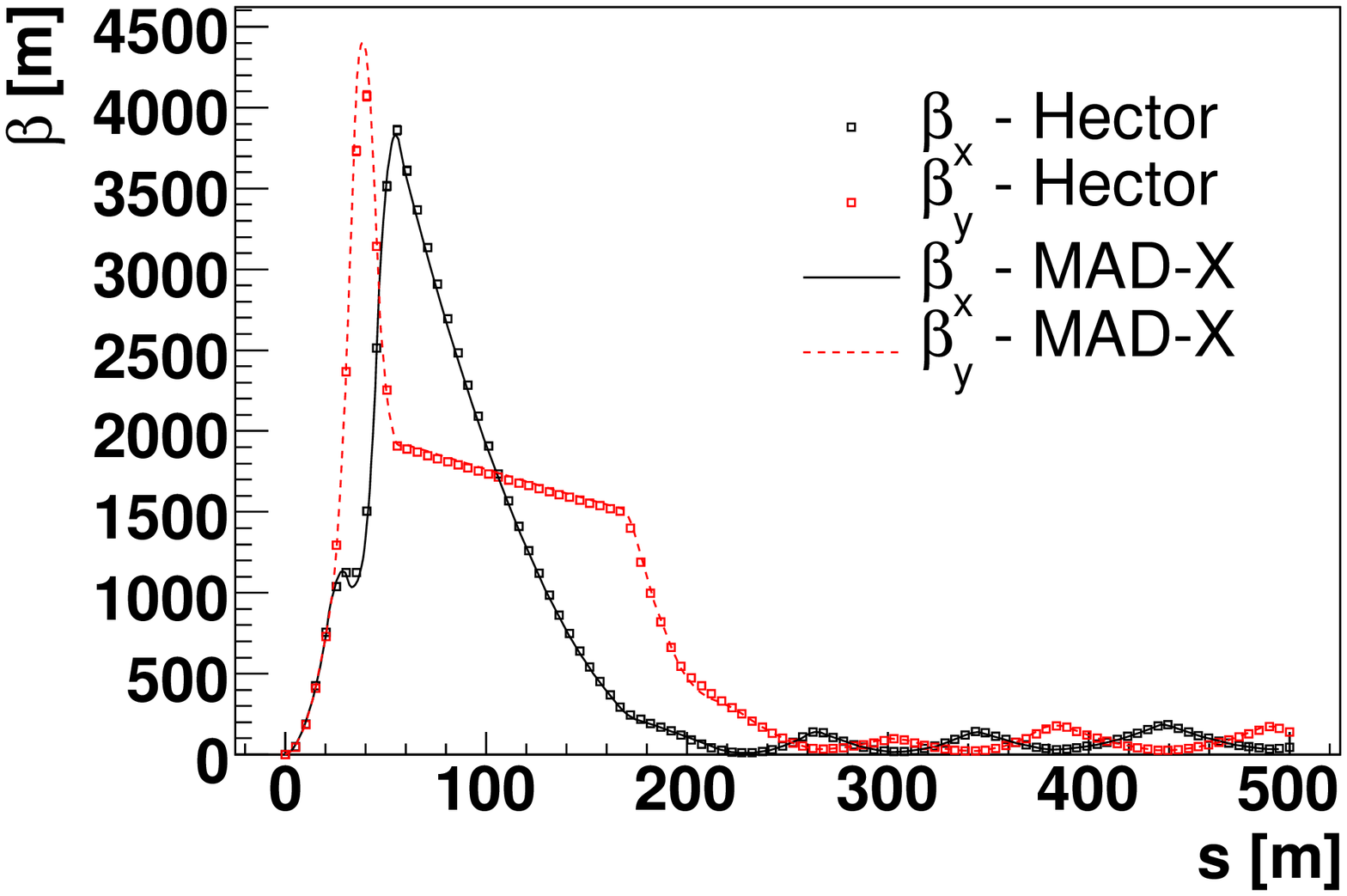}
\caption{Beta functions $\beta_x$ (horizontal) and $\beta_y$ (vertical) for
LHC beam 1 (left) and beam 2 (right) calculated by MAD-X (lines) and HECTOR (squares).}
\label{fig:MADXHector}
\end{figure}

\begin{figure}[htpb]
\centering
\includegraphics[width=8cm]{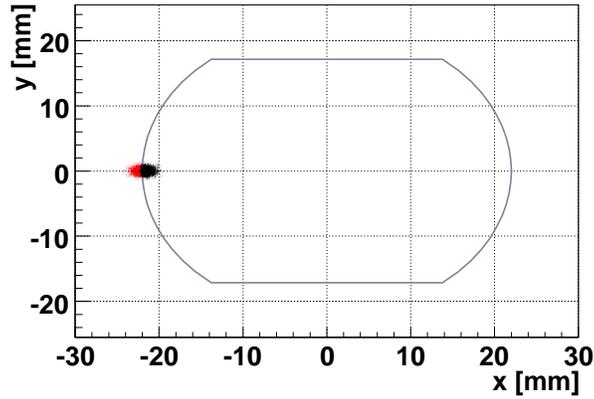}
\caption{Example of aperture check for a typical Main Bending dipole
at 338m from IP5, for a set of protons with a mean energy loss of
110~GeV. The protons which exit the aperture are shown in black, and
those which hit the walls are shown in red.}
\label{fig:app}
\end{figure}

The FP420 Collaboration has written two independent proton tracking
programs, FPTrack~\cite{fptrack} and HECTOR~\cite{hector}, and
implemented a model of the LHC beamline into the package
BDSIM~\cite{ref:bdsim} in order to simulate machine-induced
backgrounds. The BDSIM model is described in detail in section
\ref{sec:machineinduced}.  The three simulations, FPTrack, HECTOR and
BDSIM are in good agreement with each other and with MAD-X, the
standard LHC beam transport program used at CERN.
Figure~\ref{fig:MADXHector} shows the $\beta$ functions for beams 1
and 2 as computed by HECTOR and compared to MAD-X. (Comparison with
MAD-X strictly verifies the tracking programs only for 7 TeV protons.) 
HECTOR has also been verified for protons above 80\% of the nominal
beam energy (i.e. all protons within the acceptance of 220~m and 420~m
detectors) by direct comparison to numerical
calculations~\cite{hector}. All the programs perform aperture checks
through each of the LHC optical elements.  Figure~\ref{fig:app}
illustratively shows the losses occurring for a set of protons with
mean energy loss of 110~GeV in the MB.B9R5.B1 dipole at 338 m from IP5
using LHC beam 1 optics. It is aperture restrictions of this kind that
chiefly limit the high-mass acceptance of FP420.

Unless otherwise stated, we use the ExHuME Monte Carlo~\cite{exhume}
to generate outgoing protons from the central exclusive production
of a SM Higgs boson, although the results apply for any
centrally-produced system of the same mass. Version 6.500 of the LHC
optics files have been used with: $\beta^* = 0.55$ m; angular
divergence $\sigma_{\theta} = 30.2\;\mu$rad at the IP; crossing
angle = 142.5 $\mu$rad in the vertical (horizontal) plane at IP1
(IP5); beam energy spread $\sigma_E = 0.77$~GeV. Full details can be
found in~\cite{hector}. The energy spread of the 7000~GeV beam is an
irreducible limiting factor on the mass resolution that can be
obtained by proton tagging detectors at the LHC.  We show
acceptances below for both 420~m alone and for 420~m and 220~m stations
operating together.

\begin{figure}[htpb]
\centering
\includegraphics[width=6.5cm]{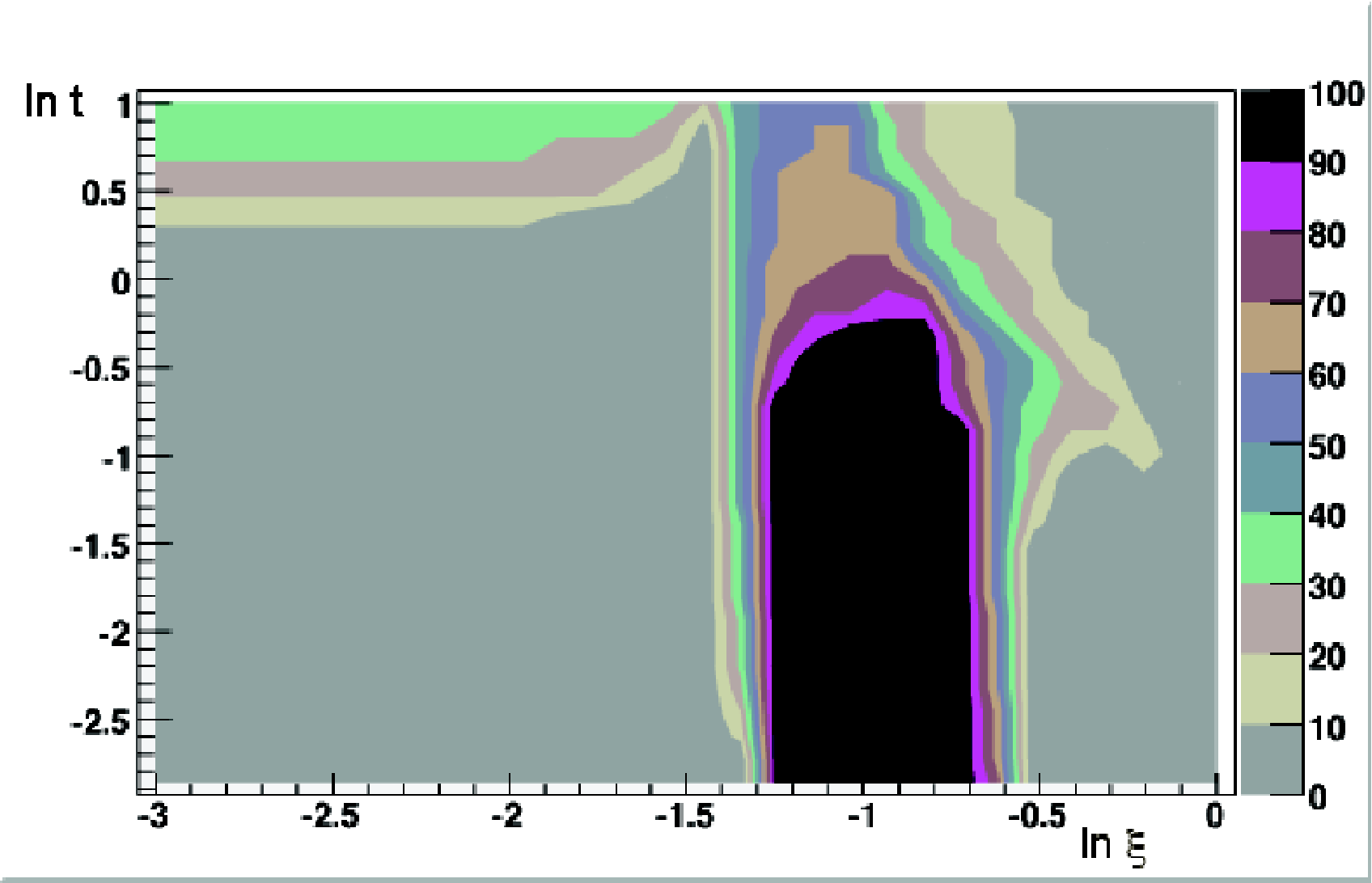}
\includegraphics[width=6.5cm]{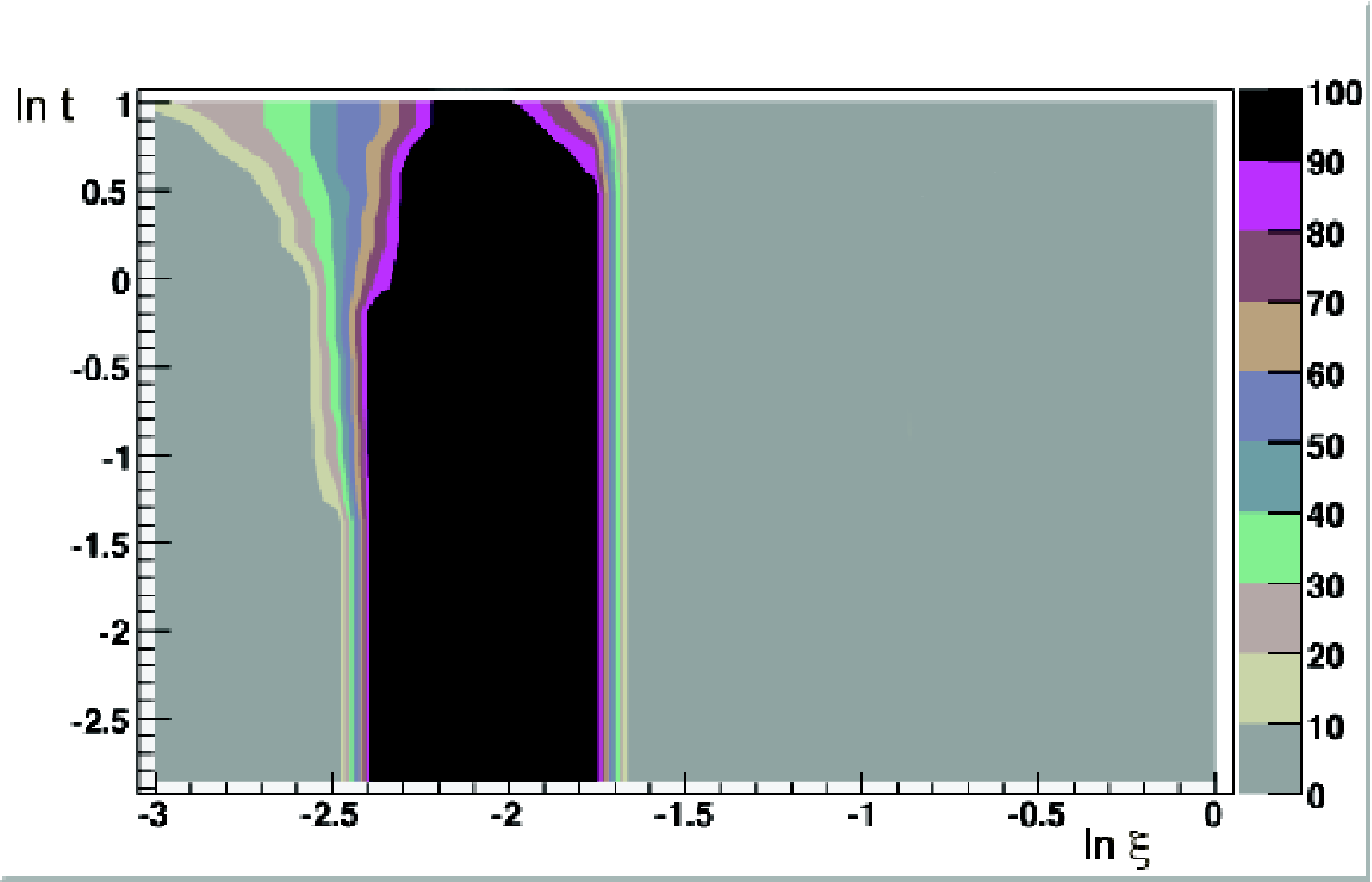}\\
\includegraphics[width=6.5cm]{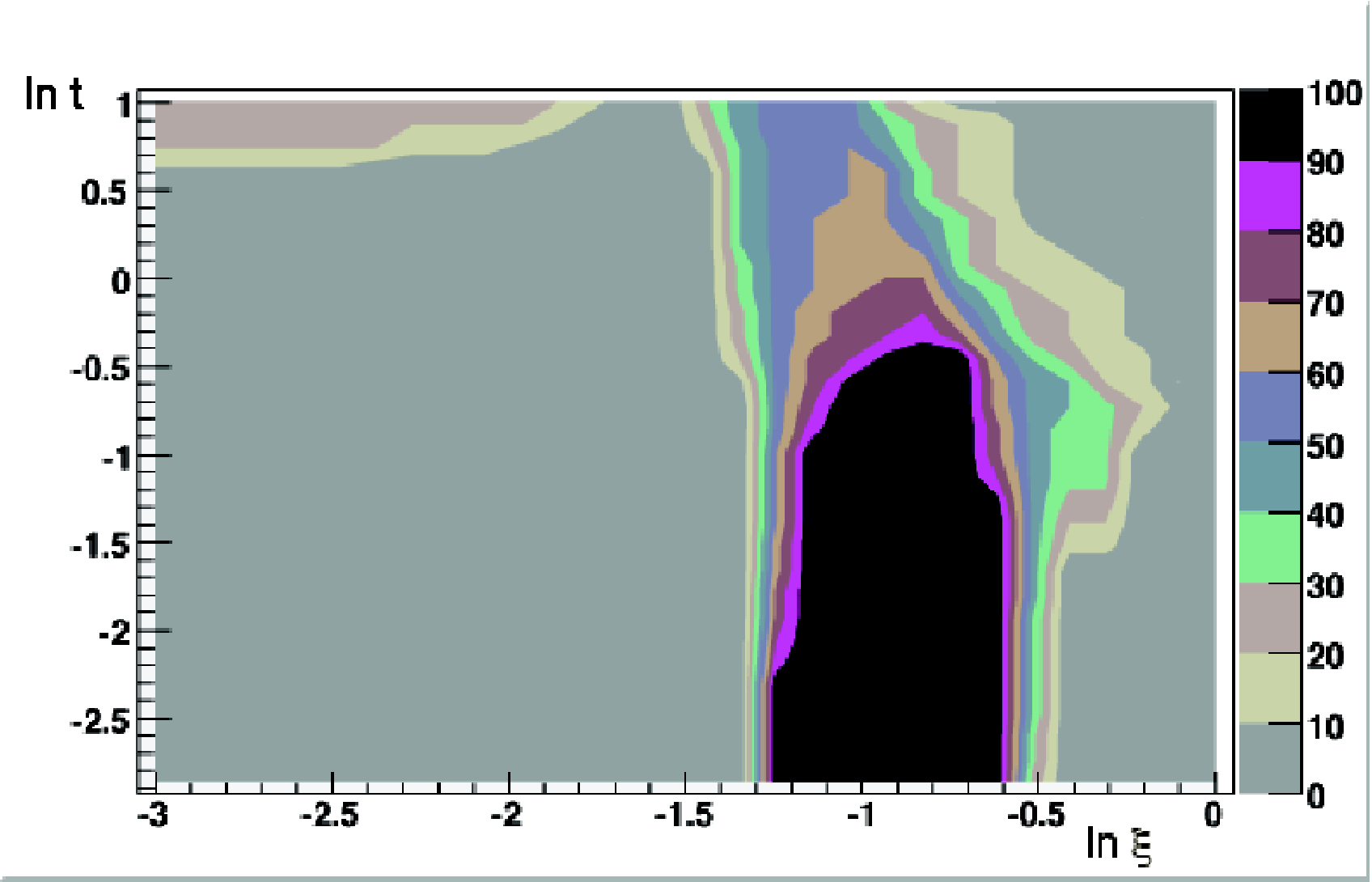}
\includegraphics[width=6.5cm]{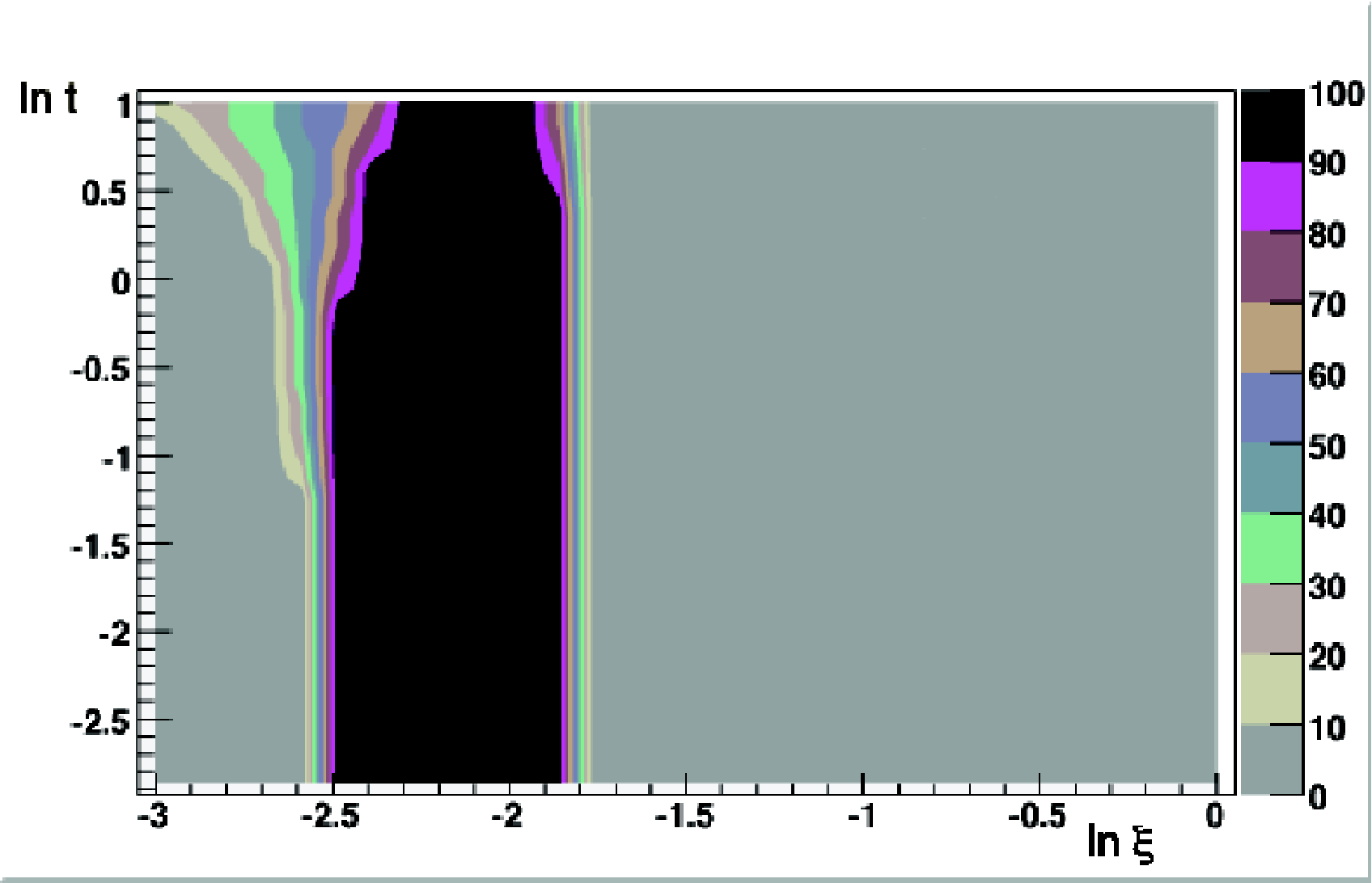}
\caption{Acceptances in the $\xi,\,t$ plane for protons to reach planes at 220~m 
(left) and 420~m (right) for beam-1 (top) and beam-2 (bottom) around IP1 (ATLAS)
computed with FPTrack. 
The variables plotted are $t$, the modulus of the squared momentum transfer to the 
proton at the IP, and $\xi$, its fractional energy loss. No detector effects are included here.}
\label{fptrackbeam}
\end{figure}

\subsection{Detector acceptance}

\begin{figure}[htpb]
\centering
\includegraphics[width=6.5cm]{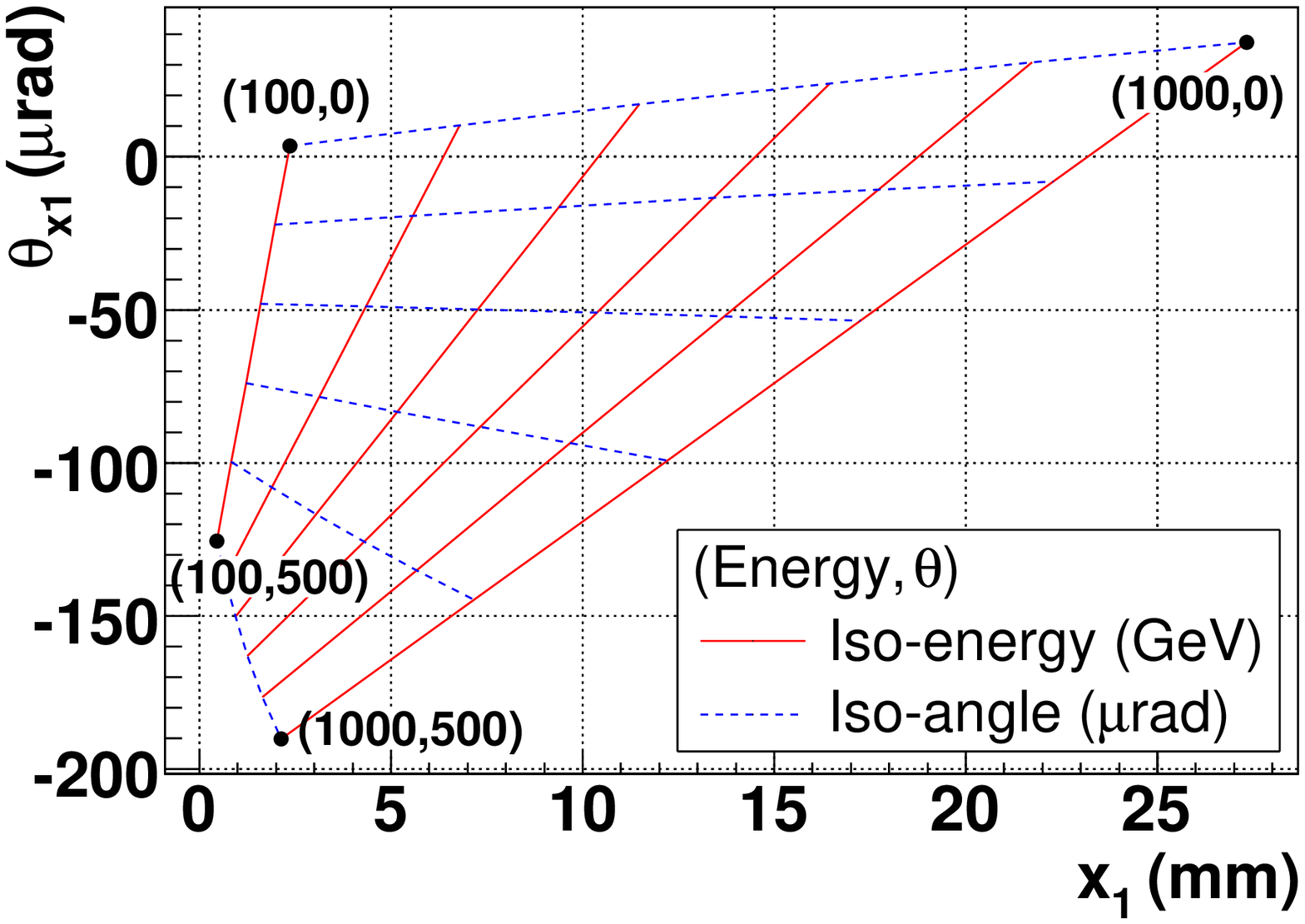}
\includegraphics[width=6.5cm]{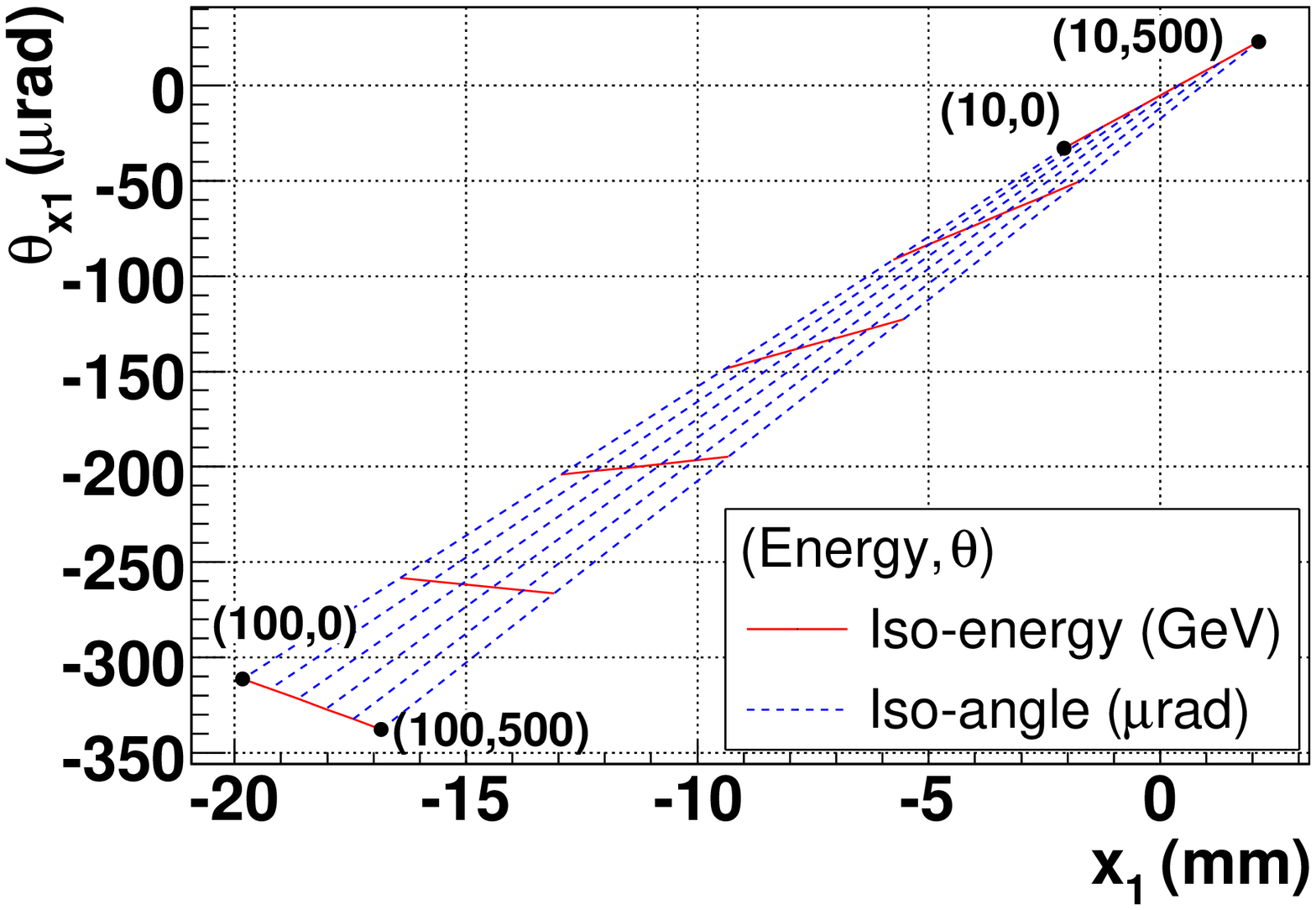}
\caption{Chromaticity grids: iso-energy and iso-angle lines for 220~m (left) and
420~m (right) detectors at IP5, beam 1 computed with HECTOR. The vertical axis
$\theta_{x1}$ is the angle of the scattered proton relative to the
beam at 220 or 420~m, and the horizontal axis $x_1$ is the horizontal
displacement of the scattered proton relative to the beam. The solid lines
are iso-energy lines, ranging from proton energy loss 0~GeV to 1000~GeV 
in steps of 200~GeV at 220~m, and from 0~GeV to 100~GeV in steps
of 20~GeV at 420~m. The dotted lines are lines of constant proton emission
angle at the interaction point, and range from 0~$\mu$rad to 500~$\mu$rad 
in steps of 100~$\mu$rad.}
\label{fig:chroma}
\end{figure}

The position and direction of a proton in the 220~m and 420~m
detectors (for a given LHC optics) depend on the energy and
scattering angle of the outgoing proton and the $z$-vertex position of
the collision. The energy and scattering angle are directly related
to the kinematic variables $\xi$, the fractional longitudinal
momentum loss of the outgoing proton, and $-t$, the square of the
four-momentum transfer.  Figure~\ref{fptrackbeam} shows the
acceptance in the $\xi$-$t$ plane for the 220~m and 420~m regions
for beam 1 and beam 2 respectively, around IP1 (ATLAS), as
calculated by FPTrack. The mapping of the energy loss and outgoing
angle of a proton at the IP to a position and angular measurement in
the detector at 220~m or 420~m can be visualised using chromaticity
grids. Figure~\ref{fig:chroma} shows iso-energy and iso-angle curves
for protons with energy loss ranging from $0$ to $1000$~GeV in steps
of $200$~GeV at 220~m (left), and from $0$ to $100$~GeV in steps of
$20$~GeV at 420~m (right). The angle of the outgoing proton at the
IP ranges from $0$ to $500$ $\mu$rad in steps of 100~$\mu$rad.  The
angle of the proton track measured at the detector $\theta_{x}$ is
shown on the vertical axis, and the horizontal position from the
beam, $x$, is shown on the horizontal axis. The non-linear nature of
the grids is due to the energy dependence of the transfer matrices,
without which the grid would be a parallelogram.
The chromaticity grids show that the measurement of the energy of the
outgoing proton requires good position and angle measurements in the
detector stations.  A measurement of the angle of the outgoing protons
from the IP, and hence $p_T$, at 420~m requires a far better spatial
resolution than the energy ($\xi$) measurement. This can be seen, for
example, by noting that the separation in $x_1$ of the (10~GeV, 0~$\mu$rad) 
and (10~GeV, 500~$\mu$rad) fixed-energy points is much
smaller than that of the (10,0) and (100,0) fixed-angle points. We
return to this issue below when discussing the required measurement
precision. We expect to achieve $\sim 1 \mu$rad precision on $\theta_x$.

\begin{figure}[htpb]
\centering
\includegraphics[width=6.5cm]{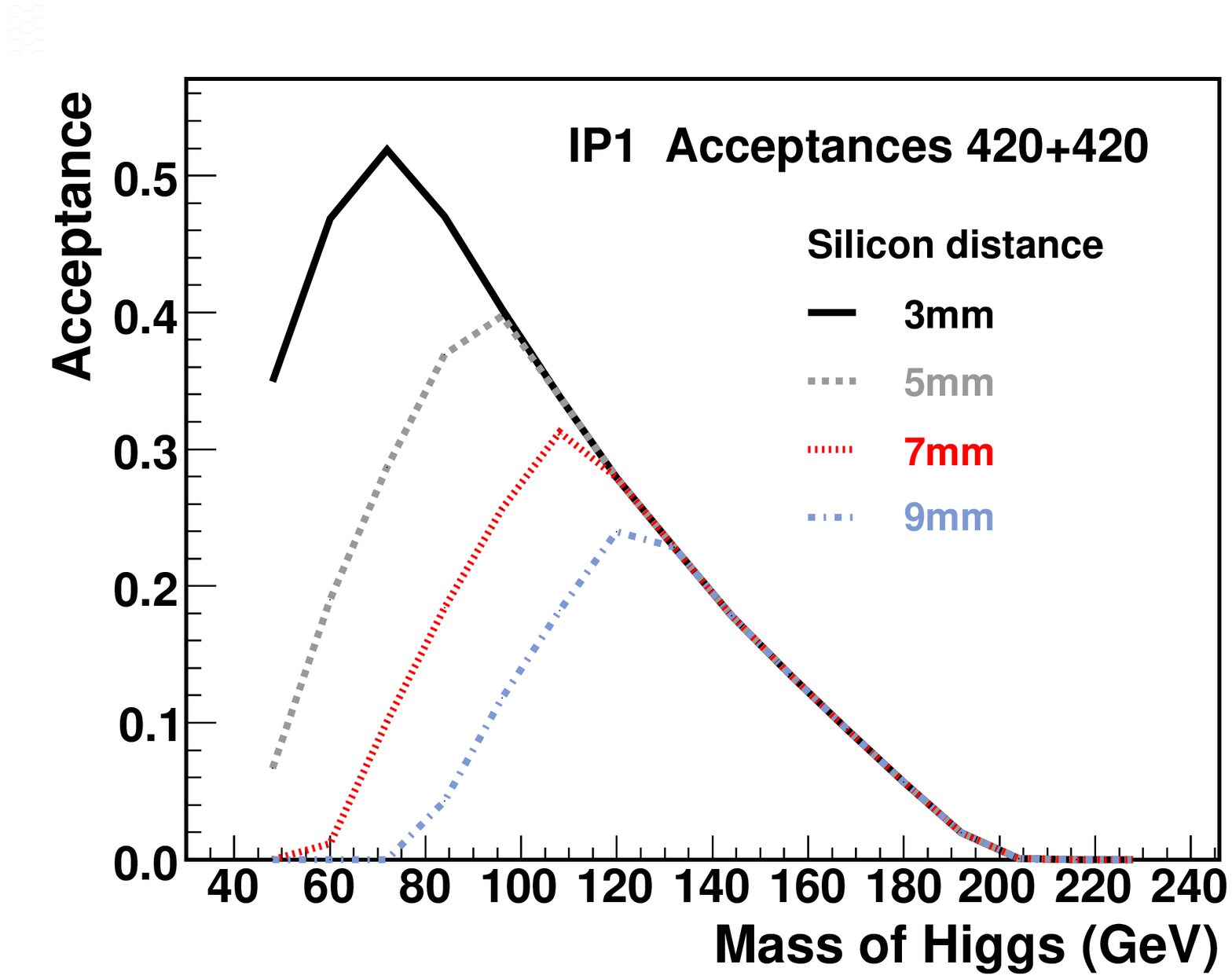}
\includegraphics[width=6.5cm]{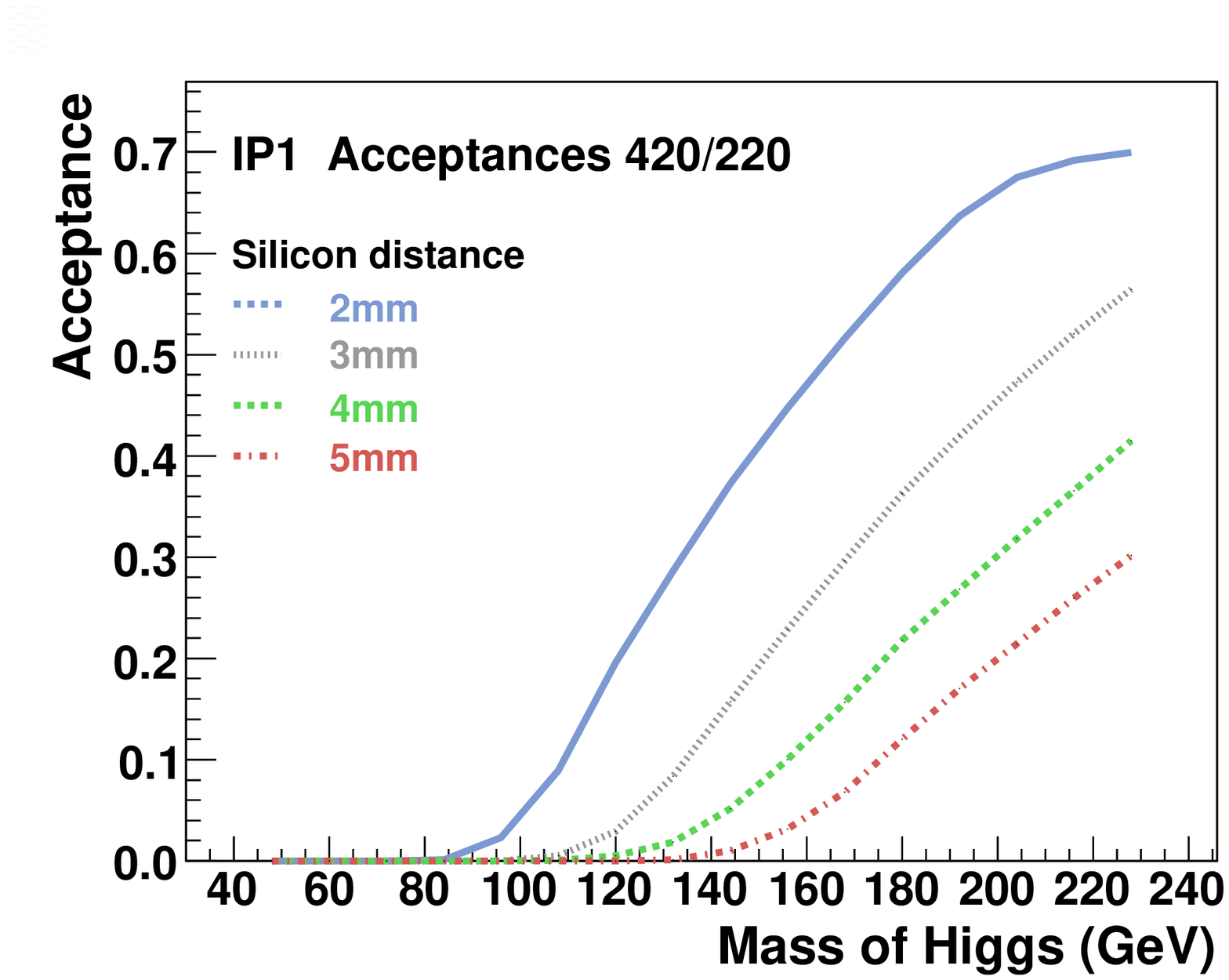}
\caption{Acceptance as a function of centrally produced
 mass for (left) 420 + 420~m proton tags for the silicon detector active edge
 positioned at different distances from the beam;
 (right) for 220 + 420~m proton
 tags with the 420~m silicon at 5 mm from the beam and the 220~m at different
 distances from the beam.  The small upward deviation at high mass for the 2 mm
 silicon positions, show the additional acceptance from 220 + 220~m coincidences.}
\label{distance_beam}
\end{figure}

The low-$\xi$ (and therefore low mass) acceptance depends critically
on the distance of approach of the active area of the silicon
sensors from the beam. This is shown in Fig.~\ref{distance_beam} for
proton tags at 420~+~420~m and 420~+~220~m.  It is clear from the
left hand plot in Fig.~\ref{distance_beam} that operating as far
away as 7.5~mm does not compromise the acceptance for central masses
of 120~\GeVcc\ and above, for 420~+~420~m tagged events. Acceptance at
higher masses requires the 420~m detectors to be used in conjunction
with 220~m detectors. For this configuration, however, the
acceptance becomes more sensitive to the distance of approach for
masses in the 120~\GeVcc\ range (Fig.~\ref{distance_beam} right). This
is because the 220~m detectors have acceptance only at relatively
high $\xi$ (Fig.\ \ref{fptrackbeam}), forcing the proton detected at
420~m to have low-$\xi$, and therefore to be closer to the beam. As
we shall see in the following sections, the possible distance of
approach depends on the beam conditions, machine-induced backgrounds
and collimator positions, and the RF impact of the detector on the
LHC beams. Such studies have been performed by us only for 420~m
stations -- for further details on the current and proposed 220~m
designs see Refs.~\cite{ATLAS220m,Albrow:2006xt}. It is envisaged that
the 220~m detectors will be able to operate as close as 1.5 mm from
the LHC beams~\cite{ATLAS220m}. At 420~m the nominal operating position
is assumed to be between 5 mm and 7.5 mm, depending on beam
conditions. This is discussed further in Sections
\ref{sec:backgrounds} and \ref{sec:RF}. For central masses above 150~\GeVcc\ 
or so, the inclusion of 220~m detectors becomes increasingly
important.

\begin{figure}[htpb]
\centering
\includegraphics[width=6.5cm]{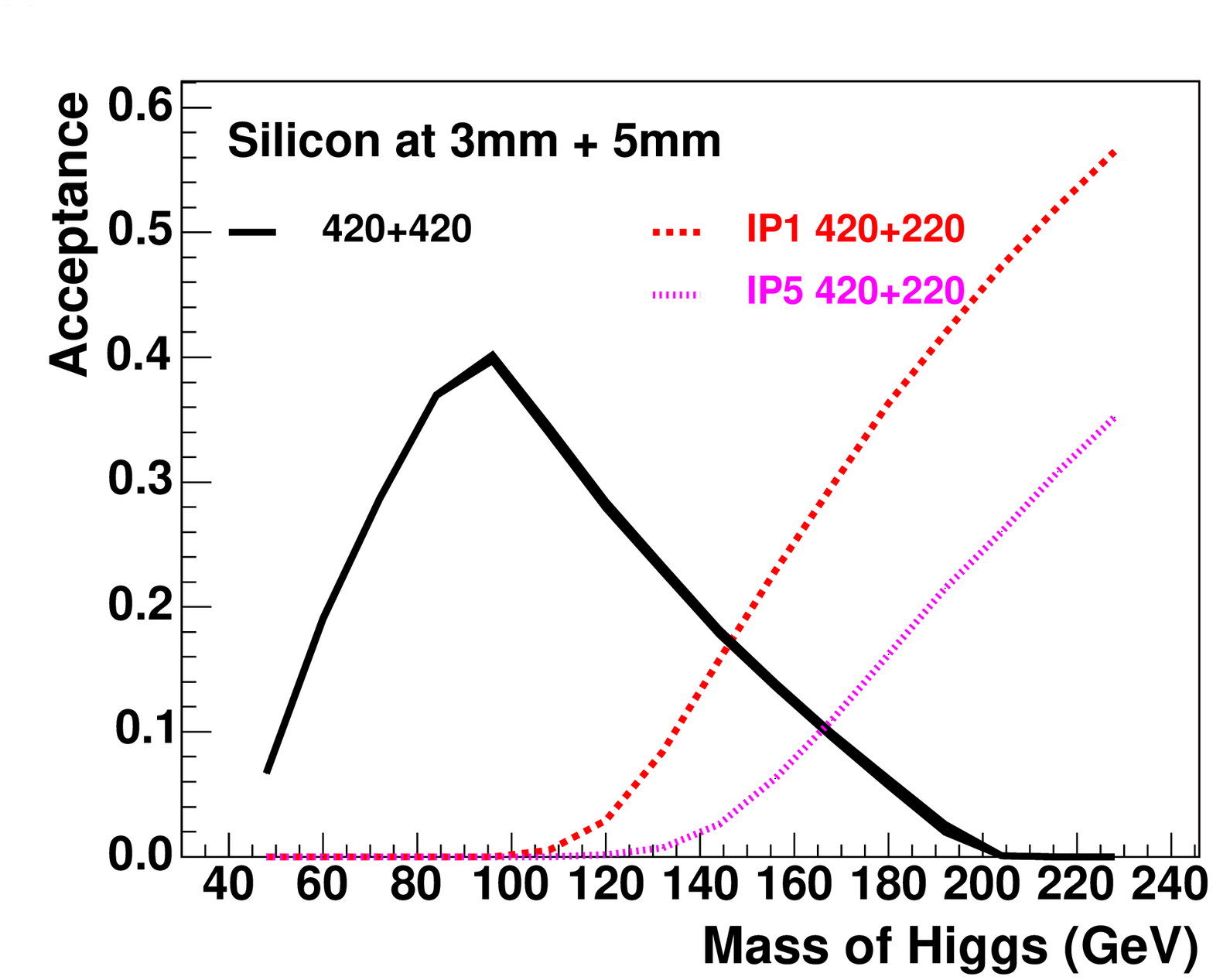}
\includegraphics[width=6.5cm]{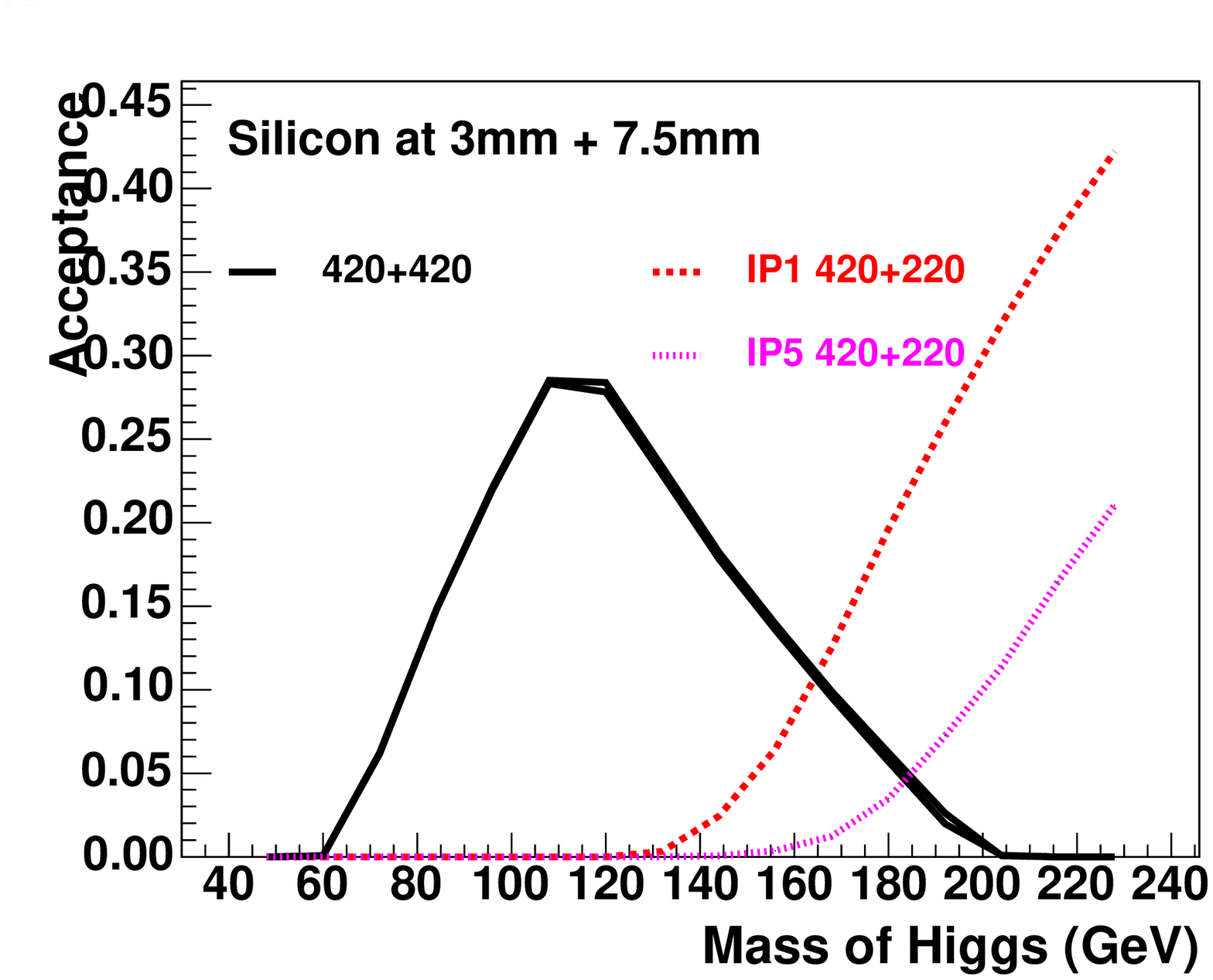}
\includegraphics[width=6.5cm]{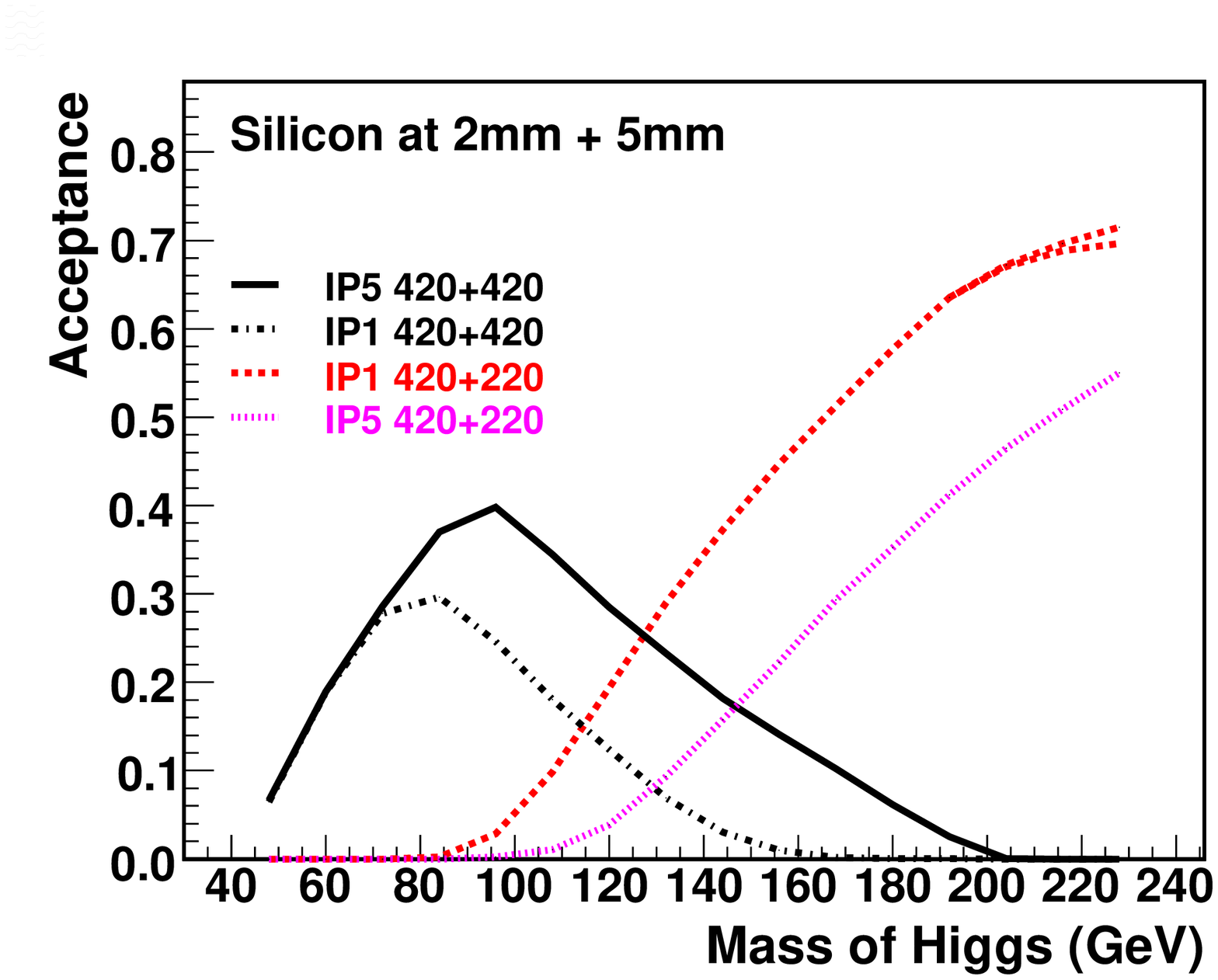}
\includegraphics[width=6.5cm]{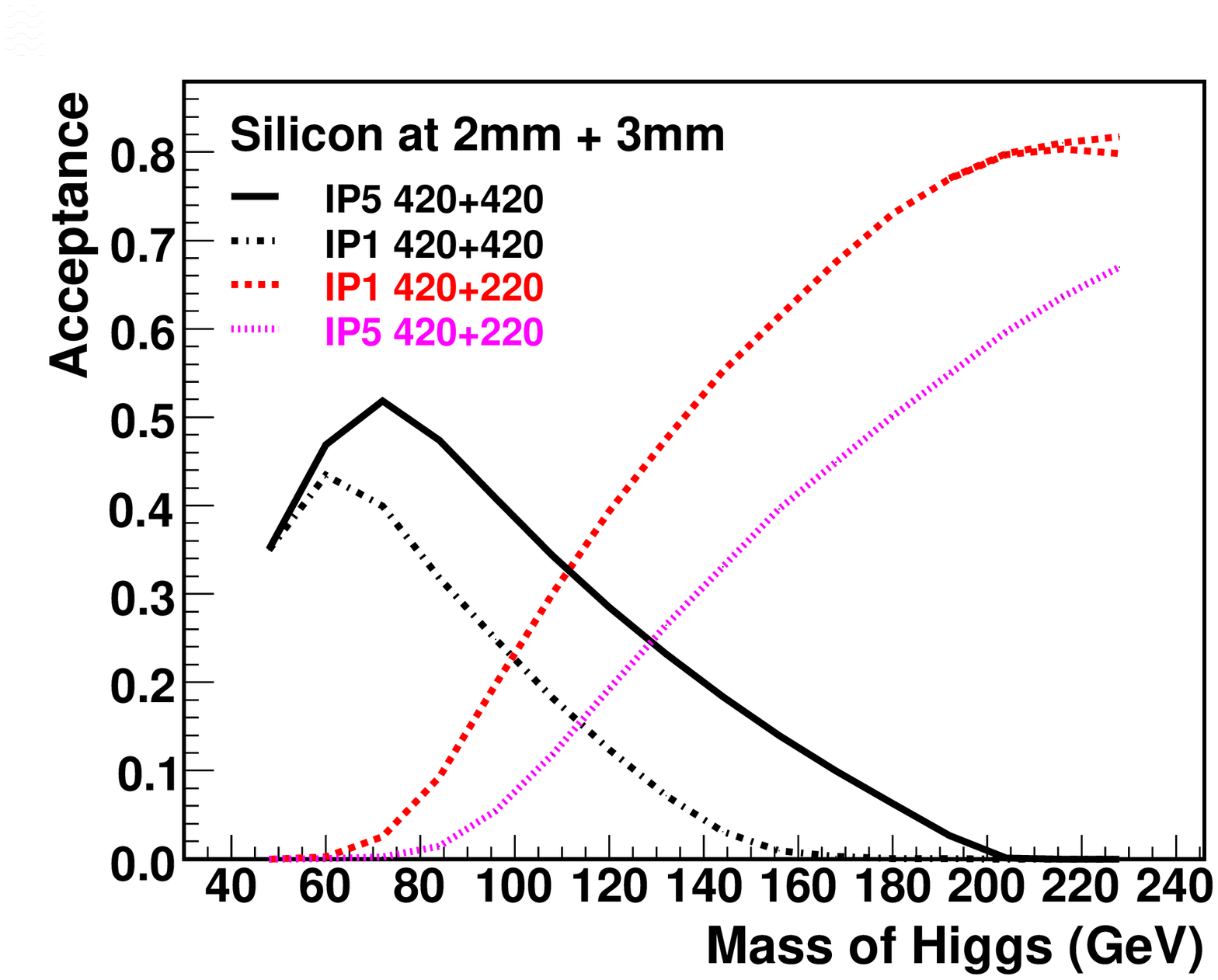}
\caption{Acceptances as a function of Higgs boson mass with
detector active edge at various distances from the beam centre at
420~m for IP1 (dotted line) and IP5 (dashed line). Also shown 
is the acceptance for events with one proton detected at 220~m 
and one proton at 420~m (or also 220~m, upper branch) .
The smaller distance in the legend is always the 220~m distance.}
\label{f5_2}
\end{figure}

Figure~\ref{f5_2} shows several interesting features of the
acceptance, including differences between the IP1 (ATLAS) and IP5
(CMS) regions. The upper plots show that if the 220~m detectors are
sufficiently far from the beam (3 mm in this case) then there is
negligible difference in 420 +420~m acceptance between IP1 and IP5,
and beam 1 and beam 2. The fact that the crossing angle is in the
vertical plane at IP1 and the horizontal plane at IP5, however,
results in a higher acceptance at IP1 than IP5 for 420 + 220~m
event\footnote{Right now the different choice of crossing plane at the IPs 
leads to a reduced acceptance for IP5, but it would be possible to use the same
crossing plane at both IPs with some minor modifications to the LHC 
around IP5.}, as shown in Fig.~\ref{f5_2}. The bottom plots show that for
closer insertions at 220~m (2 mm in this case), there is a decrease
in the 420 +420~m acceptance for the IP1 region, due to the dead
region (from the thin vacuum window) of the 220~m detectors
intercepting protons that would otherwise be detected at 420~m. This
dead region is taken as 0.7 mm in the acceptances shown in the
figure, and has negligible affect for clearances of more than 2 mm
from the beam line at 220~m.  The accuracy of the proton momentum
measurement (see next section) is higher at 420~m than at 220~m, so
the operating conditions at 220~m must be chosen so as to achieve an
optimum balance between the mass resolution and acceptance.

\subsection{Mass resolution} 
\label{sec:resolutions}

\begin{figure}[htpb]
\centering
\includegraphics[width=7.5cm]{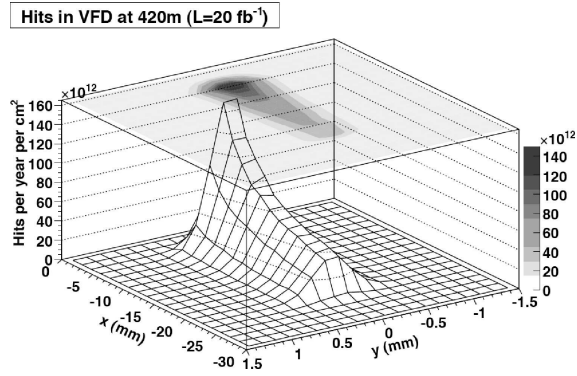}
\caption{Number of proton hits 
due to the process pp $\rightarrow$ pX for 20~fb$^{-1}$ integrated luminosity. Protons
were generated with PYTHIA 6.2.10 (single diffraction process 93) 
and tracked through the beam lattice with HECTOR.}
\label{p_rates}
\end{figure}

\begin{figure}[htpb]
\centering
\includegraphics[width=6.5cm]{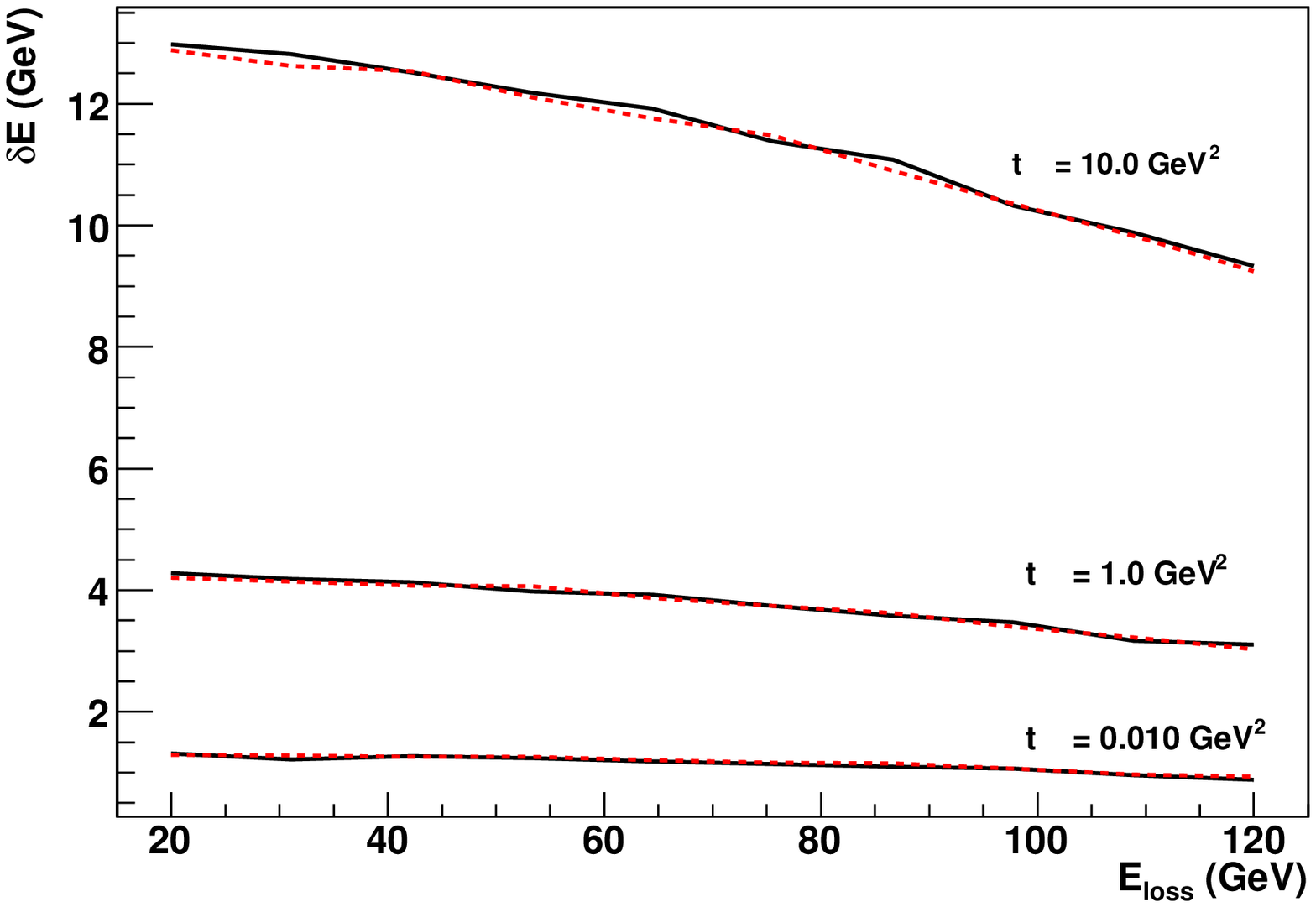}
\includegraphics[width=6.5cm]{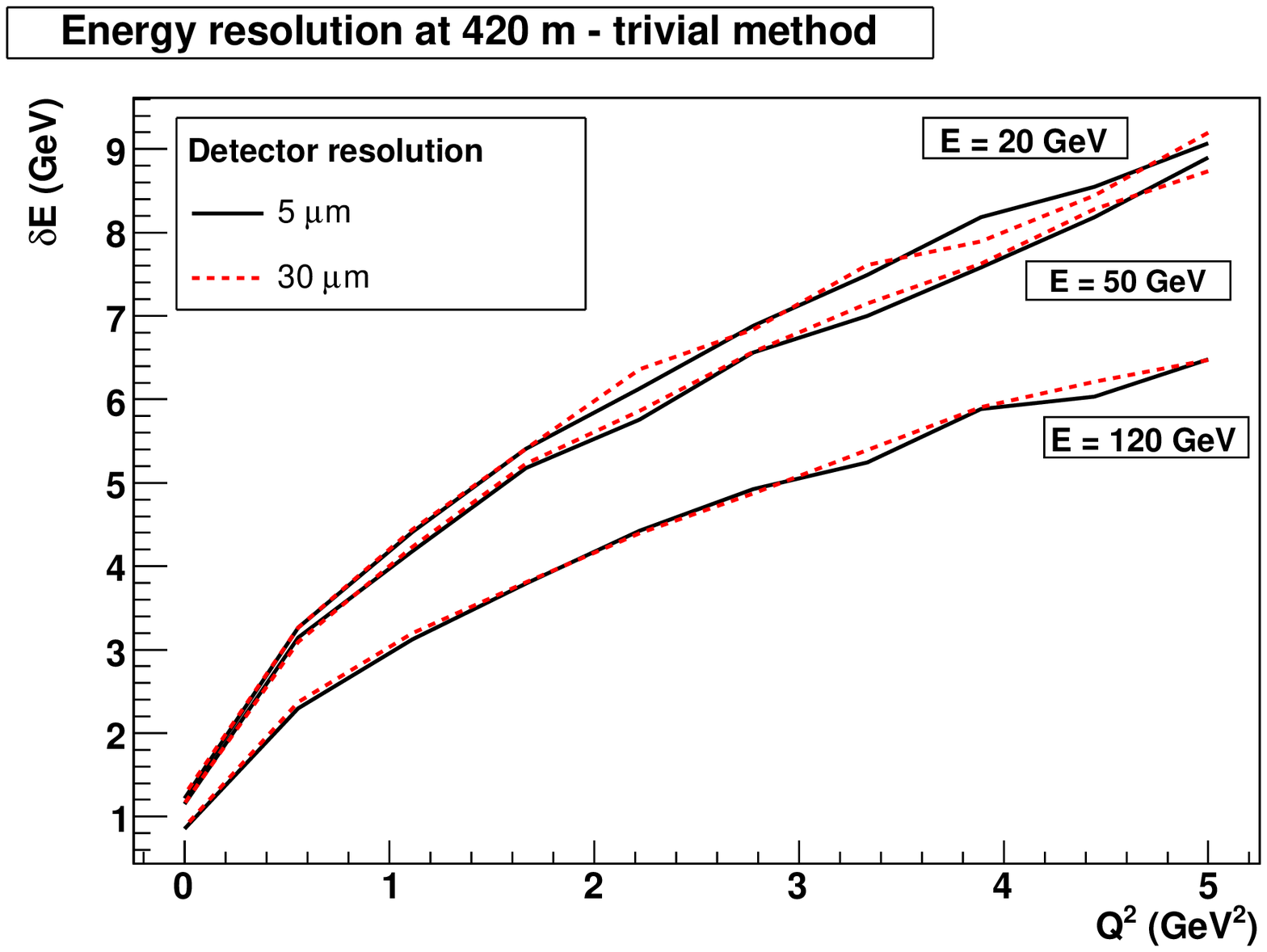}
\caption{ Energy resolution for the simple
reconstruction method described in the text for protons at 420~m. The
resolution is shown as a function of energy loss (left) and $Q^2
(=-t)$ (right). Also shown is the effect of varying the error on the
positional measurement on the detectors from 5~$\mu$m to 30~$\mu$m.}
\label{trivial}
\end{figure}

Typical $x-y$ distributions of hits in a detector at 420~m are shown in Fig.~\ref{p_rates}.
The distribution extends over the full horizontal width of the detector
but is narrowly confined vertically. Note that the detector sensitive area need only be $\sim$2~mm 
(V) $\times$ 20~mm (H). In practice we will use a larger vertical area to allow for beam 
position variations. From measurements in a minimum
of two stations in each region, the mean position and direction of
the scattered protons can be determined.  The position and angle in
the $x$-$y$ plane of a proton at any point along the beam-line can
be used to measure its energy loss and $p_T$ at the interaction
point. A simple reconstruction method for the energy of the detected
proton has been studied which takes account only of the dispersion;
here a polynomial fit is performed for the proton energy as a
function of the horizontal position at the detector (Fig.~\ref{trivial}). 
As seen in Fig.~\ref{fig:chroma}, however, an
angular measurement in the horizontal plane $\theta_x$ is required
to give good momentum reconstruction accuracy; this must be
particularly precise at 420~m because the iso-angle lines are highly
compressed.  A precision of at least $\pm1\,\,\mu$rad is necessary
and appears attainable (see Section~\ref{sec:silicon_perform}); the tracks are 
measured over an 8~m lever arm with $<$80~$\mu$m precision at front and back.

For optimal results, polynomial-based parametrization formulae have
been developed in order to evaluate the proton momenta from the
measured parameters in the silicon detectors.  The formulae are
based on fits to the calculated positions and angles, using the
generated values of the momentum and emission angle at the IP, and
averaging over the width of the beam-beam interaction region. From
the momenta of the pair of oppositely emerging protons in an event,
the mass of the centrally produced system can then be calculated by
a missing-mass formula~\cite{alb}. Using these parametrizations we
have evaluated the resolution achievable on the missing mass of a
diffractively produced object. Minimizing this resolution is
critical to the physics capabilities of the proposed new detectors.
We present results for a vertex at $z = 0$, but there is no significant 
dependence on the $z$ vertex within the interaction region. 
To allow for any dependence on $x$ we note that this will be well measured
by the central detector for every event, and is expected to be quite stable within a run. 
Thus offline corrections for these variations are easily applied. The
residual event-by-event variation of the $x$ position is taken into
account below in the mass resolution calculation.

The following factors affect the measured resolution of a narrow
object produced in the exclusive double diffraction process:
\begin{itemize}
\item The Gaussian width of the momentum distribution of the circulating
proton beam.  This is specified as 0.77~GeV/c.
\item The lateral uncertainty of the position of the interaction
point. This is taken to be 11.8 $\mu$m  from the intrinsic beam width,
but could be improved if the central silicon detector system provides
a better measurement on an event-by-event basis.
\item The position measurement uncertainty in the RP system
\item The angular measurement uncertainty in the RP system.
\end{itemize}

\begin{figure}[htpb]
\centering
\includegraphics[width=6.5cm]{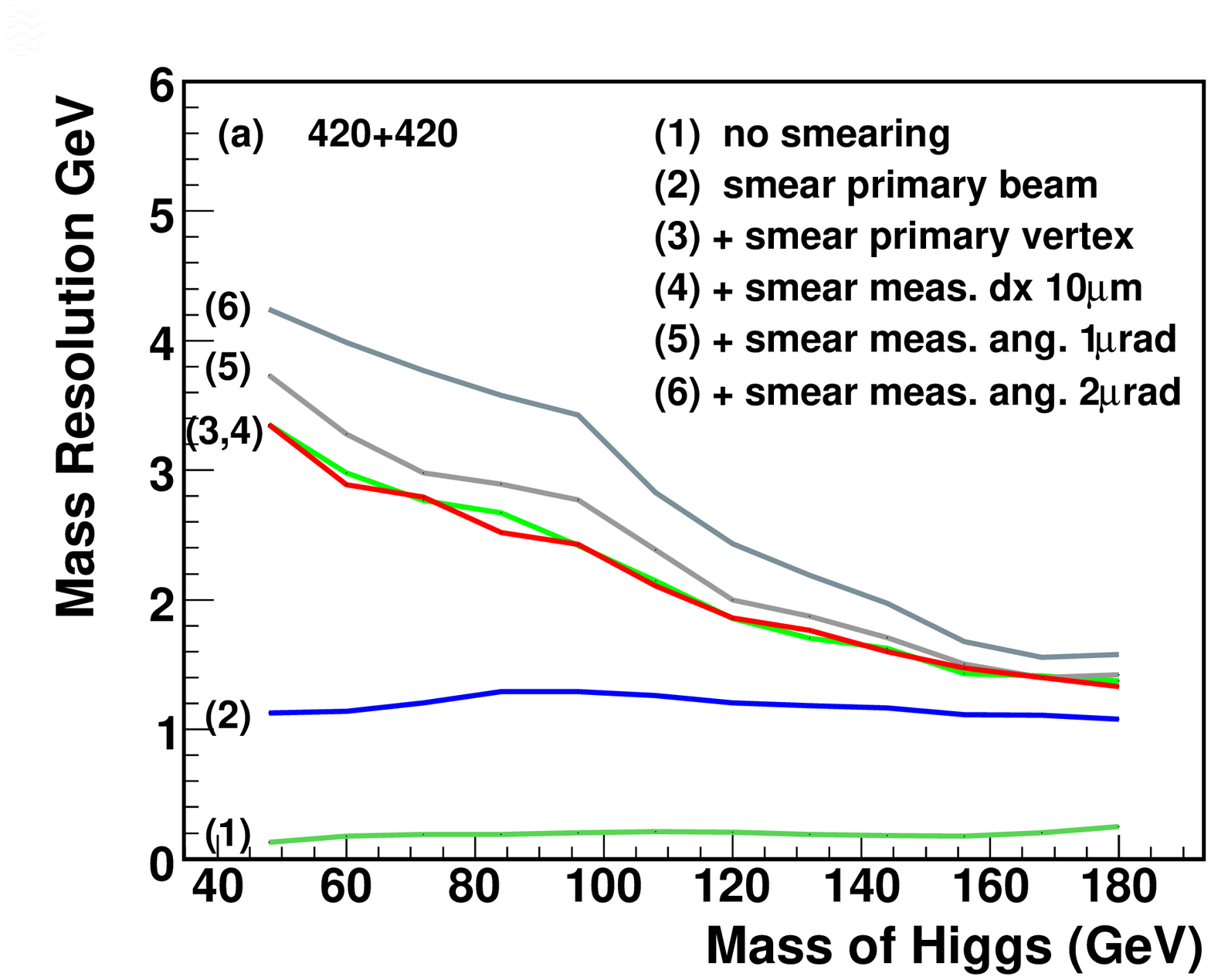}
\includegraphics[width=6.5cm]{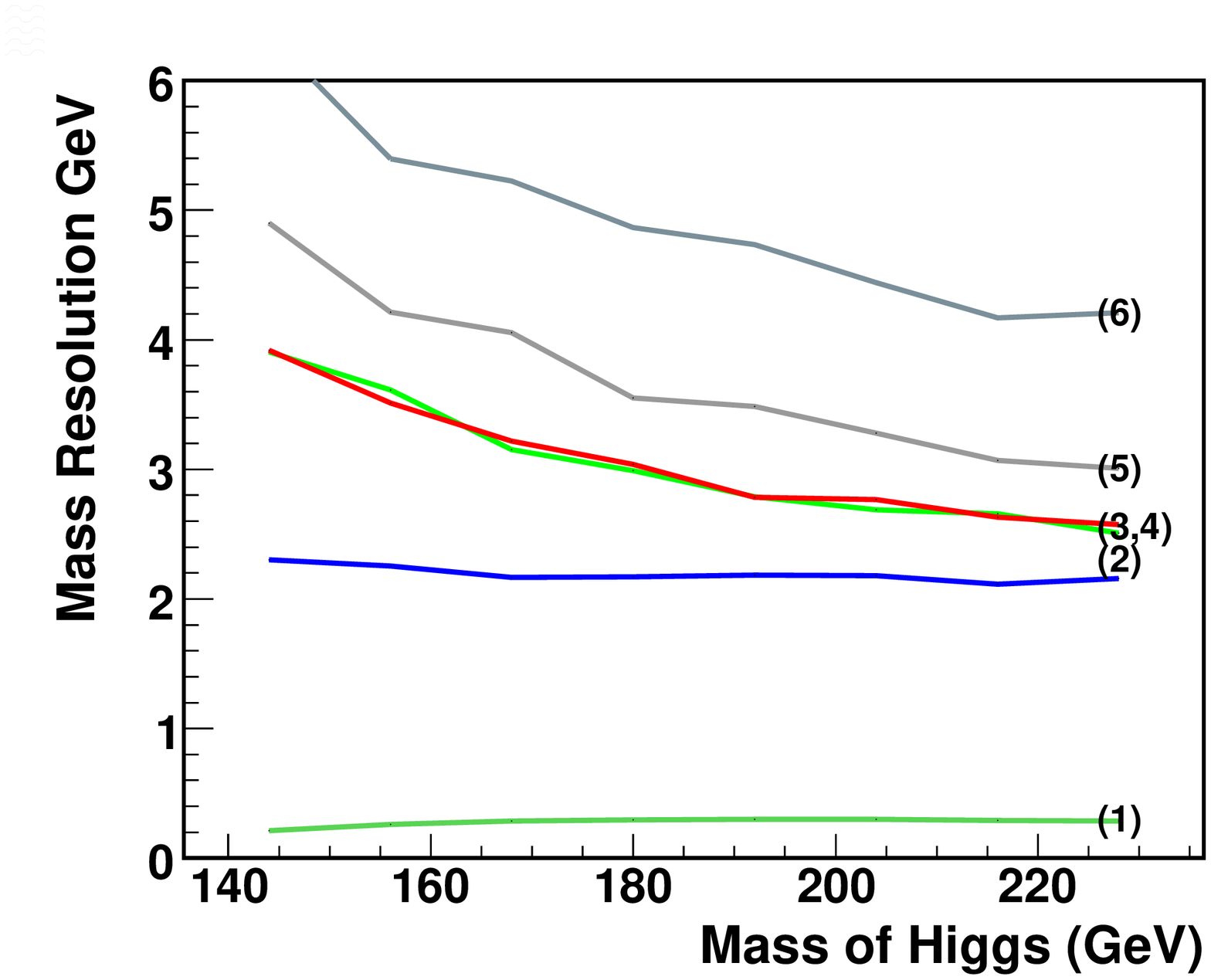}
\includegraphics[width=6.5cm]{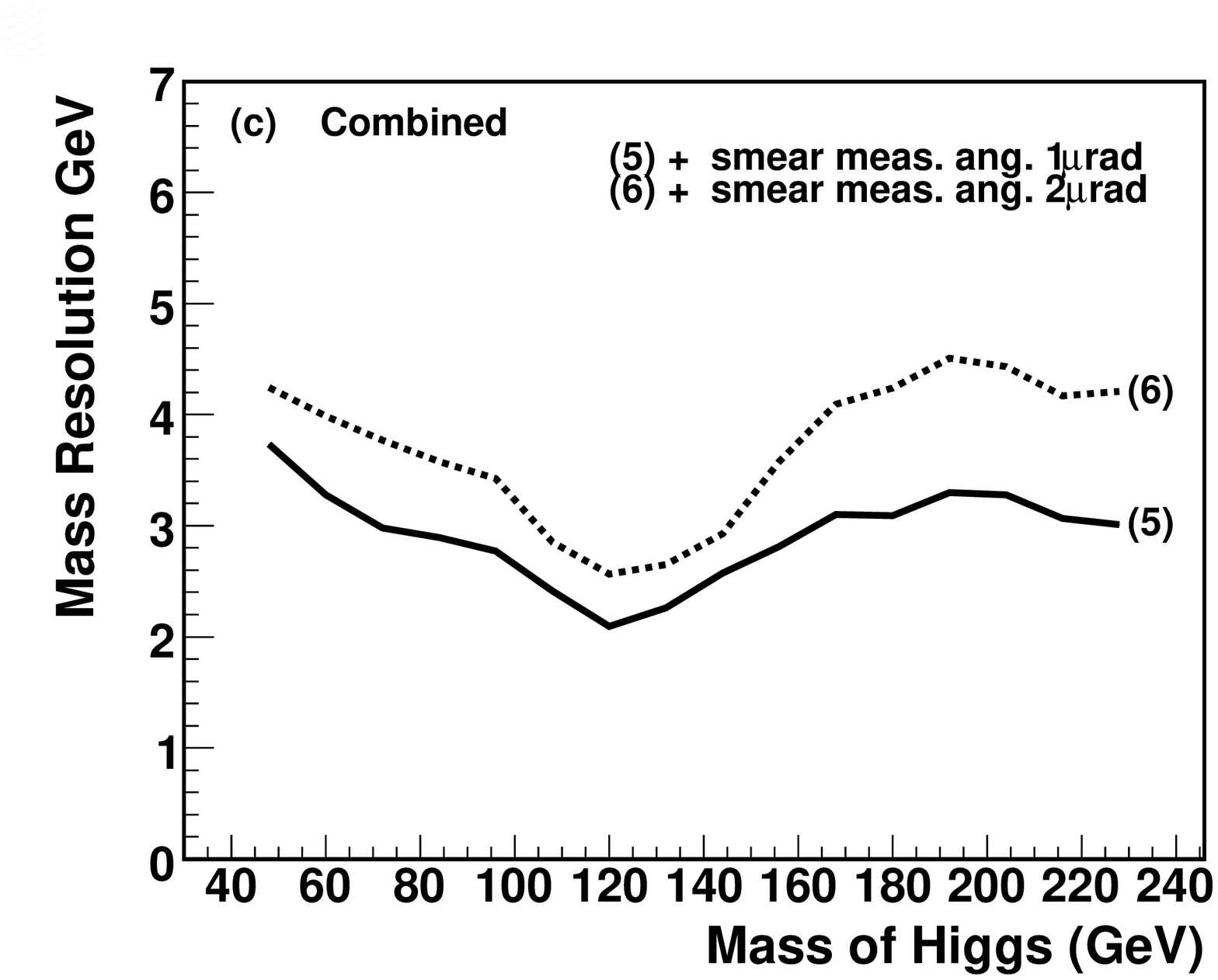}
\caption{Mass resolutions obtainable in ATLAS 
(a) for 420~+~420~m measurements, (b) for
420~+~220~m measurements, (c) combined. 
The curves have different amounts of
smearing applied as explained in the text.}
\label{f5_6}
\end{figure}

Figure \ref{f5_6} shows the effect of each one of the above factors on the
mass resolution. Full sets of curves are presented for 420~+~420~m
measurements up to 180~\GeVcc\ (left) and 420~+~220~m measurements above 
140~\GeVcc\ (right). The two top curves which are given in both figures
indicate a combination of the two measurements.  Resolutions were
determined by applying a chosen combination of Gaussian smearings and
fitting the resulting histograms of reconstructed minus true mass with
a Gaussian function, whose width is plotted here. The sets of curves
represent the resolutions obtained: (1) with no physical smearing
applied, indicating the precision of the reconstruction algorithm, (2)
applying smearing due to the 0.77~GeV/c Gaussian distribution of the
primary proton beam momentum (3), also including a 10~$\mu$m lateral
smearing of the interaction vertex within incident beam spot, (4) also
including a 10~$\mu$m smearing of the measured position $x$ in the
silicon system, (5,6) also including 1, 2 $\mu$rad smearing, respectively,
 of the $dx/dz$ measurement in the silicon system. The curves in
(c) give the overall mass
resolution under the conditions of (5) and (6) for all events in both
regions combined. The effects of a small smearing of the $x$
measurement in the silicon system are seen to be small in comparison
with the other effects.  The overall resolution is as low as 2~\GeVcc\ in
the central mass range of interest, using the expected 1$\mu$rad
angular uncertainty in the $dx/dz$ measurement.  It should be noted
that the 2 $\mu$rad curve could be considered an upper limit to the
resolution, as a comparable resolution can be obtained by simply
constraining the angle of the emitted proton to be along the beam
direction at the interaction point.


It is possible to measure the transverse momentum of the proton as
it emerges from the interaction point, again by means of
polynomial-based parametrization formulae using the measurements in
the detector stations. Both $x$ and $y$ measurements are required to
determine the full transverse momentum of the proton.  The
measurement is degraded by two factors. The angular beam spread at
the interaction points  is equivalent to a $\pm$ 0.21~GeV/c transverse
momentum spread, both horizontally and vertically, and the poorer
measurement uncertainty in the $y$ direction increases the overall
uncertainty on $p_T$ significantly. Studies are continuing to
determine the requirements for particular physics studies and
whether they can be achieved.

\subsection{Optics summary}

The beam optics at LHC allows protons that have lost momentum in a
diffractive interaction to emerge from the beam envelope at regions
220~m and 420~m from the interaction point.  By placing silicon
detector arrays in these locations we can detect the protons and
obtain good acceptance for diffractively produced objects with a wide
range of masses above 60~\GeVcc, the precise acceptances depending on how
close it is possible to place the detectors relative to the beam. Even
under cautious assumptions, the mass range above 100~\GeVcc\ is well
covered, but to obtain good acceptance for masses above 150~\GeVcc\ the
220~m system is essential. The expected position and angle resolutions
for the protons obtained in the silicon stations are expected to yield
a mass resolution reaching values of 2 to 3~\GeVcc\ per event.
\newpage

\section{Machine Induced Backgrounds}
\label{sec:backgrounds}

\subsection{Introduction}
\label{sec:machineinduced}

A precise evaluation of the particle flux environment at 420~m
caused by machine operation provides critical input to the FP420
project in several areas. It is necessary for the determination of
expected FP420 operating parameters such as the the minimum safe
distance of approach to the beams, and also for assessment of the
level of radiation exposure of  the detectors and associated
electronics. Moreover, machine-induced background entering the
detectors may result in fake proton tracks, which will contribute to
the pile-up background described in Section~\ref{sec:pilko} and
also result in increased occupancy in the silicon sensors, which must
be considered in the tracking algorithm performance. The assessment
of machine-induced backgrounds relies on detailed simulations of the
machine geometry, the LHC collimation scheme and cleaning
efficiency, the beam optics, the bunch structure and the residual
gas density. In this section, the status of the estimates of all
contributions to the background are presented, and preliminary
conclusions discussed. Unless otherwise stated, all results are
calculated for the case of full instantaneous 
luminosity of 10$^{\mathrm{34}}$~cm$^{\mathrm{2}}$s$^{\mathrm{-1}}$.\\

\noindent The background in the FP420 region is comprised of the following components:

\begin{itemize}
\item
{\bf interaction point (IP) particles}: generic proton-proton
collisions at the interaction point produce a great number of
particles dominantly in the forward direction, some of which reach
the 420~m region. The control of this so-called overlap background
is discussed in Section~\ref{sec:pileup}.
\item
{\bf beam-gas particles}: elastic and inelastic proton-nucleus collisions
between the beam protons and residual gas molecules
produce shower particles, which represent a direct background when
the collisions occur close to the FP420 detector stations. This is referred
to as the \emph{near beam-gas background}.
\item
{\bf beam halo particles}: 
\emph{distant} beam-gas interactions 
(occurring around the whole ring and not only in vicinity of the detectors), 
%
%
various beam instabilities and a limited dynamic aperture 
lead to beam protons leaving their 
design orbit and impact on the collimation system. 
This builds up beam halo particles circulating in the machine. 
\item
{\bf secondary interactions}: beam-halo particles or particles
resulting from proton-proton or beam-gas collisions can interact
with the machine elements creating {\bf secondary showers} that can
irradiate the detector region with a potentially large flux of
charged and neutral particles. Showers can also originate in the
detector structure itself.

\end{itemize}

The following sections consider each of these background sources.

\subsection{Near beam-gas background}

The beam-gas contribution arises from the interaction of beam
particles with residual gas in the beam pipe region immediately
before 420~m. These elastic and inelastic proton-nucleon collisions
perturb the angular (large-angle scattering) and momentum phase
space distribution of the primary protons, and cause secondary
production in the vicinity of the detector. Study of this background
requires a detailed model of the beam line, coupled with gas
pressure profiles and computation of proton/gas interactions. The
Protvino group have started performing simulations 
using~\cite{ref:protvinostudies} for the 
forward detectors at 220~m and 240~m from the IP1 (ATLAS) and IP5 (CMS and TOTEM) interaction points, 
based on estimated pressure profiles in the IR1 and IR5 straight sections. These calculations will be extended to 420~m, and
normalised to beam lifetime. Furthermore, the beam-gas pressure
profiles can be used within the BDSIM~\cite{ref:bdsim} (see
Sec.~\ref{sec:secondary}) simulations of the beamline, to complement
and cross-check the Protvino calculations and also to assess the
integrated beam-transport/beam-gas background spectrum at FP420.\\

Until these studies are completed, a rough estimate of the number of
beam-gas interactions per bunch in the 420~m detector region can be
extrapolated from the results obtained for the straight section regions~\cite{ref:prot_bgpresent}. 
Such simulations include protons that are lost after scattering with the gas nuclei and secondary 
particles produced due to proton losses upstream the considered scoring plane. 
In a scoring plane set at 240~m from IP1, the total number of charged hadrons assessed 
by the simulations is $\mathrm{np_{240~m} = 2.4\,s^{-1}}$. This result is obtained 
considering an average residual gas density along IR1~\cite{ref:arossi,ref:prot_bgpresent}  
$\mathrm{\rho_{240~m} = 3.4\cdot 10^{11}\,molecules\cdot m^{-3}}$ 
(converted in hydrogen-equivalent species). The dynamic residual pressure at 420~m
is expected to be higher than the straight sections, due to synchrotron radiation.
As a very conservative upper limit, we can consider a residual hydrogen density of
about $\mathrm{\rho_{420~m}=1\cdot 10^{15}\,molecules\cdot m^{-3}}$, which is compatible with a
beam-gas lifetime of 100 hours. If the level were any
higher than this the energy deposition per meter in the LHC arcs
would exceed the cooling power needed to avoid magnet
quenches~\cite{ref:lhcreport_vacuum}.  With such a gas density, the
total number of expected hadrons per bunch, due to near beam-gas  events, is around
\begin{equation}
np_{420~m} =  \frac{np_{240~m}}{N_{bs}} \cdot \frac{\rho_{420~m} }{\rho_{240~m}}= 1.8\cdot 10^{-4} 
\end{equation}
where $\mathrm{N_{bs}=4\cdot 10^{7}}$ is the number of bunches per second that 
will circulate in the LHC with nominal conditions.\\

This estimate predicts a very low background rate contribution, especially taking into account that 
it refers to the full mechanical beam pipe aperture and only a small fraction of those hadrons 
will hit the FP420 detectors. In addition, after the LHC startup phase, the residual gas density 
in the arcs is expected to correspond to a beam-gas lifetime larger than 100 hours. 
However, such an approximation has to be validated with dedicated simulations and eventually with real data.

\subsection{Beam halo}

During standard LHC operation a so-called \emph{primary halo} will
be filled continuously due to beam dynamics processes. These
particles are lost by the limitations of the mechanical aperture at
various places around the LHC ring, resulting in a finite beam
lifetime. Given the high intensity of the LHC beam, it will be
unacceptable to lose even a small amount of the particles populating
the halo in the super-conducting magnets. The collimation system has
been designed to clean the beam halo without inducing magnet
quenches due to beam losses~\cite{ref:lhcreport_coll}. The system is based on
a set of movable primary, secondary and tertiary collimators that can be
adjusted during the different phases of a physics run. They are
divided into two categories: \emph{betatron cleaning collimators} (located at IR7)
that clean particles performing large betatron oscillations, and
\emph{momentum cleaning collimators} (located at IR3) that clean particles with large
momentum offset. The two systems will be always adjusted such that they comprise
the limiting transverse and longitudinal machine apertures.

The collimation system is designed
mainly to protect the machine,
but it reduces also
the experimental backgrounds related to the
primary halo. However, the unavoidable cleaning inefficiency of the
multi-stage collimation process generates \emph{secondary} and
\emph{tertiary} halos populated by protons scattered at the
collimators. Such particles can circulate for many turns before
being removed by the cleaning/absorbing elements or in other
locations depending on the phase advance of their betatron
oscillation. Tertiary collimators are located in all experimental straight sections to protect super-conducting magnets
from the tertiary halo. Additional devices (\emph{absorbers})  are designed to protect from  
showers generated by the cleaning insertions and physics debris from the interaction points.   

The fact that primary collimators are not distributed around the LHC ring, but are
concentrated in IR7 and IR3 results in a beam secondary halo distribution that will be
different for the four potential FP420 locations around ATLAS and
CMS. 

Although FP420 is in the shadow of the
collimators,
this will not be sufficient to completely avoid hits from beam halo
particles. In particular conditions  (linked to the betatron phase
advance between the collimators and FP420, the dispersion functions,
and the particles momenta), elements with apertures larger than the
collimators may be hit by halo particles.

In the following sections, we review the LHC collimators settings
and the expected beam parameters at FP420.  We also address in more
detail the beam halo generated at the two collimation systems
elements (Sec.~\ref{sec:momclean} and~\ref{sec:betaclean}) and
around the whole LHC ring due to small scattering angles beam-gas
interactions (Sec.~\ref{sec:dbeamgas}).

%
\subsubsection{Collimator settings and beam parameters}
 
%

During high luminosity running at 7\,TeV, it is foreseen to set the
collimator position as shown in Table~\ref{tab:coll_settings}.

\begin{table}[htbp]
\centering
\begin{tabular}{| l |c|c|c|}
\hline
System              & Name & Location & Half Gap\\
                &&&[$\sigma_{\beta}$]\\
                \hline
Primary betatron cleaning  &     TCP 		& IR7 &6\\
Secondary betatron cleaning  &     TCSG 	& IR7  &7\\
\hline
Primary momentum cleaning  &     TCP 		& IR3	&15\\
Secondary momentum cleaning  &     TCSG 	& IR3	&18\\
\hline
Tertiary collimators 			&TCT&IR1,IR2,IR5,IR8 &8.3\\
\hline
Absorbers 	&TCLA &IR3	&20\\
(for showers in cleaning insertions)		&  & IR7	&10 \\
\hline
Absorbers (for physics debris) 	&TCLP &	IR1,IR5 & 10\\
\hline
\end{tabular}
\caption{\label{tab:coll_settings}
LHC collimator settings for nominal optics at 7\,TeV. Details about the 
collimator exact number, materials, location and orientation (horizontal, vertical or skew) 
can be found in~\cite{ref:coll_home,ref:g_thesis}.}
\end{table}%

Such values are expressed in terms of the radial distance from the
beam envelope evaluated at one $\sigma_\beta$. It must be noted
that at each location $s$ in the ring, the actual transverse beam
size is defined by  the particle betatron oscillations and by the closed orbit
offset due to the particle momentum error.
Considering the
horizontal plane relevant for FP420:
\begin{equation}
\sigma_x(s)=\sqrt{\frac{\epsilon^{*}_x \beta_x(s)}{(\beta\gamma)}+\left[D_x(s)\cdot\delta\right]^2}=\sqrt{\sigma^2_{\beta x}(s)+\sigma^2_{\delta_x}(s)},
\end{equation}
where $\beta_x$ and $D_x$ are the betatron and dispersion functions,
$\epsilon^{*}_x$ the normalised emittance (at 1$\sigma$),
$\delta=dp/p_0$ the particle's r.m.s momentum spread, $\gamma=E/E_0$
the Lorentz factor and $\beta=v/c$.
As a baseline, in order to guarantee safe operation conditions, experimental insertions like FP420  
will be allowed to approach the beam as close as $15\,\sigma_{x,y}$. 
Smaller distances will need to be discussed and approved by the concerned LHC committees.\\
%
%

\begin{figure}[htbp]
\centering
\includegraphics[width=0.48\columnwidth]{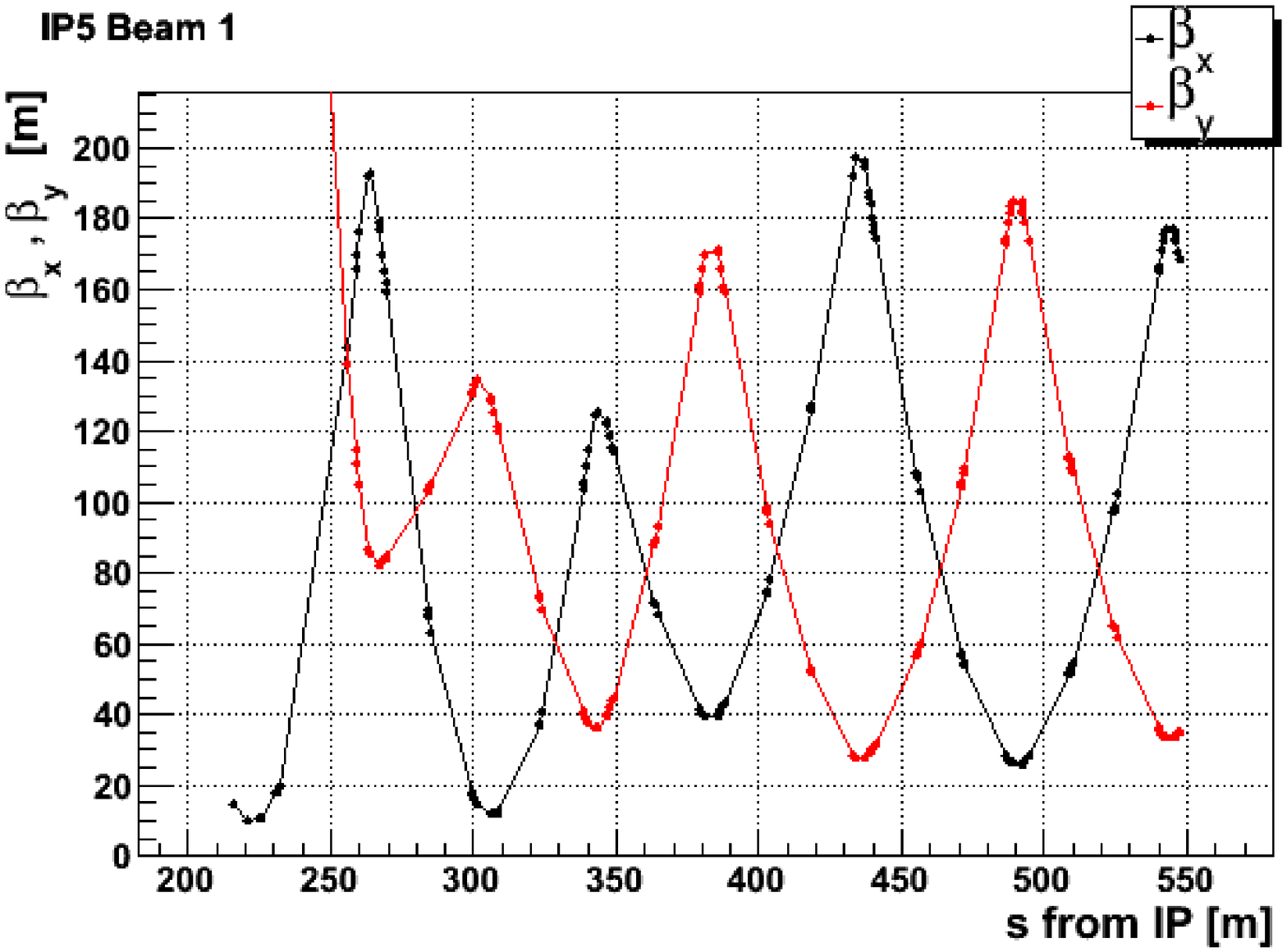}
\includegraphics[width=0.48\columnwidth]{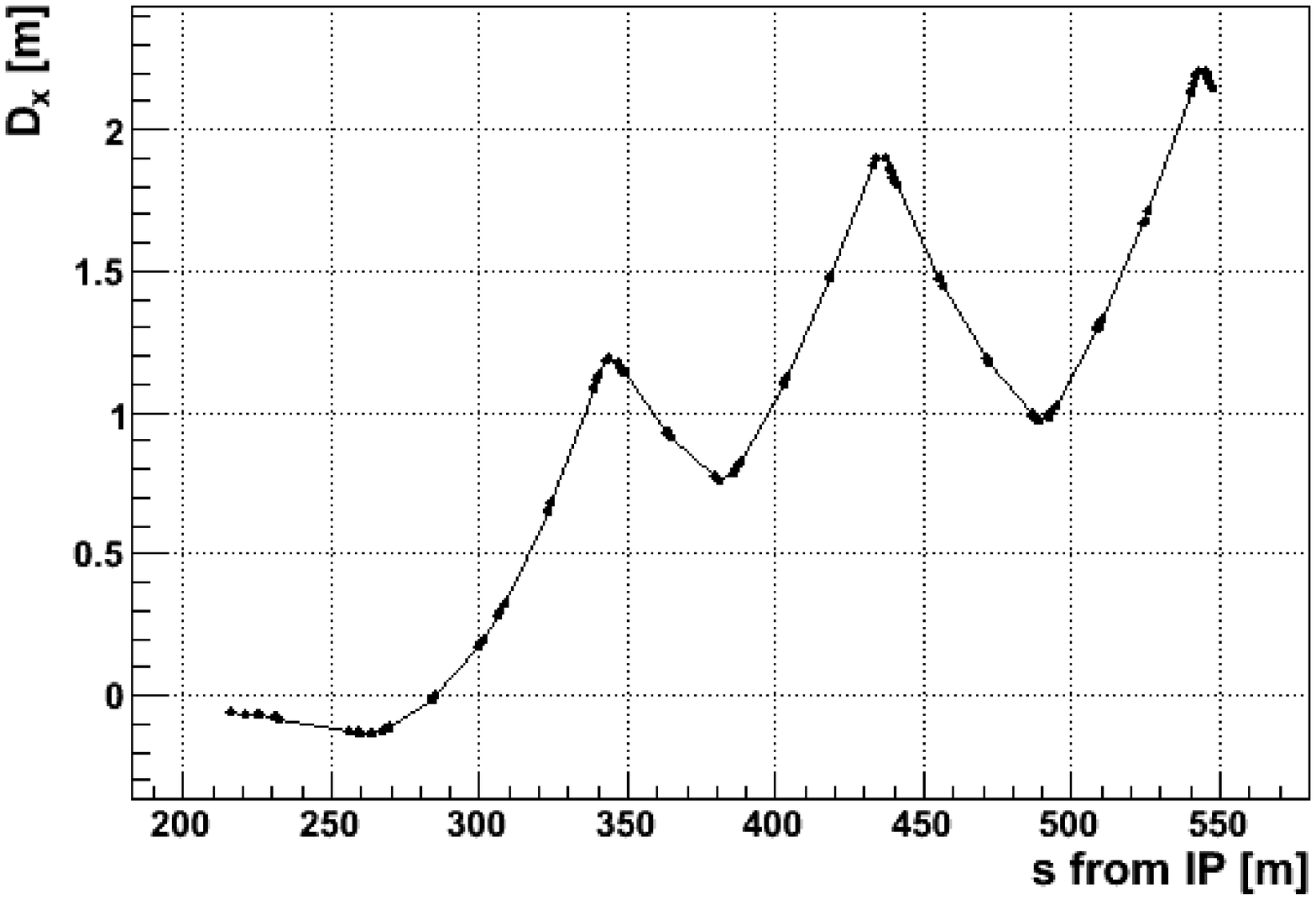}
\caption{The betatron functions and horizontal dispersion in the FP420 region for Beam 1 downstream of IP5.}
\label{fig:twiss}
\end{figure}

The optics parameters at the entrance of FP420 are summarised in
Table~\ref{tab:fp420par}. Also shown is the horizontal beam size for
the nominal values $\epsilon^{*}_x=3.75\,\rm {\mu m}$ and
$\delta=1.13\cdot10^{-4}$.
\begin{table}[htbp]
\centering
\begin{tabular}{| c | c | c | c | c | c |}
\hline
                   & distance from IP & $\beta_x$ & $D_x$ & $\sigma_x=\sigma_{\beta x}+ \sigma_\delta$ &$15 \sigma_x$\\
                   & [m]          & [m]             & [m]     & [mm]  & [mm]\\
\hline
IP1 Beam 1 &    418.5           &   127.1   &   1.51    &0.305  & 4.573\\
IP1 Beam 2 &    418.8           &   106.9   &   2.02    &0.325  &4.873\\
\hline
IP5 Beam 1 &    418.5           &   127.1   &   1.47    &0.302  &4.534\\
IP5 Beam 2 &    418.2           &   106.9   &   1.96    &0.321  &4.808\\
\hline
\end{tabular}
\caption{Beam parameters at the end of the last element before FP420 using LHC optics V5.0}
\label{tab:fp420par}
\end{table}
The betatron functions and the horizontal dispersion in the
dispersion suppressor region for one of these combinations (Beam 1
downstream IP5) are shown in Fig.~\ref{fig:twiss}.

\subsubsection{Beam halo induced by momentum cleaning collimators}
\label{sec:momclean}

During the physics runs, the momentum cleaning system is designed
mainly to protect the machine from protons leaving the RF bucket
because of energy loss due to synchrotron radiation. Off-momentum
protons can potentially perturb the operation of the FP420 detectors
due to the closed orbit displacement caused by the high dispersion
function (Table~\ref{tab:fp420par}). For this reason, a series of
simulations has been carried out in order to characterise the beam
halo populated by such protons and the effect of the cleaning system
settings on the FP420 background. The simulations have as input $2
\cdot 10^6$ protons belonging to the primary halo hitting the
momentum cleaning primary collimators in IR3. A multi-turn tracking
routine follows the protons emerging from the collimator surface
until they are absorbed by the cleaning system or lost in other
aperture limitations of the machine (not including the FP420
detectors). At each turn, the proton distribution is recorded at the
420~m locations. Two separate sets of simulations have been carried
out for Beam 1 and Beam 2 using STRUCT~\cite{ref:struct}.\\

%

\begin{figure}[htbp]
\centering
\includegraphics[width=0.48\columnwidth]{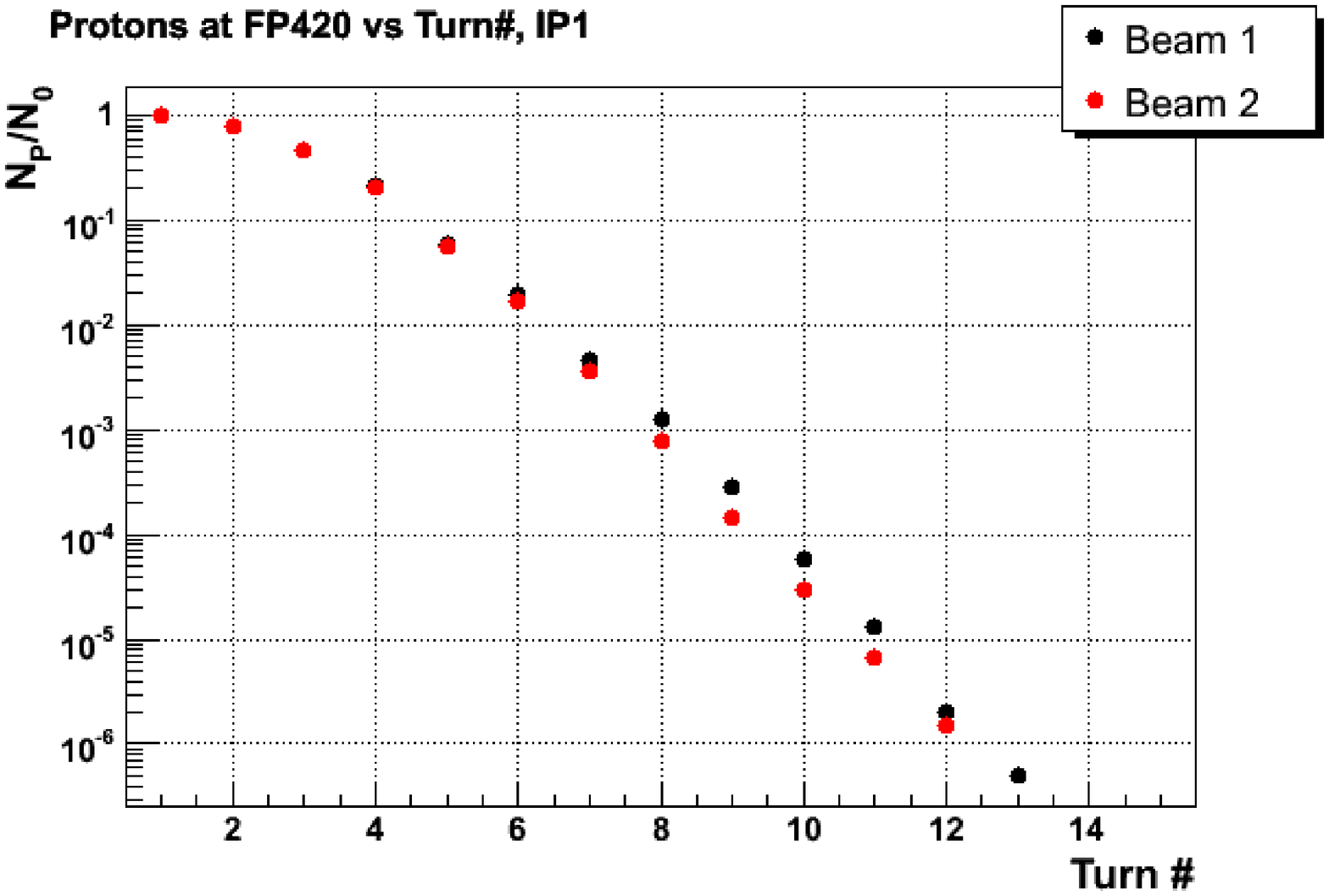}
\includegraphics[width=0.48\columnwidth]{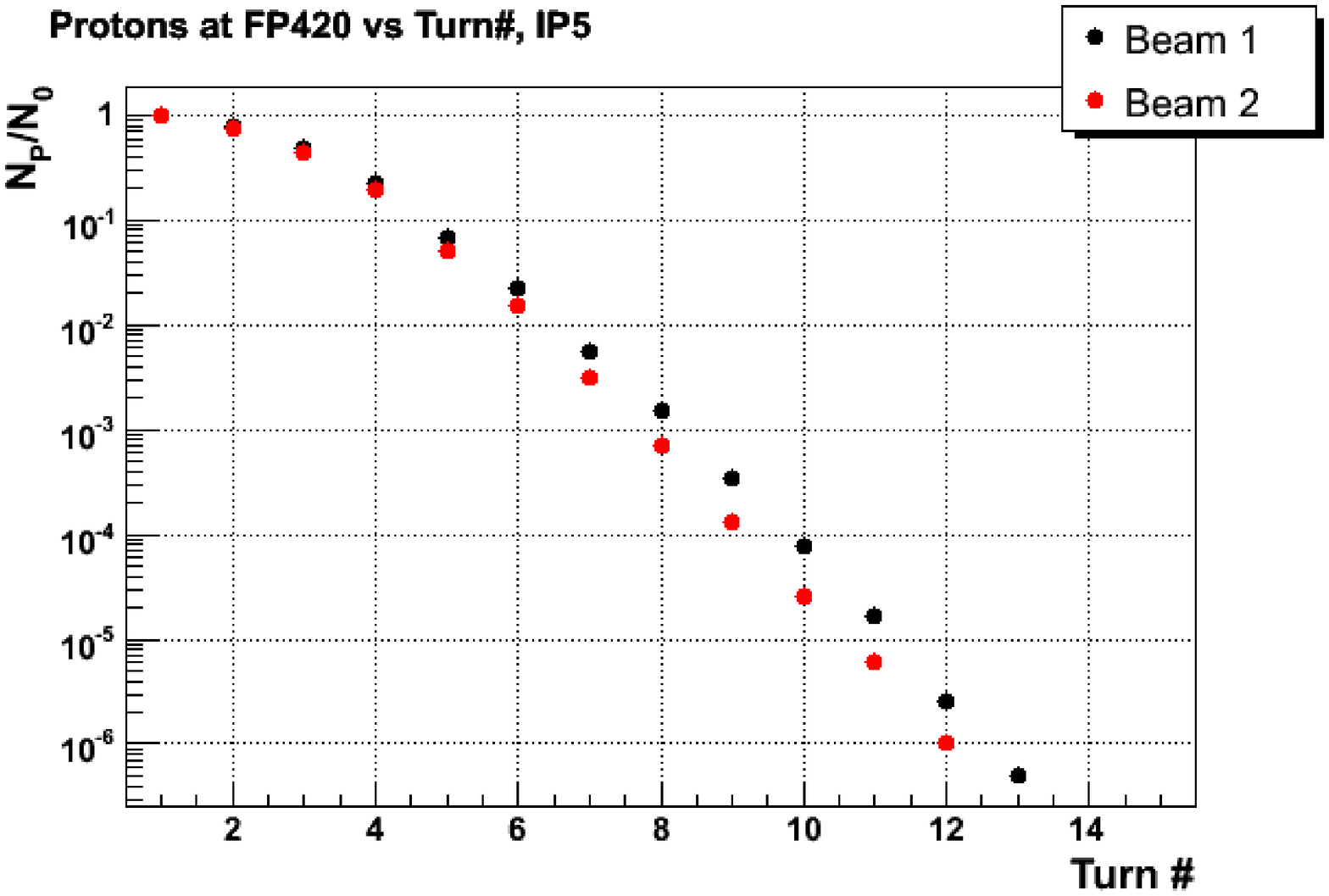}
\caption{Total number of particles at FP420 as a function of the
turn number after scattering on the momentum cleaning collimators.}
\label{fig:turns}
\end{figure}

The fraction of the initial protons reaching 420~m as function of
the number of turns after their interaction with the collimators is
shown in Fig.~\ref{fig:turns}. The plots confirm the multi-turn
nature of the cleaning process, as almost 100\%  of the

protons hitting the collimators reach 420~m at ``turn 1'' (when the
particles have only traversed the distance from IR3 to 420~m) and
almost 90\,\% of them survive the first full turn. Therefore, for
background considerations, all the primary halo off-momentum protons
that continuously hit the momentum cleaning collimators and fill the
secondary and tertiary halos, must be considered at 420~m. Of course
reasonable operating positions will be chosen to avoid the bulk of
this halo.

\begin{figure}[htbp]
\centering
\includegraphics[width=0.48\columnwidth]{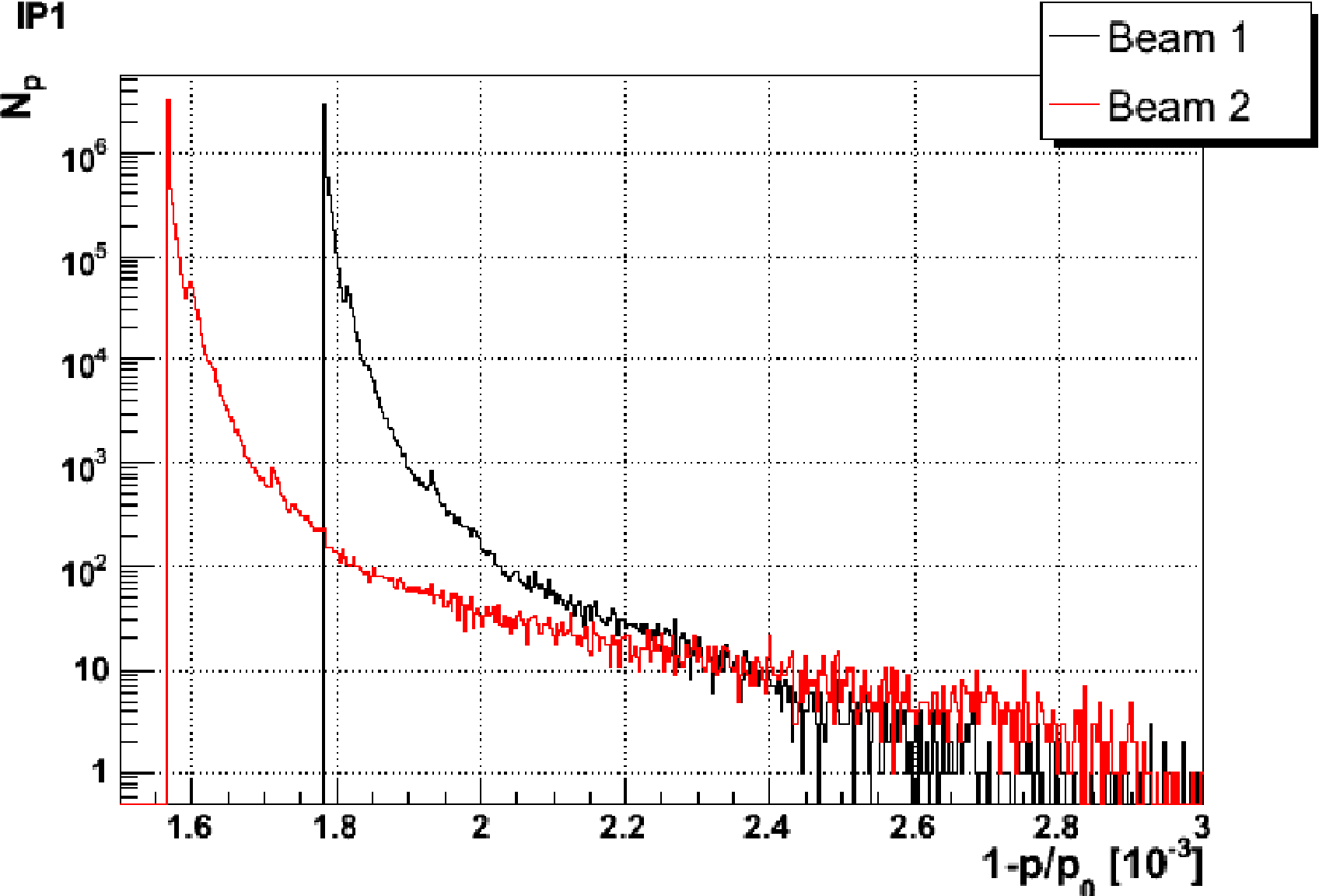}
\includegraphics[width=0.48\columnwidth]{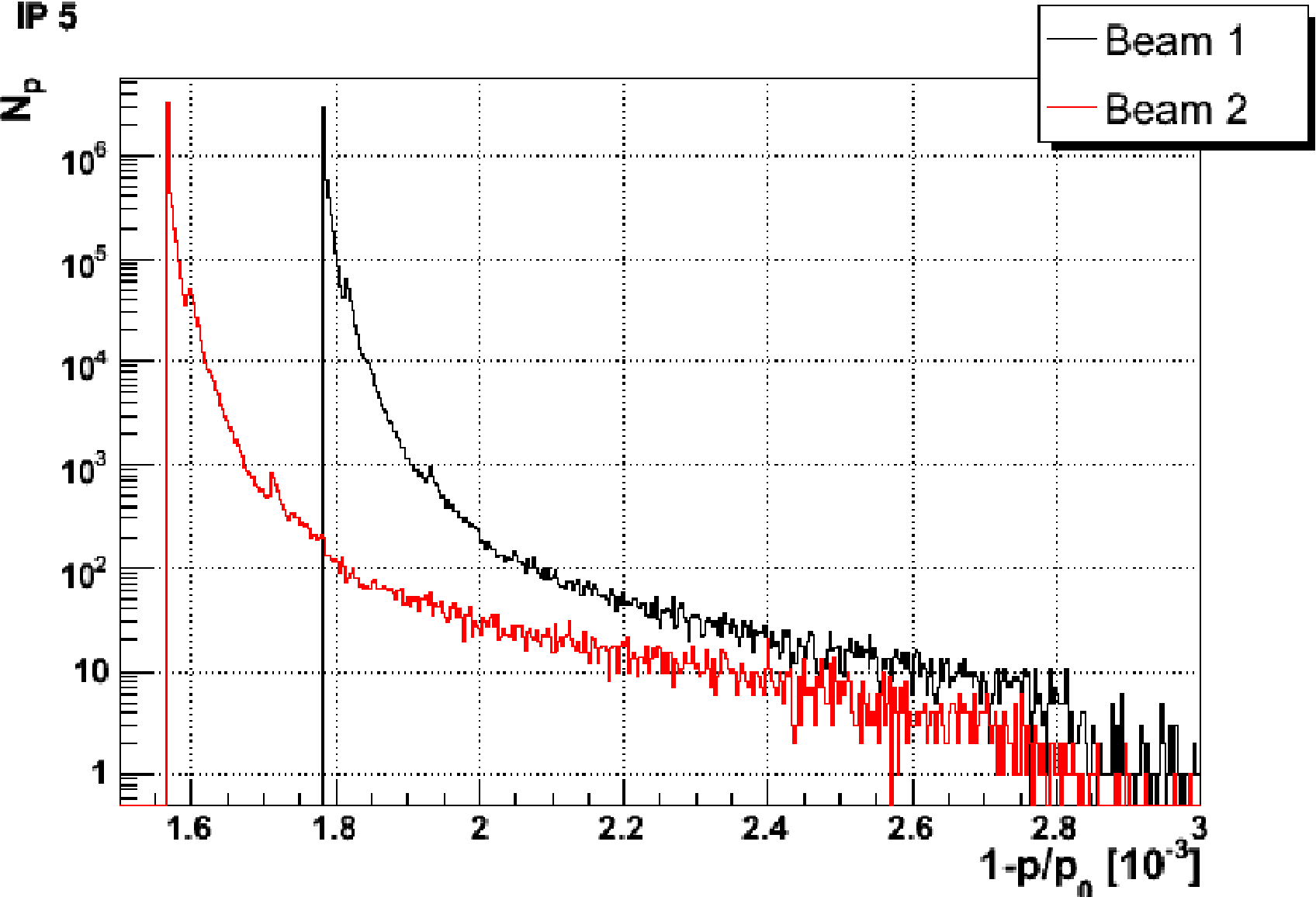}
\caption{Momentum distribution of the protons emerging from the momentum cleaning collimators.}
\label{fig:momcut}
\end{figure}

\begin{figure}[htbp]
\centering
\includegraphics[width=0.49\columnwidth]{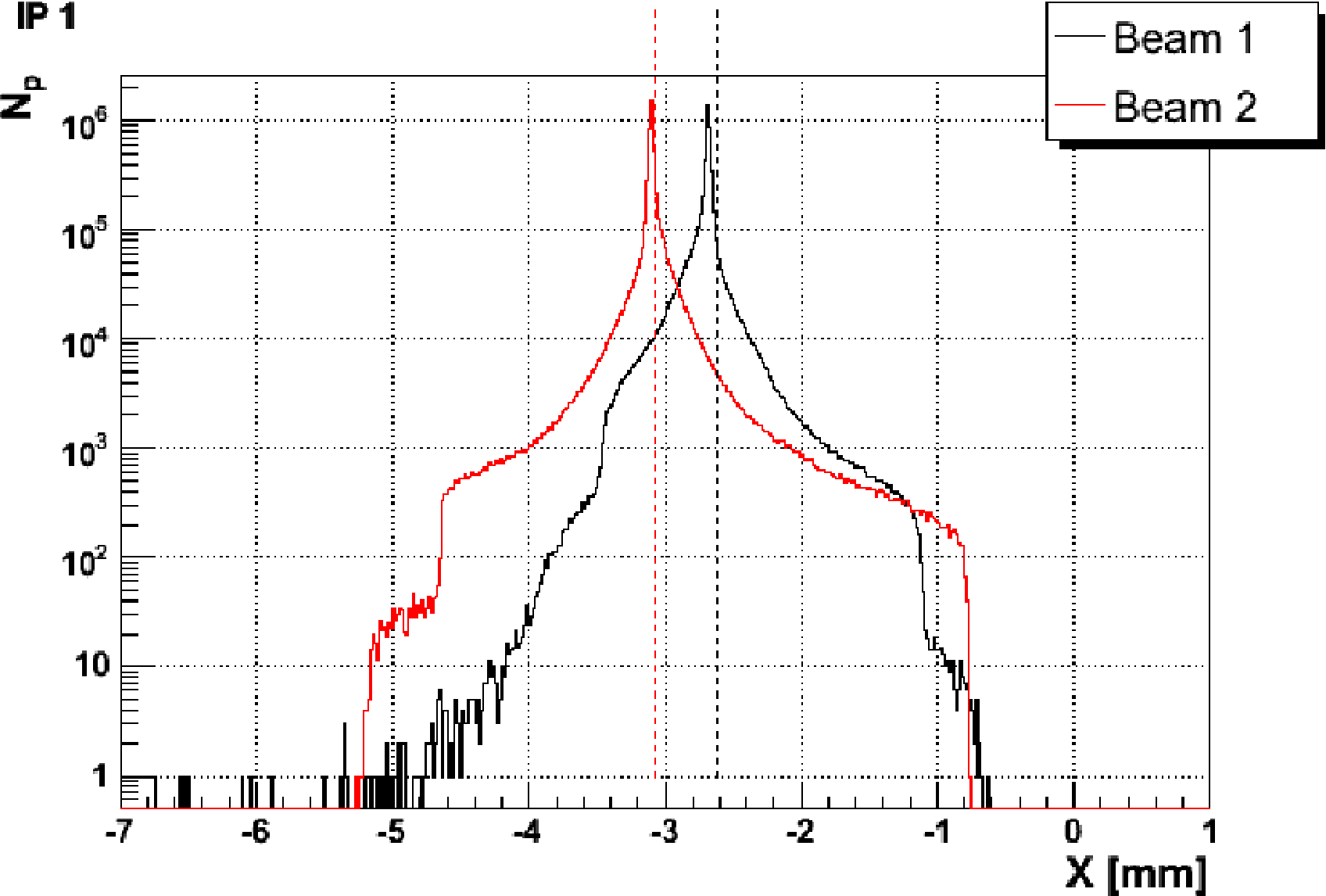}
\includegraphics[width=0.49\columnwidth]{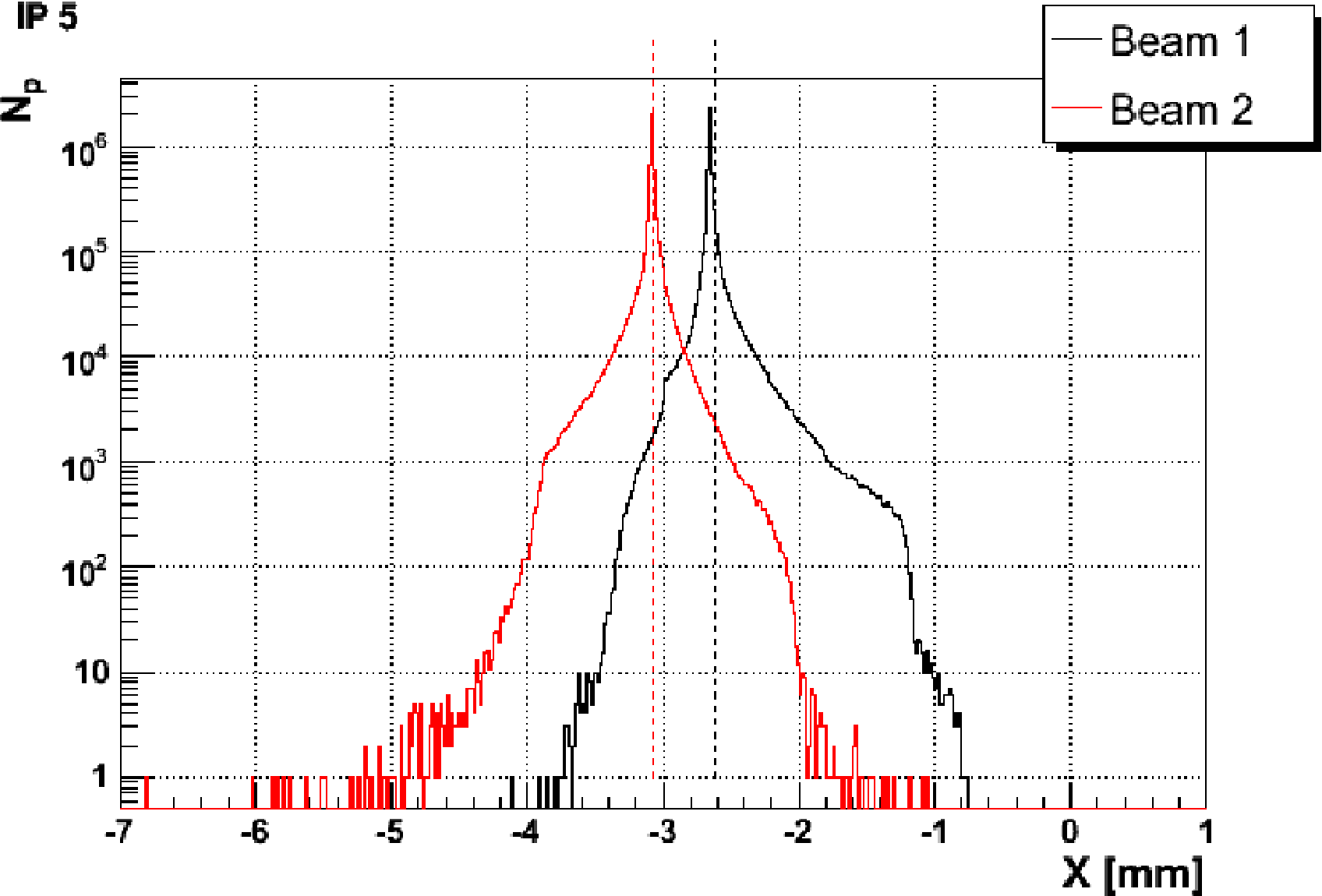}
\caption{Horizontal distribution of the protons emerging from the momentum 
cleaning collimators as observed at FP420 after the simulations of $2\cdot 10^6$ 
proton interactions with the primary momentum cleaning collimators.}
\label{fig:halodist}
\end{figure}

If the collimators in IR3 are set at $x(s_{c})$ and the dispersion
function at that location is $D_x(s_{c})$, all particles with
\begin{equation}
\delta \equiv 1 - \frac{p}{p_0} \leq \frac{x(s_{c})}{D_x(s_{c})} \equiv \delta_c
\end{equation}
hit the collimator at every turn. Given $D^{b1}_x(s_c)=2.20$ m and
$D^{b2}_x(s_c)=2.46$ m for Beam 1 and Beam 2 respectively, and the
collimator positioning at $15 \sigma_{\beta}$, the cut in
momenta for the two distributions is expected to be at
$\delta^{b1}_c=1.78\cdot10^{-3}$ and
$\delta^{b2}_c=1.57\cdot10^{-3}$. This is confirmed by
Fig.~\ref{fig:momcut}.
In addition, at the FP420 locations, the proton horizontal distribution is expected to be centred around
\begin{equation}
x^{FP420}_{cut} = -D_x(s_{420}) \cdot \delta_c,
\end{equation}
as confirmed by the simulation results in Fig.~\ref{fig:halodist}
which show the horizontal halo distributions with the expected peak
values (dashed vertical lines).
The shape of the distributions depends on the betatron phase advance
between the collimators and the detectors.

In order to estimate the absolute background level, the distributions
must be normalised for the number of protons that will interact with
the momentum cleaning collimators during normal LHC operations.
Assuming:
\begin{description}
\item[--] the nominal LHC beam intensity for high luminosity runs $I_0=3.2\cdot 10^{14} $ protons,
\item[--] an exponential decay of the beam current due to off-momentum proton losses \begin{equation*} I(t)=I_0\cdot e^{-t/\tau}\end{equation*}
\item[--] a beam lifetime accounting for losses of off-momentum particles
$\tau=150$ hours,
\end{description}
then the corresponding maximum proton loss rate is:
\begin{equation}
r(t=0)=\left.-\frac{dI}{dt}\right |_{t=0}=\frac{I_0}{\tau}\approx 5.9\cdot 10^{8}\ [p/s].
\end{equation}
Hence, the loss rate at FP420 as a function of transverse position can
be calculated by normalizing the histograms of
Fig.~\ref{fig:halodist} according to:
\begin{eqnarray}
\label{equ:norm}
r(t,\Delta x)&=&N_p\cdot \frac{r(t_0)}{N_0} [p\cdot s^{-1} \cdot (\Delta x)^{-1}]\\
 N_0&=&2\cdot 10^{6} \text{\ \ (simulation input)}\nonumber \\
 \Delta x &=&\text{bin-width.}\nonumber
\end{eqnarray}
The normalised distributions are shown in Section~\ref{sec:dbeamgas}
together with the distributions generated by beam-gas
interactions.

%
%

\subsubsection{Beam halo induced by betatron cleaning collimators}
\label{sec:betaclean}

The impact of the primary beam halo protons on the betatron cleaning
collimators will also generate secondary and tertiary halos. Given
the collimator settings shown in Table~\ref{tab:coll_settings},
however, it is clear that halo generated by these collimators will
be negligible at reasonable 420~m detector positions ($x=10$ to
$15 \sigma_x$). If this were not the case high luminosity LHC
operation and the protection of superconducting magnets would be
extremely problematic, so we can neglect this halo term in our
considerations.

\subsection{Halo from distant beam-gas interactions}
\label{sec:dbeamgas}

The LHC beam halo will be populated also by protons that experience
scattering with the residual gas nuclei. When the resulting proton
momentum loss and scattering angle are small, the protons remain
within the machine momentum and transverse acceptance and can
circulate for several turns. Therefore the scattering is elastic or
inelastic, provided the momentum loss is small enough for multi-turn
survival.

A series of simulations was carried out by the Protvino group using
STRUCT~\cite{ref:struct}. Ten million protons (for each LHC beam) were
generated at the location of the collimator labelled as TCP.6L3
(at 177 m upstream of IP3), with momentum equal to 7\,TeV and distributed
according to the nominal transverse phase space. Each proton was 
tracked around the LHC ring model while assuming a uniform gas
density in the LHC arcs and dispersion suppressor regions. After a
proton-gas interaction, all protons that are scattered with a small
angle and momentum loss are tracked around the machine until they
are either lost in a machine aperture limitation or rescattered in a
collimator. In the latter case, the scattering process proceeds as
for the momentum halo simulations in Section~\ref{sec:momclean}. At
each turn, all protons with transverse position $|x|>7 \sigma_x$
or $|y|>7 \sigma_y$ are recorded at the entrance of the FP420
regions\footnote{As for the tracking simulations related to momentum cleaning collimators, 
the FP420 detectors are considered transparent for the beam.}.

\begin{figure}[htbp]
\centering
\includegraphics[width=0.49\columnwidth]{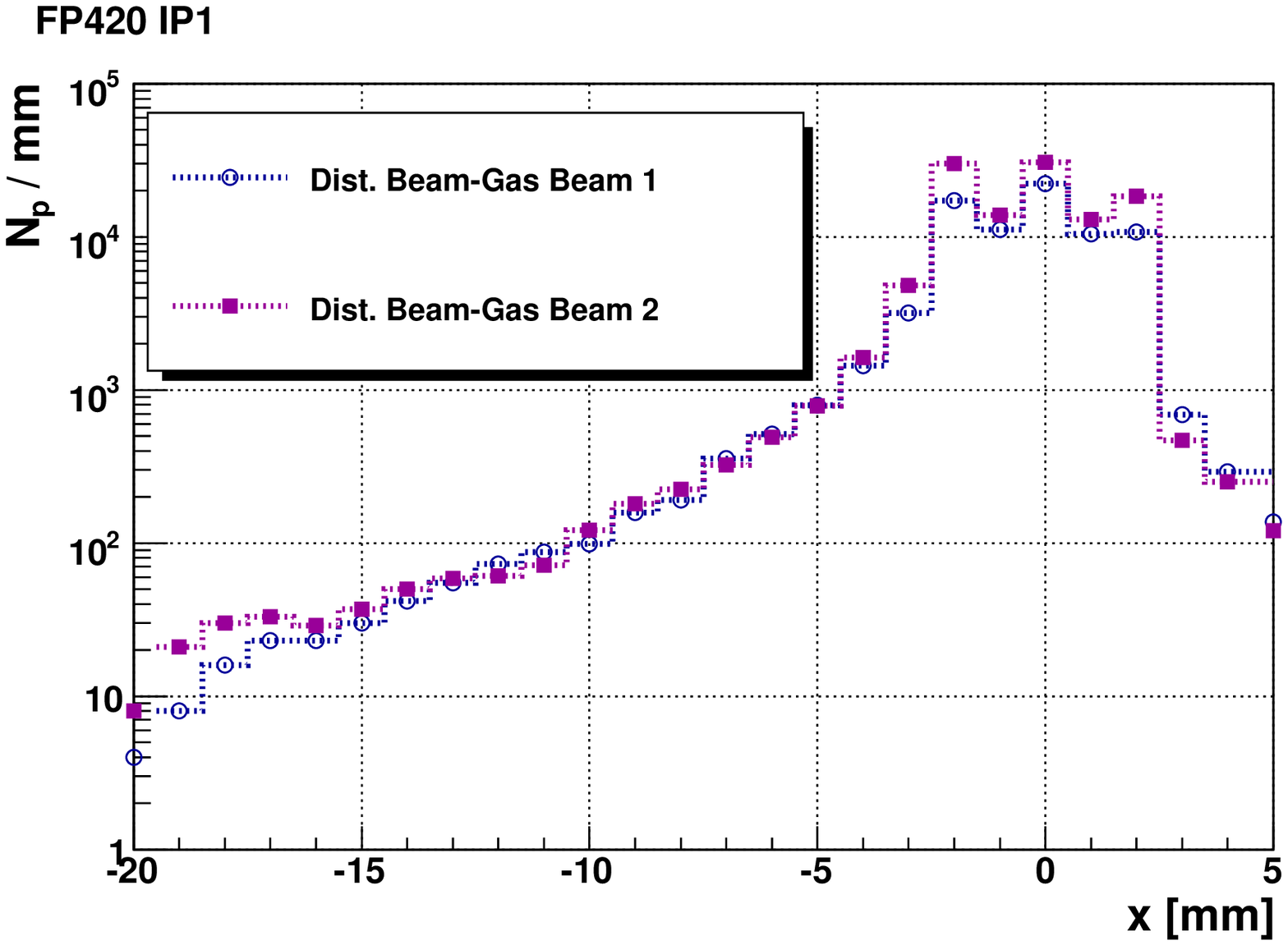}
\includegraphics[width=0.49\columnwidth]{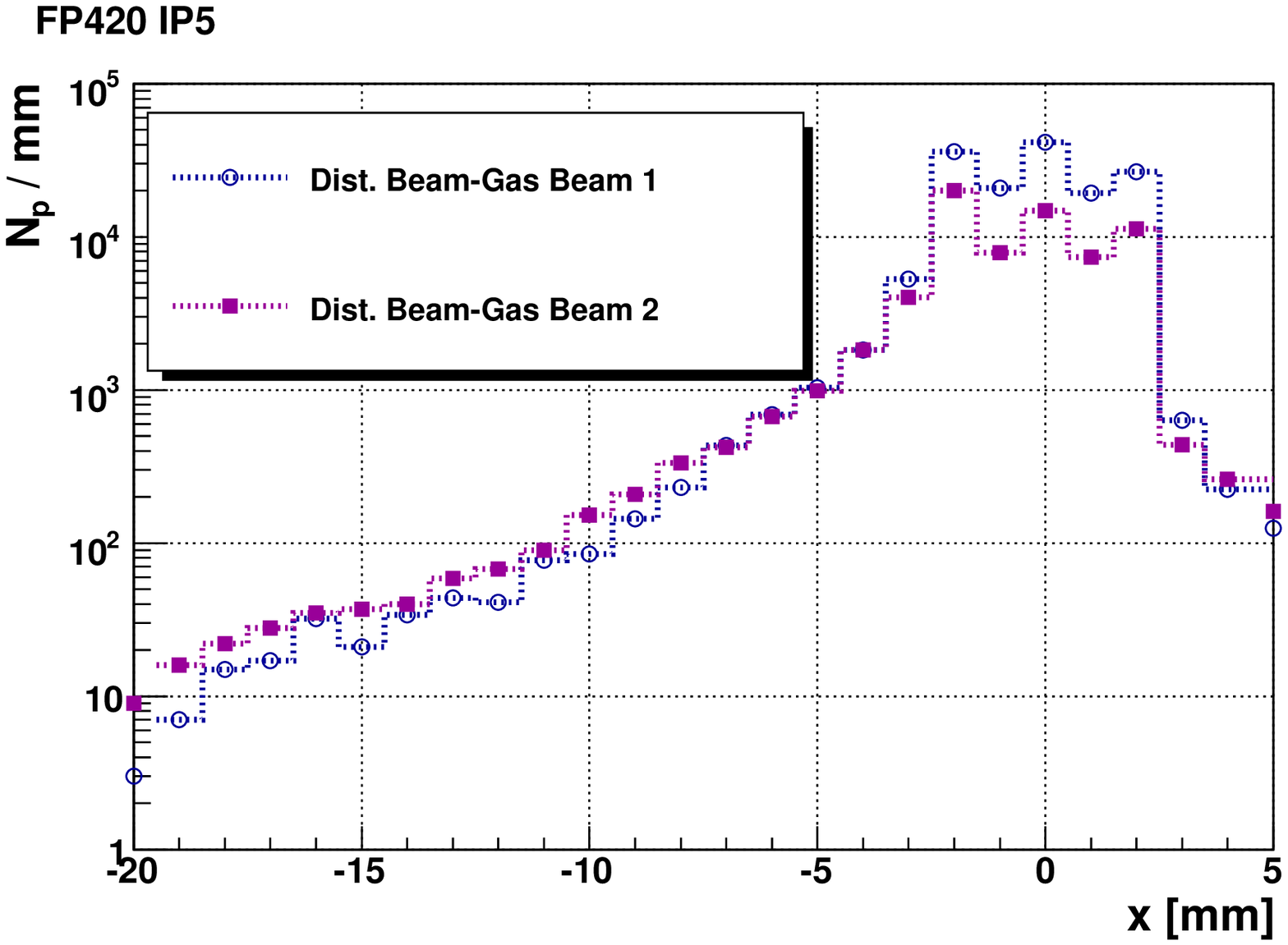}
\caption{Horizontal distribution of the protons scattered due to
beam-gas interaction, as observed at FP420 after the simulation of
$1\cdot 10^7$ proton-gas nuclei events per LHC beam.}
\label{fig:halodist2}
\end{figure}

The horizontal distribution of the beam halo protons at FP420, after
the simulation of $1\cdot 10^7$ proton-beam gas interactions per
beam are shown in Fig.~\ref{fig:halodist2}. These distributions are
normalised for the expected beam lifetime $\tau_{bg}$ related to beam-gas
interaction, as shown by Eq.~\ref{equ:norm}. 

\begin{figure}[htbp]
\centering
\includegraphics[width=0.49\columnwidth]{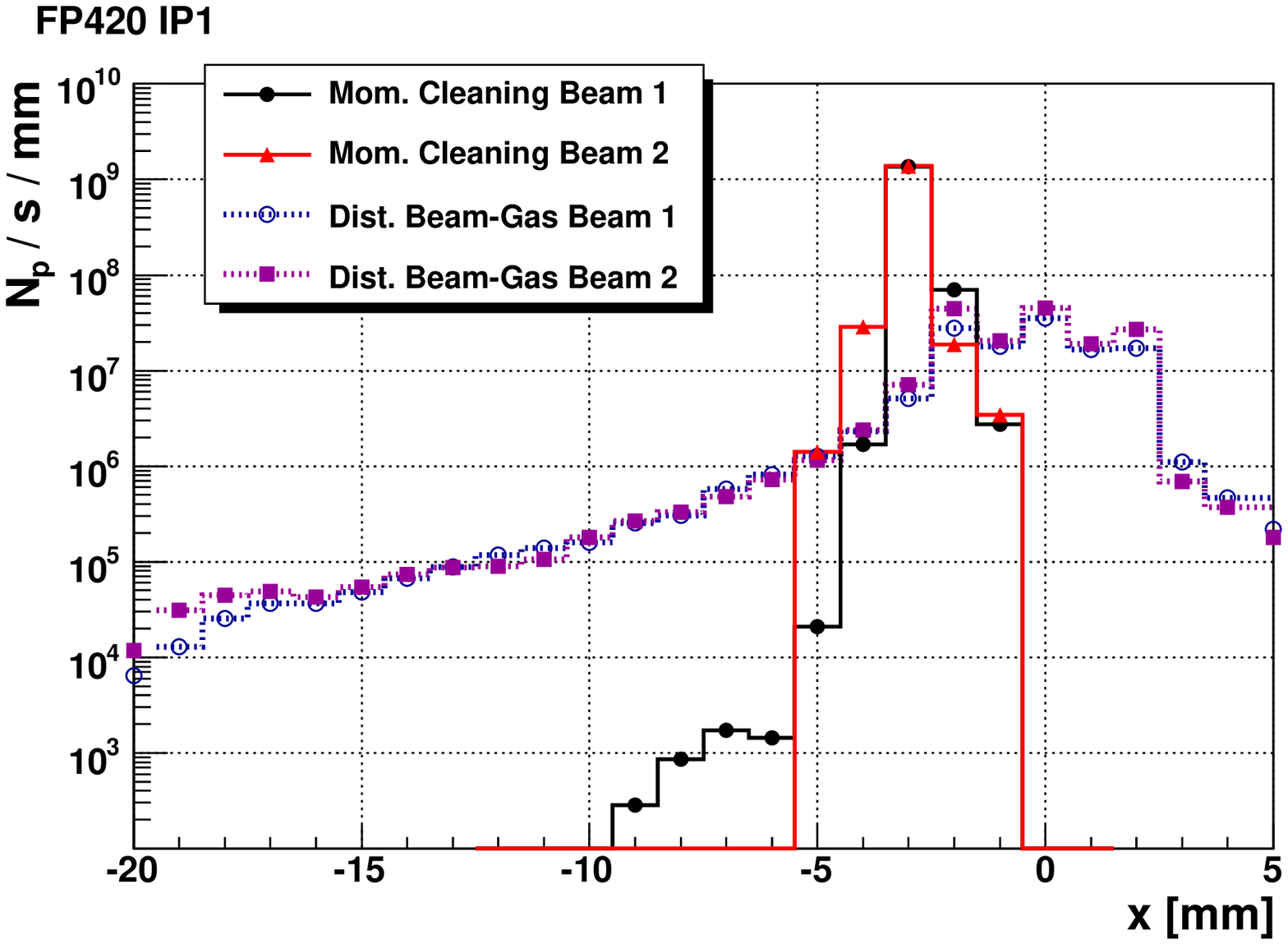}
\includegraphics[width=0.49\columnwidth]{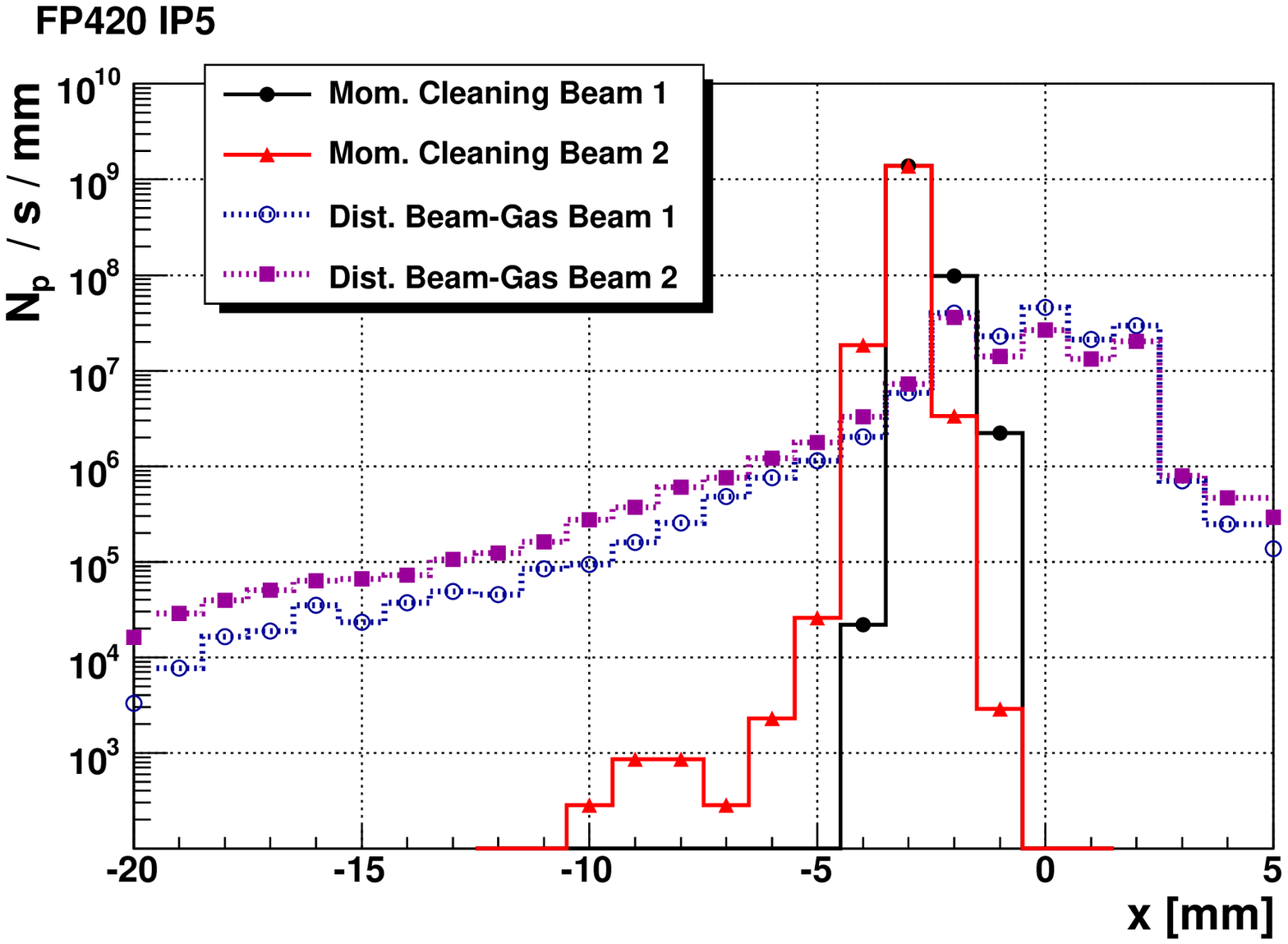}
\caption{Horizontal distribution of the protons emerging from the
momentum cleaning collimators and scattered due to beam-gas
interaction, as observed at 420~m, after normalization for the beam
lifetime as described in text.} \label{fig:halodist_norm}
%
\centering
\includegraphics[width=0.49\columnwidth]{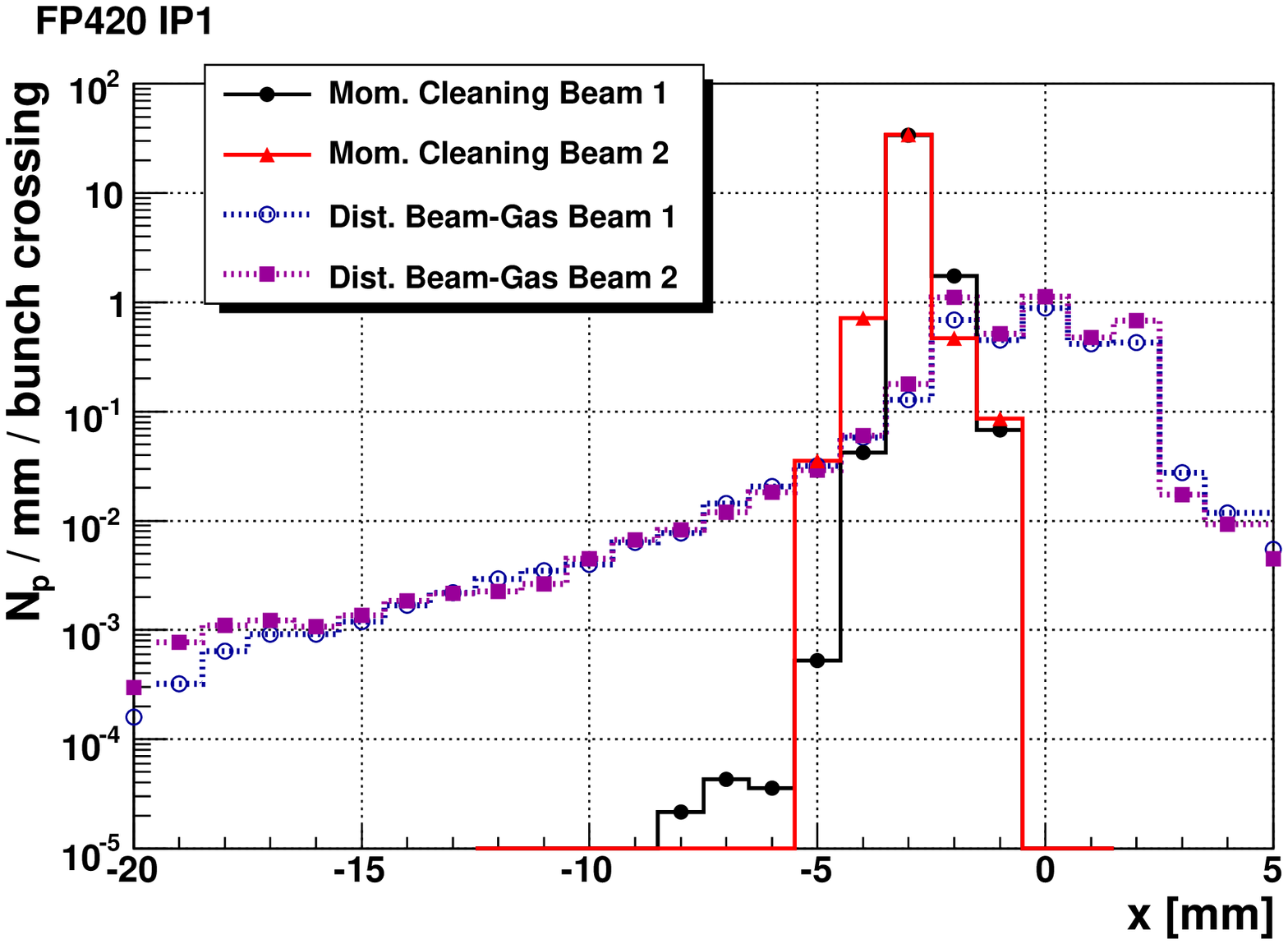}
\includegraphics[width=0.49\columnwidth]{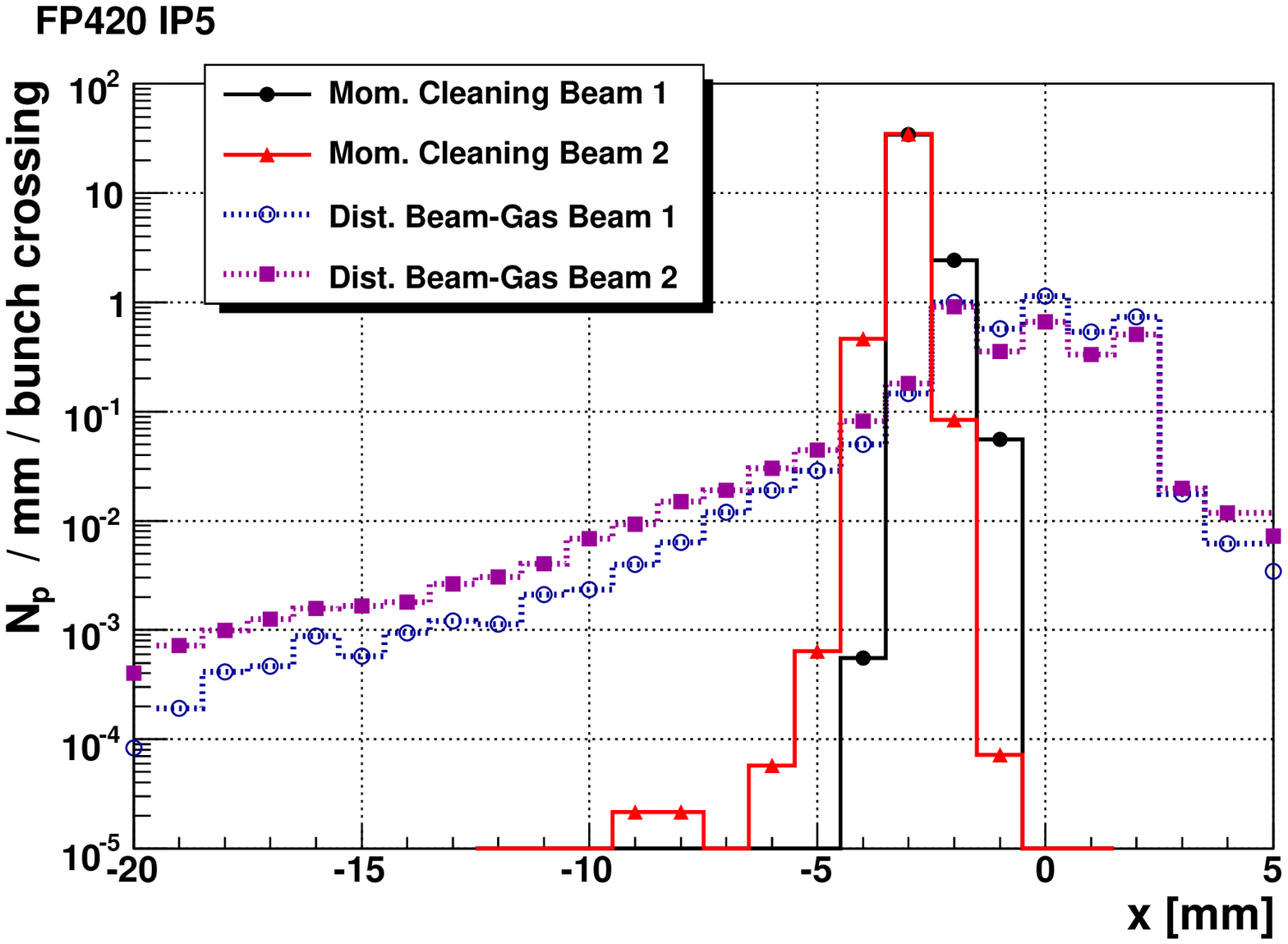}
\caption{Renormalised versions of Fig.~\ref{fig:halodist_norm}
yielding  the number of halo protons per bunch crossing.}
\label{fig:halodist_norm_bcrossing}
%
\centering
\includegraphics[width=0.49\columnwidth]{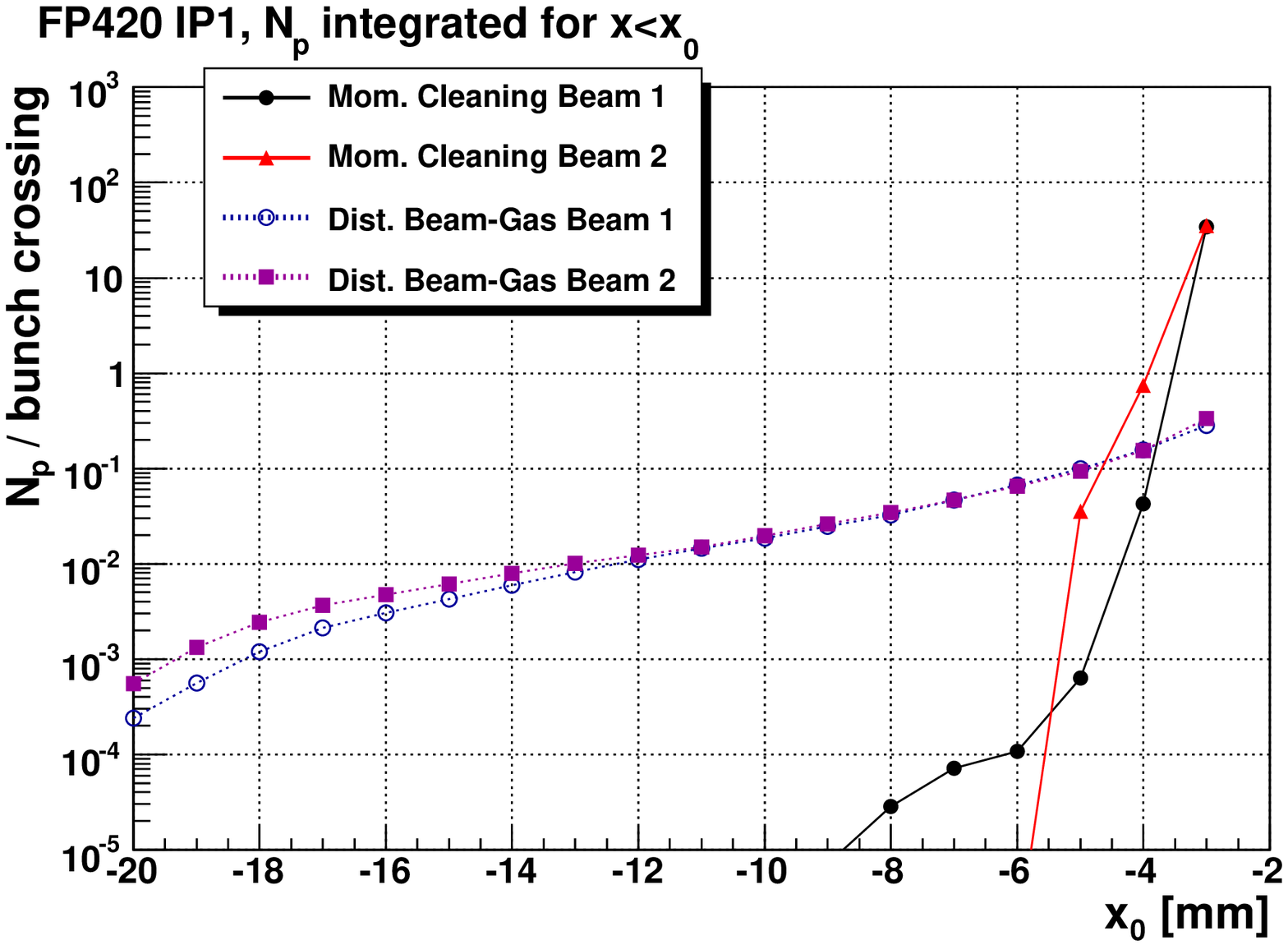}
\includegraphics[width=0.49\columnwidth]{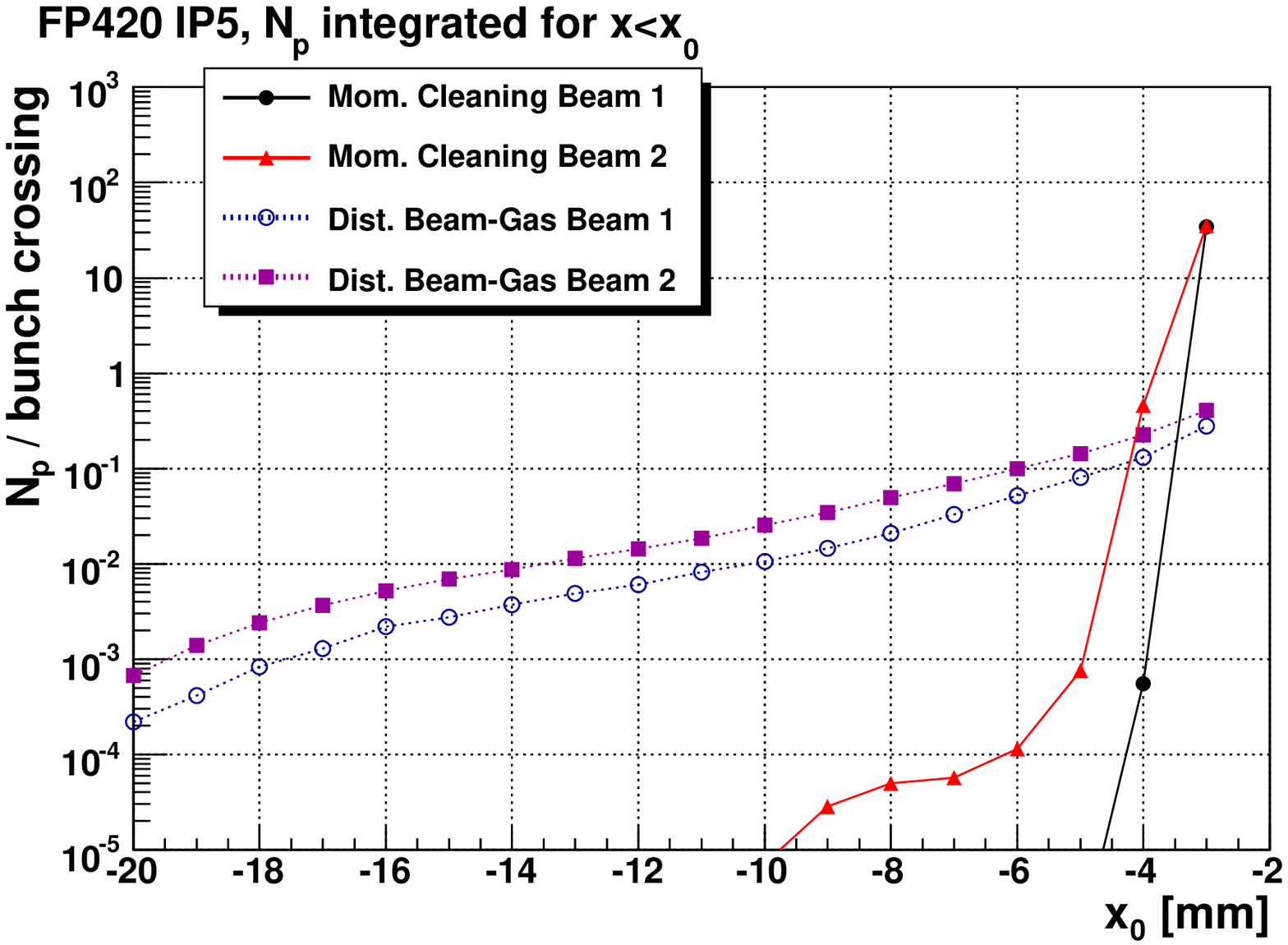}
\caption{Total amount of beam halo protons predicted at the 420~m
regions for different FP420 detector horizontal positions.}
\label{fig:halo_integrated}
\end{figure}

%
During the LHC startup period, $\tau_{bg}$, averaged over the all LHC ring, 
is expected to be around 100 hours. Later, during the LHC operation at high luminosity 
(after the so-called "beam pipe conditioning" by the beam itself), such value is expected 
to be higher and here we use $\tau_{bg}=500$ hours.   
%
The normalised profiles are shown in Fig.~\ref{fig:halodist_norm}, where
the resulting number of protons per second and per millimeter is 
compared to the simulated distribution (and normalised to the
relevant lifetime $\tau=150$ hours)  for the momentum cleaning
collimators beam halo (Sec.~\ref{sec:momclean}).

Also here, it is instructive to apply another normalization factor
$\mathrm{
N_{bs} = 4\cdot 10^7}$ (number of bunches per second), 
to obtain the beam halo distributions associated with each bunch
crossing, as shown in Fig.~\ref{fig:halodist_norm_bcrossing}. The
same data can be used to calculate the total number of beam halo
protons that will enter the 420~m regions, for different horizontal
positioning of the detectors (i.e. the number of protons integrated
from the outer beam halo edge to the detectors inner edge.). This
has been calculated in the plots of Fig.~\ref{fig:halo_integrated}.

The peak of the beam-halo distribution (see
Fig.~\ref{fig:halodist_norm} or
Fig.~\ref{fig:halodist_norm_bcrossing}) is determined by the LHC
momentum cleaning collimator settings. For nominal collimator
settings FP420 detectors located 5 mm from the beam would be well
away from this peak. To operate closer than 5 mm an adjustment of
the collimator positions would likely be required. Furthermore, for
detector distances greater than 5 mm, this background is dominated by
distant beam gas and the background rate is low.

\subsection{Secondary interactions}
\label{sec:secondary}

The transport of a proton bunch with an energy distribution will
result in proton losses when the protons interact with physical
elements of the beamline. This process results in electromagnetic
and hadronic showers, causing deposited energy and the production of
background particle species. The assessment of the effects of these
showers along the beam line and in the detector regions requires

\begin{figure}[htbp]
\centering
\includegraphics[width=0.58\columnwidth]{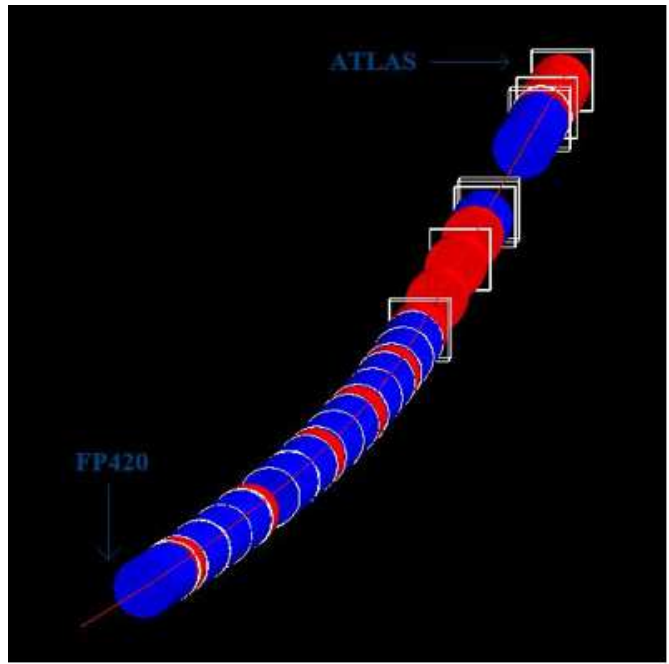}
\caption{Beam line model created with BDSIM, showing the beamline from Atlas to the position
of the FP420 detector (red denotes a
quadrupole element and blue donates a bending element).}
\label{fig:bdsim}
\end{figure}

\begin{description}
\item
- modelling of the beamline and detector regions, to correctly describe the type and distribution of matter;
\item
- simulation of the proton transport through the beam line optics;
\item
- simulation of the interaction of the beam particles with the beam line
apertures and the detectors.
\end{description}
To obtain a full simulation of secondary production along the beam
line the toolkit BDSIM~\cite{ref:bdsim} (Beam Delivery System
Simulation) has been used. This code, developed to study this class
of problem combines fast vacuum tracking of particles in the
beampipe with GEANT4~\cite{ref:g4}, which models the interaction of
beam particles with matter and is used whenever particles leave the
beampipe and enter solid parts of the machine. Hence BDSIM allows a
seamless integration of the optical properties of the beamline with
a full particle-matter interaction model. Figure~\ref{fig:bdsim}
shows the 3D volumes included in the BDSIM model of the beamline
from the ATLAS detector to FP420; red denotes a quadrupole element
and blue denotes a bending element.

\begin{figure}[htbp]
\centering
\includegraphics[width=0.68\columnwidth]{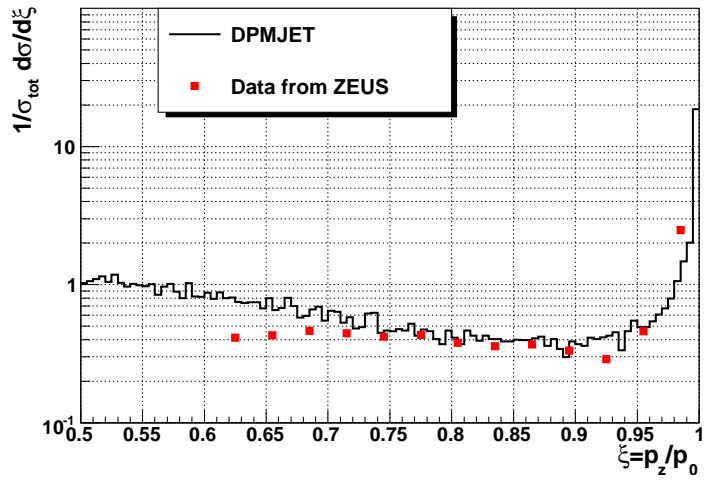}
\caption{The leading proton spectrum as a function of fractional momentum loss $\xi$ 
predicted by DPMJET and measured by the ZEUS Collaboration~\cite{Chekanov:2002yh}.
}
\label{fig:xllp}
\end{figure}

The input particle phase space from proton-proton collisions at
the interaction point was generated with the Monte Carlo
program DPMJET~\cite{ref:dpmjet}. 
DPMJET is the reference program for most of the
background studies for the LHC, and was chosen to produce the
final state proton spectra for this reason.  Figure~\ref{fig:xllp} shows a
comparison between the leading proton spectrum as a function of
fractional momentum loss $\xi$ generated by DPMJET and used in this
analysis compared to the $\xi$ distribution measured by the ZEUS
Collaboration at HERA~\cite{Chekanov:2002yh}.\\ 




%

\begin{figure}[htbp]
\centering
\includegraphics[width=0.68\columnwidth]{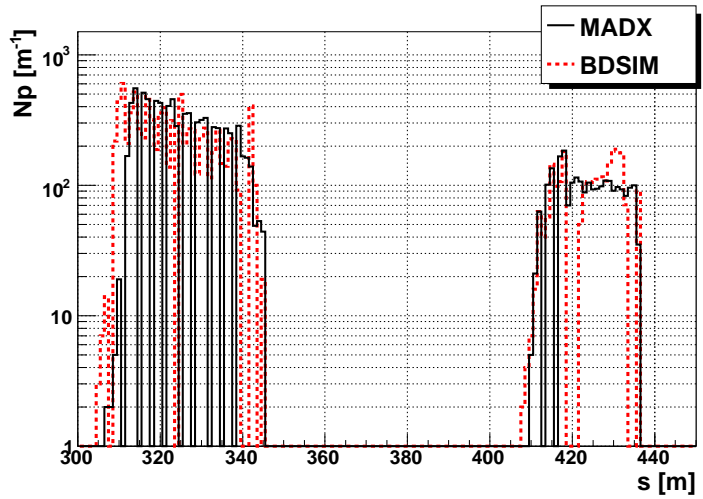}
\caption{Loss maps produced for the IP5 beamline with a DPMJET phase space sample using
MADX and BDSIM.}
\label{fig:eloss}
\end{figure}

The following simulations were performed to check the consistency of loss maps between BDSIM
and the code MADX~\cite{ref:madx}. They were performed for the IP5 beamline, for the LHC Beam 1, 
starting from the same proton sample generated with DPMJET at IP5 and consisting of 50000 protons 
with $\mathrm{dp/p<0.05}$ with respect to the nominal momentum $\mathrm{p_0=7 TeV}$.
The resulting number of protons lost as a function of the distance from the IP, in the region from 
300 to 420~m is shown in Figure~\ref{fig:eloss}. The figure shows a very good consistency of found 
loss locations between the two codes. Studies aiming at understanding the differences are ongoing. 


\subsubsection{Background particle fluxes and detector modeling} 
\label{sec:bgfluxes}

\begin{figure}[htbp]
\centering
\includegraphics[width=0.68\columnwidth]{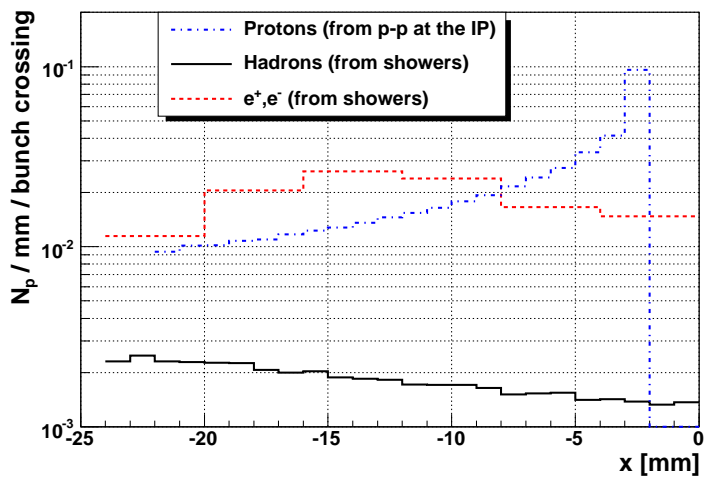}
\caption{Secondary particles flux at the entrance of the 420~m region downstream IP5 (preliminary results). 
The shower source is diffractive protons, generated with DPMJET, %
lost on the last bending magnet before FP420. The surviving diffractive protons are shown too.}
\label{fig:protvino_shower}
\end{figure}

The loss of protons shown in Fig.~\ref{fig:eloss} results in the
production of secondaries and the subsequent irradiation of the
FP420 detector region.
%
The electromagnetic and hadronic showers resulting from the transport of
the DPMJET phase space sample was calculated using BDSIM, and the
number and properties of the particle spectra estimated at 420~m.
These calculations were done using a subset of a DPMJET events with
565,000 final state protons on one forward side, which caused
proton loss and showering in the beamline immediately preceding 420~m.
The LHC total proton-proton cross section gives about 35
proton-proton collisions per bunch crossing, of which approximately 1/3 give forward protons, 
and BDSIM estimates the neutron rate to be 0.11 neutrons per bunch crossing at 420~m. This is equivalent to an
integrated rate of 44.4$\cdot$10$^{3}$ neutrons per cm$^2 \cdot$s,
with a time structure similar to the bunch structure
with a slight smearing to later times. The distribution of in-time  backgrounds is important for
time-of-flight analysis.
Hadronic models uncertainties in GEANT4 and uncertainties in the number of
events per bunch crossing imply that the numbers quoted here
are preliminary, and may result in a suppression of hadronic rates.
These numbers are currently being used to estimate the effect on the
detector signal-to-noise ratio and long-term damage, through
equivalent neutrons, and the systematic errors are under study.
In addition, the background contribution from charged secondary
particles generated by proton losses in the accelerator elements
immediately upstream of 420~m is under investigation. Preliminary
results of the Protvino group simulations
~\cite{ref:prot_bgpresent} (accounting for diffractive proton losses as source of secondary showers) 
are shown in Fig.~\ref{fig:protvino_shower}. 
%
These results have to be confirmed and crosschecked with BDSIM.

The detector region has also been simulated with
GEANT4~\cite{ref:g4}, and one aim of the background analysis is to
integrate the BDSIM model of the beamline with the GEANT4 model of
the detector. This simulation of the complete chain will allow
studies beginning with proton interactions at the IP, and ending
with the production and reconstruction of tracks in the detector
stations.

\begin{figure}[htbp]
\centering{\epsfig{file=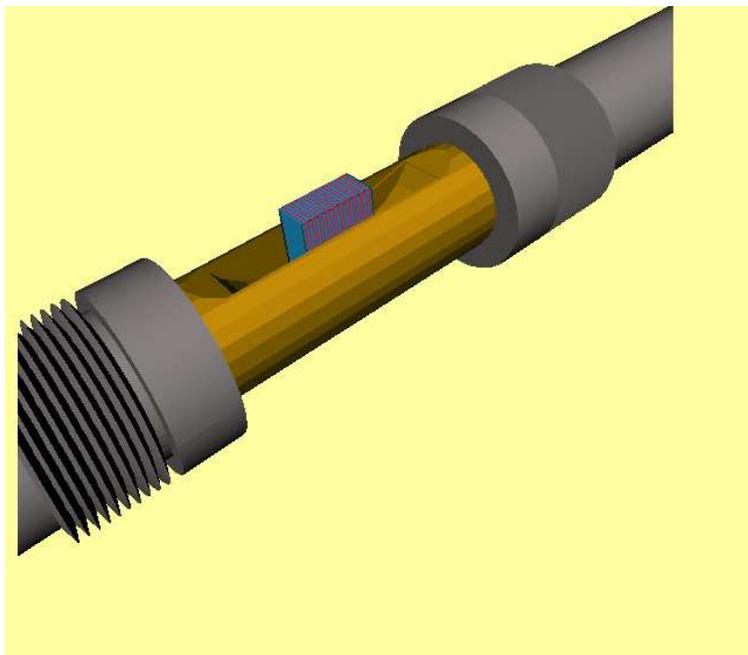,width=10cm,clip=}}
\caption{An example of the GEANT 4 geometry of the pocket hosting the
detectors. This model was used to study the interaction rate of 7 TeV protons, 
which is described in Section~\ref{sec:silicon_perform}.}
\label{fig:pocket}
\end{figure}

The GEANT 4 geometry of the detector pockets  is shown in
Fig.~\ref{fig:pocket}. In the GEANT4 simulations to-date, different
layout of the detector stations and surrounding pockets have been
considered, along with different numbers of sensitive planes. In all
cases the rate of secondary interactions of 7 TeV protons traversing
the full detector region was studied as a function of the materials
used and their thickness. The results of these studies are described
in Section~\ref{sec:silicon_perform}, 
where their impact on the
design of the layout of the detector region is discussed.

\subsection{Machine background summary}
\label{sec:bgconclusion}

The machine-induced background contribution at 420~m from
near beam-gas and the betatron cleaning collimation is expected to
be small, due to the arguments given in this chapter. However, there
is a contribution to the background rate arising from far beam-gas,
the momentum cleaning collimators and proton loss in the beamline.
The first two of these contributions give a proton background
which is described by a peak determined by the momentum cleaning
collimator settings, and a tail dominated by far beam-gas halo
protons. The combined distribution is shown in
Fig.~\ref{fig:halo_integrated}. At detectors transverse distance of 5 mm or
greater, the expected integrated number of protons from beam halo is
expected to be less than 1 per bunch crossing. The impact of a rate
of less than 1 proton per bunch crossing on the FP420 physics signal
in a pixel detector requires further study and comparison of the
background and signal spatial, angular and temporal distributions.
This may allow some degree of background rejection.

The proton loss background contribution is a mixture of charged and neutral
particles produced immediately upstream of 420m. The BDSIM estimate of the neutron
rate is 0.11 neutrons per bunch crossing at 420~m. The impact of these
preliminary neutral background rates will be assessed in term of
detector performance and survivability.

In summary, the preliminary proton and neutron background rates at
420~m have been estimated and need to be combined with detailed
detector and signal studies to understand the impact on the FP420
experiment.

\newpage

\section{A new connection cryostat at 420 m}
\label{sec:cryostat}

The LHC beamline layout downstream of an interaction point (IP)
consists of a triplet of low-beta quadrupole magnets, two beam
separation dipoles and a matching section of quadrupoles up to
quadrupole Q7. This is followed by a dispersion suppressor region of
standard dipoles and quadrupoles and finally the periodic lattice of
the arc. In the dispersion suppressors there is a 14 m drift space,
sometimes called the ``missing magnet'' drift space, which is
approximately 420 m downstream of the IP. In the LHC it was decided,
mainly for cost reasons, to place the dispersion suppressors and arc
magnets in one continuous cryostat from Q7, all the way to the
symmetric Q7 quadrupole upstream of the next IP ~\cite{lhcreport}.
At the position of the missing magnet, 420 m downstream of each IP,
there is a 14 m long Connection Cryostat (CC) which contains cold
beam-pipes, the 2K heat exchanger, or X-line, and various cryo-lines
which run throughout the continuous cryostat, as well as the
superconducting busbars and nearly 100 superconducting cables of the
main bending magnets and corrector magnets. There are sixteen CCs in
the LHC, each made to be as similar as possible to a standard arc
cryostat, as far as interconnection and handling are concerned. At
this 420 m point, the dispersion function D, with the standard high
luminosity optics, is approximately 2 m and hence protons from the
IP which have lost around 1\% of their momentum are well separated
from the circulating beam, as described in Sec.~\ref{sec:optics}.
Placing detectors directly inside the 1 m diameter cryostat at a
temperature of 2K was considered, but ultimately dismissed,
primarily because of the inevitable very high local heat load on the
LHC cryogenic system. The alternative is to replace the existing
connection cryostat with a warm beam-pipe section and a cryogenic
bypass. At the end of each arc cryostat of the LHC there is a
special short cryostat called an Arc Termination Module (ATM) which
includes cold to warm transitions for the beampipes and connects
cryo-lines and superconducting busbars and cables to the electrical
feed boxes. A New Connection Cryostat (NCC) with approximately 8 m
of room temperature beam-pipes has been designed using a modified
ATM at each end.
\begin{figure}
\centering\includegraphics[width=.9\linewidth]{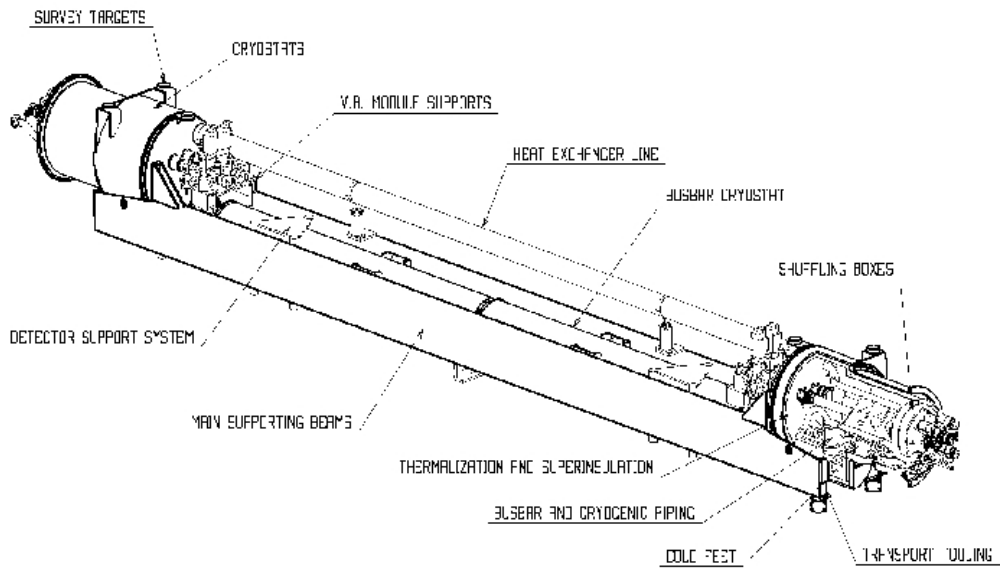}
\caption{The new connection cryostat for FP420}
\label{fig:cryostat}
\end{figure}

\begin{figure}
\centering\includegraphics[width=.6\linewidth]{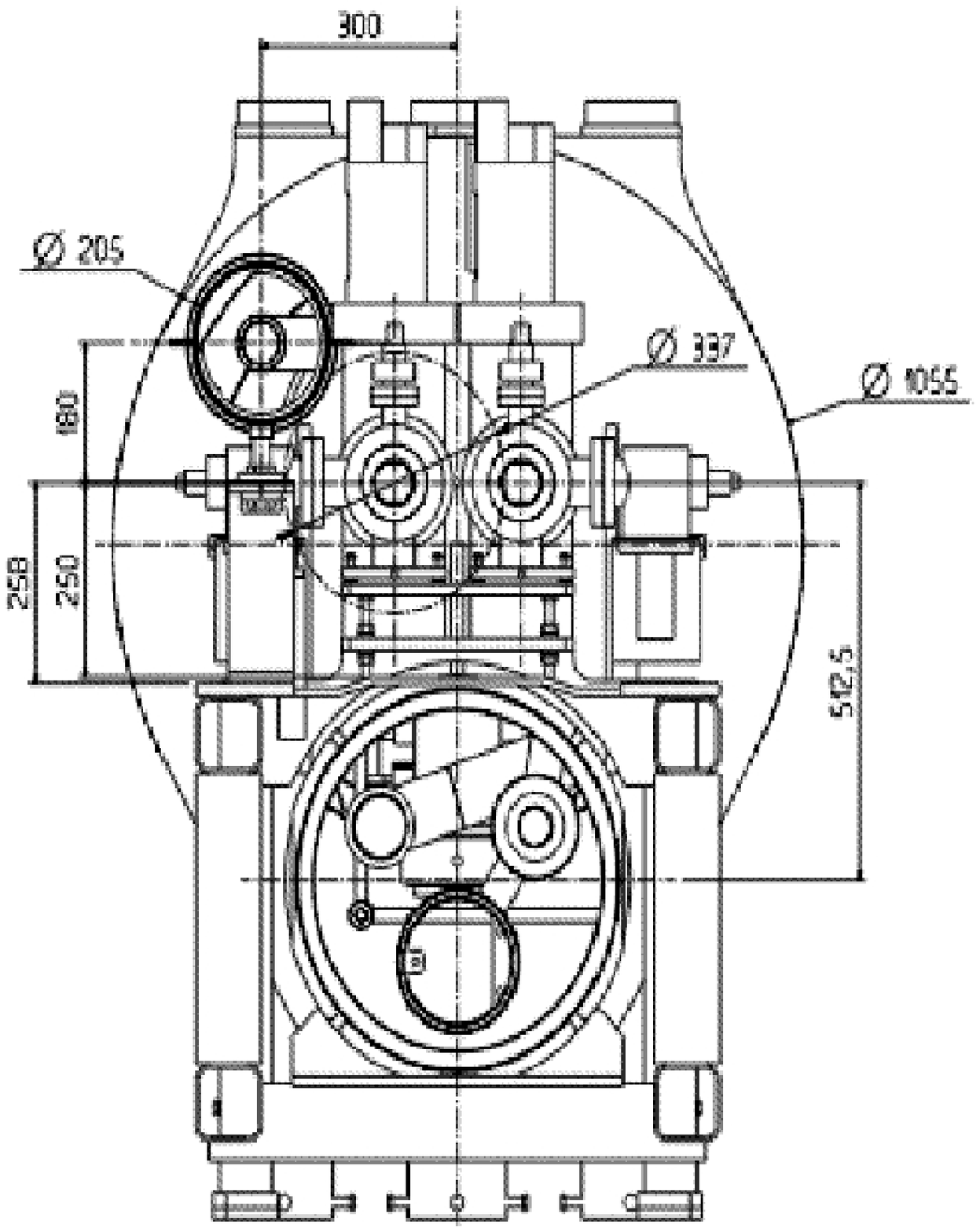}
\caption{Cross-section view of the new connection cryostat for FP420}
\label{fig:xscryostat}
\end{figure}

In addition to the two modified ATMs and warm beam-pipes, the NCC
shown in Fig.~\ref{fig:cryostat} has a small cross section cryostat
below the beam-pipes carrying all the cryo-lines and superconducting
circuits and a new specially designed cryostat for the X-line. All
this is supported by two longitudinal beams to make a single unit
which can be directly exchanged for an existing connection cryostat.
The passage of the X-line through the ATM modules is the main
modification needed to the standard ATMs, but the geometrical layout
of this passage has been arranged to be as far away as possible from
the downstream beam-pipe and hence leave adequate space for
near-beam detectors and their associated equipment. The
cross-section of the NCC, with the space around the beam-pipes
available for detectors and associated mechanics, is shown in
Fig.~\ref{fig:xscryostat}.

\begin{figure}
\centering\includegraphics[width=.9\linewidth]{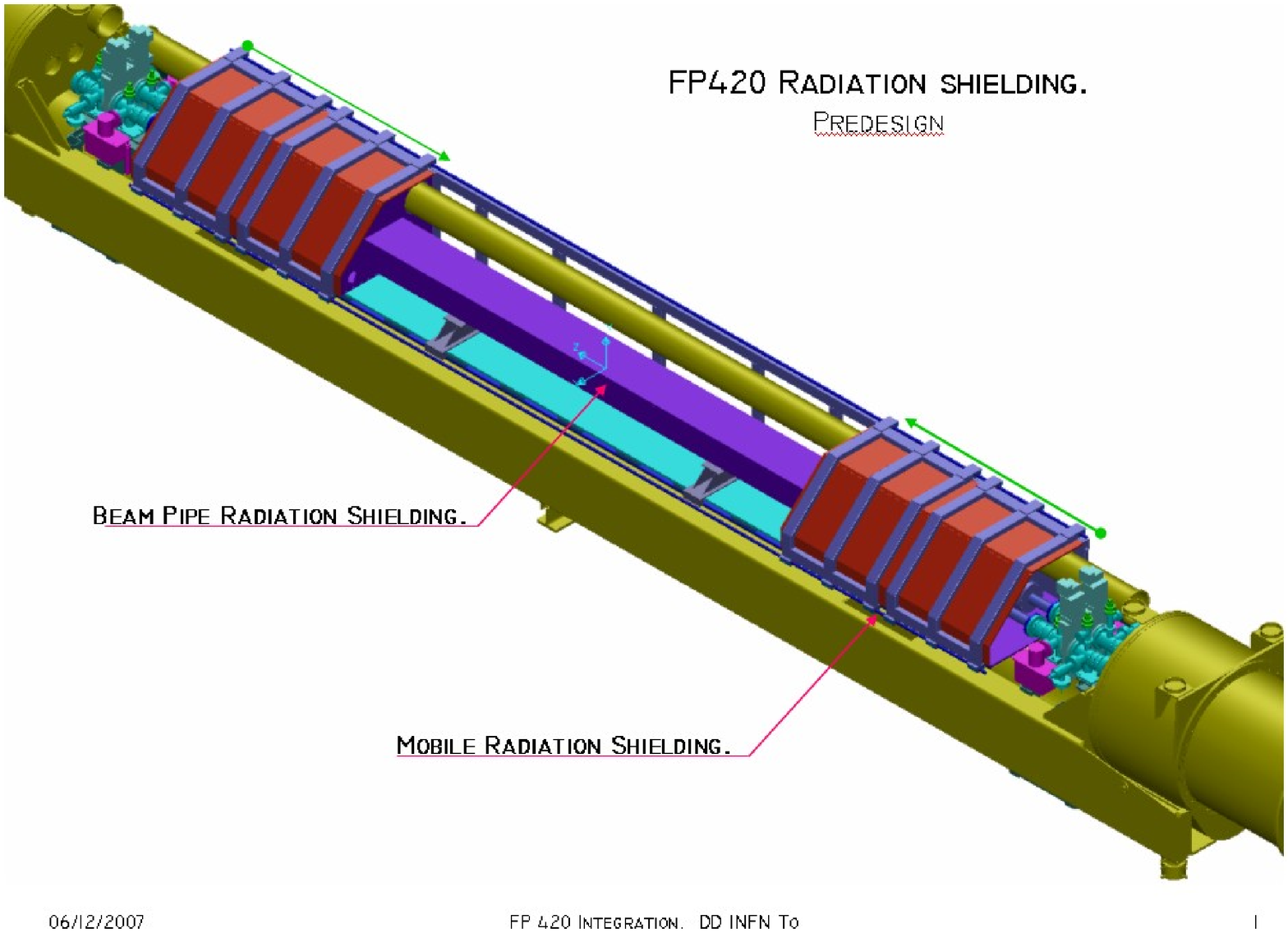}
\caption{Preliminary design for radiation shielding around the warm
beampipes and detector stations. The mobile shielding can be rolled
sideways on rails to give access to the detector stations.}
\label{fig:shielding}
\end{figure}

The existing connection cryostat contains a box structure of lead
plates of 15 mm thickness enclosing the two beam-pipes to reduce the
radiation field in the tunnel, essentially replacing the shielding
provided by the cold mass in a standard arc dipole cryostat. The
same thickness of lead shielding will be provided around the warm
beam-pipes and detector stations of the NCC. A preliminary design,
which provides a complete radiation shield while giving access to
the detector stations and passages for services is shown in
Fig.~\ref{fig:shielding}.

There are also short lengths of cylindrical shielding in the form of
collars around the beam-pipes at each end of the existing connection
cryostat to limit the risk of quenching adjacent superconducting
magnets. These same collars will be incorporated into the modified
ATM's at each end of the NCC in order to ensure that the performance
of the NCC is also equal to the existing cryostat in terms of
influence on the local radiation fields.

\begin{table}[htdp]
\begin{center}
\begin{tabular}{| c | c |}
\hline
& Normal Days \\
\hline
 Warmup from 1.9K to 4.5 K & 1\\
\hline
 Warmup from 4.5K to 300 K & 15\\
\hline
 Venting & 2 \\
\hline
Dismantling interconnection & 10\\
\hline
Removal of the connection cryostat &    2\\
\hline
Installation of the FP420 cryostat& 5\\
\hline
Realization of the interconnections&    15\\
\hline
Leak test and electrical test&  4\\
\hline
Closing of the vacuum vessel&   1\\
\hline
Evacuation/repump&  10\\
\hline
Leak test&  2\\
\hline
Pressure test & 4\\
\hline
Cool-down from 300 K to 4.5 K&   15\\
\hline
Cool-down from 4.5K to 1.9 K&    3\\
\hline
Total [days]&   89\\
\hline
\end{tabular}
\end{center}
\caption{The estimated time in days required to install one NCC}
\label{table:cryo}
\end{table}
The engineering design of the new connection cryostat is in progress
in the CERN central design office of the TS/MME group.  The design
aim is to meet or exceed the same specifications as the existing
connection cryostat, whilst providing the maximum useable space for
the FP420 detectors. The preliminary design offers acceptable
solutions for all cryogenic and mechanical engineering aspects as
well as integration into the LHC environment
\cite{Columbet1,Pattalwar1}. The final cryogenic performance will
depend on the detailed design, but it has already been established
that the additional static heat load arising from the two additional
cold to warm transitions will be tolerable for the LHC cryogenic
system. In fact, simulations show that during LHC operation the NCC
actually has a lower dynamic heat load than the existing connection
cryostat, because in the 8 m long warm section synchrotron radiation
will be absorbed at room temperature.

The detailed design of this second generation connection cryostat is
in progress and will be followed by an Engineering Change Request
(ECR) submitted to allow a detailed engineering review of all
machine aspects to be performed. Following acceptance of the ECR it
would in principle be possible to build two complete NCC's in about
a year and have them tested and ready for installation in late 2009.
Vittorio Parma of the AT Department's MCS group has accepted to take
up responsibility for the cryostat. He will lead a
working group which will verify the compatibility of the existing
conceptual design and develop the detailed design for manufacture.
As regards construction of the NCC's, the sixteen ATM
modules of the LHC were assembled at CERN in a dedicated workshop in
Building 110, under the responsibility of Ramon Folch (TS/MME). His
team has prepared a preliminary construction schedule and cost
estimate for the new cryostats \cite{Folch1}.

The cutting and removal of the existing connection cryostat and its
replacement by an NCC is very similar to the replacement of a
standard LHC dipole and has been evaluated by the group responsible
for all the LHC interconnections. As mentioned above this is the
same group that took responsibility for the design installation and
performance of the existing connection cryostat.

Table \ref{table:cryo} shows the sequence of operations and the
estimated time needed in normal working days to complete the
exchange of a connection cryostat from start of warm-up to being
ready for beam. It is thus conceivable that the installation of
FP420 modules consisting of an NCC cryostat and associated detectors
could be completed in an annual long shutdown. A preliminary study
of the transport aspects has shown that adequate tooling exists and
it can be expected that the time needed will be in the shadow of
other operations shown in Table \ref{table:cryo}. However, the
number of connection cryostats that can be replaced in a standard
annual shutdown will depend on the number of LHC magnets requiring
replacement and the work load of the interconnection teams.

\subsection{Cryostat summary}

In summary, a preliminary design for a replacement connection
cryostat that would allow detectors to be placed in the 420 m region
has been completed, and a final design is in progress. This solution
is expected to actually lower the dynamic heat load of the LHC and
have similar radiation profiles. With the appropriate approvals and
funding, two such cryostats could be built and installed in late 2009,
and in principle, two more in 2010 with negligible risk to LHC
operations. 
\newpage

\section{Hamburg beam-pipe}
\label{sec:hhpipe}

\subsection{Introduction}

Detection of diffractive protons at 420 m from the IP is
particularly challenging since it requires detectors to be placed
between the two LHC beam-pipes, the exteriors of which are separated
by about 140 mm (the distance between the pipe axes is nominally 194
mm). In addition, the nearby cryogenic lines severely limit the
available free space. Due to these space constraints the traditional
Roman Pot (RP) technique cannot be used, and another concept for near
beam detectors, pioneered at DESY is proposed. This technique of
moving sections of beam-pipe with integrated detectors is known as
``Hamburg pipes'' and was developed within the ZEUS collaboration in
1994 to measure very forward-scattered electrons as a signature of
photoproduction~\cite{piotrzkowski}. The concept was inspired by the
moving pipes used in the PETRA wiggler line to allow for beam-line
aperture changes. The ZEUS version involved small electromagnetic
calorimeters attached to the moving pipe (44 m from the interaction
point). The detectors were retracted during beam injection, but
could be positioned close to the beam axis during stable beam
conditions, and thus measure scattered electrons with reduced
energy, which exited the pipe through special thin windows. Since
the detectors were located outside of the machine vacuum, they could
be easily maintained and were successfully and routinely used for
six years, providing data essential for several publications
\cite{HHpipe2}. The detectors were positioned remotely by the HERA
shift crew, which inserted the detectors at the working position,
typically about 15 mm from the coasting electron beam, using the
HERA slow control system.

Prior to installation at HERA, the Hamburg pipe system was tested by
making several thousand displacement operations. No significant radio-frequency 
(RF) effects on the electron beam were observed due to the modified
beam-pipe geometry. It should be noted that no special RF screening
was used; it was sufficient to have the so-called RF fingers
providing good electrical contact across the connecting bellows.

The moving pipe technique has many advantages with respect to the RP
design. It allows much simpler access to detectors and provides
direct mechanical and optical control of the actual detector
positions.  In addition, unlike the Roman pot case which involves
forces from pressure differences as the detectors are inserted into
the vacuum, the Hamburg pipe maintains a fixed vacuum volume. This
results in much less mechanical stress, consequently allowing a very
simple and robust design.

\subsection{FP420 moving pipe design}

\begin{figure}[htpb]
\centering\includegraphics[width=.9\linewidth]{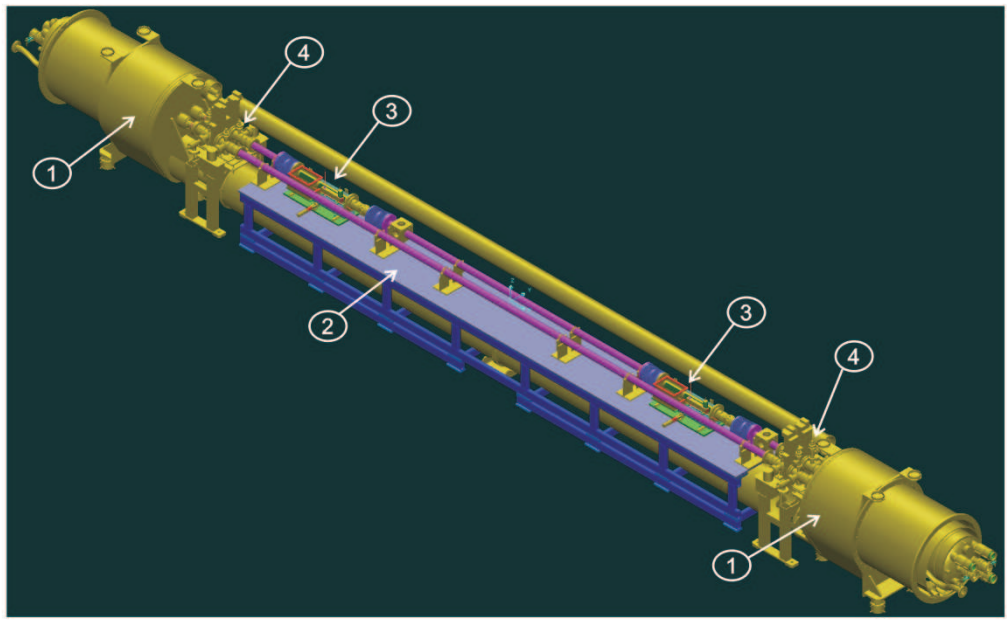}
\caption{Schematic view of the connection cryostat (1) and detector
arm with support table (2), two detector sections (3) and vacuum
pumping sections (4).} \label{hhpipe1}
\end{figure}

\begin{figure}
\centering\includegraphics[width=.9\linewidth]{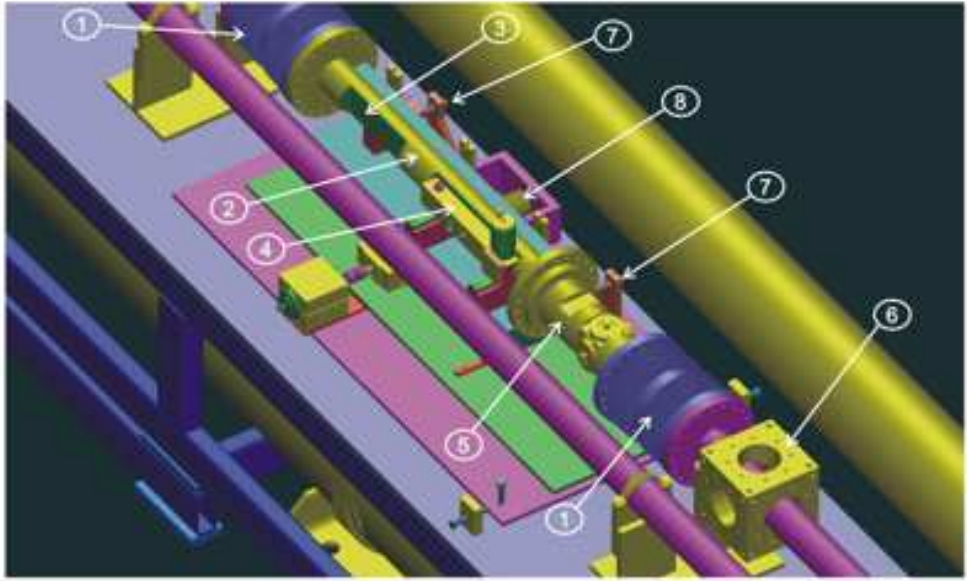}
\caption{Top view of one detector section: bellows (1), moving pipe
(2), Si-detector pocket (3), timing detector (4), moving BPM (5),
fixed BPM (6), LVDT position measurement system (7), emergency
spring system (8).} \label{hhpipe2}
\end{figure}

\begin{figure}
\centering\includegraphics[width=.9\linewidth]{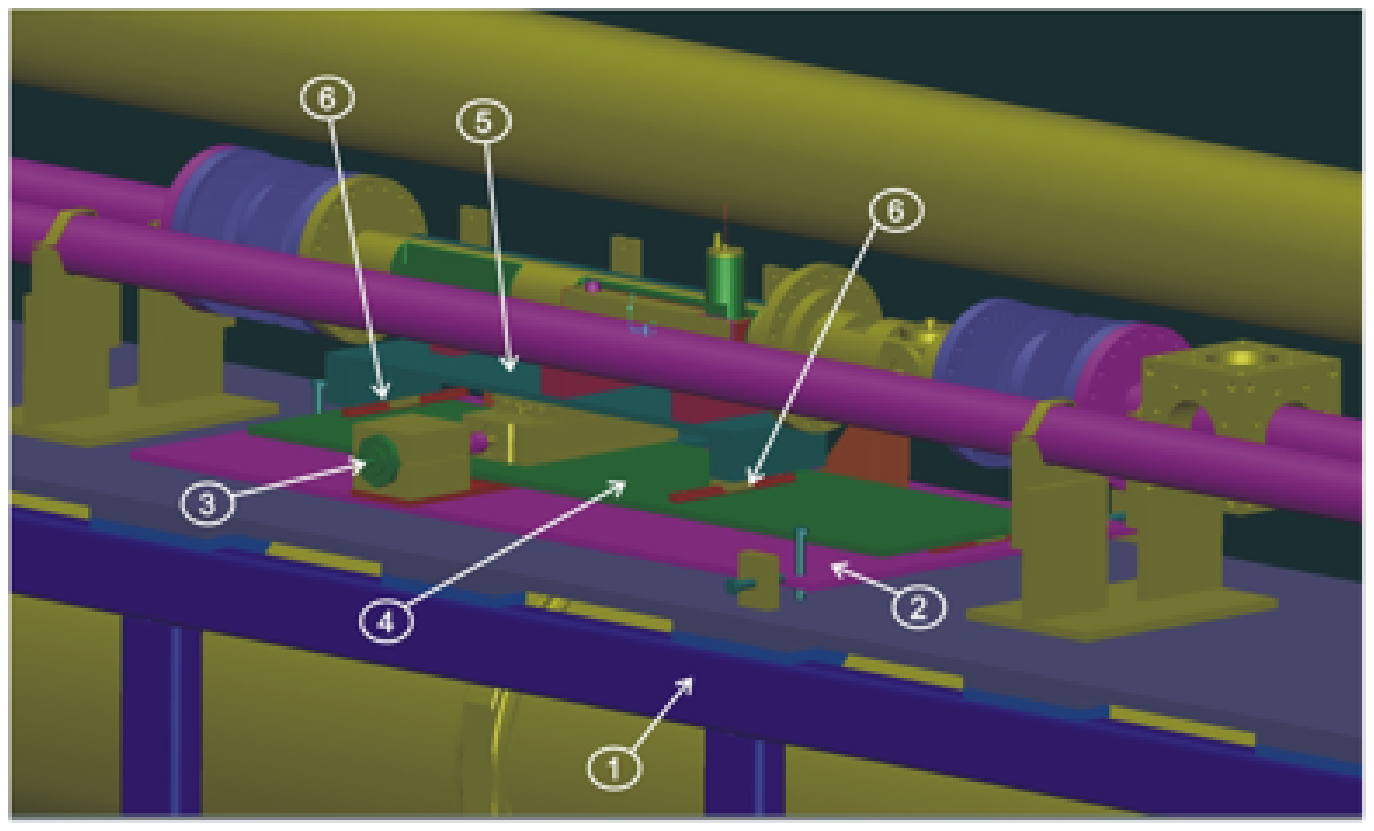}
\caption{Support table (1), drive support table with alignment
system (2), drive motor (3), intermediate table for emergency
withdrawal (4), moving support table (5), and linear guides (6).}
\label{hhpipe3}
\end{figure}

A modified connection cryostat (Section~\ref{sec:cryostat}) has been
designed with approximately eight meter long 
warm beam-pipes, providing adequate lateral space to
install the FP420 detectors. Figure~\ref{hhpipe1} shows the layout
of the connection cryostat including two detector stations and the
support table. The entire detector arm is fixed on the support
table, which is attached to the tunnel floor, independent of the
cryostat. Both ends of the detector arm are equipped with vacuum
pumping and control stations and isolation valves.
Figure~\ref{hhpipe2} shows one of the two detector stations equipped
with timing and silicon detectors, an LVDT (Linear Variable
Differential Transformer) for position measurement and one moving
and one fixed beam position monitor (BPM). The support table and
motion system are shown in Fig.~\ref{hhpipe3}.

The basic dimensions of the stations are defined by the LHC standard
beam-pipe diameter, the required lateral detector translation, and
by the longitudinal dimensions of the FP420 detectors. Each station
is composed of a beam-pipe with inner diameter of 68.9 mm, wall
thickness of 3.6 mm and a length of about 1000 mm, fixed on a
motorised drive. Rectangular thin-walled pockets are built into the
pipe to house the different detectors that must be positioned close
to the beam. The displacement between data taking position and the
retracted or parked position is about 25 mm. The ends of the moving
beam-pipes are connected to the fixed beam-pipes by a set of two
bellows. Inside, these may be equipped with moving RF-contacts to
assure electrical continuity. In general, this design allows
significant flexibility in the configuration of the detectors
stations, allowing optimization of the detector operation, scattered
proton detection,  kinematical reconstruction, and alignment.

\subsection{Pocket Design and Tests}

A key factor in the pocket design is the desire to maximise detector
acceptance, which is achieved by minimizing the distance of the
detector edge from the LHC beam. This in turn requires that the
thickness of the detector pocket wall should be minimised to limit
the dead area. Care must be taken to avoid significant window
deformation which could also limit the detector-beam distance.

\begin{figure}
\centering\includegraphics[width=.6\linewidth]{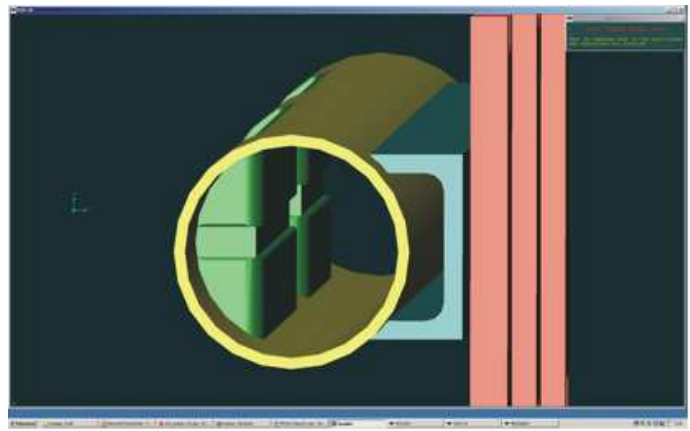}
\caption{Interior of the moving beam-pipe as seen by the particles.}
\label{hhpipe4}
\end{figure}

A rectangular shaped detector pocket is the simplest to construct,
and minimises the thin window material perpendicular to the beam
which can cause multiple scattering and degrade angular resolution
of the proton track. RF studies of the rectangular pocket have shown
(see Sec.~\ref{sec:RF}) that the effects on the beam dynamics are
minor. For reasons of mechanical stiffness, thermal stability and
fabrication of the pockets, only stainless steel beam tubes are
suitable. They will be copper coated for RF-shielding and Non-Evaporative Getter (NEG) coated
for vacuum pumping.  A rectangular slot of adequate height and
length is machined in the beam tube. A thin window is then welded in
this slot. Both welded and extruded pipes have been used in tests.
Figure~\ref{hhpipe4} shows the interior of the Hamburg pipe
including the thin vacuum window as seen by the scattered protons.

First tests using welded pipes showed excessive deformation due to
asymmetrical stresses appearing after the machining of the
cylindrical pipes. Several welding techniques for different length
pockets have been considered and two have been tested: Tungsten Inert Gas (TIG) 
welding and laser welding, the latter expected to produce somewhat
smaller deformation. A first prototype used thin (0.2 and 0.3 mm)
windows of 83 mm height and 200 mm length TIG welded in rectangular
slots, machined in a tube of diameter 89 mm. The deformation for
this setup under vacuum were unacceptable, exceeding 5 mm in the
centre. Pressure tests with this prototype have shown that the TIG
weld is quite strong, as it supported pressures of up to 7 bar.

\begin{figure}[htbp]
\centering
{\includegraphics[width=0.49\columnwidth]{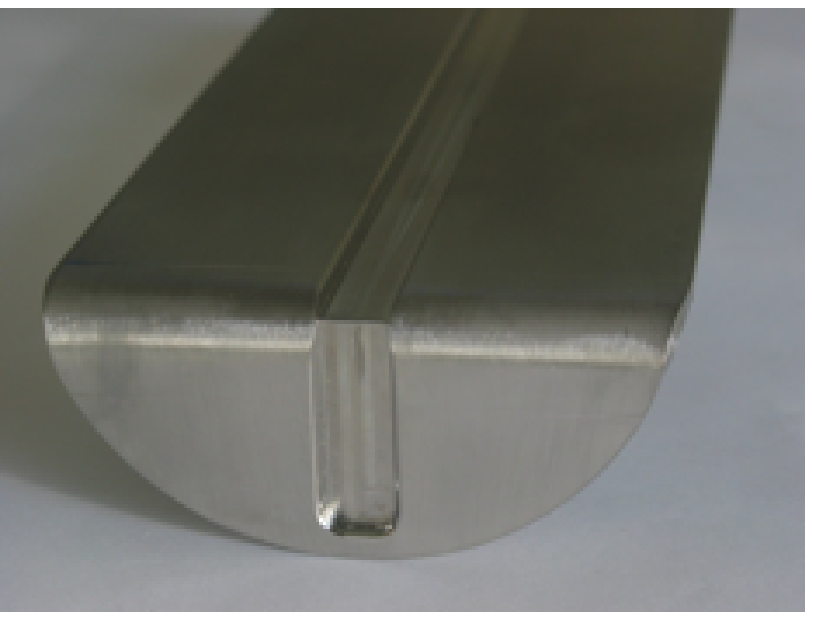}}
{\includegraphics[width=0.49\columnwidth]{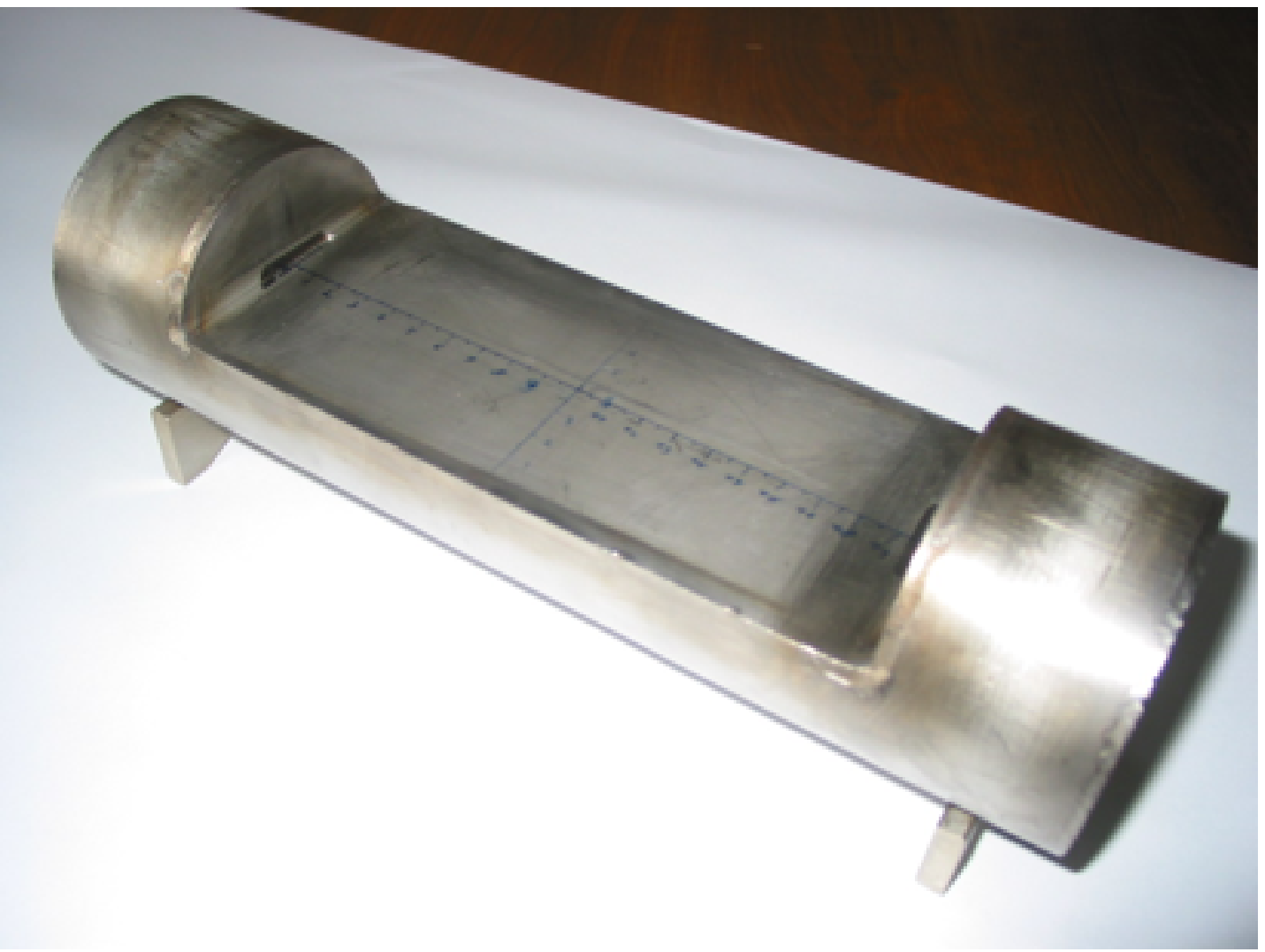}}
\caption{Hamburg pipe prototypes: (left) an end view of a machined
window before welding to the beam-pipe; (right) a 200 mm long
pocket, TIG welded in a tube without reinforcement. }
\label{hhpipe5}
\end{figure}

A first improvement was to use specially machined windows, which
have a thin wall of 0.5 mm only over 10 mm height and uses thicker,
solid walls for the remainder of the pipe (we note that the full
scale range of the scattered protons of interest is only a few mm as
shown in Sec.~\ref{sec:optics}. Figure~\ref{hhpipe5} shows this
design with the machined window TIG welded onto a long tube.

\begin{figure}[htbp]
\centering \subfigure[]
{\includegraphics[width=0.49\columnwidth]{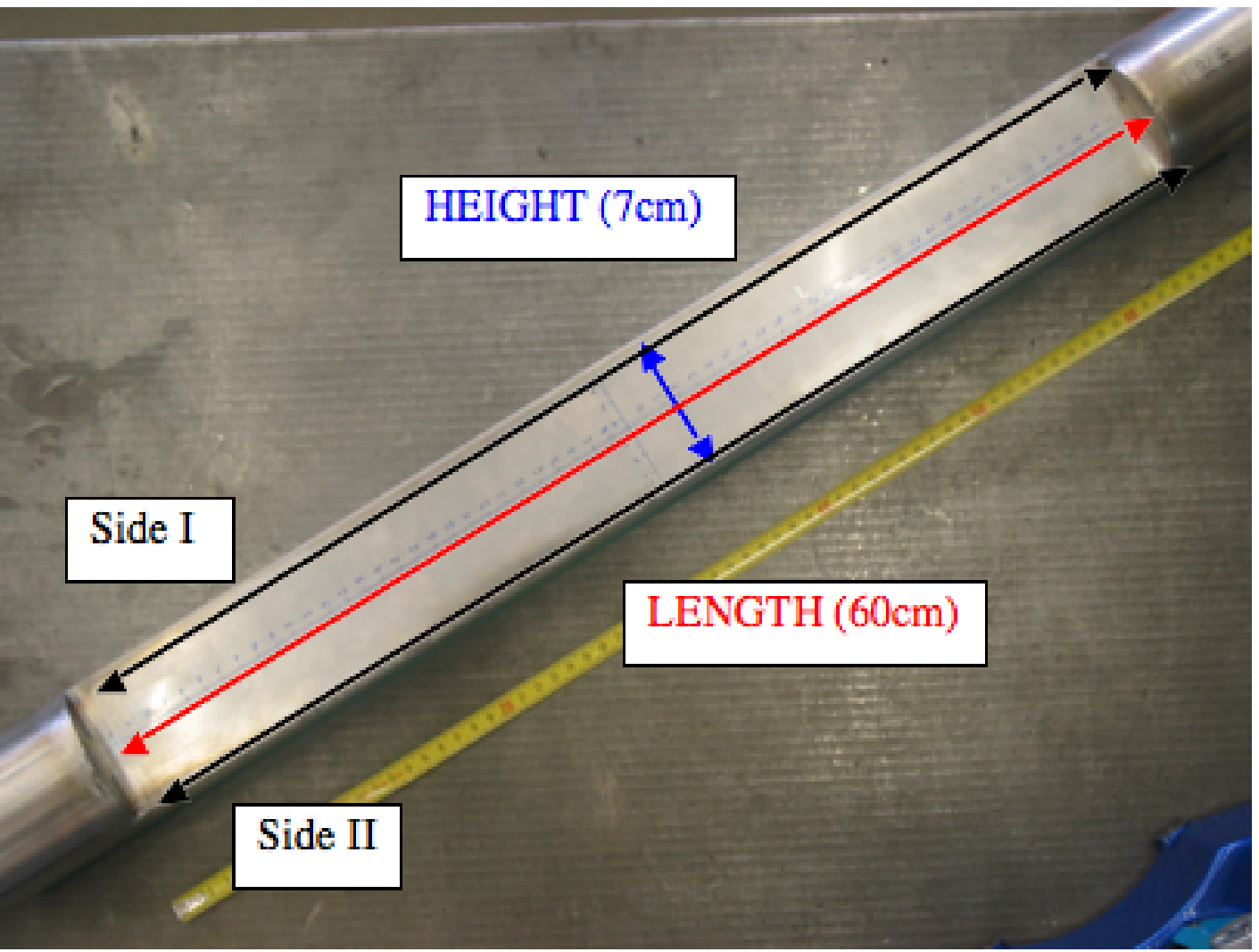}}
\subfigure[]
{\includegraphics[width=0.49\columnwidth]{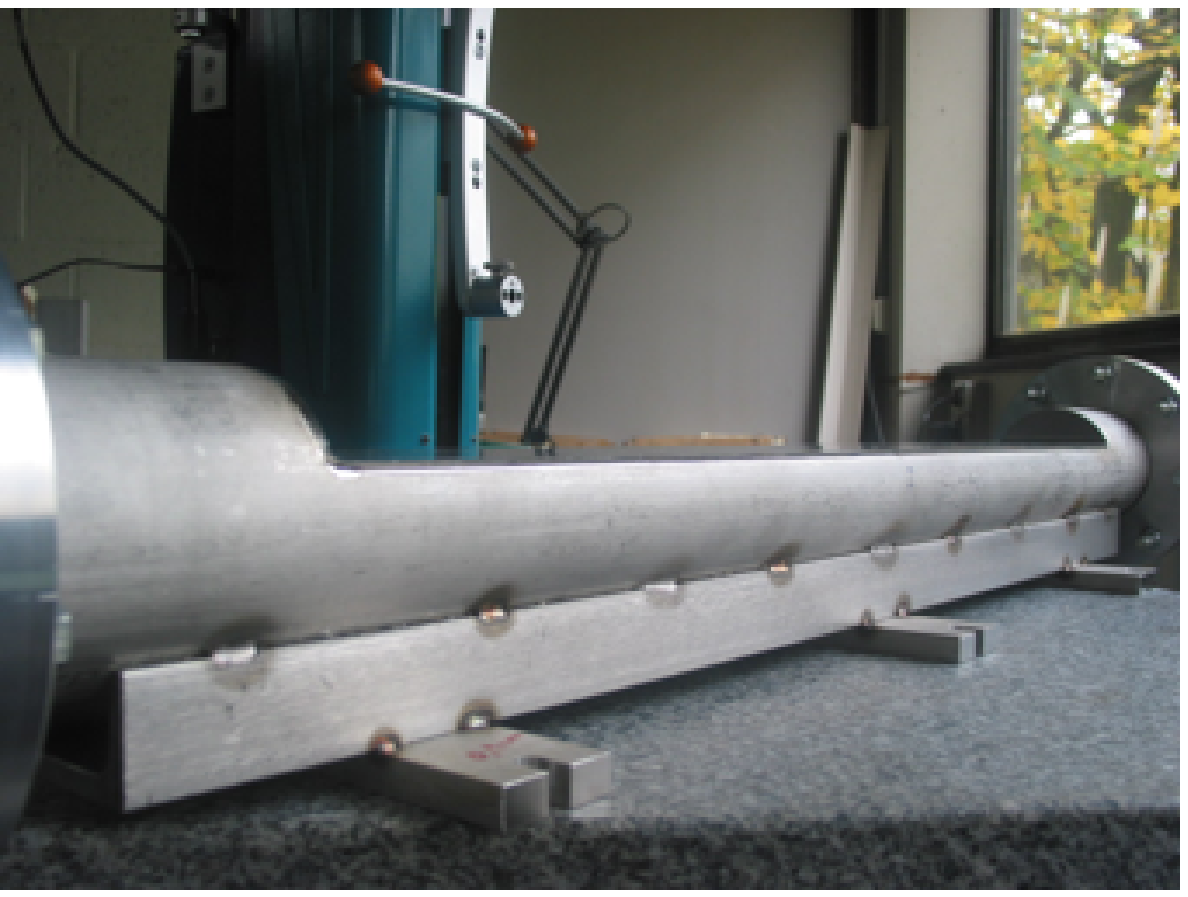}}
\caption{Hamburg pipe prototypes: (left) view of a 600 mm long
pocket, laser welded in a tube without reinforcement; (right)
picture of a 600 mm long pocket welded by laser in a reinforced
tube.}

\label{hhpipe6}
\end{figure}

A second improvement to keep the cylindrical tube from deforming was
to weld a U-shaped steel support to the back side of the pipe.
Figure~\ref{hhpipe6} shows the coordinate system used to measure the
deformation (and the locations where the measurements are made) and
the tube before and after the reinforcement is attached.

\begin{figure}
\centering\includegraphics[width=.9\linewidth]{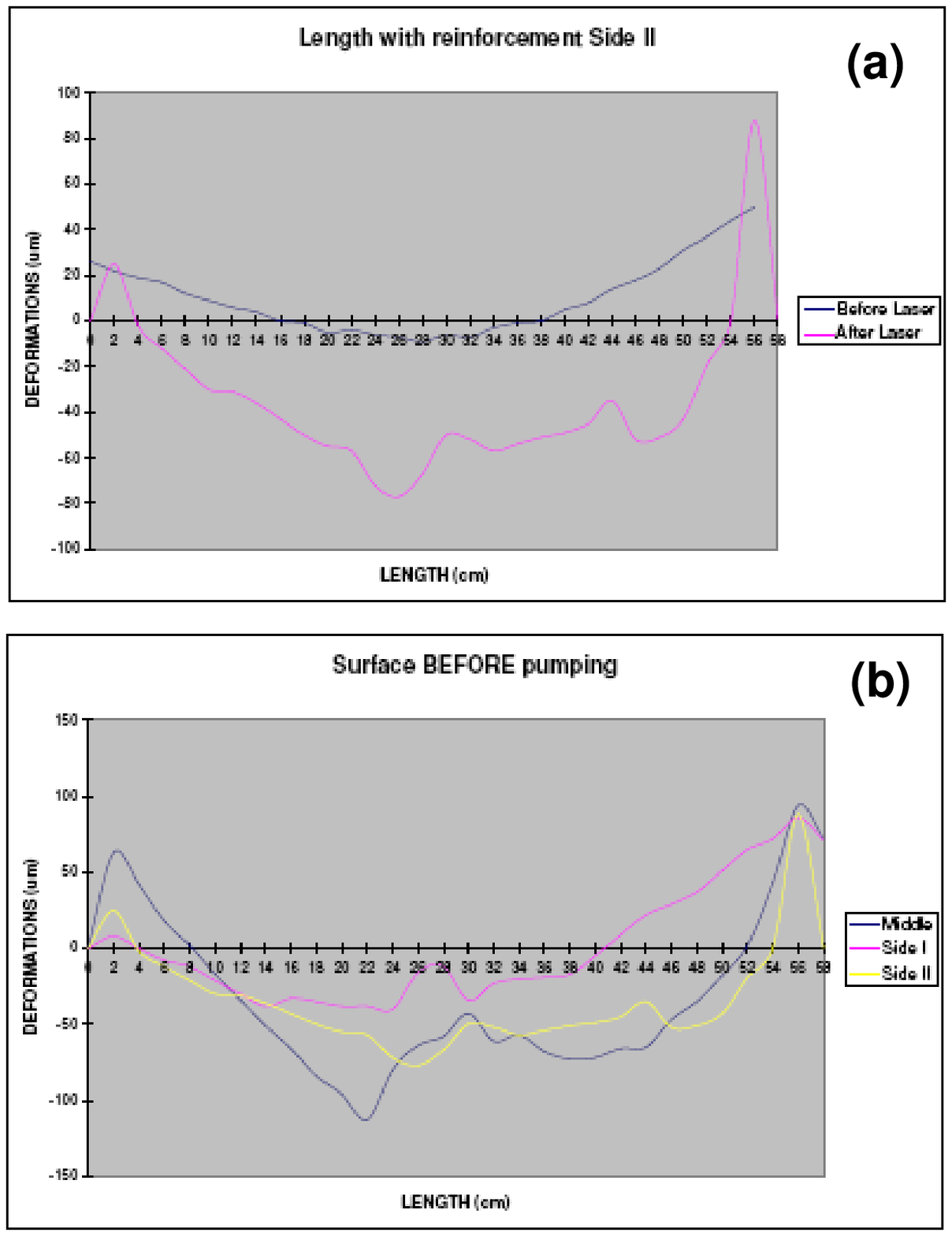}
\caption{(a) Deformation as a function of length at the ``Side 2''
location before (blue) and after (pink) laser welding. (b)
Deformation after welding but before vacuum pumping for the three
locations.} \label{hhpipe7}
\end{figure}

\begin{figure}
\centering\includegraphics[width=.9\linewidth]{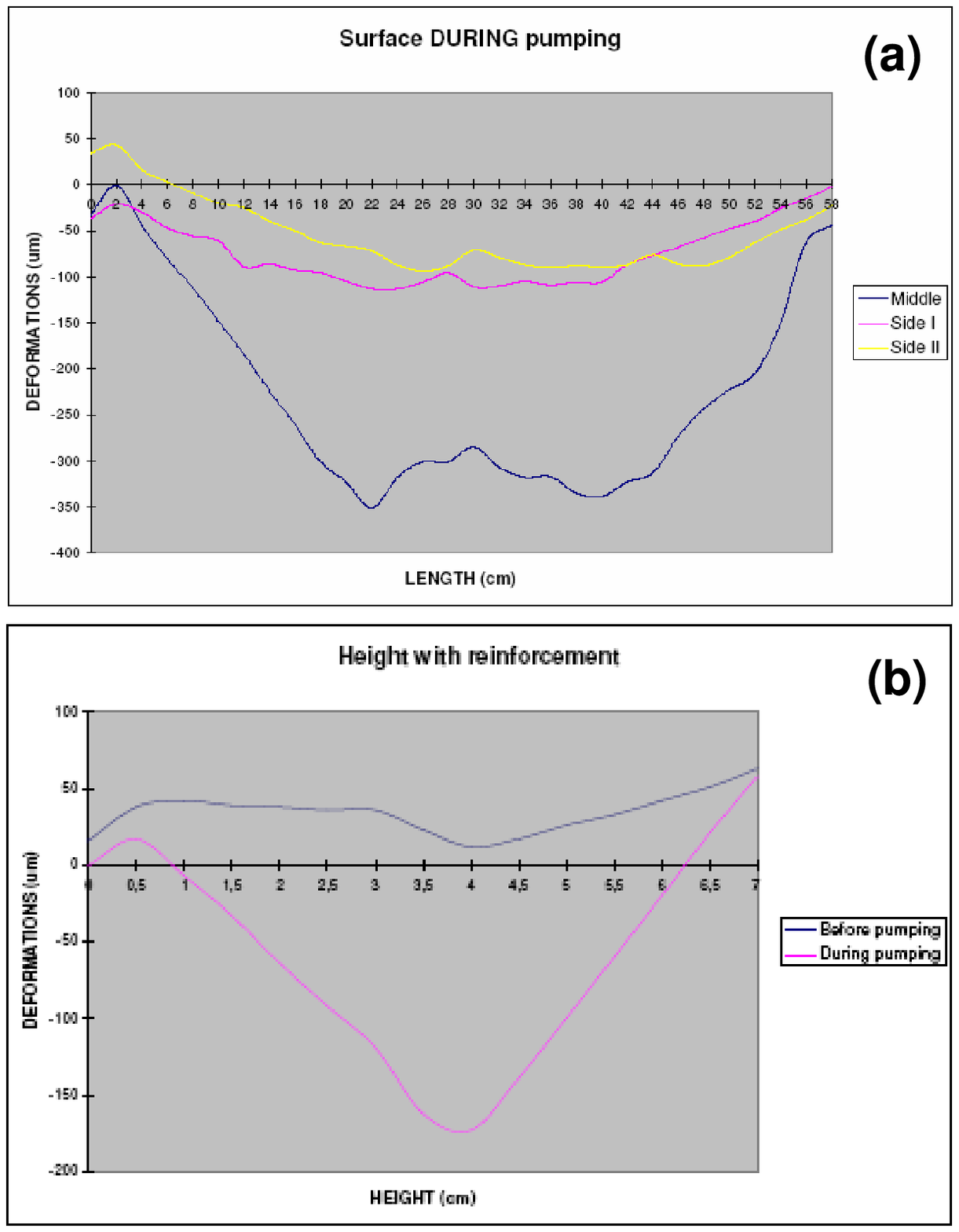}
\caption{(a) Deformation as a function of length after vacuum
pumping. (b) Height profiles after reinforcement.} \label{hhpipe8}
\end{figure}

We measured the deformation of the 600 mm pocket at different
stages. Figure~\ref{hhpipe7}(a) shows the deformation as a function
of length at the ``Side II'' (as defined in previous figure)
location before (blue) and after (pink) laser welding. Although the
magnitude of the deformation increases after welding, but is still
less than 100 $\mu$m, far superior than the TIG welding case.
Figure~\ref{hhpipe7}(b) shows the deformation after welding but
before vacuum pumping for three parallel lengths. The effect is
similar, although it is a little worse in the middle (blue) than in the
two sides, as expected, it is still less than 100 $\mu$m.
Figure~\ref{hhpipe8}(a) shows that although the deformation at the
sides (edges) is not much affected by vacuum pumping, it becomes
much larger ($>300$ $\mu$m) in the middle. After reinforcement,
however, it is reduced to acceptable levels, as shown by the
perpendicular height profile at the middle of the tube in
Figure~\ref{hhpipe8}(b). We also note that the final design will
have pockets of 1/3 to 1/2 the length, implying significantly less
deformation.

\begin{figure}
\centering\includegraphics[width=.9\linewidth]{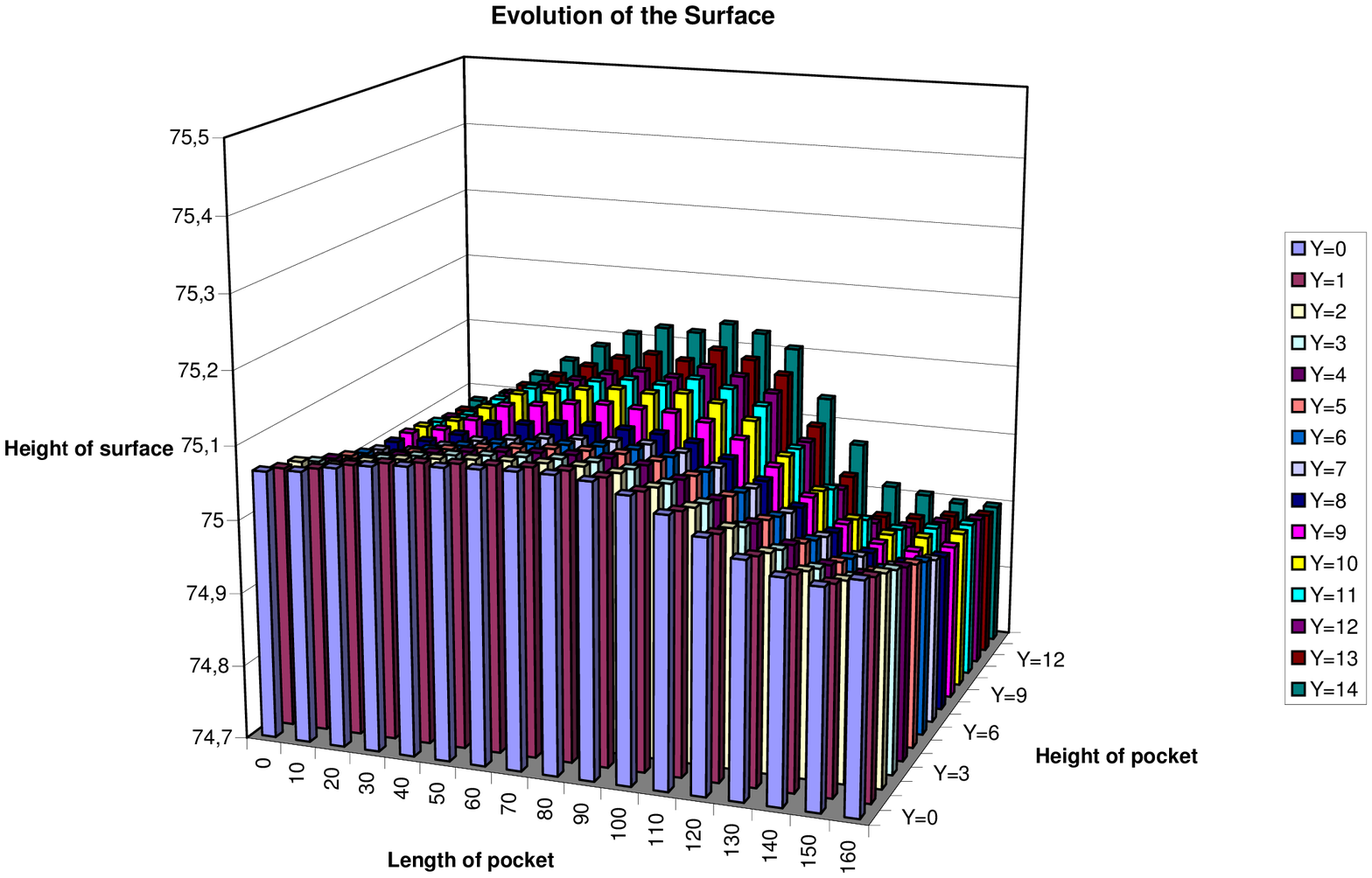}
\caption{Example of a 3D profile measurement run using the multi
sensor equipment.} \label{hhpipe9}
\end{figure}

A new 1 micron-precision 3D multisensor measuring device has been
tested on a 200 mm long TIG welded window. The result is shown in
Fig.~\ref{hhpipe9}. This device will help us fully evaluate the
final prototypes.

Two prototype beam tubes equipped with 600 mm long pockets were used
for RF measurements at the Cockcroft Lab. The results of these
measurements are described in Sec.~\ref{sec:RF}.

\subsection{Test beam prototype}

\begin{figure}
\centering\includegraphics[width=.9\linewidth]{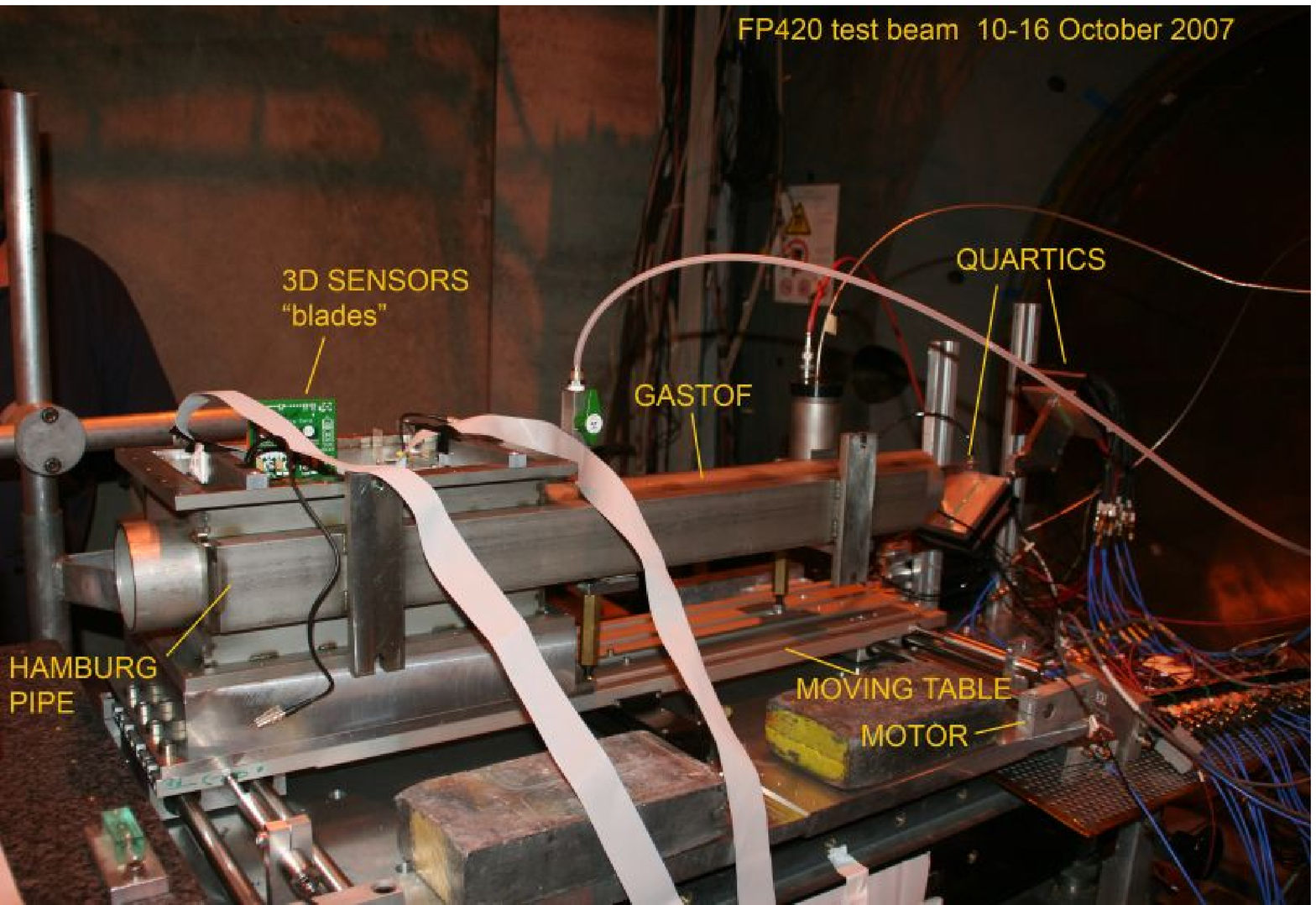}
\caption{Photograph of the prototype beam-pipe section used in the
October 2007 CERN test beam.} \label{hhtb}
\end{figure}

The baseline prototype of the moving beampipe was prepared for use
in test beam at CERN in October 2007. Figure~\ref{hhtb} shows the 1
m long beam-pipe equipped with two pockets, one of 200 mm length for
the 3D pixel detector (Section~\ref{sec:silicon}) and the other of 360 mm length for the gas
\v{C}erenkov timing detector (Section~\ref{sec:timing}) . The vacuum 
window thickness was 0.4 mm. A detector box for the 3D detectors was mounted in the first pocket.
The moving pipe was fixed on a moving table, driven by a MAXON motor
drive and guided by two high precision linear guides. The moving
table was equipped with alignment adjustments in the horizontal,
vertical, and axial (along the beam axis) directions, and was
attached to a fixed structure in the test beam area. The relative
position of the moving pipe was measured with two SOLARTRON LVDT
displacement transducers, which have 0.3 $\mu$m resolution and 0.2\%
linearity.

\subsection{Motorization and detector system positioning}

In routine operation, detector stations will have two primary
positions (1) the parked position during beam injection,
acceleration and tuning, and (2) close to the beam for data taking. The
positioning must be accurate and reproducible. Two options have been
considered: equipping both ends of the detector section with a motor
drive which are in principle moving synchronously but allowing for
axial corrections with respect to the beam axis, or a single drive
at the centre, complemented with a local manual axial alignment
system. A two motor solution in principle allows perfect positioning
of the detector station, both laterally and axially. However, it
adds complexity to the control system, reduces reliability, and
increases cost. Positioning accuracy and reproducibility are also reduced
because extremely high precision guiding systems can no longer be
used, due to the necessary additional angular degree of freedom.
Therefore, a single motor drive system has been chosen, accompanied
by two precise LVDTs. As in the LHC collimator system, no
electronics is foreseen in highly irradiated zones, close to the
motors, to limit radiation damage. For ease in integration, we are
planning to adopt the collimator stepping motor solution. As these
have never been irradiated, the stepping motors will be tested in
the high neutron flux test beamline at the Louvain-la-Neuve
cyclotron centre CRC~\cite{LouvainHHpipe}.

\subsection{System operation and safeguards}

Given the FP420 schedule, it will be possible to learn from the
experience that will be gained during the LHC commissioning by the
operation of machine elements with similar control and
surveillance aspects, namely the TOTEM~\cite{ref:totemTDR} and
ATLAS ALFA Roman Pot~\cite{ref:atlasRP} detectors and the LHC
collimators~\cite{ref:coll_controls}. Nevertheless a series of
aspects specific to FP420 need to  be addressed.
%
%
The F420 detectors will operate at all times in the shadow of the
LHC collimators in order to guarantee low background rates and to
avoid detector damage from unwanted beam losses. In addition, for
machine protection constraints, it will be unacceptable for FP420 to
interfere with the beam cleaning
system (e.g. to avoid magnet quenches downstream the 420~m region).\\
Therefore, the high-level Hamburg pipe control system will be
integrated in the collimator controls. The interface between low- and
high-level controls will be implemented with the
CERN standard Front End Standard Architecture (FESA)~\cite{ref:FESA}.\\
%
The Main Control Room will position the detectors close to the beam
after stable collisions are established. The precision movement
system will be able to operate at moderate and very low speed for
positioning the detectors near the beam. During insertion and while
the detectors are in place, rates in the timing detectors will be
monitored, as well as current in the silicon. The step motor and
LVDT's will provide redundant readback of the position of the
detectors and the fixed and moveable BPM's will provide information
on the position of the detectors with respect to the beam. Separate
mechanical alignment of the height and orientation with respect to
the beam are discussed in Sec.~\ref{sec:alignment}.

\subsection{Hamburg pipe summary and outlook}

The Hamburg moving pipe concept provides the optimal solution for
the FP420 detector system at the LHC. It ensures a simple and robust
design and good access to the detectors. Moreover, it is compatible
with the very limited space available in the modified connection
cryostat and with the expected position of the scattered protons
between the two LHC beampipes. Its reliability is linked to the
inherent absence of compensation forces and the direct control of
the actual position of the moving detectors. Finally, rather large
detectors, such as the timing devices, can naturally be incorporated
using pockets, rectangular indentations in the moving pipes. The
prototype detector pockets show the desired flatness of the thin
windows, and the first motorised moving section, with prototype
detectors inserted, has been tested at the CERN test beam. This was
a first step in the design of the full system, including assembling,
positioning and alignment aspects. A full prototype test is planned
in test beam in Fall 2008.

We want to stress that the moving pipe design development and prototyping 
has been done in direct contact with the LHC cryostat group. In particular, 
the Technical Integration Meetings (TIM), held regularly at CERN and 
chaired by K.~Potter, provided an efficient and crucial framework for 
discussions and information exchanges.
\newpage

\section{RF impact of Hamburg pipe on LHC}
\label{sec:RF}

\subsection{Motivation and introduction}

The electromagnetic interaction between the beam and its
surroundings will be one of the phenomena limiting the ultimate
performance of the LHC,  because  it can lead to single bunch and
multi-bunch beam instabilities, beam emittance growth and beam
losses.  Usually such effects are expressed in terms of \emph{wake
fields} and beam \emph{coupling impedance}. As discussed in the LHC
design report~\cite{lhcreport}, the LHC has an overall impedance
budget that requires careful design of each element of the
accelerator to minimise the total impedance. During the first years
of operation, it is expected that the maximum intensity of the LHC
colliding beams will be limited by collimation efficiency and impedance effects; 
consequently a series of studies designing upgraded configurations of the LHC was
initiated many years ago. The primary focus of these studies is the
collimation system, since it is the dominant component of the
impedance budget.

In general, the electromagnetic effects are enhanced by the use of
low electric conductivity materials, by small distances between the beam and the vacuum chamber and by any transverse cross
section variation of the vacuum chamber. In particular, the transverse
resistive wall impedance~\cite{rw,rw_PAC} increases when the beam
approaches the beam pipe wall, which will regularly occur during
FP420 detector insertion. Both the real and imaginary part of the
transverse impedance have to be controlled, in order to minimise the
effect on beam instability growth rate and betatron tune shift.

Variation of the beam pipe cross-section in the 420 m region not
only is a potential issue for LHC operations through increased
impedance, but it can also affect the FP420 detectors. Trapped modes
arising from the exchange of electromagnetic energy between the beam
and its surroundings can cause heating of the detectors which
increase their cooling requirements, since they must operate at low
temperature. Moreover, the electromagnetic fields can penetrate
through the beam pipe walls and be picked up by the detector
electronics.

We have begun a series of studies to examine the different aspects
of the FP420 impedance. Analytical calculations and numerical
simulations are underway to assess the longitudinal and transverse
impedance values.
Laboratory measurements on an FP420 station prototype have been
performed to validate the simulations and will serve to investigate
the effect of electromagnetic disturbances on the detector
electronics. These studies are also useful to suggest modifications
to the final FP420 design to minimise RF effects.

During LHC operation, the real effect of wake fields on power losses
and beam instability will be assessed by the convolution in the
frequency domain between the beam spectrum and the coupling
impedance. Therefore, the relevant upper limit on the frequency that
must be considered is assessed by the nominal LHC beam bunch length,
$\sigma_z=0.25$~ns (r.m.s.). This permits us to limit our study up
to a frequency of 3\,GHz. The following sections describe the
current status of the RF studies.

\begin{figure}[!t]
\centering{
\includegraphics[width=0.6\columnwidth]{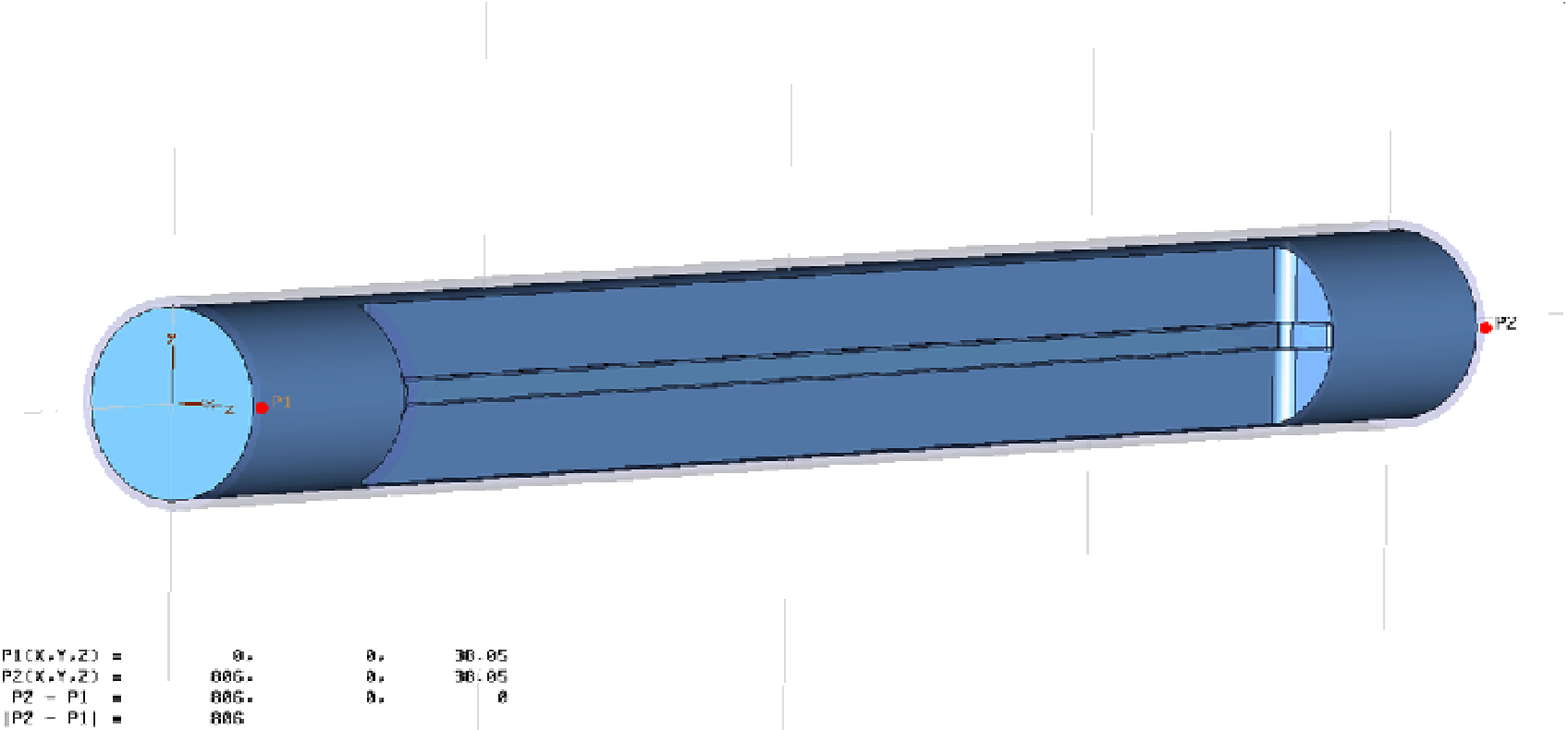}
\includegraphics[width=0.6\columnwidth]{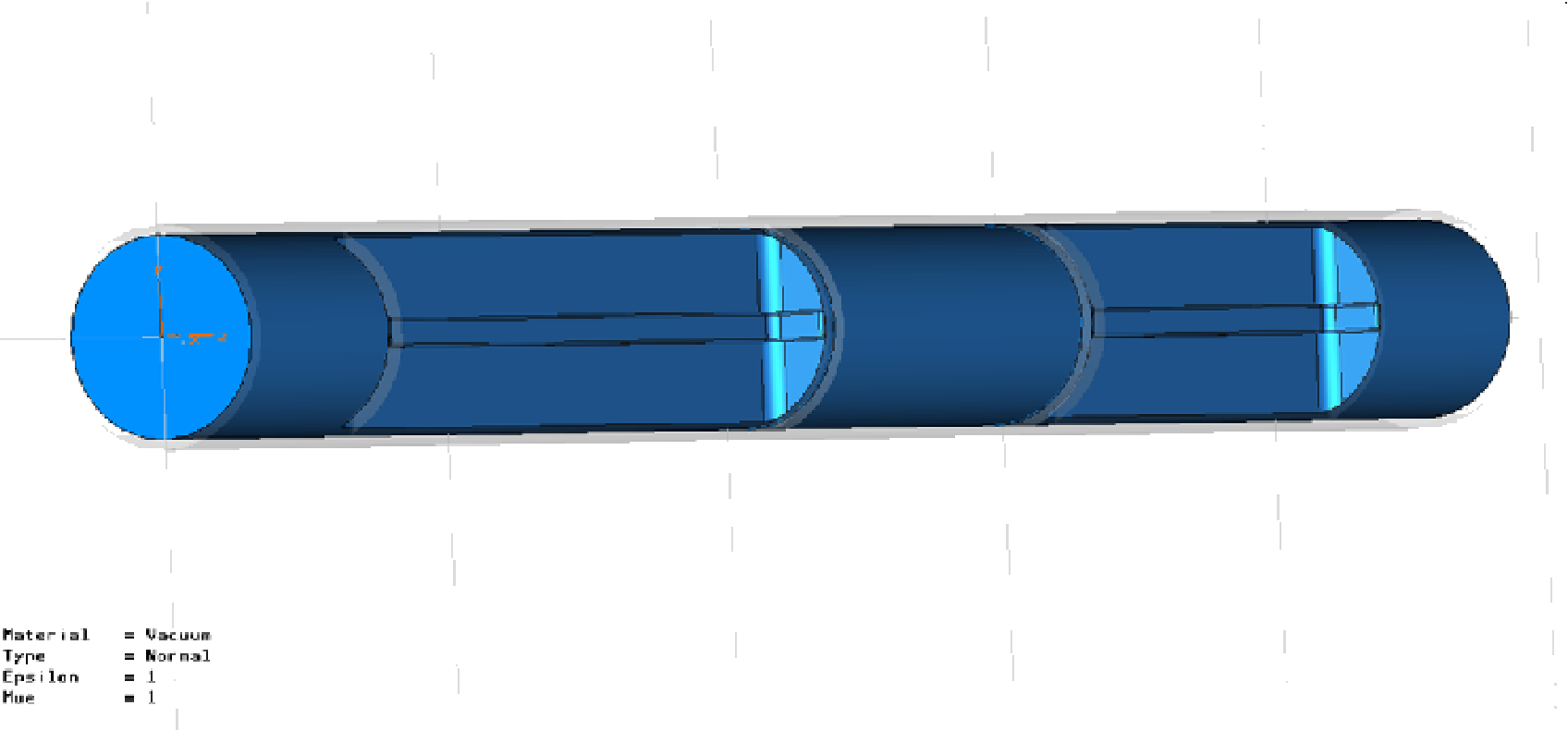}
}\caption{The two benchmark FP420 pocket designs. The upper figure
is a single pocket solution. The lower figure is a double pocket
design, which will allow separate temperature and vacuum conditions
for timing and silicon detectors} \label{fig:pockets}
\end{figure}

\subsection{Longitudinal impedance}

Most of the studies are based on the stretched wire method for
evaluating the longitudinal coupling impedance $Z_L$ through the
measurement of the scattering parameters of the network composed by
an RF source and the device under test (DUT)~\cite{ref:fritz}.
Usually, the RF source is a two ports Vector Network Analyzer that
is used to send an electromagnetic wave through the wire stretched
along the DUT.  The measurement consists in determining the
scattering parameter $S_{21}$, that is defined as the ratio of the output
of the VNA port 2 to the incident wave on port 1. With such a
method, the deviation of the impedance of the DUT from that of a
reference vessel (REF) can be modeled with a loaded transmission
line~\cite{vaccaro}.    Solving the resulting non-linear equation to
first order in impedance enables an explicit relation (as function
of frequency $f$) between the longitudinal impedance $Z_L$ and $S_{21}$
to be obtained. This is referred to as the ``log'' formula:
\begin{equation}
\label{equ:log} \vec{Z_L}(f) = -2 Z_c \;ln\frac{\vec{S}_{21}^{DUT}(f)}{\vec{S}_{21}^{REF}(f)},
\end{equation}
where $Z_c$ is the characteristic line impedance.

The results will be expressed in terms of longitudinal impedance $Z_L/n$:
\begin{equation}
\left (\frac{Z_{L}}{n}\right ) =\frac{Z_{L}(f)}{f/f_0},
\end{equation}
where $n$ = $f/f_0$ and $f_0$=11\,kHz is the beam revolution frequency in the LHC. 
This quantity can be compared to predictions and measurements of other LHC elements, 
as reported in~\cite{lhcreport}.  All the calculations and measurements refer to an FP420 
pocket made of stainless steel.
%
%
%

\subsubsection{Simulations}

Figure~\ref{fig:pockets} shows two beam pipe designs considered for
the RF simulations, a single long pocket and an alternative design
with two shorter pockets. The results of three different numerical
simulation packages are presented. Ansoft
HFSS$^\copyright$~\cite{hfss} was used to simulate the stretched
wire setup and calculate the longitudinal impedance according to
Eq.~(\ref{equ:log}), while CST Particle Studio$^\copyright$
(PST)\footnote{In performing these simulations, a beta-version of
PST has been used. }~\cite{cst} and GDFIDL~\cite{gdfidl} provide a
direct calculation of the electromagnetic field induced by a passing
bunch on the surrounding structure.

\begin{figure}[!t]
\centering{
\includegraphics[width=0.49\columnwidth]{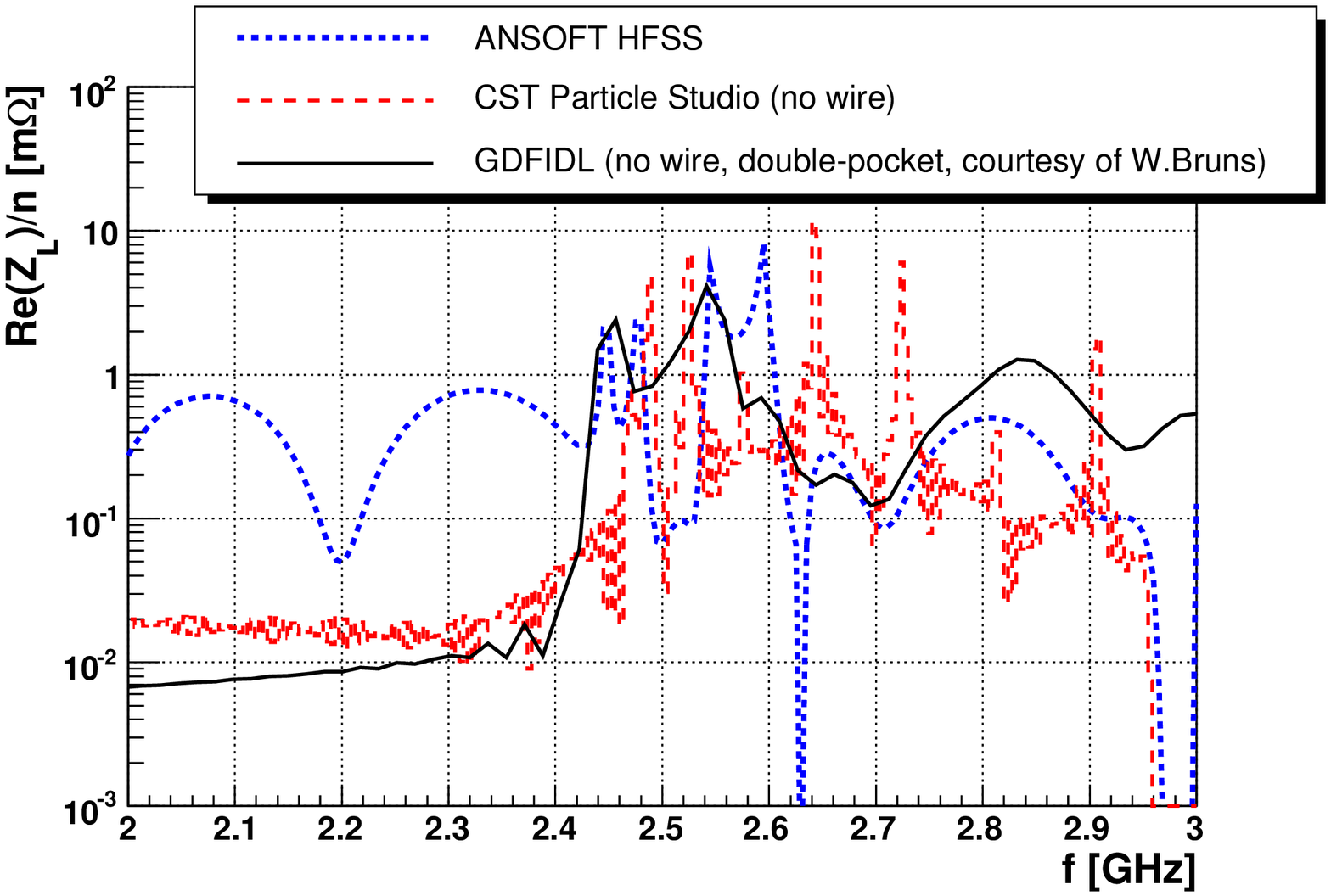}
\includegraphics[width=0.49\columnwidth]{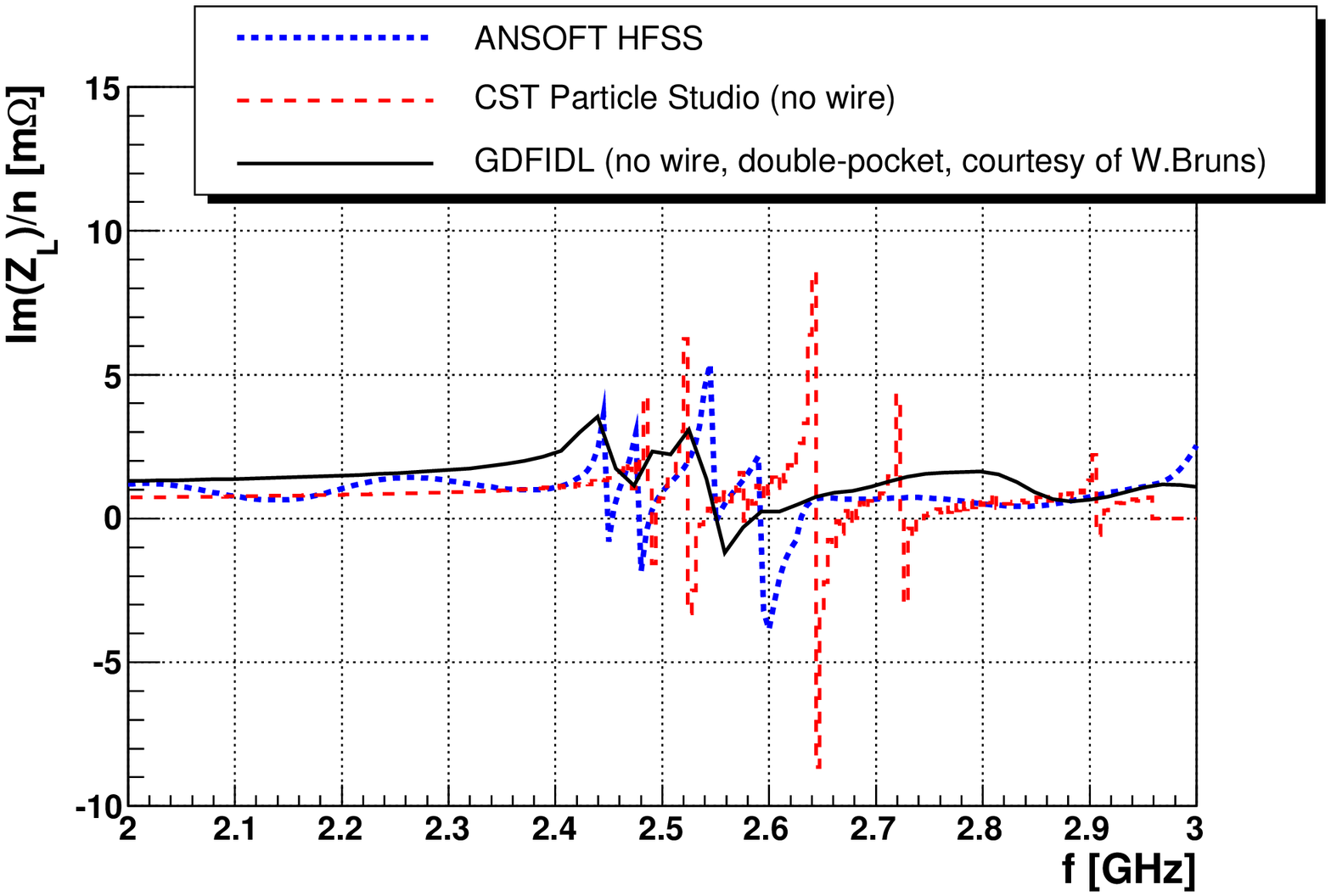}
} \caption{Real (left) and imaginary part (right) of the
longitudinal impedance for three different simulations of the single
pocket prototype, effectively assuming the pocket wall is 3 mm away from
the beam.}
\label{fig:comp_sim}
\end{figure}

Figure~\ref{fig:comp_sim} shows, for all three simulations,  the
calculations of the real and imaginary parts of the longitudinal
impedance.  For the single pocket geometry,  four narrow band
impedance peaks are observed  between 2.4 and 2.75\,GHz for the HFSS
and PST simulations. The frequency difference is attributed to the
presence of the wire in the HFSS simulation. Two of the four
resonances are significantly reduced for the double pocket
simulation with GDFIDL. The wide band resonances that we observe (in
both HFSS simulations and experiment) for  $f <2.4$\,GHz are
understood to be an artifact of the wire and do not represent a real
beam impedance effect.
Simulations of the double pocket geometry with HFSS are in progress
and preliminary results confirm the laboratory experiments that are
presented in the next section.

\vskip 0.1 cm
%
%

\subsubsection{Laboratory measurements}
\label{sec:meas} 

The laboratory setup at the Cockcroft Institute
comprises a sophisticated mechanical system equipped with micrometer
screws, in order to stretch, move and monitor the relative position
of the wire. A set of measurements in the time domain was used to
determine the absolute position of the wire with respect to the
pocket wall with an accuracy of about $100\,\mathrm{\mu}$m~\cite{PAC}.

\begin{figure}[!t]
\centering{
\includegraphics[width=0.49\columnwidth]{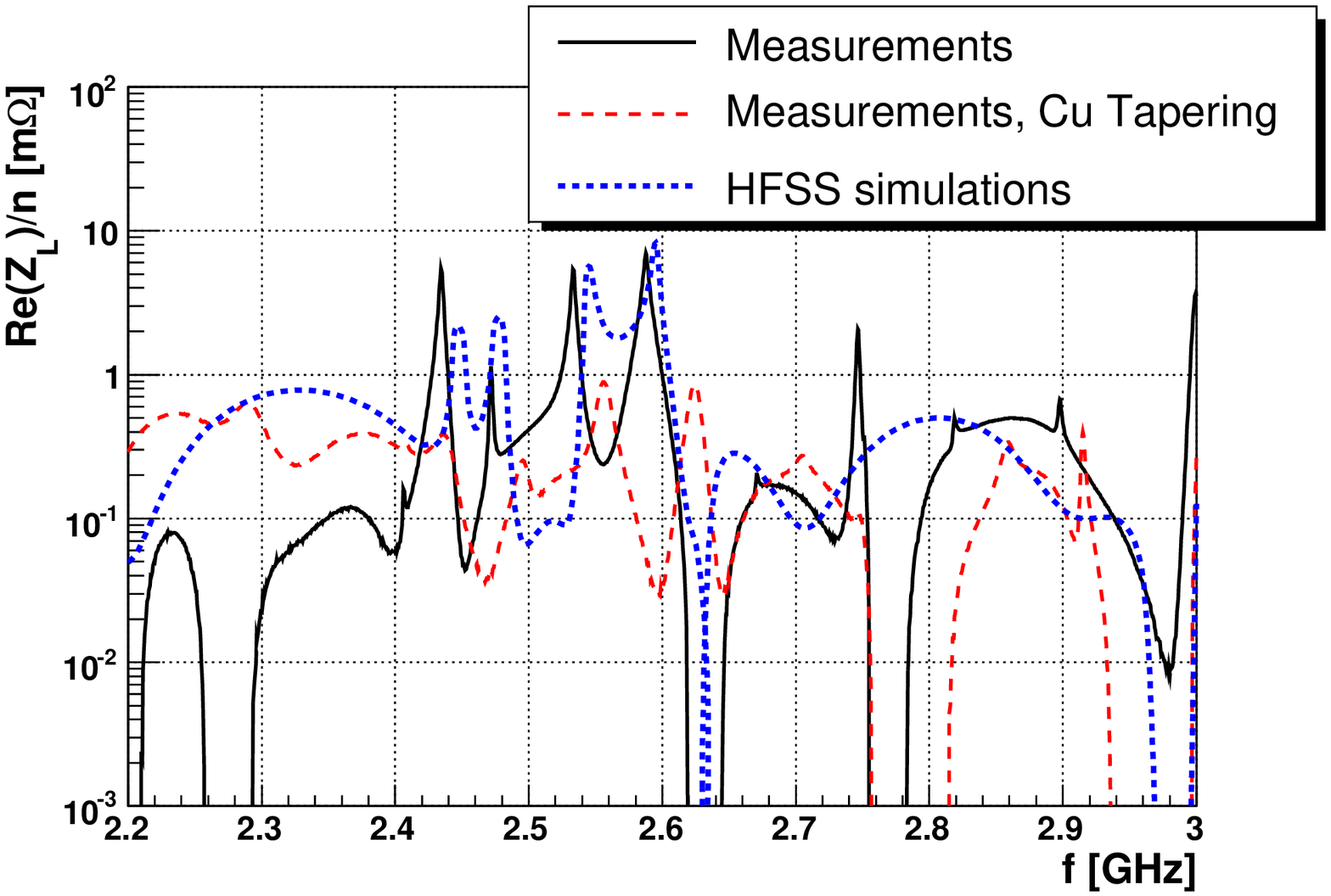}
\includegraphics[width=0.49\columnwidth]{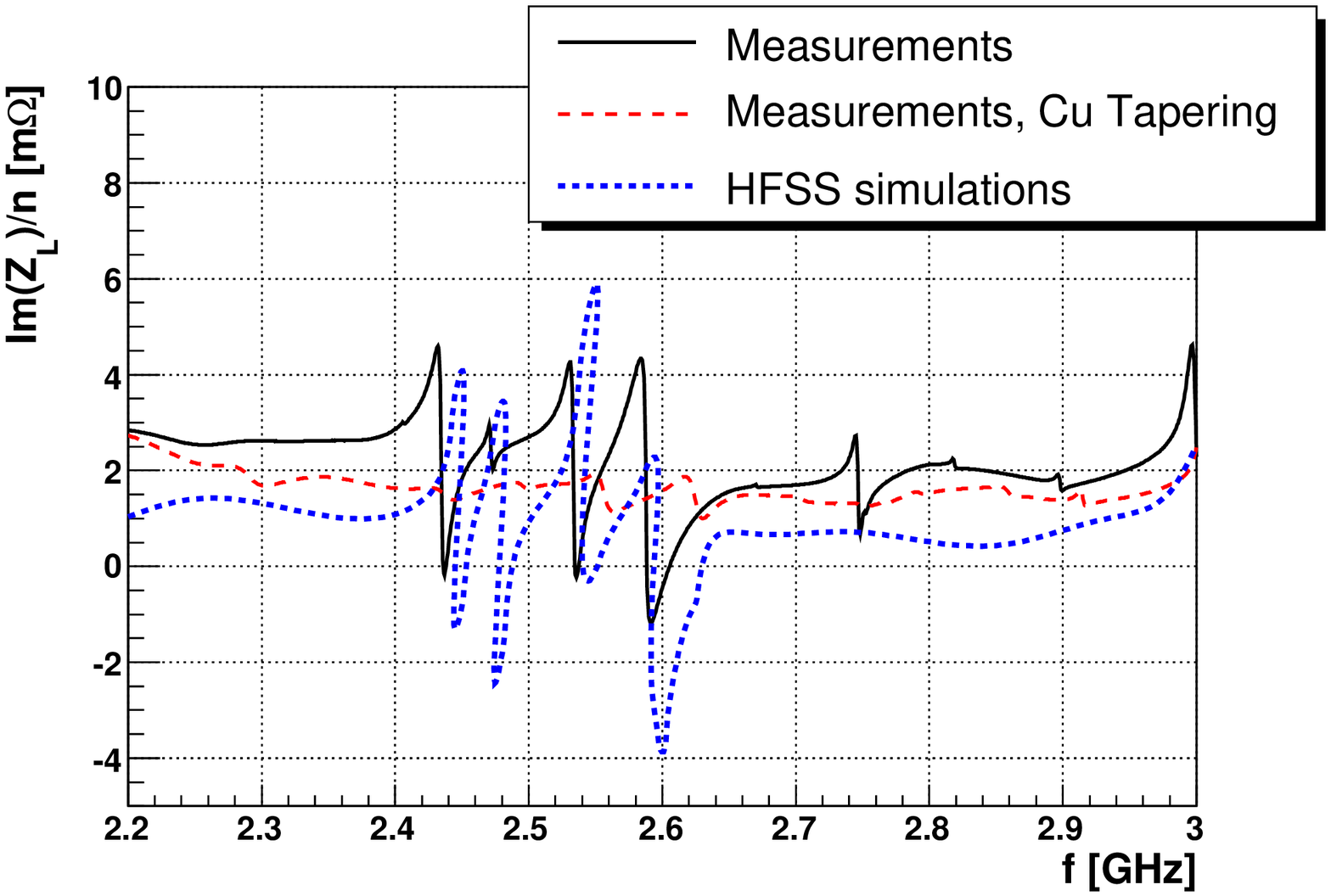}
} \caption{Measurements of real (left) and imaginary part (right) of
the longitudinal impedance with the beam 3\,mm  from the pocket wall
for the single pocket prototype; also shown are the  HFSS
simulations.} \label{fig:tapering}
\end{figure}

\paragraph*{Single pocket results}
The real and imaginary part of the FP420  longitudinal impedance
calculated from the measured $S_{21}$ parameter (black solid line)
and simulated by HFSS (blue dashed line) are shown in
Fig.~\ref{fig:tapering}.
\begin{figure}[!t]
\centering{
\includegraphics[width=0.49\columnwidth]{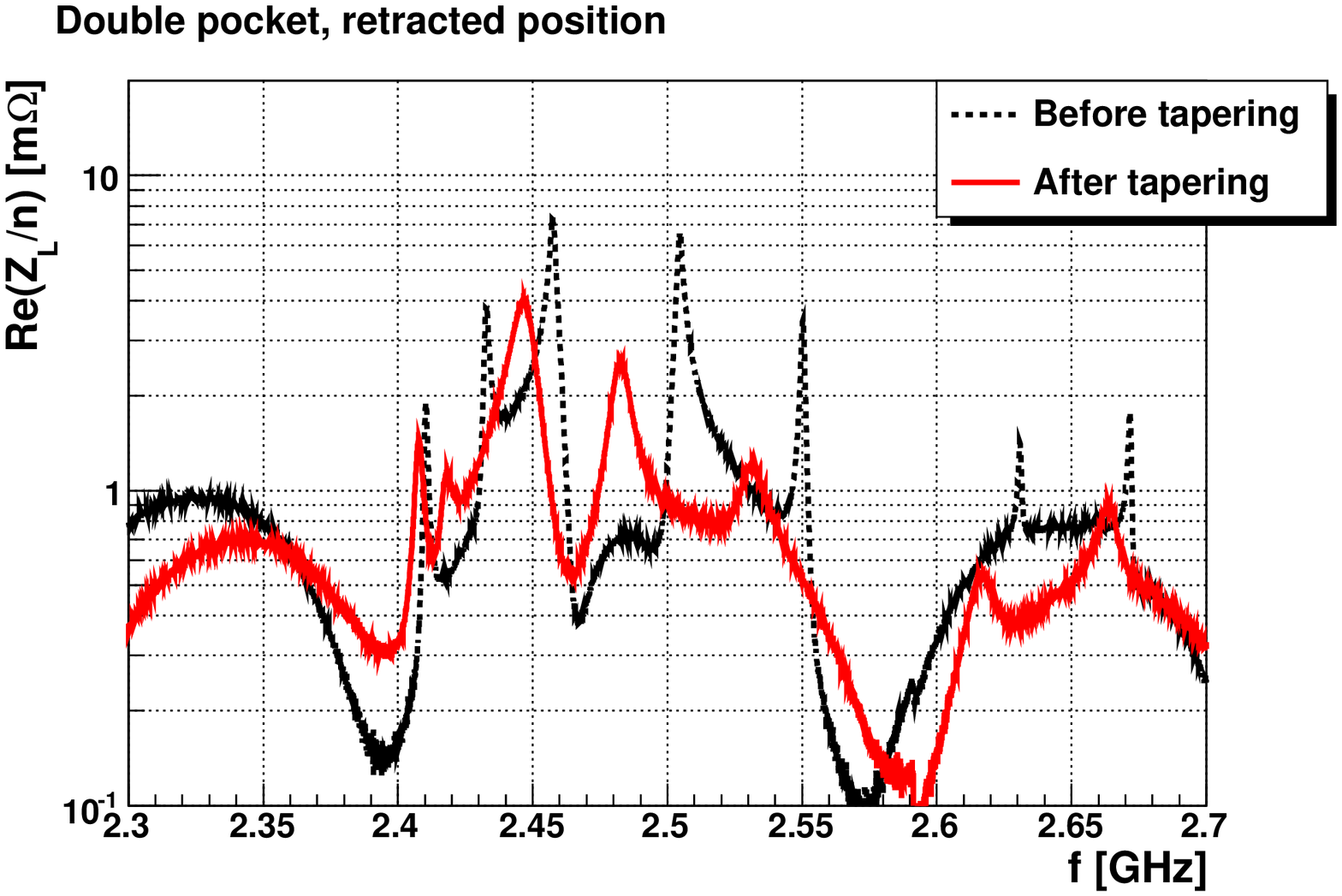}
\includegraphics[width=0.49\columnwidth]{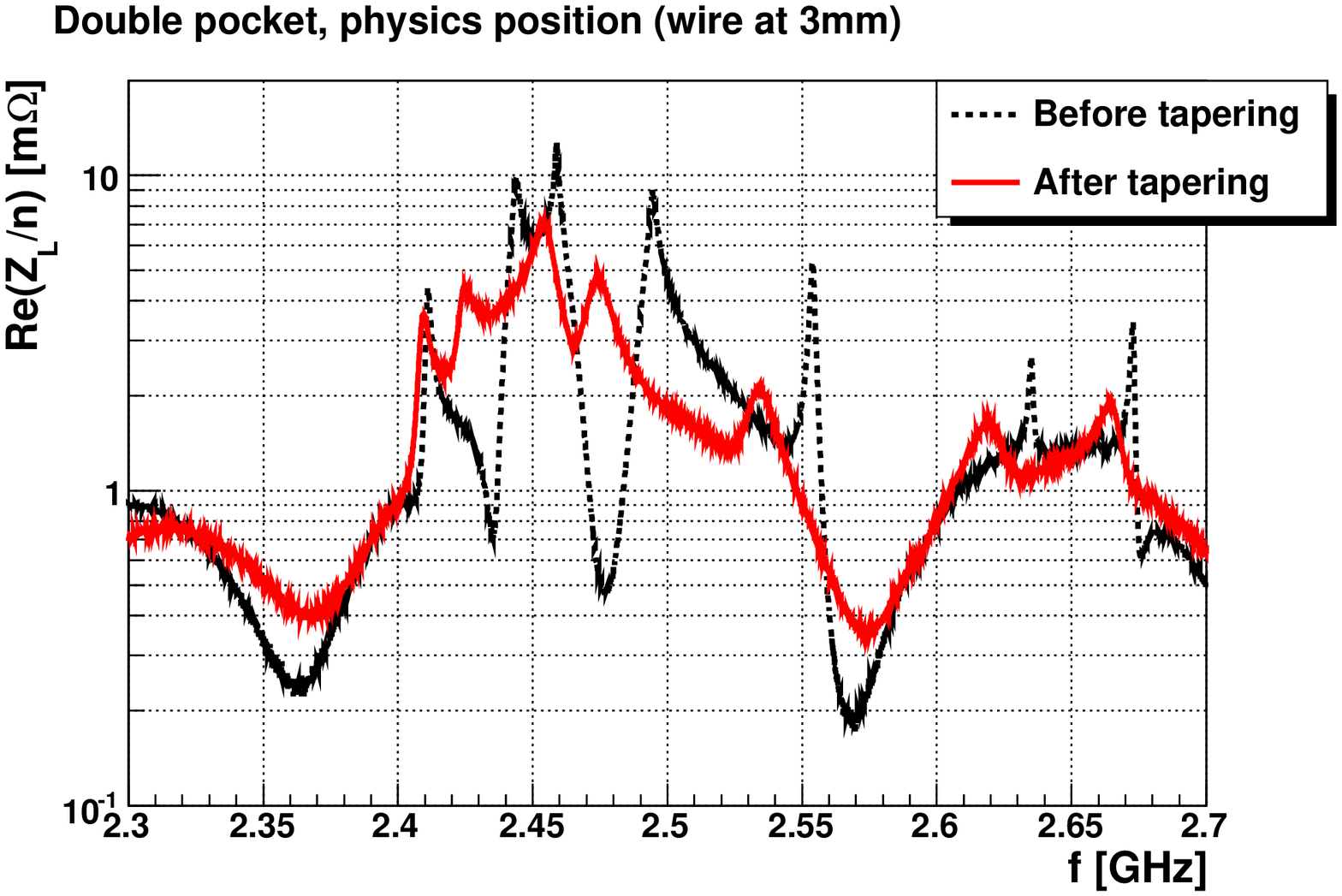}
\includegraphics[width=0.49\columnwidth]{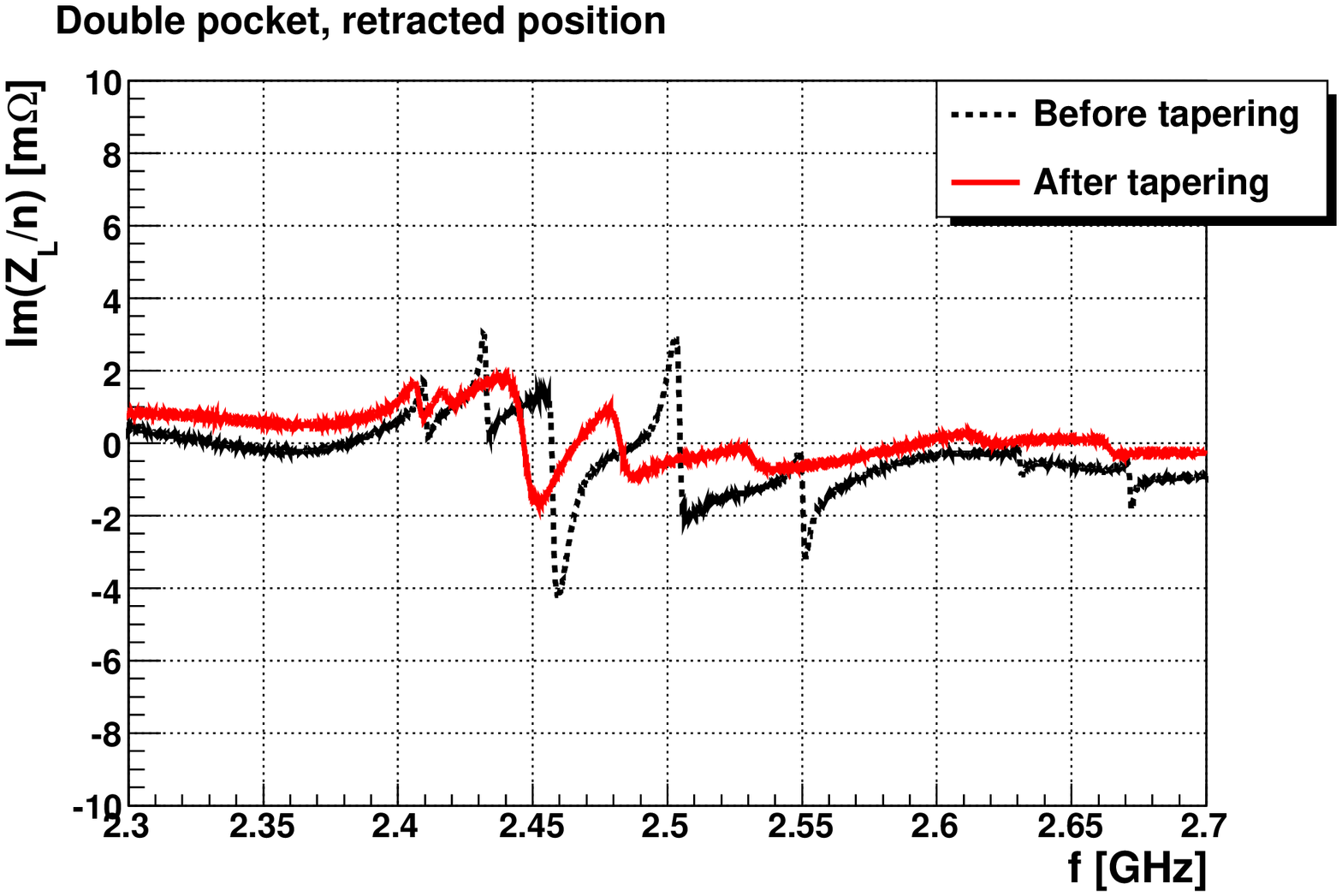}
\includegraphics[width=0.49\columnwidth]{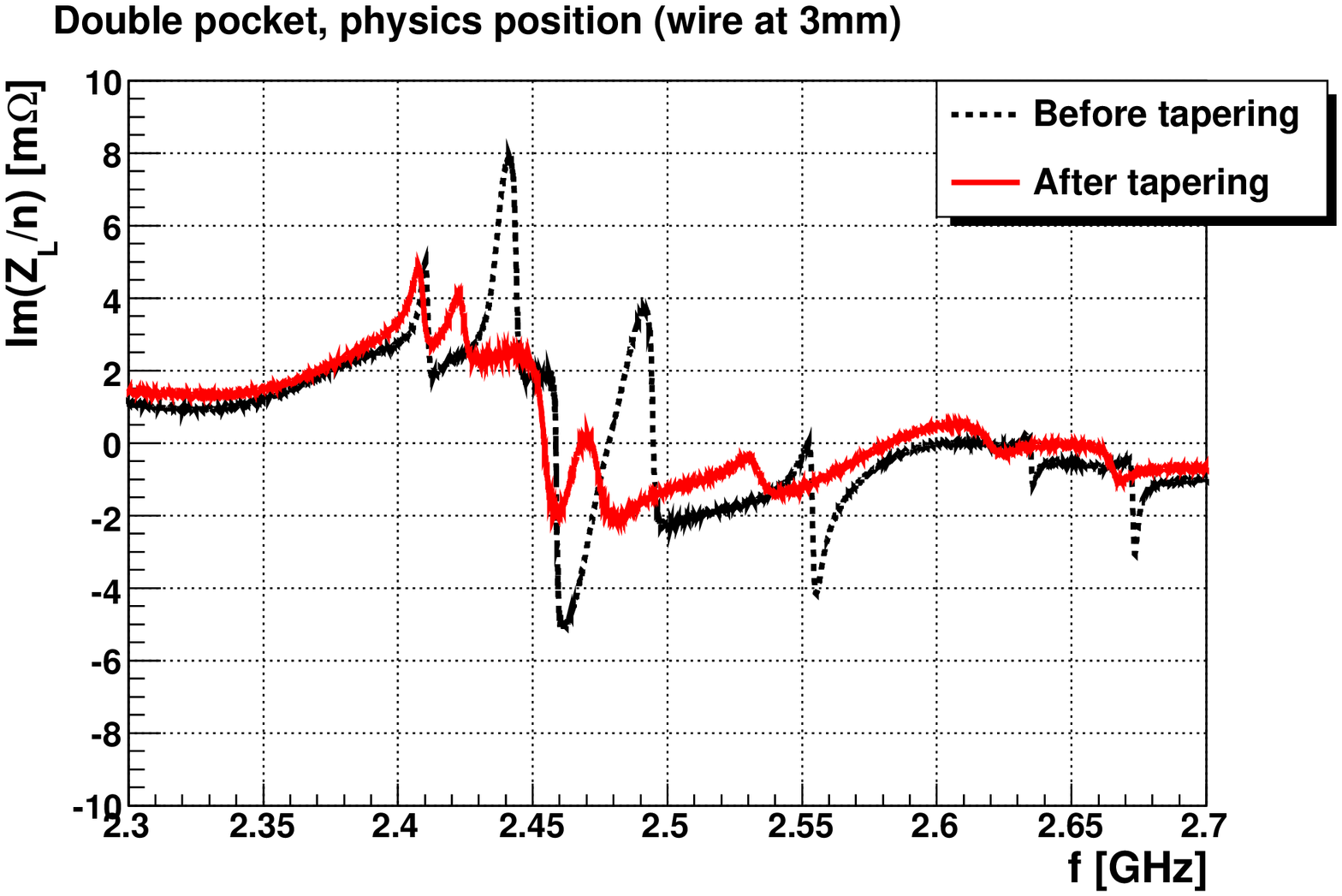}
} \caption{Real and imaginary part of the  longitudinal impedance
for the FP420 physics (inserted) and parking positions, as measured
for a double pocket prototype.} \label{fig:dpocket}
\end{figure}
This plot refers to a wire distance from the pocket wall of
$3\,\mathrm{mm}$, simulating a detector 3 mm ($10\sigma_x$) from the
beam. Measurements and simulations have been carried out for several
intermediate distances from this position of closest approach to a
retracted position  ($x > 50\sigma_x$ from the beam).

The agreement between measurements and simulations, in terms of
resonance peaks of the impedance is satisfactory as they lie within
$3\,m\Omega$ in amplitude and a few MHz in frequency. At least one
additional resonance appears in the measurements (e.g. at 2.75\,GHz)
and can be explained by a residual mismatch between the RF source
and the DUT, not considered in the simulations.

The FP420 pocket was remeasured after applying a thin copper-plated
tape at the indentation regions. The tape was placed outside the
beam orbit region (i.e. above and below the 500 $\mu$m thin window),
in order to provide a tapered transition of the beam pipe cross
section variation. The result is shown by the red solid lines in
Fig.~\ref{fig:tapering}. After tapering, the longitudinal impedance
is reduced by an order of magnitude and thus the impedance is
limited to no more than $1\,m\Omega$ over the measured frequency
band.

\paragraph*{Double pocket results}
A first set of laboratory measurements with an FP420 double pocket
prototype has been completed. The results are shown in
Fig.~\ref{fig:dpocket} in terms of the real and imaginary part of the
longitudinal impedance and for parking and physics positions. The
black dashed lines refer to the original beam pipe, whereas the
solid red lines assess the measured impedance value after applying a
copper tape at the accessible pockets indentations (i.e. for each
pocket, the indentation at the beam pipe end). The two indentations
in between the two pockets are not easily accessible after the beam
pipe fabrication and could not be tapered or connected with an RF
contact during these measurements. As for the single pocket
prototype, there are no impedance peaks for frequencies below
2\,GHz. After tapering, the real part of the longitudinal $Z_L/n$  
impedance remains above $5\,m\Omega$ at about 2.46\,GHz, when the
detectors are 3\,mm from the beam. In all the rest of the frequency
band of interest, both the real and imaginary parts of the
longitudinal impedance are below $5\,m\Omega$.

\subsection{Transverse impedance and beam instability}
\label{sec:stability} 

\begin{figure}[!t]
\centering{
 \subfigure[Real part of $Z_x$ (comparison with theory)]
{\includegraphics[width=0.49\columnwidth]{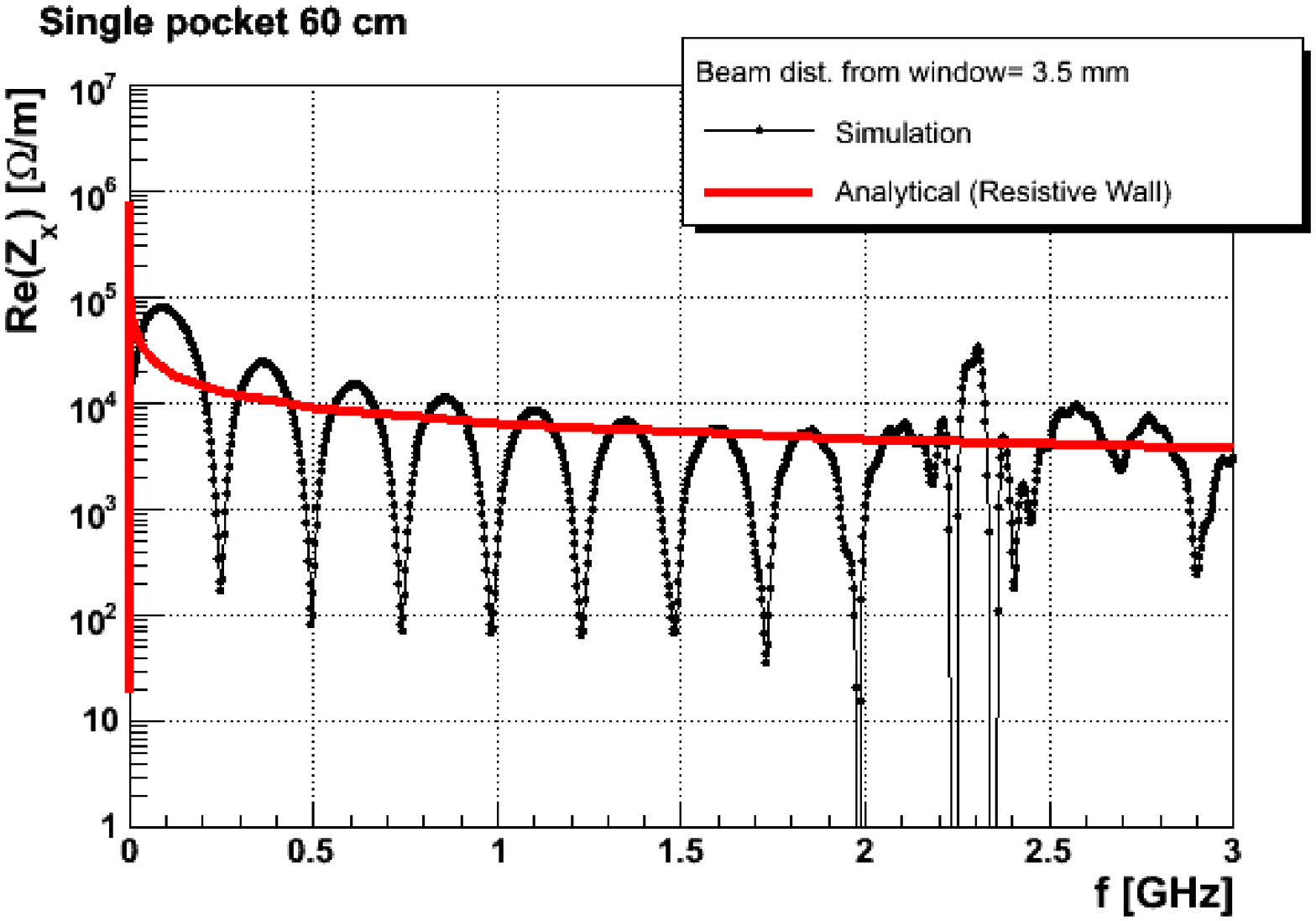}}
\subfigure[Imaginary part of $Z_x$ (comparison with theory).]
{\includegraphics[width=0.49\columnwidth]{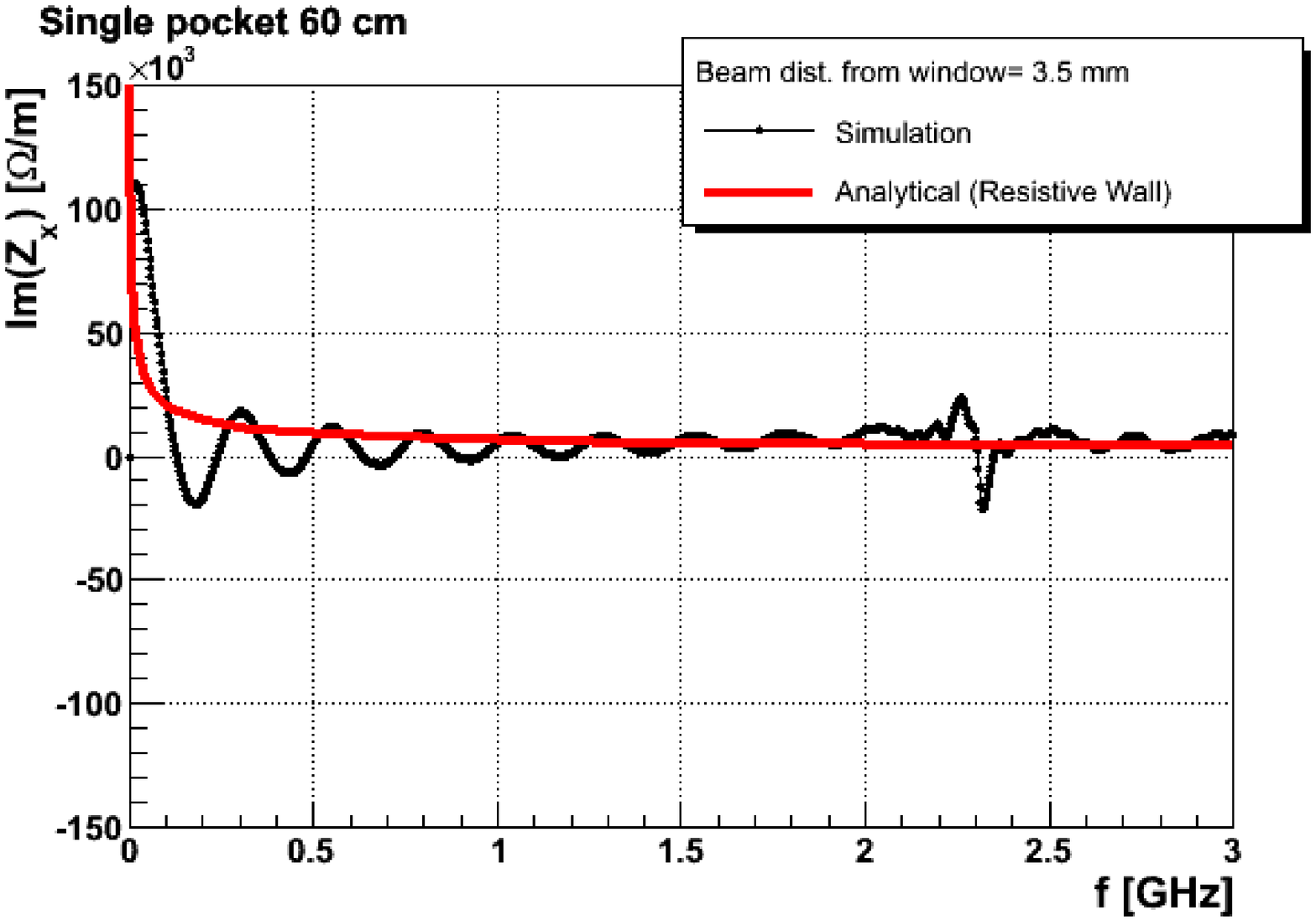}}
} \caption{Single pocket transverse impedance vs. resistive wall
theory.} \label{fig:compare_rw}
\end{figure}

\begin{figure}[!t]
\centering{
\includegraphics[width=0.8\columnwidth]{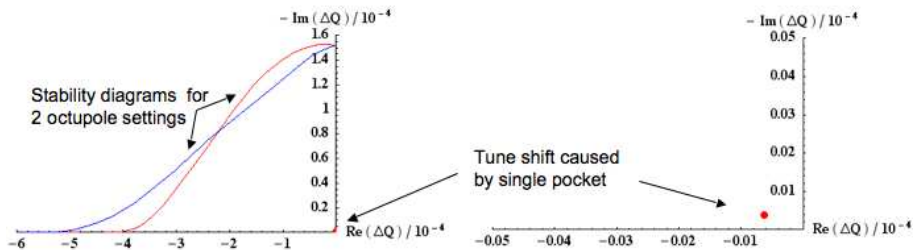}
} \caption{Tune shift induced by FP420 single pocket at 3\,mm, due
to resistive wall transverse impedance (calculations based on
B.~Zotter and E.~Metral's models). The effect is negligible when
compared to the stability diagram that assesses how the LHC
octupoles tuning can damp the instability.} \label{fig:tuneshift}
\end{figure}

The transverse impedance can be inferred by
the variation of the longitudinal impedance for different wire
(beam) transverse positions. Figure~\ref{fig:compare_rw} compares
the simulated transverse impedance with an analytical prediction
accounting for resistive wall effects only. The results are in good
agreement since the oscillation at low frequencies given by the
numerical simulations is attributed to the presence of the wire. The
resonances between 2 and 3\,GHz account for the geometric impact of
the FP420 station on the beam pipe cross section not considered by
the analytical formulas. Therefore, for frequencies below  2\,GHz,
the transverse impedance introduced by the FP420 insertion is
dominated by the resistive wall effect. The impedance values
calculated analytically can be used to predict the impact on the
beam horizontal tune shift. The effect is very small, it results in
$|\Delta Q_x| < 1\cdot 10^{-6}$, well within the stability region
defined by the available Landau damping octupoles at
LHC~\cite{elias}, as shown in Fig.~\ref{fig:tuneshift}.
%

\subsection{Coupling with detectors}

The simulation of detector signal disturbances due to
electromagnetic coupling between the beam and the surroundings is
very difficult, due to the small amount of power that could be
picked-up at the detector electronics level. A laboratory
measurement using high power spike generators and a normalization to
the real beam current is under consideration.


\subsection{RF summary}

The FP420 single pocket geometry has been
characterised in terms of coupling impedance. Numerical simulations,
analytical calculations and laboratory measurements showed
consistent results, all indicating that this design will have a
small impact on the total LHC impedance budget.

Tapering of the beam pipe indentations is recommended because it
does reduce the impedance significantly, as measured both with the
single pocket and double pocket designs. Since an effective tapering
can be done outside the beam orbit region, this design modification
can be implemented at no cost in terms of the forward proton signal
to background ratio. With a double pocket station design, the beam
pipe section between the two pockets can also be electrically
connected outside the beam orbit region, in order to provide a good
RF contact and minimise the effect of beam pipe cross section
variation. This could not be tested in the laboratory, due to the
difficulty of accessing the region after beam pipe fabrication.
Simulations and laboratory measurements of a new prototype, modified
according to the RF studies completed so far, will be continued.

The resultant effective longitudinal impedance follows from the
convolution of the results presented here with the LHC beam
spectrum. The beam harmonics at 2\,GHz are expected to be below
$10^{-2}$ of the main harmonic at 40\,MHz and well below $10^{-3}$
at 2.5\,GHz. This provides a further indication of the expected
minimal impact of a FP420 station on the LHC impedance. One of the
consequences is that, according to the available analytical models,
the horizontal tune shift induced by a FP420 station is expected to
be almost imperceptible when compared to the tune stability region
defined by the available LHC octupoles magnets.

In addition, the worst case considered in these studies, refers to
the positioning of a FP420 station at 3\,mm from the circulating
beam, whereas recent acceptance (Sec.~\ref{sec:optics}) and
background (Sec.~\ref{sec:backgrounds}) calculations indicate that
5\,mm  is a more likely distance of closest approach. This implies
that  the results are conservative in terms of disturbances to the
beam. Further studies are ongoing in order to determine the
characteristic loss factor, which will provide an estimate of the
power dissipated due to electromagnetic coupling.

\newpage

\section{Silicon Tracking Detectors}
\label{sec:silicon}

\subsection{Introduction}
\label{sec:silicon_intro}

In order to detect protons from the production of central systems of masses 
$\sim 100$~\GeVcc, the detector edge has to approach the beam axis to a
minimum distance of 5~mm (see Figure~\ref{distance_beam}). This 
represents a challenge for the radiation hardness and radio-frequency pick-up in the detector and the
nearby front-end electronics, as described in Sections~\ref{sec:backgrounds} and \ref{sec:RF}. 
The detector system has to be robust, and for satisfactory control of systematic uncertainties its
position has to be aligned and maintained to a positional accuracy of 10 ${\mu}$m 
in order to achieve the required track angular precision of $1 \mu$rad (see Section~\ref{sec:resolutions}). 

With a typical LHC beam size at 420~m of ${\sigma}_{beam}$
${\approx}$ 300 ${\mu}$m, the window surface of the Hamburg pipe
can theoretically safely approach the beam to  15 ${\times}$ ${\sigma}_{beam}{\approx}$ 4.5~mm. 
As discussed in Section~\ref{sec:backgrounds} however, this distance will ultimately be determined 
by the LHC collimator settings, since for beam 2 in particular the halo can extend to $\sim 5$~mm
with the nominal collimator positions. 
The window itself adds another 0.2~mm to the minimum possible distance of the detectors
from the beam. To maximise the acceptance for low momentum-loss protons, the detectors
should therefore be active as close to their physical edge as possible. In general, 
planar silicon detectors have a
wide (0.25~mm -- 1~mm) insensitive border region around
the sensitive area that is occupied by a sequence of guard rings. This
ring structure controls the potential distribution between the
detectors sensitive area and the cut edge to remove
leakage current. Planar silicon detectors designed for a heavy
radiation environment or generally for operation at high bias voltages,
contain multi-ring structures with typically about $\sim$20 rings. 

The key requirements for the FP420 tracking system are
\begin{itemize}
\item To track efficiently as close as possible to the sensor's physical edge. 
\item To have extreme radiation hardness. A design figure
equivalent to or better than the vertex systems used for ATLAS or CMS
will be required, i.e. better than 10$^{15}$ 1-MeV equivalent neutrons per cm$^{2}$.
\item To operate at the highest LHC luminosity
and be robust and reliable.
\item Individual detectors should have a spatial precision of $\sim$10
microns. The tracking system angular precision should be 1~$\mu$rad. 
These requirements are discussed in detail in Section~\ref{sec:optics}. 
\item At 420~m the tracking detector needs to cover an area of 25~mm
x 5~mm.
\end{itemize}

3D silicon technology has been chosen as the baseline detector technology
best equipped to meet the above requirements, although the tracking system 
has been designed such that any silicon technology  compatible with the ATLAS 
pixel readout can be used. The 3D silicon sensor R\&D is described in Section~\ref{sec:silicon_RD}.
Section~\ref{sec:silicon_support} discusses the mechanical design of the tracking detector, 
Sec.~\ref{sec:hv-lv} discusses solutions for the required high voltage and low voltage, and 
Sec.~\ref{sec:silicon_readout} discusses the infrastructure and readout.
The thermal performance of the system is described in Section~\ref{sec:silicon_thermal}. The
performance of the proposed tracking system is described in Section~\ref{sec:silicon_perform}.

\subsection{3D silicon detector development}
\label{sec:silicon_RD}

\begin{figure}
\centerline{
\epsfig{file=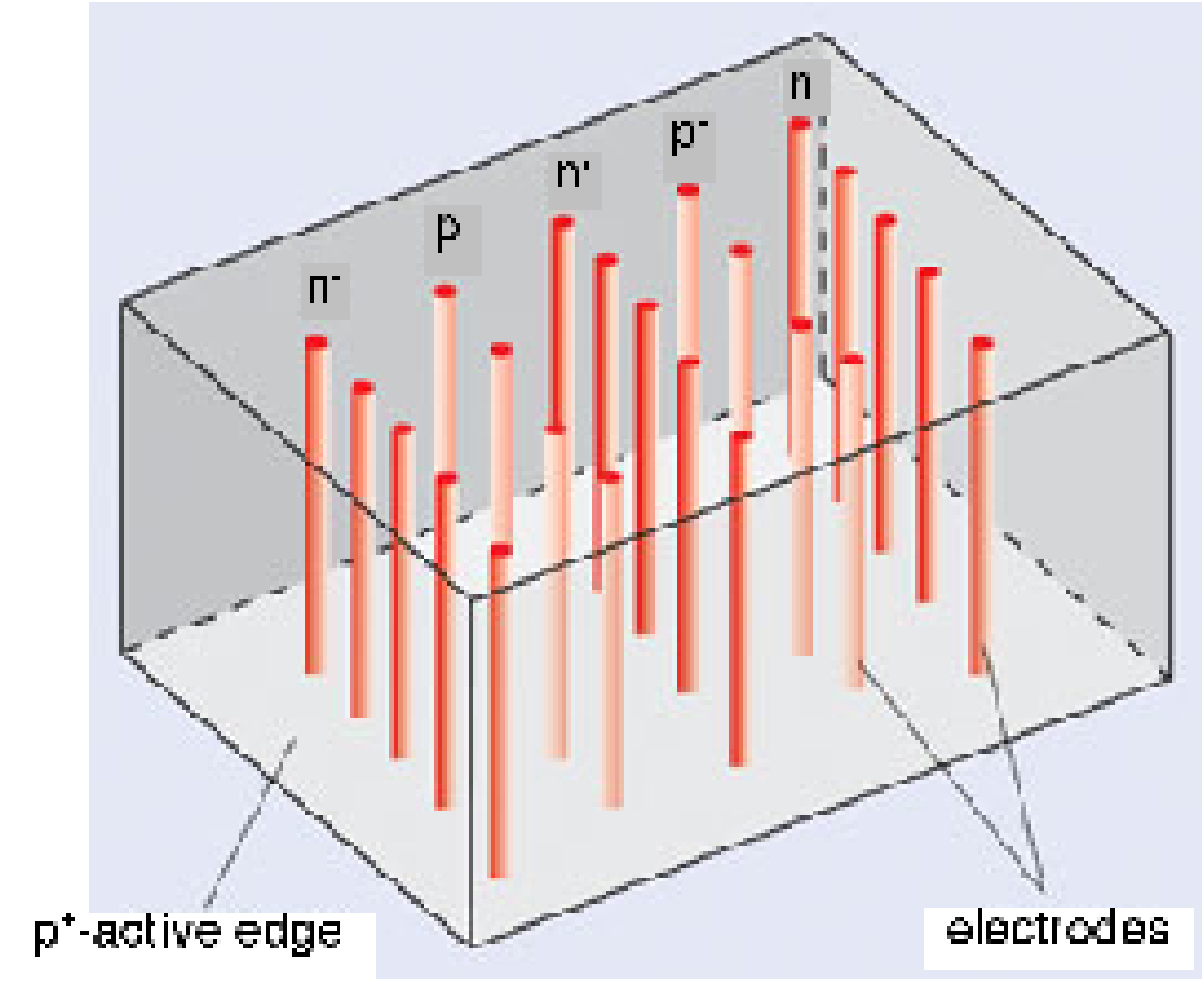,width=7cm}
\epsfig{file=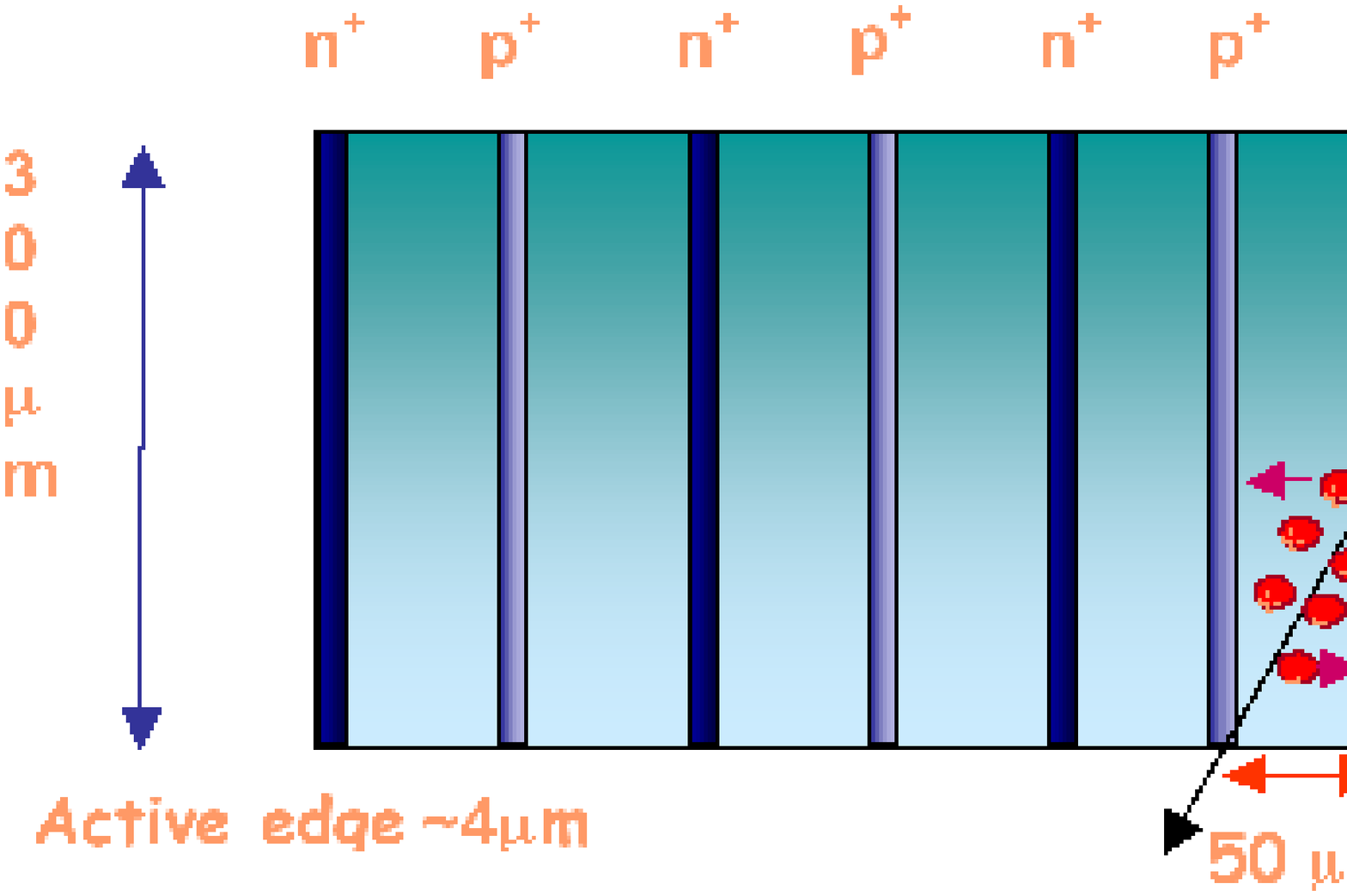,width=7cm}}
\caption{Isometric and lateral view sketches of a 3D detector
where the p+ and n+ electrodes are processed inside the silicon bulk.
The edges are trench electrodes (active edges) and surround the sides
of the 3D device making the active volume sensitive to within a few
microns of the physical edge.}
\label{fig:silicon1}
\end{figure}

\begin{figure}
\centerline{
\epsfig{file=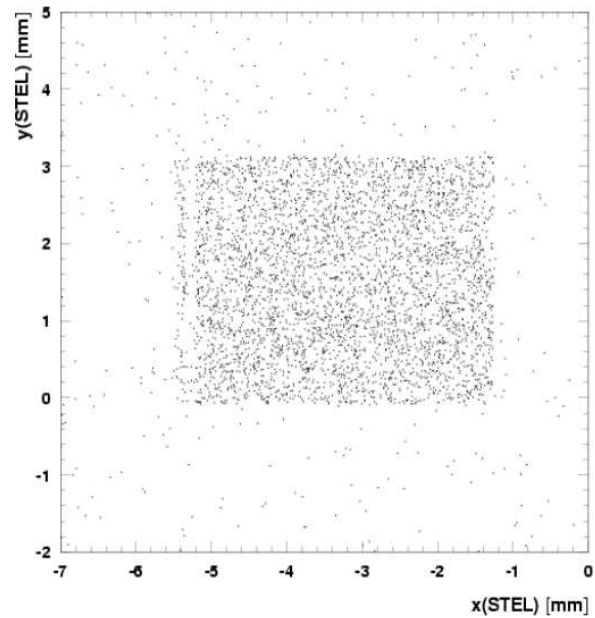,width=7.7cm}}
\caption{Two-dimensional efficiency map of a fully operational 3D detector. 
A point is plotted with respect to the position ($x,y$) predicted by a telescope, 
with a precision of 4 $\mu$m as a valid track and a hit was recorded by a 3D detector. 
The inefficient band near the lower $x$-edge was caused by the detector's bonding pads. 
The upper and lower y edges were used for active edge measurements and 
showed a device sensitivity up to 4 microns of the physical edge.}
\label{fig:3D_telescope}
\end{figure}

\begin{figure}
\centering
\epsfig{file=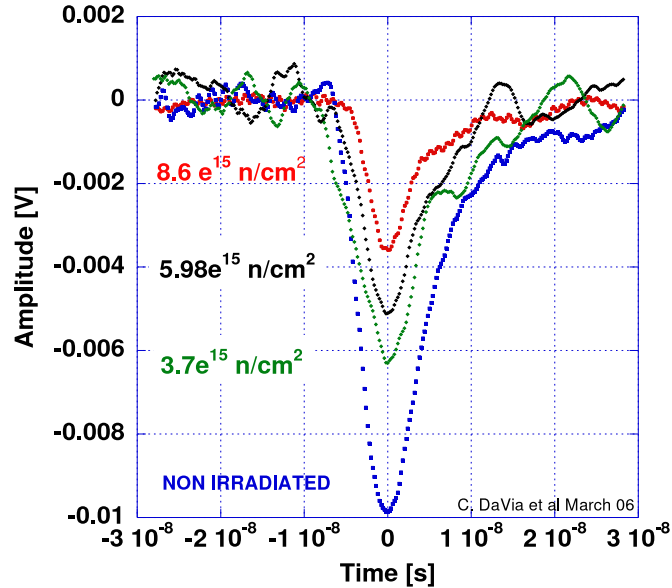,width=10cm}
\caption{Response of a 3 electrodes/pixel (3E) 3D device to a 1060 nm laser pulse after $3.7 \times 10^{15}$, 
$5.98 \times 10^{15}$ and $8.6 \times 10^{15}$ neutrons/cm$^2$.}
\label{fig:silicon3}
\end{figure}

\begin{figure}
\centering
\epsfig{file=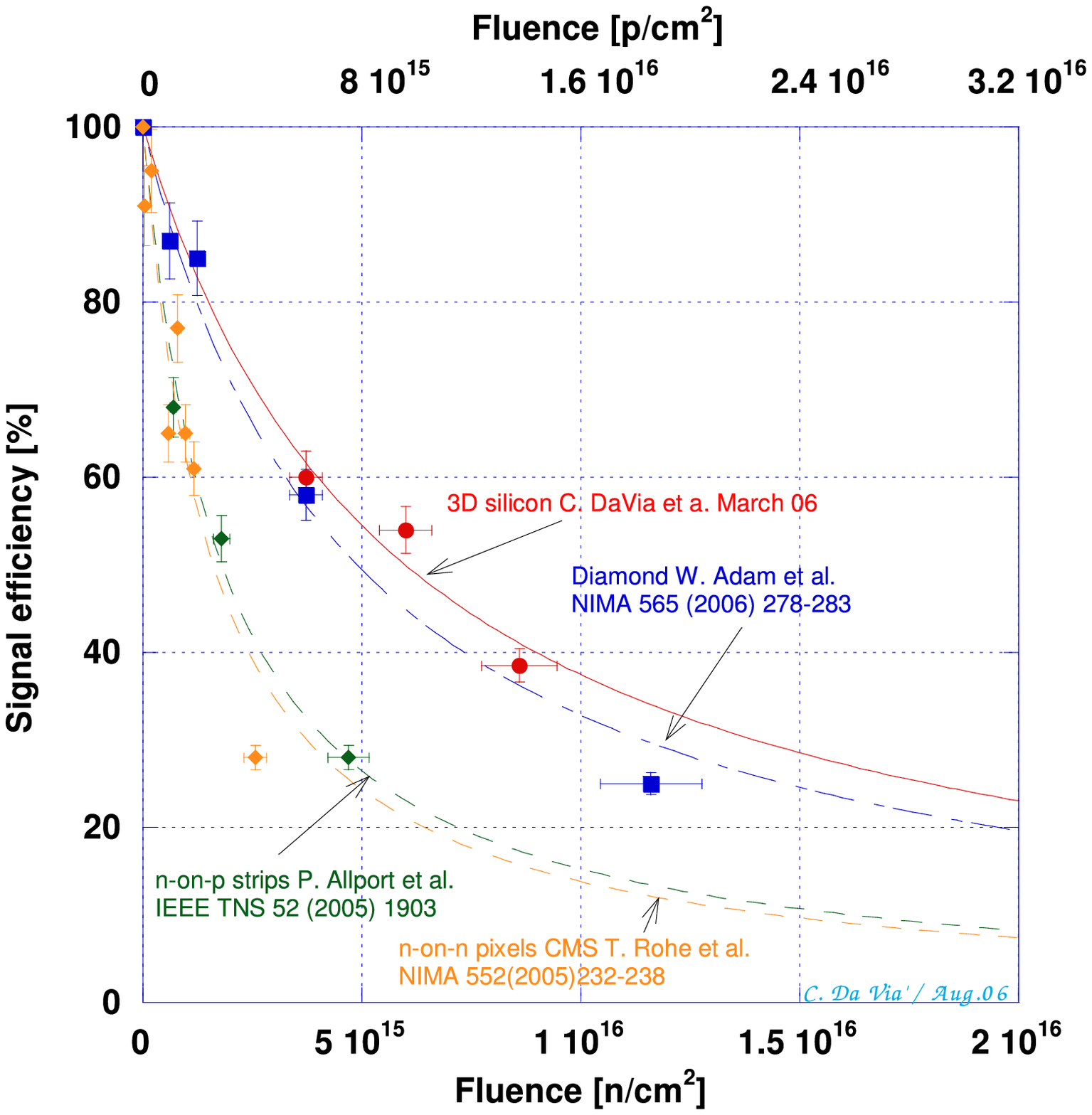,width=10cm}
\caption{Signal efficiency of 3D detector versus fluence of 1 MeV equivalent neutrons/cm$^2$. 
Data for n on p silicon strips and n-side readout pixel detectors are shown for comparison. 
Diamond detector results are also shown. Note that diamond gives a factor three less signal for a minimum ionising particle.}
\label{fig:silicon4}
\end{figure}

3D detectors are a new generation of semiconductor devices~\cite{ref:totemTDR,silref1,silref2,silref3,silref4,silref5,silref6,silref7,silref8,silref9,silref10,silref11,silref12,silref14}. 
Using micro-machining techniques, electrodes penetrate the entire
thickness of the detector perpendicular to the surface. This results in
smaller collection distances, very fast signals, and substantially
improved radiation tolerance. Figure~\ref{fig:silicon1} sketches the main features
of this novel detector design. In addition, similar micro-machining
techniques allow one to produce ``active edges'' where the amount of
dead silicon at the edge of the detector is greatly reduced. 
\begin{figure}
\centering
\epsfig{file=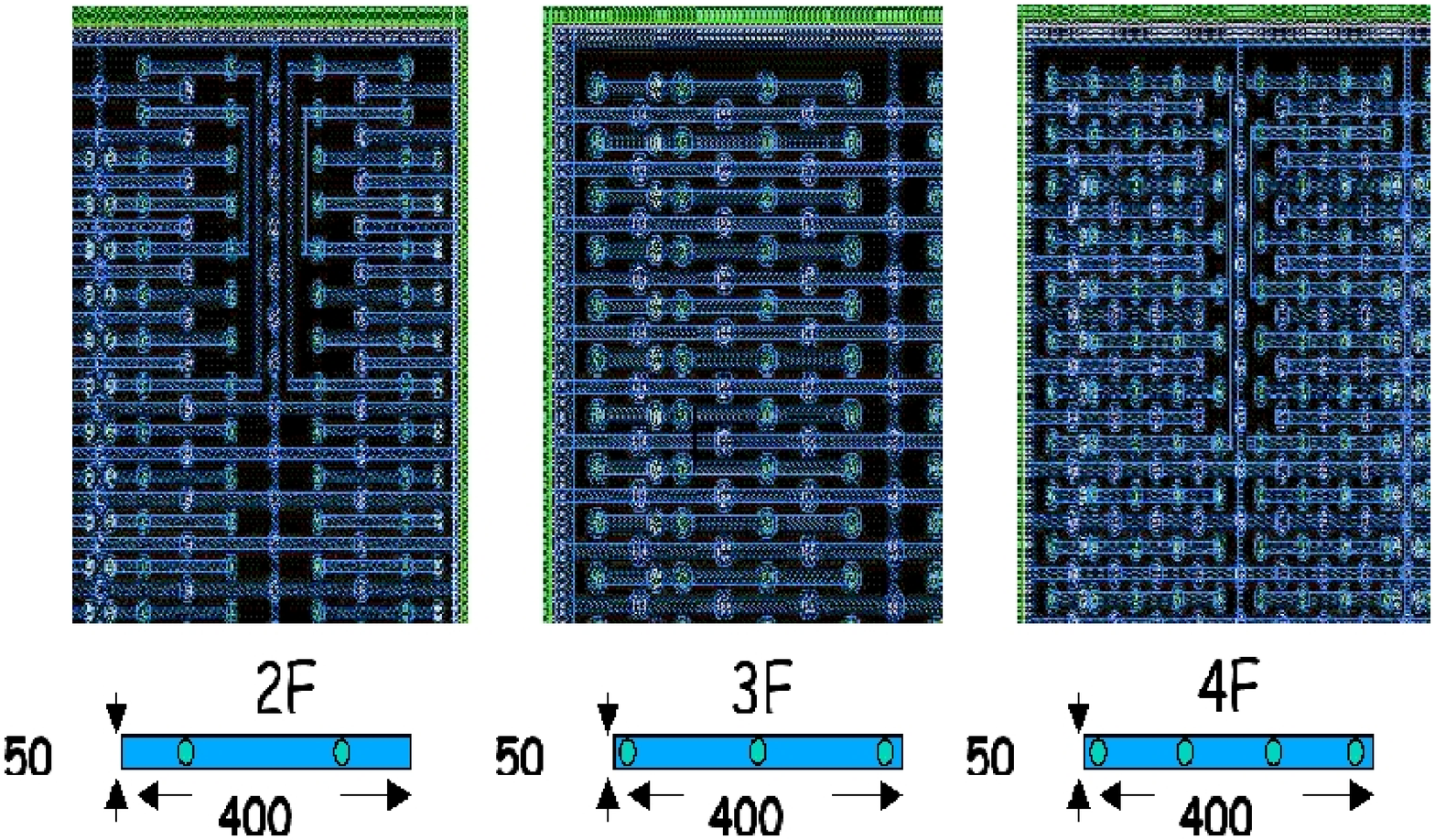,width=12cm}
\caption{Different 3D cell structures designed to be compatible with the ATLAS Pixel detector readout chip. 
The pixel size is 50~$\mu$m by 400~$\mu$m. The devices have either 2, 3 or 4 electrodes per pixel and 
are named 2E, 3E and 4E respectively. The electrodes cover 4\%, 6\% and 8\% of the total area for 2E, 
3E and 4E devices respectively.}
\label{fig:silicon5}
\end{figure}

\begin{figure}
\centerline{
\epsfig{file=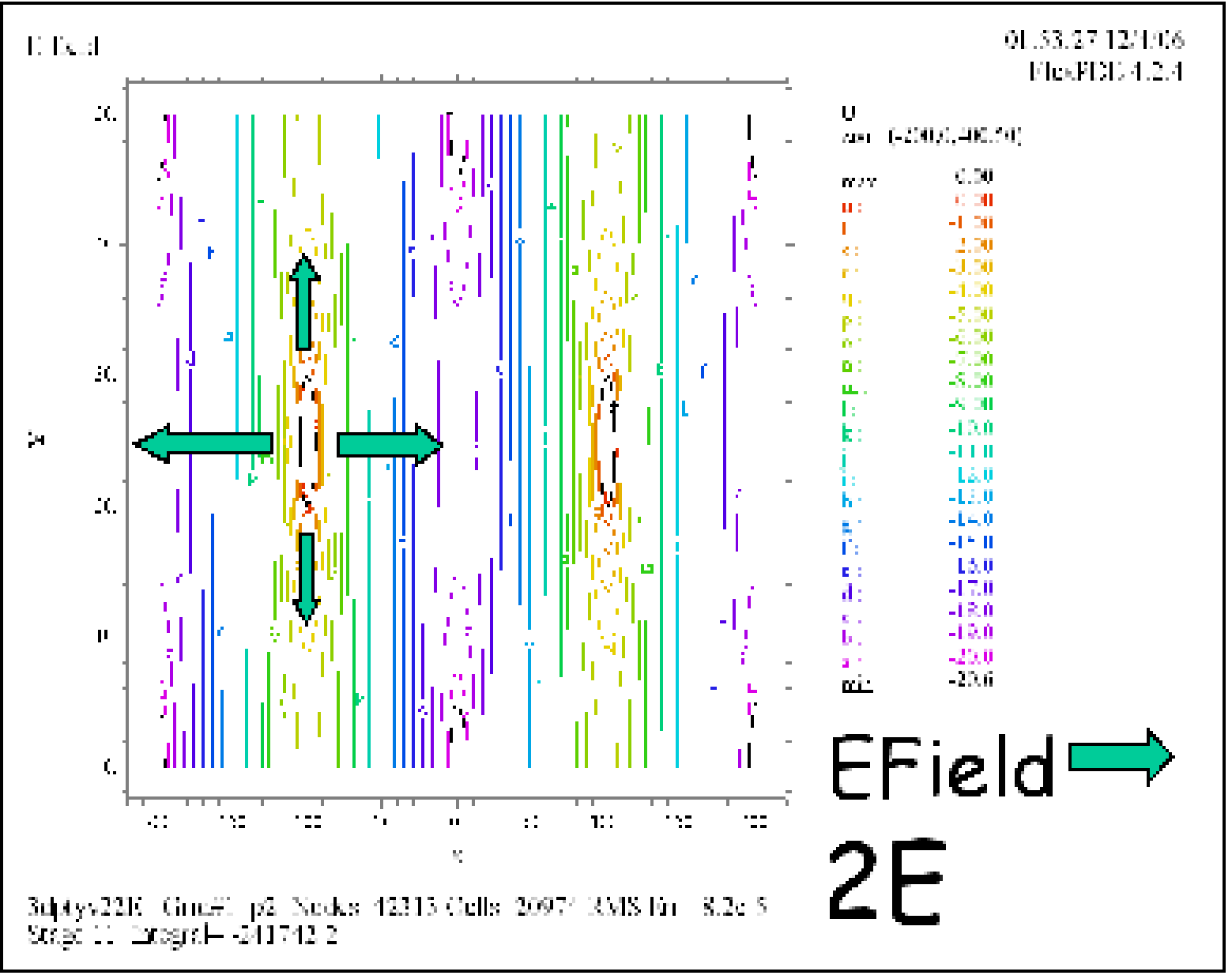,width=7cm}
\epsfig{file=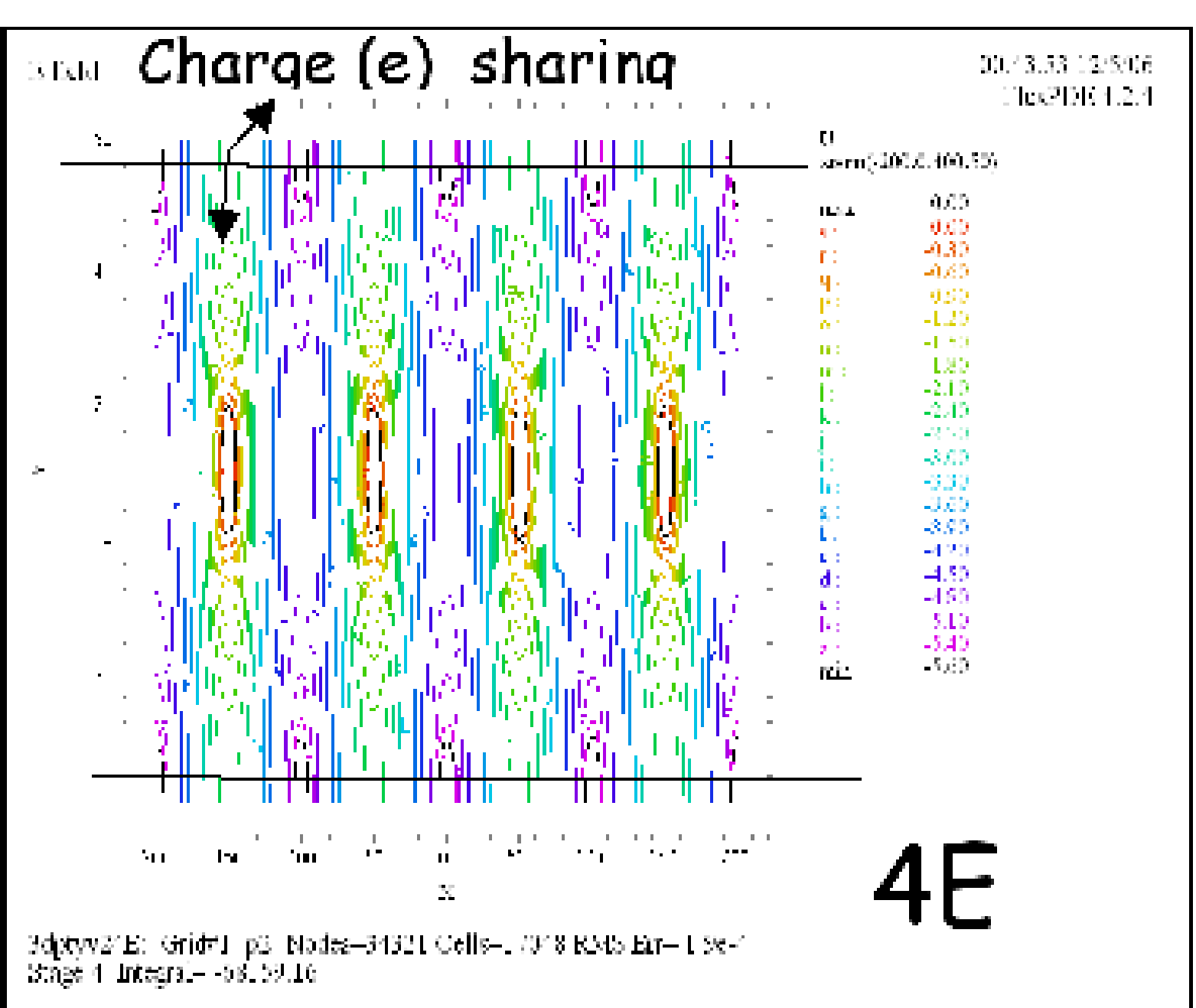,width=7cm}}
\caption{Equipotentials in a 2E and 4E 3D detector. Pixel size is 50 $\mu$m (Y) by 400 $\mu$m (X). 
The bias is 20~V and 5~V for the 2E and 4E devices respectively. The direction of the electric field is indicated. 
The n-electrodes are at the centre. A maximum field of 1~V/$\mu$m occurs at 24~V and 14~V in a 2E and 4E device 
respectively.}
\label{fig:silicon6}
\end{figure}

\begin{figure}
\centering
\epsfig{file=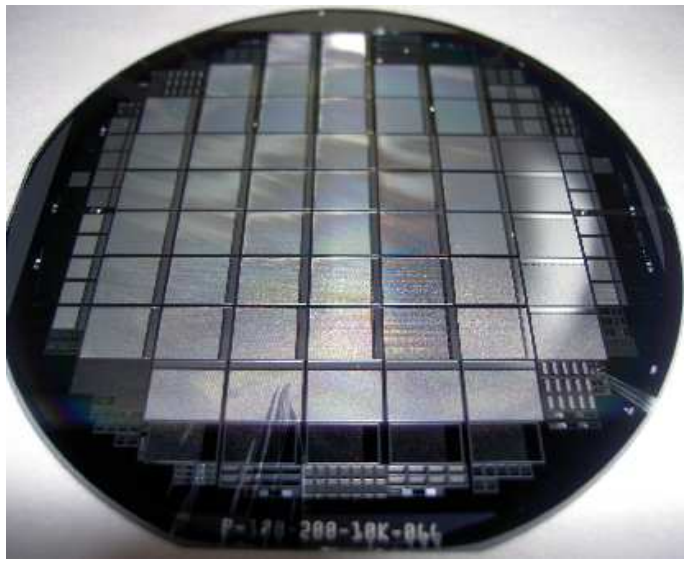,width=9cm}
\caption{Four-inch wafer processed for the FP420 project. This has 32 3E, 
6 4E and 6 2E ATLAS pixel readout compatible devices and several test structures. 
The 250 micron substrate is 12 k$\Omega$~cm p-type.}
\label{fig:silicon7}
\end{figure}

\begin{figure}
\centerline{
\epsfig{file=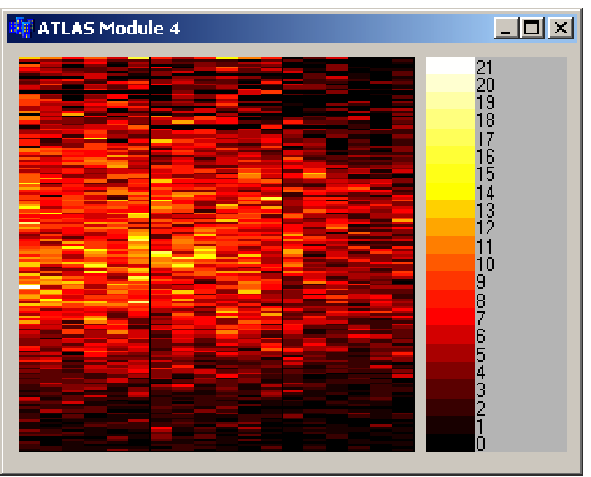,width=7cm}
\epsfig{file=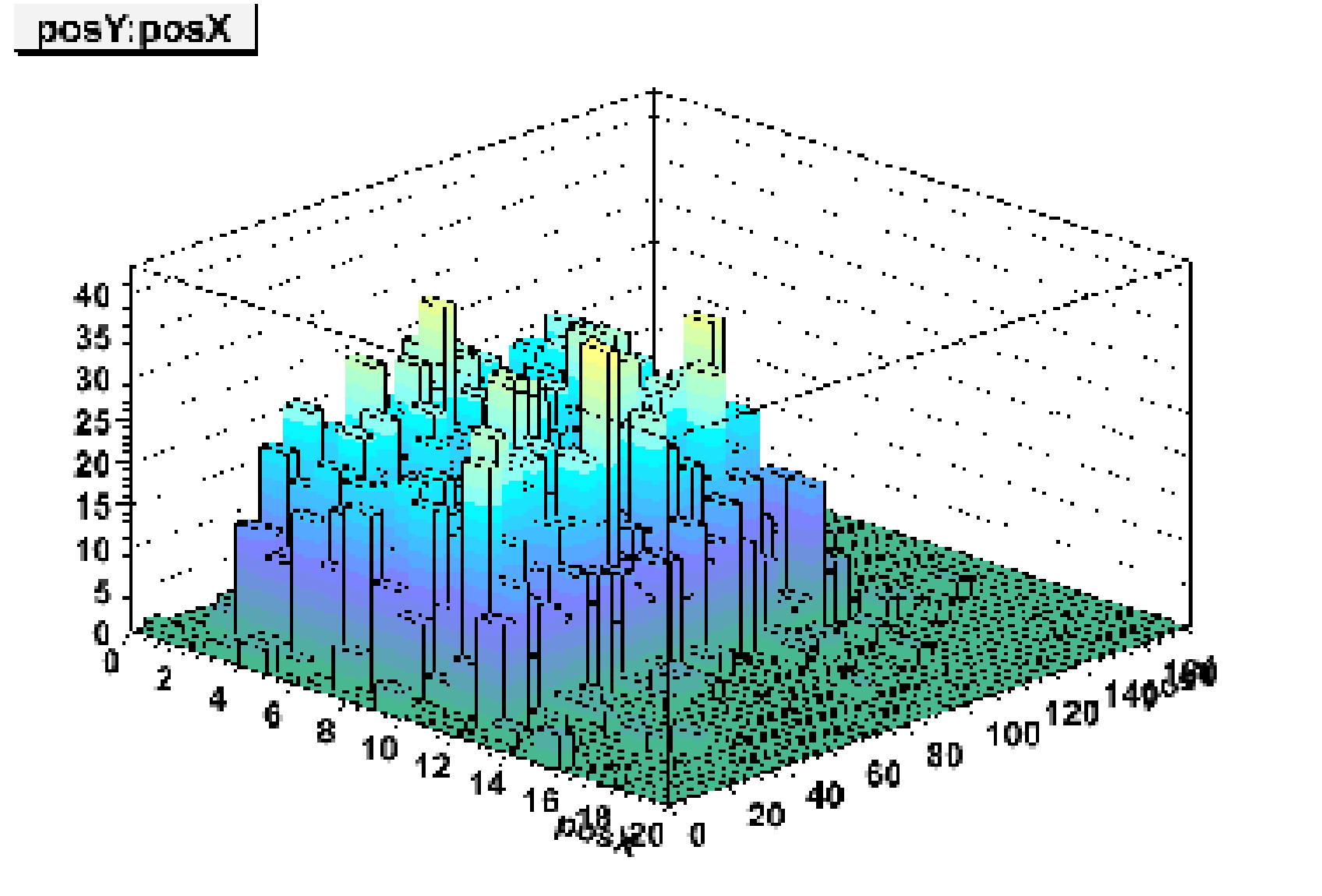,width=8cm}}
\caption{Hit-map for a $12 \times 12$~mm$^2$ (left) and legoplot for a 
$3 \times 3$~mm$^2$ (right) scintillator trigger.
Device 3D-2E-A operated at 30~V.}
\label{fig:silicon8}
\end{figure}

\begin{figure}
\centering
\epsfig{file=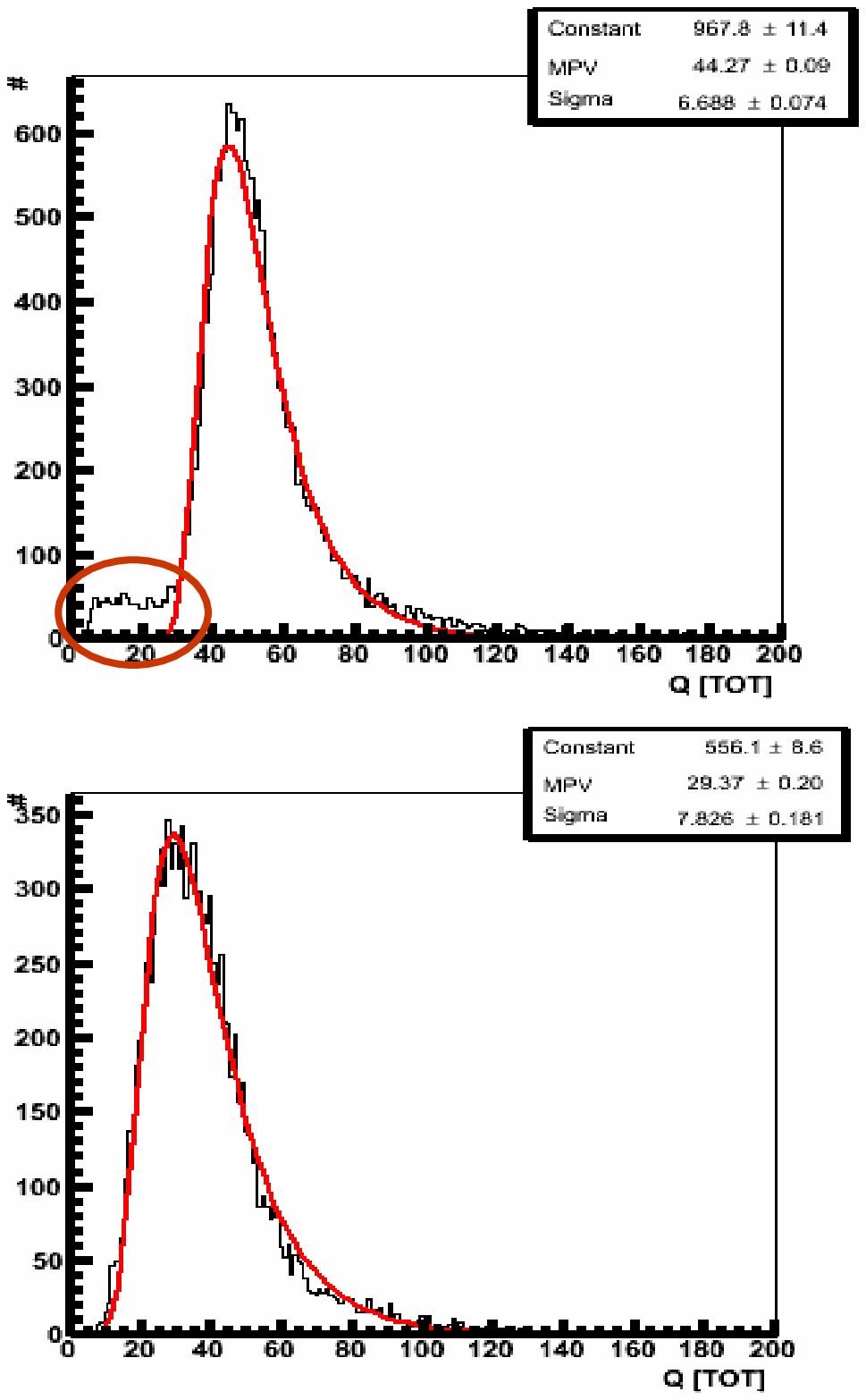,width=7cm,height=12cm}
\epsfig{file=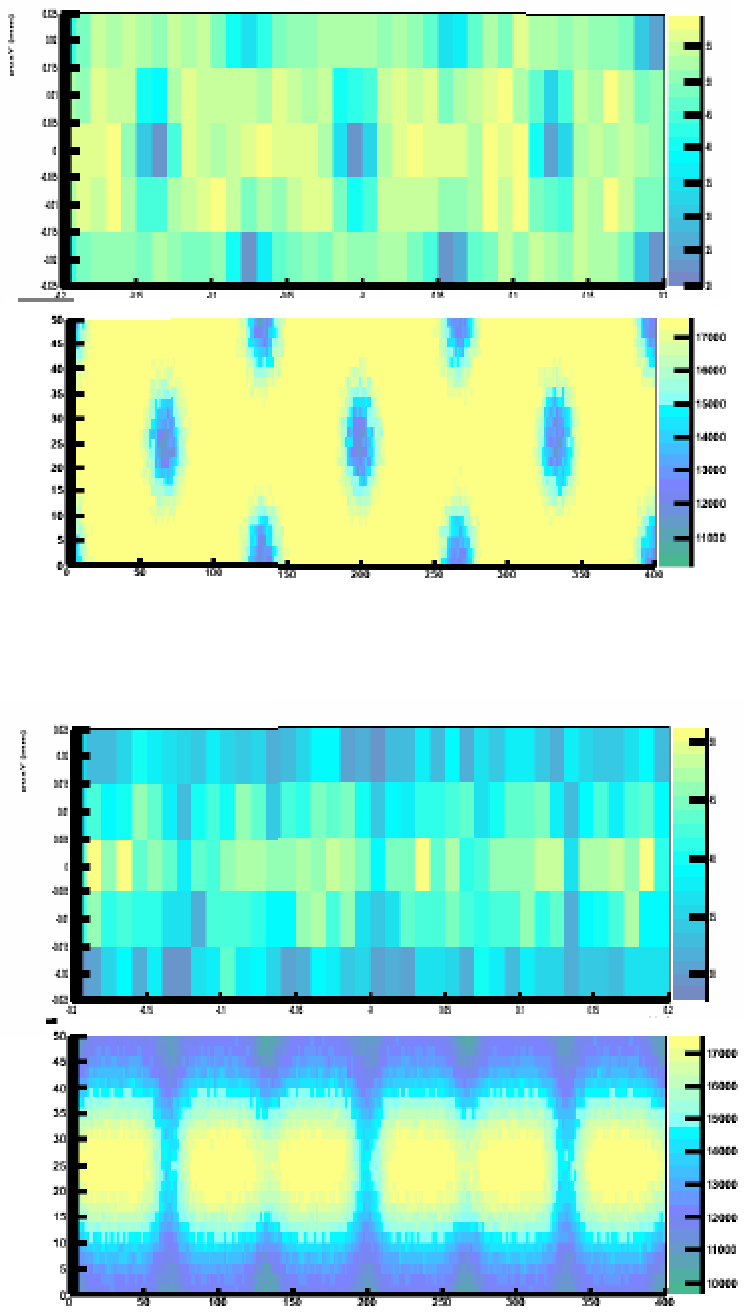,width=7cm,height=12cm}
\caption{Left: Pulse height spectrum for 100~GeV/c pions incident
perpendicularly (top) and at a 15$^{o}$ angle on a 3E-3D 
detector biased at 20~V. One Q[TOT] ADC count is 600 electrons.
The threshold was 3200 electrons. The tracking efficiencies are 95.9\% 
and 99.9\% respectively. This takes into account the partial
response of the central part of the electrodes. Right: Simulations (bottom)
are in good agreement with the experimental results (top)~[M. Mathes, Bonn].}
\label{fig:silicon9}
\end{figure}

\begin{figure}
\centering
\epsfig{file=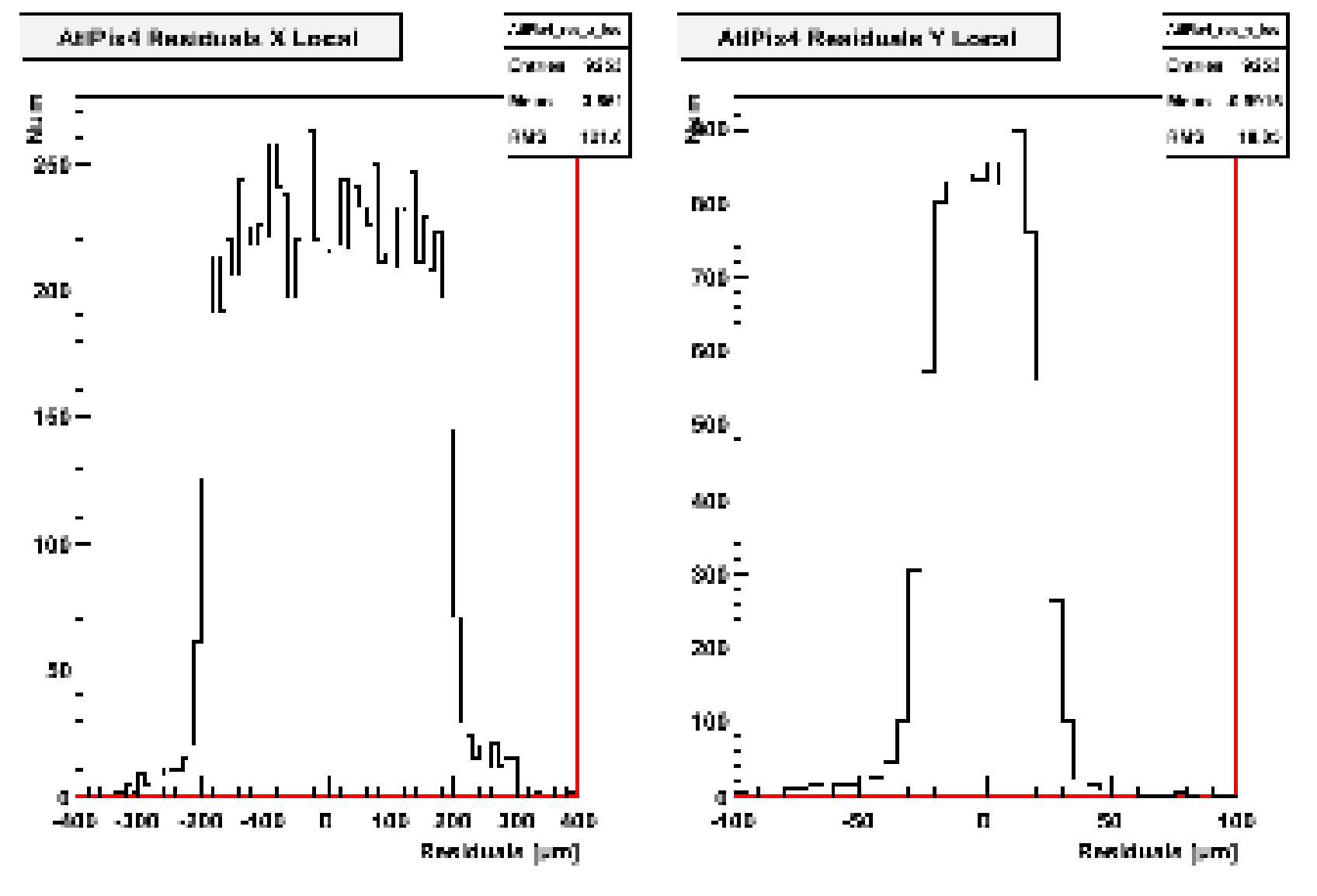,width=11cm}
\caption{Tracking residuals for 3D pixel detector [M. Mathes,Bonn].}
\label{fig:silicon10}
\end{figure}


Full-3D silicon sensors have been successfully fabricated at CIS-STANFORD 
by J. Hasi (Manchester University) and C. Kenney (Molecular Biology Consortium) 
since 2001, following the original design of Sherwood Parker, University of Hawaii 
and C. Kenney who developed active edges. The Manchester/MBC/Hawaii Collaboration 
has been working since 1999 to develop this technology for applications in particle
physics. Important results are summarised below.

The first 3D detector used 16 rows of 38 p+ electrodes spaced by 100 $\mu$m.
n+ electrodes were placed 100~mm from the p+ electrodes. The total
active area was 3.2~mm by 3.9~mm. The p+ electrodes were
connected as strips to ATLAS SCTA readout chips. After tests in the X5
beam at the CERN SPS in 2003, the efficiency was found to be around
98\% and particles were detected to within 5 $\mu$m of the physical edge,
as can be seen in Figure~\ref{fig:3D_telescope}. 
The full results of this beam
test can be found in the TOTEM TDR~\cite{ref:totemTDR} and Ref.~\cite{silref14}. 
A hybrid technology (planar/3D) detector was manufactured at Stanford and
was successfully tested by TOTEM in a prototype Roman Pot at the CERN
SPS in 2004. This uses planar technology but has a 3D active edge. This
worked well, but is a factor 100 less tolerant to irradiation than full
3D technology.

\begin{figure}
\centering
\epsfig{file=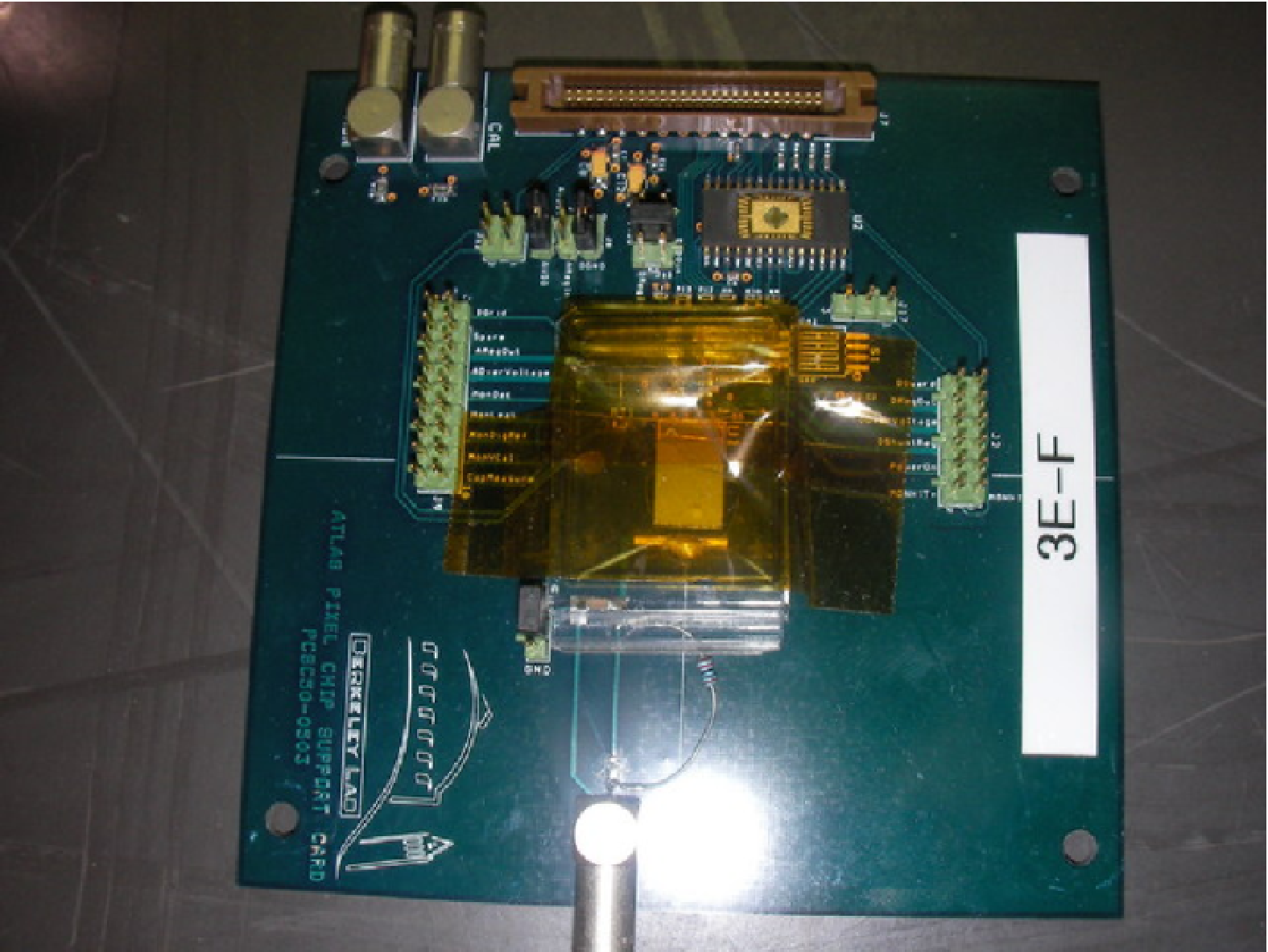,width=10cm}
\caption{Picture of the 3D-ATLAS pixel assemblies mounted
on one of a pcb testboard with a protective cover.}
\label{fig:silicon11a}
\end{figure}

\begin{figure}
\centering
\epsfig{file=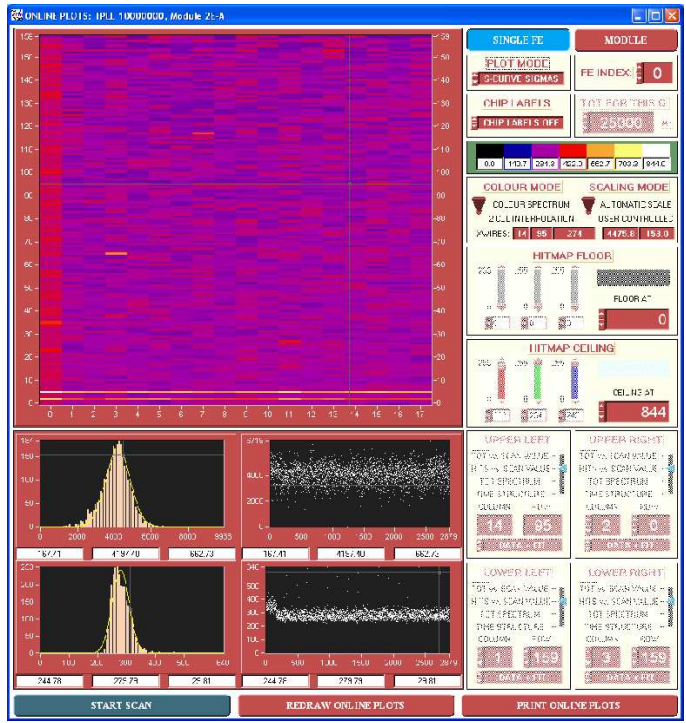,width=9.cm}
\caption{Snapshot of the online display of the TurboDaq
ATLAS pixel test setup. On the top the entire pixel matrix response to
a test pulse. At the bottom the ENC (Equivalent Noise Charge) is
measured as the sigma of the threshold distribution.}
\label{fig:silicon12}
\end{figure}

\begin{figure}
\centering
\epsfig{file=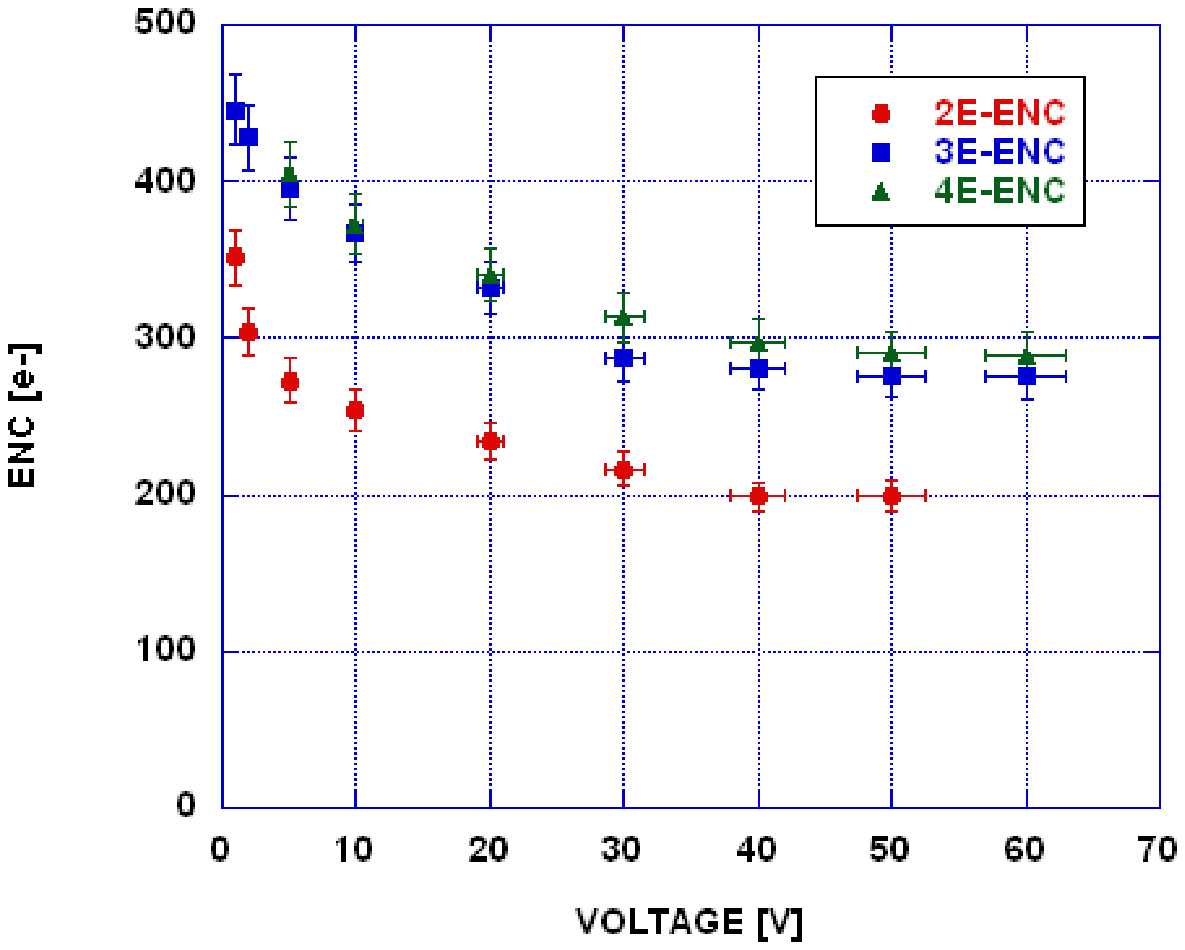,width=12cm}
\caption{The equivalent noise charge (ENC) of the 2E, 3E and
4E 3D detectors after bump-bonding with the FE-I3 ATLAS pixel
readout chip.}
\label{fig:silicon13}
\end{figure}

Initial tests on irradiated 3D samples were made in 2001~\cite{silref5}. 
The first results on the signal efficiency were obtained in 2006 using signal
generated by an infrared laser. The 3D devices were irradiated with
neutrons in Prague with an equivalent fluence of 10$^{16}$
protons/cm$^{2}$~\cite{silref12}. As expected, 3D devices can operate
at much higher fluences than conventional silicon devices. For a
minimum ionising particle, the signal size depends on the thickness.
However, the signal collection distance is determined by the
inter-electrode spacing, which can be as short as 50 microns. The
measurement is shown in Figure~\ref{fig:silicon3} for a 3E device with an
inter-electrode distance of 71 microns. This has three n-type
collection electrodes in a pixel size of 50 micron by 400 microns. Figure~\ref{fig:silicon4} 
shows the signal efficiency versus fluence for the 3D detector.
It is compared to the best that has been achieved using strip and pixel
detectors for the LHC experiments. 3D technology is about a factor five
more radiation tolerant. 
\begin{figure}
\centering
\epsfig{file=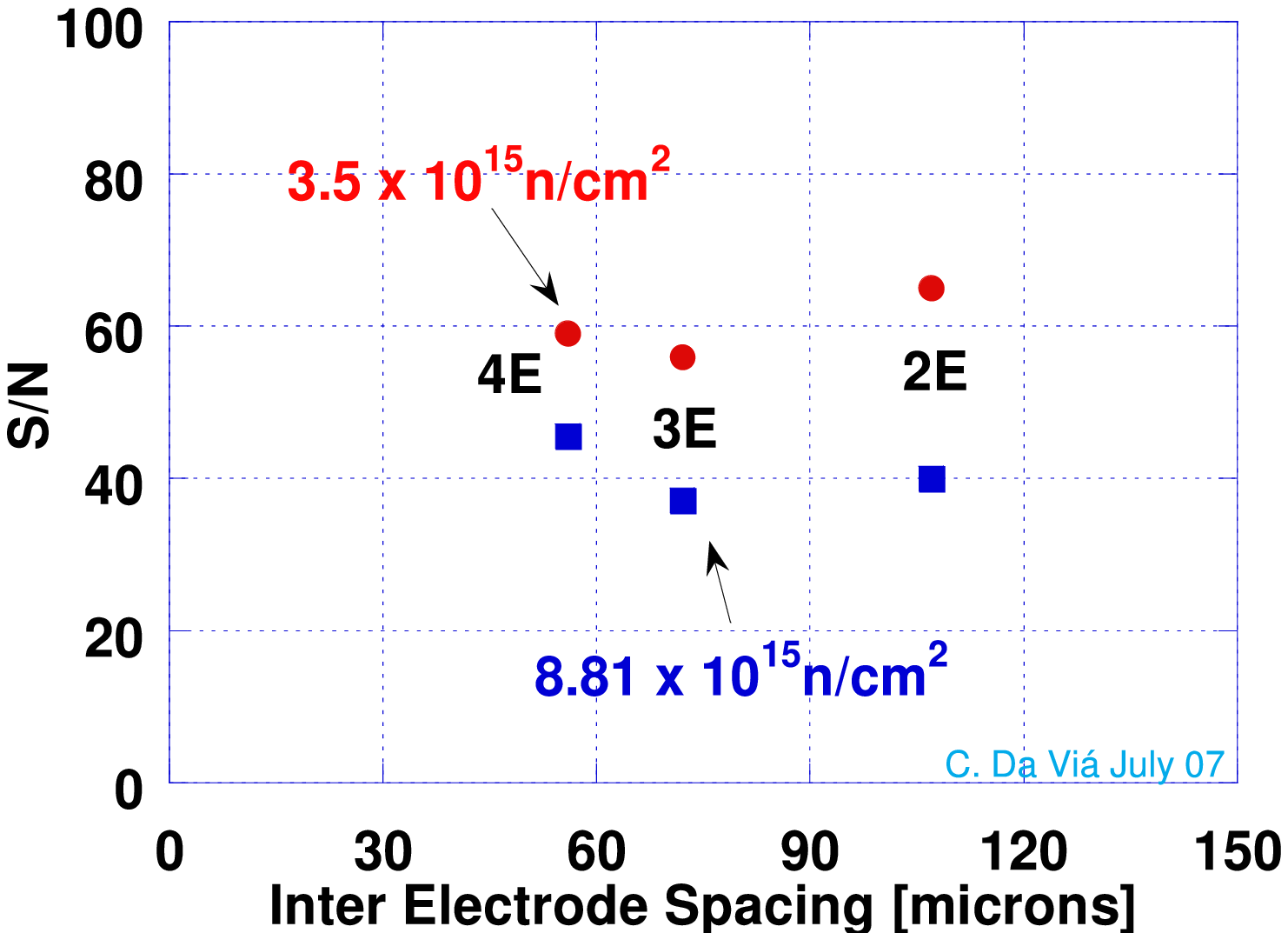,width=12cm}
\caption{Extrapolated Signal to Noise (S/N) ratios of
three 3D pixel configurations at two different irradiation fluences.}
\label{fig:silicon14}
\end{figure}

\begin{figure}
\centering
\epsfig{file=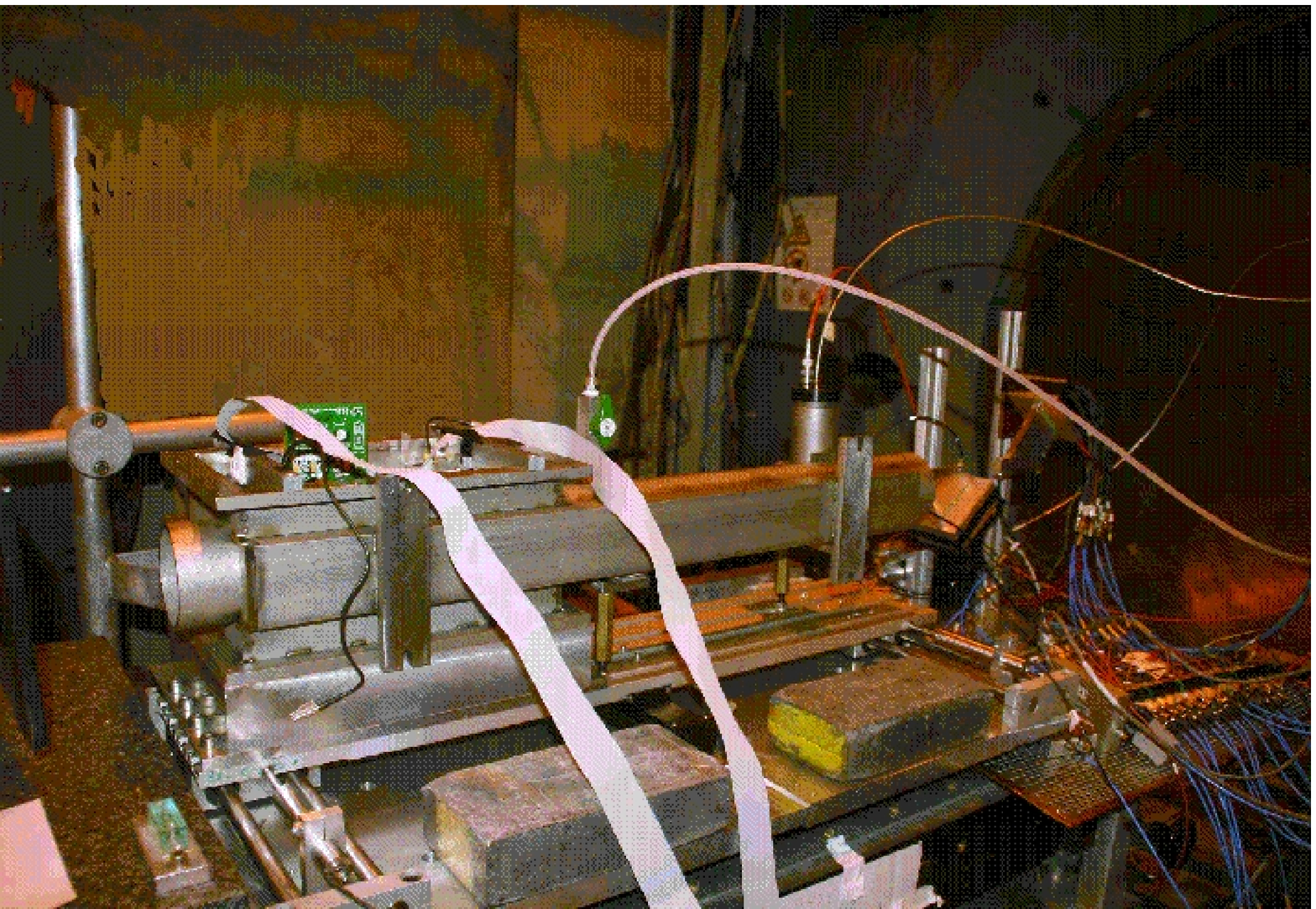,width=12cm}
\caption{FP420 2007 test beam setup including one movable station, 
two blades, and two timing detectors (one GASTOF and two QUARTICS).}
\label{fig:silicon15}
\end{figure}

For the FP420 application the ATLAS Pixel Readout chip was chosen. The
total active area is 7.2~mm by 8~mm. The pixel
structure is shown in Fig.~\ref{fig:silicon5}. The 3D detectors were bump-bonded
to the readout chip. To cover the full area, a minimum of three detectors are
required. The details of the mechanical/electronic layout required to
make a single layer with full area coverage is described in Section~\ref{sec:silicon_RD}. 
Figure~\ref{fig:silicon6} shows that the 4E device can operate at the lowest
voltage. Charge sharing only occurs very close to the pixel edge.
Operating voltages are a factor ten less than for a standard planar
device. Figure~\ref{fig:silicon7} shows a processed 3D wafer. The device yield was around 80\%.

2E, 3E and 4E devices, bump-bonded to ATLAS Pixel readout chip were
tested in the H8 beam at CERN in Autumn 2006 with support from
LBL and Bonn. Individual detectors were placed between planes of a
silicon microstrip tracking system. The beam was 100~GeV/c pions.
Figure~\ref{fig:silicon8} shows a hit map for a 12~mm~$\times$~12~mm and 
3~mm~$\times$~3~mm scintillator trigger. There were no dead or hot pixels.

Figure~\ref{fig:silicon9} shows the pulse height spectrum for a 3E detector for 
minimum ionising particles incident at zero (top) and fifteen degrees. 
The low pulse height at an ADC count of 10 is due to particles traversing the
electrode. The tracking efficiency has been measured to be 95.9\% and
99.9\% respectively using a reference telescope. In the proposed FP420 tracking system,
several planes will be used to form a track-segment. Half of the
plans will be shifted by 25${\mu}$m to improve the spatial resolution
in one dimension. This guarantees that the efficiency will not
suffer from electrode inefficiency. Figure~\ref{fig:silicon10} shows tracking
residuals. This is consistent with the pixel dimensions. The pulse
height spectrum indicates that the efficiency is very high and is
consistent with previous results. Millions of tracks have been
recorded for incident angles  between 0$^{o}$ and 90$^{o}$ 
for 2E, 3E and 4E devices. 

An extended collaboration (3DC) has been formed between Manchester,
Hawaii, Oslo, SINTEF and the Technical University of Prague, to
transfer this technology to industry and guarantee large scale
production. Variations on the full 3D detector design are also being
studied by IRST and CNM. Further developments ar discussed in  \cite{silref15}.

In order to understand the signal-to-noise performance for the
various geometry detectors, the noise performance of the 2E, 3E and
4E 3D sensors was measured after bump bonding with the FE-I3 ATLAS
pixel readout chip (Fig.~\ref{fig:silicon11a}). The equivalent noise charge (ENC)
of the entire pixel matrix was measured, for each configuration, by
injecting a variable amount of charge into each pixel front end and
looking at the threshold dispersion over the entire matrix. This
operation is possible since each front end electronics chip is equipped
with a test input capacitance. Figure~\ref{fig:silicon12} shows a snapshot of the
online display of the ATLAS pixel TurboDaq test system. The top of
the figure shows the response of the entire pixel matrix while the bottom
shows the threshold distribution before and after tuning. The noise versus bias 
voltage for all the 3D pixel configurations can be seen in Figure~\ref{fig:silicon13}.

The extrapolated signal-to-noise of the three configurations after
irradiation is shown in Figure~\ref{fig:silicon14}. The plot shows the S/N after a
fluence of 3.5 $\times$ 10$^{15}$ n cm$^{-2}$ and 8.8 $\times$ 10$^{15}$ n cm$^{-2}$ respectively.
The first set of values corresponds to the integrated fluence expected
at 4 cm from the ATLAS interaction point (i.e. the ATLAS central tracker) 
after $\sim$10 years of operation of
the LHC at nominal luminosity. The second set corresponds to the values
expected after $\sim$5 years of operation at the same distance at the
SLHC. These S/N results indicate that the lower fluences expected at the FP420 location
should not compromise the performance of the 3D pixel tracking
detectors.


In conclusion, 3D detectors readout out using the ATLAS Pixel Chip
fulfill all the requirements for use in the FP420 experiment

\subsection{Tracking detector mechanical support system}
\label{sec:silicon_support}

\begin{figure}
\centering
\epsfig{file=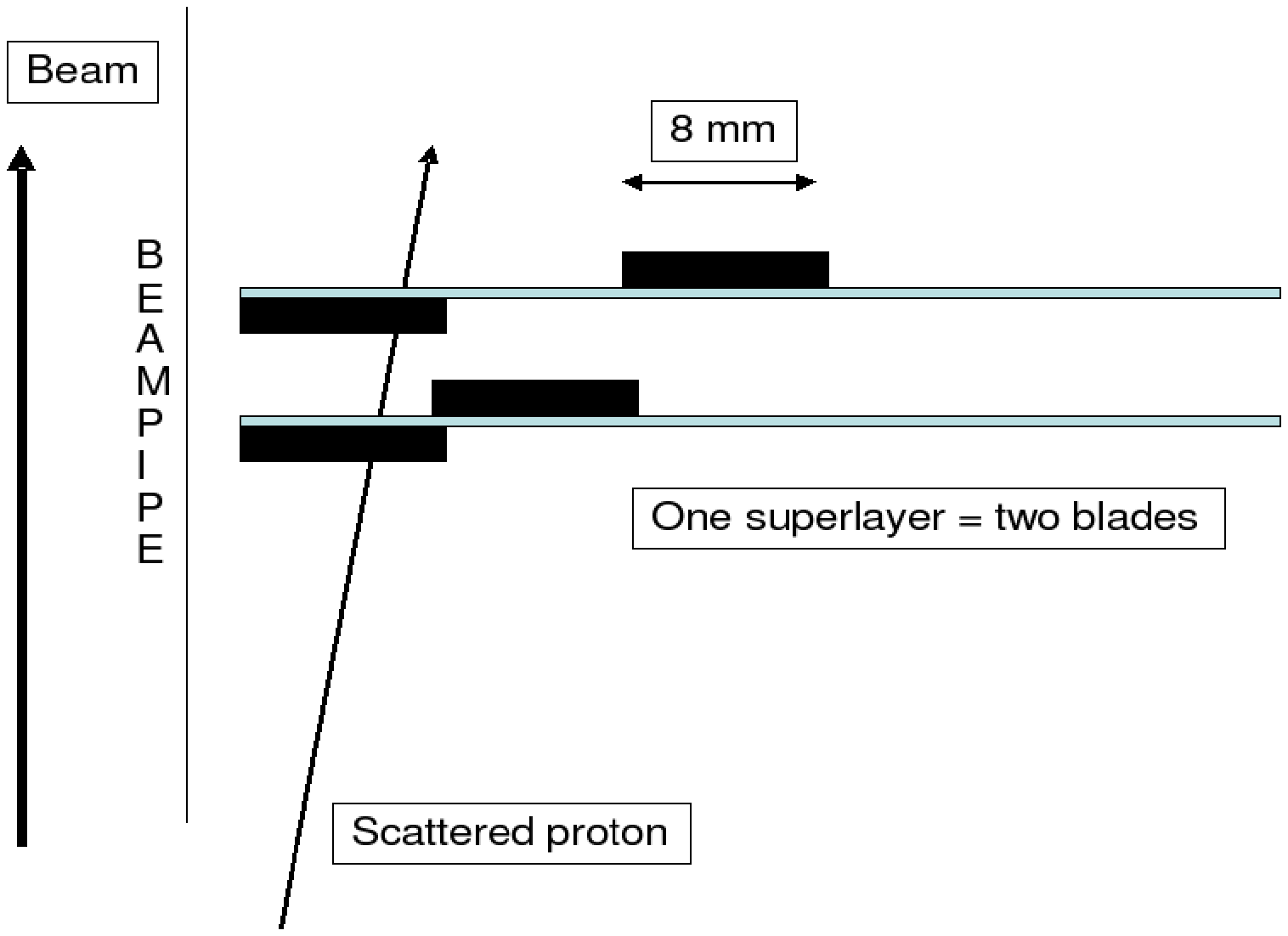,width=10cm}
\caption{Schematic of a superlayer consisting of four sensors.}
\label{fig:superplane}
\end{figure}

\begin{figure}
\centering
\epsfig{file=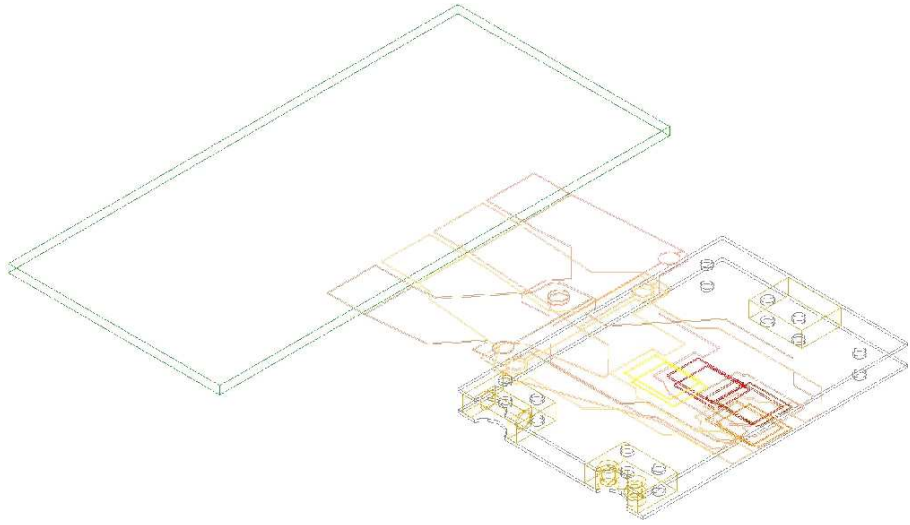,width=12cm}
\caption{A schematic drawing of a superlayer, consisting of two blades.
The flexible circuits connect the four sensors to a common control card.}
\label{fig:superlayer}
\end{figure}

\begin{figure}
\centering
\epsfig{file=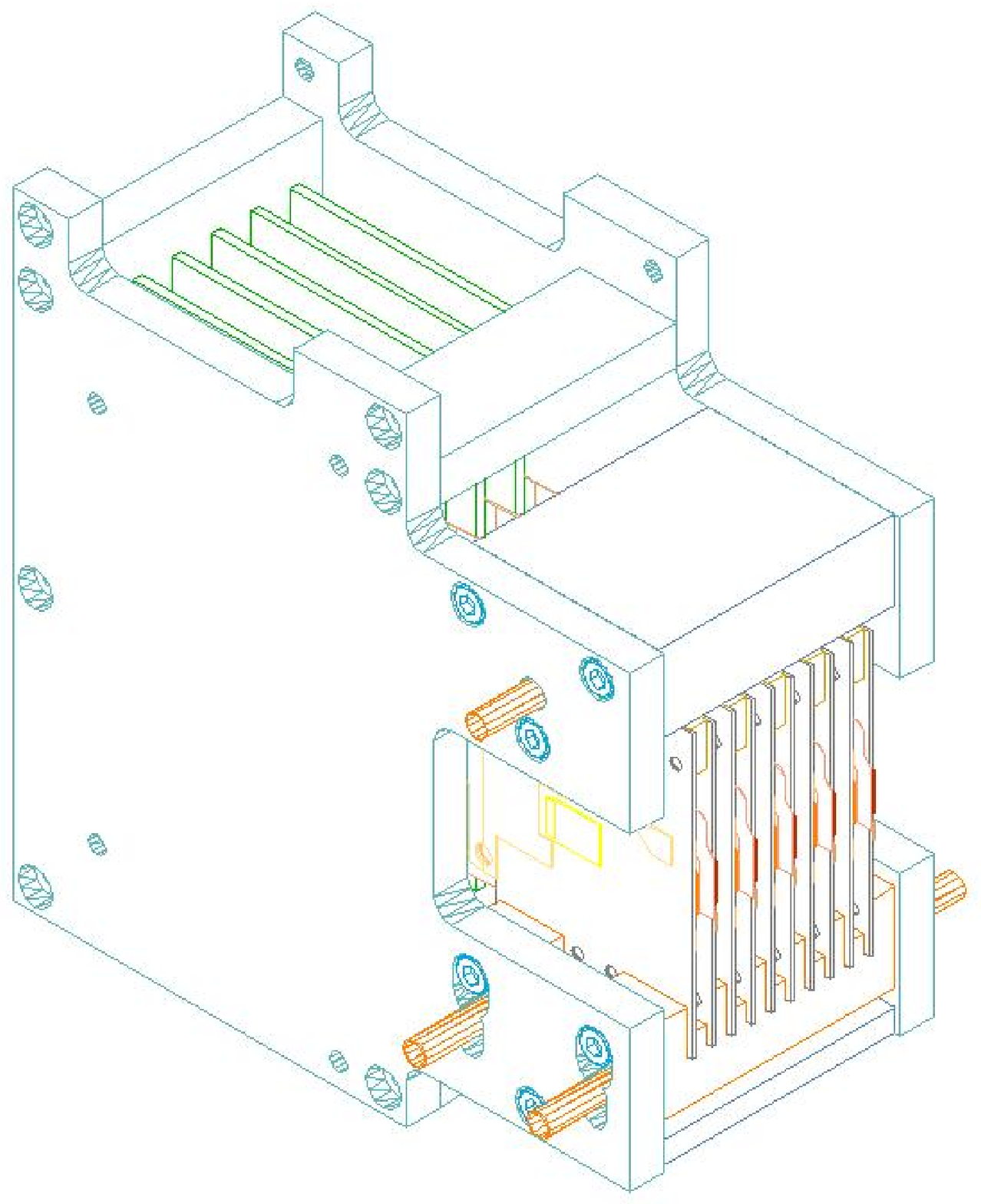,width=12cm}
\caption{A five superlayer tracking station. The mechanics supports the superlayers and also provides cooling blocks.}
\label{fig:5superlayers}
\end{figure}

The space available for the detectors is extremely limited. The baseline
design is to have two independently moving pockets, one at each end of
the 420~m region. The pockets may be sub-divided to allow different
cooling and vacuum conditions for the silicon and timing detectors. The
optimal configuration may change depending on the pile-up conditions
and the machine-induced background environment at the time of
operation. A key design goal has therefore been to allow changes in the
detector configuration to provide the optimal balance of detection
points versus traversed material, and to allow simple replacement of
failing detectors during permitted tunnel access. To achieve an active
area of 5~mm $\times$~25~mm requires a minimum of three
silicon sensors. The basic detector unit, referred to as a superlayer,
tiles the sensors to cover the required area. A superlayer is made of
two ``blades''. Figure~\ref{fig:superplane} shows a schematic of the superlayer
layout to illustrate the basic geometry and nomenclature. A single
tracking station will consist of a number of superlayers. Schematic
drawings of a superlayer and a modular tracking station consisting of
five superlayers are shown in Figs.~\ref{fig:superlayer} and~\ref{fig:5superlayers} 
respectively.

Since the 3D silicon sensors have rectangular pixels of 50 microns by
400 microns, they have better position resolution along one axis.
This means that superlayers can be designed to position the sensors
to give superior resolution in the horizontal ($x$) or vertical ($y$) plane. In the initial
phase of FP420 operations, the horizontal ($x$) deflection of the protons from the beam
is of prime importance, since this corresponds to a measurement of the energy 
loss and hence the missing mass. The vertical ($y$) position becomes important primarily
when the $p_T$ of the outgoing protons is required. Whilst there is a strong physics 
case for measuring the $p_T$ of the outgoing protons the initial priority of FP420 
is the missing mass measurement. Phase 1 will therefore be optimised for a high-precision
$x$ measurement, with $y$ measurements considered as a potential
future upgrade. Because of the modular design of the tracking stations,
superlayers optimised for enhanced $p_T$  resolution can easily be inserted in a
short tunnel access.

\subsubsection{Superlayer and blade design}

A superlayer consists of two blades, each carrying two sensors. The two
sensors closest to the beam overlap but are offset with respect to each
other by half a pixel (25 micron) to improve track resolution for low
${\xi}$ particles {--} see Section~\ref{sec:silicon_RD}. A superlayer control card is
positioned between the blades and connected by four flex circuits.
Although the 3D silicon sensor technology is edgeless, tabs required
for readout connections to the front-end ASIC, bias connections to the
sensor and edge effects imply that it is impossible to tile the
detectors in certain orientations. Even in the specific orientation
unaffected by these tabs there are residual edge effects introduced by
the front-end chip design. These constraints require detectors to be
positioned over a number of overlapping layers to provide the required
coverage. This is achieved by using both sides of the blade. 

The choice of material for the blades is critical if the design goal of
an internal mechanical alignment of 10 microns is to be achieved. The
material must be stiff but machinable, have a high thermal conductivity
and low coefficient of thermal expansion, similar to that of the
attached silicon dies. The thermal conductivity must be optimised
relative to the density to allow for extraction of heat from the
detectors without too high a thermal gradient, whilst minimising the
amount of material (radiation length) and hence multiple scattering.

\begin{table}[htdp]
\centering
\begin{tabular}[l] {|c|c|c|c|c|} \hline 
Material & Thermal conductivity & Relative Density ${\varrho}$ & K/${\varrho}$ & CTE \\
 &  K (Wm$^{-1}$K$^{-1}$) & & & (10$^{-6}$~K$^{-1}$)\\\hline
CE7 (70/30 Si/Al) &\ \ 125&\ \ \ \ 2.4&\ \ 52&\ \ 7.4\\\hline
Aluminum Nitride &\ \ 180&\ \ \ 3.26&\ \ \ 55&\ \ 5.2\\\hline
Silicon &\ \ \ 156-200&\ \ \ 2.33&\ \ \ 67-86&\ \ 2.6\\\hline
\end{tabular}
\caption{Possible blade materials.}
\label{tab:}
\end{table}

Beryllium oxide and Beryllium metal although possessing good thermal and
low mass parameters were rejected at this stage because of
difficulties due to their toxicity, which makes prototyping difficult,
time consuming and expensive.

Several blade design variants have been prototyped. We were initially
attracted to the possibilities of CE7, a hypereutectic alloy of 70/30
silicon and aluminum because its aluminum component makes it
machinable with conventional tooling, making it possible to construct
a blade as one single component. Its K/${\varrho}$ value
of 52 compares well with more conventional materials such as aluminum
nitride. The prototype blades used in the Sept 07 CERN test beam
runs were of this design with the centre sector machined down to 500
microns. However material scattering considerations are pushing the
design to be even thinner {--} 300 microns. It has proved difficult to 
machine CE7 to this tolerance due to its granular structure.
Hence we have investigated an alternate design using a CE7 frame and a
decoupled planar thin front section supporting the detector. This
allows the use of hard materials such as silicon or Al\,N whose 
thicknesses can be lapped down to 300 micron with high surface
finish. The superplane shown in Figure~\ref{fig:superlayer} has such a design. In a
planar geometry the requisite shapes can be laser cut.

\subsubsection{Thermal tests of the blades}

\begin{figure}
\centerline{
\epsfig{file=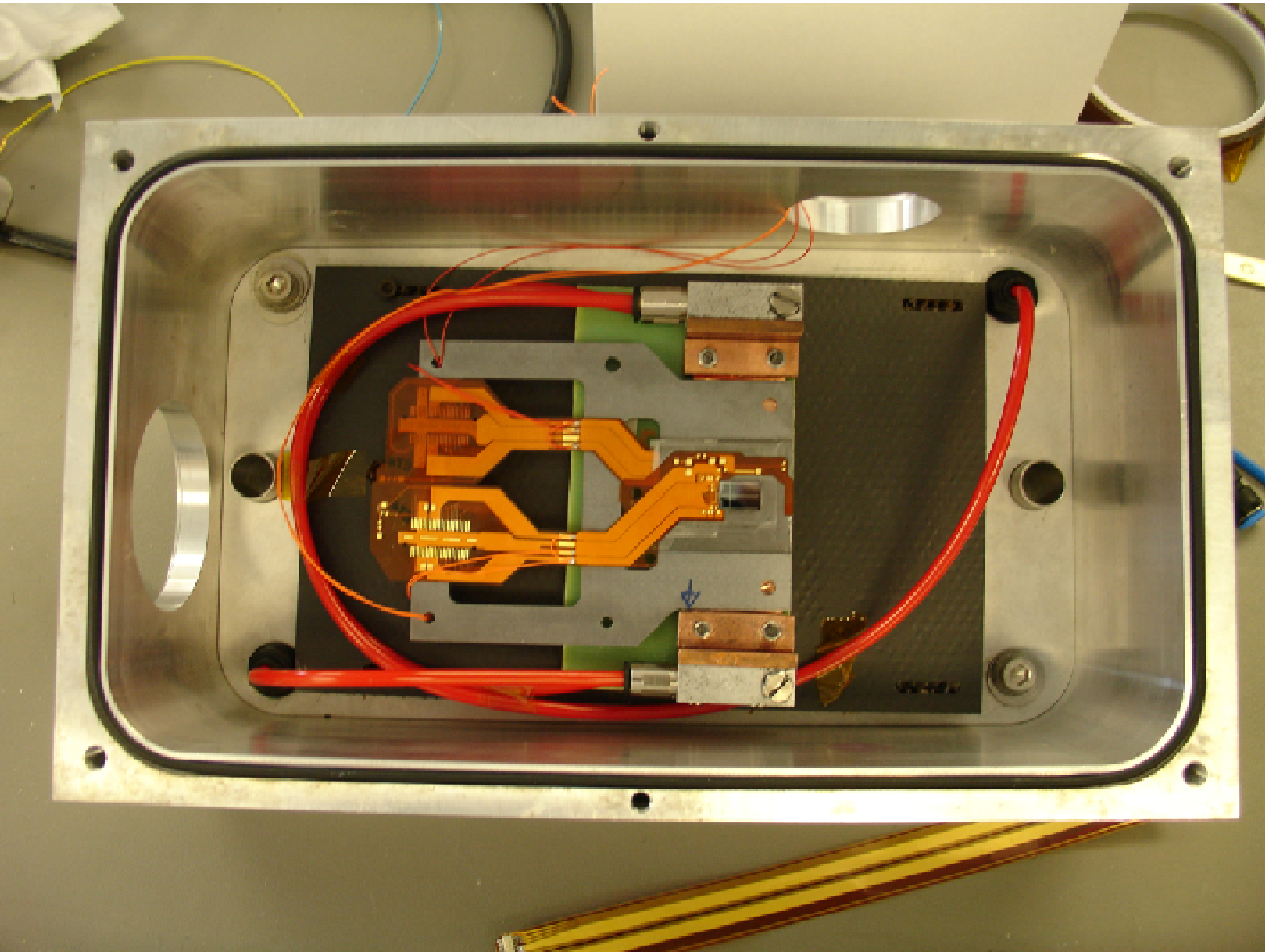,width=7cm}
\epsfig{file=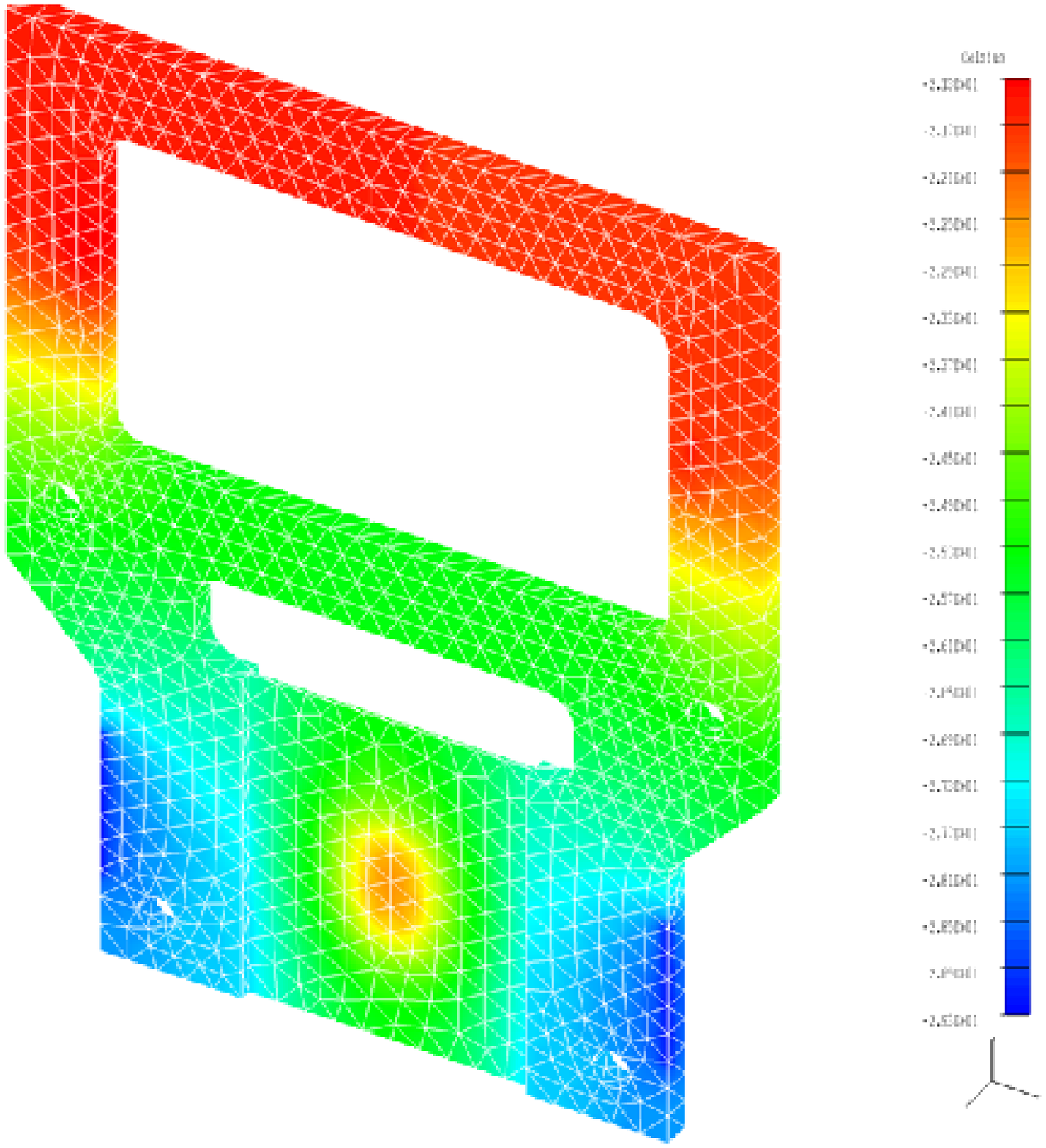,width=7cm}}
\caption{Left: A thermal model blade with the silicon resistors as described in the text. 
Right: Finite element analysis of the thermal properties of a blade.}
\label{fig:thermal1}
\end{figure}

Test blades have been built to investigate heat flow and thermal gradients and 
the resulting mechanical displacements using a thermal camera and a ``smartscope'' 
to measure the displacement. A realistic chip/glue/support interface
structure was constructed using custom silicon resistors that match the
size and power of the front end chip and have a similar bond pad
layout. The model used for thermal testing is shown in Fig.~\ref{fig:thermal1}. Also
shown is the finite element analysis of the blades performed at Mullard
Space Science Laboratory. The preliminary thermal tests indicate that
the blade design meets the required criteria of thermal and mechanical
stability at the 10 micron level.

\subsubsection{Assembly and alignment}

\begin{figure}
\centering
\epsfig{file=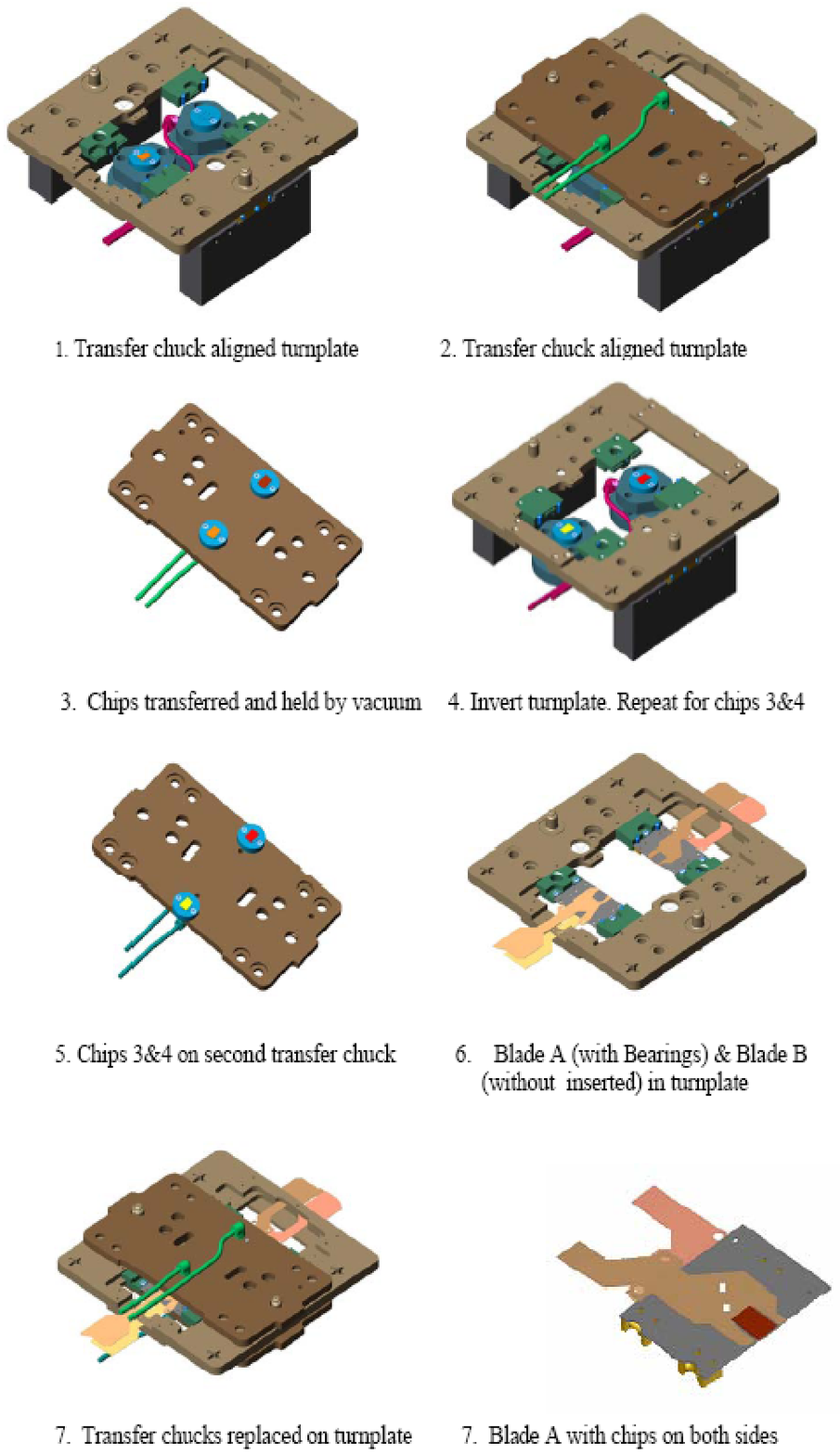,width=12.5cm}
\caption{ Blade assembly -- Positioning of chips 1 to 4 on Blades A and B.}
\label{fig:blade_assemb}
\end{figure}

\begin{figure}
\centering
\epsfig{file=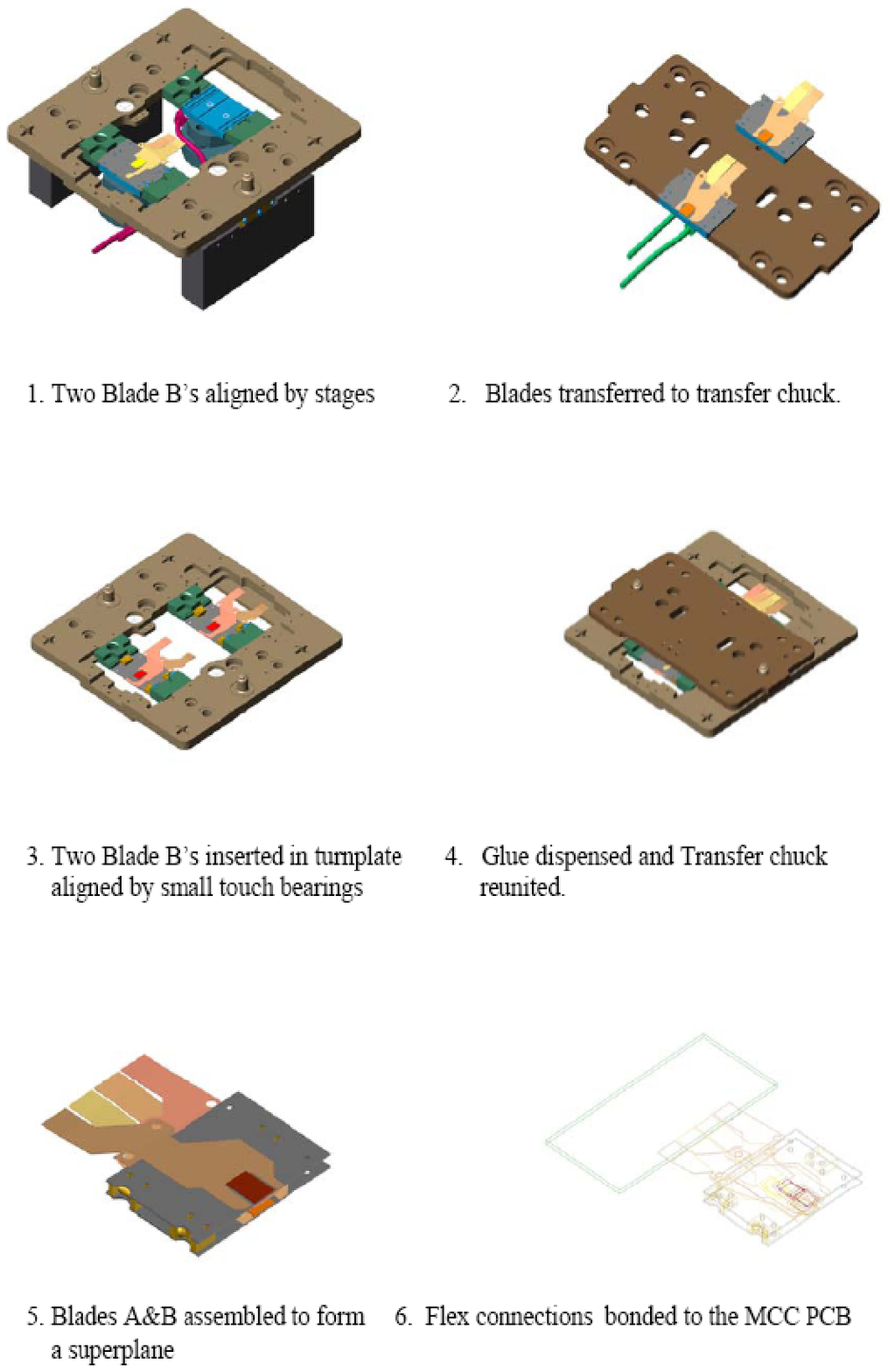,width=12.5cm}
\caption{Superplane assembly.}
\label{fig:superplane_assemb}
\end{figure}

The silicon sensors will be positioned on the blades using an
adaptation of the automated assembly stages and jigs used to construct
silicon modules for the ATLAS SCT at Manchester. The system uses
automatic pattern recognition of fiducials on the readout chip to
provide coordinates to $x,y, \theta$ motion stages which position the
detector on precision jigs. Components are glued using a Sony CastPro
dispensing robot under software control. The system allows silicon
sensors to be reliably positioned on opposite sides of a blade with an absolute position
accuracy of 5 microns. Detector blades are independently surveyed using
a Smartscope optical coordinate measuring machine capable of one micron
precision. Figs.~\ref{fig:blade_assemb}  and~\ref{fig:superplane_assemb} 
show the build sequence for a blade and then a superplane.
Once the superplanes have been manufactured, the station needs to be assembled.\\

\begin{figure}
\centering
\epsfig{file=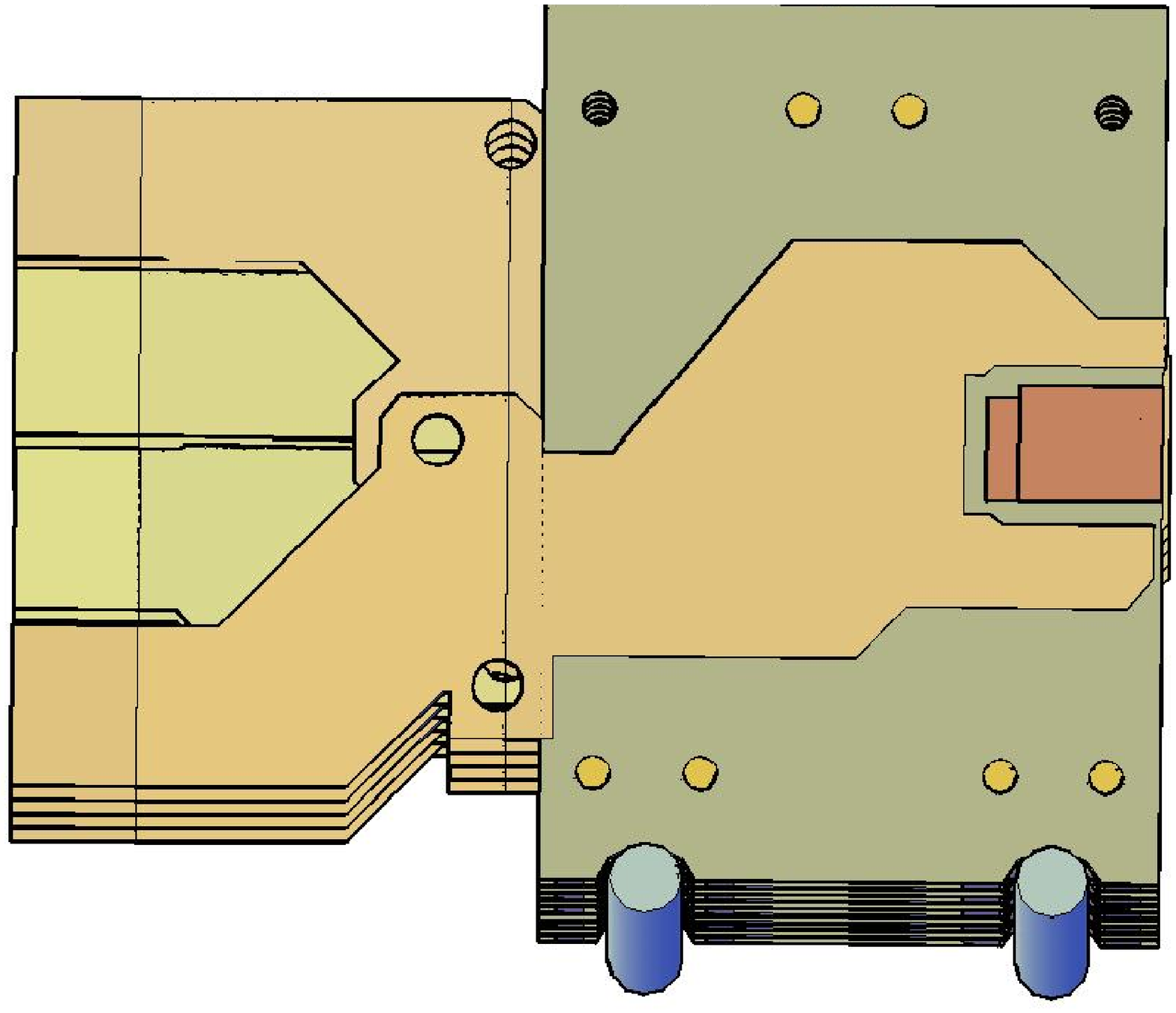,width=6.5cm}
\epsfig{file=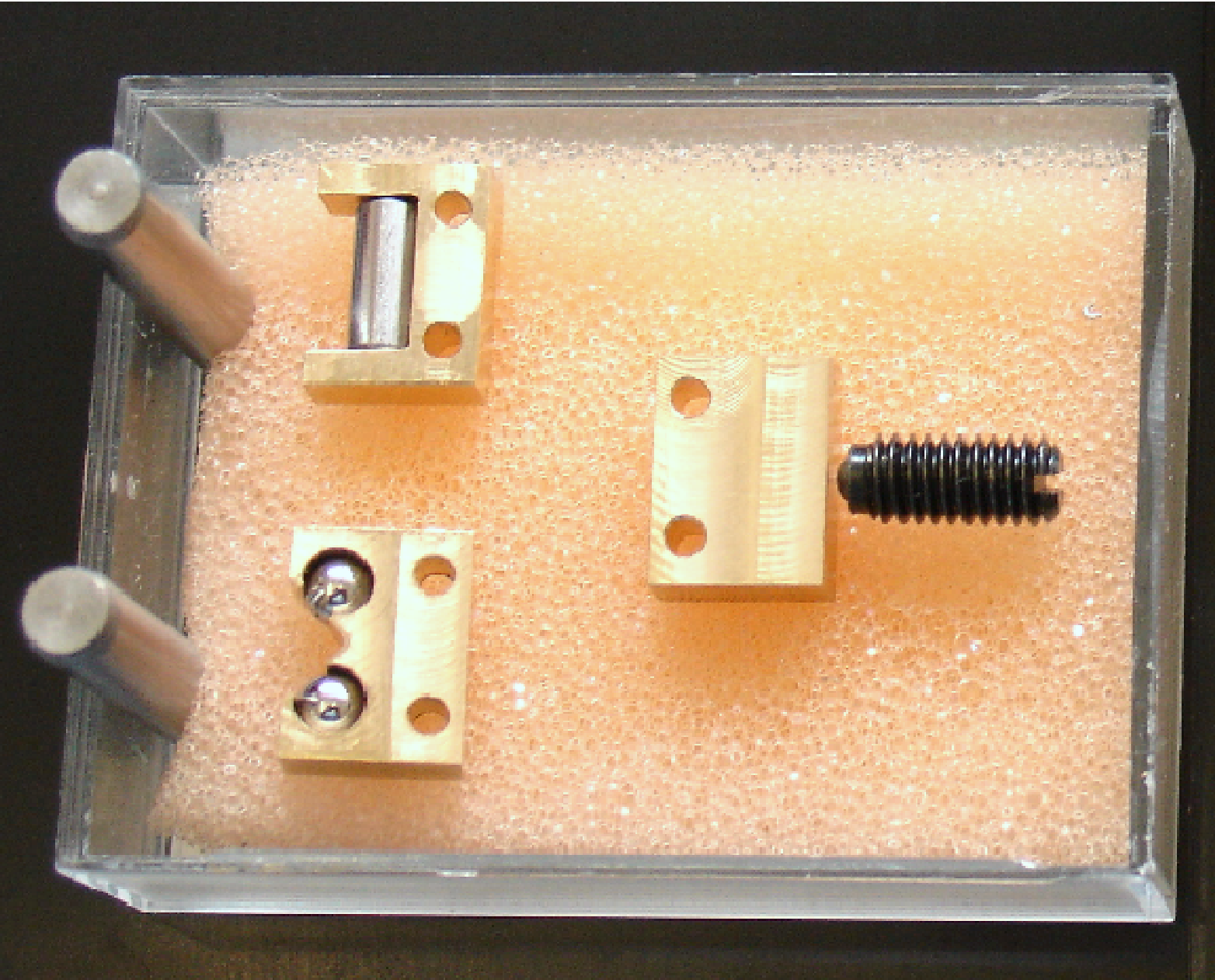,width=7cm}
\caption{Left: Several superplanes showing the bearings
without the support structure. Right: Miniature touch bearing.
The ball bearings are 3mm diameter.}
\label{fig:suplanes}
\end{figure}

Linking individual blades together, in pairs as superlayers, and then
into an entire station has some complexities. Several approaches have
been prototyped. The simplest idea is to use linking dowel rods and
precision wire cut washers stuck to each blade following the ATLAS SCT
experience. These can be manufactured to 5 micron tolerance. However,
the alignment of a stack of 10 is limited by sliding tolerances and
difficulties in maintaining dowel angular tolerances. This led us to
touch bearing designs. A touch bearing consists of a bar perpendicular
to the dowel rod, pushed against it by a spring force such that there
is a unique point contact between the two. These are arranged in a
kinematic manner, providing a V and a flat. The kinematic single point
contacts provide high reproducibility, whilst the spring force allows
easy movement to position along the dowel. The challenge is to make
such a bearing design small enough for this application. One of the
restrictions imposed by through dowels is that it is difficult to
remove an individual superplane without dismantling the entire system.
This leads us to our current baseline design: open-sided touch
bearings, one V, one flat are sandwiched between two blades as part
of the superplane assembly process. These are located against two external
dowel rods, held by a small ball spring. Figure~\ref{fig:suplanes} shows several
superplanes and their bearings without the support structure and also
the miniature touch bearings.

\begin{figure}
\centering
\epsfig{file=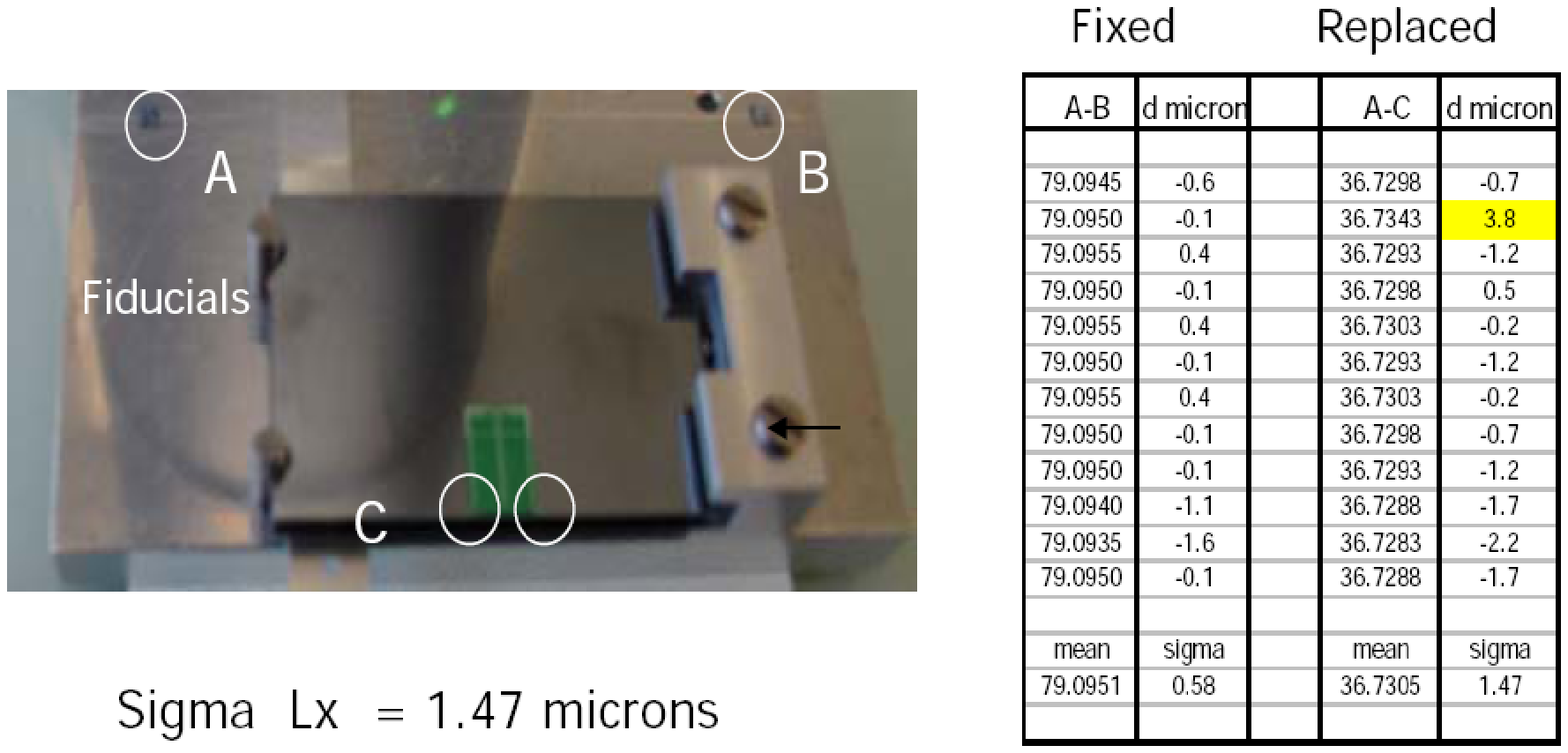,width=12cm}
\caption{Superplane positioning accuracy measurement.}
\label{fig:superplane_position}
\end{figure}

Figure~\ref{fig:superplane_position} shows results of repeatedly repositioning a superplane {--}
reproducibility at the 5 microns level or better is clearly demonstrated using the touch 
bearings. Measurements were taken using
a Smartscope optical coordinate measuring machine of fiducials on a
superplane front end compared with fixed fiducials on a base. The
superplane was repeatedly removed and repositioned against the end
bars.

To summarise:  sensor to sensor positioning on a blade element can be
achieved with an accuracy of 5 microns, and within a superplane to 10
microns.\\

The position of any sensor in the station once built can be surveyed by
the Smartscope with a single measurement accuracy of 1 micron
(several measurements may be required to link all sensors).

\subsubsection{Electrical details of the superplane}

A flex circuit situated behind and bonded to the sensor is used to connect the FEC chips 
to the power supply and data lines via wire bonds. The flex circuit is fitted with a control 
chip (the MCC) which services the FEC chips, distributing clock, control and trigger
information and collecting data for onward
transmission. Aside from some slow single-ended signals, the
connection between the FEC and MCC chips are implemented using
LVDS-style differential signaling, with lower current than LVDS
terminated into 600${\Omega}$. The two-layer flex circuit is built on a
50 micron polyimide core with a nominal track/spacing of 100 microns
falling to 60 microns in the bond region, 100 micron laser drilled
vias, and a Ni/Au finish suitable for Al wire bonding. The flex is
pre-assembled (passive decoupling components soldered) then glued to
the blade. Positioning is visual with respect to the chip and performed
with a manual placement workstation with a typical accuracy of around
20 microns. The positioning is not critical, the bonding process can
cope with many tens of microns misplacement between flex and front end
assembly. 25 micron Al wire with (99\% Al, 1\% Si) is used.
Wedge-wedge bonding has been undertaken with a manual
(semi-automatic) wire bonder during the prototyping phase; an
automatic bonder will be used in production. We plan to investigate the
benefits of plasma cleaning the flex, although our experience thus far
has shown no difficulties bonding to the flex using a slightly elevated
bonding power setting to overcome any surface contaminants. The
individual blades need to be tested before final assembly as both the
sensor assembly and their connections on the internal faces of the
blades are not accessible after the blades have been combined into a
superlayer, and it would not be practical to split and repair after
assembly. The flex circuits have sacrificial tails that bring the
signals to diagnostic headers. These connect to adapter boards allowing
connection to the ATLAS Pixel TurboDAQ system which can be used for
single chip testing. We foresee the option of potting the bonds after
successful testing. Once both sides of the blade have been processed
and all tests have been successfully completed the sacrificial tails
are cut away. Two blades are combined with a control card and fixed
together to form a superlayer. The flex circuits are glued to the
control card with solder connections between the underside of the flex
and the card for power, and data connections made by wire bond between
pads on the topside of both. The bond pitch is much more generous and
the alignment is not critical. The flex tension does not have any
impact on the sensor positioning. The control card is a hybrid based on
conventional PCB construction expected to have microvia breakout of the
high density wire-bond connections to the MCC chip. Because of a
shortage of MCC chips in this prototyping phase it will be necessary to
mount the chip in a ceramic carrier which is placed into a socket on
the board, but final production boards will be true hybrids. The power
planes of this card provide the thermal path for the heat generated by
the MCC chip. The connection from the superlayer has not been
finalised. The prototypes use a SAMTEC QTE connector that straddles the
board edge and mates to a custom made cable assembly.

Differential (LVDS like) data paths from each superlayer, together with power supply
connections, span the detector box assembly to the support crate positioned either 
inside one of the support legs of the NCC or in an overhead gantry nearby. 
At the support crate, data links are merged and passed to the optoboard. 
Each superlayer has one inward link that provides
clock/trigger/control, and one outward link for data. The MCC support
chips offer dual output links, but because of the low occupancy and
small number of FE chips associated to each MCC (1/4 of density of the
Pixel detector) only one link is required. We hope to be able to adopt
the opto-components used at ATLAS, however the multimode fibre is
susceptible to radiation which over these long distances may cause
excessive attenuation, so it may be necessary to periodically replace
the fibre. Alternatively, a rad-hard monomode based connection, as
used at CMS, may need to be developed. 

\subsubsection{Station positioning}

From an electrical point of view, a station is simply a collection of
superlayers. It is worth noting however that the station is positioned
inside a box that is welded to the beampipe and fitted with substantial
lid, and so is a good Faraday cage. The blade carrier material is
itself conductive; one point to be established therefore is whether
this should be actively tied to the ground reference (the box/ beam
pipe) or left to float. RF modeling studies together with practical
testing on the RF test rig at the Cockcroft institute will help to
determine the optimum strategy.\\

The tracking station will be loosely mounted from the lid of the vacuum
vessel by flexible supports. Services , cables and cooling feedthroughs will be on the lid.\\

\begin{figure}
\centerline{
\epsfig{file=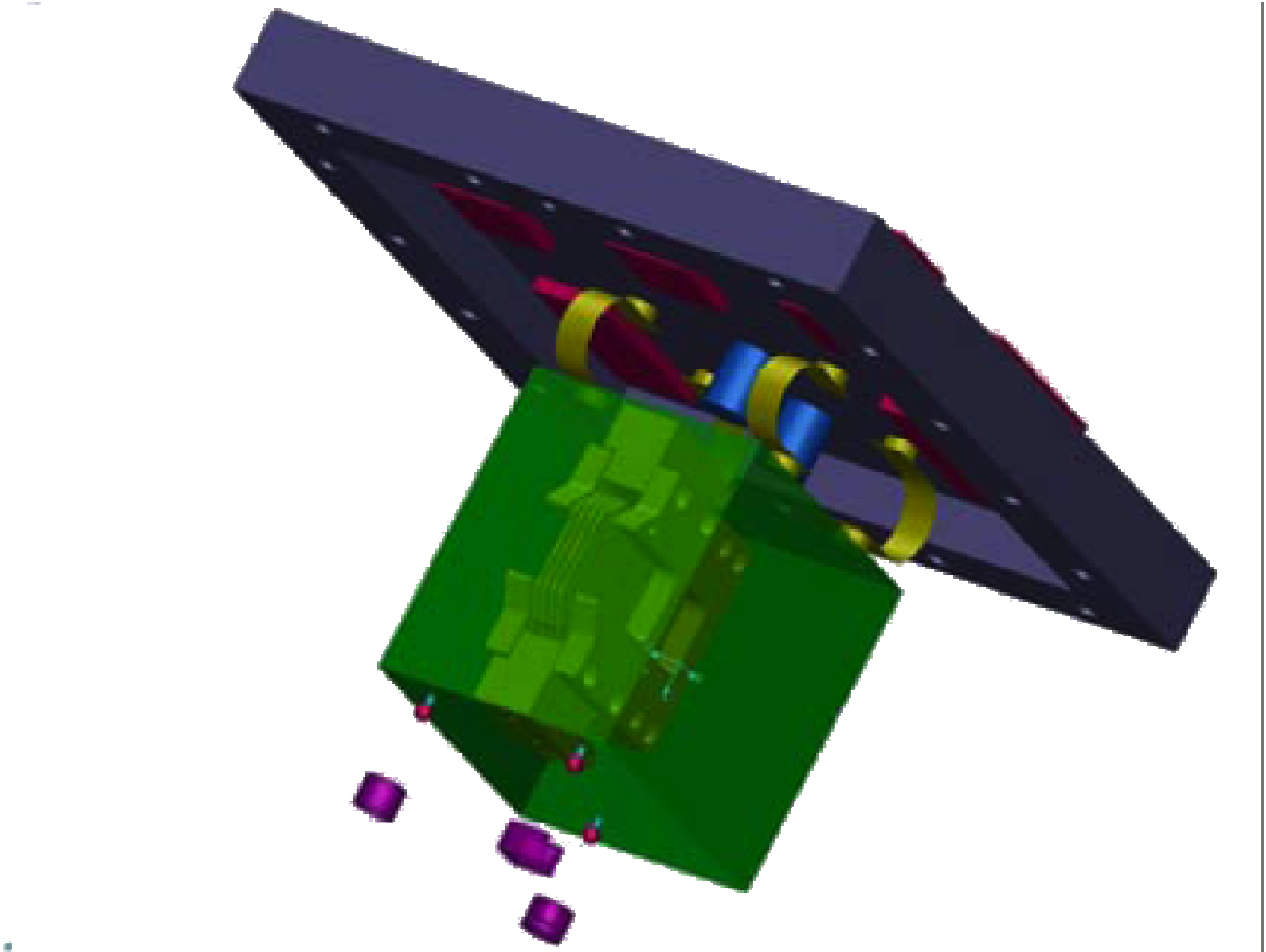,width=7cm}
\epsfig{file=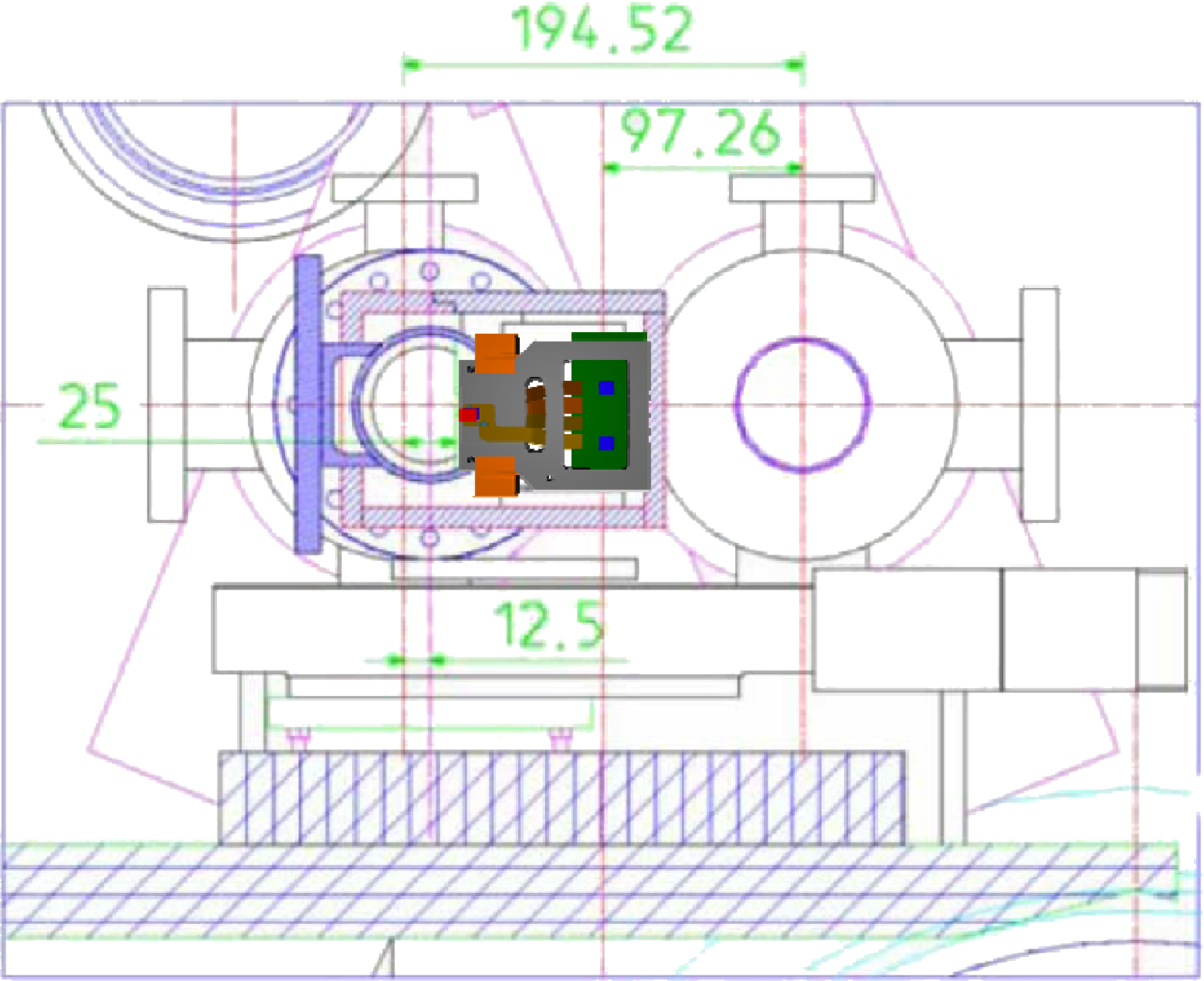,width=7cm}}
\caption{Left: A complete tracking station attached to the lid. Also shown are the 
positioning studs. Right: Schematic view of the box in position around the beam pipe.}
\label{fig:box}
\end{figure}

\begin{figure}
\centering
\epsfig{file=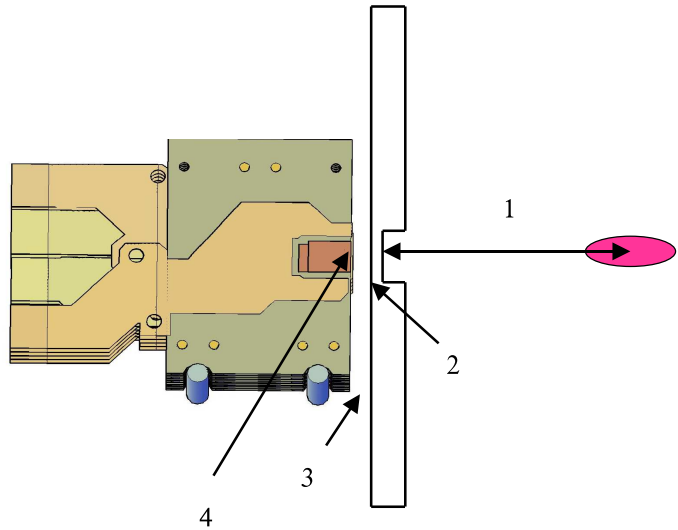,width=9cm}
\caption{Key dimensions from the beam to the edge of the first chip.
1 {--} beam to window, 2- window thickness, 3 - standoff
of detector from window depends on thermal considerations and assembly 
tolerances , 4- distance of first pixel from from edge of blade defined by
dicing considerations.}
\label{fig:key_dims}
\end{figure}

Precision alignment with respect to the beam pipe is achieved by location with  
kinematic ruby ball mounts on the base of the box. Figure~\ref{fig:box} shows a station and lid, 
and relates these to the LHC beam pipes. Figure~\ref{fig:key_dims}  defines key distances that will 
determine how closely the active silicon will be to the beam.

\nopagebreak

\subsection{High-voltage and low-voltage power supplies}
\label{sec:hv-lv}

This section outlines some of the solutions envisaged for the bias and 
low-voltage supplies.  Emphasis is on the supplies for the 3D sensors and 
their front-end chips. For more details, the reader is referred 
to Ref.~\cite{fp420-note-08-001}.

\subsubsection{Low-voltage power supplies specifications}

Each superlayer requires two low-voltage supplies, preferably floating
with minimum 1 V compliance range relative to each other, see
Table~\ref{tab:hvlv1}. The low-voltage supply for a superlayer should be
floating relative to that of any other superlayer. There will be
decoupling capacitors close to the load.

\begin{table}[htbp]
\begin{center}
\begin{tabular}{|l|c|c|c|c|}\hline
\multicolumn{1}{|c|}{One Pixel FE-I3 chip} & Voltage range& Nominal 
voltage & Current & Current limit \\ \hline
Analog AVDD & 1.6-2.0 V & 1.6 V & 5-70 mA & 100 mA \\ \hline
Digital VDD & 1.5-2.5 V & 2.0 V &  40-50~mA (1\% occ.) & 100 mA \\ 
            &           &       &  60-70~mA (10\% occ.)&  \\ \hline
\end{tabular}
\caption{Low-voltage requirements for one ATLAS FE-I3 front-end chip. 
Currents are given for both 1\% and 10\% occupancy.\label{tab:hvlv1}}
\end{center}
\end{table}

The required digital supply current depends on the detector occupancy.  
High occupancy results in higher current. The supply and its cables should
take this into account. In Table~\ref{tab:hvlv2} the requirements for one
readout controller chip MCC are listed.

\begin{table}[htbp]
\begin{center}
\begin{tabular}{|l|c|c|c|}\hline
\multicolumn{1}{|c|}{One MCC} & Voltage & Current  & Current limit \\ \hline
Digital VDD & 1.8-2.5 V & 120-150 mA & 170 mA \\ \hline
\end{tabular}
\caption{Supply requirement for one ATLAS MCC  chip.\label{tab:hvlv2}}
\end{center}
\end{table}

As each superlayer has 4 detectors and 4 FE-I3 chips plus one MCC chip 
sharing the digital supply with the front-end, we can sum up the total 
requirement per superlayer as shown in Table~\ref{tab:hvlv3}.

\begin{table}[htbp]
\begin{center}
\begin{tabular}{|l|c|c|c|c|}\hline
4 FE-I3 + 1 MCC  & Voltage range & Nominal voltage & Current  & Current limit \\ 
 Read-out driver &  &  &  &  \\ \hline
Analog (AVDD) & 1.6-2.0 V & 1.6 V& 20-280 mA & 310 mA \\ \hline
Digital (VDD) & 1.8-2.5 V & 2.0 V& 280-350 mA (1\% occ.) & 480 mA \\  
              &           &      & 360-430 mA (10\% occ.) &  \\ \hline
Monitor resolution& $<20$ mV&    & $<10$ mA & \\ \hline
\end{tabular}
\caption{Overall specification for a low-voltage supply segment for one 
superlayer consisting of 4 PixelChips FE-I3 and one MCC chip. Remote 
monitor should enable observation of the voltage and current.\label{tab:hvlv3}} 
\end{center}
\end{table}

A few comments are in order. The voltages may need adjustments in the
course of the lifetime of the system due to radiation effects. The
low-voltage supply may need to have remote-sense feedback to compensate
for the voltage drop. There must be a current limit which can be set
either locally or remotely; it would be an advantage if its value can be
set remotely as this will allow a more flexible system, capable of dealing
with changes due to, for instance, radiation damage.  The current limiting
can be either of a saturating type or a fold-back with latching action.
The latter requires some means of remote reset. Currents and voltages must
be monitored and results provided remotely with the accuracy given in
Table~\ref{tab:hvlv3}. A sample rate of the order of 1~Hz is sufficient. It
is important that each superlayer low-voltage supply can be switched
on/off individually (and remotely).

\subsubsection{High-voltage power supplies specifications}

A superlayer requires two high-voltage bias supplies, Vb1 and Vb2, with
remotely controlled voltage in the range 0 to $-120$~V.  Vb1 and Vb2
should be floating relative to each other within a superlayer with a
compliance range on the zero terminal of at least 2~V. The high voltage
bias supplies to a superlayer should be floating relative to any other
superlayer with a similar compliance range. As the bias voltage for
depleting the detector increases with radiation damage, it is an advantage
to segment the supply into two:  one for the detector pair closest to the
beam (Vb1) and one for the pair away from the beam (Vb2).  It is not
necessary to separate the ground between Vb1 and Vb2 at the superlayer.
The Vb zero line will be tied to the AVDD line. There will be passive RC
low-pass filtering close to the load.  Table~\ref{tab:hvlv4} summarises
the requirements.

\begin{table}[htbp]
\begin{center} 
\begin{tabular}{|l|c|c|c|} \hline
\multicolumn{1}{|c|}{4 detectors/2 voltages} & Voltage & Current  & 
Current limit \\ \hline
Vb1 & 0 to $-120$~GeV & $<$1~mA & 1 mA \\ \hline
Vb2 & 0 to $-120$~GeV & $<$1~mA & 1 mA \\ \hline
Monitor accuracy & $<$1~V & 1~$\mu$A $\sim 12$ bit res. & \\ \hline
Setting accuracy & $<$3~V $\sim 6$ bit res. & & \\ \hline
\end{tabular}
\caption{Specifications for the high voltage bias supplies for one superlayer 
consisting of four detectors powered by two independent 
voltages. The voltage and current should be monitored remotely with at 
least the specified accuracy. The voltage should be controllable from 
remote with a resolution of better than 3~V.\label{tab:hvlv4}} 
\end{center}
\end{table}

There must be a current-limit at the indicated value.  To increase
flexibility, it would be an advantage if its value can be remotely
adjusted. The limiting can be a simple saturating current-source type.
Currents and voltages must be monitored and results provided remotely.
Sample rate of the order of 1~Hz is sufficient. The high-voltage supply
has to be remotely controllable.  Remote-sense feedback on the wires to
the load is not required as the current-induced voltage drop is negligible
with respect to the required accuracy.

\subsubsection{Power budget}

Table~\ref{tab:hvlv5} gives the power dissipated in the front-end for a
worst case scenario where the occupancy is 10\% and the voltages are at a
maximum. For cooling design, the power from the radiation and the thermal
flux from the ambient will have to be added to this list.

\begin{table}[htbp]
\begin{center} 
\begin{tabular}{|l|c|c|c|} \hline
\multicolumn{1}{|c|}{One superlayer} & Voltage (V) & Current (A) & Power (W)\\ \hline
AVDD & 2.0 & 0.28 & 0.56 \\ \hline
VDD &  2.5 & 0.43 & 1.08  \\ \hline
Vbias1	&120	&0.001	&0.12  \\ \hline
Vbias2	& 120	&0.001	&0.12 \\ \hline\hline
Total per Superlayer& & & 1.88 \\ \hline \hline
 & & no of superlayers & \\ \hline \hline
Total per pocket	 & &  	5& 9.38 \\ \hline \hline
 & & no of pockets& \\ \hline \hline
Total per cryostat	& &  	3& 	28.13 \\ \hline
\end{tabular}
\caption{Power dissipated in the front-end electronics assuming 5 superlayers per 
pocket. Numbers are worst case values with 10\% occupancy and maximum 
voltages and currents.\label{tab:hvlv5}} 
\end{center}
\end{table}

\subsubsection{Low- and high-voltage channel count}

Table~\ref{tab:hvlv6} gives the number of channels assumed.
The final count may differ from this.

\begin{table}[htbp]
\begin{center}
\begin{tabular}{|c|c|c|c|c|}\hline
no. of channels& One superlayer	&One pocket&One cryostat&FP420 \\ \hline
 & 4 det.+ FE+1MCC & with 5 superlayers& with 3 pockets &with 4 cryostats \\ \hline
Low voltage&	2&	10&	30	&120 \\ \hline
High voltage&	2&	10&	30	&120 \\ \hline
\end{tabular}
\caption{Number of low- and high-voltage supplies channels.\label{tab:hvlv6}} 
\end{center}
\end{table}

\subsubsection{Temperature monitoring}

The temperature in the front-ends needs to be monitored. It will probably 
be necessary to have a probe on each superlayer. Temperature sensors of 
NTC type are known to be radiation tolerant and are used in other 
detectors at LHC. For instance LHCb (VELO repeater board, Low Voltage 
Card) uses NTC 103KT1608-1P from Semitec. The selection of the most 
appropriate device will require a later study. It is however sure that  
both excitation circuitry and an ADC to read the temperature values will 
be needed. It is an advantage if this excitation and measurement system 
can be integrated into the power supply crates.

\subsubsection{QUARTIC/GASTOF high- and low-voltage supplies}

The QUARTIC/GASTOF modules have different requirements than the 3D
detectors. The specifications per cryostat are for the moment rather
loosely set as described in Table~\ref{tab-hvlv-7}.

\begin{table}[htbp]
\caption{Preliminary specifications for the QUARTIC/GASTOF power 
supplies for one cryostat.
} 
\begin{center}
\begin{tabular}{|c|c|c|c|c|}
\hline
        & Number of channels	& Nominal voltage& Current & Current limit 
\\ \hline
High voltage&	4& 	$-3.5$~kV & TBD &	TBD \\ \hline
AVp12       &	1&	12~V      & TBD &	TBD \\ \hline
AVm12       &	1&	$-12$~V   & TBD &	TBD \\ \hline
DVp5        &   1&      5~V       & TBD &       TBD \\ \hline
DVp3.3      &   1&      $3.3$~V   & TBD &       TBD \\ \hline
\end{tabular}
\end{center}
\label{tab-hvlv-7}
\end{table}

\subsubsection{Discussion of the solutions considered}

All solutions discussed in the following are based on commercially 
available modules. Three conceptually different approaches have been 
studied.

\begin{enumerate}
\item Power supplies located in the tunnel next to the FP420 cryostats and
stowed underneath the adjacent magnets. The advantage is the low cable
cost combined with options for extensive remote control and monitoring.  
The major drawback is the sensitivity to radiation, combined with 
difficult access for maintenance. A study of the radiation 
tolerance~\cite{Thijs} of a solution based on CAEN supplies (see below) 
concludes that there may be 0.1 SEU (Single Event Upsets)/module/day if 
the modules are placed in the tunnel close to the cryostat. This will be 
the case from day-one of operation. This means that there will be several 
SEUs per day, in addition to the damage due to dose gradually accumulating 
over time (tens of Grays per year).
\item Power supplies in the alcove areas RR17/13 for ATLAS and RR57/53 for 
CMS. The expected level of radiation here is 0.05-0.36 Gy/year at full LHC 
luminosity. This solution is similar to that adopted for the TOTEM Roman 
Pots.
\item All critical power supply electronics in the counting room and only 
very simple linear, radiation-hard regulators in the tunnel next to the 
cryostat.
\end{enumerate}

Except for the CAEN version of solution 1. (see below), the high-voltage
supplies are always assumed to be in the counting room, which is
advantageous because radiation is thus no more a concern. The wires for
high voltage can have a small cross section due to the small current
($<1$~mA) and need no remote sense. The high-voltage cables must be well
shielded and with a noise filter at the detector.

\begin{figure}[htb]
\begin{center}
\includegraphics[width=0.6\textwidth]{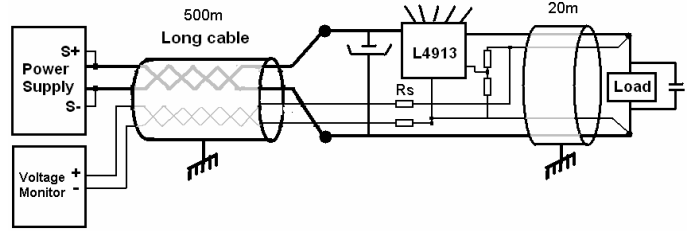}
\caption{Block diagram showing the principle of using a local 
radiation-hard linear regulator. Here for a positive voltage and the option of 
remote monitor of the load voltage via isolation resistors $R_s$.}  
\label{fig:hvlv0} \end{center}
\end{figure}

Solutions 2. and 3. with 200~m long (or longer) low-voltage cables require
local regulators next to the load. Without them it will not be possible to
maintain a stable load voltage. Cables with a length of 200-500~m would
have to have large cross section in order to limit the voltage drop to the
level required (roughly $<200$~mV). Remote sensing, the classical
way of overcoming this, is not effective due to the long delay in the
cable.  Linear regulators, albeit with much shorter cables, are used
in many LHC detector systems, such as the TOTEM Roman Pots and the LHCb
Vertex Locator (VELO). A pair of radiation-hard linear regulators have 
been
developed in the framework of RD-49. The regulators are LHC4913 for
positive voltages (SCEM:
08.57.56.011.7; 1.23~V to 9~V at 3~A) and for negative voltages LHC7913-4
(SCEM: 08.57.56.111.4;  $-1.2$~V to $-7$~V at 3~A). In other LHC
experiments using a linear regulator, a separate monitoring system for the
voltage is exploited, which has to be radiation hard. For instance, in the
CMS central tracker a system of FEC, DOHM and CCUs is used. The main issue
with this solution is that it is highly specialised for these applications
and not easily adapted to the FP420 requirements. Added to this is the
difficulty of finding the components. As an alternative solution we
suggest the following setup, which allows remote monitoring of the load
voltage (see Fig.~\ref{fig:hvlv0}). The voltage at the load is fed back to the
location of the power source via pairs in the same cable as the power
source. We propose to put isolation resistors in series with the sense
wires. As long as the ADCs at the acquisition end have high impedance and
low leakage and bias current, the average current and thus the voltage
drop across the sense resistors will be small. This means that the average
voltage measured at the acquisition end will equal the average voltage at
the load.

\subsubsubsection*{Solution 1: supplies next to the cryostat}

For this configuration we have one proposal from CAEN and two (A and B)  
from Wiener; all solutions still need refinements. The CAEN solution
envisages putting both the low- and the high-voltage supplies in the
tunnel; the Wiener solutions foresee only the low-voltage supplies in the
tunnel.

\bigskip

{\bf CAEN}

\begin{figure}[htb] 
\begin{center}
\includegraphics[angle=270,width=0.6\textwidth]{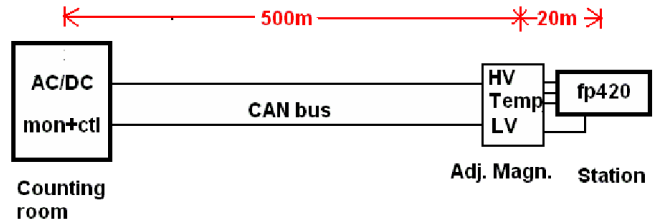}
\caption{Solution with all supplies in the tunnel, adjacent to the cryostat.
``Station" indicates the FP420 cryostat and ``Adj. Magn." the magnets 
adjacent to the FP420 cryostat.}
\label{fig:hvlv1} 
\end{center} 
\end{figure}

\begin{figure}[htb]
\begin{center}
\includegraphics[width=0.6\textwidth]{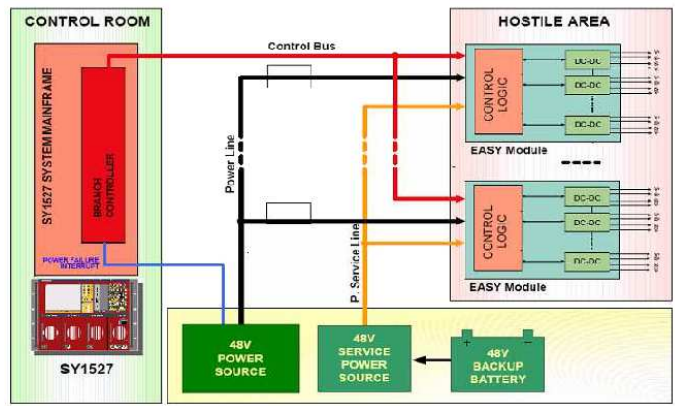}
\caption{Block diagram of the CAEN setup.}  
\label{fig:hvlv2}
\end{center}
\end{figure}

\begin{figure}[htb]
\begin{center}
\includegraphics[width=0.7\textwidth]{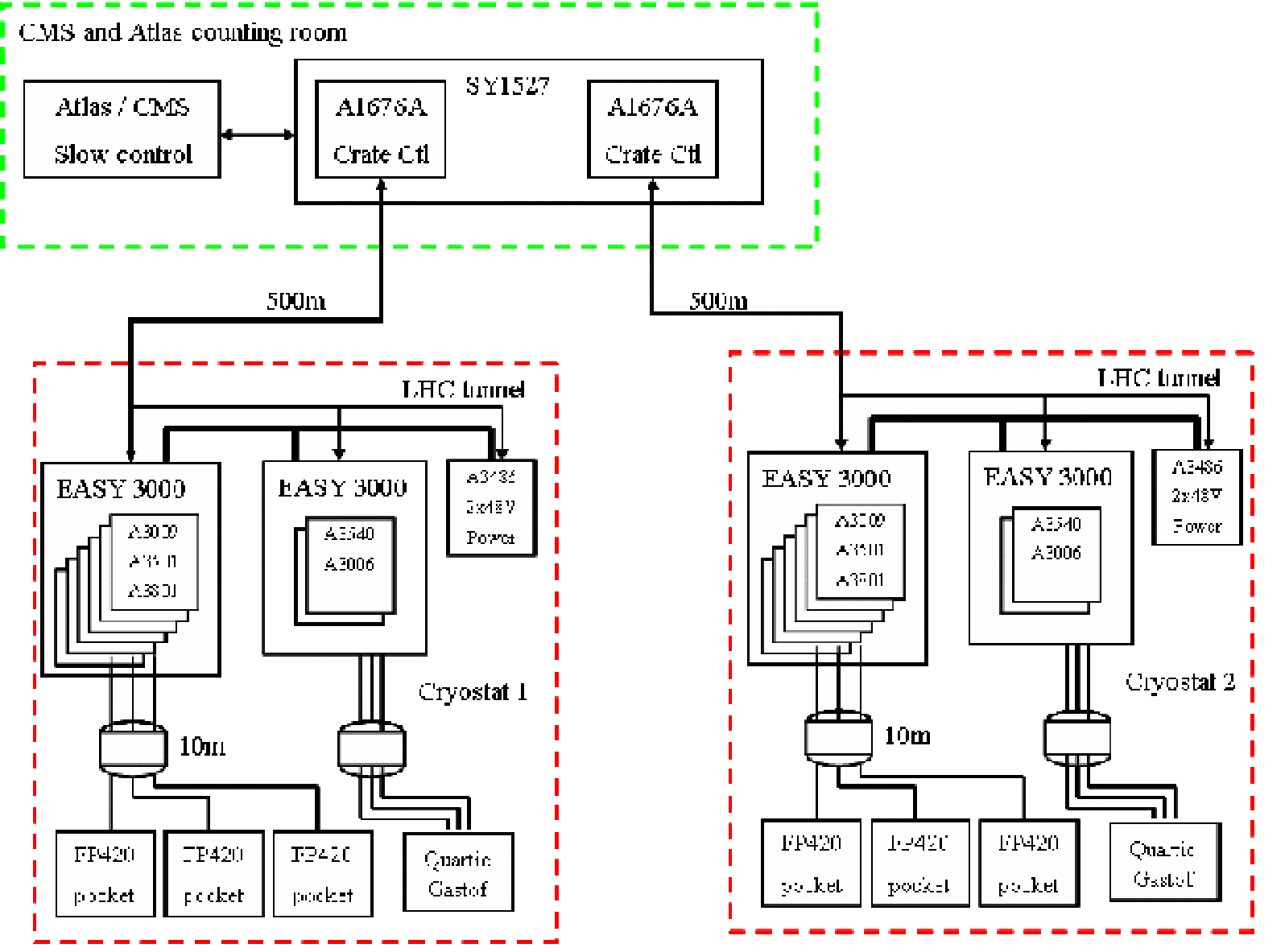}
\caption{Diagram of the solution suggested by CAEN. The number of pockets 
assumed is 3 per cryostat. Shown are also the system for temperature 
monitoring (A3801) and the supplies for QUARTIC/GASTOF detectors (A3540 
and A3006).}  
\label{fig:hvlv3} 
\end{center}
\end{figure}

The schematic layout is shown in Figs.~\ref{fig:hvlv1}-\ref{fig:hvlv3}.

The A3006 low-voltage supply is adjustable in the 4 to 16V range and may
thus not be able to cover all the way down to 3.3~V, necessary for QUARTIC
and GASTOF, without additional modifications. 

Details of the degree of radiation tolerance of the CAEN modules are given
in~\cite{fp420-note-08-001}. Modules A3009, A3486, A3540, A3801 have been
tested to work up to doses of about 140-150 Gy. The first three are
radiation-certified for ATLAS. Module A3501 has never been tested, but its
radiation behaviour should be the same as that of A3540, which has been
radiation-certified for ATLAS.

The CAEN standard-module communication is not guaranteed to work over a 
500~m cable. The CAEN CAN bus is operated at 250~kbit/s. A speed of 
250~kbit/s
has been verified to work over cable SCEM 04.21.52.140.4, without
affecting signal integrity, but, due to the cable delay, the timing
requirements of the CAN bus arbitration protocol are violated. Lowering
the bit rate to 125~kbit/s would make the 500~m cable meet the
specifications of signal integrity and arbitration protocol. CAEN has
offered, at an additional cost, to modify the modules such that they 
operate at 125~kbit/s, but the modules will then become non-standard and
will no longer be exchangeable with those used elsewhere at CERN.

In addition, the CAEN module A3501 is designed for 0 to $-100$~V, whereas
$-120$~V may be necessary, as specified above. CAEN is able to modify the
modules at an additional cost.

\bigskip
{\bf WIENER}
\bigskip

{\bf Wiener, Solution A, MPOD LV next to cryostat, MPOD HV in counting 
room}

\begin{figure}[htbp]
\begin{center}
\includegraphics[width=0.6\textwidth]{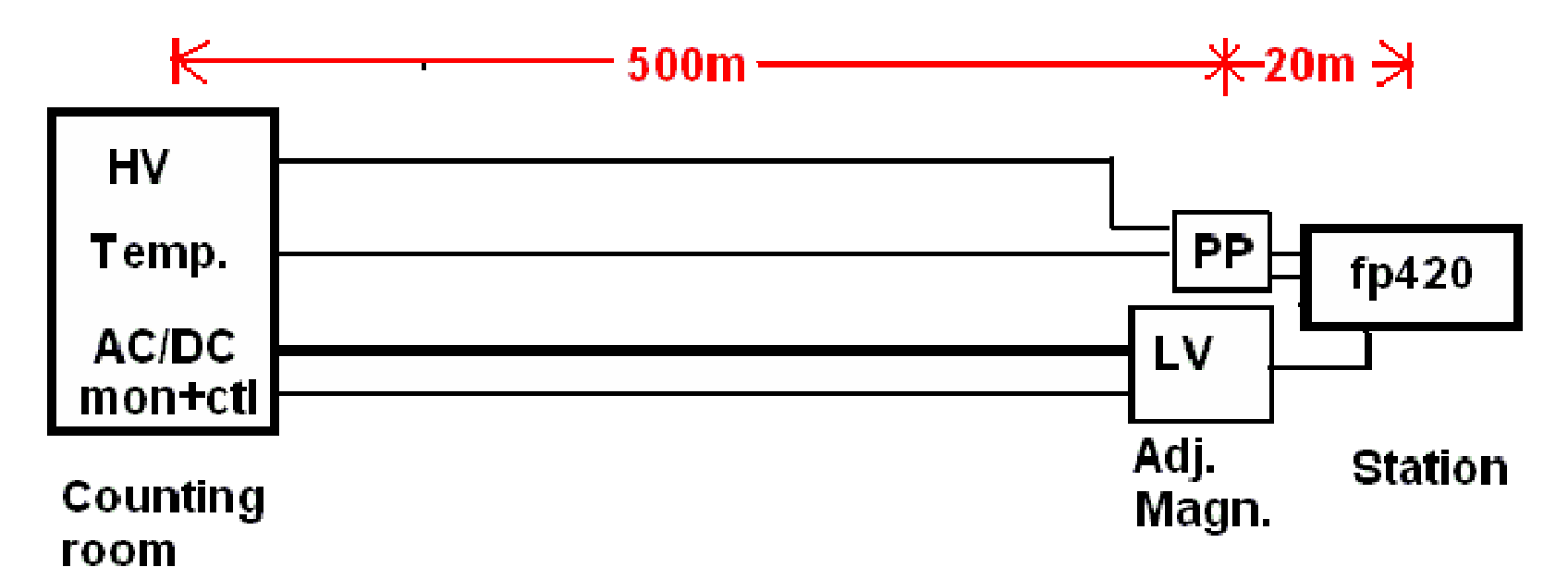}
\caption{Overview of the Wiener Solution 1, with MPOD LV next to cryostat,  
MPOD HV in the counting room. ``Station" indicates the FP420 cryostat 
and ``Adj. Magn." the magnets adjacent to the FP420 cryostat. ``PP" is a 
patch panel.}  
\label{fig:hvlv4}
\end{center}
\end{figure}

\begin{figure}[htbp]
\begin{center}
\includegraphics[width=0.8\textwidth]{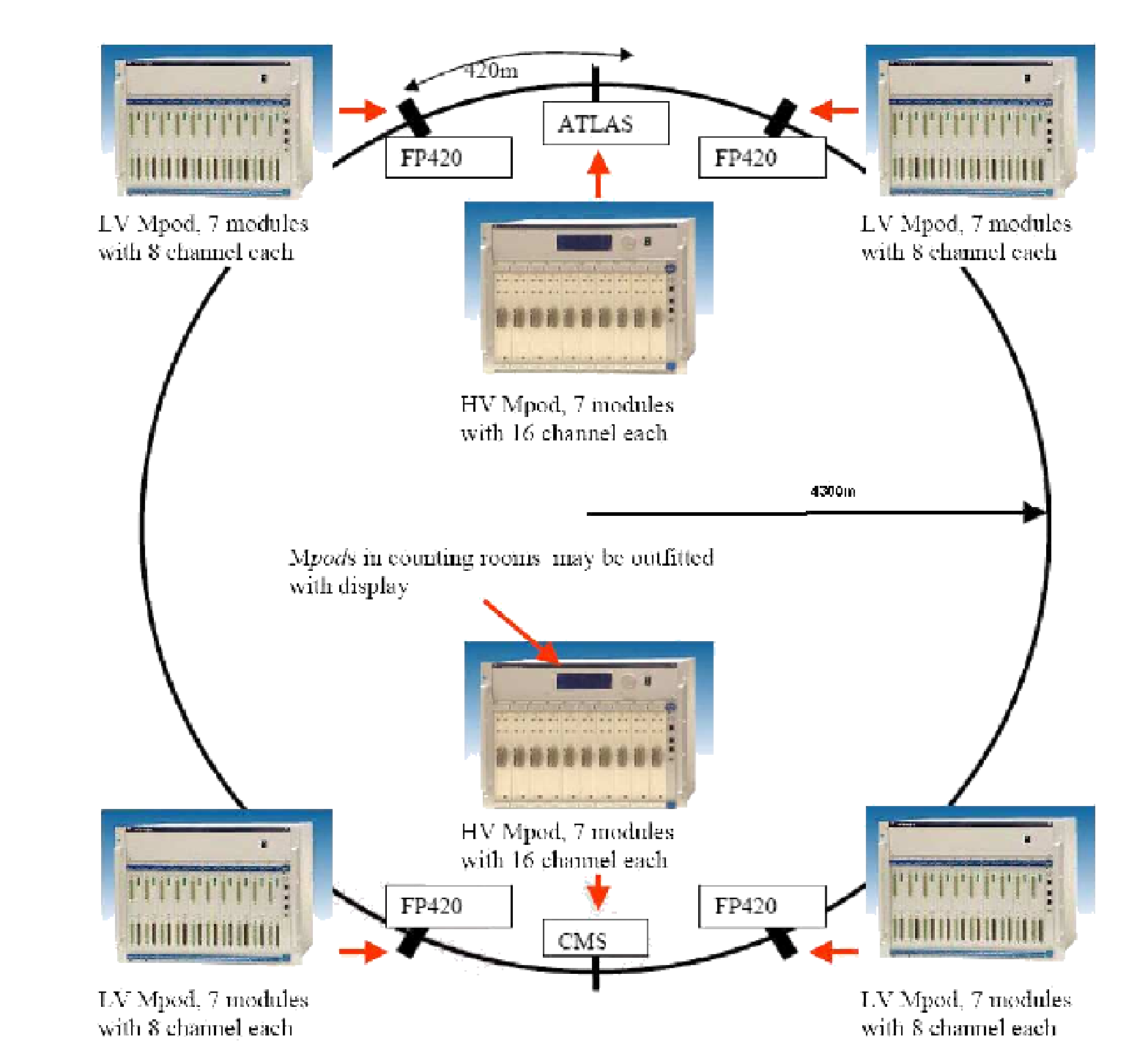}
\caption{Wiener solution with LV supplies in the tunnel and HV supplies 
in the counting room, delivering the bias via 500~m cables. The MPOD will 
require custom $-120$~V modules.}
\label{fig:hvlv5}
\end{center}
\end{figure}

Figures~\ref{fig:hvlv4}-\ref{fig:hvlv5} show the schematic layout of the
proposed system.  This solution, based on Wiener MPOD modules, has only
the low-voltage part in the tunnel. One crate at each location will be
needed for the 3D supplies. The high voltage is supplied from MPOD modules
in the counting room via a 500~m cable.  No auxiliary power crate is
needed in the tunnel, different from the CAEN solution. The MPOD modules
have never been radiation tested.  According to the company they are made
in a way which is likely to qualify them to the level we require. It will
however be necessary to test the modules in both proton and gamma fields.

\bigskip

{\bf Wiener, Solution B, Maraton LV crates next to cryostat, MPOD HV in 
counting room}

\bigskip

\begin{figure}[htbp]
\begin{center}
\includegraphics[width=0.6\textwidth]{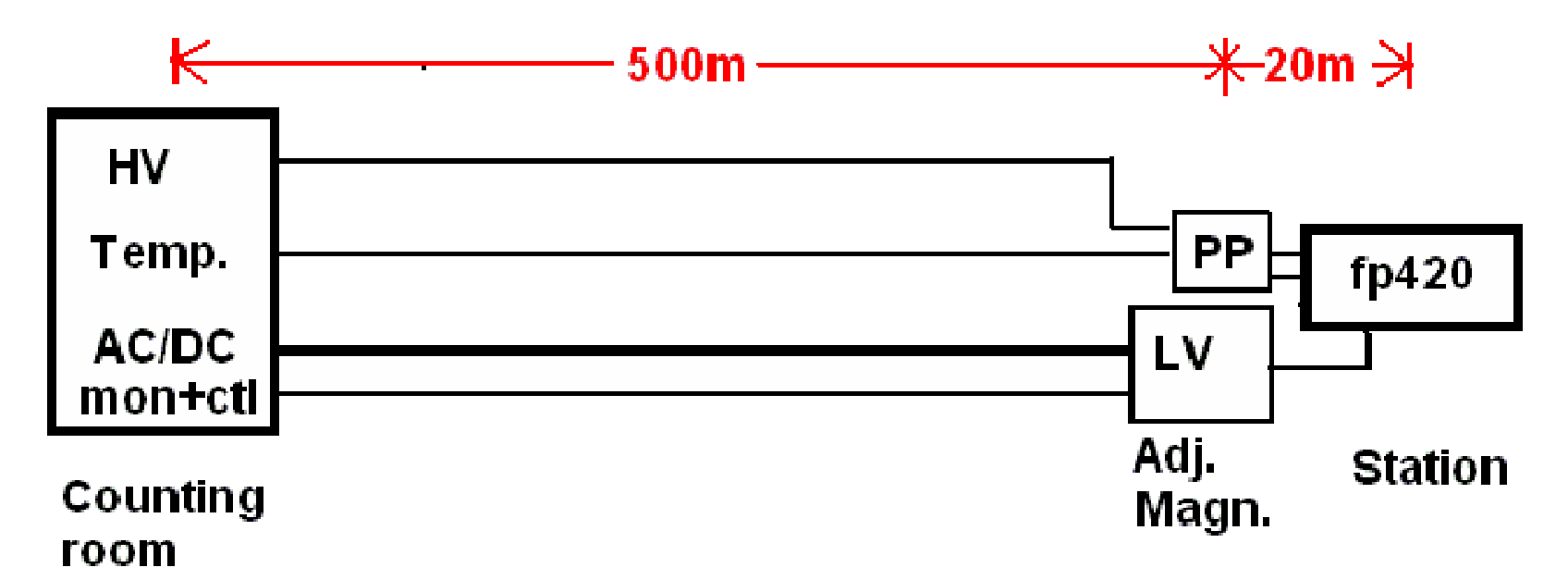}
\caption{Overview of the Wiener Solution 2, based on the Maraton modules 
next to the cryostat. ``Station" indicates the FP420 cryostat and 
``Adj. Magn." the magnets adjacent to the FP420 cryostat. ``PP" is a patch 
panel.}  
\label{fig:hvlv6}
\end{center} 
\end{figure}

\begin{figure}[htbp]
\begin{center}
\includegraphics[width=0.8\textwidth]{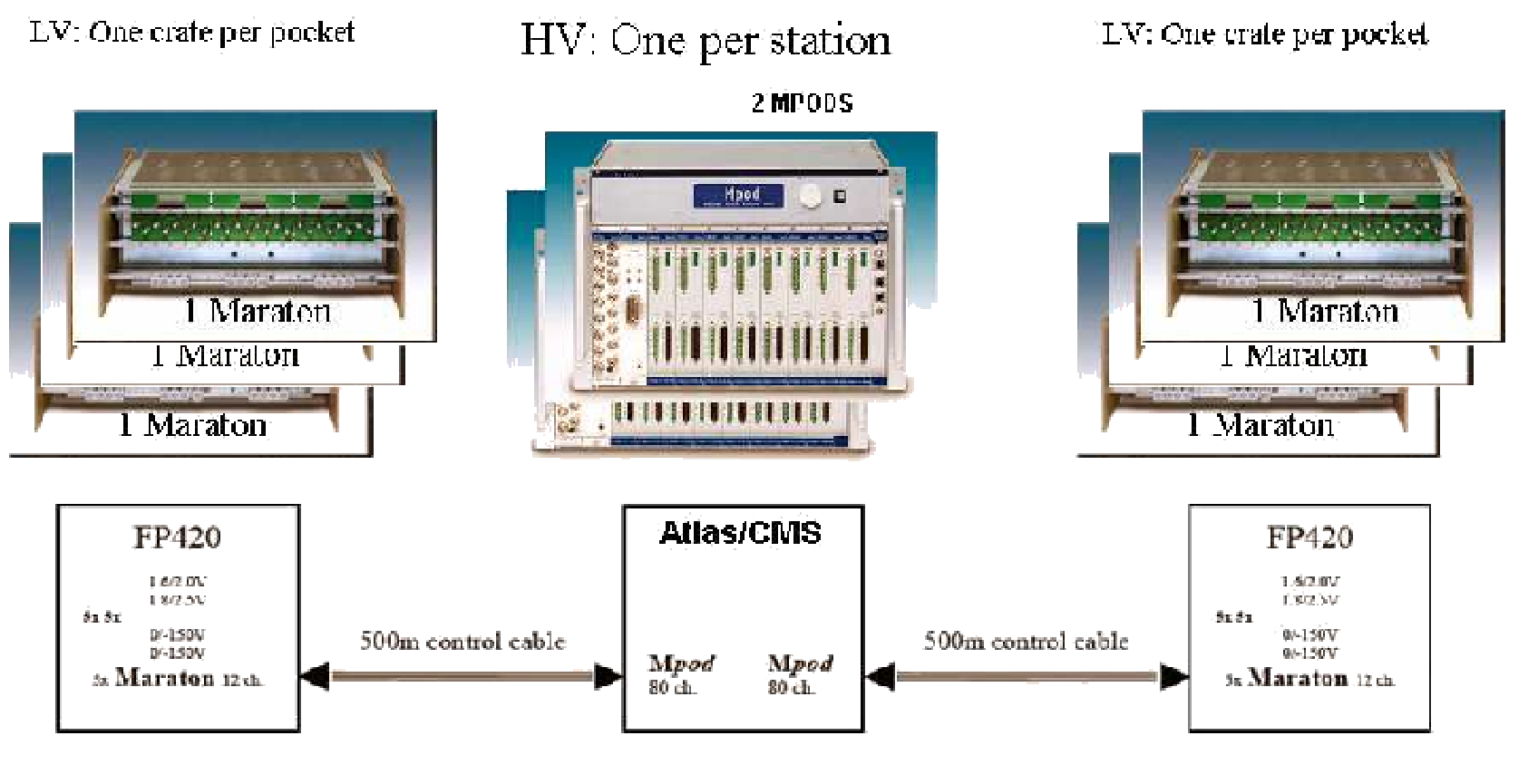}
\caption{Low-voltage Wiener Maraton supplies in the tunnel. High-voltage 
MPOD type supplies are located in the counting room. The Wiener Maraton 
system is qualified for the radiation environment expected in the tunnel 
under the magnets near the FP420 cryostat. The illustration shows the 
setup for either ATLAS or CMS. ``Station" indicates the FP420 cryostat.}
\label{fig:hvlv7} 
\end{center}
\end{figure}

Figures~\ref{fig:hvlv6}-\ref{fig:hvlv7} show the schematic layout of the 
proposed system.

This solution has the low-voltage supplies housed in Wiener Maraton crates
in the tunnel next to the cryostat. One crate will be needed per pocket.  
The high voltage is supplied over a 500~m long cable by an MPOD module in
the counting room. This solution requires a customization of Wiener
Maraton low-voltage modules in order to optimise it for low currents. The
monitoring of the Wiener Maraton is with individual twisted pairs from
each channel. The ADCs for this will need to be in a radiation-free
environment, i.e. in the counting room. The length of the monitor and
control cable of 500~m is beyond the specification in the data sheet, so
this length of cable needs further testing.

The Wiener Maraton modules have been radiation qualified to 722~Gy, and $8
\times 10^{12}$~n/cm$^2$. Their good radiation tolerance is partly
obtained by moving the digital part of the control and monitoring
circuitry away from the radiation zone. This results however in less
flexibility compared to the CAEN and the Wiener MPOD solutions. So in the
Wiener Maraton system the output voltage and current limit cannot be
adjusted from remote, and monitoring is via analogue differential wires.
One pair is required per measurement value (voltage and current) resulting
in a large amount of monitor wires. The ADCs for this will need to be in a
low-radiation environment, i.e. in the counting room. For improved
radiation tolerance, mains supply AC to DC conversion is also done in the
counting room.

The advantage of this solution is that it will fit the QUARTIC/GASTOF 
requirements without much modification. The disadvantages are the exposure 
to radiation and difficult access for maintenance. In addition, the Wiener 
Maraton only allows the voltage setting and current limits to be adjusted 
manually using potentiometers on the modules. No remote tuning is 
possible. 

\subsubsubsection*{Solution 2: low-voltage supplies in alcoves,  
high-voltage supplies and temperature monitor in counting room, local 
regulators at load}

\begin{figure}[htb]
\begin{center}
\includegraphics[width=0.6\textwidth]{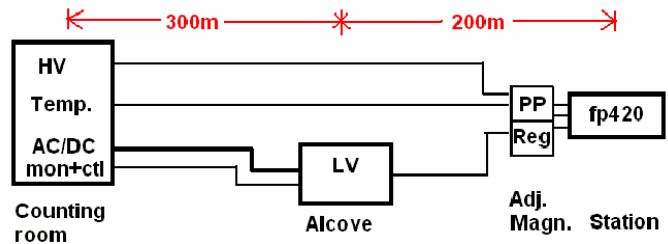}
\caption{Low voltage in alcoves, rest in counting room using 200~m cables 
from alcove to cryostat. ``Station" indicates the FP420 cryostat and 
``Adj. Magn." the magnets adjacent to the FP420 cryostat. ``PP" is a patch 
panel. ``Reg" are linear regulators next to the load.} 
\label{fig:hvlv8} 
\end{center} 
\end{figure}

\begin{figure}[htbp]
\begin{center}
\includegraphics[angle=90,width=0.8\textwidth]{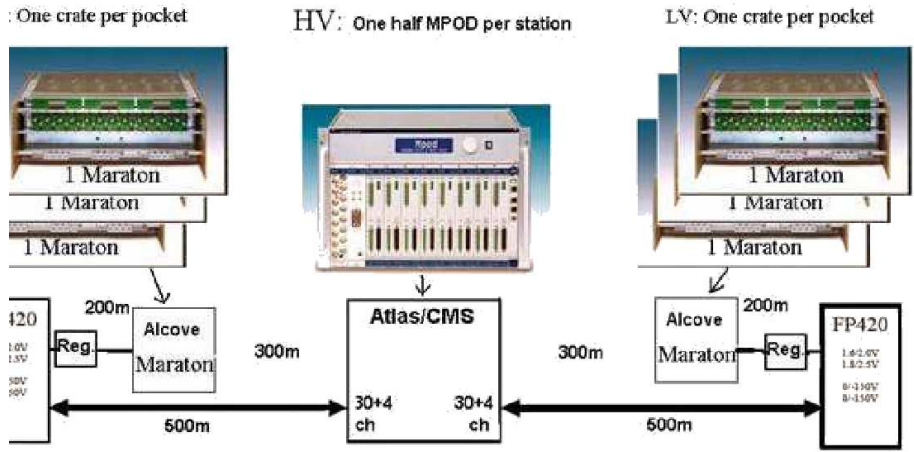}
\caption{Low-voltage Wiener Maraton supplies in the alcoves. High-voltage 
MPOD supplies in the counting room. The Wiener Maraton system is 
qualified for the expected radiation environment within a large 
margin. ``Station" indicates cryostat.} 
\label{fig:hvlv9} 
\end{center}
\end{figure}

This solution (Figs.~\ref{fig:hvlv8}-\ref{fig:hvlv9}) is based on the use
of Wiener Maraton low-voltage supplies placed in the alcoves. CAEN also
has radiation-tolerant power supplies, which could be considered. The
Wiener Maraton modules are used for the TOTEM Roman Pot detectors and are
also placed in the alcoves.

This solution requires a customization of Wiener Maraton low-voltage
modules in order to optimise it for low currents. The length of the
monitor cable of 300~m is beyond the specification in the data sheet, so
this length of cable also needs further testing, as already 
discussed. A linear voltage regulator is placed next to the front-end to 
ensure the voltage stability at the load.

\subsubsubsection*{Solution 3: low- and high-voltage supplies and temperature 
monitor in counting room, local regulators at load}

\begin{figure}[htbp]
\begin{center}
\includegraphics[width=0.6\textwidth]{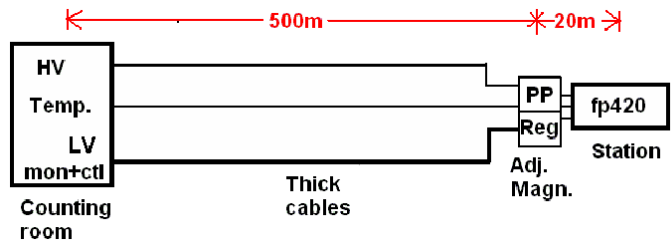}
\caption{Low- and high-voltage supplies in counting room using 500~m long 
cables to a patch panel with regulators next to the cryostat. ``Station" 
indicates the FP420 cryostat and ``Adj. Magn." the magnets adjacent to the 
FP420 cryostat. ``PP" is a patch panel. ``Reg" indicates linear regulators 
next to the load.}
\label{fig:hvlv10}
\end{center}
\end{figure}

This solution is illustrated in Fig.~\ref{fig:hvlv10}. The advantage is
that the power supplies are not exposed to radiation. This widens the
number of power supply candidates significantly, lowers their cost and
makes the system simpler to maintain.  The major drawback is the cable
cost and the need for local regulators.

The low voltage needs to be regulated at the load as discussed earlier.
The absolute maximum cable drop in the low-voltage long cables is 5.7~V.
At this limit, 6 LV cables per cryostat will be necessary. Hardware tests
will have to be done in order to determine if a voltage drop of 5.7~V is
tolerable.

\subsubsection{Summary of solutions}

\begin{figure}[htbp]
\begin{center}
\includegraphics[width=1.\textwidth]{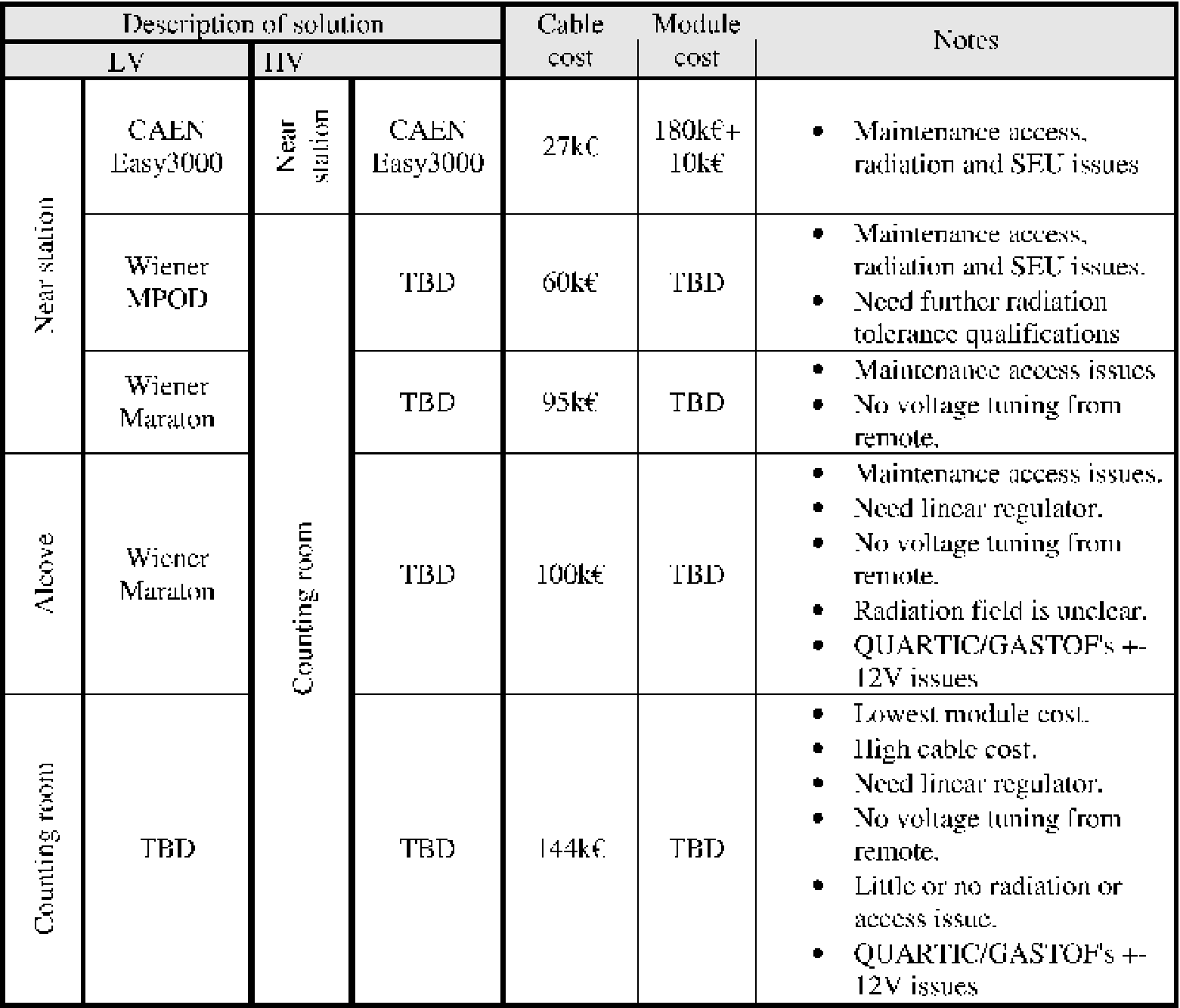}
\caption{Summary of cable and module cost for various solutions covering 
both ATLAS and CMS. Custom modules with linear regulator are estimated to 
cost a total of 6~kEU.  The cost of cable pulling and connector mounting 
is not included. ``TBD" means that no particular manufacturer stands out 
as the best choice based on the investigations done so far. 
``QUARTIC/GASTOF +-12V issues" refers to the problem that the LHC4713/ 
LHC7913 regulators will not be suitable to regulate $\pm 12$~V presumably 
required for QUARTIC/GASTOF. Other solutions will have to be found for 
that case. ``Station" indicates the FP420 cryostat.} 
\label{fig:hvlv11}
\end{center}
\end{figure}

Figure~\ref{fig:hvlv11} summarises the solutions outlined in this 
section.

\nopagebreak

\subsection{Readout and infrastructure at the host experiment}
\label{sec:silicon_readout}

\subsubsection{CMS and ATLAS Specific issues}

\begin{figure}[htb]
\centerline{
\epsfig{file=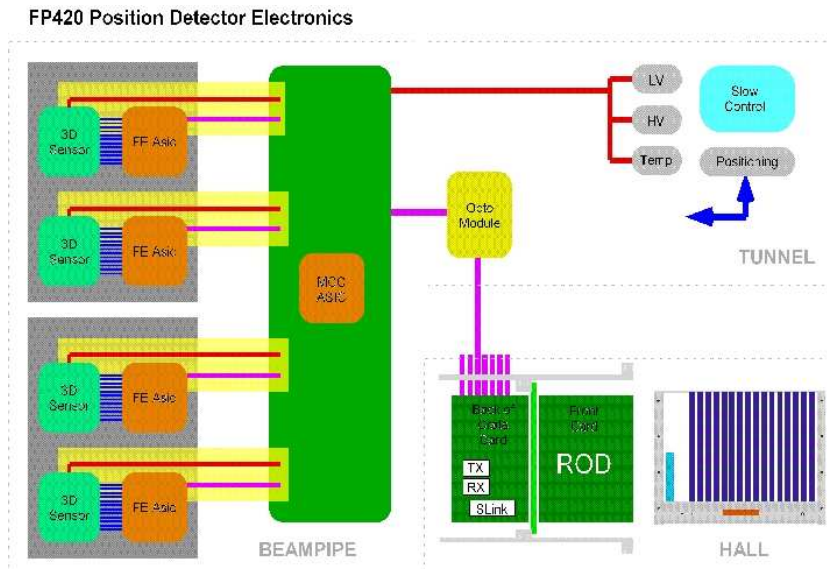,width=11cm}}
\caption{Layout of the readout and DAQ system.}
\label{fig:readout1}
\end{figure}

Readout installations at ATLAS and CMS necessarily differ, but will be
based on the same parts, which are essentially single-crate versions of
the ATLAS silicon readout. Refer to Figure~\ref{fig:readout1}. Fibre connections
from the tunnel arrive at optomodules fitted to a back of crate BOC
card. The BOC provides timing adjustments and passes the data to the
ROD where event segments are combined and DSPs can perform monitoring.
Event data are passed back through the BOC to an SLINK transmitter and
onward to the ATLAS standard ROS. Integration into CMS will require
some modification of the ROD firmware so that the output format can be
interpreted as a CMS format event stream. CMS experts describe this as
``relatively straightforward''. 
DCS and DSS requirements have not been studied, but again it is anticipated that
these will follow the example of the existing experiments.

\subsubsection{Tracker readout and downstream data acquisition}

The 3D silicon assemblies and their readout take advantage
of the significant design investment made by the ATLAS pixel groups. The bump-bonded detector assembly mimics an
ATLAS pixel element and the downstream readout of FP420 can therefore
be based very closely on the equivalent parts of the ATLAS pixel
system. Each superlayer has independent connections to a support card
situated within the support structure. LV and HV are supplied from
commercial units positioned nearby, as described in Section~\ref{sec:hv-lv}. 
Fibre optic data links to and from the central detector areas terminate on
the support cards. Each station has its own link back to a ROD card
that drives each arm of FP420. The ROD crates are easily integrated
into the ATLAS readout. Integration into CMS should require minimal
work.

\subsection{Thermal Design}
\label{sec:silicon_thermal}

\subsubsection{Overview}

Running detectors at -20{\textordmasculine}C implies that if they are
not shielded from the tunnel environment they will ice-up. In order
to prevent this from happening it is crucial to isolate these detectors
from the LHC tunnel environment. This can be achieved in various ways.
One possibility is to use a foam insulation surrounding the detectors,
another is to purge dry-nitrogen gas within the detectors to isolate
them from the air in the tunnel. A third option is to enclose the
detector block within a box. Then there are again two options, either
purge the box with dry nitrogen or keep the detector box under vacuum.

\begin{enumerate}
\item Foam insulation is not practical within the limited available space. It
would render the detectors themselves inaccessible (foam will have to
be applied between and around the detector planes) and would not
absolutely guarantee that no icing will take place at any point. This
method is considered to be cumbersome and potentially harmful to the
detectors with no guarantee it will work.
\item Purging dry nitrogen gas during operation is a viable option from an
engineering point of view. However purging nitrogen gas continuously
into the LHC tunnel is not allowed by CERN.
\item Maintaining a water vapour free environment around the detectors by
enclosing them in a box, filled with dry nitrogen is an option. It
would however compromise the cooling of the detectors themselves due to
natural convection inside the box. It would require more heat to be
pumped away compared to cooling the detectors in vacuum, which in itself
is not directly considered to be a show-stopper. There is however a
potential for icing-up of the enclosure due to the internal
convection, which could be solved by applying heaters to the
outside of the enclosure. The box would have to be gas tight, in order
not to leak nitrogen into the tunnel. The convection of the nitrogen
gas will yield larger thermal gradients over the detector plane
compared to vacuum and potentially cause an asymmetric temperature
distribution that could affect the measurements.
\item If the enclosure is kept under vacuum all drawbacks of option 3
disappear. Maintaining a vacuum around cold detectors is standard
practice in laboratories and the technology required to maintain this
vacuum is bulk-standard, off the shelf. It will minimise the cooling
requirements and it will minimise the thermal gradients. Maintaining a
gas-tight enclosure around the detectors (option 3) has the same level
of complexity as a vacuum environment.
\end{enumerate}

Option 1 is not considered to be viable and option 2 is not allowed.
Because of the advantages attached to maintaining a vacuum around the
cold detectors, compared to option 3, it was decided to go for option 4.

\subsubsection{Thermal Requirements}

The thermal requirements for the detectors are as follows;

\begin{figure}
\centering
\epsfig{file=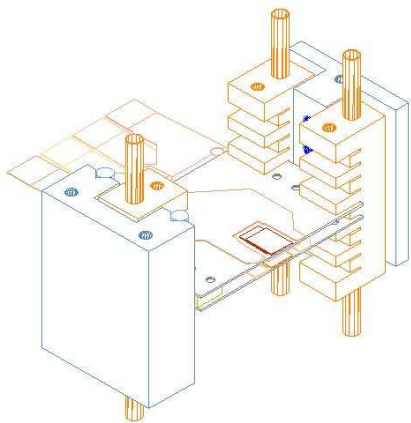,width=12cm}
\vspace{-1cm}
\caption{Front-end of a superlayer showing the cooling block arrangement.}
\label{fig:heat1}
\end{figure}

\begin{itemize}
\item Lowest allowed operating temperature is -22{\textordmasculine}C. 
\item Nominal operating temperature is -20{\textordmasculine}C. 
\item Highest allowed operating temperature is -18{\textordmasculine}C.
\item Required thermal stability during operation is better than 0.5{\textordmasculine}C per 24 hours. 
\item The maximum allowed thermal gradient over an individual detector (chip) is 0.5{\textordmasculine}C.
\item All detectors within a test setup will be operating within 2{\textordmasculine}C of each other. 
That is, the temperature of the hottest detector at any given time is no more than 2{\textordmasculine}C 
higher than that of the coldest detector.
\item The vacuum pipe, enclosing the LHC beam, will be at 30{\textdegree}C $\pm$ 5{\textdegree}C.
\item The extreme temperatures to which the detectors will be exposed when non operating 
will be the ambient temperature in the LHC tunnel and that during transport. These are expected to be in the
range of 10{\textdegree}C to 40{\textdegree}C. FP420 will not be part
of the overall beam line tube bake out.
\end{itemize}
Figure~\ref{fig:heat1} illustrates how heat will be transferred from the
superplanes to copper blocks in the station support.

\subsubsection{Heat Loads}

\begin{figure}[htpb]
\centerline{\epsfig{file=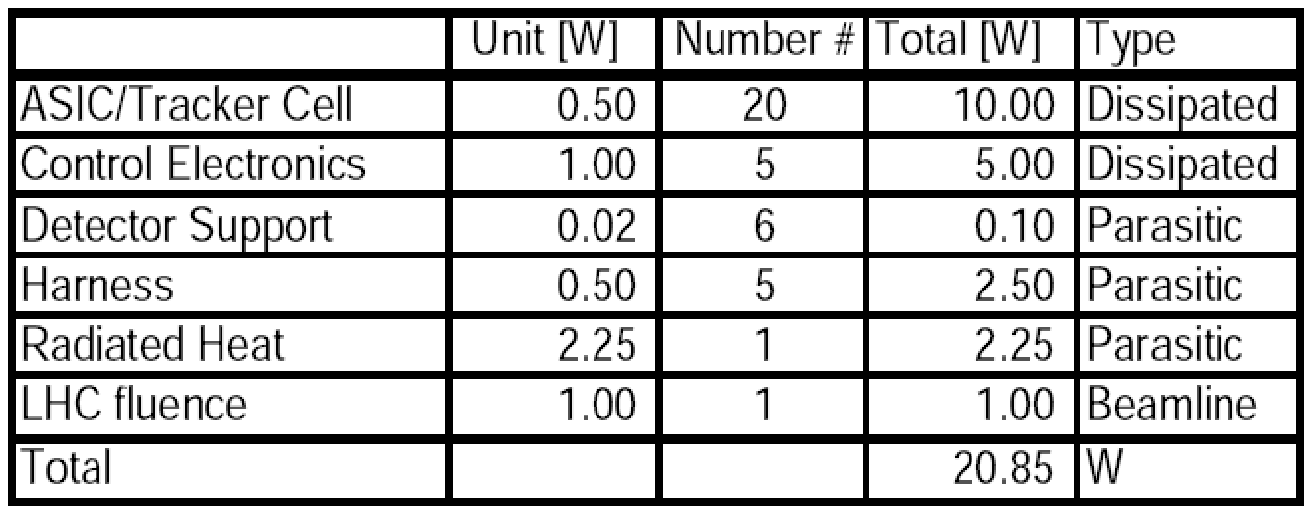,width=11cm}}
\caption{The heat loads.}
\label{fig:heat2}
\end{figure}

Heat is dissipated inside the tracker cell (mostly in the ASIC
underneath) and the control card. Other than that, heat enters the
detector block via thermal radiation (enclosure is sitting at 30{\textdegree}C) 
and parasitic conductive heat loads via the harness and the supports. Analyses 
have been carried out to size these heat loads. The results are listed in the 
table~\ref{fig:heat2}.

The dissipated heat loads are conservative estimates and make up 75\% of
all dissipated heat. The parasitic heat loads are best estimates at the
time of writing this document.
It would be prudent to put a safety factor of 2 on these numbers to quantify the 
required cooling power. Therefore the recommended cooling power for the cold sink 
should be better than 42 W. In the next sections the temperature of the cold sink
is determined.

\subsubsection{Heat Flow}

The heat flow/gradient is determined by the dissipated heat together
with the thermal resistance between the source and the cold sink. The
heat flow has been pictured schematically below in Figure~\ref{fig:heat3}. 
When the overall heat flow is known, together with the thermal resistance of
the network, gradients and overall temperature differences can be
determined.

\begin{figure}
\centerline{\epsfig{file=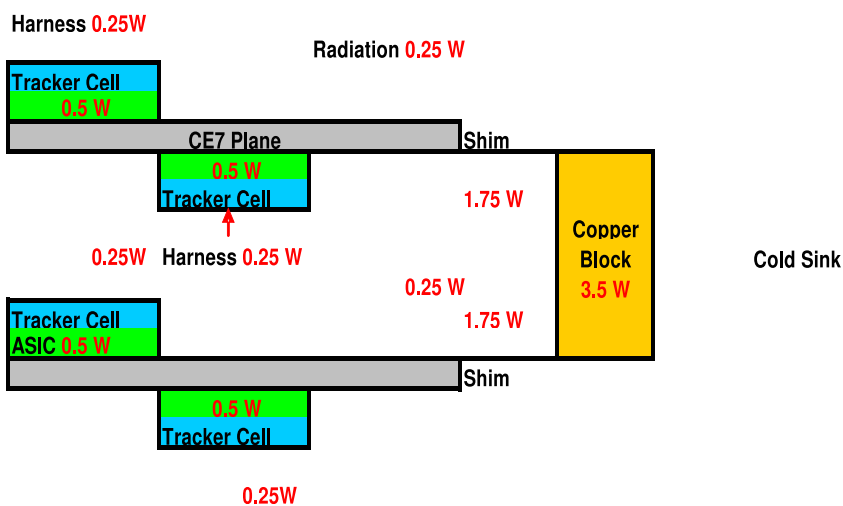,width=12cm}}
\caption{Heat flow surrounding the tracker cells (local thermal network).}
\label{fig:heat3}
\end{figure}

\begin{figure}
\centerline{\epsfig{file=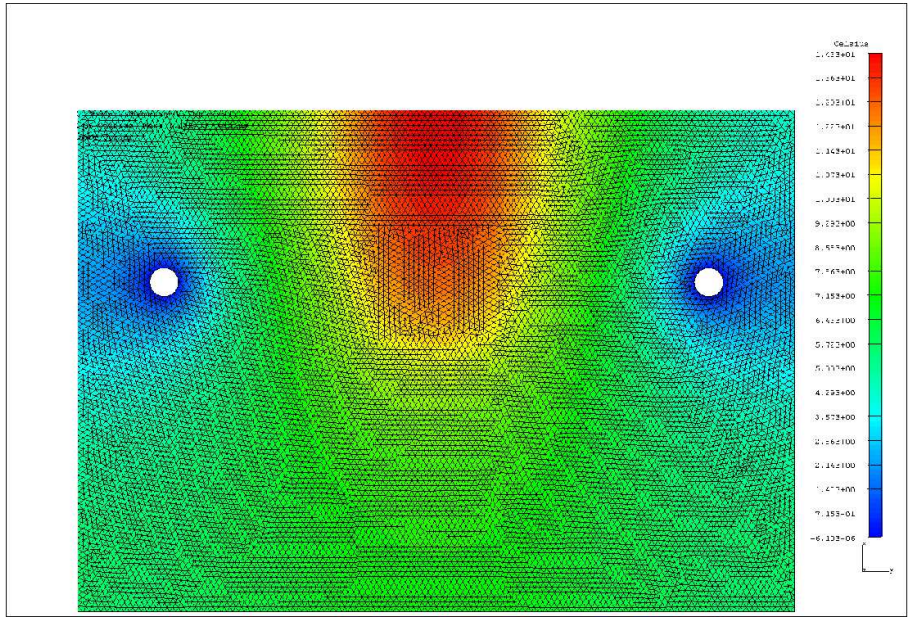,width=12cm}}
\caption{CE7 detector plane gradient (worst-case configuration).}
\label{fig:heat4}
\end{figure}

The CE7 (70/30 Si/Al) plane with two tracker cells has been analysed in some detail.
Simplified thermal models were used to assess the effective thermal
conductance between the edge and the tracker cell. In Figs.~\ref{fig:heat3} 
and~\ref{fig:heat4}  the overall temperature distribution for an
artificial load (1~W, with boundary at 0{\textdegree}C) is given. The
resulting thermal resistance towards the edges is 1/14.3 = 0.07 W/K,
which assumes heat sinks on either side of the tracker planes.

\begin{figure}
\centerline{\epsfig{file=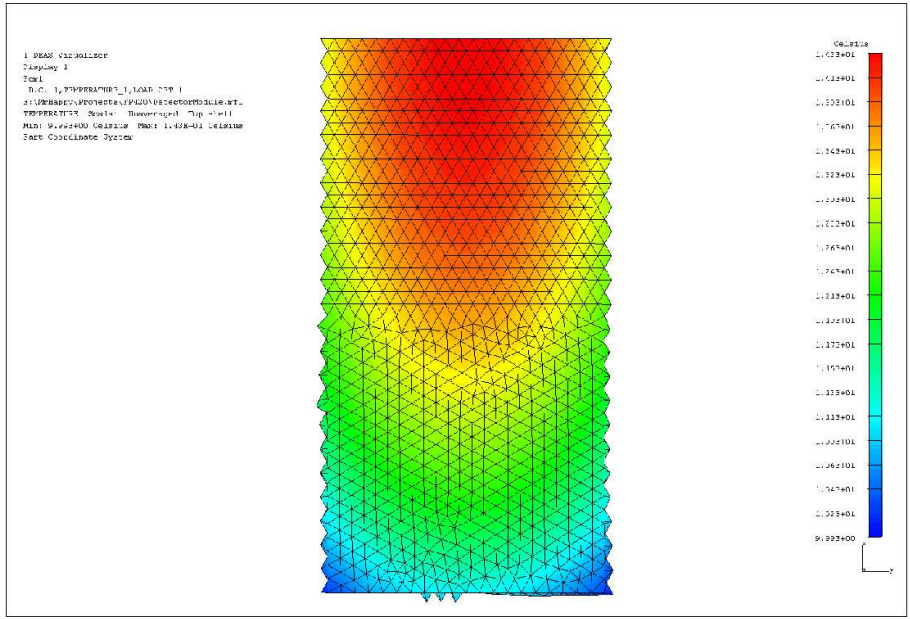,width=12cm}}
\caption{Gradient in the CE7 plane underneath the tracker cell (4.3{\textdegree}C).}
\label{fig:heat5}
\end{figure}

Figure~\ref{fig:heat5} shows the gradient between the location of the tracker
cells and two cold sinks on either side (represented by two holes). One
has to assume that not the whole edge of the CE7 plane is available for
a thermal load path (sink), hence this worst-case approach. There are
some obvious improvements that can be made, but not many will yield a
significant smaller gradient. The thermal ``choke'' as it were is the
limited thickness of the CE7 plate, assumed here to be 300 microns.

Of interest to the sensors themselves is the gradient, in the CE7
support. This gradient is shown in Figure~\ref{fig:heat5}.

\begin{figure}
\centerline{\epsfig{file=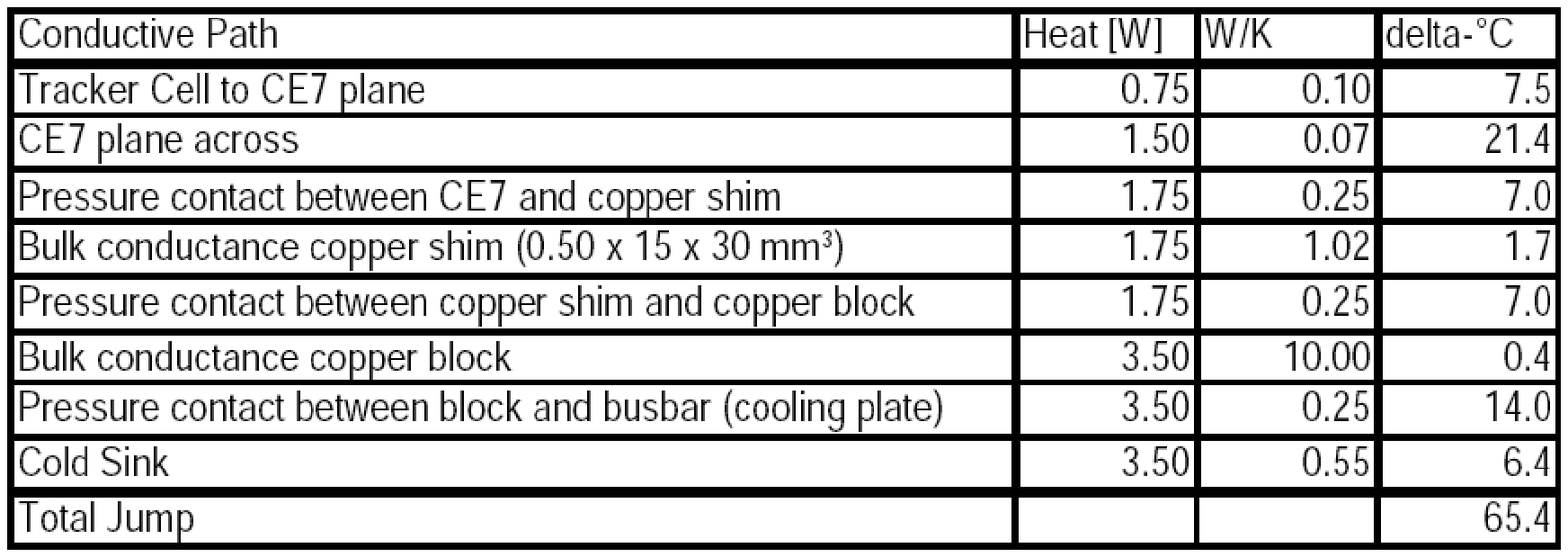,width=14cm}}
\caption{Table: Heat flow for various conductive paths.}
\label{fig:heat_flows_tab}
\end{figure}

The various thermal resistances between the actual tracker cell and the
cold plate next to the detector block have been analysed and the
results are listed in the Table~\ref{fig:heat_flows_tab}.

As can be seen from Table~\ref{fig:heat_flows_tab}, the accumulated gradient between the
tracker cell and the cold sink is 65.4{\textdegree}C. In order to gain
some extra margin with respect to temperature, the recommended cooler
temperature (at the heat sink) is -90{\textdegree}C, which gives
5{\textdegree}C margin on top of the 100\% margin on the pumping
capacity.

\subsubsection{Cold Sink}

The cold sink as shown in the table above needs to sit at
-90{\textdegree}C since the tracker cell operates at -20{\textdegree}C 
nominally with a gradient of 65{\textdegree}C down to
the cold sink (and 5{\textdegree}C extra margin). As mentioned in the
heat load section, the cold sink needs to absorb 42 W (including a
safety factor of 2). This amount of heat and the gradient excludes
the use of Peltier cooling devices. Peltier cooling devices are not
practical when they need to bridge gradients exceeding 50{\textdegree}C 
at sub-zero temperatures. At these temperatures, Peltier
devices have trouble pumping heat and they are not efficient at all
($<$ 5\%). Using Peltier coolers in stacks to bridge the gap
between -90{\textdegree}C (cold sink) and +30{\textdegree}C (ambient)
with an efficiency of less than 5\% would yield the need to dump at
least 10kW of heat into the LHC tunnel and we would still struggle to
reach the required temperatures. The alternative would be to use some
kind of fluid/vapour cooling stage; however the environment directly
surrounding the beam line is extremely limiting. Not many cooling
agents can survive the extremely intense radiation environment. 

Within CERN several cooling methods have been developed. The cooling system developed for the TOTEM
project seems appropriate to cool the FP420 detectors as it has been
designed and is acceptable for use in the LHC tunnel. It can reach the
required cold-sink temperature with margin and has sufficient cooling
power. Other options we looked into required cooling fluids with a heat
exchanger but none of the cooling fluids could be guaranteed to be
radiation hard. Due to symmetry conditions and in order to have at
least partial redundancy in the cooling system, it would be good to
operate two coolers in parallel per detector block. Envelope
restrictions or cost may however exclude this option.

\subsubsubsection*{Conclusions}

The cooling system should be able to run for 2 years next to the LHC
beam line, without servicing. It is strongly recommended to operate the
tracker cells in vacuum.
The required cooling for operation in vacuum is specified as follows: 42
W pumping power at -90{\textdegree}C.   There are significant thermal
gradients predicted across the CE7 plane and underneath the tracker
cell.

\subsubsubsection*{Recommendations}

Maintain a symmetrical cooling system, following the symmetry in each
detector plane. It will minimise gradients and provide redundancy.
A cooling system by CERN as for the TOTEM detector is recommended. The
selected cooling system needs to be subjected to significant radiation
levels during sub-system testing in preparation for the final design
to prove performance and stability. When the tracker cell design and
the flexible links have matured, together with the overall geometry,
the analysis needs to be repeated at a slightly more detailed level. 
If gradients between the different super planes have to be minimised it
would be prudent to introduce ``dummy'' planes at either end of the
stack, sitting at the same temperature as the other planes. The extra
planes would provide for a more uniform thermal radiative background.

\subsection{Performance of the tracking system}
\label{sec:silicon_perform}

The performance of the tracker has been evaluated using a simple Monte
Carlo program and also by a full GEANT4 simulation. In the GEANT 4 simulation, the
energy deposits within the sensitive detector volumes are translated
into elementary charges and their collection on the electrodes is
simulated. Capacitative coupling between closely placed channels as
well as noise contribution are taken into account. The signal collected
channel by channel is corrected by a gain factor, converted into an
integer number and fed into a cluster-finding algorithm, if above a 
threshold. Clusters typically ($\sim$90\% of the cases) include just one
channel. The efficiency to find at least one cluster per plane is 99.7\%.
A resolution on the simulated hit position close to 10$\mu$m
has been measured for each plane. A track finding/fit algorithm based on a
$\chi^2$ fit loops over the available clusters.

One feature of forward tracking that does not occur in central trackers
is that the tracks have a very small angle. This means that hits in
each tracking layer are highly correlated and one does not improve the
resolution by 1/$\sqrt{N}$, where N is the number of layers. To improve
matters, alternate layers will need to be shifted by half a pixel width to
improve the tracking precision.

\begin{figure}
\centering
\epsfig{file=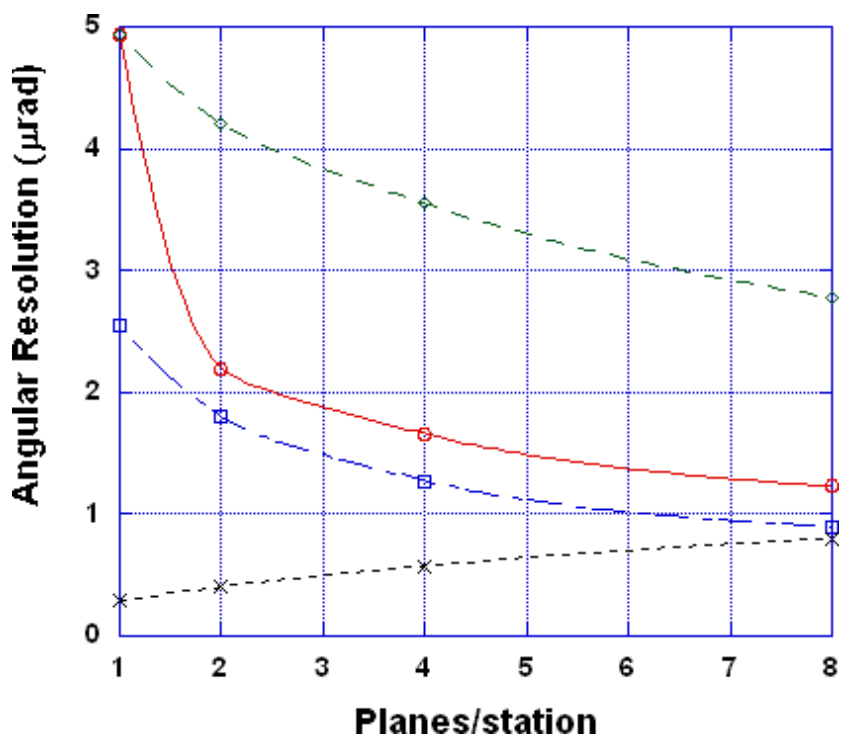,width=11cm}
\caption{Angular resolution for a tracker consisting of two stations separated by 8 metres.
Each layer has a detector with a pitch of 50 microns. The curves from top to bottom are: 
aligned tracking layers, alternate layers shifted by 25 microns, theoretical best result, 
and multiple scattering contribution. The design goal is one $\mu$rad.}
\label{fig:silicon11b}
\end{figure}

This is shown using a simple Monte Carlo model in Figure~\ref{fig:silicon11b}. The
multiple scattering angle is roughly 2 $\mu$rad $\times \sqrt{\rm thickness/X_{0}}$)
per layer at 7 TeV. If each layer corresponds to about 1\% of a radiation length, 
then one has a multiple scattering contribution of 0.2 $\mu$rad per layer. For the materials in
this model tracker, roughly 0.2\% of the protons will interact per
layer. Figure~\ref{fig:silicon11b} shows calculations for a tracker consisting of N
planes per station, with two stations placed 8 metres apart. The
spatial precision per layer is 50 microns/ sqrt(12) = 14 micron.
Shifting alternate layers by 25 micron significantly improves the
tracking performance. Multiple scattering degrades the tracking
resolution if the number of planes per station is increased beyond ten
layers. However, ten layers will give the design figure of one $\mu$rad.

\begin{figure}[htbp]
\centering
\epsfig{file=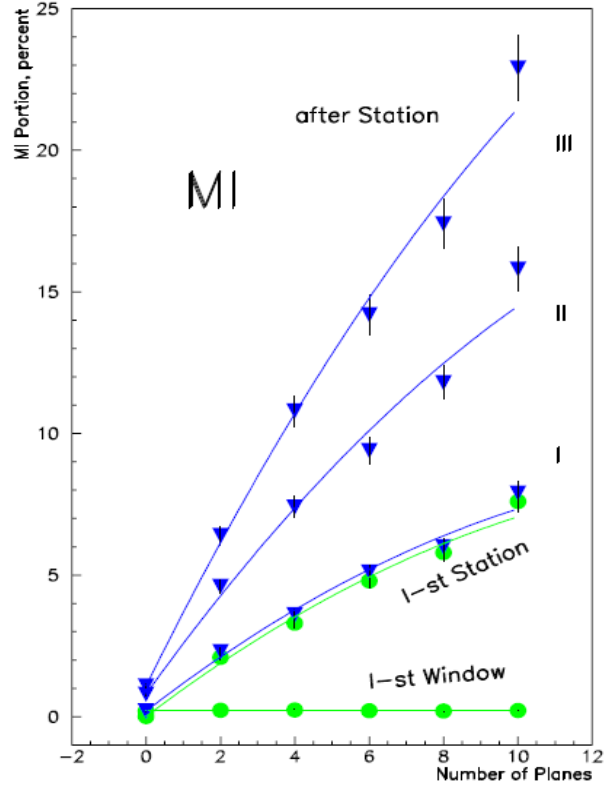,width=9cm,height=11cm}
\caption{Percentage of secondary interactions (MI) as a 
function of the number of planes and tracking stations. A revised
design improves the performance {--} the two-station tracker has a
6.8\% secondary interaction rate. See text.}
\label{fig:mi-rate}
\end{figure}

In a full GEANT4 simulation, different layouts of the detector stations
with different numbers of planes were simulated and their impact in
terms of secondary interactions of 7 TeV protons was assessed.
Moreover, the impact of a middle (3$^{rd}$) station was
evaluated.

The secondary interaction rate (Multiple Interaction, or MI in the
figures) was evaluated as the fraction of proton tracks which have an
inelastic interaction anywhere along the spectrometer before the last
plane of the last station. It was found that in 1~mm of stainless
steel, ceramic, and silicon the secondary interaction rates are 1\%, 0.5\%
and 0.4\%, respectively. Figure~\ref{fig:mi-rate} shows the rate of secondary
interactions as a function of the number of planes per station for a three-station layout.
Contributions to the 20\% rate resulting after the third station come
mainly ($\sim$15\%) from the 1~mm ceramic support structure of the
silicon detectors. Note that this is much larger than the model tracker
discussed above. The GEANT4 results led us to consider CE7, a 70\%/30\% 
Si-Al compound support as an alternative. The contribution of a 250
${\mu}$m stainless steel window, one for each station, turned out to be
negligible. Consequently a more reliable secondary interaction
estimate, based on an analysis of hits in the detector using realistic
materials and a three-station layout is 10.1\%. For a two-station
layout, this drops to 6.8\%. It should be noted that if an interaction
takes place in the third station some tracks can nevertheless be well
reconstructed with a $\chi^{2}$/NDF less than 1.5.
With this cut, the contamination of events with secondary interactions
in the signal sample is negligible {--} around 0.5\%. Losses of events
are comparable to the secondary interaction rates, and are 10.4\% and
7.1\% for the three- and two-station layouts respectively.

\begin{figure}
\centerline{\epsfig{file=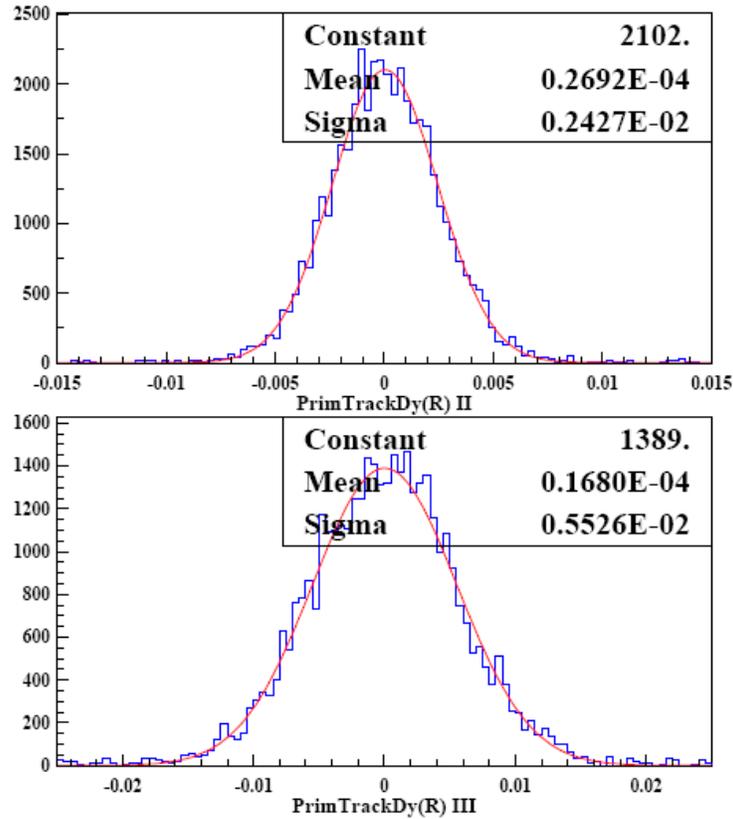,width=10cm}}
\caption{GEANT4 estimate of the multiple scattering (in mrads) in the middle (top) and 
at the end (bottom) of the three-station tracker.}
\label{fig:MI-2_3_stations}
\end{figure}

\begin{figure}
\centerline{\epsfig{file=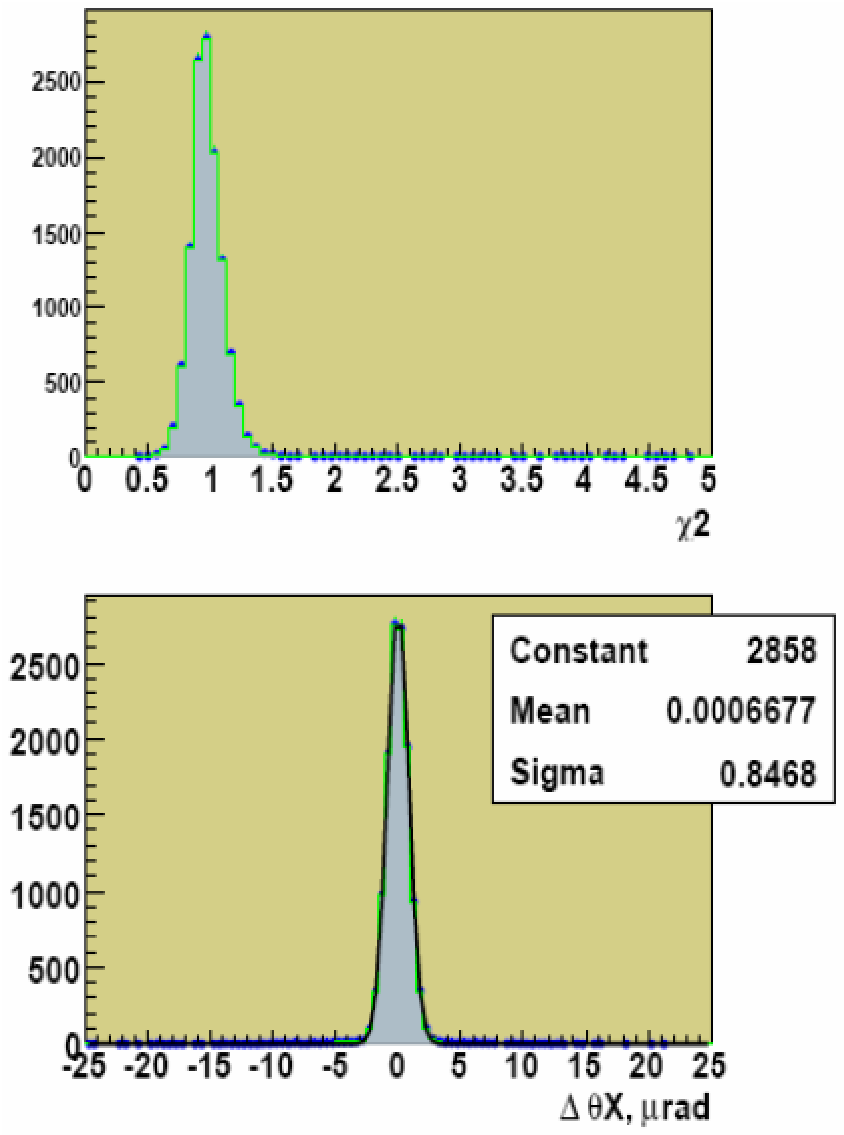,width=9cm}}
\caption{Track $\chi^{2}$/NDF (top) and angular resolution (bottom) for a 
two-station tracker. The angular resolution is 0.85 ${\mu}$rad if the 
$\chi^{2}$/NDF is selected to be less than 1.5.}
\label{fig:ang_resol1}
\end{figure}

\begin{figure}
\centerline{\epsfig{file=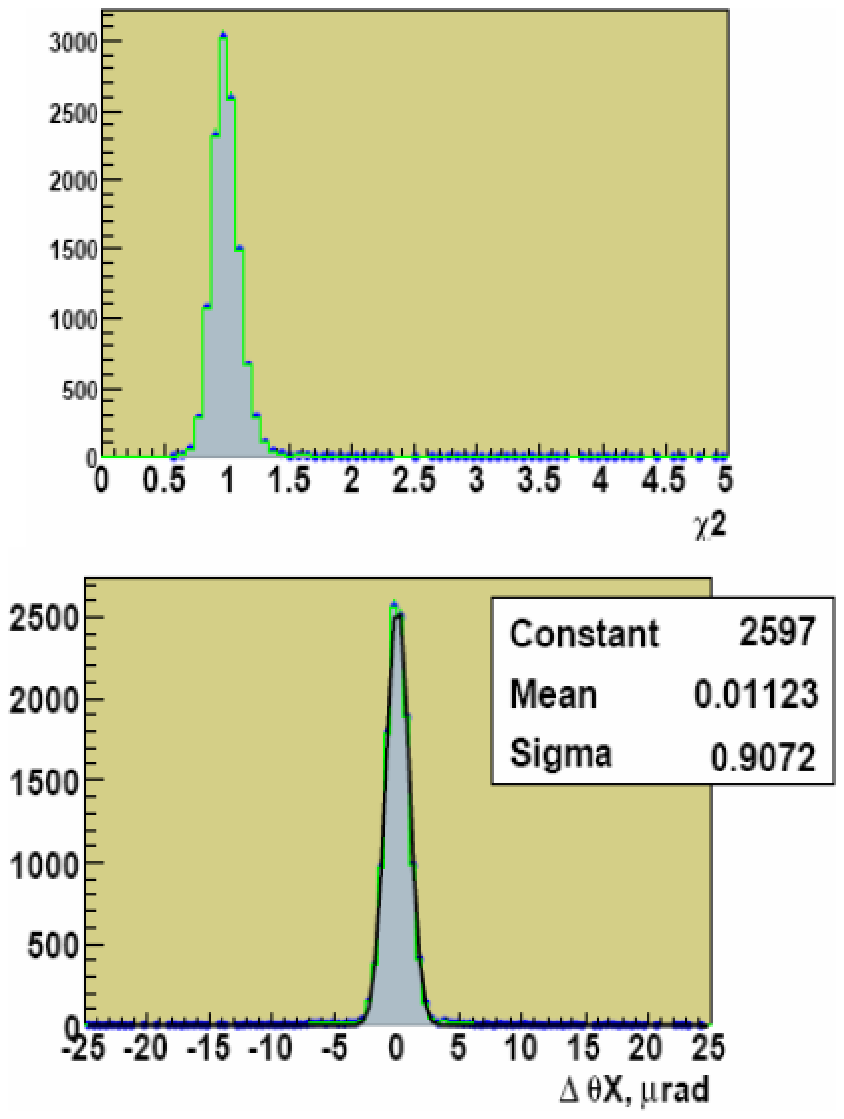,width=9cm}}
\caption{Track $\chi^{2}$/NDF (top) and angular resolution (bottom) for a 
three-station tracker. The angular resolution is 0.91 ${\mu}$rad 
if the $\chi^{2}$/NDF is selected to be less than 1.5.}
\label{fig:ang_resol2}
\end{figure}

An estimate of the multiple scattering for the two- and three-station
layouts is shown in Figure~\ref{fig:MI-2_3_stations}. Figures~\ref{fig:ang_resol1} 
and~\ref{fig:ang_resol2} show the $\chi^{2}$/NDF and angular resolution for the
two-station (0.85 ${\mu}$rad) and three-station (0.91 ${\mu}$rad)
layouts. These are both within the design specification. Finally, the
efficiency of two-track reconstruction has been found to be 86\% and
80\% respectively for the two- and three-station layouts.


\newpage

\section{Fast Timing Detectors}
\label{sec:timing}

\subsection{Overlap background and kinematic constraints}
\label{sec:pileup}

\begin{figure}[htbp]
\centering\includegraphics[width=.8\linewidth]{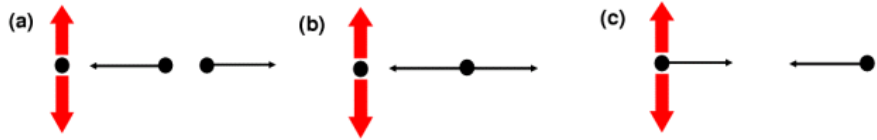}
    \caption{A schematic diagram of overlap  backgrounds
 to central exclusive production: (a) [p][X][p]: three interactions,
 one with a central system, and two with opposite direction single
 diffractive protons (b) [pp][X]: two interactions,
 one with a central system, and the second with two opposite direction protons
 (c) [p][pX]: two interactions, one with a central system and a proton,
 the second with a proton in the opposite direction. \label{pxp} }
\end{figure}

   The FP420 detectors must be capable of operating at the LHC design luminosity $\mathcal{L} \approx 10^{34}$~cm$^{-2}$s$^{-1}$
   in order to be sensitive to femtobarn-level cross sections in the central exclusive channel [pXp].
    At these luminosities overlap background from two single diffractive events superimposed with
    a central hard scatter ([p][X][p]), as shown in Fig.~\ref{pxp}(a), becomes a significant
   concern, especially in dijet final states. The 2-fold overlap coincidence backgrounds, shown in Fig.~\ref{pxp}(b) and (c), also must be considered,
   however; as they scale with $\mathcal{L}$$^2$
   instead of  $\mathcal{L}$$^3$ they are less of a concern in
   the high luminosity limit.  Fortunately, there are a number of techniques we can employ to reduce this
   overlap background. It can be substantially reduced at the
   high level trigger stage, or offline, by employing kinematic constraints.
   These factors, discussed in detail in the physics overlap discussion (Section~\ref{sec:pilko}),
   include consistency between the central system and the protons in rapidity
   and mass, and also use the fact that the number of particle tracks associated with the
   event vertex is much smaller for exclusive than generic collisions.
   Even after the
   significant background rejection afforded by these
   constraints, overlap backgrounds are still expected to
   dominate the signals without the additional rejection provided
   by precision timing of the protons, as detailed below.

\subsection{Timing}

High-precision time of flight (ToF) detectors at 420 m can be used
to obtain a large reduction in overlap (or pile-up)
backgrounds~\cite{alb}. We need only measure the \emph{relative}
arrival time of the two protons, $\Delta t = t_L - t_R$. Under the
assumption that they originate from the same event, the $z$-position
of that event can be calculated as $z_{pp} = \frac{1}{2}\Delta t
\times c$. The uncertainty on $z_{pp}$ is $\delta z_{pp} =\frac{c}{\sqrt{2}}\delta t$, 
where $\delta t$ is the (r.m.s.) time
resolution of the proton measurement. For example, $\delta t$ = 10
ps implies  $\delta z_{pp}$ = 2.1 mm. We then require a match
between $z_{pp}$ and the vertex position from the central detector,
$z_{vertex}$, which is known with extremely good precision
($\approx$ 50~$\mu$m)~\cite{atlas_tdr}.

In the case of the overlap backgrounds, the protons do not originate
from the same event as the hard scatter and so the vertex
reconstructed from time-of-flight information will, in general, not
match the vertex observed in the central detector, which implies
that a large rejection factor can be obtained. This rejection factor
depends on four parameters; the timing resolution $\delta t$, the
spread in interaction points $\sigma_z$, the vertex window size
(i.e. the degree to which the vertices are required to match) and
the luminosity. As the luminosity increases, the probability of
there being more than one proton in an arm of FP420 increases. If
any of the subsequent timing measurements results in a vertex that
coincides with the central vertex, then these protons would be chosen
as the `correct' protons. Hence the rejection factor degrades
slightly with increasing luminosity. The vertex window size is a
trade-off between high signal efficiency and high background
rejection. Clearly a smaller vertex window results in a higher
background rejection but will also lead to more signal events
failing the vertex matching requirement. Common choices are that the
vertices must coincide to within 1, 1.5 or 2$\times \delta z_{pp}$,
which corresponds to a signal efficiency of 68\%, 87\% and 95\%
respectively. Finally, the rejection factor increases if the spread
in vertices increases and is also approximately linear with $\delta
t$.

The prototype detectors described below have a timing resolution of
$\delta t \leq 20$~ps.  As the luminosity grows, better timing
resolution is required. We envisage a program of detector upgrades
to match this requirement, eventually attaining resolutions smaller
than 10~ps, as discussed in Section~\ref{sec:timsum}. The relatively
small and inexpensive precision ToF detectors discussed here make
this approach viable.

We have calculated the background rejection for the three overlap
cases shown in Fig.~\ref{pxp} (a) [p][p][X] (b) [pp][X] and (c)
[pX][p]. For example, if $\delta t$ = 20~ps ($\delta z_{pp}=4.2$~mm)
and the spread in interaction points is $\sigma_z \approx 50$
mm~\cite{atlas_tdr}, we obtain a rejection factor of 21 for the
first two cases and 15 for the third if the vertex measurement from
proton time-of-flight is  required to fall within $\pm4.2$~mm ($\pm
1\times \delta z_{pp}$) of the vertex measured by the central
detector. Case (a) dominates at high luminosity and consequently for
$\delta t$ = 10~ps, we would be able to obtain a rejection factor of
greater than 40 (for a $\pm 1\times \delta z_{pp}$ vertex window),
enabling FP420 to effectively cope with the large overlap
backgrounds at the design luminosity. Note that the rejection
factors presented in Table~\ref{tab:cuteff} in
Sec.~\ref{sec:results} are smaller than those presented here due to
a larger vertex window ($\pm2\delta z_{pp}$), which maximises the
signal efficiency, and also a narrower spread in interaction points
of 44.5~mm. This pessimistic vertex distribution is based on  a
large crossing angle scenario and results in a reduced background
suppression power using the ToF detectors. For the nominal crossing
angle of 250~$\mu$rad, the vertex spread exceeds 5~cm, and in
addition, the expected growth in $\sigma_z$ would result in an
improved rejection. The final choice of vertex window will be
optimised based on the analysis goals and instantaneous luminosity.
For example, a discovery measurement would likely maximize signal to
background, while a measurement of a state's properties, might
demand very low background at the expense of signal efficiency.

In addition to detector performance, there are other factors that
could impact the overall timing precision.  If the path length of
protons detected in FP420 were to vary significantly, this could
degrade the vertex measurement accuracy. We have determined that
even for the largest energy loss for protons in our acceptance
compared to the beam protons,  the path difference amounts to less
than 30~$\mu$m, corresponding to a 100~fs time difference (an even
smaller effect is expected from proton velocity differences). A
second concern is that a precise measurement of the arrival time
difference between deflected protons in the ToF detectors requires a
reference timing signal at each detector with a $t_L-t_R$ jitter
that is small enough not to contribute significantly to the overall
time resolution. The large ToF detector separation of about 850 m
makes this a challenging requirement. Our reference timing system,
designed to yield an r.m.s. jitter of $\sigma_{LR} \approx$ 5 ps, is
described in Sec.~\ref{sec:reftime}.

The absolute calibration of the ToF detectors $z$-coordinate measurement 
will be determined and monitored with double pomeron exchange (DPE) 
physics events to correlate the vertex position measured with the central trackers 
with the vertex measured by the FP420 timing detectors. Since it is not possible to 
trigger on the protons at Level-1, it will be necessary to add a double Pomeron filter 
at the High-Level-Trigger to the highest cross section candidate DPE processes 
that pass the Level-1 trigger, dijets and dileptons, for example, to select an 
adequate sample of events. Given the high cross section for DPE dijets 
(1.2 nb for $E_T>50$~GeV, see Table~\ref{tab:hard-diffr-crossections}),
it will be possible to collect hundreds of such events per hour. 



\subsection{Timing detectors}

For quite a while the standard for time of flight detectors has been
in the 100~ps range. Recently, there has been an explosion of
interest in fast timing for medical purposes in addition to high
energy physics detectors, and the idea of a detector with a few~ps
resolution is no longer considered unreasonable~\cite{UCtiming}. The
ALICE collaboration has developed a time of flight system that has
achieved a time resolution of about 20~ps~\cite{alice}. A time
resolution of $\sigma$~=~6.2~ps (with $\sigma \approx$ 30~ps for a
single photoelectron) was recently achieved by a group from
Nagoya~\cite{akatsu} utilizing prompt \v{C}erenkov radiation. A beam
of 3~GeV/c pions was passed through a quartz radiator in line with a
micro-channel plate photomultiplier tube (MCP-PMT). MCP-PMTs consist
of a quartz faceplate and a photocathode followed by two
back-to-back chevroned microchannel plates read out by a single
anode or multi-anode pads. They are compact (only about 3~cm in
depth) and provide a gain of about $10^6$ for a typical operating
voltage of 2.5 to 3.0 kV. Our requirements of an edgeless detector
to measure particles within several mm of the beam combined with the
very high beam energy renders the Nagoya geometry unusable, but
alternate detector concepts described below are likely to be capable of
10~ps or better resolution.

Three main factors affect the time resolution of \v{C}erenkov
detectors: (1) the spread in arrival time of photons at the
photocathode, (2) the time resolution of the MCP-PMT, dominated by
the transit time spread (TTS) of the electrons from emission at the
photocathode to arrival at the anodes, and (3) the downstream
electronics, including signal dispersion in cables. The first factor
is minimised using \v{C}erenkov light and optimised geometrical
designs. The MCP-PMTs we are considering have a small TTS, about 30
ps for a single photoelectron, or better, leading to a resolution of
30~ps/$\sqrt{n_{pe}}$. The major manufacturers of MCP-PMTs are
Hamamatsu~\cite{hama}, Photek~\cite{photek}, and
Photonis~\cite{burle}. Hamamatsu and Photek have concentrated on
single channel tubes typically of small active area (11~mm
diameter), for which the TTS may be as low as approximately 15~ps.
Photonis' tubes are larger (48~mm~$\times$~48~mm) and include a 64
pixel version that is well matched to one of our detector concepts.

We are developing two types of ToF counters for FP420, GASTOF (Gas
Time Of Flight) and QUARTIC (QUARtz TIming \v{C}erenkov). Prototypes
of both types of detector have been built and tested.

\begin{figure}[htbp]
\centering
\includegraphics[width=.6\linewidth]{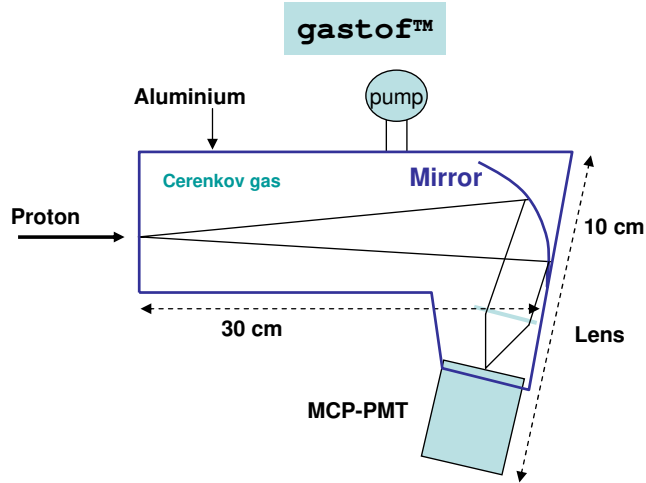}
\caption{Schematic of GASTOF, a gas-based \v{C}erenkov counter proposed by Louvain-la-Neuve, as described in the text.}
\label{gastof3}
\end{figure}

A schematic diagram of the GASTOF detector developed at UC Louvain
is shown in Fig.~\ref{gastof3}. It has a gas radiator at 1.1~--~1.4 bar in 
a rectangular box of 20~--~30 cm length, with a very thin wall adjacent to 
a specially designed flat pocket in the Hamburg beam pipe 
(Section~\ref{sec:hhpipe}). The protons are all essentially parallel 
to the axis. A thin 45$^\circ$ concave mirror at the back reflects 
the light to a MCP-PMT. The gas used in the tests, and which we 
propose to use in FP420, is C$_4$F$_{10}$, which is non-toxic and 
non-flammable, and has a refractive index $n=1.0014$  between 200~nm
and 650~nm, giving  a \v{C}erenkov angle ($\beta$~=~1) of about
3.0$^\circ$.  C$_4$F$_{10}$ is used in the RICH1 detector of the LHCb 
experiment. 


   The in-line material in a GASTOF (thin windows, mirror and gas) is minimal and does not cause
   significant multiple scattering. It can therefore be placed before the final tracking detectors.
   The GASTOF is intrinsically radiation hard, the only sensitive element being the
   MCP-PMT. Lifetime tests on gain, transit time spread, and quantum
   efficiency under laser light irradiation were carried out on Hamamatsu and Budker Institute tubes by
   the Nagoya group~\cite{kishi}.
   At 2.8$\times 10^{14}$ photons/cm$^2$ some gain decrease occurred, recoverable by increasing the HV,
   but the TTS was not affected. However, a significant deterioration of the photocathode quantum efficiency (QE) 
   was observed. Such an effect could be remedied by increasing the gas pressure and thus the number of \v{C}erenkov 
   photons -- for a 30 cm GASTOF and the nominal QE, the mean number of photoelectrons can well exceed 10.

\begin{figure}
\centering
\includegraphics[width=.6\linewidth]{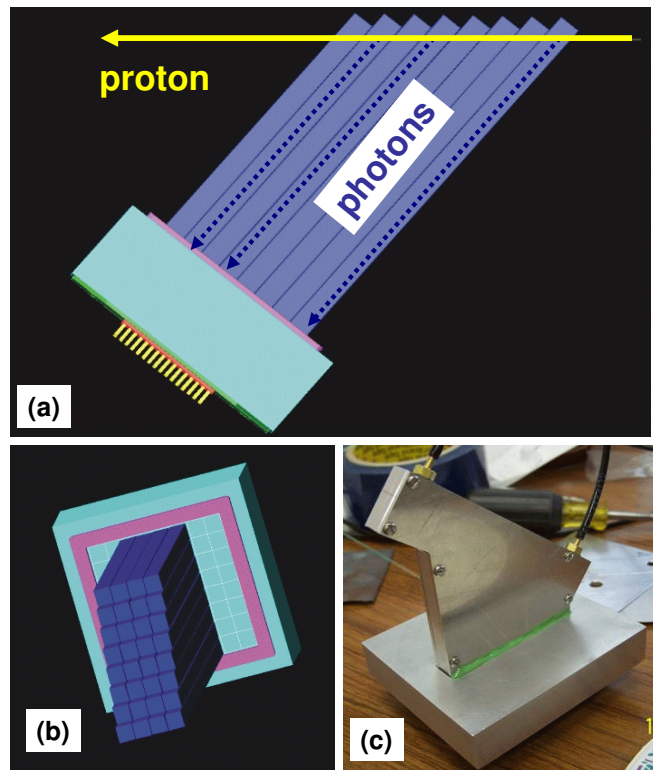}
  \caption{Conceptual drawings of
    a QUARTIC detector, (a) showing the proton passing through eight bars of one row in $x$
    providing eight measurements of the proton time (b) showing the
    $4 \times 8$ layout of QUARTIC bars
    (c) A photograph of the prototype detector used in the June 2008 CERN
    test beam.}
\label{quartic}
\end{figure}

The QUARTIC detector, which utilises fused silica (artificial
quartz) bars as radiators, is being developed by the University of
Alberta, Fermilab, and the University of Texas, Arlington and Louvain-la-Neuve groups.
Figure~\ref{quartic}(a) shows the concept: a proton passing through
the silica bars radiates photons which are measured by the MCP-PMT.
Figure~\ref{quartic}(b) shows the 4  $\times$ 8 array of bars with a
6 mm  $\times$ 6 mm cross section and length ranging from about 
110~mm for the first bar hit by the proton to 70 mm for the last, and
will be flush with the surface of the MCP-PMT. The 
bars are oriented at the average \v{C}erenkov angle, $\theta_c 
\approx$~48$^{\circ}$, which serves to minimise the number of 
reflections as the light propagates to the MCP-PMT. 
Figure~\ref{quartic}(c) shows a third generation single-row 
prototype used in the June 2008 CERN test beam. The final four-row 
version will have a very thin wall adjacent to the beam pipe, 
matching the dead area of the silicon detectors, to ensure full 
acceptance for all measured tracks. It will also use short tapered 
aluminised air light guides to channel the light into the respective 
pixels of the MCP-PMT.

Since the GASTOF and QUARTIC detectors have complementary features
as discussed below, we are proposing to use both: one GASTOF
detector will be located in its own beam pipe pocket after the first
silicon detector pocket and two QUARTIC detectors will be located in
their own pocket after the final silicon tracking detector, to
mitigate the impact of multiple scattering.

\subsection{Detector simulations}

Ray tracing and GEANT4 simulations of the propagation, absorption,
reflection, and arrival time (at the MCP-PMT face) of \v{C}erenkov
photons have been performed using the GASTOF and QUARTIC detector
designs. These simulations provide an important aid to our
understanding of the proposed detectors.

\begin{figure}
\centerline{ \epsfig{figure=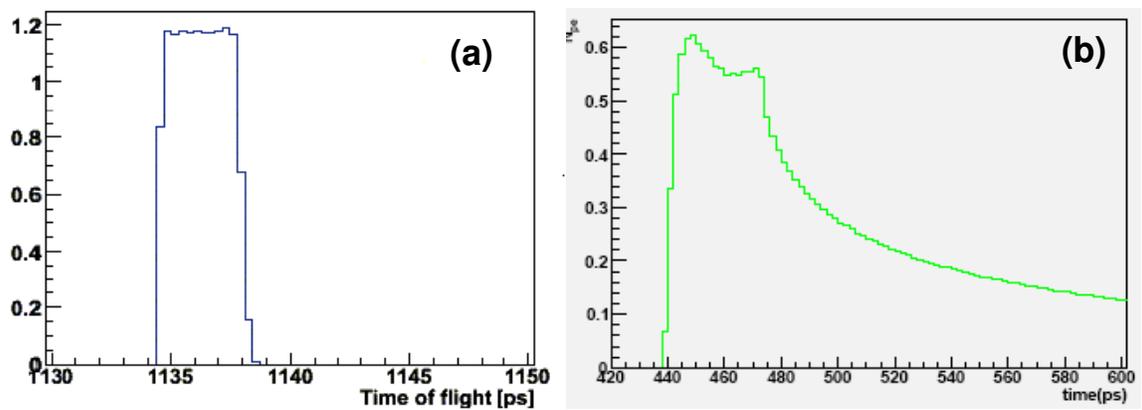,width=6.0in}}
  \caption{Simulated
    time of arrival of \v{C}erenkov photons at the MCP-PMT photocathode
    for
    (a) GASTOF (b) QUARTIC.}
\label{gqtime}
\end{figure}

Figure~\ref{gqtime}(a) shows GEANT simulation results for the
distribution of arrival time of photons at the MCP-PMT face for  a
30 cm long GASTOF. Due to GASTOF's optimised geometry and small
\v{C}erenkov angle all the photons arrive within a few picoseconds,
consequently the time resolution is dominated by transit time jitter
in the MCP-PMT and the subsequent electronics.

Figure~\ref{gqtime}(b) shows simulation results for the distribution
of arrival time of photons for a 90 mm long QUARTIC bar. For the
QUARTIC bar the bulk of the photoelectrons arrive within 40 ps, and
there is a long tail to much larger times.  The width of the peak is
due to the time dispersion from the wavelength dependence of the
index of refraction in silica, while the long tail is due to photons
emitted at different azimuthal angles, which leads to a variable
path length. Both the GASTOF and the QUARTIC simulations show an
expected yield after quantum efficiency of about 10 photoelectrons.
Although the time spread for a QUARTIC bar is much larger than the
GASTOF, the philosophy of the QUARTIC detector is to compensate for
the inferior resolution of a single channel by performing multiple
measurements. A proton traverses eight bars in each of the two
QUARTIC detectors, giving 16 measurements with up to a 4-fold
improvement in resolution over that of a single bar. The QUARTIC
detector also has $x$-segmentation that will be useful to time
multiple protons in the same bunch crossing.

Ideas for combining the advantages of the two detectors to yield a
segmented detector with superior resolution for 2012 are being
pursued and are discussed in Section~\ref{sec:timsum}.

\subsection{Performance in test-beam measurements}

   Test beam measurements over the last few years have validated the
   detector concepts.  Two of the test beam runs, March 2007 at Fermilab and June 2008
   at CERN were particularly successful and some of the results are reported here.

   In the March 2007  test-beam we used a 120 GeV proton beam to test two GASTOFs, G1 and G2,
   and several 15 mm long QUARTIC bars. G1 was an initial prototype which ganged together four central
   channels of the ($8 \times 8$ array of 6 mm
   $\times$ 6 mm pixels) Burle 85011-501 MCP-PMT with 25~$\mu$m
   pores. G2 was a second-generation prototype using an 11 mm diameter single channel Hamamatsu
   R3809U-50 MCP-PMT with 6~$\mu$m pores. The QUARTIC detector used a Burle 85011-501 with 10~$\mu$m pores.
   The signal for the MCP-PMT's was amplified using a GHz amplifier,
   passed through a constant fraction discriminator (CFD), and read
   out by a Phillips 7186 TDC.  Several types of amplifiers were tested: ORTEC 9306, Phillips
   BGA2712, Hamamatsu C5594,
   and Mini-Circuits ZX60-14012L. Several different CFD's were also used: ORTEC 934, ORTEC 9307, and a
   Louvain-made CFD circuit (LCFD).
   We used a CAMAC-based data-acquisition system triggered by  scintillator tiles located
   on either end of the detector setup.  Multiwire proportional
   chambers provided track position information.

   While the data-acquisition system provided a wealth of data allowing us to compare the performance of the
   different components and multiple channels, the most useful results
   for evaluating the detector performance
   were derived from an analysis of waveforms recorded
   from four channels (G1, G2, Q1 and Q4) using a Tektronix DPO70404 4 GHz digital oscilloscope.
   Offline we applied fixed threshold discrimination and constant fraction
   discrimination algorithms.

\begin{figure}[htbp]
\centerline{ \epsfig{figure=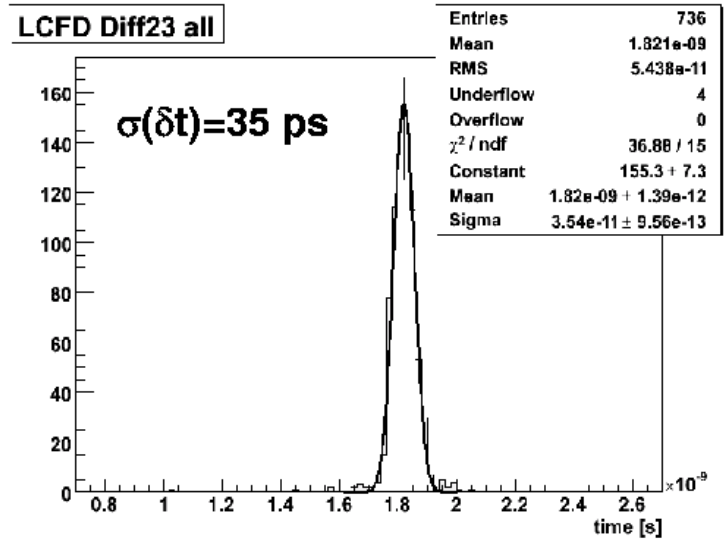,width=4.0in}}
\caption{The time difference between the first (G1) and second
generation (G2)
 GASTOF prototype detectors from the March 2007 test beam.} \label{f:g1g2}
\end{figure}

Figure~\ref{f:g1g2} shows the time difference between the two GASTOF
detectors, with $\delta t(G1-G2)$ = 35$\pm$1 ps (r.m.s.). If the two
detectors were identical this would imply a 25 ps resolution for
each.  The G2 detector is superior to G1, however, due to an
improved design resulting in a larger number of photoelectrons, and
the use of the 6 $\mu$m pore Hamamatsu MCP-PMT resulting in a
reduced transit time jitter. From the time differences between all
pairs of channels we infer the individual resolutions obtaining
$\delta t(G1)$ = 32~ps and $\delta t(G2)$ = 13~ps.
   The G2 detector is expected to be superior due to a better mirror
   and a faster MCP-PMT. Unfolding the resolutions of QUARTIC bars, we find $\delta t$
   $\approx$ 60~ps for the 15 mm long bars.  The G1 efficiency was measured to
   be about $\sim$ 98\%, while the G2 and QUARTIC bar efficiencies were measured to be about
   80\%, but due to limited statistics and concerns about the tracking
   alignment, we repeated these measurements in the subsequent June
   2008 test beam.

The June 2008 test beam at CERN used a beam of 180 GeV pions with
the same GASTOF detectors as the Fermilab test beam, but a new
QUARTIC prototype that had 90 mm long bars. Mini-Circuit amplifiers
were used, with and without the LCFD.  LeCroy 7300 and 8620
oscilloscopes, 3 and 6 GHz respectively, were used in segmented
memory mode for the data acquisition. Preliminary measurements of
the oscilloscope data confirm that the G2 resolution is 13 to 15 ps.
A silicon strip tracking system with 50 $\mu$m pitch was used to
provide tracking information. The trigger for the silicon telescope
was used also to trigger the oscilloscopes for some portion of the
data, allowing synchronization of the tracking readout with the
oscilloscopes. The tracking allowed a detailed study of the
efficiency as a function of position, yielding an efficiency
measurement of about 90\% near the edge of the GASTOF detector
rising over a few mm to 99\% at the center of the detector.

\begin{figure}
\centerline{ \epsfig{figure=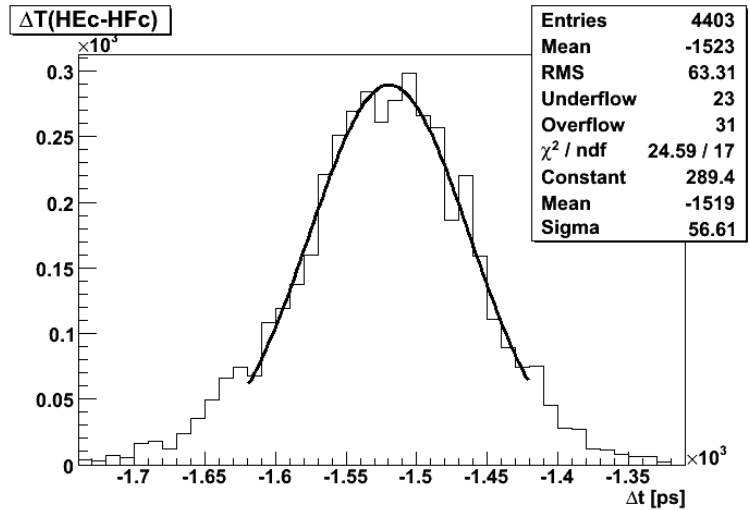,width=4.0in}}
\caption{The time difference between two 90 mm long QUARTIC bars as
described in text.} \label{f:q1q2}
\end{figure}

During this test beam, signals from several of the long QUARTIC bars
were passed through the LCFD, and the resulting NIM pulses were
recorded with the LeCroy 8620a oscilloscope. Figure~\ref{f:q1q2}
shows the time difference between two of the long QUARTIC bars, with
$\delta t(Q1-Q2)$ = 56 ps (r.m.s.). 
This implies a QUARTIC bar resolution of 40 ps including the LCFD. 
Measurements of the LCFD resolution alone give about 
25~ps, implying an intrinsic bar resolution of 30~ps. Preliminary 
studies show a uniform efficiency of near 90\% across the bar.  It 
is likely possible to increase the efficiency, by lowering the 
threshold of the constant fraction discriminator, and such studies 
are in progress using the raw waveforms. Given a 90\% efficiency we 
would expect about 14 measurements per event for two 8-bar QUARTIC 
detectors, implying an overall resolution of about 11~ps for the 
QUARTIC detectors alone, including the effect of a 20~ps TDC.  This 
$\sqrt{N}$ dependence must still be demonstrated using a full 
detector row with the final detector and TDC system.

The single-channel GASTOF detector has an intrinsic detector/MCP-PMT
resolution on the order of 10~ps, so requires a different
electronics strategy to maintain this superior resolution. As
discussed in the next section, we envisage using a single photon
counter with fast oscilloscope technology to maintain an overall
timing resolution of 15~ps or better, even without further
improvements, such as offline corrections to position of the tracks
through the detector (which will be known to $\sigma(x,y) \approx$
5~$\mu$m).

\subsection{Electronics and data acquisition}

The fast readout electronics must  provide a timing resolution
compatible with the baseline design of   the ToF detectors. The
Alberta and Louvain groups have extensive experience in this area
and are responsible for the design and prototyping of the readout
electronics. Both groups  have developed fast amplifier boards and
CFD boards for use in the beam tests. Independent tests with a fast
laser have verified that the performance of these boards is
comparable to commercial units, but the custom boards have the
advantages of being much more compact and less expensive.

The largest single contribution to the timing resolution in our
first test-beam run in Summer 2006 was the ORTEC 934 constant
fraction discriminator. For the March 2007 test-beam run, we employed 
the LCFD board described above. This new unit was designed to work 
with rise times as short as 150~ps, and to be insensitive to the 
non-linearity and saturation of the amplifier. Based on test beam 
results, this unit meets the needs of the QUARTIC detector.

The Alberta board consists of an integrated amplifier and CFD,
providing an alternative approach to the separate amplifiers and
CFD's developed at UC Louvain. The amplifier uses the Phillips BGA2717
chip, while the CFD is based on one developed by Alberta for the
GlueX Experiment. It has also recently been adopted by the ATLAS
LUCID detector. The circuit has been upgraded to use the most recent
comparators and logic. Laser tests at SLAC gave a preliminary
measured resolution of 19~ps for the Alberta CFD  (ACFD) board.

For beam tests we used the Phillips 7186 25~ps least-bit TDC and
fast oscilloscopes to measure the time of flight. The final readout
for QUARTIC will use the HPTDC (High Precision Time to Digital
Converter) chip which forms the basis of the CAEN V1290A TDC VME
board and is employed in the ALICE ToF detector readout system. In
addition, the HPTDC chip is radiation hard and has been designed for
use at the LHC, including a 40 MHz clock and appropriate buffering.
We have tested the HPTDC chips using CAEN 1290 VME boards. The
Alberta group has designed a custom readout system, comprised of
ACFD and HPTDC boards, that will interface with the ATLAS ROD
readout system. We plan to test a vertical slice of the FP420
readout chain in a 2009 test beam.

We are also exploring other TDC options for when the TDC performance
becomes a limiting factor. The development of a sub-10~ps TDC now
seems to be possible, and is somewhat simplified by the limited
dynamic range of $\sim$~500~ps required for our application. New
ideas such as sampling the waveform to replace the CFD/TDC
functionality are also being pursued~\cite{UCtiming}.

The amplifier/CFD combination ideally would be located close to the
detector to minimise time dispersion in the cables. We are exploring
the possibility of locating this front-end electronics in a shielded
compartment at the base of the cryostat support connected to the
detector via SMA 18~GHz cable. The length of the cable run to the TDC 
is not critical, so a mini-VME crate can be located nearby in a
shielded area. If the radiation hardness of the CFD comparator
becomes an issue, we may use the Louvain amplifier solution near the
detector with a longer cable run to the Louvain CFD, which would be
located near the TDC. We will be testing these options and 
effects of radiation in 2009. Low-voltage and high-voltage power 
supplies are standard units (described in Sec.~\ref{sec:services}) 
and will follow the same specifications as the silicon detector 
power supplies.

For the GASTOF detectors, a single photon counter, such as Boston
Electronics SPC-134, can be used to replace the amplifier, CFD and
TDC. This device has a timing resolution of 5~ps r.m.s., but is
expensive (\$10K per channel), making it impractical for 
use with the 32-channel QUARTIC detectors.

\subsection{Reference time system}
\label{sec:reftime}

A reference time signal without significant jitter and skew between the East (E) and West (W)
ToF detectors is an essential component of the ToF system. The goal of the design is to deliver
a pair of synchronized trigger pulses to the East and West ToF detectors, which are physically
separated by 420+420 meters, to a timing accuracy of $\sim$5~ps rms. This design uses long length
of optical fibers to deliver pairs of low timing jitter trigger pulses to both ToF detectors. To
minimize modal and group velocity dispersions, single-mode communication optical fibers and
reliable industrial 10-Gbit Ethernet optical backbone components operating at the wavelength of
1.3~$\mu$m are used.

We rely on the high stability master bunch clock at CERN to derive our reference timing system.
This master bunch clock is frequency divided by 10 from the 400~MHz superconducting RF cavity
frequency. The divider is synchronized to the revolution frequency. The delay between the edge
of the master bunch clock and the passage of a bunch is fixed from beam run to run. This master bunch
clock is a stable constant frequency square wave at 40.078879~MHz with a stability of 50~ppm
corresponding to 1.25~ps phase variation. Although this encoded bunch clock signal is broadcast by
the Timing Trigger and Control (TTC) transmitters around the CERN complex via single mode fibers
(1.31 $\mu$m) to various LHC instrumentation and experiment areas, to minimize the added timing
jitter and skew we design our reference time system starting at the stabilized 40.078879~MHz reference
electrical square pulse.

\begin{figure}[htbp]
\centering\includegraphics[width=.9\linewidth]{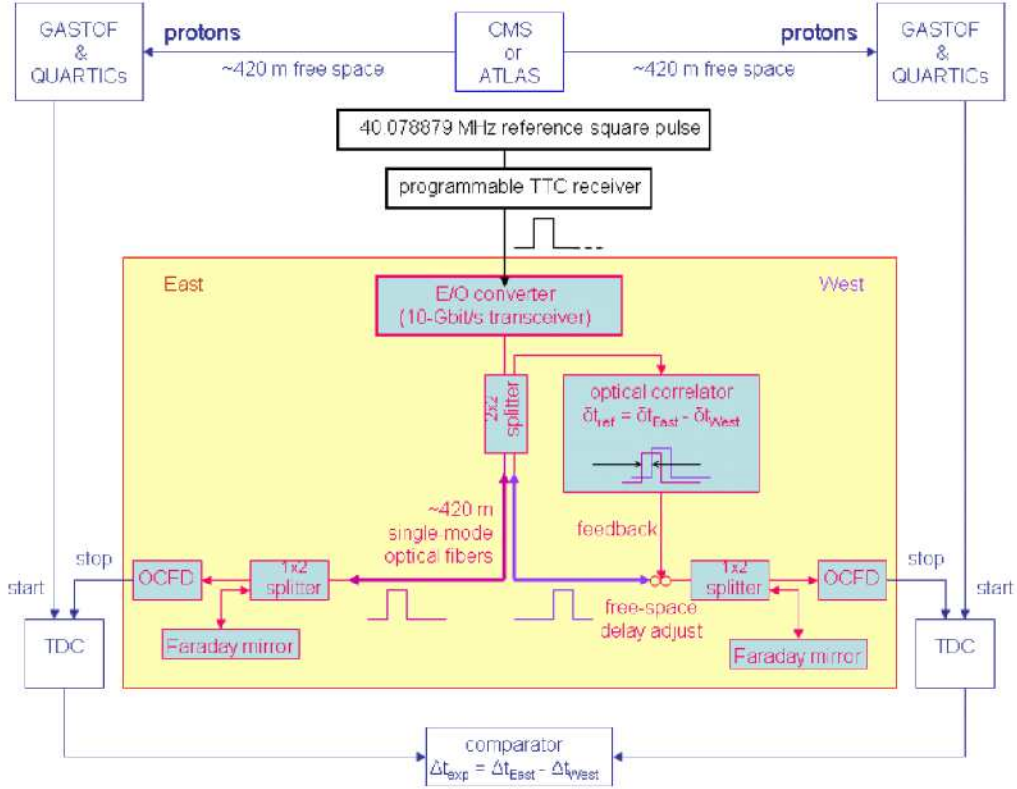}
\caption{ A schematic of the reference time system. Components labeled in black are expected to come from
CERN TTC system, components labeled in blue are associated with the ToF detectors, and the components in
red (highlighted in the yellow box) are the optical components related to the reference timing system.
\label{reftime} }
\end{figure}

A block diagram depicting the various elements of this system is shown in Figure~\ref{reftime}. A stable
signal pulse is derived from the 40.078879~MHz bunch clock provided by the LHC TTC system~\cite{baron}.
This RF signal is received by a programmable TTC receiver and to provide programmable post-skewing or
de-skewing of the clock period in steps of 104~ps. Because the coincident RF trigger pulse arrives
$\sim$700~ns later than the proton-proton interaction, the post-skewing feature allows
the triggers for the ToF detectors to be associated with the corresponding bunch crossings.

A time drift between both arms is expected to be dominated by environmental effects, primarily caused by the long length of 420~meter
single-mode fibers. Based on the thermal optical coefficient ($dn/dT$) of $\sim$1.28$\times$10$^{-5}$/$^o$C
and the thermal expansion coefficient ($dl/dT$) of $\sim$5.5$\times$10$^{-5}$/$^o$C for typical communication
optical fibers, a total time drift of $\sim$20~ps/$^o$C is anticipated for a $\sim$420 meter long single-mode fiber.
Hence, all fibers and optoelectronics need to be housed in a temperature stabilized environment.

This 40~MHz electrical pulse is converted to 1.3~$\mu$m optical pulse through an E/O converter. This E/O converter
can be a 1.3~$\mu$m single-mode Vertical cavity surface emitting laser (Vcsel) based optical front end driver for the
10 Gbit/s (IEEE 802.3ae standard) Ethernet transceiver which has a low timing jitter of 10~ps peak-to-peak
(such as Finisar FTLX1411D3). A typical 10-Gbit/s return-to-zero optical pulse has a width of $\sim$100 ps with
a rise-time of $\sim$20 ps and a peak optical power of 2~mW containing $\sim$10$^6$~photons/pulse. The optical
pulse is then split to two replica in a non-polarization single-mode 2$\times$2 splitter and sent along each of the
$\sim$420~meter long single-mode fibers to both East and West ToF detector stations. At each end of the optical fiber,
the light pulse is further split to two, one to trigger the stop of the Time-to-Digital Converter (TDC), while the other is
fiber butt-coupled to a Faraday mirror and retro-reflected back to the source point through the 2$\times$2 splitter. The
returned pulse typically contains $\sim$10$^5$~photons/pulse and is fiber-coupled to an optical correlator or an optical
oscilloscope. At the input end of the optical correlator an amplified fast photodiode or a single-photon detector
convert the optical pulses and are displayed on an electronic sampling oscilloscope with a time resolution of $<$5 ps.
Alternatively, a 500 GHz optical sampling oscilloscope (Alnair Labs Eye-Checker) can be used which has temporal
resolution of $<$2 ps and a timing jitter of $<$0.1~ps. The time different between the rising edges of the two pulses
$\delta t_{ref}$ retro-reflected from both arms is used as a feedback signal to drive a free-space delay scan stage deployed
on one arm (West) of the $\sim$420~meter long optical path. The error signal from $\delta t_{ref}$ is automatically
adjusted to less than $<$5~ps rms (1.5~mm free space) for every bunch crossing. Prior to trigger the TDC, an Optical
Constant Fraction Discriminator (OCFD), such as OFC-401, is used to suppress the electronic trigger delay shifts due to
optical signal pulse amplitude variation. With this reference timing system, the two path of the trigger pulse for
the ToF detectors can be synchronized to an absolute time difference of $<$5~ps rms. It is worth to note that this
optical reference timing system is used to compensate mostly the slow environmental drift on the two arms of the
trigger system.

\subsection{Central detector timing}

To this point, we have been focussing on relative timing of the
forward protons to provide a vertex position measurement for
comparison with the position of the central vertex. In
Ref.~\cite{snw} the space-time distribution of the luminosity
profile for design beam parameters was calculated, and it was found
that the position and time distributions of the vertex factorise.
This implies that an absolute timing of the central detector portion
of the event (two 60~GeV $b$-quark jets, for example) to significantly better 
than 170~ps would in principle provide a further overlap reduction
factor for [p][X][p] events discussed earlier.

From test-beam results the ATLAS Electromagnetic Calorimeter (ECAL)
was found to have a noise term of 500~ps/E(GeV) and a constant term
of 70~ps. Similar test-beam results were obtained by the CMS ECAL. 
Reductions in the clock jitter could result in a smaller
constant term during standard data taking.  We have begun
simulations to determine what central detector time resolution is
possible for ATLAS and CMS. A 70~ps event resolution already would
provide an additional factor of two in overlap rejection, and if it
is eventually possible to reduce this to 10~ps this factor grows to 12.

\subsection{Timing summary and future plans}
\label{sec:timsum}

We are in the process of developing an ultra-fast TOF detector
system which will have a key role in the FP420 project by helping to
reject overlap background that can fake our signal. Tests of the
current prototype detector design imply an initial detector
resolution of $\delta t \leq$ 20~ps, including the full electronics
chain, with an upgrade path to resolutions better than 10~ps
matching the need for improved rejection as the luminosity
increases. For a luminosity of $\mathcal{L}$ $\approx
\,2\,10^{33}$~cm$^{-2}$s$^{-1}$, a 30~ps detector would be
sufficient to keep the overlap background to the level of other
backgrounds for the dijet channels, and render it negligible for
other final states. For $\mathcal{L}$ $\approx \,
5\,10^{33}$~cm$^{-2}$s$^{-1}$, a 10~ps detector (still with loose
vertex cuts to maximise signal efficiency) would be desirable to
keep overlap backgrounds totally under control, without any loss in
signal efficiency. For $\mathcal{L}$ $\approx\, 7 \,10^{33}$ -- 10$^{34}$~cm$^{-2}$s$^{-1}$
to the design luminosity, we would
control the background by (i) developing timing detectors in the
5~ps range, or (ii) adding extra rejection from central timing, or
(iii) tightening the vertex window or other background cuts (a
factor of several in rejection is possible with a modest loss of
efficiency), or more likely a combination of all of the above.

In addition to further analysis and beam tests to fully evaluate the
current prototypes, we are continuing a program of simulation,
development and testing of the detector concepts and electronics to
provide a fully optimised robust timing solution. The simplest
approach to achieving faster timing is in upgrades to the
existing detectors.  Further optimisation is possible through
advances in MCP-PMT technology.  The transit time spread of the
MCP-PMT dominates the GASTOF resolution, so a pore size of  3 $\mu$m
(offered by Photek) would already be expected to yield a time
resolution better than 10 ps. 
For the QUARTIC detector a next generation
MCP-PMT with smaller pixel sizes would allow finer $x$ segmentation
for improved multi-proton timing.  A smaller pore size would not be
expected to dramatically improve the time resolution, since the
largest component is the intrinsic detector resolution, but might
yield modest improvements on the 10-20\% level.  Better electronics
could also give an improvement, with the combination
resulting in a sub 10~ps QUARTIC detector as well.


Each arm of FP420 will only contain three MCP-PMTs (in the present baseline design)
and, unlike the central detectors elements, they could simply be replaced every long shutdown
if necessary. The mechanical design will allow for quick replacement.


At maximum luminosity the proposed detectors will have rates in the
10~MHz range and see an integrated charge of a few to tens of
coulombs per year depending on the exact details of the detectors
and the gain at which the phototubes are operated. The current
commercially available MCP-PMT's will not sustain such high rates
and will not have an adequate lifetime. We are pursuing developments
with the phototube vendors to address these issues. There are many
rate and lifetime improvement proposals in various stages of
development that will improve the rate and lifetime capabilities of
these phototubes, including an increased pore angle to reduce ions
striking and damaging the photo-cathode, adding a thin aluminum film
at the front face or between the two microchannel plates to suppress the back-scattered
ions, gating the cathode during dead-time periods, and new
photocathode and MCP materials. We plan to make our own rate and
lifetime measurements using newly operational laser test stands at
UC-Louvain and UT-Arlington. The UCL group has already demonstrated that they
can successfully operate the GASTOF with the Hamamatsu tube at a
gain at least five times lower than the standard $10^6$ gain
(reducing both the lifetime and rate issues), and similar studies
are in progress at UTA for the Photonis tubes. We anticipate that
the GASTOF detectors will be able to operate to intermediate
luminosities $\mathcal{L}$ $\approx 2 \times
10^{33}$~cm$^{-2}$s$^{-1}$, with little or no improvements to the
MCP-PMT's, while the QUARTIC detectors likely need modest MCP-PMT
improvements to survive Phase I (intermediate luminosity) operations.  In order to have
satisfactory performance under Phase II (high luminosity) conditions, upgrades will be
needed to both improve the timing resolution of the detectors and
the rate and lifetime characteristics of the MCP-PMT's. These
modifications seems plausible on the timescale in which they must be
developed.  We also note that there have been interesting advances in APD,
silicon PMT, fast streak cameras, and other fast photo-sensitive
devices in the past few years, any of which could be adapted
for use with our current detectors if their performance surpasses
that of the MCP-PMT technology.

Hybrid detectors which combine the advantages of GASTOF and QUARTIC
are also being developed. Fermilab is  developing a new \v{C}erenkov
concept using conical quartz radiators. Saclay is investigating a
design that would give  a precise knowledge of the Cerenkov photon
origin, and therefore a better timing accuracy. This design is
comprised of a blade of MgF2 (refractive index 1.39), a focusing
mirror, and a segmented MCP-PMT. Alberta and UTA are investigating a
quartz fiber version of QUARTIC.  
A most promising new idea is a multi-anode GASTOF detector being developed by Louvain,
that uses the QUARTIC idea of distributing photons among many pixels
to make several $\sim 40$ ps measurements instead of a single 10 ps
measurement. This has the advantage of a much lower rate/$cm^2$,
improved lifetime due to spreading the charge over a larger area,
and the added benefit of multi-proton timing, since the photons from
two protons will typically be distributed among different pixels.
Simulations of these designs are in progress, and beam tests are
planned for 2009. We are also collaborating with other
groups~\cite{stof} who have long-term plans to develop large-area
timing detectors with ps-level resolution, and have semi-annul fast
timing workshops to keep in contact with developments in the rapidly
evolving field of fast timing.

The radiation environment of the detectors remains a concern that
has not been fully evaluated.  Simulations are in progress to
determine the radiation levels at the detector location and the
composition of the radiation, especially with respect to soft
particles that could cause background in the timing detectors. The
issue of radiation hardness of certain electronics components is
also a concern and different options are being explored depending on
these levels as discussed above. Radiation exposure tests of the
electronics are planned. The detectors are small with relatively few
channels and can be upgraded or replaced on a one-year time scale if
significant technological improvements are made or if the MCP-PMT
performance is significantly degraded. Since the intrinsic
\v{C}erenkov detector resolution is only a couple of picoseconds,
eventual timing detector performance at the 2~ps level is
conceivable with improvements in the electronics. The development of
central detector timing also provides a path towards better
background rejection and is being pursued in parallel.

At high luminosity it is desirable to have the ability to measure
multiple protons per bunch. Currently the GASTOF detector can only
measure one proton per bunch (the first one), while the QUARTIC
detector can measure two protons if they pass through different rows
(about 2/3 of the time for 6 mm width bars). At design luminosity
this will result in about a 10\% efficiency loss. An upgrade to
better determine the time of more than one proton per bunch is
conceivable, either through a multi-anode GASTOF detector, or by
reducing the pixel size in the $x$-direction for the QUARTIC
detectors.

As the reference timing is also an important component of the timing
resolution, we are also exploring other options for this, including
interferometrically stabilised fibre optic links, where the standard
is in the 10 femtosecond range.

In conclusion,  we believe we have a fast timing concept that will
work for moderate luminosities (Phase I), and an R\&D plan that will
provide a next generation of detectors by the time that high
luminosities are achieved (Phase II).

\newpage

\section{Alignment and calibration}
\label{sec:alignment}


Precise measurement of the momentum loss of the
outgoing protons will be achieved in FP420 by measuring the
proton-beam displacement and relative direction (slope) as accurately as possible.
To avoid significant degradation of
the intrinsic uncertainty arising from physics processes and beam optics,
FP420 must be aligned internally and relative
to the beam to an accuracy of at most a few tens of microns.

In this section we discuss (1) internal alignment of the track detectors within the 420 m arm; 
(2) determination of the displacement of the detectors with respect
to the passing beam, and their relative angle; (3) calculation of the proton momentum vector using 
the known LHC field elements (transfer matrices); (4) ``on-line'', real-time checks of the 
beam-track separation from data and (5) measurement
of the $M_X$ scale \emph{and its resolution} from a known physics process, in particular
exclusive $\mu^+\mu^-$ production.

\subsection{Alignment requirements}

``Internal alignment'' is the issue of knowing the \emph{relative} positions of all the tracking elements,
with respect to a fiducial entrance point [$x_{in},y_{in},z_{in}$] at 420 m and an exit point [$x_{out},y_{out},z_{out}$] at about 428 m.
The mechanical construction of the detector mountings on the moving pipe, and precision control of the motions (described
below) will give us these relative positions to an accuracy $\sim 10\mu$m. Any fine corrections can be obtained from
the straight-track fits, as the high energy protons are not significantly affected by remnant magnetic fields. It
remains to measure the entrance and exit points $x_{in},y_{in},x_{out},y_{out}$ with respect to the beam ($z_{in}$ and
$z_{out}$ do not need to be very precisely known).

\begin{figure}[htbp]
   \begin{center}
   \includegraphics[width=11cm]{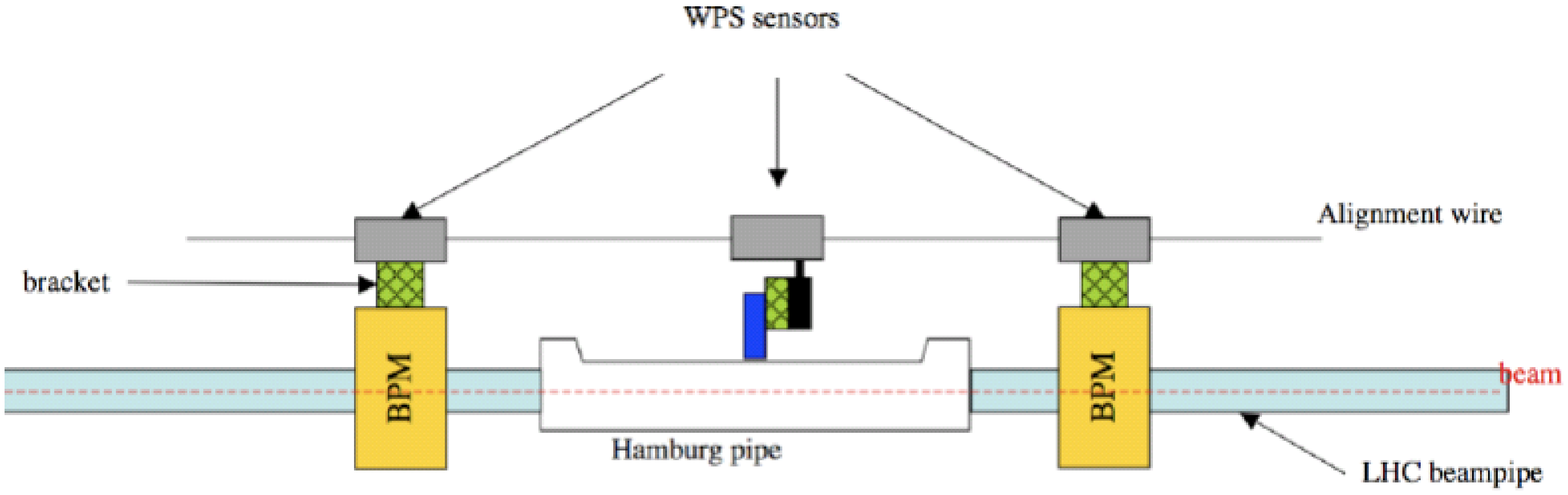}
   \\*
   \vspace*{0.25in}
   \includegraphics[width=11cm]{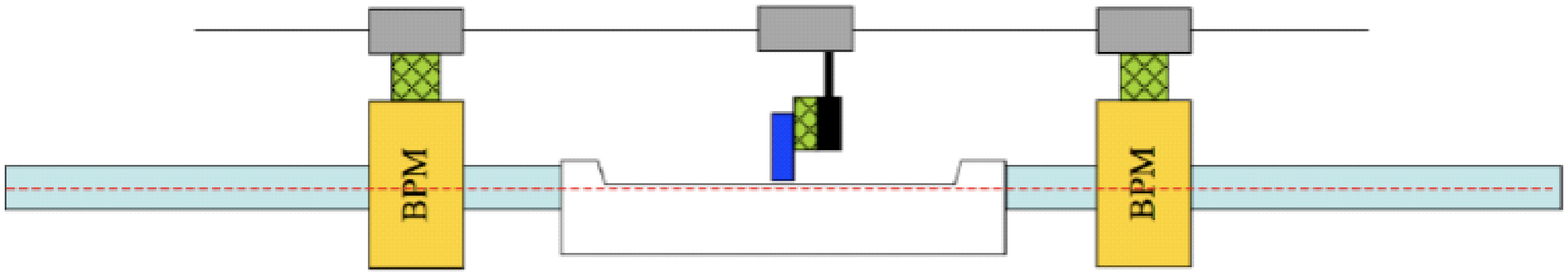}
   \caption{The proposed overall alignment system, shown with
     detectors in garage position (top picture) and in operating
     position (bottom picture).}
   \label{fig:alignsys}
   \end{center}
\end{figure}

For this we plan to build an independent
real-time alignment system into the detector, for on-line knowledge of positions and also as it will be needed for safety while moving FP420 into its
working position.
Two options, both based on Beam Position Monitors (BPMs), are being considered:
a `local' system consisting of a large-aperture BPM
mounted directly on the moving beampipe and related to the position of
the silicon detectors by knowledge of the mechanical structure of the
assembly, and an `overall' system consisting of BPMs mounted on the
(fixed) LHC beampipe at both ends of FP420, with their position and
the moving silicon detectors' positions referenced to an alignment
wire using a Wire Positioning Sensor (WPS) system. Figure~\ref{fig:alignsys} shows a schematic diagram
of the proposed `overall' alignment subsystem. To simplify the illustration only one moving beam pipe section
is shown, although there may actually be more than one. Note that the larger aperture BPMs for the
`local' alignment system are not shown (one would be mounted on each moving beam pipe section),
although it is likely that both the local and overall BPM alignment schemes will be implemented.

Sources of uncertainty in such a system include the intrinsic
resolution of the WPS system, the intrinsic resolution (and calibration) of the BPMs,
and the mechanical tolerances between the components. The mechanical uncertainties are expected to be affected by
temperature fluctuations and vibrations in the LHC tunnel, and measurement
is complicated by the fact that the detectors move with respect to the
beam.  The individual components of the system, with comments on their
expected accuracy, are described in the following sections.

\subsubsection{Beam Position Monitors (BPMs)}

A direct measurement of the beam position at FP420 can be obtained
with beam position monitors (BPMs). Although there are several pickup
techniques available, an obvious choice would be the type used in
large numbers in the LHC accelerator itself.  The precision and
accuracy of these electrostatic button pickups~\cite{LHC-BPMs} can be optimised
through the choice of electrode geometry and readout
electronics (for a description of the LHC electronics, see~\cite{LHC-BPM-electronics}.)
While BPMs can be made with precision geometry, an important issue is
balancing the gain of the right and left (or up and down) electronics;
one can have a time-duplexed system such that the signals from opposing electrodes
are sent through the same path on a time-shared basis, thus cancelling any gain differences.
Multiplexing of the readout chain will avoid systematic errors due to
different electrical parameters when using separate channels and
detuning through time and temperature drift.
Preliminary tests with electrostatic BPMs designed for
the CLIC injection line have shown promising behavior on the test
bench, even when read out with general purpose test equipment.

The LHC button-electrode pickups have been designed for best integration
within the accelerator and its environment. Specially designed
semi-rigid cables allow the front-end electronics to be moved to locations
with lower radioactive exposure. However, less specific cables
providing a sufficient bandwidth can be envisaged for FP420 since the
BPM will be at room-temperature and therefore not subject to large
temperature variations.

Although the requirements are not as demanding for the LHC as for FP420,
it has been estimated that the necessary level of precision,
resolution and acquisition speed can be  obtained. It should be
emphasised that the precision will depend to a large extent on the
mechanical tolerances which can be achieved. Tests of these BPMs will begin soon
on an alignment bench.
Several strategies and optimizations have been proposed to reach
precision and resolution of a few microns, and to achieve bunch-by-bunch
measurement. The effect of the intrinsic non-linearity of button
electrodes can be reduced if the particle beam passes close to the
centre of the pickup in the operating position. In the case where only two
electrodes are required the linearity of the signal could possibly be
further improved by larger electrodes. While the detectors are in the
parking position, away from the beam, the beam position measurement is
also less critical.

Multi-turn integration will improve the resolution at least by a
factor 10. This should still allow bunch/bunch measurements since the
bunches in LHC can be tagged. In this case measurements of each bunch
will be integrated over a number of turns. The variation of one
specific bunch between turns is expected to be small.
The estimated maximum orbit offset among bunches
is 0.2$\sigma$ and only subject to ``slow'' orbit drifts~\cite{LHC502}.

Wide band amplifiers could be envisaged to obtain single shot measurements, 
whereas narrow band amplifiers should allow a better resolution and
signal-to-noise ratio.

Shortly before the installation of each complete
420 m section (with trackers and BPMs) a test-bench survey using a
pulsed wire to simulate the LHC beam
will provide an initial calibration of the BPMs. Further, in-situ calibration
could be done by moving each BPM in turn and comparing
its measured beam position with that expected from the measurements in the other
BPMs in the system; the potential for success of such an online BPM calibration
scheme has been demonstrated with cavity-style BPMs intended for use in linear colliders~\cite{ILC-BPM-systest,PerfHiResBPM}.
Such calibration may even be possible at the beginning and end of data-taking runs when the BPMs
are being moved between garage and operating positions,
and therefore should not require dedicated calibration runs.

\subsubsection{Wire Positioning Sensors (WPSs)}

\begin{figure}[htbp]
   \begin{center}
   \includegraphics[width=2in]{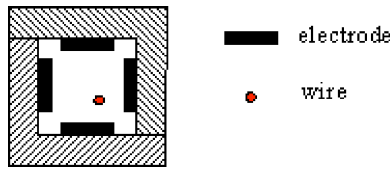}
   \caption{A cross-sectional schematic of a WPS sensor and alignment wire.}
   \label{fig:WPSprin}
   \end{center}
\end{figure}

\begin{figure}[htpb]
\centering
\includegraphics[width=0.48\columnwidth]{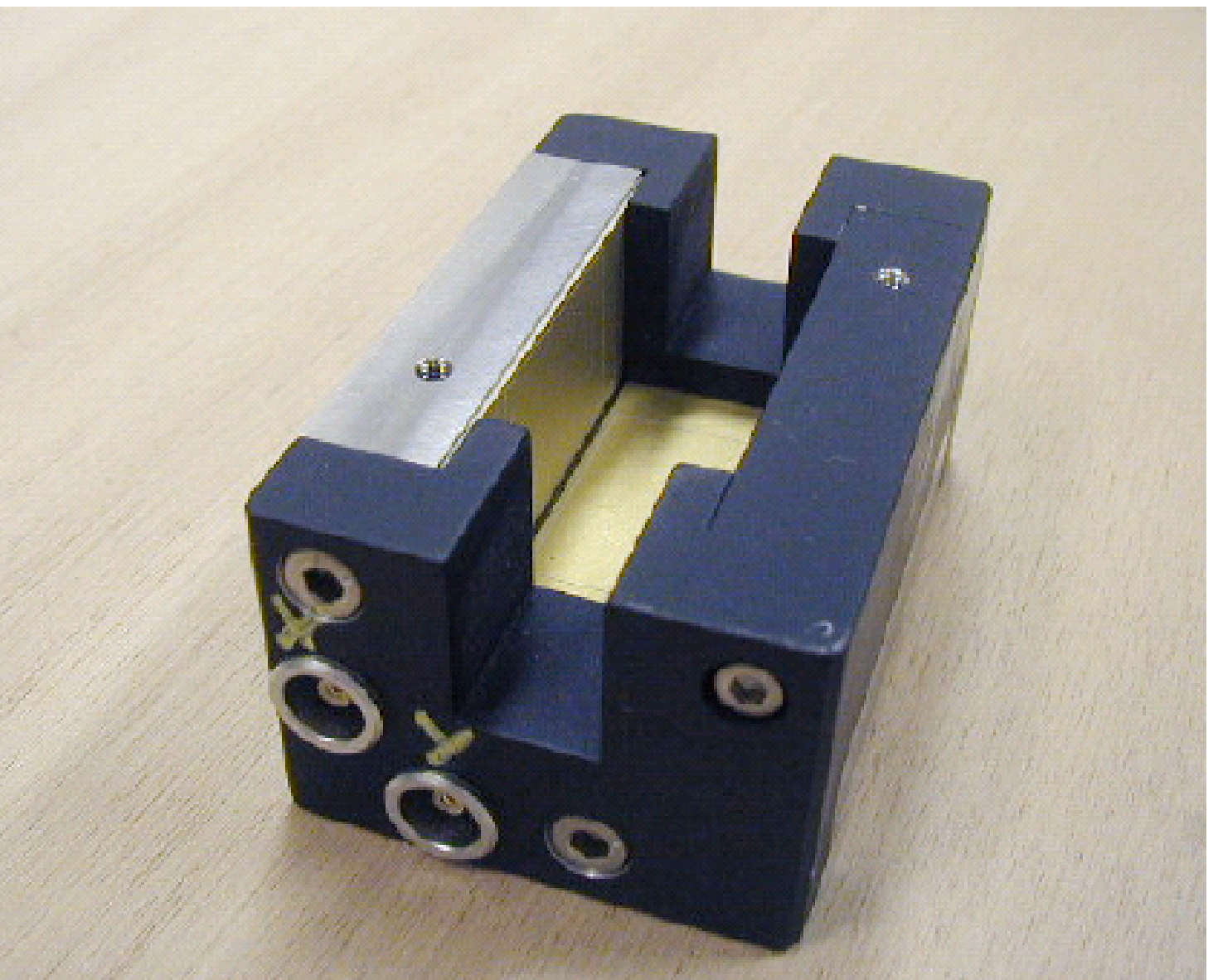}
\includegraphics[width=0.48\columnwidth]{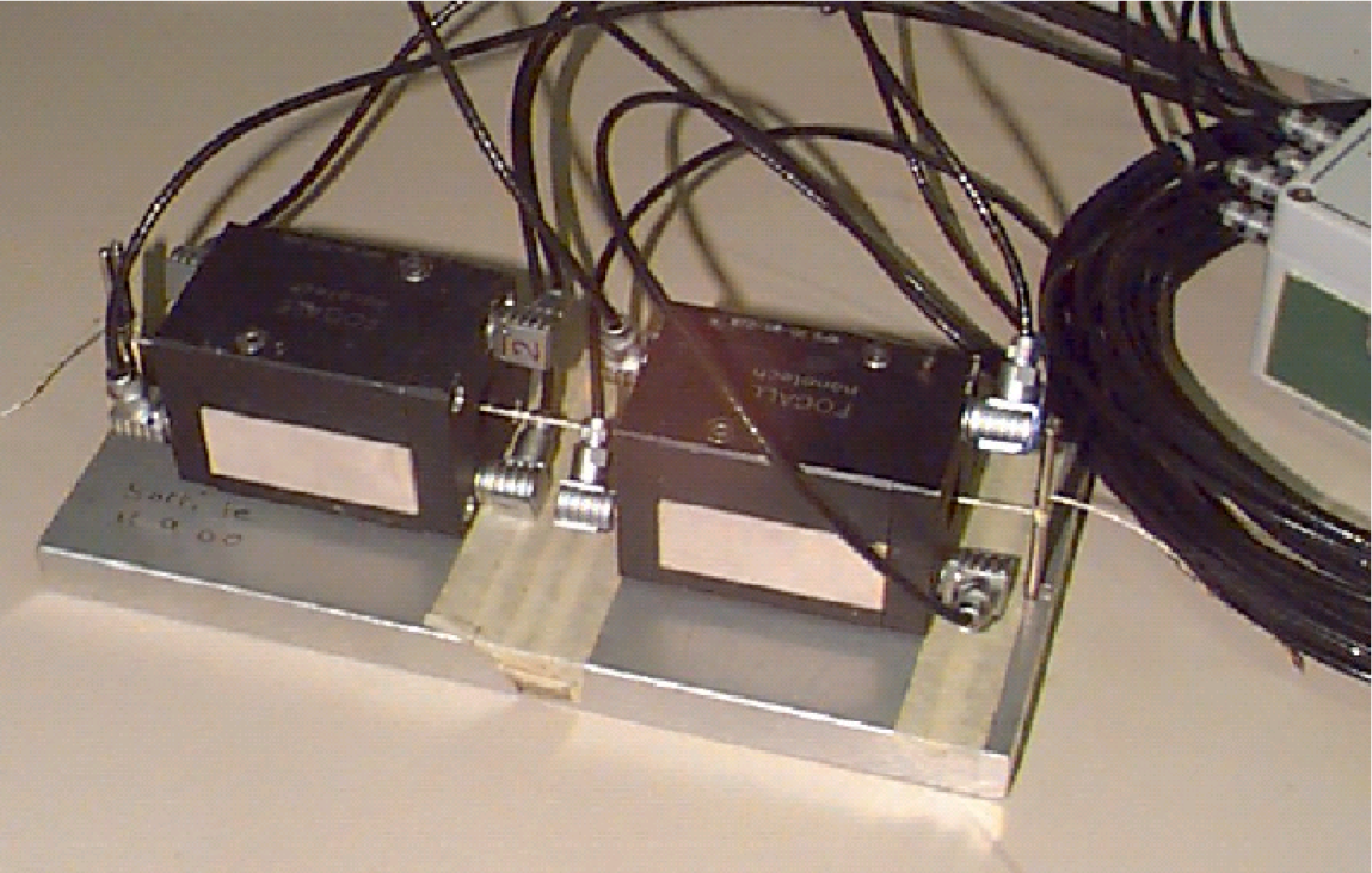}
\caption{A WPS sensor with lid removed (left), showing the
  electrodes. The aperture is 1cm square. Also shown are two WPS
  sensors on the test bench (right).}
\label{fig:WPSsensors}
\end{figure}

Wire Positioning Sensor systems use a capacitive measurement
technique to measure the sensors' positions, along two perpendicular
axes, relative to a carbon-fibre alignment wire. Such systems have
been shown to have sub-micron resolution capability in accelerator
alignment applications~\cite{WPS-LEP} and will be used in LHC
alignment. The principle of operation is shown in
Fig.~\ref{fig:WPSprin}. Photographs of a sensor (with cover removed)
and of two end-to-end sensors are shown in
Fig.~\ref{fig:WPSsensors}.

\subsubsection{The moving detectors}

FP420's silicon detectors will be mounted on moving beampipes. The
detectors however must be referenced to the fixed WPS sensors. One way
to achieve this is to use an LVDT or similar mechanical
displacement-measurement device. However, ``off-the-shelf'' examples
with long enough stroke length to accommodate the motion of the moving
beampipe tend not to have sufficient accuracy, and they (particularly their readout electronics) are not generally
guaranteed to be radiation-hard at the level needed by FP420. However
Schaevitz${\ooalign{\hfil\raise .00ex\hbox{\scriptsize R}\hfil\crcr\mathhexbox20D}}$
have made special LVDTs for aligning the LHC
collimators~\cite{Schaevitz}. These devices are by design sufficiently
radiation-hard for our purposes, and although they are longer
and less accurate than required for FP420, initial discussions with
the company have resulted in the expectation that a similar device to meet
FP420's needs can be manufactured; currently prototypes are being
designed. In the event that this fails, there are potential fallback
solutions, including a combination of a long stroke-length LVDT for
the garage position with a shorter, more accurate device for the
operating system; or an optical positioning (e.g. laser-based) system.

\subsection{Beam and proton transfer calculations}

\begin{figure}[htbp]
   \begin{center}
   \includegraphics[width=5in]{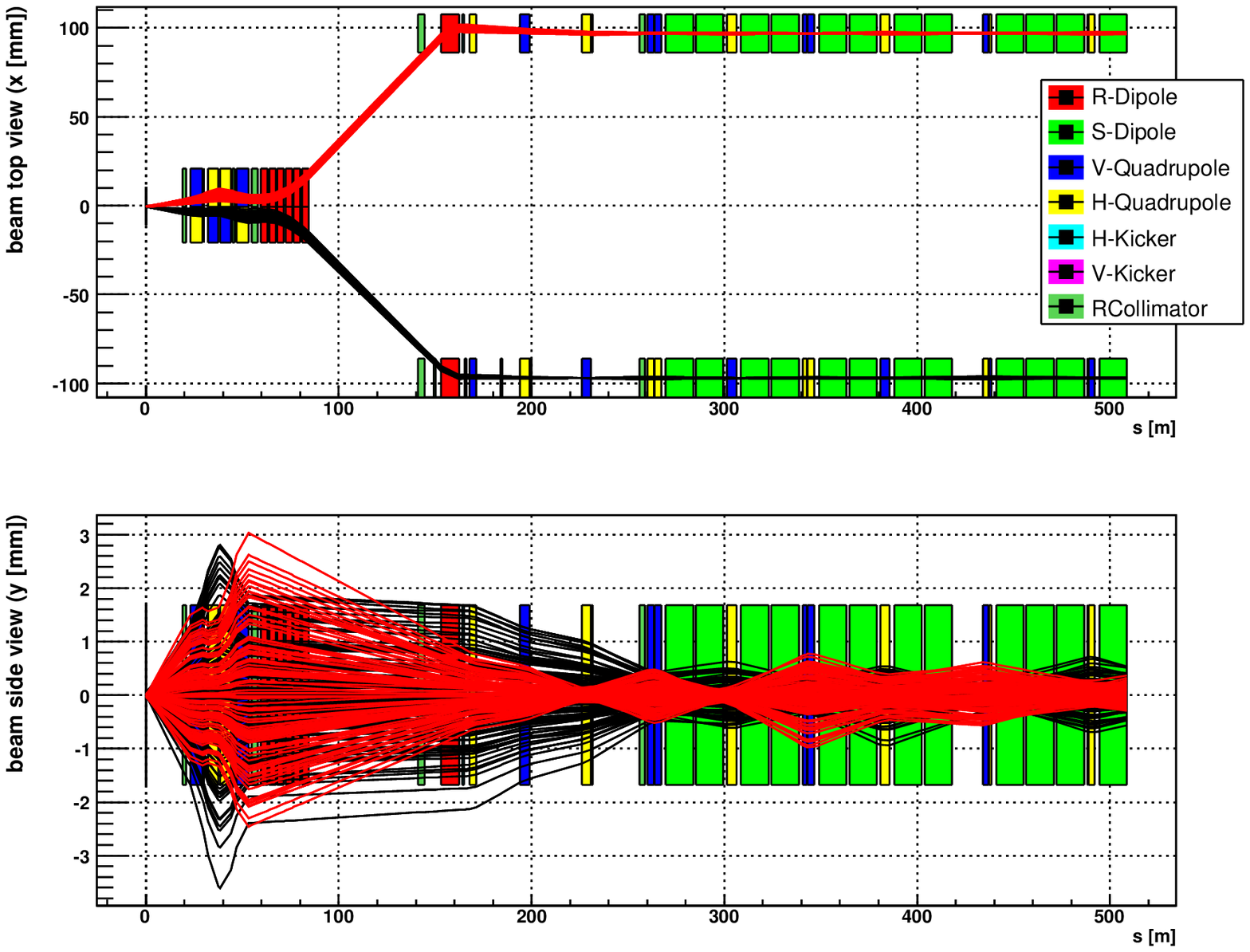}
   \includegraphics[width=5in]{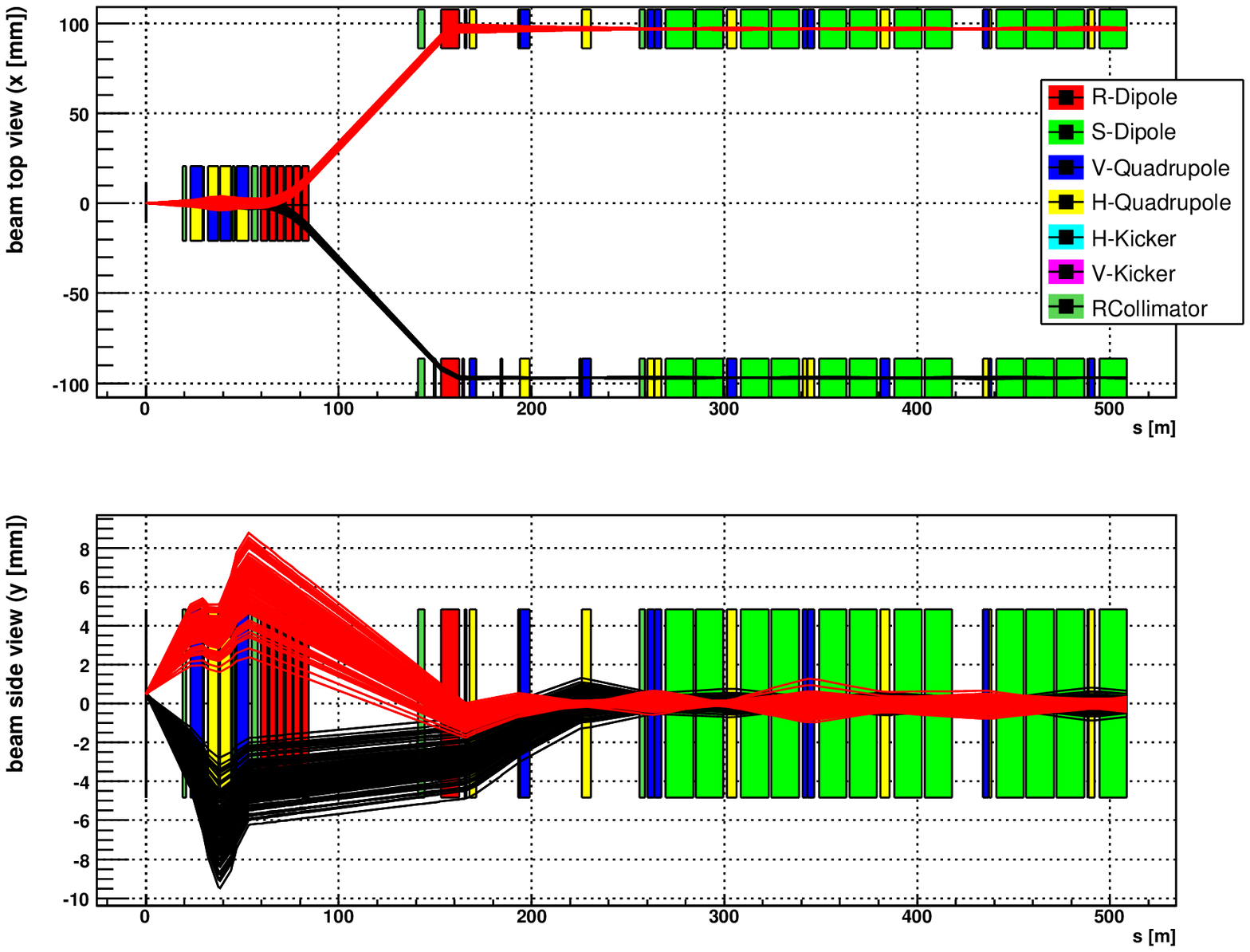}
   \caption{Beam particle paths calculated by \textsc{HECTOR}~\cite{hector}, 
   around CMS (top two figures) and ATLAS (bottom two figures). The beam line has
   been artificially straightened through the dipole region $z >$~250~m.}
   \label{fig:beampaths}
   \end{center}
\end{figure}

There are several simulations of proton transport through the LHC machine elements. We have developed a fast simulator,
\textsc{HECTOR}~\cite{hector}. Each generated proton is represented by a phase space vector $(x,x',y,y',E)$ at its point of origin.
This vector is rotated in phase space by a product $M_{beamline}$ of matrices, each corresponding to a machine element (drifts, quadrupoles,
dipoles, etc.). Aperture limitations are included. \textsc{HECTOR} has been validated by comparison with \textsc{MAD-X}~\cite{madx} and found to be very
accurate, providing the lateral positions of particles with inelastic protons with a precision at the few micron level.
Figure~\ref{fig:beampaths} shows top and side views, in CMS and ATLAS (which are quite different due to the orthogonal crossing planes) of the beam
protons. The bending of the dipoles has been switched off for display purposes, straightening the beam line after 250 m.

\begin{figure}[htbp]
   \begin{center}
   \includegraphics[width=5in]{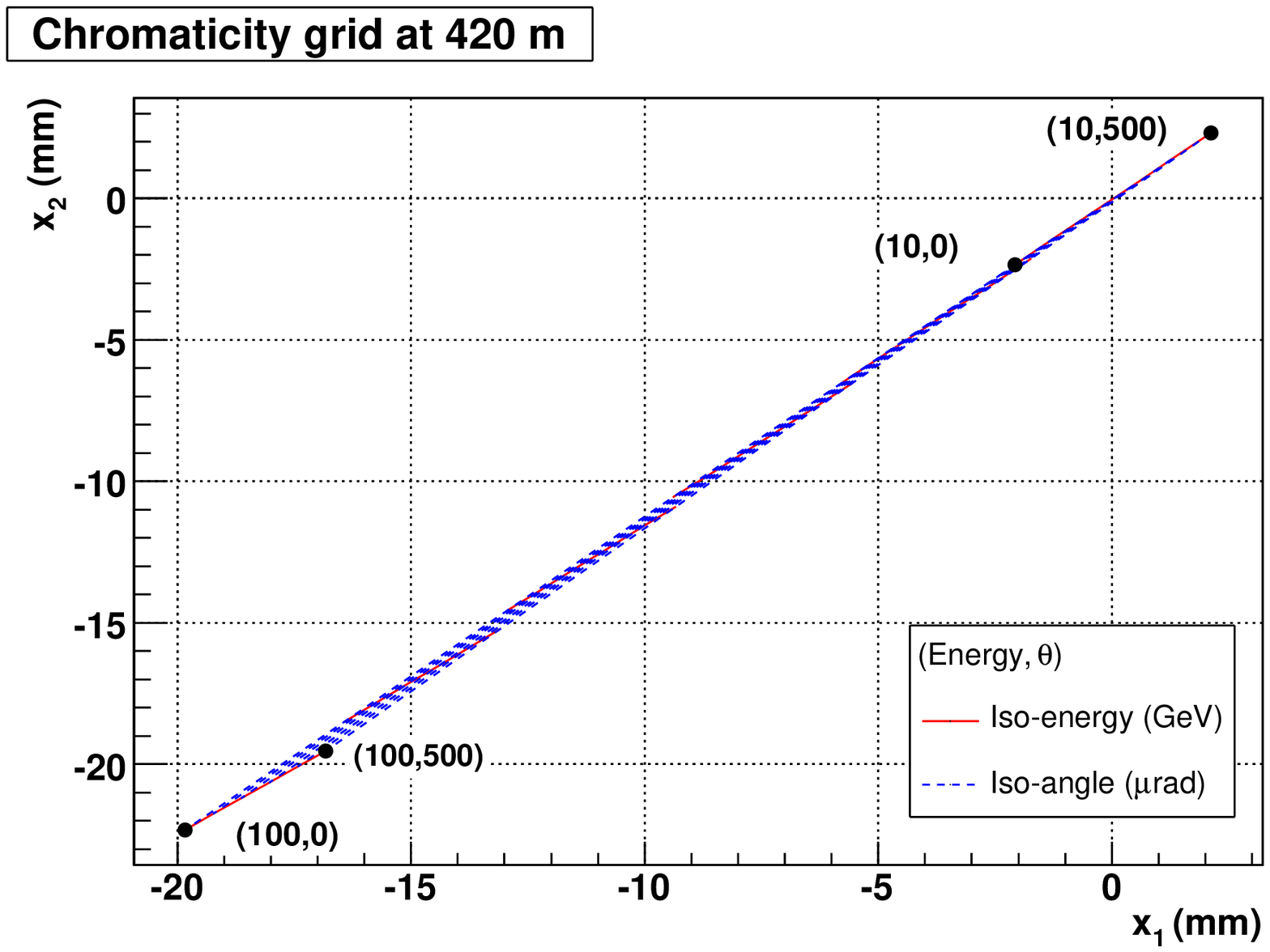}
   \includegraphics[width=5in]{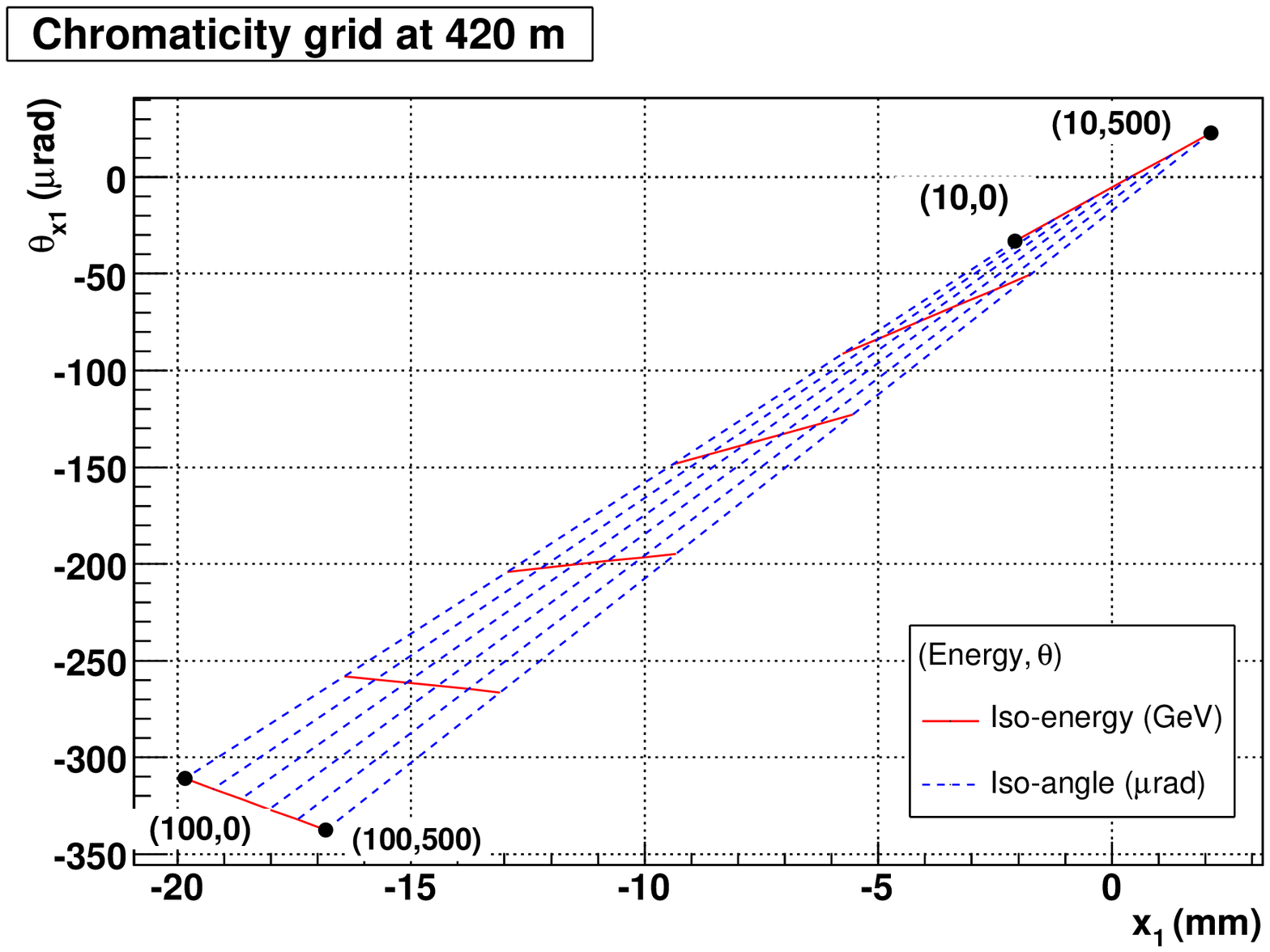}
   \caption{Chromaticity grids: iso-energy loss and iso-production angle lines, 
   from 10~GeV to 100~GeV and 0 to 500~$\mu$rad~\cite{hector}.}
   \label{fig:chrom}
   \end{center}
\end{figure}

Figure~\ref{fig:chrom} (top) shows the close correlation between
scattered protons at $x_1$ ($z=420$~m) and $x_2$ ($z=428$~m).
Numbers in parentheses are the energy loss (in~GeV) and the
production angle (in $\mu$rad). The bottom plot of $\theta_{x1}$ vs
$x_1$ opens up the angular dependence and demonstrates that for good
resolution it is not enough to measure the displacement of the
proton from the beam; the angle is also crucial.

\subsection{Machine alignment}

\begin{figure}[htbp]
   \begin{center}
   \includegraphics[width=5in]{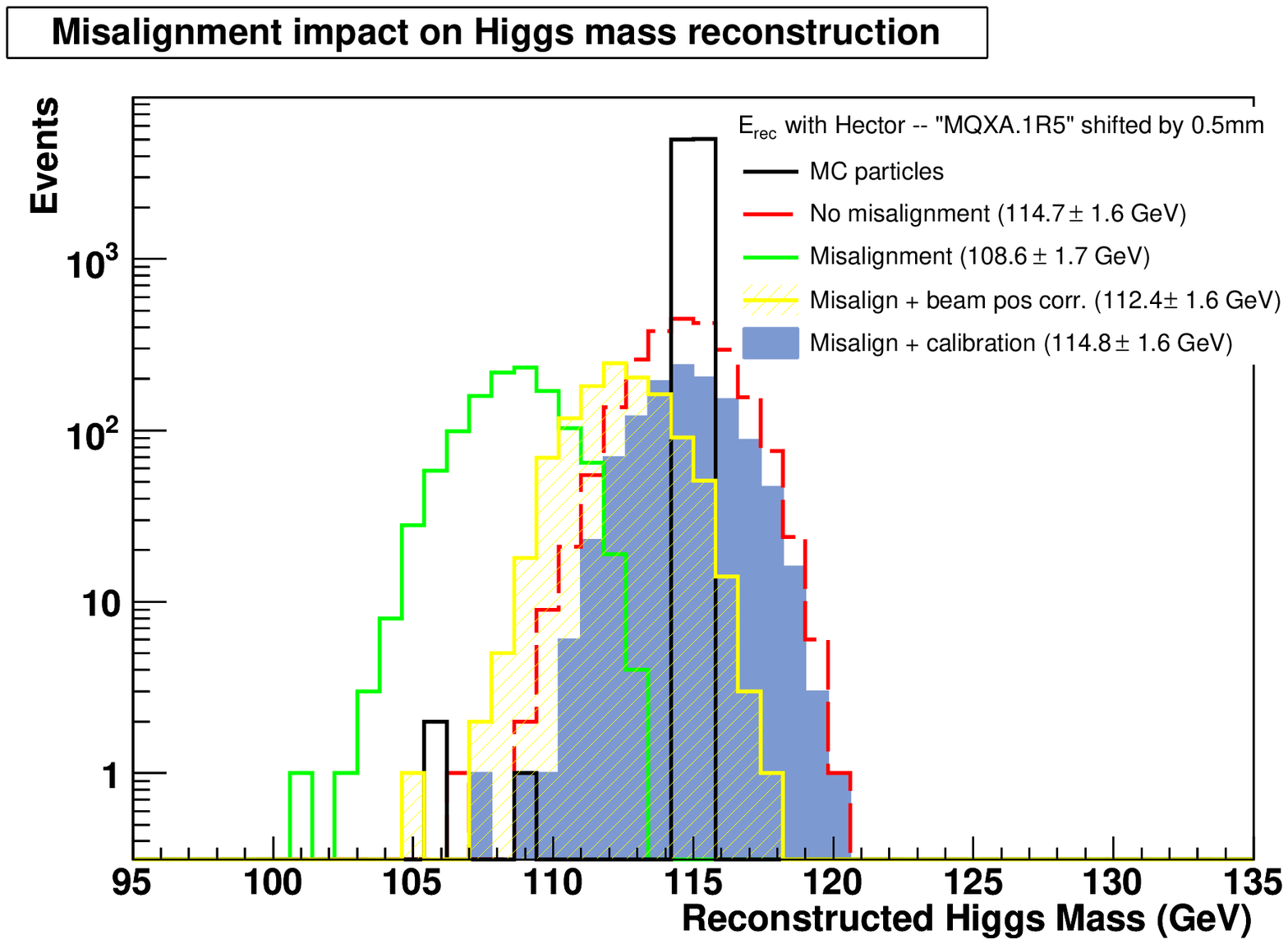}
   \caption{Illustration of the effect on the missing mass reconstruction due to a misalignment
of LHC quadrupoles. In this example a quadrupole (MQXA1R5) has been misaligned by
500~$\mu$m. Data from BPMs cannot fully recover the mass shift, while the exclusive
dimuon calibration recovers it fully, and the resolution is not affected as
long as the misalignment is stable~\cite{hector}.}
   \label{fig:misalign}
   \end{center}
\end{figure}

Primarily because of the quadrupoles, the spectrometer performance
is degraded by small misalignments of the LHC elements. We have
studied these with \textsc{HECTOR}. One example in
Figure~\ref{fig:misalign} shows reconstructed 115~\GeVcc\ Higgs boson
masses with no misalignment (central value 114.6~\GeVcc\, 
$\sigma(M)$~=~1.6~\GeVcc), and with 500 $\mu$m misalignment 
of the MQXA1R5 quadrupole at 29~m~\cite{hector}. The resolution 
is little changed but the central value shifts to 108.6~\GeVcc. 
A partial correction can be applied using BPM
information, and a full correction using exclusive dimuon
calibration, see below. This assumes stability on a week- or
month-time scale; it will be difficult to correct more frequent
shifts in alignment, especially of the quadrupoles.

\subsection{Mass scale and resolution measurement with physics processes}
\label{sec:align_physics}

The study of exclusive Higgs boson production in FP420 demands not
only good missing mass resolution, but also a means of calibrating
the mass scale and measuring the mass resolution $\sigma(M)$. The
width of a state can only be determined from an observed width by
unfolding the resolution $\sigma(M)$. While a perfect knowledge of
the machine, the central vertex and the FP420 tracking tells us
this, in principle, a verification using data is very important. The
production of \emph{exclusive dileptons}, $p+p\rightarrow p+e^+e^- +
p$ and $p+p\rightarrow p + \mu^+\mu^- +p$  is almost an ideal
calibration reaction: a measurement of the central dilepton gives
both forward proton momenta with very high precision. (One does not
need to detect both protons.) The exclusive $\mu^+\mu^-$ will be
easier to trigger on and will have less background. There are two
contributing processes. Two photon production $\gamma\gamma
\rightarrow \mu^+\mu^-$ is a purely QED reaction with a precisely
known cross section, such that it has been proposed as a means of
calibrating the LHC luminosity. The dimuon mass $M(\mu^+\mu^-)$ is a
continuum; there are no significant resonances in the mass region
considered here. While two-photon production of lepton pairs is well
known at $e^+e^-$ and $ep$ colliders, it has only recently been
observed (by CDF~\cite{cdfee,cdfmumu}) at a hadron collider. The
other important process is vector meson $V$ photoproduction: $\gamma
I\!\!P \rightarrow \Upsilon \rightarrow \mu^+\mu^-$ (muons from the
$J/\psi,\psi'$ family have too low $p_T$). The $\Upsilon$
photoproduction cross section ($\times$ branching ratio) is larger
in the mass region 9 -- 11~\GeVcc\ than the two-photon continuum, so a
trigger that includes this region is desirable, and achievable. In
the FP420 detectors, protons with energy loss as low as 20~GeV, $\xi
=$~20/7000 = 0.0029 are accepted. For a pair of exclusive muons each
with transverse momentum $p_T$ (in these processes the muons' $p_T$
are approximately equal) and equal pseudorapidity $\eta$, we have
$\xi_{1(2)} = \frac{2}{\sqrt{s}} p_T e^{+(-)\eta}$. So for $p_T = 4$
GeV/c and $\eta$~= 2.0 (2.5), $\xi_1 = 0.0042 (0.0070)$, inside the
acceptance. The other proton is at much lower $\xi$. The exclusive
events can be selected in the presence of pile-up, by requiring no
other tracks on the dimuon vertex, and $\Delta\phi_{\mu\mu} \approx
\pi$ with $p_T(\mu^+) \approx p_T(\mu^-)$, or $p_T(\mu^+\mu^-)\lesssim$~1~GeV/c, 
with a coincident consistent track in FP420. The dominant uncertainty of the forward proton momentum comes from the
incoming beam spread ($\frac{\delta p}{p} \approx 10^{-4} = 700$ MeV), as the central dimuons are measured with a
better resolution~\cite{fp420}. The two-photon cross section for central
``large''-$p_T$ muons is small; we expect about 300 events/fb$^{-1}$
with $p_T(\mu) \geq$~5~GeV/c and $|\eta(\mu)| < 2.5$. If the
threshold can be as low as 4~GeV/c, to include the $\Upsilon$, the
number of events is approximately doubled. With such good resolution
on the predicted proton momentum, combinatorial background can be
tolerated and a momentum scale calibration is achieved with very few
(tens of) events, i.e. on a daily basis. However a good measurement
of the resolution will require more events.
A potentially important use of the dimuon events is not only to measure the 
spectrometer performance, but to optimise it. For example, different tracking 
procedures can be tried and their resolution measured. While the two-photon dimuon 
events calibrate the missing mass scale in FP420, it cannot be used for frequent 
re-calibrations of detectors at 220~m as the process cross section is much smaller at 
such high $\xi$. The $\Upsilon$ photoproduction is important not only for improvement 
of calibration statistics but also for checking the resolution and bias of the muon 
$p_T$ reconstruction in the central detectors. In addition, the $\Upsilon$ events 
can be used to check the forward proton angular reconstruction.\\

Other reactions and forward instrumentation can provide information that can be used to 
calibrate the forward detectors, not as well as exclusive muon pairs but in almost real time. 
One can make use for example of the Zero Degree Calorimeter, ZDC, installed at 140 m from 
both the ATLAS~\cite{zdc_atlas} and CMS~\cite{zdc_cms} IPs.
The bremsstrahlung process $p+p \rightarrow p + p + \gamma$ with the photons emitted 
into a very forward cone has a cross section of about 10 nb for $E_\gamma >$~100~GeV. 
The photon is detected in the ZDCs at 140~m, and the proton at 420~m. This allows a 
cross-calibration; the proton spectrometer is only calibrated as well as the ZDC. The background level (e.g.
from forward $\pi^\circ$ production) remains to be seen. The angular distribution of bremsstrahlung photons is very forward peaked
(typically with $\theta(\gamma) \lesssim$~150~$\mu$rad) which helps with background reduction. The flow of neutral particles 
measured at the ZDC is huge and will allow the monitoring of the beam direction (tilt) at the IP with high precision. 
The LHC luminosity monitors (BRAN)~\cite{bran} will also be capable of fast online tilt measurement 
(also bunch-by-bunch) with a resolution better than 10~$\mu$rad. This information together with the precise 
control of the lateral position of the proton collisions at the IP, provides very 
good and independent on-line monitoring of the actual proton-beam trajectory.

A measurement of the relative positions of the beam and the track detectors can also come from the distributions of 
single diffractive protons. The less critical vertical ($y$)
distribution peaks at $y = y_{beam}$. (This allows a bunch-by-bunch monitor with time of $y_{beam}$.) Suppose the horizontal position to have
a poorly known offset $\delta x$. Most of the tracks of protons from the intersection region will be from single diffraction, $p+p
\rightarrow p + X$, which has an exponential $t$-distribution (at least in the low $|t|$-range), $\frac{d\sigma}{dt} \approx e^{bt}$. The intercept
of the distribution at $t$=0 has a maximum when the beam-detector distance is correct, so by applying offsets offline one can
find the actual distance. One can also vary the offset to find the maximum slope $\frac{d\sigma}{dt}$. This has been successfully applied
in CDF~\cite{cdfslope,gallinaro}; the accuracy on the offset was (at the Tevatron) approximately $\pm 30 \mu$m in $x$ and $y$.

Note that protons with $t=0$ (more strictly $\theta = 0$) are inside our acceptance; this is where the diffractive 
cross section has a maximum. An improvement on
the method could come from a measurement of the diffractive mass from the central detector, which would allow this technique to be used
selecting bands of $\xi$; however that can not be done in the presence of pile-up\footnote{One or two bunch crossings with deliberately low
luminosity, such that $<n> \approx$~1, could be useful for this and many other reasons.}.

\subsection{Alignment summary}

    While alignment and calibration issues are crucially important for FP420 tracking, we have viable solutions to all the issues: within the
    FP420 arm, over its motion towards the beam, with respect to the passing beam, and through the 420 m spectrometer. We have on-line checks of
    the proton energy using bremsstrahlung and displacement from the beam using diffractively scattered protons. Finally, and very important, we
    will use the $p+p\rightarrow p+\mu^+\mu^- + p$ reaction to calibrate, offline, both the mass scale and its resolution, and to optimise the
    latter. We hope to continue to push the mass resolution towards the limit given by the incoming beam momentum spread. It is important to
    miminise instabilities (bunch-to-bunch, store-to-store, week-to-week etc.) of LHC elements (especially quadrupole alignments), and to monitor
    any residual instabilities to allow for off-line corrections to be applied.

\newpage

\section{Near detector infrastructure and detector services}
\label{sec:services}

\begin{figure}[htpb]
\centerline{
\epsfig{file=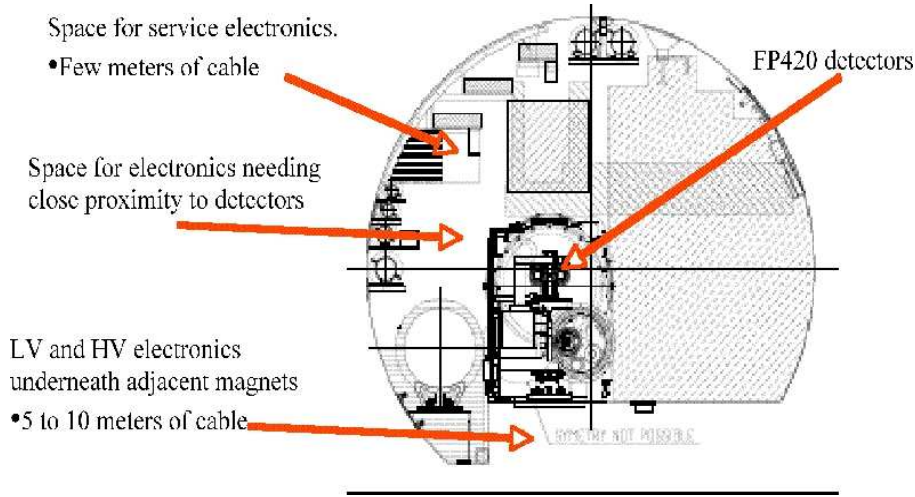,width=12cm}}
\caption{Plan view of the tunnel area around FP420 and the available space for 
equipment and services, as detailed in table~\ref{tab:services}.}
\label{fig:tunnel}
\end{figure}

The tunnel region at the location of the FP420 detectors presents a number of constraints 
for the installation of instrumentation such as FEE, HV and LV power supplies, cooling and 
detector gas supplies. A plan view of the tunnel area around FP420, showing the available 
space for services and electronics, is shown in figure~\ref{fig:tunnel}. 

Simulations have shown~\cite{silref10} that the zone will be exposed to a reasonably high 
radiation dose. A careful assessment of the possible locations 
and type of instrumentation is therefore required. Provisionally space has already been reserved 
underneath adjacent magnets upstream and downstream from the NCC where the LV and HV 
supplies can be placed [$yy$ = L0762023PL, L0762024PL, L0762045PL, L0762046PL, L0722023PL, 
L0722024PL, L0722045PL, L0722046PL]. Most of the machine electronics is already placed in 
this volume and such reducing the available space. The dose in this position under the 
arc magnets is expected to be between 10 and 20 Gy/year. FEE like trigger electronics, alignment 
and detector positioning control can be installed in cavities of the support beams of the NCC. 
In order to  limit the radiation load on the environment  at most to the level estimated with the 
present configuration, an envelop of adequate shielding (Pb plates) will be placed around the 
beam pipes and detectors as described in section~\ref{sec:cryostat}. It will also be feasible to 
place some equipment along the LHC tunnel wall underneath the cable trays which run above 
the QRL line but the radiation level will be somewhat higher than under the dipole magnets.\\

\begin{figure}[htpb]
\centerline{
\epsfig{file=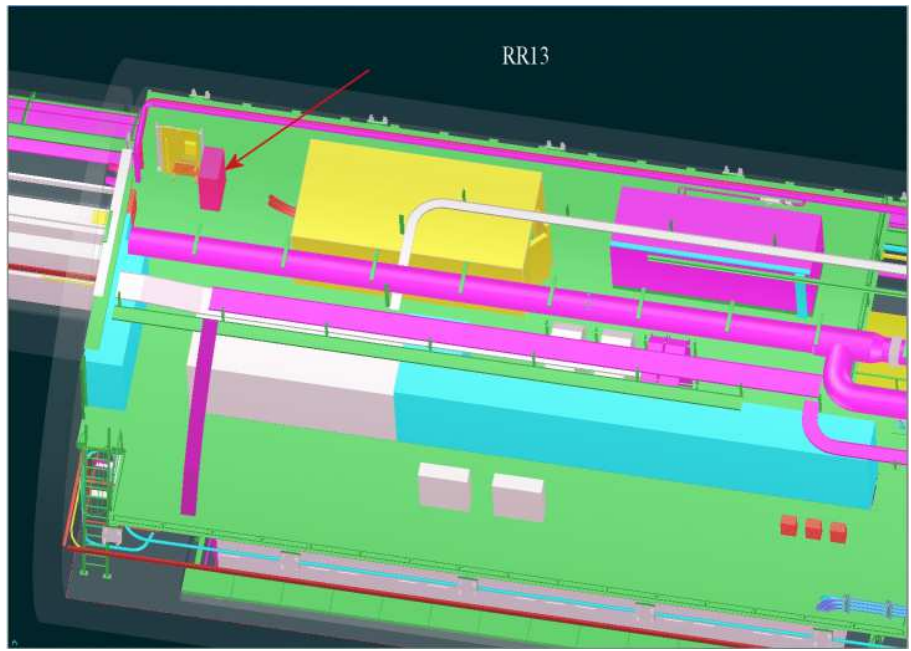,width=12cm}}
\caption{Reserved rack space in the LHC alcove areas (RRs).}
\label{fig:RRs}
\end{figure}

Currently, active radiation monitoring instrumentation is being installed in the 420m region of the 
LHC tunnel. These monitors will be operational at LHC startup and thus provide valuable data to 
assess the real radiation levels in the area. At present no general services are provided in the LHC 
tunnel at the FP420 region. Power- and controls- cables as well as tubes for fibre-optic (FO) will therefore need 
to be pulled from the corresponding experimental area to this location. It is therefore proposed to 
install additional cable trays next to the overhead rail of the monorail. This strategy has already been 
used for the routing of the services for LHCf to both sides of point 1. FP420 can reuse these trays 
which will have to be extended from the present 150m to 420m. The FP420 power supplies and 
detector controls instrumentation could be connected to the LHC machine power. This would assure 
the electricity supply as long as the LHC machine power is available and reduce the cabling impact in 
the tunnel since additional power cables would only need to be routed from the nearest RR alcoves at 
a distance of about 200m. 
Space has also been reserved in the RR13, RR17, RR53, RR57 alcoves on the 2nd floor level above the 
LHC power converters (Figure~\ref{fig:RRs}). At this location the expected annual fluence of hadrons 
($E >$~20~MeV) is in the order of 10$^8$, corresponding to a dose rate in the order of 0.3 Gy per 
annum, and considered suitable for installation of detector power supplies and electronics for the 
alignment monitoring system.

The assessment of types
and quantities of services needed for the FP420 detectors is still rather difficult 
at this stage of the project. Those requirements which correspond to the present 
estimation for each sub-system are summarised in Table~\ref{tab:services}.

\begin{table}[htbp]
\centering
\begin{tabular}{|c|c|c|c|}
\hline
Sub-system & Requirements & Location & Comment \\
\hline
GasTOF & gas & Detector station & Secondary vacuum \\
       & FEE, DAQ& Cryostat support beams & \\
       & elec. power & see Fig.~\ref{fig:hvlv4} & \\
\hline
Quartic & FEE, DAQ & Cryostat support beams & \\
        & elec. power & see Fig.~\ref{fig:hvlv4} &\\
\hline
3D (Silicon) & Cooling & Detector station & \\ 
                 & FEE, DAQ & Cryostat support beams & \\
                 & elec. power & see Fig.~\ref{fig:hvlv4} & \\
\hline
General Cooling & power, control & & \\
\hline
Alignment & BPM, BLM,  & Detector station, & \\
          &     WPS, other    & Detector support table RRs & \\
\hline
Positioning & Movement drive, & Detector support table & \\
            & control & see Fig.~\ref{fig:hvlv4} & \\
\hline
Timing & TTC, BST & Counting rooms & \\
\hline
Interlocks & Injection, & Machine IF rack in & \\
& Dump & exp. counting room& \\
\hline
Electrical power & 400~V AC & RE alcoves & UPS for controlled \\
& 230~V AC & & shutdown ? \\
& 48~V DC & & \\
\hline
HV & 16 ch. HV 4kV & see Fig.~\ref{fig:hvlv4} & Control from IP (CAN bus ?) \\
& (timing) & & \\
& 36 ch. HV & & \\
& (tracking) & & \\
\hline
LV & 6 ch. LV & see Fig.~\ref{fig:hvlv4} & Control from IP (CAN bus ?) \\
& (timing) & & \\
& 36 ch. LV & & \\
& (tracking) & & \\
\hline
Communication, & FO, Field bus & ECR $\leftarrow \rightarrow$ FP420 & Space available for \\
RH-diodes: & RH-diodes: & & $\ge 2 \times 24$ fibres to each \\
& MITSUBISHI & & station \\
& FU-427SLD-F1 & & Use BLM / BPM FEE ? \\
\hline
Miscellaneous & Cameras, lights & FP420 & \\
\hline
Instrumentation & VME crate & Cell 12L/R at IP1 \& IP5 & \\
space & equivalent & each (5); & \\
& & 13.8m tunnel wall; & \\
& & Call 11 L/R (5) & \\
 & each (5); & \\
\hline
\end{tabular}
\caption{Summary of detector services required for FP420. Abbreviations: 
BPM - Beam Position Monitors, BLM - Beam Loss Monitors, 
FEE - Front End Electronics, FO - Fibre optics,  
WPS - Wire Positioning System.}
\label{tab:services}
\end{table}
\newpage

\section{Conclusions}
\label{sec:conclusions}

The FP420 project proposes to install silicon tracker and fast timing detectors
in the LHC tunnel at 420~m from the interaction points of the ATLAS and CMS experiments for the detection of very 
forward protons as a means to study Standard Model (SM) and New Physics signals.
The FP420 detector system is a magnetic spectrometer consisting of a moveable silicon 
tracking system which measures the spatial position of protons scattered by a few hundreds $\mu$rads 
relative to the LHC beam line and their arrival times at several points in a 12 m region around 420 m. 
The measurement of the displacement and angle of the outgoing protons relative to the beam 
allows the reconstruction of their momentum loss and transverse momentum. 
The combined detection of both outgoing protons 
and the associated centrally produced system using the current ATLAS and/or CMS detectors
gives access to a rich programme of studies in QCD, electroweak, Higgs and Beyond the 
Standard Model physics. The addition of such detectors will add the capability to make measurements 
which are currently unique at the LHC, and difficult even at a future linear collider.

A prime process of interest for FP420 is Central Exclusive Production (CEP), $pp \rightarrow p + \phi + p$, 
in which the outgoing protons remain intact and the central system $\phi$ may be a single particle 
such as a Higgs boson. Observation of new particle production in the CEP channel benefits from 
(i) enhanced signal over backgrounds (giving access to the difficult Higgs fermionic decay channels
for example), and allows one to directly measure  (ii) its quantum numbers (the central system has an 
approximate $J^{PC} = 0^{++}$  selection rule) as well as (iii) its mass with very good resolution,
$\mathcal{O}$(2~GeV/c$^2$) irrespective of the decay channel of the particle. In some 
beyond-SM scenarios, the FP420 detectors may be the primary means of discovering new 
particles at the LHC.
Section~\ref{sec:physics} of this document has presented an overview of the physics case for FP420 
including the current theoretical status of CEP predictions. The state-of-the-art calculations of the 
production cross section for a 120~GeV/c$^{2}$ Standard Model Higgs boson via the CEP process 
at the LHC yields a central value of 3~fb, with a factor of 4 uncertainty. Supersymmetric extensions
of the SM, yield Higgs boson cross sections 10 or 100 times larger and would allow the 5$\sigma$ discovery 
of all CP-even scalar bosons in practically the whole $M_A - \rm{tan}\beta$ plane with $\mathcal{O}$(100 fb$^{-1}$). 
Section~\ref{sec:pilko} has presented a detailed study of the trigger strategy, expected acceptance, 
reconstruction efficiencies, signal over backgrounds and final mass resolutions and yields for a particular 
$p\,p \to p\, H\, p$ measurement with Higgs boson decay in the $b\bar{b}$  mode. The Higgs boson 
line shape in this channel can be reconstructed with a $3\sigma$ or better significance with an integrated 
luminosity of 60 fb$^{-1}$.\\

A summary of various interesting photon-photon and photon-proton processes accessible to FP420 is presented
(Section~\ref{sec:photon_phys}). Photon-induced reactions tagged 
with forward protons can provide a very clean environment for the study of various signals such as 
anomalous top or associated $WH$ production in $\gamma\,p$ interactions; as well as anomalous 
gauge boson couplings, exclusive dileptons or supersymmetric pair production in $\gamma\,\gamma$ interactions.
Hard diffraction studies (single-diffractive and double-Pomeron production of $B$-mesons, $W$, $Z$ bosons or di-jets),
sensitive to generalised parton distributions, are discussed in Section~\ref{sec:diffraction}.\\

The beam optics at LHC (Section~\ref{sec:optics}) allows protons that have lost momentum in a
diffractive interaction to emerge from the beam envelope at regions 220~m and 420~m from the interaction point.  
The acceptance of silicon detector arrays in these locations placed at distances 3 -- 9 mm from the beam centre
allows for the detection of both outgoing protons from centrally produced 
objects with a wide range of masses above 60~\GeVcc. However, to obtain good acceptance 
for masses above 150~\GeVcc, the 220~m system is essential. The expected position and angle resolutions
for the protons obtained in the silicon stations yield a mass resolution reaching values of 2 to 3~\GeVcc.\\

The expected machine-induced backgrounds at 420 m such as beam-halo and beam-gas backgrounds
are discussed in Section~\ref{sec:backgrounds}. Contributions at 420 m from {\it near} 
beam-gas and the betatron cleaning collimation are found to be small. For  transverse separations between
the detectors and the beam centre above 5 mm, the integrated number of protons, photons and neutrons 
from beam halo is expected to be less than 1, 0.16 and 0.003 per bunch crossing respectively. The impact of these
estimated background rates needs to be assessed in term of detector performance and survivability.\\

Section~\ref{sec:cryostat} describes the new 420 m connection cryostat which will allow moving 
near-beam detectors with no effects on LHC operations. A preliminary design for a replacement connection
cryostat that would allow detectors to be placed in the 420 m region has been completed, and a 
final design is in progress. Such a solution is expected to actually lower the dynamic heat load of 
the LHC and have similar radiation profiles. With the appropriate approvals and funding, two such 
cryostats could be built and installed in late 2009 (installation time is around 90~days), 
and in principle, two more in 2010 with negligible impact on LHC operations.\\

The design of the beam pipe in the FP420 region and the movement mechanism are
discussed in Section~\ref{sec:hhpipe}. The Hamburg moving-pipe concept provides the optimal 
solution for the FP420 detector system as it ensures a simple and robust
design and good access to the detectors. Moreover, it is compatible with the very limited space 
available in the modified connection cryostat and with the expected position of the scattered protons
between the two LHC beampipes, and it permits the incorporation of rather large detectors, 
such as the timing devices, using pockets, i.e. rectangular indentations in the moving pipes. 
The prototype detector pockets show the desired flatness of the thin windows, and the first motorised 
moving section, with prototype detectors inserted, has been tested at the CERN test beam. 
A full prototype test, including assembling, positioning and alignment aspects, is planned in test beam 
in Fall 2008.\\

The studies of the radio-frequency impact of the design on the LHC are described in 
Section~\ref{sec:RF}. Numerical simulations, analytical calculations and laboratory measurements 
have showed consistently that the proposed FP420 design will have a small impact on the total LHC 
impedance budget, even for transverse distances of the stations from the beam centre as small as 3~mm.
Tapering of the beam pipe indentations is recommended because it reduces the impedance significantly, 
as measured both with the single pocket and double pocket designs. The beam harmonics at 2\,GHz 
are expected to be below $10^{-2}$ of the main harmonic at 40\,MHz and well below $10^{-3}$
at 2.5\,GHz, and the horizontal tune shift induced by a FP420 station is expected to be almost 
imperceptible when compared to the tune stability region defined by the available LHC octupoles 
magnets.\\

In Section~\ref{sec:silicon} we present a detailed description of the design of the FP420 3D silicon sensors
including mechanical support system, superlayer and blade design and thermal tests, assembly and alignment, 
high- and low- voltages, tracker readout, downstream data acquisition and 
infrastructure at the host experiment. The performance of the tracker has been evaluated using a simple
Monte Carlo program as well as a full GEANT4 simulation. Estimates of the multiple scattering for 
the three (two) station layouts indicate that the expected angular resolution is 0.85~${\mu}$rad 
(0.91~${\mu}$rad), well within design specifications. The efficiency of two track reconstruction 
has been found to be 86\% and 80\% respectively for the two and three station layouts.\\

Since the cross sections for CEP of the SM Higgs boson and other new physics scenarios are 
relatively small (few fb), FP420 must therefore be designed to operate at the highest LHC instantaneous 
luminosities of $10^{34}$cm$^{-2}$s$^{-1}$.
A measurement of the relative time of arrival of the protons at FP420 in the 10 picosecond range is required 
for matching of the detected protons with a central vertex within $\sim$2 mm, which will enable the 
rejection of a large fraction of the pile-up overlap background. Section~\ref{sec:timing} describes 
two complementary fast timing detector designs: GASTOF (GAS Time Of Flight) and QUARTIC 
(QUARtz TIming \v{C}erenkov). 
 The {\it prototype} detector design is approaching a resolution of 20 ps. An upgrade to determine the 
time of more than one proton per bunch is conceivable, either by reading out individual pixels in the
GASTOF MCP-PMT to resolve separate, but overlapping, \v{C}erenkov discs, or by reducing the 
pixel size in the $x$-direction for the QUARTIC detectors. We are also developing a promising new type of 
focusing quartz \v{C}erenkov detector. As the reference timing is also an important 
component of the timing resolution, we are also exploring interferometrically stabilised fibre optic links, 
where the standard is in the 10 femtosecond range.\\

In Section~\ref{sec:alignment} we describe the alignment and calibration strategy, 
using both physics and beam position monitor techniques. Alignment and calibration 
is guaranteed for all experimental conditions: within the FP420 arm, over its motion 
towards the beam, with respect to the passing beam, and through the 420 m spectrometer. 
We have on-line checks of  the proton energy using bremsstrahlung and of displacement from the 
beam using diffractively scattered protons. We will use the $p+p\rightarrow p+\mu^+\mu^- + p$ 
reaction to calibrate, offline, both the mass scale and its resolution, and to optimise the latter. 
It is important to miminise instabilities (bunch-to-bunch, store-to-store, week-to-week etc.) 
of LHC elements (especially quadrupole alignments), and to monitor any residual instabilities 
to allow for off-line corrections to be applied. Chapter~\ref{sec:services} outlines the near 
detector infrastructure and detector services required for the FP420 project.\\

The studies presented in this document have shown that it is possible to install detectors in the 420 m 
region with no impact on the operation or luminosity of the LHC. These detectors can be aligned and 
calibrated to the accuracy required to measure the mass of the centrally produced system to between 
$2$ and $3$~GeV/c$^{2}$. This would allow an observation of new particles in the $60 - 180$~GeV/c$^{2}$ 
mass range in certain physics scenarios during 3 years of LHC running at instantaneous luminosities of 
$2 \times 10^{33}$ cm$^{-2}$ s$^{-1}$, and in many more scenarios at instantaneous luminosities 
of up to $10^{34}$ cm$^{-2}$ s$^{-1}$. Events can be triggered using the central detectors alone 
at Level 1, using information from the 420 m detectors at higher trigger levels to reduce the event rate. 
Observation of new particle production in the CEP channel would allow a direct measurement of the 
quantum numbers of the particle and an accurate determination of the mass, irrespective of the decay 
channel of the particle. In some scenarios, these detectors may be the primary means of discovering 
new particles at the LHC, with unique ability to measure their quantum numbers. The FP420 opens, 
moreover, the possibility to develop an extensive, high-rate $\gamma \gamma$ and $\gamma p$ 
physics program. The addition of the FP420 detectors will thus, for a relatively small cost, 
significantly enhance the discovery and physics potential of the ATLAS and CMS experiments.

\section{Costing}

A preliminary estimate of the costing of the major components of FP420 detectors is given 
here as an indication. A detailed costing evaluation is  still being performed.

\begin{itemize}
\item Two new cryostats per experiment, amounting to a total of 1.5~MCHF/experiment
\item The silicon tracker including the electronics and mechanical parts:
0.7 to 1.0~MCHF/ experiment, depending on the purchasing of equipment
\item Quartic timing detectors, including electronics, 100~kCHF/experiment for 4 detectors.
\item GASTOF timing detectors, including electronics, DAQ, slow controls and cables: 145~kCHF
\item BPMs and beampipe mechanics: 380~kCHF/experiment
\item High voltage/Low Voltage: 160~kCHF/experiment
\end{itemize}

This leads to a approximate grand total of 3.5~MCHF/experiment for 
equipping both sides with FP420 detectors.
\newpage

\section*{Acknowledgments}
\label{sec:acknowledgments}

We thank many people in the accelerator (AB and AT) and technical (TS) Departments of CERN
for their valuable help and continuing support in the FP420 design study (beam vacuum, electrical specifications, 
RF effects ...). In particular, we want to thank Thierry Renaglia, Sebastien Marques, Thierry Colombet, 
Alain Poncet and Vittorio Parma for their expertise and generous help with the cryostat studies.
This work was supported in the UK by grants from the STFC and the Royal Society; 
in the USA by the Department of Energy (including Fermilab and Brookhaven National Lab funding, 
UT-Arlington base funding and the Advanced Detector Research program), and the Texas Advanced 
Research Program;
in Belgium by FNRS and its associated fund (IISN), by FWO-Vlaanderen, IIKW, and by the 
Inter-University Attraction Poles Programme subsidised by the Belgian Federal Science Policy;
in Italy by the Italian Istituto Nazionale di Fisica Nucleare (INFN) and by the 
Italian Ministry for Education, University and Scientific Research under the programme 
``Incentivazione alla mobilit\`a di studiosi stranieri e italiani residenti all'estero".



\end{document}